\PassOptionsToPackage{unicode}{hyperref}
\PassOptionsToPackage{hyphens}{url}
\PassOptionsToPackage{dvipsnames,svgnames,x11names}{xcolor}
\documentclass[
  10pt,
]{report}
\usepackage{xcolor}
\usepackage[margin=1in]{geometry}
\usepackage{amsmath,amssymb}
\usepackage{iftex}
\ifPDFTeX
  \usepackage[T1]{fontenc}
  \usepackage[utf8]{inputenc}
  \usepackage{textcomp} % provide euro and other symbols
\else % if luatex or xetex
  \usepackage{unicode-math} % this also loads fontspec
  \defaultfontfeatures{Scale=MatchLowercase}
  \defaultfontfeatures[\rmfamily]{Ligatures=TeX,Scale=1}
\fi
\usepackage{lmodern}
\ifPDFTeX\else
\fi
\IfFileExists{upquote.sty}{\usepackage{upquote}}{}
\IfFileExists{microtype.sty}{% use microtype if available
  \usepackage[]{microtype}
  \UseMicrotypeSet[protrusion]{basicmath} % disable protrusion for tt fonts
}{}
\makeatletter
\@ifundefined{KOMAClassName}{% if non-KOMA class
  \IfFileExists{parskip.sty}{%
    \usepackage{parskip}
  }{% else
    \setlength{\parindent}{0pt}
    \setlength{\parskip}{6pt plus 2pt minus 1pt}}
}{% if KOMA class
  \KOMAoptions{parskip=half}}
\makeatother
\usepackage{color}
\usepackage{fancyvrb}

\DefineVerbatimEnvironment{Highlighting}{Verbatim}{commandchars=\\\{\}}
\usepackage{framed}
\definecolor{shadecolor}{RGB}{245,247,250}
\newenvironment{Shaded}{\begin{snugshade}}{\end{snugshade}}

\newcommand{\BaseNTok}[1]{\textcolor[rgb]{0.54,0.31,0.07}{#1}}
\newcommand{\BuiltInTok}[1]{\textcolor[rgb]{0.60,0.20,0.14}{#1}}

\newcommand{\CommentTok}[1]{\textcolor[rgb]{0.36,0.42,0.48}{\textit{#1}}}

\newcommand{\ControlFlowTok}[1]{\textcolor[rgb]{0.12,0.31,0.47}{\textbf{#1}}}
\newcommand{\DataTypeTok}[1]{\textcolor[rgb]{0.60,0.20,0.14}{#1}}
\newcommand{\DecValTok}[1]{\textcolor[rgb]{0.54,0.31,0.07}{#1}}

\newcommand{\ErrorTok}[1]{\textcolor[rgb]{1.00,0.00,0.00}{\textbf{#1}}}

\newcommand{\FloatTok}[1]{\textcolor[rgb]{0.54,0.31,0.07}{#1}}
\newcommand{\FunctionTok}[1]{\textcolor[rgb]{0.02,0.16,0.49}{#1}}
\newcommand{\ImportTok}[1]{\textcolor[rgb]{0.12,0.31,0.47}{\textbf{#1}}}

\newcommand{\KeywordTok}[1]{\textcolor[rgb]{0.12,0.31,0.47}{\textbf{#1}}}
\newcommand{\NormalTok}[1]{\textcolor[rgb]{0.10,0.10,0.10}{#1}}
\newcommand{\OperatorTok}[1]{\textcolor[rgb]{0.33,0.33,0.33}{#1}}
\newcommand{\OtherTok}[1]{\textcolor[rgb]{0.43,0.25,0.70}{#1}}
\newcommand{\PreprocessorTok}[1]{\textcolor[rgb]{0.12,0.31,0.47}{\textbf{#1}}}

\newcommand{\SpecialCharTok}[1]{\textcolor[rgb]{0.33,0.33,0.33}{#1}}

\newcommand{\StringTok}[1]{\textcolor[rgb]{0.18,0.49,0.27}{#1}}
\newcommand{\VariableTok}[1]{\textcolor[rgb]{0.61,0.23,0.42}{#1}}

\usepackage{longtable,booktabs,array}
\usepackage{calc} % for calculating minipage widths
\usepackage{etoolbox}
\makeatletter
\patchcmd\longtable{\par}{\if@noskipsec\mbox{}\fi\par}{}{}
\makeatother
\IfFileExists{footnotehyper.sty}{\usepackage{footnotehyper}}{\usepackage{footnote}}
\makesavenoteenv{longtable}
\providecommand{\tightlist}{%
  \setlength{\itemsep}{0pt}\setlength{\parskip}{0pt}}
\usepackage{xcolor}
\definecolor{accent}{HTML}{1F4E79}     % deep blue
\definecolor{codebg}{HTML}{F5F7FA}     % very light blue-grey for code
\definecolor{rulegrey}{HTML}{C9D2DC}
\definecolor{zebra}{HTML}{EEF3F8}      % faint blue for alternate table rows
\usepackage{colortbl}                  % per-row \rowcolor zebra (injected by callout.lua)

\usepackage{microtype}

\newlength{\anebodyparskip}
\AtBeginDocument{\setlength{\parindent}{0pt}\setlength{\parskip}{\anebodyparskip}}
\AddToHook{para/begin}{\setlength{\parskip}{\anebodyparskip}}

\usepackage[most]{tcolorbox}
\newtcolorbox{summarybox}{%
  enhanced, breakable, sharp corners, boxrule=0pt, frame hidden,
  colback=accent!6, borderline west={3pt}{0pt}{accent},
  left=12pt, right=10pt, top=8pt, bottom=8pt,
  before skip=10pt, after skip=12pt,
  before upper={{\bfseries\scriptsize\textcolor{accent}{SUMMARY}}\par\vspace{2pt}}}
\newtcolorbox{calloutbox}[1]{%
  enhanced, breakable, sharp corners, boxrule=0.4pt, colframe=accent!60,
  colback=accent!6, coltitle=white, colbacktitle=accent,
  fonttitle=\bfseries\small, title={#1},
  left=10pt, right=10pt, top=6pt, bottom=6pt,
  before skip=10pt, after skip=12pt}

\usepackage{fvextra}
\fvset{breaklines=true,breakanywhere=true,fontsize=\footnotesize,%
  breakautoindent=true,breakindent=1.4em,%
  breaksymbolleft={\footnotesize\color{accent!55}\ensuremath{\hookrightarrow}\;},%
  breaksymbolsepleft=0pt}
\usepackage{framed}
\colorlet{shadecolor}{codebg}
\renewenvironment{Shaded}{%
  \MakeFramed{\hsize\dimexpr\linewidth-2pt-16pt\relax\FrameRestore}}%
  {\endMakeFramed}

\usepackage{etoolbox}
\AtBeginEnvironment{longtable}{\footnotesize}

\arrayrulecolor{accent!55}

\usepackage{graphicx}
\usepackage{float}
\usepackage{caption}
\usepackage{newfloat}
\DeclareFloatingEnvironment[fileext=lol,placement=H,name=Listing]{listing}
\AfterEndEnvironment{Shaded}{\setlength{\parskip}{\anebodyparskip}}
\AfterEndEnvironment{listing}{\setlength{\parskip}{\anebodyparskip}}
\AfterEndEnvironment{longtable}{\setlength{\parskip}{\anebodyparskip}}

\makeatletter
\let\origtexttt\texttt
\protected\def\texttt#1{{\origtexttt{\ANEallowbreaks{#1}}}}
\def\ANEallowbreaks#1{\@tfor\ANEtk:=#1\do{%
  \ANEtk\ifx\ANEtk\_\penalty\z@\else\penalty2800\fi}}
\makeatother

\usepackage{fancyhdr}
\renewcommand{\headrulewidth}{0.6pt}
\renewcommand{\headrule}{\hbox to\headwidth{\color{rulegrey}\leaders\hrule height \headrulewidth\hfill}}
\fancypagestyle{plain}{%
  \fancyhf{}\fancyfoot[C]{\small\thepage}\renewcommand{\headrulewidth}{0pt}}

\usepackage{sectsty}
\sectionfont{\bfseries\color{accent}}
\subsectionfont{\bfseries\color{accent!85!black}}
\subsubsectionfont{\bfseries\color{accent!85!black}}
\partfont{\bfseries\color{accent}}
\newif\ifANEinappendix
\makeatletter
\def\@makechapterhead#1{%
  \vspace*{10\p@}%
  {\parindent\z@\raggedright\bfseries\color{accent}\Huge
   \ifANEinappendix Appendix~\thechapter\quad\else\thechapter\quad\fi #1\par\nobreak
   \vskip 4\p@{\color{rulegrey}\hrule height 1.2pt}\vskip 26\p@}}
\def\@makeschapterhead#1{%
  \markboth{#1}{}% set the running mark, since \chapter* otherwise leaves the
  \vspace*{10\p@}%
  {\parindent\z@\raggedright\bfseries\color{accent}\Huge #1\par\nobreak
   \vskip 4\p@{\color{rulegrey}\hrule height 1.2pt}\vskip 26\p@}}
\makeatother

\AtBeginDocument{\hypersetup{colorlinks=true,linkcolor=accent,urlcolor=accent,citecolor=accent}}

\AtEndPreamble{%
  \usepackage{cleveref}%
  \crefname{figure}{figure}{figures}\Crefname{figure}{Figure}{Figures}%
  \crefname{table}{table}{tables}\Crefname{table}{Table}{Tables}%
  \crefname{listing}{listing}{listings}\Crefname{listing}{Listing}{Listings}%
}

\newcommand{\aneeyebrow}[1]{#1}
\providecommand{\anecurrentversion}{v2}
\providecommand{\anebuilddate}{June 2026}
\providecommand{\aneorigrelease}{June 21, 2026}
\IfFileExists{anedate.tex}{\def\anecurrentversion{v3}
\def\anebuilddate{October 14, 2026}
}{}
\renewcommand{\maketitle}{%
  \begin{titlepage}
  \centering
  \vspace*{0.13\textheight}
  {\color{accent}\rule{0.72\linewidth}{0.8pt}}\par
  \vspace{0.85cm}
  {\color{accent}\fontsize{32}{38}\selectfont\bfseries Apple Neural Engine\par}
  \vspace{0.55cm}
  {\Large\color{black!75} Architecture, Programming, and Performance\par}
  \vspace{16pt}
  {\color{accent}\rule{0.72\linewidth}{0.8pt}}\par
  \vspace{1.0cm}
  {\large\color{black!80} Spencer H. Bryngelson\par}
  \vspace{8pt}
  {\footnotesize\color{black!55}
   School of Computational Science \& Engineering\\
   Georgia Institute of Technology, Atlanta, GA 30332, USA\\
   ORCID 0000-0003-1750-7265\par}
  \vfill
  {\color{accent!70}\footnotesize\aneeyebrow{A11--A18 \;\textbullet\; M1--M5}\par}
  \vspace{0.7cm}
  {\small\color{black!70}
   Current edition: \anecurrentversion\ (\anebuilddate)\\[3pt]
   Original release: v1 (\aneorigrelease)\par}
  \vspace{0.6cm}
  \end{titlepage}
  {\thispagestyle{empty}\footnotesize\color{black!70}\raggedright
   \vspace*{\fill}
   \noindent Apple Neural Engine: Architecture, Programming, and Performance.\par
   \vspace{8pt}
   \noindent Copyright \textcopyright\ 2026 Spencer H. Bryngelson.\par
   \vspace{8pt}
   \noindent This work is licensed under a Creative Commons Attribution 4.0
   International License (CC~BY~4.0). You are free to share and adapt the material
   for any purpose, including commercially, provided you give appropriate credit.
   The full license is at \texttt{creativecommons.org/licenses/by/4.0/}.\par
   \vspace{8pt}
   \noindent Apple, Apple silicon, Core ML, and the Apple Neural Engine are
   trademarks of Apple Inc. This is an independent reference work and is not
   affiliated with, authorized by, or endorsed by Apple Inc. The private
   interfaces it describes are undocumented, unsupported, and may change without
   notice.\par
   \vspace{10pt}
   \noindent\textbf{Use of AI tools.}\par
   \noindent Claude Code assisted in drafting and editing the prose of this guide.
   The reverse engineering it reports is the author's own, and the author is
   responsible for the text. See the statements at the end for the full
   declaration.\par
   \vspace{10pt}
   \noindent\textbf{How to cite.}\par
   \noindent Spencer H. Bryngelson. \textit{Apple Neural Engine: Architecture,
   Programming, and Performance.} arXiv:2606.22283, 2026.\par
   \vspace{16pt}}
  \clearpage\pagenumbering{roman}}

\makeatletter
\newcommand{\anepartopen}[2]{%
  \clearpage\thispagestyle{empty}%
  \def\ANE@eb{#1}%
  \addcontentsline{toc}{part}{\ifx\ANE@eb\@empty #2\else #1. #2\fi}%
  \vspace*{0.20\textheight}%
  {\parindent\z@\raggedright
   \ifx\ANE@eb\@empty\else
     {\color{accent!75}\large\aneeyebrow{#1}}\par\vskip 9\p@
   \fi
   {\color{accent}\hrule height 2pt}\vskip 16\p@
   {\color{accent}\bfseries\fontsize{30}{34}\selectfont #2\par}%
   \vskip 13\p@{\color{rulegrey}\hrule height 1pt}\vskip 24\p@}%
  \begingroup\parindent\z@\raggedright}
\newcommand{\anechap}[3]{%
  \noindent{\color{accent}\bfseries #1}\hspace{1.1em}{\color{black!85}#2}\par
  \def\ANE@cb{#3}\ifx\ANE@cb\@empty\else
    \nobreak\noindent\hspace*{2.7em}{\color{black!55}\small #3}\par
  \fi
  \vskip 9\p@}
\newcommand{\anepartclose}{\endgroup\clearpage}
\makeatother
\usepackage{bookmark}
\IfFileExists{xurl.sty}{\usepackage{xurl}}{} % add URL line breaks if available
\makeatletter
\@ifundefined{xmpquote}{}{}
\makeatother
\hypersetup{
  pdftitle={Apple Neural Engine},
  pdfauthor={Spencer H. Bryngelson},
  colorlinks=true,
  linkcolor={Maroon},
  filecolor={Maroon},
  citecolor={Blue},
  urlcolor={Blue},
  pdfcreator={LaTeX via pandoc}}

\title{Apple Neural Engine}
\usepackage{etoolbox}
\makeatletter
\providecommand{\subtitle}[1]{% add subtitle to \maketitle
  \apptocmd{\@title}{\par {\large #1 \par}}{}{}
}
\makeatother
\subtitle{Architecture, Programming, and Performance}
\author{Spencer H. Bryngelson}
\date{June 2026}

\begin{document}
\maketitle
\begin{abstract}
The Apple Neural Engine (ANE) is Apple\textquotesingle s fixed-function
matrix accelerator. It has shipped in every Apple system-on-chip since
the A11 in iPhones and iPads and the M1 in Macs. Applications reach it
only through Core ML, which documents almost nothing below its own
surface. This work reverse-engineers the engine. We reach it directly,
below Core ML, from an ordinary user process, and that access opens the
rest to measurement. We measure on Apple silicon and decompile the
private runtime, compiler, kernel driver, and firmware, checking each
against the other. We give the datapath and the roofline that bounds
throughput and energy. Below that are the compiler and the program
format it emits, the weight-compression scheme, the kernel driver, the
firmware, and the command protocol. We decode compiler targets for the
A11 through A18 and M1 through M5 families, with per-chip tables and an
operation-by-device matrix. We measured silicon on the M1 and M5 only,
and we mark every claim measured, decompile-derived, or predicted. We
record how we reached the engine and what we could not measure. The
route is callable from user space, but Apple does not document or
support it, and it breaks across operating-system updates. It serves
measurement and on-device research. Core ML remains the path for shipped
software.
\end{abstract}

{
\setcounter{tocdepth}{0}
\tableofcontents
}
\clearpage\pagenumbering{arabic}

\chapter*{Introduction}\label{introduction}
\addcontentsline{toc}{chapter}{Introduction}

The Apple Neural Engine (ANE) is among the most widely deployed
machine-learning accelerators, and among the least documented. Apple
built the ANE into every Apple system on chip since the A11 in 2017 and
the M1 in 2020, so nearly every iPhone and iPad active today has one.
The count of active Apple devices surpasses 2.5 billion (measured early
2026)
\hyperref[ref-appleactivedevices2026]{{[}AppleActiveDevices2026{]}}. On
these devices it largely runs the vision, speech, and language models
used by the operating system and its applications. Of the programmable
engines on an Apple chip (CPU, GPU, and ANE), the ANE is the most
opaque. There is no public instruction set, no driver interface, and no
documented way for a program to confirm that a computation ran on it. An
application reaches the ANE indirectly, through the lightly documented
Apple Core ML framework, which treats the engine as one option behind a
placement hint. Core ML is at least partly documented. Nothing below it
is: the datapath and numerics, the performance and energy envelope, the
compiler and program format, the kernel driver, the firmware, and the
command protocol.

Here we report what was not previously available about the ANE,
recovered by reverse engineering from the silicon datapath to the system
interface. Our evidence is of two kinds, checked against each other. We
measure on Apple silicon. We decompile the private runtime, compiler,
kernel driver, and firmware, and attach to them at run time. We find the
engine reachable directly from an ordinary user process, below Core ML.
The ANE compiler lowers a graph to the engine's own program format, the
runtime loads and dispatches it without the model framework, and the
operations the compiler accepts need no special entitlement. This direct
route is what opens the rest of the device to measurement. Some of the
quantities we report, and some of the bytecodes we emit, are likely to
break across operating-system updates. Nothing here should be built into
software that has to keep working.

We stage the performance results around a roofline. On the M1 Max, the
engine reaches about 12 fp16 TFLOP/s of compute, bounded above by a
DRAM-bandwidth ceiling. The roofline has a ridge point near 141 FLOP per
byte, a 2 MB working-set threshold, a 0.23 ms floor under any single
dispatch, and efficiency near 0.37 picojoules per FLOP at peak compute.
On a 256-channel 3x3 convolution it runs about 3.8 times faster than the
same chip's GPU and is 9 times more energy-efficient. We take the power
side of those figures from shipped device measurement tooling.

Reaching the engine is not the same as running an arbitrary
computational graph on it. The operations the engine executes are
distinct from the ones a capability bit advertises. We count a feature
only once a compile-and-run confirms it. Attestation in the hardware
tables, or acceptance by the compiler frontend, is not enough. Several
advertised operations never lower to the engine at all,
three-dimensional convolution among them. Weight compression on the
direct path cuts bandwidth, not only stored size. On the unentitled
engine, int4 lookup-table weights run about 2.37 times faster than fp16,
and structured sparsity 1.55 to 1.64 times faster at 0.43 times the
bytes.

The engine runs on a private stack beneath the datapath. It contains the
compiler and its backend dialect, the on-disk program and container
format, the kernel-driver IOKit ABI, the unencrypted firmware and its
ninety-three-command host protocol, and the address-translation path
that maps host buffers into the engine. We decode twenty-eight compiler
targets, spanning the A11 through A18 and M1 through M5 families. The
decode yields the rule that maps each M-series part to its internal
H-series identity, the per-family operation floors, and an
operation-by-device matrix. Our cross-generation predictions held when
we checked them on a second physical chip, and a seeded training run
reproduced across generations to within 0.001 in final accuracy.

\section*{Related work}\label{related-work}
\addcontentsline{toc}{section}{Related work}

The closest precedent is \hyperref[ref-orion2026]{{[}Orion2026{]}},
concurrent work that characterizes and programs the engine for
large-language-model training and inference. A line of community reverse
engineering precedes it. The tinygrad project recovered the HWX program
format and the AppleH11ANEInterface IOKit path
\hyperref[ref-tinygrad]{{[}tinygrad{]}}, and Yoon's \texttt{ane} project
built a reverse-engineered Linux driver and the \texttt{anecc} compiler
\hyperref[ref-eilnane]{{[}eilnANE{]}}. Singh's recent series decodes the
M4 engine \hyperref[ref-singh2026]{{[}Singh2026{]}}, and Handley
reconstructs the engine's hardware architecture from Apple's patent
filings \hyperref[ref-handley]{{[}Handley{]}}.

The \texttt{libane} native runtime exposes the engine to ordinary
programs \hyperref[ref-libane]{{[}libane{]}}, and whisper.cpp reaches it
through the Core ML-routed path
\hyperref[ref-whispercpp]{{[}whispercpp{]}}. The long-maintained catalog
of Hollemans \hyperref[ref-hollemans]{{[}Hollemans{]}} and Apple's own
engineering note on deploying Transformers
\hyperref[ref-appleanetransformers]{{[}AppleANETransformers{]}} collect
what is publicly known. Several further repositories document direct
access and runtime APIs \hyperref[ref-communityane]{{[}CommunityANE{]}},
and an earlier thesis treats decoupling on-device intelligence from the
application on IoT hardware
\hyperref[ref-plyenkov2019]{{[}Plyenkov2019{]}}.

The performance treatment is organized around the roofline model
\hyperref[ref-williams2009]{{[}Williams2009{]}} and its later
refinements: the energy roofline
\hyperref[ref-choi2013]{{[}Choi2013{]}}, the cache-aware roofline
\hyperref[ref-ilic2014]{{[}Ilic2014{]}}, the instruction roofline
\hyperref[ref-ding2019]{{[}Ding2019{]}}, hierarchical roofline analysis
\hyperref[ref-yang2020]{{[}Yang2020{]}}, and its application to
machine-learning accelerators
\hyperref[ref-verhelst2025]{{[}Verhelst2025{]}}. It belongs to a wider
body of work on neural accelerators and edge inference. That work
includes the in-datacenter analysis of the first tensor processing unit
\hyperref[ref-jouppi2017]{{[}Jouppi2017{]}}, smartphone deep-learning
benchmarks \hyperref[ref-ignatov2019]{{[}Ignatov2019{]}}, and
edge-platform inference benchmarking
\hyperref[ref-jayanth2024]{{[}Jayanth2024{]}}. It also includes
energy-and-time roofline studies on edge accelerators
\hyperref[ref-prashanthi2025]{{[}Prashanthi2025{]}}, NPU energy
efficiency on microcontrollers
\hyperref[ref-fanariotis2025]{{[}Fanariotis2025{]}}, LLM inference
trade-offs across mobile NPU and GPU under sustained load
\hyperref[ref-tummalapalli2026]{{[}Tummalapalli2026{]}}, and on-device
LLM roofline benchmarking \hyperref[ref-bi2026]{{[}Bi2026{]}}. The work
closest in workload studies heterogeneous on-device LLM inference: fast
NPU inference \hyperref[ref-xu2025]{{[}Xu2025{]}}, GPU-NPU hybrid
serving for long context \hyperref[ref-moon2025]{{[}Moon2025{]}}, and
mobile-SoC characterization for heterogeneous execution
\hyperref[ref-chen2025]{{[}Chen2025{]}}. On Apple silicon specifically,
Benazir and Lin study mixture-of-experts inference on the NPU
\hyperref[ref-benazir2026]{{[}Benazir2026{]}}, Hübner and colleagues
evaluate the M-series for HPC efficiency
\hyperref[ref-hubner2025]{{[}Hubner2025{]}}, and the ML.ENERGY project
\hyperref[ref-zeus2025]{{[}Zeus2025{]}} treats the programmatic energy
measurement the power figures here depend on.

We do not claim to be first to reach the engine. The direct route we
measure is released as the open-source ANEForge runtime
\hyperref[ref-aneforge2026]{{[}ANEForge2026{]}}, which this guide
accompanies. Our account differs from prior work in scope and in method,
since the coverage runs the full stack, from the fp16 datapath down to
the firmware and command protocol. Beyond the access result we add a
reachable-operation census, unentitled weight streaming, a prediction of
an operation's reachability from a callable compiler validator, and an
account of how the compiler fuses a whole graph into one program.
Earlier Apple-silicon rooflines are GPU-only; here we cover the engine,
pair it with measured power, and give a batched-serving energy
crossover.

\section*{How to read this guide}\label{how-to-read-this-guide}
\addcontentsline{toc}{section}{How to read this guide}

We have written the two halves to be taken independently. The front
half, Parts I through V, covers using the engine: what the hardware is,
how to reach it, how it performs, and how to fit real workloads to it.
The back half, Parts VI through IX, covers the architecture and system
internals beneath that surface, down to the firmware and command
protocol. The appendices collect the operation-by-device matrix, decoded
reference tables, glossary, and provenance record, and the references
close the guide.

Every substantive claim has an evidentiary mark. A measured claim was
observed directly on Apple silicon, primarily the M1 and M5. We read
decompile-derived claims out of the disassembled runtime, compiler,
kernel driver, or firmware. Predicted claims we inferred from a model or
a per-chip table; silicon has not confirmed them yet. Appendix
\hyperref[appendix-e.-provenance]{E} records the mark on every claim.
The methodology describes how we reached the engine, and the open
questions name what we were unable to measure.

\anepartopen{Part I}{The Machine}
\anechap{01}{What the ANE is}{The fixed-function fp16 datapath, its wide accumulator, and the route below Core ML.}
\anechap{02}{Execution model}{Compile once and dispatch many; the per-call latency budget and the program cache.}
\anechap{03}{Numerics}{fp16 end to end, the wide accumulator, saturation, and where precision breaks down.}
\anechap{04}{Capability surface}{Which operations the engine runs, against which a capability bit only advertises.}
\anepartclose

\chapter{What the ANE is}\label{what-the-ane-is}

\begin{summarybox}

The Apple Neural Engine is a fixed-function fp16 matrix accelerator with
a wide accumulator, reachable directly below Core ML. That pairing
explains most of its numerical behavior and much of its efficiency. It
leads on compute-bound vision and convolution: 3.8 times faster than the
GPU on a 256-channel 3x3 convolution, at 9 times the efficiency. The GPU
leads only on bandwidth-bound decode. The overhead-isolated compute
slope nears 12 fp16 TFLOP/s against 85 GB/s of DRAM, and a 2 MB on-chip
working set is the binding limit.

\end{summarybox}

\begin{figure}[H]\centering
\includegraphics[width=4.737in]{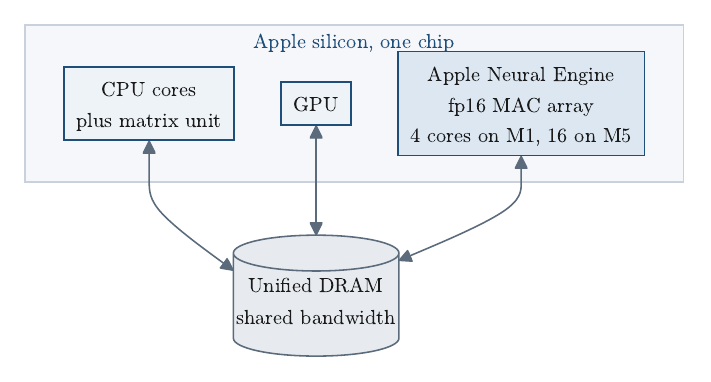}
\caption{The Apple Neural Engine beside the CPU and GPU on one chip, sharing unified DRAM.}
\label{fig:soc}
\end{figure}

The Apple Neural Engine is a fixed-function matrix accelerator built
into every recent Apple system on chip, from the A11 and the M1 onward.
It runs the feed-forward neural networks of on-device perception, vision
and speech, at low power, and leaves the CPU and the GPU free for the
rest of the system. Apple distributes it in volume and documents almost
none of it. The engine is beside the CPU and the GPU on the same chip,
where the three share one pool of DRAM, as \cref{fig:soc} shows.

\section{Reachable surface}\label{reachable-surface}

The public way to use the engine is Core ML
\hyperref[ref-applecoreml]{{[}AppleCoreML{]}}, sketched in
\cref{lst:c1-coreml-predict}.

\begin{listing}[H]
\caption{The Core ML load-and-predict path, where the compute-unit field is a placement hint rather than a guarantee.}\label{lst:c1-coreml-predict}

\begin{Shaded}
\begin{Highlighting}[]
\CommentTok{// The public path: ask for the engine, but Core ML still places the work.}
\KeywordTok{let} \VariableTok{config} \OperatorTok{=}\NormalTok{ MLModelConfiguration}\OperatorTok{()}
\NormalTok{config}\OperatorTok{.}\NormalTok{computeUnits }\OperatorTok{=} \OperatorTok{.}\NormalTok{cpuAndNeuralEngine  }\CommentTok{// a hint, not a guarantee}
\KeywordTok{let} \VariableTok{model} \OperatorTok{=} \ControlFlowTok{try}\NormalTok{ MLModel}\OperatorTok{(}\NormalTok{contentsOf}\OperatorTok{:}\NormalTok{ url}\OperatorTok{,}\NormalTok{ configuration}\OperatorTok{:}\NormalTok{ config}\OperatorTok{)}
\KeywordTok{let} \VariableTok{out} \OperatorTok{=} \ControlFlowTok{try}\NormalTok{ model}\OperatorTok{.}\NormalTok{prediction}\OperatorTok{(}\NormalTok{from}\OperatorTok{:}\NormalTok{ input}\OperatorTok{)}  \CommentTok{// planner splits ops across CPU, GPU, ANE}
\CommentTok{// The caller is not told which device ran each segment.}
\end{Highlighting}
\end{Shaded}

\end{listing}

A cost-driven placement planner segments a model handed to Core ML
across the CPU, GPU, and ANE. An eligibility check decides which
operations the engine can accept, and a roofline and transfer cost model
decides where each segment runs. The planner is opaque. A caller does
not choose the device and is not told which device ran the work.

A direct route reaches the engine without Core ML. The same private
Espresso runtime that Apple's own dispatchers use is callable from
ordinary user space. It compiles a network to the engine's program
format, loads it, and drives an execution stream, with no placement
planner in the path and no special entitlement for the operations the
compiler accepts. This route makes the engine a target a developer can
address on purpose. It is not a supported or App-Store-safe path; see
the status note in the front matter. Part II covers it in full.

The stack from a graph to the silicon is a fixed set of layers, and the
direct route enters one level below Core ML at the runtime
(\cref{tbl:c1-stack}).

\begin{longtable}[]{@{}
  >{\raggedright\arraybackslash}p{(\linewidth - 2\tabcolsep) * \real{0.5000}}
  >{\raggedright\arraybackslash}p{(\linewidth - 2\tabcolsep) * \real{0.5000}}@{}}
\caption{The software and firmware stack from a compute graph to the
silicon, with the role of each
layer.}\label{tbl:c1-stack}\tabularnewline
\toprule\noalign{}
\begin{minipage}[b]{\linewidth}\raggedright
\textbf{Layer}
\end{minipage} & \begin{minipage}[b]{\linewidth}\raggedright
\textbf{Role}
\end{minipage} \\
\midrule\noalign{}
\endfirsthead
\toprule\noalign{}
\begin{minipage}[b]{\linewidth}\raggedright
\textbf{Layer}
\end{minipage} & \begin{minipage}[b]{\linewidth}\raggedright
\textbf{Role}
\end{minipage} \\
\midrule\noalign{}
\endhead
\bottomrule\noalign{}
\endlastfoot
Core ML, the GPU graph framework, the direct runtime route & three
independent front ends \\
\rowcolor{zebra}Espresso and the runtime & the universal compile, load,
and execution-stream layer \\
the engine framework and its daemon & program lifecycle, brokered over
cross-process messaging \\
\rowcolor{zebra}the kernel driver & the coprocessor endpoint, mailbox
transport, signed program load \\
the firmware & the engine's own real-time operating system, the dispatch
loop, the data-movement and power control \\
\rowcolor{zebra}the silicon & the multiply array, the on-chip memory,
and the address-translation unit \\
\end{longtable}

The compile path requires the modern intermediate language, which the
compiler accepts only from the A13 and M1 generation up. The pre-A13
parts have no path through this toolchain, which is why the M1 is the
practical floor for addressing the engine directly.

\section{One fact that explains the
machine}\label{one-fact-that-explains-the-machine}

The products are fp16 while the accumulator is wide, of fp32 class, and
that property accounts for most of what the engine does. The datapath
multiplies in fp16 end to end: fp16 inputs, fp16 weights, fp16 outputs.
The frontend accepts fp32, int32, and bf16 as type annotations, but the
backend does not implement them. The datapath reconstructs compressed
weights to fp16 before they reach the multiplier.

The running sum is not fp16. Input tiles round to fp16 on the way in,
the output port rounds to fp16 on the way out, and the accumulation
between those two points is held in a wide accumulator. Any sum the fp16
grid can represent comes back near exact. Sixteen thousand ones reduce
bit exact; a naive fp16 running sum stalls near two thousand instead,
once the partial total exceeds the spacing of its own increments. A
cancellation probe settles the question: a sum of a large value, its
negation, and one, taken near sixteen thousand, returns the one intact,
which an fp16 running sum would have swallowed. The sum is physically
wider than fp16.

The wide accumulator is what suits the engine to vision, audio, and
encoders. Those workloads are convolutions, matrix multiplies, and
normalizations whose partial sums stay in range and whose results are
representable, so the accumulator keeps their precision. The same
arithmetic explains where a transformer decoder loses precision in fp16.
The loss comes from the per-product fp16 rounding of the inputs and
weights under heavy cancellation, in the down-projection in particular.
The accumulator is wide enough; the inputs to it are already quantized.
A cancellation-heavy step has no fp16-safe form on the engine and needs
a wider anchor on the CPU or the GPU.

The fp16 datapath also accounts for the power efficiency. A narrow
multiply and a fixed-function pipeline move and compute far fewer bits
per result than a general-purpose vector unit.

\section{Two core counts}\label{two-core-counts}

The engine reports two different core counts, and only one is the
throughput unit. The advertised 16-core figure is the count Apple
markets, 16 on the M1, exposed at runtime as the device property for the
number of engine cores. The architectural figure is the core count the
compiler tiles work across and the cost model scales by, read from the
hardware-abstraction-layer offset \texttt{0x238}, which is 4 on the M1.
The core count is 4 on the M1 base part, 8 on the M1 Pro and Max, and 16
on the M5. Each core emits up to 4 output channels per cycle in the
default fp16 path. On the M1 that works out to about 16 output-channel
multiply-accumulates per cycle in fp16, and int8 doubles the lanes to
reach about 32, consistent with the overhead-isolated compute slope near
12 fp16 TFLOP/s.

\section{Where it leads}\label{where-it-leads}

On compute-bound work the engine is faster than the GPU outright. A 3x3
convolution at 256 channels runs about 3.8 times faster than the same
work on the GPU and about 9 times more energy efficient, and batched
matrix multiply is more efficient at every batch size
(\cref{tbl:c1-gpu}).

\begin{longtable}[]{@{}
  >{\raggedright\arraybackslash}p{(\linewidth - 4\tabcolsep) * \real{0.3333}}
  >{\raggedright\arraybackslash}p{(\linewidth - 4\tabcolsep) * \real{0.3333}}
  >{\raggedright\arraybackslash}p{(\linewidth - 4\tabcolsep) * \real{0.3333}}@{}}
\caption{Engine versus GPU speed and efficiency on the convolution and
batched matrix multiply workloads.}\label{tbl:c1-gpu}\tabularnewline
\toprule\noalign{}
\begin{minipage}[b]{\linewidth}\raggedright
\textbf{workload}
\end{minipage} & \begin{minipage}[b]{\linewidth}\raggedright
\textbf{engine vs GPU speed}
\end{minipage} & \begin{minipage}[b]{\linewidth}\raggedright
\textbf{engine vs GPU efficiency}
\end{minipage} \\
\midrule\noalign{}
\endfirsthead
\toprule\noalign{}
\begin{minipage}[b]{\linewidth}\raggedright
\textbf{workload}
\end{minipage} & \begin{minipage}[b]{\linewidth}\raggedright
\textbf{engine vs GPU speed}
\end{minipage} & \begin{minipage}[b]{\linewidth}\raggedright
\textbf{engine vs GPU efficiency}
\end{minipage} \\
\midrule\noalign{}
\endhead
\bottomrule\noalign{}
\endlastfoot
3x3 convolution (256 channels) & 3.8x faster & 9x more efficient \\
\rowcolor{zebra}batched matrix multiply & faster below N of 2048 & more
efficient at every batch size \\
\end{longtable}

The case where the GPU leads is real but limited: it holds for
bandwidth-bound autoregressive decode, where the work is moving weights.
The measured envelope on the M1 fixes the scale. The overhead-isolated
compute slope reaches nearly \(P \approx 12\) fp16 TFLOP/s against a
DRAM bandwidth of \(B \approx 85\) GB/s, at about \(0.5\) pJ per FLOP
sustained, near \(0.37\) at the compute optimum. The roofline ridge
point, the arithmetic intensity above which a kernel is compute-bound,
is at \(I^{*} = P / B \approx 141\) FLOP per byte. A hard 2 MB on-chip
working-set threshold is the primary design limit: a kernel whose live
tiles exceed it stalls on off-chip streaming. Newer chips scale the core
count and the clock, but the form of the roofline and the fp16 datapath
apply across the family.

\section{Reference: the M1 envelope}\label{reference-the-m1-envelope}

\Cref{tbl:c1-envelope} collects the M1 envelope figures.

\begin{longtable}[]{@{}llr@{}}
\caption{The M1 envelope figures, the two distinct core counts, and the
family core-count span.}\label{tbl:c1-envelope}\tabularnewline
\toprule\noalign{}
\textbf{Quantity} & \textbf{Symbol} & \textbf{M1/H13 value} \\
\midrule\noalign{}
\endfirsthead
\toprule\noalign{}
\textbf{Quantity} & \textbf{Symbol} & \textbf{M1/H13 value} \\
\midrule\noalign{}
\endhead
\bottomrule\noalign{}
\endlastfoot
Compute roof & \(P\) & 12 fp16 TFLOP/s \\
\rowcolor{zebra}DRAM bandwidth roof & \(B\) & 85 GB/s \\
Energy per FLOP & & 0.5 pJ sustained, 0.37 pJ optimum \\
\rowcolor{zebra}Roofline ridge point & \(I^{*}\) & 141 FLOP/byte \\
On-chip working-set threshold & & 2 MB \\
\rowcolor{zebra}fp16 maximum finite magnitude & & 65504 \\
Advertised 16-core figure, M1 & & 16 \\
\rowcolor{zebra}Core count, M1 (HAL \texttt{0x238}) & & 4 \\
Core count, M1 to M5 & & 4 to 16 \\
\rowcolor{zebra}Output channels per cycle per core, fp16 & & 4 \\
\end{longtable}

\chapter{Execution model}\label{execution-model}

\begin{summarybox}

The engine is an autonomous coprocessor: a network compiles once into
its program format, then dispatches many times against that one program.
Compiling is costly, so keep it out of the hot loop; dispatch binds
operands, posts one mailbox command, and waits. The compiled program is
a static graph the hardware walks, so control flow is fixed at compile
time and cannot read runtime values. A buffer can stay resident across
dispatches, so a key-value cache or optimizer state persists without a
host round-trip.

\end{summarybox}

\section{Compile once, dispatch many}\label{compile-once-dispatch-many}

Work reaches the engine in two phases whose costs are far apart. The
public surface exposes only a load-and-predict view of this split
\hyperref[ref-applecoreml]{{[}AppleCoreML{]}}; the two phases below are
the mechanism beneath it. The compile phase turns a network into the
engine's program format: the compiler lowers the operation graph, lays
out weights for the streaming datapath, and produces a loadable program.
The dispatch phase runs that program against a set of operands and reads
back the output.

The compile phase is costly. It runs the full lowering and layout
pipeline, writes a program to a content-addressed cache on disk, and the
first dispatch of a freshly compiled program pays a further one-time
cost to produce the loadable hardware form. The dispatch phase binds
operand buffers, hands the program to the engine, and waits for
completion.

\Cref{tbl:exec-phases} sets the two phases against the runtime call
surface.

\begin{longtable}[]{@{}
  >{\raggedright\arraybackslash}p{(\linewidth - 6\tabcolsep) * \real{0.2500}}
  >{\raggedright\arraybackslash}p{(\linewidth - 6\tabcolsep) * \real{0.2500}}
  >{\raggedright\arraybackslash}p{(\linewidth - 6\tabcolsep) * \real{0.2500}}
  >{\raggedright\arraybackslash}p{(\linewidth - 6\tabcolsep) * \real{0.2500}}@{}}
\caption{The compile and dispatch phases, when each runs, its cost, and
the runtime calls in each.}\label{tbl:exec-phases}\tabularnewline
\toprule\noalign{}
\begin{minipage}[b]{\linewidth}\raggedright
\textbf{Phase}
\end{minipage} & \begin{minipage}[b]{\linewidth}\raggedright
\textbf{When it runs}
\end{minipage} & \begin{minipage}[b]{\linewidth}\raggedright
\textbf{Cost}
\end{minipage} & \begin{minipage}[b]{\linewidth}\raggedright
\textbf{Example calls}
\end{minipage} \\
\midrule\noalign{}
\endfirsthead
\toprule\noalign{}
\begin{minipage}[b]{\linewidth}\raggedright
\textbf{Phase}
\end{minipage} & \begin{minipage}[b]{\linewidth}\raggedright
\textbf{When it runs}
\end{minipage} & \begin{minipage}[b]{\linewidth}\raggedright
\textbf{Cost}
\end{minipage} & \begin{minipage}[b]{\linewidth}\raggedright
\textbf{Example calls}
\end{minipage} \\
\midrule\noalign{}
\endhead
\bottomrule\noalign{}
\endlastfoot
Compile & once, ahead of the loop & full lowering and layout, written to
disk & compile the network, open the program library, retain the program
function, load the function for execution \\
\rowcolor{zebra}Dispatch & once per frame, token, or request & bind
operands, post command, wait & encode the operation, execute the stream,
read the output \\
\end{longtable}

The runtime exposes the two phases as distinct call families, the
compile-side calls running once (\Cref{lst:exec-compile}) and the
dispatch-side calls on every request (\Cref{lst:exec-dispatch}).

\begin{listing}[H]
\caption{The one-time compile-side calls that turn a network into a loaded program.}\label{lst:exec-compile}

\begin{Shaded}
\begin{Highlighting}[]
\CommentTok{/* once, out of the hot loop */}
\NormalTok{e5rt\_e5\_compiler\_compile}\OperatorTok{(}\NormalTok{compiler}\OperatorTok{,}\NormalTok{ model\_path}\OperatorTok{,}\NormalTok{ options}\OperatorTok{,} \OperatorTok{\&}\NormalTok{library}\OperatorTok{);}
\NormalTok{e5rt\_program\_library\_retain\_program\_function}\OperatorTok{(}\NormalTok{library}\OperatorTok{,}\NormalTok{ name}\OperatorTok{,} \OperatorTok{\&}\NormalTok{function}\OperatorTok{);}
\NormalTok{e5rt\_program\_function\_load\_for\_execution}\OperatorTok{(}\NormalTok{function}\OperatorTok{);}
\end{Highlighting}
\end{Shaded}

\end{listing}

\begin{listing}[H]
\caption{The per-call dispatch-side calls that encode and run the loaded program.}\label{lst:exec-dispatch}

\begin{Shaded}
\begin{Highlighting}[]
\CommentTok{/* once per frame, token, or request */}
\NormalTok{e5rt\_execution\_stream\_encode\_operation}\OperatorTok{(}\NormalTok{stream}\OperatorTok{,}\NormalTok{ op}\OperatorTok{);}
\NormalTok{e5rt\_execution\_stream\_execute\_sync}\OperatorTok{(}\NormalTok{stream}\OperatorTok{);}
\end{Highlighting}
\end{Shaded}

\end{listing}

Chapter \hyperref[dispatching-without-core-ml]{6} gives the full
create-bind-encode-execute sequence, including the compute operation and
the input and output port binding.

The compile phase produces three program forms, not one, lowered at
successive layers (\cref{tbl:exec-forms}).

\begin{longtable}[]{@{}
  >{\raggedright\arraybackslash}p{(\linewidth - 4\tabcolsep) * \real{0.3333}}
  >{\raggedright\arraybackslash}p{(\linewidth - 4\tabcolsep) * \real{0.3333}}
  >{\raggedright\arraybackslash}p{(\linewidth - 4\tabcolsep) * \real{0.3333}}@{}}
\caption{The three program representations a compiled network passes
through, from the cached bundle to the firmware load
format.}\label{tbl:exec-forms}\tabularnewline
\toprule\noalign{}
\begin{minipage}[b]{\linewidth}\raggedright
\textbf{Representation}
\end{minipage} & \begin{minipage}[b]{\linewidth}\raggedright
\textbf{What it is}
\end{minipage} & \begin{minipage}[b]{\linewidth}\raggedright
\textbf{Where it is}
\end{minipage} \\
\midrule\noalign{}
\endfirsthead
\toprule\noalign{}
\begin{minipage}[b]{\linewidth}\raggedright
\textbf{Representation}
\end{minipage} & \begin{minipage}[b]{\linewidth}\raggedright
\textbf{What it is}
\end{minipage} & \begin{minipage}[b]{\linewidth}\raggedright
\textbf{Where it is}
\end{minipage} \\
\midrule\noalign{}
\endhead
\bottomrule\noalign{}
\endlastfoot
the bundle & a flat-buffer container whose fused graph collapses to a
three-op chain, a cast in, the engine inference, a cast out, with a
parametric per-op descriptor whose size does not grow with the tensor &
the on-disk content-addressed cache \\
\rowcolor{zebra}the program image & a signed executable the kernel
loader parses, magic \texttt{0xbeefface}, with a register-write text
section, a weight section, constants, and scratch & materialized below
the host boundary at load \\
the firmware container & the firmware's load format, a three-level
container keyed by program identity, with generic, kernel, text,
operation, and procedure sections & resident on the engine \\
\end{longtable}

The loader expands the parametric descriptor in the bundle into the
explicit register-write program below the host boundary, so the
shape-specific program appears in no host buffer. This is why the
compile cost is paid once on disk and a further one-time cost on the
first dispatch, when the loadable hardware form is produced.

\section{Host drives an autonomous
coprocessor}\label{host-drives-an-autonomous-coprocessor}

The engine is on the system on chip with its own controller and its own
local memory. The host never reads or writes the engine's compute
registers and never steps it instruction by instruction. It hands over a
compiled program and the operand buffers, signals the engine through a
command mailbox, and waits for a completion notification.

The mailbox is a ring buffer of command records shared between the host
and the engine's controller. To start work, the host writes a command
that names the program and its operands and rings a doorbell. The
controller picks the command up, drives the datapath through the work,
and signals completion back across the same channel. Operand buffers are
mapped through the engine's own address translation unit, so the engine
reads inputs and writes outputs directly in memory the host prepared.
The host waits on the completion signal; it does not supervise the
computation.

Once the command is posted, the host CPU is idle with respect to that
work and can prepare the next operands or post more commands. Several
programs from several processes can have work outstanding at once; the
engine time-shares itself across them without host involvement.

Each inference is a procedure call that contains its operand buffers,
its optional wait, signal, and shared events, and a set of
task-descriptor partitions. The firmware pushes one engine request per
partition onto a task queue, and the engine preempts at task-queue
granularity with a mid-flight abort, so a higher-priority program does
not wait for a lower-priority one to drain. The host posts a command
through a header that names the command, its size, a priority in the
range 0 to 7, and the program, process, and procedure identities. A
secure mode can claim the engine exclusively by quiescing and
power-cycling around the boundary, which is how protected-content work
is isolated.

A single dispatch stream keeps one operation in flight reliably, but
overlapping two or more streams in one process is the unfinished path on
the M1. The completion event for the first stream signals and its waiter
returns, while the completion notification for a second concurrently
overlapped stream does not fire, so its waiter blocks. The runtime has
the controls that would change this, a low-latency event path and a
submit call with a timeout, but the default serialized path is the sound
one. A caller that needs aggregate throughput runs independent streams,
since sequential decode cannot overlap with itself in any case.

\section{What one dispatch costs}\label{what-one-dispatch-costs}

A single dispatch is governed by the cost of getting to the engine and
back, not by the engine compute itself. Measured live on the M1 with a
read-only trace, a tiny graph of a 3-by-3 convolution from 8 channels to
8 with padding 1, then a relu, then a mean, runs in a hot loop of about
2000 iterations. Each call costs about 190 microseconds of wall-clock
time. About 98 percent of that is software and firmware dispatch
overhead (\cref{tbl:exec-budget}).

\begin{longtable}[]{@{}
  >{\raggedright\arraybackslash}p{(\linewidth - 2\tabcolsep) * \real{0.5000}}
  >{\raggedright\arraybackslash}p{(\linewidth - 2\tabcolsep) * \real{0.5000}}@{}}
\caption{The per-call latency budget of a single small dispatch on the
M1, from the user-space binding through the firmware round trip to
kernel-side completion.}\label{tbl:exec-budget}\tabularnewline
\toprule\noalign{}
\begin{minipage}[b]{\linewidth}\raggedright
\textbf{Stage}
\end{minipage} & \begin{minipage}[b]{\linewidth}\raggedright
\textbf{Cost}
\end{minipage} \\
\midrule\noalign{}
\endfirsthead
\toprule\noalign{}
\begin{minipage}[b]{\linewidth}\raggedright
\textbf{Stage}
\end{minipage} & \begin{minipage}[b]{\linewidth}\raggedright
\textbf{Cost}
\end{minipage} \\
\midrule\noalign{}
\endhead
\bottomrule\noalign{}
\endlastfoot
User-space binding, runtime, and host fp16 input and output copy & about
25 microseconds \\
\rowcolor{zebra}Building the firmware request & about 16 microseconds \\
Firmware kick: the doorbell, which returns asynchronously & about 2 to 3
microseconds \\
\rowcolor{zebra}Firmware round trip & about 130 microseconds \\
Kernel-side completion processing & about 10 microseconds \\
\end{longtable}

The kernel user-client submit, the \texttt{ANE\_ProgramSendRequest}
external method, takes about 163 microseconds from entry to return.
Completion is interrupt-driven: an interrupt handler fires about twice
per inference, and the asynchronous-message completion path is not used
on this synchronous small-model path. During that firmware round trip
the firmware wakes, picks the command off its queue, executes, and
signals back. It is not sub-splittable from user space with read-only
tools, because the firmware per-task-descriptor latency profiler is
gated.

\section{Submissions serialize at one in
flight}\label{submissions-serialize-at-one-in-flight}

The driver keeps at most one firmware command in flight at a time, a
single-pending-queue scheduler. Two concurrent submission threads
serialize, measured at 1.04 times, so the round trip is not hidden by
overlapping requests.

\section{A walked graph, not a decoded
stream}\label{a-walked-graph-not-a-decoded-stream}

The compiled program is a static graph of work segments that the
hardware walks, with no instruction stream for a processor to decode.
There is no program counter to read and no microcode to dump. The
compiler fixes the order and shape of every segment ahead of time, and
the engine's controller advances through that fixed structure,
programming the data movement engines and the multiply array for each
segment in turn.

A direct consequence is that control flow must be static. The graph has
a shape decided at compile time, so the work the engine performs cannot
depend on values computed during the run. Data-dependent branching does
not execute on the engine: no path through the walked graph selects
itself from a runtime value. A network that needs such a branch must
resolve it on the host or restructure it so the branch becomes a fixed
computation, for example a mask applied to both sides rather than a
choice between them. A loop with a fixed trip count unrolls into a fixed
graph and is admissible; a loop whose length depends on the data is not.

The same property explains the absence of a readable instruction trace.
The unit the engine executes is a precompiled segment of data movement
and multiply work, parameterized by operand addresses and shapes. The
fine-grained register program that drives the silicon is materialized
below the host boundary at load time and never appears in a host buffer.
The program format is the subject of a later chapter.

\section{State kept resident across
dispatches}\label{state-kept-resident-across-dispatches}

A dispatch does not have to round-trip every tensor through the host.
The engine can keep a buffer resident in its working set across calls,
so a value produced by one dispatch is available to the next without a
copy back to the host and a copy forward again. The mechanism aliases an
output buffer of one call to an input buffer of the following call, so
the data persists in place between dispatches.

The aliasing reuses the same port-binding calls that chapter
\hyperref[dispatching-without-core-ml]{6} uses for ordinary I/O. One
buffer object is bound to the output port of the operation and to the
input port of the next dispatch, so the dispatch that writes the held
tensor and the dispatch that reads it name the same memory. The binding
appears call by call in \cref{lst:exec-resident}.

\begin{listing}[H]
\caption{Keeping a state buffer resident across dispatches by binding it to both an output port and the next step’s input port.}\label{lst:exec-resident}

\begin{Shaded}
\begin{Highlighting}[]
\CommentTok{/* One buffer object holds the resident state (a KV{-}cache or optimizer state). */}
\NormalTok{e5rt\_buffer\_object\_alloc}\OperatorTok{(\&}\NormalTok{state\_buf}\OperatorTok{,}\NormalTok{ nbytes}\OperatorTok{,} \CommentTok{/*type=*/}\DecValTok{0}\OperatorTok{);}

\CommentTok{/* Bind it to BOTH the output port that writes the new state ... */}
\NormalTok{e5rt\_execution\_stream\_operation\_retain\_output\_port}\OperatorTok{(}\NormalTok{op}\OperatorTok{,} \StringTok{"state\_out"}\OperatorTok{,} \OperatorTok{\&}\NormalTok{out\_port}\OperatorTok{);}
\NormalTok{e5rt\_io\_port\_bind\_buffer\_object}\OperatorTok{(}\NormalTok{out\_port}\OperatorTok{,}\NormalTok{ state\_buf}\OperatorTok{);}

\CommentTok{/* ... and the input port that reads it on the next step: same buffer object. */}
\NormalTok{e5rt\_execution\_stream\_operation\_retain\_input\_port}\OperatorTok{(}\NormalTok{op}\OperatorTok{,} \StringTok{"state\_in"}\OperatorTok{,} \OperatorTok{\&}\NormalTok{in\_port}\OperatorTok{);}
\NormalTok{e5rt\_io\_port\_bind\_buffer\_object}\OperatorTok{(}\NormalTok{in\_port}\OperatorTok{,}\NormalTok{ state\_buf}\OperatorTok{);}

\ControlFlowTok{for} \OperatorTok{(}\DataTypeTok{int}\NormalTok{ step }\OperatorTok{=} \DecValTok{0}\OperatorTok{;}\NormalTok{ step }\OperatorTok{\textless{}}\NormalTok{ n}\OperatorTok{;}\NormalTok{ step}\OperatorTok{++)} \OperatorTok{\{}
    \CommentTok{/* Send only the small per{-}step input (a token or a minibatch). */}
    \CommentTok{/* schematic: write the step input into its own bound port, not state\_buf */}
\NormalTok{    e5rt\_execution\_stream\_encode\_operation}\OperatorTok{(}\NormalTok{stream}\OperatorTok{,}\NormalTok{ op}\OperatorTok{);}
\NormalTok{    e5rt\_execution\_stream\_execute\_sync}\OperatorTok{(}\NormalTok{stream}\OperatorTok{);}
    \CommentTok{/* state\_buf now holds the updated state; it is never re{-}sent from the host. */}
\NormalTok{    e5rt\_execution\_stream\_reset}\OperatorTok{(}\NormalTok{stream}\OperatorTok{);}
\OperatorTok{\}}
\end{Highlighting}
\end{Shaded}

\end{listing}

Any multi-step computation that stores state uses this mechanism. An
autoregressive decoder keeps its key and value cache resident, so each
step appends the new entry in place, and no step restreams the whole
cache through the host. A training loop keeps its optimizer state
resident across steps for the same reason. The host sends only the small
per-step inputs, a new token or a minibatch, and reads the resident
buffers back at a checkpoint. The large held tensor stays on the engine
instead of crossing the host boundary twice per step.

The output-to-input aliasing is the reachable face of the firmware's
data-chaining subsystem, which keeps an output set resident and chains
it as the next call's input. This is the route on the M1. The M1 task
descriptor has no in-place resident-state data-movement engine: the
encoders for that path are stubbed with the message that the
data-movement form is not supported on this architecture, and the
engine's native state operations are rejected when the program compiles.
The held tensor is written as one call's output and read as the next
call's input through one bound buffer, with the positional write done as
a standard masked update against a small position vector sent each step.
A resident accumulator built this way returns 1, 2, 3, 4 over four
dispatches with no host re-send, and a resident cache fills each slot in
place across steps, both confirmed on the M1.

\section{Compile out of the loop, dispatch inside
it}\label{compile-out-of-the-loop-dispatch-inside-it}

The cost split dictates the loop structure. A content hash keys the
compiled program, so recompiling an unchanged network is a cache hit.

The loaded program is dispatched against fresh operands per frame, per
token, or per request (\cref{lst:exec-loop}).

\begin{listing}[H]
\caption{The compile-once, dispatch-many loop, with compile and load above the loop and bind and dispatch inside it.}\label{lst:exec-loop}

\begin{Shaded}
\begin{Highlighting}[]
\CommentTok{/* Once, out of the hot loop: compile and load. A cache hit skips the lowering. */}
\NormalTok{e5rt\_e5\_compiler\_compile}\OperatorTok{(}\NormalTok{compiler}\OperatorTok{,}\NormalTok{ model\_path}\OperatorTok{,}\NormalTok{ options}\OperatorTok{,} \OperatorTok{\&}\NormalTok{library}\OperatorTok{);}
\NormalTok{e5rt\_program\_library\_retain\_program\_function}\OperatorTok{(}\NormalTok{library}\OperatorTok{,}\NormalTok{ fn\_name}\OperatorTok{,} \OperatorTok{\&}\NormalTok{function}\OperatorTok{);}
\NormalTok{e5rt\_program\_function\_load\_for\_execution}\OperatorTok{(}\NormalTok{function}\OperatorTok{);}
\NormalTok{e5rt\_precompiled\_compute\_op\_create\_options\_create\_with\_program\_function}\OperatorTok{(\&}\NormalTok{op\_opts}\OperatorTok{,}\NormalTok{ function}\OperatorTok{);}
\NormalTok{e5rt\_execution\_stream\_operation\_create\_precompiled\_compute\_operation\_with\_options}\OperatorTok{(\&}\NormalTok{op}\OperatorTok{,}\NormalTok{ op\_opts}\OperatorTok{);}
\NormalTok{e5rt\_execution\_stream\_operation\_retain\_input\_port}\OperatorTok{(}\NormalTok{op}\OperatorTok{,} \StringTok{"x"}\OperatorTok{,} \OperatorTok{\&}\NormalTok{in\_port}\OperatorTok{);}
\NormalTok{e5rt\_io\_port\_bind\_buffer\_object}\OperatorTok{(}\NormalTok{in\_port}\OperatorTok{,}\NormalTok{ in\_buf}\OperatorTok{);}   \CommentTok{/* bound once, refilled per frame */}
\NormalTok{e5rt\_execution\_stream\_create}\OperatorTok{(\&}\NormalTok{stream}\OperatorTok{);}

\ControlFlowTok{for} \OperatorTok{(}\DataTypeTok{int}\NormalTok{ frame }\OperatorTok{=} \DecValTok{0}\OperatorTok{;}\NormalTok{ frame }\OperatorTok{\textless{}}\NormalTok{ n\_frames}\OperatorTok{;}\NormalTok{ frame}\OperatorTok{++)} \OperatorTok{\{}
    \CommentTok{/* Inside the loop: write the next frame, then encode, execute, reset. */}
\NormalTok{    e5rt\_execution\_stream\_operation\_prepare\_op\_for\_encode}\OperatorTok{(}\NormalTok{op}\OperatorTok{);}
\NormalTok{    e5rt\_execution\_stream\_encode\_operation}\OperatorTok{(}\NormalTok{stream}\OperatorTok{,}\NormalTok{ op}\OperatorTok{);}
\NormalTok{    e5rt\_execution\_stream\_execute\_sync}\OperatorTok{(}\NormalTok{stream}\OperatorTok{);}
\NormalTok{    e5rt\_execution\_stream\_reset}\OperatorTok{(}\NormalTok{stream}\OperatorTok{);}
\OperatorTok{\}}
\end{Highlighting}
\end{Shaded}

\end{listing}

\chapter{Numerics}\label{numerics}

\begin{summarybox}

The datapath is fp16 end to end with a wide accumulator of fp32 class,
so representable sums come back near exact. Convolution, matrix
multiply, and normalization keep their precision. A cancellation-heavy
step like the transformer down-projection loses it to per-product fp16
input rounding, not to the accumulator, and needs a wider anchor
off-engine. The fp16 ceiling is 65504, and a width-axis slice with a
nonzero begin offset applies a fixed gain of sixteen. A fill above 4094
therefore overflows silently to infinity on the M1 and the A14.
Activation lookup tables are accurate to about half a unit in the last
place. They coerce a NaN to a clamp value and have a small origin bias
that a bit-exact oracle must model.

\end{summarybox}

Computation on the engine is fp16 with a wide accumulator. That
property, stated in chapter \hyperref[what-the-ane-is]{1}, decides which
computations keep their precision on the engine and which lose it.

\section{fp16 datapath}\label{fp16-datapath}

The multiply array is fp16 end to end. An fp16 value decodes from a sign
bit, five-bit exponent, and ten-bit fraction as

\[x = (-1)^s \, 2^{e-15} \left(1 + \frac{m}{2^{10}}\right),\]

with a maximum finite magnitude of \(65504\). Inputs are fp16, weights
are fp16, outputs are fp16. The frontend accepts fp32, int32, and bf16
as type annotations, but the backend does not implement them, so those
annotations do not reach the silicon as wider arithmetic. Bias is the
single exception in the descriptor: it may be held as fp32, but it is
added in the fp16 datapath. The datapath reconstructs compressed weights
to fp16 before they reach the multiplier; it dequantizes an int8, int4,
or palettized weight to fp16 on the way in, and the multiply that
follows is the same fp16 multiply as for an uncompressed weight. The
dequantization is affine, \(w = s \, (q - z)\), with the scale \(s\)
scalar or per-output-channel, and agrees with the documented
conversion-tool quantization relation
\hyperref[ref-applecoremltools]{{[}AppleCoreMLTools{]}}. On the M1
generation the quantization is symmetric, so \(z = 0\) and
\(w = s \, q\); the inverse encode is \(q = \mathrm{round}(x/s) + z\).

The order of rounding in one multiply-accumulate pass is fixed. A weight
is dequantized to fp16, the fp16 multiply runs, the products accumulate
in the wide register, an optional per-channel scale and bias apply in
fp16, an optional activation table applies, and the result stores to
memory as fp16. Two rounding points bracket the accumulation: the inputs
round to fp16 going in, and the output rounds to fp16 going out.

\section{Wide accumulator}\label{wide-accumulator}

The running sum between those two rounding points is not fp16. The input
port rounds tiles to fp16 on the way in and the output port rounds to
fp16 on the way out, but the reduction in between is held in a wide
register of fp32 class on every ANE device. The accumulator width is a
fixed hardware property.

Representable sums thus come back near exact. A reduction of sixteen
thousand ones is bit exact, where a naive fp16 running sum would stall
near two thousand once the partial total passes the spacing of its own
increments. The spacing of fp16 at a partial of magnitude \(p\) is
\(\mathrm{ulp}(p) = 2^{\lfloor \log_2 p \rfloor - 10}\), so once
\(p > 2048\) the spacing exceeds \(1\) and an added unit increment falls
below half a step and is swallowed.

The worked case that shows the accumulator is wider than fp16 yet
supplied by fp16-rounded inputs is the sum of one value of 4096 followed
by 1024 ones. The exact sum is

\[[4096] + [1]\times 1024 = 5120,\]

and the engine returns 5116, between the naive-fp16 result \(4096\) and
the exact total. A naive fp16 running sum returns 4096, because each
added one falls below half the spacing at 4096 and is swallowed. The
engine instead keeps the 1024 ones that fp16 would drop, and the small
deficit from the exact 5120 is the fp16 rounding of the single input
tile that contains the 4096.

The discriminating test is a cancellation triple. A sum of a large
value, its negation, and a one, repeated sixteen times near a magnitude
of four thousand, returns all sixteen ones intact. An fp16 running sum
holding four thousand has a spacing of two there, so each one would
round away. The ones survive, so the accumulator is physically wider
than fp16. The first reduction stage groups input lanes into tiles of
four, fp16-rounded, which then supply the one wide accumulator. Swept
across magnitudes, the survivor count is layout-independent: the same
result whether the one precedes or follows the negation, so the hardware
re-associates over a fixed lane lattice rather than the source order.
\Cref{tbl:c3-cancel-sweep} records the survivor count against the
cancellation magnitude and the fp16 spacing there.

\begin{longtable}[]{@{}
  >{\raggedleft\arraybackslash}p{(\linewidth - 4\tabcolsep) * \real{0.3333}}
  >{\raggedleft\arraybackslash}p{(\linewidth - 4\tabcolsep) * \real{0.3333}}
  >{\raggedleft\arraybackslash}p{(\linewidth - 4\tabcolsep) * \real{0.3333}}@{}}
\caption{The cancellation-threshold survivor sweep, M1/H13 measured,
which fixes the first-stage tile width at
four.}\label{tbl:c3-cancel-sweep}\tabularnewline
\toprule\noalign{}
\begin{minipage}[b]{\linewidth}\raggedleft
\textbf{Cancellation magnitude}
\end{minipage} & \begin{minipage}[b]{\linewidth}\raggedleft
\textbf{fp16 spacing at that magnitude}
\end{minipage} & \begin{minipage}[b]{\linewidth}\raggedleft
\textbf{Survivors of sixteen ones}
\end{minipage} \\
\midrule\noalign{}
\endfirsthead
\toprule\noalign{}
\begin{minipage}[b]{\linewidth}\raggedleft
\textbf{Cancellation magnitude}
\end{minipage} & \begin{minipage}[b]{\linewidth}\raggedleft
\textbf{fp16 spacing at that magnitude}
\end{minipage} & \begin{minipage}[b]{\linewidth}\raggedleft
\textbf{Survivors of sixteen ones}
\end{minipage} \\
\midrule\noalign{}
\endhead
\bottomrule\noalign{}
\endlastfoot
1024 & 1 & 16 \\
\rowcolor{zebra}3000 & 2 & 16 \\
4090 & 2 & 16 \\
\rowcolor{zebra}4096 & 4 & 4 \\
8000 & 4 & 4 \\
\rowcolor{zebra}16000 & 8 & 4 \\
30000 & 16 & 4 \\
\end{longtable}

The survivor count holds at sixteen up to a magnitude of 4090, then
drops to a hard floor of exactly four at 4096 and stays at four out to
30000. The threshold is at 4096 because that is the first magnitude
where the fp16 spacing reaches four, so a one sharing a four-lane tile
with a rounded partial at or above 4096 falls below the in-tile rounding
threshold and vanishes. The flat floor of four is the signature of a
four-lane tile: a three-element triple period beating against a
four-lane tile leaves exactly four unaffected lanes out of sixteen. A
single one added next to one large resident value cannot expose the
tile, because that value forces the whole wide partial onto its own
coarse spacing and the one rounds away regardless of which tile it is
in.

Both behaviors are reproducible on the engine by reducing an fp16
vector. The cancellation probe routes a sum through a matmul against a
ones vector so the reduction is in the wide accumulator, and the slice
trigger exercises the width-axis crop gain (\cref{lst:c3-accum-probe}).

\begin{listing}[H]
\caption{Probes for the wide accumulator’s cancellation behavior and the width-slice saturation that overflows to infinity above the fp16 ceiling.}\label{lst:c3-accum-probe}

\begin{Shaded}
\begin{Highlighting}[]
\CommentTok{\# Pseudocode for the engine\textquotesingle{}s wide accumulator and the width{-}slice saturation.}
\CommentTok{\# Each reduce\_via\_matmul(v) runs on the engine: v is fp16, summed against a}
\CommentTok{\# fp16 ones{-}vector so the reduction accumulates in the wide (fp32{-}class) register;}
\CommentTok{\# the result is read back as fp16.}

\CommentTok{\# (a) Cancellation probe: a large value, its negation, and ones near the threshold.}
\CommentTok{\#     A naive fp16 running sum at 4000 has ulp = 2 and swallows every one.}
\NormalTok{big }\OperatorTok{=} \FloatTok{4000.0}
\NormalTok{v }\OperatorTok{=}\NormalTok{ ([big, }\OperatorTok{{-}}\NormalTok{big, }\FloatTok{1.0}\NormalTok{] }\OperatorTok{*} \DecValTok{16}\NormalTok{)               }\CommentTok{\# 16 ones survive the cancellation}
\ControlFlowTok{assert}\NormalTok{ reduce\_via\_matmul(v) }\OperatorTok{==} \FloatTok{16.0}       \CommentTok{\# engine keeps all 16; fp16 loop {-}\textgreater{} 0}

\CommentTok{\#     The accumulator is wide, but the inputs still round to fp16 going in:}
\NormalTok{v }\OperatorTok{=}\NormalTok{ [}\FloatTok{4096.0}\NormalTok{] }\OperatorTok{+}\NormalTok{ [}\FloatTok{1.0}\NormalTok{] }\OperatorTok{*} \DecValTok{1024}               \CommentTok{\# one tile of 4096, then 1024 ones}
\ControlFlowTok{assert}\NormalTok{ reduce\_via\_matmul(v) }\OperatorTok{==} \FloatTok{5116.0}     \CommentTok{\# between naive{-}fp16 4096 and exact 5120}

\CommentTok{\# (b) Width{-}axis slice saturation: a nonzero begin offset applies a x16 crop gain.}
\CommentTok{\#     The fp16 ceiling is 65504, so 4094*16 == 65504 passes but 4096*16 overflows.}
\KeywordTok{def}\NormalTok{ width\_slice\_with\_offset(value):       }\CommentTok{\# slice begin offset != 0 on the width axis}
    \ControlFlowTok{return}\NormalTok{ value }\OperatorTok{*} \FloatTok{16.0}                   \CommentTok{\# the fixed crop{-}DMA gain, then fp16 store}

\ControlFlowTok{assert}\NormalTok{ width\_slice\_with\_offset(}\FloatTok{4094.0}\NormalTok{) }\OperatorTok{==} \FloatTok{65504.0}          \CommentTok{\# at the ceiling, finite}
\ControlFlowTok{assert}\NormalTok{ width\_slice\_with\_offset(}\FloatTok{4096.0}\NormalTok{) }\OperatorTok{==} \BuiltInTok{float}\NormalTok{(}\StringTok{"inf"}\NormalTok{)     }\CommentTok{\# 65536 \textgreater{} 65504 {-}\textgreater{} inf}
\end{Highlighting}
\end{Shaded}

\end{listing}

\section{Compressed weights and the streaming
gate}\label{compressed-weights-and-the-streaming-gate}

The datapath reconstructs a compressed weight to fp16 before it reaches
the multiply array, so the arithmetic is fp16 regardless of the storage
format. What differs by format and by chip is whether the compressed
bytes stream to the engine compressed and decompress on the way in, or
fold to dense fp16 in memory first. A format that streams moves fewer
bytes across the DRAM boundary; a format that folds yields the storage
saving but not the bandwidth. Which formats stream and which fold varies
across three chip generations (\cref{tbl:c3-weight-stream}).

\begin{longtable}[]{@{}llll@{}}
\caption{Which compressed-weight formats stream natively versus fold to
dense fp16, by generation, M1 measured with A14 and M5 rows from the
per-format gate.}\label{tbl:c3-weight-stream}\tabularnewline
\toprule\noalign{}
\textbf{Format} & \textbf{M1 / A13} & \textbf{A14 / M2} & \textbf{A16 /
M5} \\
\midrule\noalign{}
\endfirsthead
\toprule\noalign{}
\textbf{Format} & \textbf{M1 / A13} & \textbf{A14 / M2} & \textbf{A16 /
M5} \\
\midrule\noalign{}
\endhead
\bottomrule\noalign{}
\endlastfoot
int4 palette lookup table & stream & stream & stream \\
\rowcolor{zebra}int8 affine & fold & stream & stream \\
sparse & stream & stream & stream \\
\rowcolor{zebra}blockwise & fold & fold & stream \\
\end{longtable}

On the M1 the int4 palette form streams natively, measured at about 2.37
times the bandwidth of the dense weight, and a sparse weight with at
least half its values zero streams as a one-bit keep-mask with the
packed fp16 nonzeros. The int8 affine and blockwise forms fold to dense
fp16 before the data-movement step. The fold expands the int8 weight to
a dense fp16 constant in DRAM before the multiply, so the bytes that
cross the DRAM boundary are full-width fp16 and the layer gets no
bandwidth gain on the M1. The int8 fold is a stored-size saving only on
the M1. The weight is half the size on disk, but it is reconstructed to
fp16 in DRAM before the data-movement step, so a weight-streaming-bound
matmul moves the same bytes as fp16 and runs at the fp16 latency. The
int8 form first streams as int8 on the A14 and M2 generation, where it
is dispatched as int8 and dequantized at the multiply port so it moves
half the bytes of fp16. A bandwidth-bound matmul reaches about 0.52
times the fp16 latency at a weight of 8192 by 8192. The
compressed-weight quantization adds error only at the input rounding: an
int8 conv weight tracks the fp16 result at a cosine near 1.0, with a
relative error near 0.6 percent against an fp32 reference, against about
0.02 percent for fp16.

\section{Where precision holds and where it is
lost}\label{where-precision-holds-and-where-it-is-lost}

Convolution, matrix multiply, and normalization keep their precision
when their partial sums stay in fp16 range and their results are
representable. The reduction happens in the wide accumulator, and the
only quantization is the fp16 rounding of the inputs and the output.
Vision, audio, and encoder workloads are in this regime, and run on the
engine without a precision penalty.

Precision is lost on cancellation-heavy steps. The transformer decoder
down-projection is the case that fails: a large positive and a large
negative contribution nearly cancel, and the result is a small
difference between two large numbers. The loss does not come from the
accumulator, which is wide enough to hold the partial. It comes from the
per-product fp16 rounding of the inputs and weights before they enter
the accumulator. Once the operands are quantized to fp16, the
cancellation amplifies that quantization into the result. There is no
fp16-safe form for such a step on the engine; the anchor has to be
computed wider, on the CPU or the GPU.

\section{Activation functions}\label{activation-functions}

Lookup tables evaluate the nonlinear activations. Identity and plain
ReLU are not table-driven, since ReLU is a max, but sigmoid, tanh, gelu,
swish, erf, exp, and the rest route through a piecewise-linear table of
fixed knots. The table is a 33-knot piecewise-linear curve, not a dense
sample grid. The input maps affinely onto the 32 segments between 33
knots, the bracketing segment evaluates as a slope times the input plus
an intercept in fp16, and the value clamps to the end-knot asymptote
past the table domain. Accuracy comes from the piecewise fit and the
per-function domain. A user sigmoid does not lower to the plain sigmoid
table. It lowers to a high-precision sigmoid table by default, which
pushes the domain to a wider range with a finer subdivision near the
linear region, so a sigmoid-heavy or attention-gate-heavy model stays
accurate under fp16.

On the standard set, measured on device, the worst absolute error is at
the level of the fp16 storage floor: sigmoid 0.0034, tanh 0.0017, gelu
0.0059, each under 0.4 percent of the function range. The on-device
value matches the fp16-rounded exact function to about half a unit in
the last place. The table adds nothing measurable on top of fp16 storage
rounding. The exceptions are sin, cos, and atan, which have up to about
0.04 to 0.12 absolute error near the seams of their argument reduction;
a model that evaluates trig directly near a magnitude of pi should fold
the range upstream.

\section{Edge behavior a correctness oracle must
model}\label{edge-behavior-a-correctness-oracle-must-model}

Several edge behaviors of the fp16 datapath depart from a host IEEE
reference, and a model that reproduces engine results bit for bit has to
encode them. The engine coerces a NaN to positive infinity at the input
boundary, and never produces a NaN anywhere. A NaN sent in and echoed
through the identity \(x + 0\) returns positive infinity, with the same
bits as sending positive infinity directly, and every downstream op then
behaves as if the value had been positive infinity. A NaN into relu
returns infinity, a NaN into sigmoid or tanh returns 1.0, a NaN into erf
returns 1.0, and a NaN into exp returns infinity. A NaN into the maximum
of two values returns infinity, and a softmax of a lane holding a NaN
puts all the mass on that lane. The one case where a NaN is not
bit-identical to positive infinity is the variance reduction inside
layer normalization, where the NaN enters as a large finite magnitude
rather than as literal infinity. Rms normalization gives identical
results for a NaN input and an infinity input. An upstream NaN through
any gate leaves the engine as a finite or infinite value and does not
surface as a NaN to downstream code.

The engine flushes to positive zero all the indeterminate forms that
produce a NaN under IEEE. The infinity minus infinity case returns
positive zero, the zero times infinity case returns positive zero,
\(\mathrm{sqrt}(-1)\) returns positive zero, \(\mathrm{rsqrt}(-1)\)
returns positive zero, and \(\log(-1)\) returns positive zero.

The engine preserves denormals elementwise but flushes them inside the
multiply-accumulate. An fp16 denormal down to \(2^{-24}\) (about
\(5.96 \times 10^{-8}\)) echoes bit-exactly through \(x + 0\) and
\(x \times 1\) on the elementwise path, and scales correctly, so the
common assumption that the engine flushes denormals globally is false
for that path. A pair of denormals summed inside a matmul, with a
representable denormal total, returns positive zero, so the
flush-to-zero is a property of the accumulator stage and not of the
datapath as a whole. The M5 accumulator instead preserves denormals, so
a denormal input and a product of two denormals both survive a matmul
reduction unflushed. The flush is an M1-generation property, and the M2
through M4 parts between them were not measured, so the generation at
which it changed is unknown.

Signed zero loses its sign before a reciprocal or a reciprocal square
root. The reciprocal of negative zero returns positive infinity where
IEEE returns negative infinity, and the reciprocal square root of
negative zero returns positive infinity where IEEE returns negative
infinity, while \(\mathrm{sqrt}\) of negative zero stays positive zero
per IEEE. A negative zero echoes as positive zero through \(x + 0\), so
the engine drops the sign bit of a zero before the reciprocal path.

A few activation tables collapse a large input: softplus and softsign of
positive infinity return positive zero, where the values should be
positive infinity and a unit magnitude. The logarithm of positive zero
returns a finite sentinel of \(-45440\) rather than negative infinity.
These are properties of the table approximations, distinct from the NaN
coercion above.

Softmax subtracts a hardware maximum before the exponential, so it does
not overflow even when a raw exponential would. A softmax of
\([1000, 1, 2, 3]\) returns \([1, 0, 0, 0]\) and matches a wide
reference, despite \(\exp(1000)\) being far past the fp16 range, and a
softmax of four equal values returns four quarters. A bare \(\exp\) with
no max-subtraction overflows to infinity at an input near
\(11.094 = \ln(65504)\), so the stable route is the fused softmax rather
than a hand-rolled exponential over a sum.

Rounding on the fp16 output grid is round half to even. A tie at the
midpoint between two representable fp16 values rounds to the value with
an even last bit, not away from zero. A partial of \(2049\) at a grid
spacing of \(2\) rounds to \(2048\), the even neighbor, and a trailing
half above the \(2048\) threshold rounds the \(L + 0.5\) accumulation to
even. The M5 returns \(2050\) for this case, rounding up where the M1
rounds to even. Whether the wide accumulator presents a value just above
\(2049\) rather than an exact tie, or the rounding differs by
generation, is unresolved.

Some activation tables have a small constant bias at the origin. The
decoded gelu table returns \(-0.000543\) at \(x = 0\) where the exact
gelu is \(0\), and the swish table returns \(-0.001259\) at \(x = 0\)
where the exact swish is \(0\). The bias is below the fp16 storage floor
and does not affect the per-op accuracy figures, but a bit-exact oracle
has to hold it.

\section{Saturation hazard}\label{saturation-hazard}

The fp16 maximum finite value is 65504. The compute datapath has no fp32
margin, so a value above that saturates to infinity silently.

The multiply-accumulate output stage saturates earlier than the storage
format, at exactly \(2^{15} = 32768\), just above half the fp16 ceiling.
This is a different axis from the width-slice gain below: it is a
property of the accumulator output port, and it fires on matrix
multiply, linear, and any convolution that accumulates two or more taps,
whatever the number of accumulation terms. The threshold is pinned to
the bit: the largest fp16 value below \(2^{15}\), which is \(32752\),
passes through a linear, and the next fp16 value, \(32768\), returns
infinity. It holds for \(K = 1\), for \(K = 2\), and for a two-channel
convolution, so it tracks the would-be output magnitude at the
accumulator port. An interior partial that exceeds \(2^{15}\) overflows
to infinity even when a later cancellation would have brought the final
result back into range. The paths that drive a dedicated reduction or a
single elementwise multiply have the full fp16 range instead. A
reduce-sum rounds to 65504 first and then overflows, an elementwise
multiply overflows at the true fp16 limit of 65504, and a pointwise
one-by-one convolution with a single input channel passes a fill of
60000. This earlier accumulator ceiling is consistent with the
multiply-accumulate datapath having about one bit less margin than the
fp16 storage format on the M1, so the guard for a matrix multiply or a
multi-tap convolution is to keep the output magnitude below \(2^{15}\).

The slice path has a second saturation. A width-axis slice with a
nonzero begin offset routes through a crop DMA that applies a fixed gain
of sixteen. A value at or below 4094 passes through bit exact, because
\(4094 \times 16 = 65504\) is the fp16 ceiling. A value of 4095 or above
becomes infinity, because it rounds to 4096 on the fp16 grid and
\(4096 \times 16 = 65536 > 65504\), past the ceiling. The control case,
a slice with a zero begin offset, is free of the saturation at the same
fill values, so the trigger is the nonzero width-axis offset and not the
magnitude alone. The hazard is on the width axis only: a nonzero begin
offset on the height, channel, or batch axis stays finite at the same
fill values. The saturation is measured on both the M1 generation (H13)
and the A14 generation (H14), where a fill at width offset 4094 stays
finite (\(4094 \times 16 = 65504\)) and a fill at 4096 overflows to
infinity (\(4096 \times 16 = 65536\)). The non-saturating route arrives
on the A15 generation and later, so the hazard is not fixed by the
immediate next family. The guard is to avoid nonzero last-axis begin
offsets on a width slice, or to route the offset onto a different axis.

\section{Determinism}\label{determinism}

For a fixed graph and a fixed input the M1 engine is bit-deterministic:
the raw fp16 output bytes are identical across reruns, across an
independent recompile, and across a fresh process. The lowered program
fixes the accumulation order, so there is no run-to-run drift to round
away.

Re-executing one compiled program on the same input returns the same
fp16 bytes every time, measured at zero units in the last place over 200
repeats for a matrix multiply, two-layer convolution, and long
reduction. The same holds over 50 repeats for softmax, reduce-mean, and
a large matrix multiply. Compiling the same graph twice at optimization
level zero, the byte-identical lowering path, produces two programs
whose outputs agree to the bit, including a genuinely tiled
\([128, 1024] @ [1024, 1024]\) matrix multiply. Running the same graph
in a fresh subprocess, with a new dispatch client, returns the same
output digest as the in-process run, so daemon and process state do not
perturb the result.

The result is also independent of batch size and batch position. A row
computed alone is bit-identical to the same row computed inside a batch,
and the same row placed at every position of a batch of sixteen gives
sixteen identical outputs. Row zero is invariant across batch sizes of
one, two, eight, thirty-two, and one hundred twenty-eight. The engine
computes each batch element identically regardless of its neighbors.

The one thing that changes the bits is changing the math. Writing a
graph with a different association order, \((a + b) + c\) against
\(a + (b + c)\), gives different fp16 rounding. About 31 percent of the
elements differ, by about one fp16 unit in the last place at that
magnitude, with a maximum absolute difference near
\(7.8 \times 10^{-3}\). Each ordering is itself perfectly reproducible,
so this is ordinary fp16 non-associativity of the graph as written and
not hardware nondeterminism. Two high-level ops that lower to the same
kernel agree to the bit. A dot product written as a matrix multiply and
the same dot product written as an elementwise multiply followed by a
reduce-sum return identical bytes here, because the compiler lowered
both to the same accumulation order. Reassociation bites only when the
graph dictates a different order.

\section{Keeping a graph inside the fp16
envelope}\label{keeping-a-graph-inside-the-fp16-envelope}

Two numeric decisions belong before a graph is compiled. A developer
anchors a cancellation-heavy reduction off-engine, on the CPU or the
GPU. A width-axis slice avoids a nonzero begin offset, because the fixed
gain of sixteen overflows a fill above 4094 to infinity.

The cancellation-heavy step routes to a wider unit, and the rest stays
on the engine, with the three rules worked through one graph in
\cref{lst:c3-safe-graph}.

\par\addvspace{\medskipamount}\noindent\captionof{listing}{The three rules that keep a graph inside the fp16 envelope before compile.}\label{lst:c3-safe-graph}\nopagebreak\par\vspace{1pt}

\begin{Shaded}
\begin{Highlighting}[]
\CommentTok{\# Keep a graph numerically safe BEFORE compiling it, by applying three rules}
\CommentTok{\# to every reduction and every cropped tile in the graph.}

\CommentTok{\# fp16 limits the whole datapath uses:}
\NormalTok{fp16\_max         }\OperatorTok{=} \DecValTok{65504}        \CommentTok{\# largest finite fp16 value; above this {-}\textgreater{} infinity}
\NormalTok{accum\_out\_max    }\OperatorTok{=} \DecValTok{32768}        \CommentTok{\# multiply{-}accumulate output port saturates here}
\NormalTok{width\_slice\_gain }\OperatorTok{=} \DecValTok{16}           \CommentTok{\# a width{-}axis crop with a nonzero begin offset}
                                \CommentTok{\#   multiplies its values by this fixed gain}

\CommentTok{\# RULE 1: route a reduction through a matmul against a ones{-}vector, so the}
\CommentTok{\#         sum accumulates in the wide (fp32{-}class) accumulator and stays near exact.}
\CommentTok{\#         A plain elementwise running sum rounds at every step in narrow fp16.}
\NormalTok{function safe\_reduce\_sum(vector v):                  }\CommentTok{\# v is fp16}
\NormalTok{    ones }\OperatorTok{=}\NormalTok{ vector\_of\_ones(length(v))                 }\CommentTok{\# fp16, same length as v}
    \ControlFlowTok{return}\NormalTok{ matmul(v, ones)                           }\CommentTok{\# one dot product, wide accumulator}

\CommentTok{\# RULE 2: a cancellation{-}heavy step (a large positive nearly canceling a large}
\CommentTok{\#         negative) has no fp16{-}safe form on the engine, because the operands}
\CommentTok{\#         were already rounded to fp16 before the subtract. Scale it up so the}
\CommentTok{\#         small difference clears the fp16 grid, OR split it out to a wider}
\CommentTok{\#         unit (CPU or GPU) and feed the result back as a graph input.}
\NormalTok{function place\_cancellation\_step(step):}
    \ControlFlowTok{if}\NormalTok{ is\_cancellation\_heavy(step):}
        \ControlFlowTok{return}\NormalTok{ compute\_off\_engine(step)              }\CommentTok{\# wider anchor, fed back as input}
    \ControlFlowTok{else}\NormalTok{:}
        \ControlFlowTok{return}\NormalTok{ keep\_on\_engine(step)                  }\CommentTok{\# representable sums stay near exact}

\CommentTok{\# RULE 3: keep every tile value under the limit that applies to its path,}
\CommentTok{\#         so nothing saturates silently to infinity mid{-}graph.}
\NormalTok{function check\_tile\_value(value, path):}
    \ControlFlowTok{if}\NormalTok{ path }\OperatorTok{==} \StringTok{"matmul\_or\_multitap\_conv"}\NormalTok{:}
\NormalTok{        require value }\OperatorTok{\textless{}=}\NormalTok{ accum\_out\_max               }\CommentTok{\# output{-}port ceiling, \textasciitilde{}half fp16\_max}
    \ControlFlowTok{if}\NormalTok{ path }\OperatorTok{==} \StringTok{"width\_slice\_with\_nonzero\_offset"}\NormalTok{:}
\NormalTok{        require value }\OperatorTok{*}\NormalTok{ width\_slice\_gain }\OperatorTok{\textless{}=}\NormalTok{ fp16\_max }\CommentTok{\# 4094*16 == 65504 passes; 4096*16 overflows}
\NormalTok{    otherwise:}
\NormalTok{        require value }\OperatorTok{\textless{}=}\NormalTok{ fp16\_max                    }\CommentTok{\# plain elementwise / single reduction}

\CommentTok{\# Assemble the safe graph: representable conv/matmul/norm stay on the engine,}
\CommentTok{\# reductions go through safe\_reduce\_sum, the cancellation step is anchored wide.}
\NormalTok{graph G:}
    \BuiltInTok{input}\NormalTok{  x                                         }\CommentTok{\# fp16 activations}
\NormalTok{    feats  }\OperatorTok{=}\NormalTok{ conv(x, weights }\OperatorTok{=}\NormalTok{ W)                    }\CommentTok{\# representable sums, stays on engine}
\NormalTok{    pooled }\OperatorTok{=}\NormalTok{ safe\_reduce\_sum(feats)                  }\CommentTok{\# RULE 1: wide{-}accumulator reduction}
\NormalTok{    output pooled}
\NormalTok{program P }\OperatorTok{=} \BuiltInTok{compile}\NormalTok{(G, target }\OperatorTok{=}\NormalTok{ H13)                 }\CommentTok{\# fp16 datapath, wide accumulator}
\NormalTok{result }\OperatorTok{=}\NormalTok{ dispatch(P, features)}
\CommentTok{\# The cancellation{-}heavy down{-}projection (RULE 2) is computed off{-}engine and}
\CommentTok{\# fed back as an input, never lowered into this graph.}
\end{Highlighting}
\end{Shaded}

\addvspace{\medskipamount}

\section{Reference: fp16 numeric
constants}\label{reference-fp16-numeric-constants}

\Cref{tbl:c3-constants} collects the fp16 numeric constants, the
saturation thresholds, and the activation-table error figures.

\begin{longtable}[]{@{}
  >{\raggedright\arraybackslash}p{(\linewidth - 2\tabcolsep) * \real{0.4286}}
  >{\raggedleft\arraybackslash}p{(\linewidth - 2\tabcolsep) * \real{0.5714}}@{}}
\caption{The fp16 numeric constants, M1/H13 measured, with the
saturation reproduced on
A14/H14.}\label{tbl:c3-constants}\tabularnewline
\toprule\noalign{}
\begin{minipage}[b]{\linewidth}\raggedright
\textbf{Constant}
\end{minipage} & \begin{minipage}[b]{\linewidth}\raggedleft
\textbf{Value}
\end{minipage} \\
\midrule\noalign{}
\endfirsthead
\toprule\noalign{}
\begin{minipage}[b]{\linewidth}\raggedright
\textbf{Constant}
\end{minipage} & \begin{minipage}[b]{\linewidth}\raggedleft
\textbf{Value}
\end{minipage} \\
\midrule\noalign{}
\endhead
\bottomrule\noalign{}
\endlastfoot
fp16 maximum finite magnitude & 65504 \\
\rowcolor{zebra}Multiply-accumulate output ceiling & 32768 \\
Width-slice crop-DMA gain & 16 \\
\rowcolor{zebra}Width-slice finite fill ceiling & 4094 \\
Width-slice overflow fill & 4096 \\
\rowcolor{zebra}Wide-accumulator bit-exact reduction & 16000 ones \\
Worked sum, \([4096] + [1] \times 1024\) & 5116 \\
\rowcolor{zebra}First reduction-stage tile width & 4 lanes \\
Sigmoid worst absolute error & 0.0034 \\
\rowcolor{zebra}Tanh worst absolute error & 0.0017 \\
Gelu worst absolute error & 0.0059 \\
\rowcolor{zebra}Gelu origin bias at \(x = 0\) & \(-0.000543\) \\
Swish origin bias at \(x = 0\) & \(-0.001259\) \\
\rowcolor{zebra}Trig seam absolute error (sin, cos, atan) & 0.04 to
0.12 \\
Activation table knot count & 33 \\
\rowcolor{zebra}Sigmoid table domain clamp & \([-9.938,\ +8.320]\) \\
Exp input where output first reaches infinity & 11.094 \\
\rowcolor{zebra}Square input where output first reaches infinity &
256 \\
int4 palette native stream speedup, M1 & 2.37x \\
\rowcolor{zebra}int8 fold on M1 (stored-size saving, no stream gain) &
1.0x fp16 latency \\
int8 weight-stream latency, A14/M2 (8192 weight) & 0.52x fp16 \\
\end{longtable}

\chapter{Capability surface}\label{capability-surface}

\begin{summarybox}

The engine natively runs what on-device perception networks are built
from: convolution, matrix multiply, fused attention, the normalizations,
the activations, and data movement, all from the M1 onward. Another set
has no hardware path on any family, among them reduce\_prod, the scatter
family, and the recurrent cells. A network that needs one of those
rewrites it. Some operations are family-gated: the texture-engine
sampler arrives on the A14, sin and cos on the A15. A capability
attested at one layer is not a reachable operation. Three-dimensional
convolution has a capability byte yet fails backend lowering on every
device, so only a compile-and-run confirms an operation.

\end{summarybox}

What the engine will run is a fixed vocabulary of tensor operations. The
full operation-by-device table is Appendix
\hyperref[appendix-a.-operation-by-device-matrix]{A}; this chapter gives
its shape and the rules for fitting work to the engine.

\section{What fits}\label{what-fits}

The engine runs the operations that on-device perception networks are
built from. Convolution and matrix multiply are at the center, the
normalizations and activations around them, and data-movement operations
between the compute steps.

Two-dimensional convolution is native, including the transpose
(deconvolution) and the dilated, depthwise, and grouped forms, with
kernel and tensor sizes bounded by the per-chip limits in Appendix
\hyperref[appendix-a.-operation-by-device-matrix]{A}. The compiler
selects a Winograd path for eligible 3x3 stride-1 convolutions without
any caller action. Matrix multiply and fully-connected layers are native
and fold into the convolution datapath when the operand fits the on-chip
working set, tiling when it does not. Attention runs as a fused
operation: scaled dot-product attention runs on the matrix-multiply and
softmax path, native on every family from the M1 onward and not gated
behind the texture engine.

The common normalizations all run: layer, instance, group, and L2
normalization, plus batch normalization folded to an affine at
inference. Pooling (average, max, and L2) is native, as are the
elementwise arithmetic and comparison operations. The activation set
covers ReLU and its variants, sigmoid, tanh, gelu, swish, softmax, erf,
exp, and log, which evaluate through the lookup tables described in
chapter \hyperref[numerics]{3}; that table is a programmable
piecewise-linear curve, and chapter
\hyperref[hidden-layers-and-direct-netplist-authoring]{26} covers
reaching an arbitrary pointwise function through it. Data movement is
native for reshape, flatten, expand, squeeze, transpose, concat, split,
stack, constant pad, and slice, all descriptor edits or DMA operations.
A reduction or transpose over a large axis switches to a tiled route at
a per-chip threshold, at no change to the result.

The backend represents these operations in its own dialect,
\texttt{anec.*}, one level below the frontend intermediate language.
\Cref{lst:c4-anec-dialect} gives the signatures for the core operations.

\begin{listing}[H]
\caption{The backend dialect signatures for convolution, matrix multiply, and normalization, with the operands and attributes the compiler expects.}\label{lst:c4-anec-dialect}

\begin{Shaded}
\begin{Highlighting}[]
\CommentTok{// 2D conv: weights and bias fold in; rank 4 or 5; M1 kernel at most 29x29 (13x13 fp16).}
\FunctionTok{anec.convolution}\NormalTok{(}\VariableTok{\%input}\NormalTok{) \{}
\NormalTok{    strides, dilation\_rates, groups, kernel\_sizes,}
\NormalTok{    explicit\_padding, padding\_style, weights\_layout}
\NormalTok{\}}

\CommentTok{// matrix multiply: contraction direction is set by the transpose flags, not by operand order.}
\CommentTok{// depth D must be 1 on both operands ("depth \textgreater{} 1 is not supported for MatMult inputs").}
\FunctionTok{anec.matmul}\NormalTok{(}\VariableTok{\%lhs}\NormalTok{, }\VariableTok{\%rhs}\NormalTok{) \{ transpose\_lhs, transpose\_rhs \}}

\CommentTok{// layer and group normalization are one atom: group\_norm is layer\_norm with num\_groups \textgreater{} 1.}
\CommentTok{// channels must be divisible by num\_groups; the output format must be Float.}
\FunctionTok{anec.layer\_norm}\NormalTok{(}\VariableTok{\%input}\NormalTok{) \{ groups, epsilon, axis \}   }\CommentTok{// gamma and beta fold in}
\end{Highlighting}
\end{Shaded}

\end{listing}

The tensor layout the compiler reasons about is five-dimensional,
\texttt{N} batch, \texttt{D} depth, \texttt{C} channel, \texttt{H},
\texttt{W}, with a maximum rank of five, and transforms act only on the
last four dimensions.

\section{Hard limits}\label{hard-limits}

Some operations have no hardware path on any current family. They do not
run on the engine and must be computed off-engine or reformulated. The
confirmed cases, unsupported on every family from the M1 through the M5,
are: reduce\_prod, the scatter family (scatter, scatter along axis,
scatter ND), mod, one\_hot, non\_zero, band\_part, reverse\_sequence,
shape, sliding\_windows, the logical and, or, and xor operations, the
recurrent gru, lstm, and rnn cells, the inverse and hyperbolic
trigonometric functions, and the random-sampling operations other than
the uniform generator. A network that needs one of these rewrites it: a
product reduction through log-sum-exp, logical-and through a minimum,
recurrent cell through an unrolled fixed-trip-count graph.

A second class is family-gated, and Appendix
\hyperref[appendix-a.-operation-by-device-matrix]{A} gives the chip that
enables each. Gather on the M1 takes a software path valid only for a
batch of one, depth of one, and three-element index channel, and is
rejected outside it. The texture-engine operations (resize as a hardware
sampler, crop-resize, resample, affine) are absent on the M1 and arrive
on the A14 generation. Trigonometric sin and cos are native only from
the A15 generation; the M1 and the A14 reject them and decompose on the
host.

\Cref{tbl:c4-op-classes} groups the operation classes by status.

\begin{longtable}[]{@{}
  >{\raggedright\arraybackslash}p{(\linewidth - 4\tabcolsep) * \real{0.3333}}
  >{\raggedright\arraybackslash}p{(\linewidth - 4\tabcolsep) * \real{0.3333}}
  >{\raggedright\arraybackslash}p{(\linewidth - 4\tabcolsep) * \real{0.3333}}@{}}
\caption{Operation classes by status, from native on every family
through limited or family-gated paths to those with no hardware
path.}\label{tbl:c4-op-classes}\tabularnewline
\toprule\noalign{}
\begin{minipage}[b]{\linewidth}\raggedright
\textbf{Class}
\end{minipage} & \begin{minipage}[b]{\linewidth}\raggedright
\textbf{Examples}
\end{minipage} & \begin{minipage}[b]{\linewidth}\raggedright
\textbf{Status}
\end{minipage} \\
\midrule\noalign{}
\endfirsthead
\toprule\noalign{}
\begin{minipage}[b]{\linewidth}\raggedright
\textbf{Class}
\end{minipage} & \begin{minipage}[b]{\linewidth}\raggedright
\textbf{Examples}
\end{minipage} & \begin{minipage}[b]{\linewidth}\raggedright
\textbf{Status}
\end{minipage} \\
\midrule\noalign{}
\endhead
\bottomrule\noalign{}
\endlastfoot
Convolution & 2D, transpose, dilated, depthwise, grouped & Native, M1
onward \\
\rowcolor{zebra}Matrix multiply, fully-connected & \texttt{matmul},
\texttt{linear} & Native, M1 onward \\
Fused attention & scaled dot-product attention & Native, M1 onward \\
\rowcolor{zebra}Normalization & layer, instance, group, L2, batch-fold &
Native, M1 onward \\
Pooling & average, max, L2 & Native, M1 onward \\
\rowcolor{zebra}Elementwise, activation & add, mul, ReLU, sigmoid, tanh,
gelu, swish, softmax & Native, M1 onward \\
Data movement & reshape, transpose, concat, split, pad, slice & Native,
M1 onward \\
\rowcolor{zebra}Product reduction & \texttt{reduce\_prod} & No path on
any family \\
\end{longtable}

\section{Type limits}\label{type-limits}

The datapath is fp16, as chapter \hyperref[numerics]{3} shows. An fp32,
int32, or bf16 annotation is accepted at the frontend and implemented
nowhere below it, so it never reaches the silicon as wider arithmetic.
The compiler rejects a cast to int32 on the M1, and bf16 is not usable
as a program input or output dtype. Requesting a wider type gains no
precision: the multiply, activation tables, and output port are fp16
regardless, and only the accumulator is wide.

\begin{figure}[H]\centering
\includegraphics[width=4.127in]{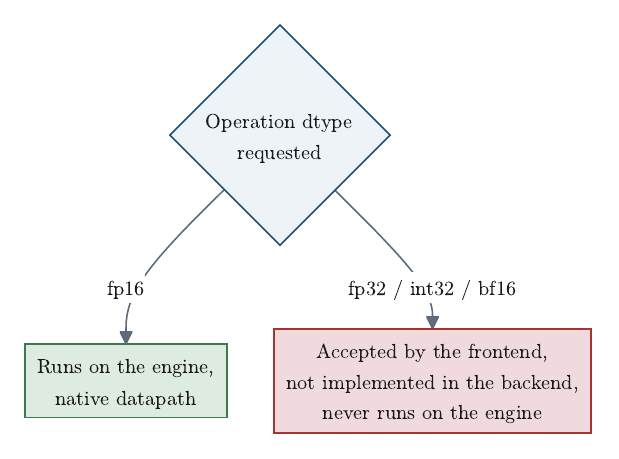}
\caption{How a requested operation dtype routes on the engine.}
\label{fig:c4-dtype-route}
\end{figure}

The fp16 and wider-type paths part by requested dtype
(\cref{fig:c4-dtype-route}).

\section{Attested is not reachable}\label{attested-is-not-reachable}

A hardware capability bit, or an operation the compiler appears to
accept, does not guarantee a reachable, correct operation on the direct
path. Capability is attested at one layer and must be confirmed at the
layer that runs the work.

The case that fixes the rule is three-dimensional convolution. It is
advertised by a hardware capability byte and recognized by the compiler
frontend, yet it fails backend lowering on every device mask: the
operation exists in the attestation and does not run. The gap appears in
the other direction too. On the M1 the top-k, sort, and dynamic-slice
validators are all callable, yet the code generator rejects all three. A
capability advertised in a table, recognized by a frontend, or validated
by a checker is a claim about one layer; only a compile-and-run on the
target confirms the operation at the layer that executes it.

The guide reports a reachable surface rather than the advertised one.
Every operation marked native in Appendix
\hyperref[appendix-a.-operation-by-device-matrix]{A} was compiled and
run on the M1. The advertised surface is the larger, and the difference
is the operations that pass an earlier check and fail at code generation
or backend lowering. This account extends and partly corrects the
documented convertible operation set, which lists operations the public
converter accepts rather than the operations that lower to the engine
\hyperref[ref-applecoremltools]{{[}AppleCoreMLTools{]}}.

\section{Two compile routes}\label{two-compile-routes}

The engine has two compile routes with different capability gates, and
an operation's availability can differ between them on the same chip.
The intermediate-language route is gated by the per-operation family
floor: an operation is native when the chip family is at or above the
operation's floor, and below the floor the compiler decomposes or
rejects it. The bridge route authors a fused layer directly and is gated
by the hardware-abstraction-layer feature bytes rather than the family
floor (\cref{tbl:c4-gate-bytes}). The clear example is the whole-tensor
argument-maximum. On the intermediate-language route it is gated to the
A15 generation and is rejected on the M1, yet on the bridge route it
runs correctly on the M1, because the feature byte that gates it is
already set on the A13. The same structure explains the other direction.
Native sin and cos have no bridge route, so they stay on the
intermediate-language route and are rejected on the M1, while the rank
and sort operations have a bridge route that is rejected at code
generation on the M1.

\begin{longtable}[]{@{}llllll@{}}
\caption{The hardware-abstraction-layer feature bytes that gate the
bridge-route capabilities, with the per-generation value of
each.}\label{tbl:c4-gate-bytes}\tabularnewline
\toprule\noalign{}
\textbf{Capability} & \textbf{Gate byte} & \textbf{older} & \textbf{A13
/ M1} & \textbf{A14} & \textbf{A15} \\
\midrule\noalign{}
\endfirsthead
\toprule\noalign{}
\textbf{Capability} & \textbf{Gate byte} & \textbf{older} & \textbf{A13
/ M1} & \textbf{A14} & \textbf{A15} \\
\midrule\noalign{}
\endhead
\bottomrule\noalign{}
\endlastfoot
softmax & \texttt{0x815} & 0 & 1 & 1 & 1 \\
\rowcolor{zebra}instance normalization & \texttt{0x816} & 0 & 1 & 1 &
1 \\
argument-maximum, hardware form & \texttt{0x4f2} & 0 & 1 & 1 & 1 \\
\rowcolor{zebra}square-after-reduction fusion & \texttt{0x494} & 0 & 0 &
1 & 1 \\
texture engine & \texttt{0x81d} & 0 & 0 & 1 & 1 \\
\rowcolor{zebra}dropout and random & \texttt{0x4a9} & 0 & 0 & 0 & 1 \\
kernel-memory streaming select & \texttt{0x48f} & 0 & 1 & 1 & 1 \\
\end{longtable}

\section{Confirming an operation on the
target}\label{confirming-an-operation-on-the-target}

Because attestation is not reachability, an operation is confirmed by
compiling the graph to the chip target and running it. A graph that
lowers and runs on the target is reachable there; a graph that the
compiler rejects names the operation and the layer that refused it.

The graph compiles against the target and runs before an operation is
relied on, one operation at a time (\cref{lst:c4-confirm-op}).

\begin{listing}[H]
\caption{Confirming one operation on a target by compiling a single-op graph and reading the compiler’s verdict.}\label{lst:c4-confirm-op}

\begin{Shaded}
\begin{Highlighting}[]
\CommentTok{\# Confirm one operation on a target by compiling a graph that contains ONLY}
\CommentTok{\# that operation, then reading the compiler\textquotesingle{}s verdict. Do not trust a}
\CommentTok{\# capability byte: a byte can be set for an op that still fails to lower.}

\CommentTok{\# Build the smallest legal graph that exercises the operation under test.}
\NormalTok{function one\_op\_graph(operation, target):}
\NormalTok{    graph G:}
        \BuiltInTok{input}\NormalTok{  x                                  }\CommentTok{\# a dummy input of a legal shape}
\NormalTok{        output operation(x)                       }\CommentTok{\# the single op we want to confirm}
    \ControlFlowTok{return}\NormalTok{ G}

\CommentTok{\# Ask the compiler to lower the graph against the target. Two outcomes only:}
\CommentTok{\#   {-} it lowers and loads        {-}\textgreater{} the op is REACHABLE on this target}
\CommentTok{\#   {-} the compiler rejects it    {-}\textgreater{} it names the op and the layer that refused}
\NormalTok{function confirm\_op(operation, target):}
\NormalTok{    G }\OperatorTok{=}\NormalTok{ one\_op\_graph(operation, target)}
    \ControlFlowTok{try}\NormalTok{:}
\NormalTok{        P }\OperatorTok{=} \BuiltInTok{compile}\NormalTok{(G, target }\OperatorTok{=}\NormalTok{ target)           }\CommentTok{\# lower + load against this chip}
        \ControlFlowTok{return}\NormalTok{ NATIVE                             }\CommentTok{\# reached the silicon: reachable}
\NormalTok{    catch reject }\ImportTok{as}\NormalTok{ r:}
        \CommentTok{\# r holds the rejecting layer and message, e.g.}
        \CommentTok{\#   "Some ops are not supported on any of the specified backends"  (no path)}
        \CommentTok{\#   "\textless{}op\textgreater{} requires family \textgreater{}= N"                                    (family{-}gated)}
\NormalTok{        report r.layer, r.message                 }\CommentTok{\# the reject string IS the signal}
        \ControlFlowTok{return}\NormalTok{ REJECTED}

\CommentTok{\# Worked checks against the M1 target (family H13):}
\NormalTok{status\_conv }\OperatorTok{=}\NormalTok{ confirm\_op(conv\_2d,     target }\OperatorTok{=}\NormalTok{ H13)   }\CommentTok{\# NATIVE: 2D conv lowers from M1 onward}
\NormalTok{status\_sin  }\OperatorTok{=}\NormalTok{ confirm\_op(sin,         target }\OperatorTok{=}\NormalTok{ H13)   }\CommentTok{\# REJECTED: sin/cos are gated to A15+}
\NormalTok{status\_prod }\OperatorTok{=}\NormalTok{ confirm\_op(reduce\_prod, target }\OperatorTok{=}\NormalTok{ H13)   }\CommentTok{\# REJECTED: no hardware path on any family}

\CommentTok{\# An op gated to a later family is rejected when compiled against an earlier}
\CommentTok{\# target, so the target argument is how a caller checks the family floor}
\CommentTok{\# before building anything on top of the operation.}
\end{Highlighting}
\end{Shaded}

\end{listing}

\section{Reference: the per-chip shape
limits}\label{reference-the-per-chip-shape-limits}

The operations that run have per-chip shape and kernel limits, read from
the compiler's hardware-abstraction-layer tables. The headline limits
separate the M1 from the M5 by a factor of four on the tensor extent and
by a step on the kernel size; the full table is Appendix
\hyperref[appendix-a.-operation-by-device-matrix]{A}.
\Cref{tbl:c4-shape-limits} gives the per-generation kernel and tensor
shape limits.

\begin{longtable}[]{@{}lrrrrr@{}}
\caption{The per-generation kernel and tensor shape limits, decoded from
the compiler's hardware-abstraction-layer tables, M1
measured.}\label{tbl:c4-shape-limits}\tabularnewline
\toprule\noalign{}
\textbf{Limit} & \textbf{older} & \textbf{A13 / M1} & \textbf{A14} &
\textbf{A15} & \textbf{A16+ (M5)} \\
\midrule\noalign{}
\endfirsthead
\toprule\noalign{}
\textbf{Limit} & \textbf{older} & \textbf{A13 / M1} & \textbf{A14} &
\textbf{A15} & \textbf{A16+ (M5)} \\
\midrule\noalign{}
\endhead
\bottomrule\noalign{}
\endlastfoot
max kernel width, default format & 29 & 29 & 32 & 32 & 32 \\
\rowcolor{zebra}max kernel width, fp16 & 13 & 13 & 16 & 16 & 16 \\
min kernel width, large mode & 16 & 16 & 1 & 1 & 1 \\
\rowcolor{zebra}max kernel depth & 1 & 16 & 16 & 16 & 16 \\
max tensor width, height & 16384 & 16384 & 16384 & 16384 & 65536 \\
\rowcolor{zebra}max tensor depth & 1 & 16384 & 16384 & 16384 & 65536 \\
max tensor batch & 4096 & 65536 & 65536 & 65536 & 65536 \\
\rowcolor{zebra}reduction-to-transpose threshold & none & 192 & 192 &
384 & 384 \\
matrix-multiply working set & 2 MB & 2 MB & 2 MB & 2 MB & 2 MB \\
\rowcolor{zebra}has texture engine & no & no & yes & yes & yes \\
\end{longtable}

A caller meets two per-operation envelopes early, both hard limits. The
matrix-multiply operation requires the depth axis to be 1 on both
operands, and the fully-connected operation rejects an input of rank 5
or more. The software gather path on the M1 is valid only inside a small
envelope: the data batch and depth must be 1, the index channel must be
3, the index width and depth must be 1, and the gather-axis count must
be 3. Outside it the compiler aborts rather than falling back.

\section{Reference: the compile-legal
envelope}\label{reference-the-compile-legal-envelope}

The shape limits above bound the operations that run; a second set of
rules bounds the programs the compiler accepts at all. These were
measured compile-only on the M1, building each graph and calling the
compiler inside a guard, never running the result, so a rejected program
names its layer without provoking the firmware. Two layers refuse a
program: a Python-side check in the frontend, before anything reaches
the device, and the on-device compiler, which validates and returns an
Espresso exception.

A tensor has a maximum rank of five, the \texttt{N,\ D,\ C,\ H,\ W}
layout of the per-chip limits, and the frontend rejects a graph that
builds a rank-6 tensor. Per-operation support is narrower than the
generic rank cap: a rank-5 batched matrix multiply builds yet the
on-device matrix-multiply backend rejects it, and convolution is pinned
to four-dimensional \texttt{NCHW} input. Each axis is capped at
\(2^{14}\), the 16384 tensor extent of the M1, applied per-axis rather
than to the last axis alone: the matrix-multiply output and contraction
dims, both elementwise axes, and the rank-1 length all accept 16384 and
reject 16385. The convolution input-channel and output-channel dims
escape this cap, both \texttt{Cin} and \texttt{Cout} of 16385 compile,
while the convolution input spatial axis obeys it. Size-1 dims and the
rank-0 scalar are legal and compile; a zero-size dim is the one
degenerate shape the on-device validator rejects, as a type mismatch,
since the frontend does not pre-screen it.

Program inputs are \texttt{fp16} or \texttt{uint8} only. The
\texttt{fp16} input is the compute type; the \texttt{uint8} input
supplies only the in-graph dequantization path for an integer image
input, not an \texttt{add} or \texttt{matmul} directly, and the frontend
rejects any other input dtype. Weight constants accept any float numpy
dtype, with \texttt{fp32} constants down-cast to \texttt{fp16}, while
the compiler rejects integer-typed weights. Convolution pre-screens the
kernel width in the frontend, rejecting a width above 15 on the M1,
while the fp16 datapath caps it lower at 13 as the limit table above
shows, so an fp16 kernel of width 14 or 15 still rejects at lowering.
The kernel height is unconstrained, so a 16x3 kernel compiles and a 3x16
kernel does not. Stride, dilation, grouped and depthwise forms, and
padding wider than the input all compile; the on-device compiler rejects
an indivisible group count.

\Cref{tbl:c4-compile-envelope} records each compile-legal probe with its
result, the layer that ruled on it, and the message the rejecting layer
returns.

\begin{longtable}[]{@{}
  >{\raggedright\arraybackslash}p{(\linewidth - 6\tabcolsep) * \real{0.2500}}
  >{\raggedright\arraybackslash}p{(\linewidth - 6\tabcolsep) * \real{0.2500}}
  >{\raggedright\arraybackslash}p{(\linewidth - 6\tabcolsep) * \real{0.2500}}
  >{\raggedright\arraybackslash}p{(\linewidth - 6\tabcolsep) * \real{0.2500}}@{}}
\caption{The compile-legal envelope on the M1, separating the frontend
rejects from the on-device compiler
rejects.}\label{tbl:c4-compile-envelope}\tabularnewline
\toprule\noalign{}
\begin{minipage}[b]{\linewidth}\raggedright
\textbf{Probe}
\end{minipage} & \begin{minipage}[b]{\linewidth}\raggedright
\textbf{Result}
\end{minipage} & \begin{minipage}[b]{\linewidth}\raggedright
\textbf{Layer}
\end{minipage} & \begin{minipage}[b]{\linewidth}\raggedright
\textbf{Compiler message}
\end{minipage} \\
\midrule\noalign{}
\endfirsthead
\toprule\noalign{}
\begin{minipage}[b]{\linewidth}\raggedright
\textbf{Probe}
\end{minipage} & \begin{minipage}[b]{\linewidth}\raggedright
\textbf{Result}
\end{minipage} & \begin{minipage}[b]{\linewidth}\raggedright
\textbf{Layer}
\end{minipage} & \begin{minipage}[b]{\linewidth}\raggedright
\textbf{Compiler message}
\end{minipage} \\
\midrule\noalign{}
\endhead
\bottomrule\noalign{}
\endlastfoot
tensor rank 1 through 5 & accept & & \\
\rowcolor{zebra}tensor rank 6 & reject & frontend &
\texttt{tensor\ rank\ 6\ exceeds\ the\ ANE\ maximum\ of\ 5} \\
batched matmul rank 5 & reject & on-device &
\texttt{Some\ ops\ are\ not\ supported\ on\ any\ of\ the\ specified\ backends} \\
\rowcolor{zebra}any axis = 16384 & accept & & \\
any axis = 16385 & reject & frontend &
\texttt{exceeds\ ANE\ family\ 2\textquotesingle{}s\ max\ dimension\ 16384} \\
\rowcolor{zebra}conv \texttt{Cin} or \texttt{Cout} = 16385 & accept &
& \\
size-1 dims, rank-0 scalar & accept & & \\
\rowcolor{zebra}zero-size dim \texttt{(0,\ 8)} & reject & on-device &
\texttt{Expected\ tensor\textless{}fp16,{[}1,8{]}\textgreater{};\ got\ tensor\textless{}fp16,{[}0,8{]}\textgreater{}} \\
\end{longtable}

\anepartopen{Part II}{Reaching the ANE}
\anechap{05}{Software stack}{The layers from a graph to the engine, and the daemon that brokers access.}
\anechap{06}{Dispatching without Core ML}{The five-step direct route: build, compile, load, bind, and dispatch.}
\anechap{07}{Weights and compression}{The int4, sparse, int8, and blockwise forms, and which stream natively.}
\anechap{08}{Entitlement boundary}{The signed-load gate and the entitlements that bound the reachable surface.}
\anepartclose

\chapter{Software stack}\label{software-stack}

\begin{summarybox}

Four user-space layers separate an application from the engine. The
execution runtime beneath the model framework is reachable directly,
with no placement planner and no entitlement for accepted operations.
The model framework segments work across the processor, graphics
processor, and engine by a shortest-path solve over a per-operation cost
graph. The public model-plan read-out reports that solve. A system
daemon has the single privileged device gate, a content-hashed program
cache, and one time-shared request queue that arbitrates every client.

\end{summarybox}

Part I treated the engine as a machine: a fixed-function matrix
accelerator with an fp16 datapath, driven as an autonomous coprocessor
through a command mailbox. Part II treats it as a target a developer can
address. Between an application and the silicon are several layers of
system software, each with a defined job. This chapter maps those layers
and marks the one this Part enters.

\section{Layered stack}\label{layered-stack}

\begin{figure}[H]\centering
\includegraphics[width=4.167in]{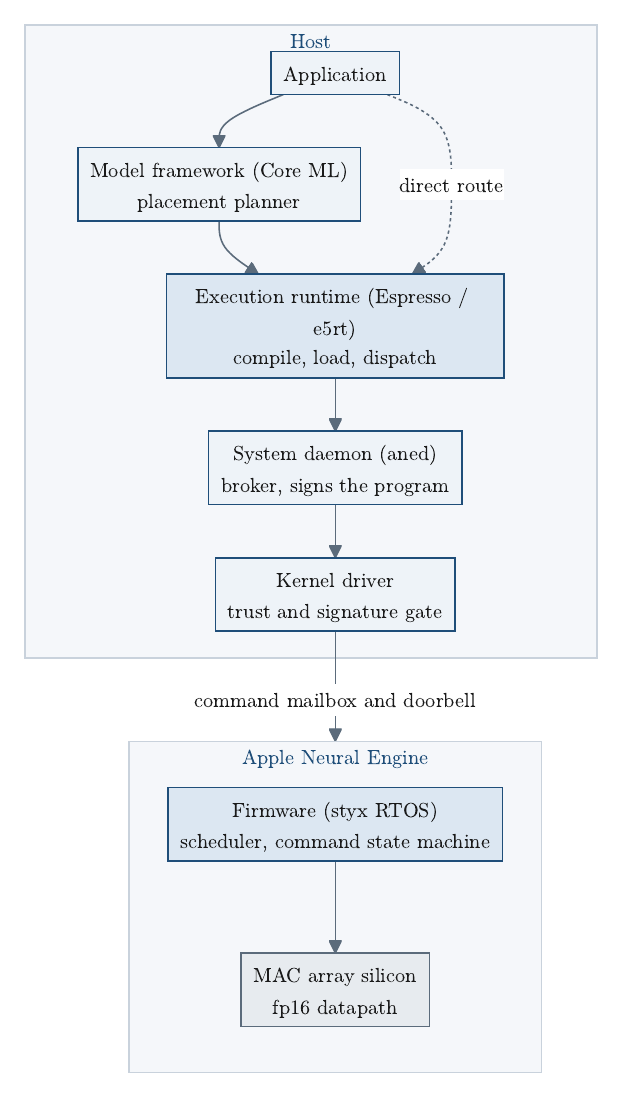}
\caption{The full stack from an application down through the kernel driver, the engine firmware, and the silicon, with the two routes to the runtime.}
\label{fig:c5-stack}
\end{figure}

The full stack runs from the application down to the silicon, with two
routes reaching the runtime (\cref{fig:c5-stack}). A system daemon and a
kernel driver are below the four user-space layers, out of the dispatch
band.

The top layer is the model framework. It accepts a trained network,
decides which operations the engine can accept, and segments the work
across the central processor, graphics processor, and engine under a
cost-driven placement planner. Device choice is not the caller's at this
layer, and the placement is never reported back. Chapter
\hyperref[what-the-ane-is]{1} describes this public surface.

Beneath it is the execution runtime, the layer every path passes
through. It owns compilation: it turns a network in the intermediate
representation into the engine's program format, writes the result to a
content-addressed cache on disk, and produces a loadable program. It
owns the program library and the callable functions inside it. It owns
the execution stream, the queue against which a compiled program is
encoded and dispatched, and the descriptors that bind operands to a
program's named ports. The same runtime runs the central-processor and
graphics-processor executors as alternative backends, so a single stream
can contain operations placed on different devices. The engine is one
backend among those three.

Beneath the runtime, on the path to the engine, is the client layer. It
splits into two duties. The client layer hands program lifecycle,
meaning compile, load, instantiate, cache, and purge, to the system
daemon over an interprocess channel. Per-inference dispatch goes
directly to the kernel driver once a program instance exists. The daemon
has the privileged device handle and the on-disk program cache; the
client process keeps the connection it uses for the hot path.

The lowest user-space layer is a thin shim over the kernel driver. Its
calls translate into driver method invocations on the engine's user
client. Below that is the kernel driver itself, then the firmware and
the silicon.

\Cref{tbl:c5-layers} lists the four user-space layers and their
reachability from ordinary user space.

\begin{longtable}[]{@{}
  >{\raggedright\arraybackslash}p{(\linewidth - 4\tabcolsep) * \real{0.3333}}
  >{\raggedright\arraybackslash}p{(\linewidth - 4\tabcolsep) * \real{0.3333}}
  >{\raggedright\arraybackslash}p{(\linewidth - 4\tabcolsep) * \real{0.3333}}@{}}
\caption{Layers of the engine software stack, each layer's role, and
whether a caller can reach it
directly.}\label{tbl:c5-layers}\tabularnewline
\toprule\noalign{}
\begin{minipage}[b]{\linewidth}\raggedright
\textbf{Layer}
\end{minipage} & \begin{minipage}[b]{\linewidth}\raggedright
\textbf{Role}
\end{minipage} & \begin{minipage}[b]{\linewidth}\raggedright
\textbf{Reachable directly}
\end{minipage} \\
\midrule\noalign{}
\endfirsthead
\toprule\noalign{}
\begin{minipage}[b]{\linewidth}\raggedright
\textbf{Layer}
\end{minipage} & \begin{minipage}[b]{\linewidth}\raggedright
\textbf{Role}
\end{minipage} & \begin{minipage}[b]{\linewidth}\raggedright
\textbf{Reachable directly}
\end{minipage} \\
\midrule\noalign{}
\endhead
\bottomrule\noalign{}
\endlastfoot
Model framework & Cost-driven placement planner; segments work across
processor, graphics, and engine & No: device is a scheduling outcome,
not a choice \\
\rowcolor{zebra}Espresso runtime & Network loading and operator graph
beneath the framework & Yes, but the framework drives it \\
Execution runtime & Owns compile, program library, execution stream, and
operand descriptors & Yes: names the engine as target, no entitlement
for accepted ops \\
\rowcolor{zebra}Client / user-client layer & Splits program lifecycle to
the daemon from per-inference dispatch to the driver & Yes, for the hot
dispatch path \\
System daemon (out of band) & Holds the privileged device gate and the
content-hashed program cache; arbitrates clients & No: reached through
its interprocess channel \\
\rowcolor{zebra}Kernel-driver shim (out of band) & Thin user-space
translation into driver method invocations on the engine's user client &
Yes, as the lowest user-space call \\
\end{longtable}

Each layer has a distinct symbol family in the binaries, and the family
name identifies the layer a call is on. The execution runtime exports a
flat C facade over a C++ core in the \texttt{E5RT::} namespace, with
every entry point prefixed \texttt{e5rt\_}. The client layer is
Objective-C, every class prefixed \texttt{\_ANE}. The daemon
communicates over a private interprocess protocol whose selectors name
the arguments verbatim. The shim into the kernel driver is the
\texttt{ANEServices} C and C++ entry points, which lower to the IOKit
\texttt{IOConnectCall*} calls on the engine's user client. Each family
appears beside its layer in \cref{tbl:c5-symbols}.

\begin{longtable}[]{@{}
  >{\raggedright\arraybackslash}p{(\linewidth - 2\tabcolsep) * \real{0.5000}}
  >{\raggedright\arraybackslash}p{(\linewidth - 2\tabcolsep) * \real{0.5000}}@{}}
\caption{The entry-point symbol family that marks each layer of the
engine software stack, from the execution runtime down to the
kernel-driver shim.}\label{tbl:c5-symbols}\tabularnewline
\toprule\noalign{}
\begin{minipage}[b]{\linewidth}\raggedright
\textbf{Layer}
\end{minipage} & \begin{minipage}[b]{\linewidth}\raggedright
\textbf{Symbol family (entry points)}
\end{minipage} \\
\midrule\noalign{}
\endfirsthead
\toprule\noalign{}
\begin{minipage}[b]{\linewidth}\raggedright
\textbf{Layer}
\end{minipage} & \begin{minipage}[b]{\linewidth}\raggedright
\textbf{Symbol family (entry points)}
\end{minipage} \\
\midrule\noalign{}
\endhead
\bottomrule\noalign{}
\endlastfoot
Execution runtime (\texttt{Espresso.framework}) &
\texttt{e5rt\_e5\_compiler\_compile} (compile MIL to \texttt{.e5}),
\texttt{e5rt\_program\_library\_retain\_program\_function},
\texttt{e5rt\_program\_function\_load\_for\_execution},
\texttt{e5rt\_execution\_stream\_execute\_sync} /
\texttt{\_submit\_async}; C++ core \texttt{E5RT::E5Compiler},
\texttt{E5RT::ExecutionStream} \\
\rowcolor{zebra}Client layer (\texttt{AppleNeuralEngine}) &
\texttt{\_ANEClient}, \texttt{\_ANEModel} / \texttt{\_ANEInMemoryModel},
\texttt{\_ANERequest}, \texttt{\_ANEIOSurfaceObject},
\texttt{evaluateWithQoS:options:request:error:} \\
System daemon (\texttt{aned}, NSXPC \texttt{com.apple.aned}) &
\texttt{createProgramInstanceForModel:modelToken:...:statsMask:memoryPoolID:enableLateLatch:...:error:} \\
\rowcolor{zebra}Kernel-driver shim (\texttt{ANEServices.framework}) &
\texttt{\_ANEServicesProgramCreate},
\texttt{ANE::ANEServicesDevice::ANE\_ProgramSendRequest(...)} to
\texttt{IOConnectCallAsyncMethod(selector=2)} on \texttt{H11ANE} \\
\end{longtable}

\section{Runtime is reachable below the
framework}\label{runtime-is-reachable-below-the-framework}

The same execution runtime that the system's own dispatchers use is
callable from ordinary user space, below the model framework. A caller
can compile a network to the engine's program format, open the resulting
program library, load a function for execution, bind operand buffers to
its ports, encode the operation onto a stream, and submit it. No
placement planner is in this path, and the operations the compiler
accepts require no special entitlement. The caller names the engine as
the target.

\subsection{Runtime surface and its tunable
dictionary}\label{runtime-surface-and-its-tunable-dictionary}

The execution runtime is a flat C facade over a refcounted C++ core in
the \texttt{E5RT::} namespace. The binary exports 292 \texttt{e5rt\_}
entry points, organized into five object families: a compiler and its
configuration, program library and the functions inside it, precompiled
compute operation, execution stream with its buffer and port objects,
and asynchronous event for fences. Each family pairs a \texttt{\_create}
constructor with a \texttt{\_release} destructor, and each has a
create-options object whose typed get/set accessors are the runtime's
tunable dictionary.

The dictionary spans four objects. The compile-time options object
(\texttt{e5rt\_e5\_compiler\_options\_}) exposes 21 keys: the backend
bitmask (\texttt{compute\_device\_types\_mask}, where \texttt{0x4} names
the engine), the cache controls \texttt{force\_recompilation} and
\texttt{force\_fetch\_from\_cache}, the \texttt{segmenter} selection,
and a set of backend-preference and experimental toggles. A separate
compiler-configuration object
(\texttt{e5rt\_e5\_compiler\_config\_options\_}) has the 2 cache keys:
\texttt{cache\_bundle\_location}, the directory the daemon writes the
compiled bundle into, and \texttt{bundle\_cache\_apfs\_purgeable}, which
marks that bundle reclaimable under storage pressure. The per-operation
options object
(\texttt{e5rt\_precompiled\_compute\_op\_create\_options\_}) contains 14
keys, including \texttt{operation\_name},
\texttt{allocate\_intermediate\_buffers}, the resident-weight path
\texttt{mutable\_mil\_weight\_paths}, and a cross-process IOSurface pool
binding. The execution-stream configuration object has 3 keys:
\texttt{enable\_concurrent\_sync\_execution},
\texttt{enable\_low\_latency\_async\_events}, and
\texttt{skip\_io\_fences}. A neutral all-engine compile sets only a
small subset of these: the engine bitmask,
\texttt{force\_recompilation}, and the \texttt{"graph"} segmenter,
leaving every other key at its runtime default.

The on-disk program cache the daemon writes is keyed by content, not by
source filename. The runtime assembles a \texttt{cacheURLIdentifier}
from a per-segment key computed over each segment's network structure
and weight blob containers, combined with the resolved source URL,
options dictionary, and platform. Two compiles of structurally identical
graphs with identical weights and options resolve to the same identifier
and hit the cache, while changing any weight, shape, operation, the
device mask, or the segmenter changes the key. The
\texttt{force\_recompilation} key bypasses the cache fetch and rewrites
the bundle unconditionally, which is the documented inverse of
\texttt{force\_fetch\_from\_cache}.

\subsection{Dispatch selector that splits the
path}\label{dispatch-selector-that-splits-the-path}

The split between the daemon and the client described above resolves to
specific kernel-driver selectors, measured by read-only tracing of
\texttt{IOConnectCall*Method} on this M1. Per-inference submit is
\texttt{IOConnectCallAsyncMethod} selector 2, the
\texttt{ANE\_ProgramSendRequest} handler in the H11ANE dispatch table,
and the unentitled client issues it directly: a freshly compiled program
issued selector 2 exactly once per execute, observed in the client
process and never in the daemon. The daemon issues only the lifecycle
selectors over its own connection during the same compile and prepare,
the synchronous selectors 3 through 6 for program destroy, status,
instance create with unprepare, and program create. The division of
labor is precise: the daemon compiles the network and creates, prepares,
and destroys the program object over IOKit, while the client submits
each inference over selector 2 itself. A warm re-execute on an
already-submitted program issued no additional \texttt{IOConnectCall}
from the client, since the runtime reuses the armed async submit ring
and signals the engine without a fresh external method per call. This is
consistent with the warm-eval cost being firmware round-trip bound
rather than dispatch-call bound.

\section{Placement segmenter above the
runtime}\label{placement-segmenter-above-the-runtime}

The model framework adds one stage the direct route does not have: it
decides, per operation, which of the three backends runs it. When the
framework places a model, it segments the operation graph across the
central processor, graphics processor, and engine by solving a shortest
path (Dijkstra) over a cost graph with one node per
operation-and-backend pair. The per-operation cost on each backend comes
from a set of learned regression decision trees, several hundred of
them, keyed by operation and by backend. Two coarse compute and
bandwidth anchors order the backends, and the trees supply the
calibrated per-operation cost the solver minimizes. A fixed launch
penalty per segment and a transfer penalty at every backend boundary
bias the solution toward fewer and larger engine segments, for the
reason given with the cost equation below. An operation the engine
cannot accept has no engine node in the cost graph, so the minimum-cost
path routes around the engine through the central or graphics processor.
This mechanism drives the framework's automatic fallback, and the direct
route skips it by authoring a single all-engine graph.

A developer can read this placement decision without running inference,
through the supported public model-plan API. \texttt{MLComputePlan.load}
parses a compiled model and reports the segmenter's choice per
operation. \texttt{deviceUsage(for:)} returns the set of devices that
could run an operation and the one device the planner preferred, with a
\texttt{Reason} describing why an operation is or is not supported on a
device. \texttt{estimatedCost(of:)} returns the per-operation cost
weight the segmenter compared. These are documented at
developer.apple.com/documentation/coreml
\hyperref[ref-applecoreml]{{[}AppleCoreML{]}}.

The segmenter is a named pipeline inside the ahead-of-time compiler. The
compiler canonicalizes the intermediate program for the platform,
expands the requested compute-unit mask into a set of backend
identifiers, then runs one of three segmenters over the graph. Its
stages run from eligibility through validity and cost to segmentation
(\cref{tbl:c5-segmenter}).

\begin{longtable}[]{@{}
  >{\raggedright\arraybackslash}p{(\linewidth - 4\tabcolsep) * \real{0.3333}}
  >{\raggedright\arraybackslash}p{(\linewidth - 4\tabcolsep) * \real{0.3333}}
  >{\raggedright\arraybackslash}p{(\linewidth - 4\tabcolsep) * \real{0.3333}}@{}}
\caption{The named stages of the placement segmenter inside the
ahead-of-time compiler.}\label{tbl:c5-segmenter}\tabularnewline
\toprule\noalign{}
\begin{minipage}[b]{\linewidth}\raggedright
\textbf{Stage}
\end{minipage} & \begin{minipage}[b]{\linewidth}\raggedright
\textbf{Component}
\end{minipage} & \begin{minipage}[b]{\linewidth}\raggedright
\textbf{Role}
\end{minipage} \\
\midrule\noalign{}
\endfirsthead
\toprule\noalign{}
\begin{minipage}[b]{\linewidth}\raggedright
\textbf{Stage}
\end{minipage} & \begin{minipage}[b]{\linewidth}\raggedright
\textbf{Component}
\end{minipage} & \begin{minipage}[b]{\linewidth}\raggedright
\textbf{Role}
\end{minipage} \\
\midrule\noalign{}
\endhead
\bottomrule\noalign{}
\endlastfoot
Eligibility & \texttt{ComputeUnitsToBackends} & expands the compute-unit
mask into concrete backend identifiers \\
\rowcolor{zebra}Validity & \texttt{AneValidator},
\texttt{BnnsValidator}, \texttt{MpsGraphValidator} & per-backend
per-operation legality, one validator per backend \\
Cost & \texttt{EstimatorMILDecisionTree}, \texttt{ConstCostEstimator} &
the per-operation cost the solver minimizes \\
\rowcolor{zebra}Segmentation & \texttt{SegmenterCoarse},
\texttt{SegmenterGraph}, \texttt{SegmenterShortestPath} & greedy,
graph-grouped, or shortest-path placement \\
\end{longtable}

The cost the shortest-path segmenter minimizes is the larger of compute
and bandwidth time plus penalties. The per-operation compute cost is the
larger of the compute time and the bandwidth time, drawn from the engine
compute rate and bandwidth primitives \texttt{GetEngineGflopsPerS} and
\texttt{GetEngineBwGbPerS}. The placement adds a fixed launch cost per
segment and a transfer cost at each backend boundary.

\[\mathrm{cost} = \max\!\left(\frac{\mathrm{flops}}{\mathrm{gflops}},\ \frac{\mathrm{bytes}}{\mathrm{bw}}\right) + \mathrm{launch} + \mathrm{transfer}\]

The transfer cost is why the minimum-cost solution favors long
single-backend runs: each boundary charges a tensor repack between the
engine channel-interleaved fp16 layout and the host layout, so one
engine segment is cheaper than several. The placement audit trail is
itself a queryable compile output: the program library exposes a
post-compile analytics dictionary, keyed by \texttt{selected\_backend},
\texttt{backend\_support}, \texttt{estimated\_runtime},
\texttt{op\_type}, \texttt{op\_path}, and \texttt{validation\_messages},
so the chosen backend and predicted runtime for each operation are
readable without running inference.

\section{Broker}\label{broker}

A system daemon mediates access to the engine. Only that daemon and its
per-user sibling have the kernel gate that opens the device, so every
other process reaches the engine through it. A client proves itself to
the daemon over the interprocess channel, and the daemon performs the
privileged device open on the client's behalf, returning a program
handle the client then drives.

The daemon arbitrates across clients. There is one physical engine, so
concurrent demand resolves by time-division on a single request queue.
The daemon owns the shared program cache, keyed by a content hash of
each compiled network, so a network already compiled by one client is a
cache hit for the next. It sets a quality-of-service value on every
request and adjusts a client's requested queue depth under contention,
which keeps one client from depriving another of service. A request from
a developer-signed program and a request from a system dispatcher arrive
at the same broker, and the same queue arbitrates both.

The daemon is the one process holding the kernel gate, and its
interprocess surface is the program-lifecycle protocol every client
drives. The program-instance method lists the per-instance parameters
verbatim in its selector, the residency, power, and statistics arguments
among them (\cref{lst:c5-instance-selectors}).

\begin{listing}[H]
\caption{The daemon’s program-instantiation selectors, with the per-instance residency, power, and statistics parameters held as named arguments.}\label{lst:c5-instance-selectors}

\begin{Shaded}
\begin{Highlighting}[]
\PreprocessorTok{\# the daemon\textquotesingle{}s central program{-}instance method (recovered selector, NSXPC \_ANEDaemonProtocol)}
\NormalTok{createProgramInstanceForModel}\OperatorTok{:}\NormalTok{modelToken}\OperatorTok{:}\NormalTok{modelFilePath}\OperatorTok{:}\NormalTok{qos}\OperatorTok{:}\NormalTok{isPreCompiled}\OperatorTok{:}
\NormalTok{  enablePowerSaving}\OperatorTok{:}\NormalTok{skipPreparePhase}\OperatorTok{:}\NormalTok{statsMask}\OperatorTok{:}\NormalTok{memoryPoolID}\OperatorTok{:}\NormalTok{enableLateLatch}\OperatorTok{:}
\NormalTok{  modelIdentityStr}\OperatorTok{:}\NormalTok{owningPid}\OperatorTok{:}\NormalTok{cacheUrlIdentifier}\OperatorTok{:}\NormalTok{aotCacheUrlIdentifier}\OperatorTok{:}
\NormalTok{  optOutOfModelMemoryUnwiring}\OperatorTok{:}\NormalTok{error}\OperatorTok{:}

\PreprocessorTok{\# the weights{-}streaming variant: one resident base, a thin adapter per call}
\NormalTok{createProgramInstanceWithWeights}\OperatorTok{:}\NormalTok{modelToken}\OperatorTok{:}\NormalTok{qos}\OperatorTok{:}\NormalTok{baseModelIdentifier}\OperatorTok{:}
\NormalTok{  owningPid}\OperatorTok{:}\NormalTok{numWeightFiles}\OperatorTok{:}\NormalTok{error}\OperatorTok{:}
\end{Highlighting}
\end{Shaded}

\end{listing}

The on-the-wire instance is a C struct the daemon returns to the client,
two opaque pointers, two 512-byte name buffers, the input and output
port tables, a procedure record, and counters, with its recovered type
encoding in \cref{lst:c5-instance-struct}.

\begin{listing}[H]
\caption{The program-instance struct the daemon returns over the interprocess channel.}\label{lst:c5-instance-struct}

\begin{Shaded}
\begin{Highlighting}[]
\PreprocessorTok{\# ANEProgramInstanceStruct (recovered NSXPC type encoding)}
\OperatorTok{\^{}\{}\NormalTok{ANEProgramInstanceStruct}\OperatorTok{=\^{}}\NormalTok{v}\OperatorTok{\^{}}\NormalTok{vQQC}\OperatorTok{\^{}[}\DecValTok{512}\NormalTok{c}\OperatorTok{]\^{}}\NormalTok{Q}\OperatorTok{\^{}\{}\NormalTok{ANEProgramIOInfoStruct}\OperatorTok{\}}
  \OperatorTok{\^{}\{}\NormalTok{ANEProgramIOInfoStruct}\OperatorTok{\}}\NormalTok{C}\OperatorTok{\^{}[}\DecValTok{512}\NormalTok{c}\OperatorTok{]\^{}}\NormalTok{QC}\OperatorTok{\^{}\{}\NormalTok{ANEProgramProcedureStruct}\OperatorTok{\}}\NormalTok{QIQQcQQiC}\OperatorTok{\}}
\end{Highlighting}
\end{Shaded}

\end{listing}

\section{Access-control model}\label{access-control-model}

The device has one user-space entry point. The kernel driver denies
opening the user client without the entitlement
\texttt{com.apple.ane.iokit-user-access}, and exactly two binaries on
the system have it: the daemon and its per-user sibling. Every other
process proves itself to the daemon over the interprocess channel
instead, gated by a \texttt{com.apple.aned.private.*} entitlement family
that the daemon checks per connection and per method.
\Cref{tbl:c5-entitlements} gives that entitlement family and what each
member authorizes.

\begin{longtable}[]{@{}
  >{\raggedright\arraybackslash}p{(\linewidth - 4\tabcolsep) * \real{0.3000}}
  >{\raggedright\arraybackslash}p{(\linewidth - 4\tabcolsep) * \real{0.3000}}
  >{\raggedleft\arraybackslash}p{(\linewidth - 4\tabcolsep) * \real{0.4000}}@{}}
\caption{The entitlement family that gates the engine, with the count of
system binaries holding each in this
build.}\label{tbl:c5-entitlements}\tabularnewline
\toprule\noalign{}
\begin{minipage}[b]{\linewidth}\raggedright
\textbf{Entitlement}
\end{minipage} & \begin{minipage}[b]{\linewidth}\raggedright
\textbf{What it authorizes}
\end{minipage} & \begin{minipage}[b]{\linewidth}\raggedleft
\textbf{Holders}
\end{minipage} \\
\midrule\noalign{}
\endfirsthead
\toprule\noalign{}
\begin{minipage}[b]{\linewidth}\raggedright
\textbf{Entitlement}
\end{minipage} & \begin{minipage}[b]{\linewidth}\raggedright
\textbf{What it authorizes}
\end{minipage} & \begin{minipage}[b]{\linewidth}\raggedleft
\textbf{Holders}
\end{minipage} \\
\midrule\noalign{}
\endhead
\bottomrule\noalign{}
\endlastfoot
\texttt{com.apple.ane.iokit-user-access} & the hard kernel gate:
privileged device open, compile, cache & 2 \\
\rowcolor{zebra}\texttt{com.apple.aned.private.allow} & baseline:
compile, load, and instantiate models through the daemon & 18 \\
\texttt{com.apple.aned.private.ANEAccess.allow} & the inference-client
access grant & 14 \\
\rowcolor{zebra}\texttt{com.apple.aned.private.adapterWeight.allow} &
stream adapter weights onto a shared resident base model & 5 \\
\texttt{com.apple.aned.private.processModelShare.allow} & share one
resident model across processes & 4 \\
\rowcolor{zebra}\texttt{com.apple.aned.private.secondaryANECompilerServiceAccess.allow}
& use the longer-duration secondary compiler service & 1 \\
\end{longtable}

A privileged subset of 27 binaries also has a sandbox exception for the
class \texttt{H11ANEInDirectPathClient}, which lets a latency-sensitive
client open the low-latency user client and drive per-inference
submission on its own connection rather than round-tripping each
inference through the daemon. The exception grants no device access by
itself: the daemon still performs the privileged open and returns the
program handle. A developer-signed binary cannot assert the kernel gate,
since it names a restricted user-client class that ad-hoc and
development signing cannot claim. The direct route reaches the engine
the same way a sanctioned application does, through the daemon. It
authors its work at the model and program layer.

\section{Compiler is its own gated
service}\label{compiler-is-its-own-gated-service}

Compilation does not happen in the calling process. The service that
turns a network into the engine's program format is a separate sandboxed
interprocess service, \texttt{com.apple.ANECompilerService}, reached
through an \texttt{NSXPCConnection} named for it. That service vends a
single entry point, the method
\texttt{compileModelAt:csIdentity:sandboxExtension:options:tempDirectory:...:withReply:},
and admits a connection only through an entitlement gate. The service's
\texttt{listener:shouldAcceptNewConnection:} delegate calls
\texttt{valueForEntitlement:} on the connection against the string
returned by \texttt{+compilerServiceAccessEntitlement}, then logs
whether the client has the entitlement or is missing it.

This is the concrete reason compilation leaves the calling process even
for a hand-authored or self-compiled program. The service has two
entitlements an ordinary process does not: it writes the
system-protected compile cache under
\texttt{rootless.storage.ane\_model\_cache}, and it decrypts under
\texttt{coreml.decypt\_allowed}. A caller hands its network to the
service and receives the compiled bundle back, the same separation the
daemon imposes on device access.

The service keeps a warm-start cache so a repeat compile of the same
network is not re-paid. The cache is a nested directory tree of the form
\texttt{cache/com.apple.e5rt.e5bundlecache/\textless{}os-build\textgreater{}/\textless{}hash\textgreater{}/},
keyed by the model hash \texttt{model.anehash}, which is a double
SHA-256 over the program. A repeat compile of the same program reuses
the cached bundle. The \texttt{force\_recompilation} option defeats the
warm start and compiles from scratch.

\section{Entering the runtime below the
framework}\label{entering-the-runtime-below-the-framework}

A caller that names the engine as the target enters at the
execution-runtime layer rather than the model framework, so no placement
planner is in the path. The runtime compiles a graph to the engine
program format, loads it, and dispatches it.

The neutral workflow builds a graph, compiles it to a chip target, and
dispatches, with no segmenter between the graph and the engine.

\begin{Shaded}
\begin{Highlighting}[]
\CommentTok{/* Enter at the execution{-}runtime layer: compiler {-}\textgreater{} library {-}\textgreater{} function {-}\textgreater{} stream. */}
\NormalTok{e5rt\_e5\_compiler\_create\_with\_config}\OperatorTok{(\&}\NormalTok{compiler}\OperatorTok{,}\NormalTok{ config}\OperatorTok{);}            \CommentTok{/* the runtime object */}
\NormalTok{e5rt\_e5\_compiler\_compile}\OperatorTok{(}\NormalTok{compiler}\OperatorTok{,}\NormalTok{ model\_path}\OperatorTok{,}\NormalTok{ options}\OperatorTok{,} \OperatorTok{\&}\NormalTok{library}\OperatorTok{);} \CommentTok{/* to the engine format */}

\CommentTok{/* The program library vends the callable function, below any model framework. */}
\NormalTok{e5rt\_program\_library\_retain\_program\_function}\OperatorTok{(}\NormalTok{library}\OperatorTok{,}\NormalTok{ fn\_name}\OperatorTok{,} \OperatorTok{\&}\NormalTok{function}\OperatorTok{);}
\NormalTok{e5rt\_precompiled\_compute\_op\_create\_options\_create\_with\_program\_function}\OperatorTok{(\&}\NormalTok{op\_opts}\OperatorTok{,}\NormalTok{ function}\OperatorTok{);}
\NormalTok{e5rt\_execution\_stream\_operation\_create\_precompiled\_compute\_operation\_with\_options}\OperatorTok{(\&}\NormalTok{op}\OperatorTok{,}\NormalTok{ op\_opts}\OperatorTok{);}

\CommentTok{/* The execution stream is the runtime\textquotesingle{}s dispatch queue, with no segmenter above it. */}
\NormalTok{e5rt\_execution\_stream\_operation\_retain\_input\_port}\OperatorTok{(}\NormalTok{op}\OperatorTok{,} \StringTok{"x"}\OperatorTok{,} \OperatorTok{\&}\NormalTok{in\_port}\OperatorTok{);}
\NormalTok{e5rt\_io\_port\_bind\_buffer\_object}\OperatorTok{(}\NormalTok{in\_port}\OperatorTok{,}\NormalTok{ in\_buf}\OperatorTok{);}
\NormalTok{e5rt\_execution\_stream\_create}\OperatorTok{(\&}\NormalTok{stream}\OperatorTok{);}
\NormalTok{e5rt\_execution\_stream\_encode\_operation}\OperatorTok{(}\NormalTok{stream}\OperatorTok{,}\NormalTok{ op}\OperatorTok{);}
\NormalTok{e5rt\_execution\_stream\_execute\_sync}\OperatorTok{(}\NormalTok{stream}\OperatorTok{);}
\end{Highlighting}
\end{Shaded}

\chapter{Dispatching without Core ML}\label{dispatching-without-core-ml}

\begin{summarybox}

Five steps reach the engine from ordinary user space: build a graph,
compile it to the program format, load it, bind operand buffers, and
dispatch. No placement planner is in the path, and the operations the
compiler accepts need no entitlement. One submission can drive several
on-engine steps with no host round-trip per step. It runs at the same
rate as the steps issued one call at a time, so the unentitled path is
performance-complete for dispatch.

\end{summarybox}

\section{Five steps}\label{five-steps}

Build a network description first. This is the layer-and-wiring graph
for the work: the operations, their parameters, the weight blobs, and
the input and output ports. The runtime accepts the description
directly, so the graph does not have to come from a trained model file
or pass through a converter.

Compile that description to the engine's program format. The runtime
drives the compiler, which lowers the graph, lays the weights out for
the streaming datapath, validates every operation against the
native-layer catalog, and writes a loadable program to a
content-addressed cache on disk. The cost is here, and a network pays it
once.

Load the compiled program. The runtime opens the program from the cache
and instantiates the function it exposes, preparing the program for
execution on the device. A freshly compiled program pays a one-time cost
on this first load to produce the loadable hardware form; a program
already in the cache is a hit.

Bind operand buffers. Each external input and output of the program is a
named port. Binding attaches a buffer to each port so the engine reads
its inputs and writes its outputs in memory the caller prepared.

Dispatch last. The caller encodes the loaded program as an operation
into an execution stream and submits the stream, synchronously or
asynchronously, then reads the output buffers back when the work
completes. Dispatch is the low-cost phase, and it runs in the hot loop.

The runtime that runs these steps is the engine's own C dispatch layer,
the \texttt{e5rt\_*} family exported from the Espresso framework. Every
entry point returns an \texttt{int64\_t} error code, zero on success,
and an entry point that creates an object returns it through its first
out-parameter, while the compile and retain calls return through their
last. The five steps map onto the call sequence in
\cref{lst:c6-direct-route}.

\begin{listing}[H]
\caption{The direct route, from compiling a network to dispatching it on an execution stream.}\label{lst:c6-direct-route}

\begin{Shaded}
\begin{Highlighting}[]
\CommentTok{/* Compile (step 2): the runtime drives the out{-}of{-}process compiler. */}
\NormalTok{e5rt\_e5\_compiler\_create\_with\_config}\OperatorTok{(\&}\NormalTok{compiler}\OperatorTok{,}\NormalTok{ config}\OperatorTok{);}
\NormalTok{e5rt\_e5\_compiler\_compile}\OperatorTok{(}\NormalTok{compiler}\OperatorTok{,}\NormalTok{ model\_path}\OperatorTok{,}\NormalTok{ options}\OperatorTok{,} \OperatorTok{\&}\NormalTok{library}\OperatorTok{);}

\CommentTok{/* Load (step 3): retain the callable function the program exposes. */}
\NormalTok{e5rt\_program\_library\_retain\_program\_function}\OperatorTok{(}\NormalTok{library}\OperatorTok{,}\NormalTok{ fn\_name}\OperatorTok{,} \OperatorTok{\&}\NormalTok{function}\OperatorTok{);}
\NormalTok{e5rt\_precompiled\_compute\_op\_create\_options\_create\_with\_program\_function}\OperatorTok{(\&}\NormalTok{op\_opts}\OperatorTok{,}\NormalTok{ function}\OperatorTok{);}
\NormalTok{e5rt\_execution\_stream\_operation\_create\_precompiled\_compute\_operation\_with\_options}\OperatorTok{(\&}\NormalTok{op}\OperatorTok{,}\NormalTok{ op\_opts}\OperatorTok{);}

\CommentTok{/* Bind (step 4): attach a CPU buffer object to each named I/O port. */}
\NormalTok{e5rt\_buffer\_object\_alloc}\OperatorTok{(\&}\NormalTok{buf}\OperatorTok{,}\NormalTok{ nbytes}\OperatorTok{,} \CommentTok{/*type=*/}\DecValTok{0}\OperatorTok{);}
\NormalTok{e5rt\_execution\_stream\_operation\_retain\_input\_port}\OperatorTok{(}\NormalTok{op}\OperatorTok{,} \StringTok{"x"}\OperatorTok{,} \OperatorTok{\&}\NormalTok{in\_port}\OperatorTok{);}
\NormalTok{e5rt\_io\_port\_bind\_buffer\_object}\OperatorTok{(}\NormalTok{in\_port}\OperatorTok{,}\NormalTok{ buf}\OperatorTok{);}

\CommentTok{/* Dispatch (step 5): encode the op into a stream and submit, in the hot loop. */}
\NormalTok{e5rt\_execution\_stream\_create}\OperatorTok{(\&}\NormalTok{stream}\OperatorTok{);}
\ControlFlowTok{for} \OperatorTok{(;;)} \OperatorTok{\{}
    \CommentTok{/* fill input buffer via e5rt\_buffer\_object\_get\_data\_ptr(buf, \&ptr) */}
\NormalTok{    e5rt\_execution\_stream\_operation\_prepare\_op\_for\_encode}\OperatorTok{(}\NormalTok{op}\OperatorTok{);}
\NormalTok{    e5rt\_execution\_stream\_encode\_operation}\OperatorTok{(}\NormalTok{stream}\OperatorTok{,}\NormalTok{ op}\OperatorTok{);}
\NormalTok{    e5rt\_execution\_stream\_execute\_sync}\OperatorTok{(}\NormalTok{stream}\OperatorTok{);}   \CommentTok{/* or async\_submit */}
    \CommentTok{/* read output buffer back via its data pointer */}
\NormalTok{    e5rt\_execution\_stream\_reset}\OperatorTok{(}\NormalTok{stream}\OperatorTok{);}
\OperatorTok{\}}
\end{Highlighting}
\end{Shaded}

\end{listing}

A network expressed for the compiler is a netplist, the
\texttt{.espresso.net} representation the compiler accepts alongside
\texttt{.mil}. Its layers are dictionary entries the binary calls Units,
keyed under a network body, with a \texttt{ProcedureList} of callable
entry points naming the \texttt{InputList}, \texttt{OperationList}, and
\texttt{OutputList}. A single matmul lowers to the native
\texttt{MatrixMultLayerDesc}, whose backend operation is
\texttt{anec.matmul} with \texttt{transpose\_lhs} and
\texttt{transpose\_rhs} attributes.

The netplist keys are dictionary string constants the compiler parses
verbatim, recovered from the runtime binary. They name the network body
(\texttt{Networks}, \texttt{NetworkName}, \texttt{ProcedureList},
\texttt{Units}, \texttt{Weights}), the per-unit wiring (\texttt{Name},
\texttt{OperationName}, \texttt{OperationList}, \texttt{Bottom},
\texttt{Top}), the external ports (\texttt{InputList},
\texttt{OutputList}, \texttt{InputName}, \texttt{OutputName},
\texttt{InputType}, \texttt{OutputType}), and the port shape, a
five-tuple plus an interleave factor and a dtype (\texttt{BatchSize},
\texttt{InputChannels}, \texttt{InputDepth}, \texttt{InputHeight},
\texttt{InputWidth}, \texttt{InputInterleave}, \texttt{OutputChannels},
\texttt{OutputInterleave}). Chapter \hyperref[compiler]{22} lists each
key with its role.

A Unit is one layer. It has a \texttt{Type} tag, a \texttt{Bottom}
wiring list naming its input symbols, the per-bottom and output dtypes,
and a \texttt{Params} sub-dictionary of typed attributes. The
directed-graph wiring is by symbol name, a Unit's \texttt{Bottom}
referencing another Unit's \texttt{Name} or an external
\texttt{InputName}. The execution order is the network body's
\texttt{Units} array. \Cref{tbl:c6-call-families} maps the
execution-runtime call families onto the phases of the direct route.

\begin{longtable}[]{@{}
  >{\raggedright\arraybackslash}p{(\linewidth - 4\tabcolsep) * \real{0.3333}}
  >{\raggedright\arraybackslash}p{(\linewidth - 4\tabcolsep) * \real{0.3333}}
  >{\raggedright\arraybackslash}p{(\linewidth - 4\tabcolsep) * \real{0.3333}}@{}}
\caption{The execution-runtime call families, mapped onto the compile,
load, bind, and dispatch phases of the direct
route.}\label{tbl:c6-call-families}\tabularnewline
\toprule\noalign{}
\begin{minipage}[b]{\linewidth}\raggedright
\textbf{Family}
\end{minipage} & \begin{minipage}[b]{\linewidth}\raggedright
\textbf{Phase}
\end{minipage} & \begin{minipage}[b]{\linewidth}\raggedright
\textbf{Representative entry points}
\end{minipage} \\
\midrule\noalign{}
\endfirsthead
\toprule\noalign{}
\begin{minipage}[b]{\linewidth}\raggedright
\textbf{Family}
\end{minipage} & \begin{minipage}[b]{\linewidth}\raggedright
\textbf{Phase}
\end{minipage} & \begin{minipage}[b]{\linewidth}\raggedright
\textbf{Representative entry points}
\end{minipage} \\
\midrule\noalign{}
\endhead
\bottomrule\noalign{}
\endlastfoot
\texttt{e5rt\_e5\_compiler\_*} & compile &
\texttt{create\_with\_config}, \texttt{compile},
\texttt{is\_new\_compile\_required} \\
\rowcolor{zebra}\texttt{e5rt\_program\_library\_*},
\texttt{e5rt\_program\_function\_*} & load & \texttt{create},
\texttt{retain\_program\_function}, \texttt{load\_for\_execution} \\
\texttt{e5rt\_precompiled\_compute\_op\_*} & load &
\texttt{create\_options\_create\_with\_program\_function} \\
\rowcolor{zebra}\texttt{e5rt\_buffer\_object\_*},
\texttt{e5rt\_io\_port\_*} & bind & \texttt{alloc},
\texttt{get\_data\_ptr}, \texttt{bind\_buffer\_object} \\
\texttt{e5rt\_execution\_stream\_*} & dispatch &
\texttt{encode\_operation}, \texttt{execute\_sync},
\texttt{submit\_async}, \texttt{reset} \\
\end{longtable}

\begin{figure}[H]\centering
\includegraphics[width=5.293in]{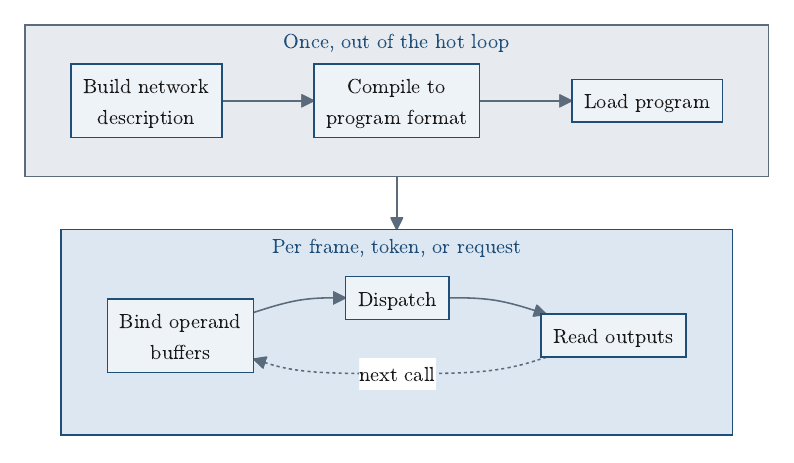}
\caption{The compile-once, dispatch-many split between one-time setup and the hot loop.}
\label{fig:c6-once-loop}
\end{figure}

One network is compiled and loaded a single time, and the same program
is dispatched against fresh operands on every call
(\cref{fig:c6-once-loop}).

\section{Compiler and runtime
options}\label{compiler-and-runtime-options}

Compiler and runtime options are not a fixed-layout struct. They are a
string-keyed dictionary of \texttt{std::any} values, which is why
setting an option changes no fixed byte offset in the option handle.
Each key has its value type in the name, for example
\texttt{forceRecompilation\textless{}bool\textgreater{}},
\texttt{segmenter\textless{}std::string\textgreater{}}, and
\texttt{computeDeviceTypesAllowed\textless{}std::vector\textless{}ComputeDeviceType\textgreater{}\textgreater{}},
and the compute-device mask value \texttt{0x4} selects the engine.

Most of the runtime is reachable but unused by an ordinary dispatch. The
runtime exposes 292 exported entry points, of which about 275 are
callable and about 52 are exercised by a normal dispatch. The largest
reachable but unused capabilities are zero-copy input and output through
an image surface or a graphics buffer, firmware-side data chaining, a
warm-start cache, a quality-of-service control, mutable weights, and an
asynchronous submission with a timeout.

\section{No planner and no
entitlement}\label{no-planner-and-no-entitlement}

The direct route has no placement planner. The public framework path
loads a model and selects engine eligibility through a compute-unit
setting, segments a model across the processors, and decides per segment
where the work runs, and the caller is not told the result
\hyperref[ref-applecoreml]{{[}AppleCoreML{]}}. On the direct route the
caller has already decided: the network is compiled for the engine and
dispatched to the engine, with no segmentation step in between.

The operations the compiler accepts require no special entitlement. The
runtime, the compiler, and the dispatch path are all reachable from
ordinary user space. The set of operations is the catalog the compiler
validates: an operation the compiler accepts compiles and dispatches
without a privileged handle, and the compiler rejects an operation
outside that set at compile time. Reachable here means no entitlement is
required, not that the path is supported: these interfaces are private,
unsupported, version-fragile across operating-system updates, and not
App-Store-safe. The boundary that does require an entitlement is a
separate matter, taken up in chapter \hyperref[entitlement-boundary]{8}.

\section{Unentitled path is
performance-complete}\label{unentitled-path-is-performance-complete}

Host dispatch is not the limiting cost on the direct path. A single
submission can drive several on-engine steps without a host round-trip
per step. The mechanism is the resident state from chapter
\hyperref[execution-model]{2}: a buffer stays in the engine's working
set across steps, so the output of one step is the input of the next in
place. The submission advances through the steps without returning to
the host between them. The host supplies only the small per-step inputs
and reads the held buffers back at a checkpoint.

The same steps issued one host call at a time run at the same rate, so
removing the per-step host calls does not speed the work up. The
unentitled direct path is performance-complete for dispatch: it reaches
the same throughput as any path with more privilege.

The stream offers one synchronous and two asynchronous submission forms.
The synchronous form blocks the caller until the encoded stream
completes. The lightweight asynchronous form returns a submit and a
complete identifier, and the full asynchronous form delivers an error
object and accepts a timeout, the timeout existing because an
asynchronous completion can hang. Single-stream pipelining is sound:
keep several operations in flight, each with its own completion event,
and drain by waiting on the events. Overlapping two or more streams at
once is the unsound path, where the completion event of the second and
later streams never notifies and the waiter blocks; the
low-latency-async-event stream option and the timeout argument are the
controls that break that wait. The caller keeps resident state across
steps by binding a buffer object once and reusing it: the output port of
one step is the input port of the next in the same buffer. The caller
re-encodes the prepared operation against fresh small inputs without
rebinding the held buffer.

\section{Compiling once, dispatching
many}\label{compiling-once-dispatching-many}

The five real steps collapse to two phases against the neutral API:
compile and load once out of the hot loop, then bind and dispatch
against fresh operands on every call. The caller reuses the compiled
program across calls, so the costly compile phase runs a single time.

Build and compile are above the loop; bind and dispatch are inside it.

\begin{Shaded}
\begin{Highlighting}[]
\CommentTok{/* Once, above the loop: compile to a program library, then retain the callable function. */}
\NormalTok{e5rt\_e5\_compiler\_compile}\OperatorTok{(}\NormalTok{compiler}\OperatorTok{,}\NormalTok{ model\_path}\OperatorTok{,}\NormalTok{ options}\OperatorTok{,} \OperatorTok{\&}\NormalTok{library}\OperatorTok{);}
\NormalTok{e5rt\_program\_library\_retain\_program\_function}\OperatorTok{(}\NormalTok{library}\OperatorTok{,}\NormalTok{ fn\_name}\OperatorTok{,} \OperatorTok{\&}\NormalTok{function}\OperatorTok{);}
\NormalTok{e5rt\_precompiled\_compute\_op\_create\_options\_create\_with\_program\_function}\OperatorTok{(\&}\NormalTok{op\_opts}\OperatorTok{,}\NormalTok{ function}\OperatorTok{);}
\NormalTok{e5rt\_execution\_stream\_operation\_create\_precompiled\_compute\_operation\_with\_options}\OperatorTok{(\&}\NormalTok{op}\OperatorTok{,}\NormalTok{ op\_opts}\OperatorTok{);}
\NormalTok{e5rt\_execution\_stream\_create}\OperatorTok{(\&}\NormalTok{stream}\OperatorTok{);}      \CommentTok{/* ports bound once via e5rt\_io\_port\_bind\_buffer\_object */}

\ControlFlowTok{for} \OperatorTok{(;;)} \OperatorTok{\{}                                  \CommentTok{/* the hot loop: dispatch the same op per call */}
\NormalTok{    e5rt\_execution\_stream\_operation\_prepare\_op\_for\_encode}\OperatorTok{(}\NormalTok{op}\OperatorTok{);}
\NormalTok{    e5rt\_execution\_stream\_encode\_operation}\OperatorTok{(}\NormalTok{stream}\OperatorTok{,}\NormalTok{ op}\OperatorTok{);}
\NormalTok{    e5rt\_execution\_stream\_execute\_sync}\OperatorTok{(}\NormalTok{stream}\OperatorTok{);}
\NormalTok{    e5rt\_execution\_stream\_reset}\OperatorTok{(}\NormalTok{stream}\OperatorTok{);}    \CommentTok{/* read outputs back, then reset for the next call */}
\OperatorTok{\}}
\end{Highlighting}
\end{Shaded}

\chapter{Weights and compression}\label{weights-and-compression}

\begin{summarybox}

The engine reconstructs four compressed weight forms at the multiplier
input. On the unentitled direct route a form either streams its
compressed bytes for a bandwidth gain or folds to dense fp16 for none.
The target chip decides which. The M1 streams only int4 lookup-table, at
2.37 times fp16, and structured sparsity at 1.55 to 1.64 times; it folds
int8 and blockwise affine. The M5 streams all four at 1.6 to 1.8 times.
Choose the form by what streams on the target. On the M1, fall back to
int8 only where its halved stored size pays, since the int8 fold expands
to fp16 in DRAM and yields no bandwidth there.

\end{summarybox}

A compiled network stores its weights as a separate stream that the
engine reads from DRAM on every dispatch. The weight stream is the
primary cost of any layer whose arithmetic intensity is low, which
covers most decode-shaped and projection-heavy work. Compressing the
weights moves fewer bytes across that stream; it does more than shrink a
stored file.

\section{Compression is a bandwidth feature on the direct
route}\label{compression-is-a-bandwidth-feature-on-the-direct-route}

Compressed weights reach the engine through the direct runtime described
in chapter \hyperref[software-stack]{5}. They are not gated behind an
entitlement, and the operations that reconstruct them are accepted by
the compiler without special privilege. The reconstruction happens at
the multiplier input, consistent with the fp16 datapath of chapter
\hyperref[numerics]{3}: a compressed weight is turned back into fp16
before it reaches the multiply array, and the multiply that follows is
the same fp16 multiply as for an uncompressed weight.

The word compression covers two distinct outcomes. A form that streams
reaches the engine in its compressed bytes and is decompressed on chip,
so fewer bytes cross DRAM and the layer runs faster when it was
bandwidth bound. A form that folds is reconstructed to a dense fp16
constant before the dispatch, so the bytes that cross DRAM are
full-width fp16 and the layer gets no bandwidth gain. The target chip
sets which outcome applies to a form, not the frontend.

\section{Compression forms}\label{compression-forms}

The engine reconstructs four weight-compression forms, the same linear
quantization, palettization, and pruning the conversion tools document
as a model-size feature
\hyperref[ref-applecoremltools]{{[}AppleCoreMLTools{]}}. Per-tensor and
per-channel int8 keep the weight as a single byte per element with an
fp16 scale, one scale for the whole tensor or one per output channel,
reconstructed as the scale times the byte. The reconstruction is the
affine dequantization

\[w = s\,(q - z)\]

with stored quantized byte \(q\), fp16 scale \(s\), and zero point
\(z\), the encode being \(q = \mathrm{round}(x / s) + z\). The zero
point folds to zero on the M1 generation, so the int8 form there is
symmetric and the relation reduces to \(w = s\,q\) with \(z = 0\)
(\cref{lst:c7-int8-dequant}).

\begin{listing}[H]
\caption{The affine int8 dequantization evaluated at the multiplier input, with the symmetric form the M1 reduces to.}\label{lst:c7-int8-dequant}

\begin{Shaded}
\begin{Highlighting}[]
\CommentTok{\# affine int8 dequant, evaluated at the multiplier input}
\NormalTok{w\_fp16 }\OperatorTok{=}\NormalTok{ scale }\OperatorTok{*}\NormalTok{ (q }\OperatorTok{{-}}\NormalTok{ zero\_point)        }\CommentTok{\# general affine form}
\NormalTok{w\_fp16 }\OperatorTok{=}\NormalTok{ scale }\OperatorTok{*}\NormalTok{ q                       }\CommentTok{\# M1: zero\_point folds to 0, symmetric only}
\CommentTok{\# q is int8 in [{-}127, 127]; scale is fp16, scalar or one per output channel.}
\end{Highlighting}
\end{Shaded}

\end{listing}

Each compressed form is declared to the compiler as a single
\texttt{constexpr\_*} reconstruction op that folds into the weight
descriptor, one per form (\cref{lst:c7-constexpr-ops}). The op count per
form is fixed, and a violation is a hard validation error.

\begin{listing}[H]
\caption{The single reconstruction op that declares each compressed weight form to the compiler.}\label{lst:c7-constexpr-ops}

\begin{Shaded}
\begin{Highlighting}[]
\CommentTok{\# the constexpr reconstruction op per form (one MIL op each, folded into the conv/linear weight)}
\NormalTok{constexpr\_affine\_dequantize(q, scale, zero\_point)   }\CommentTok{\# int8 affine:  w = scale * (q {-} zero\_point)}
\NormalTok{constexpr\_lut\_to\_dense(indices, lut)                }\CommentTok{\# int4 palette: w = lut[indices]}
\NormalTok{constexpr\_sparse\_to\_dense(mask, nonzeros)           }\CommentTok{\# sparsity:     scatter into mask positions}
\NormalTok{constexpr\_blockwise\_shift\_scale(q, scale)           }\CommentTok{\# blockwise:    one scale per contiguous block}
\end{Highlighting}
\end{Shaded}

\end{listing}

The int4 lookup-table form uses a four-bit index per element into a
sixteen-entry fp16 codebook, and reconstruction is a table lookup with
no arithmetic since the codebook is already fp16. Structurally it is a
palette rather than an arithmetic dequantization, which is why it
behaves differently from int8 on the streaming question.

A worked example shows the packing and the decode. The four weights
\texttt{{[}1.0,\ 0.0,\ 0.0,\ 1.0{]}} index a sixteen-entry fp16 codebook
whose first two slots contain \texttt{0.0} (entry 0, \texttt{0x0000})
and \texttt{1.0} (entry 1, \texttt{0x3c00}). The index stream
\texttt{{[}1,\ 0,\ 0,\ 1{]}} packs two four-bit indices per byte, low
nibble first, so the four weights occupy the two bytes \texttt{0x01} and
\texttt{0x10}. Decoding reads each nibble as a table index, recovering
\texttt{{[}1.0,\ 0.0,\ 0.0,\ 1.0{]}}.

Structured sparsity has a one-bit mask marking the nonzero positions
plus the packed fp16 values of those nonzeros. The mask costs one bit
per element and the values cost two bytes per surviving element, so a
weight that is half zeros or more stores well below its dense size.
Reconstruction scatters the values back into the masked positions
exactly apart from the fp16 rounding of the kept values. Blockwise
affine gives a separate scale to each contiguous block of elements,
finer than a per-channel scale and so lower in quantization error.

\section{What streams on each family}\label{what-streams-on-each-family}

\Cref{tbl:c7-streams} gives the per-form outcome on each chip
generation.

\begin{longtable}[]{@{}
  >{\raggedright\arraybackslash}p{(\linewidth - 10\tabcolsep) * \real{0.1667}}
  >{\raggedright\arraybackslash}p{(\linewidth - 10\tabcolsep) * \real{0.1667}}
  >{\raggedright\arraybackslash}p{(\linewidth - 10\tabcolsep) * \real{0.1667}}
  >{\raggedright\arraybackslash}p{(\linewidth - 10\tabcolsep) * \real{0.1667}}
  >{\raggedright\arraybackslash}p{(\linewidth - 10\tabcolsep) * \real{0.1667}}
  >{\raggedright\arraybackslash}p{(\linewidth - 10\tabcolsep) * \real{0.1667}}@{}}
\caption{Whether each compressed weight form streams or folds to dense
per chip generation, with the measured
speedup.}\label{tbl:c7-streams}\tabularnewline
\toprule\noalign{}
\begin{minipage}[b]{\linewidth}\raggedright
\textbf{Form}
\end{minipage} & \begin{minipage}[b]{\linewidth}\raggedright
\textbf{M1/H13}
\end{minipage} & \begin{minipage}[b]{\linewidth}\raggedright
\textbf{A14/M2}
\end{minipage} & \begin{minipage}[b]{\linewidth}\raggedright
\textbf{A15/M3 and later}
\end{minipage} & \begin{minipage}[b]{\linewidth}\raggedright
\textbf{M5/H17s}
\end{minipage} & \begin{minipage}[b]{\linewidth}\raggedright
\textbf{Speedup}
\end{minipage} \\
\midrule\noalign{}
\endfirsthead
\toprule\noalign{}
\begin{minipage}[b]{\linewidth}\raggedright
\textbf{Form}
\end{minipage} & \begin{minipage}[b]{\linewidth}\raggedright
\textbf{M1/H13}
\end{minipage} & \begin{minipage}[b]{\linewidth}\raggedright
\textbf{A14/M2}
\end{minipage} & \begin{minipage}[b]{\linewidth}\raggedright
\textbf{A15/M3 and later}
\end{minipage} & \begin{minipage}[b]{\linewidth}\raggedright
\textbf{M5/H17s}
\end{minipage} & \begin{minipage}[b]{\linewidth}\raggedright
\textbf{Speedup}
\end{minipage} \\
\midrule\noalign{}
\endhead
\bottomrule\noalign{}
\endlastfoot
int8 (per-tensor / per-channel) & fold & stream & stream & stream & M5
1.6-1.8x; M1 folds, no stream gain \\
\rowcolor{zebra}int4 lookup-table & stream & stream & stream & stream &
M1 2.37x; M5 1.6-1.8x \\
structured sparsity & stream & stream & stream & stream & M1 1.55-1.64x
at 0.43x dense bytes; M5 1.6-1.8x \\
\rowcolor{zebra}blockwise affine & fold & fold & stream & stream & M5
1.6-1.8x; M1 and M2 fold, no stream gain \\
\end{longtable}

On the M1, of the H13 generation, only the int4 lookup-table form
streams natively. A bandwidth-bound stack of one-by-one convolutions
runs 2.37 times faster with int4 weights than with fp16, measured on the
M1, because the four-bit indices move at a quarter of the fp16 byte
count. Structured sparsity also streams natively on the M1: a
convolution stack that is about sixty-three percent zeros runs 1.55 to
1.64 times faster than the same weights stored dense, at 0.43 times the
dense weight bytes. The output is bit-faithful to the dense reference
apart from fp16 rounding.

The int8 and blockwise forms fold on the M1. The accuracy cost is the
per-output int8 quantization: the int8 weight tracks the fp16 result at
a cosine near 1.0, with a relative error near one percent against an
fp32 reference; fp16 against the same reference is near two parts in ten
thousand.

From the A14 generation the int8 and sparse forms also stream, alongside
int4, measured on the M2 of the H14 generation at 0.64 and 0.54 times
fp16 on a bandwidth-bound matmul. On the A14 and M2 the int8 weight is
dispatched as int8 and reconstructed from half the bytes. The measured
latency runs from 0.85 times fp16 at a two-thousand-wide weight down to
0.52 times fp16 at an eight-thousand-wide weight, deeper as the weight
grows and the stream dominates. Blockwise affine does not stream on the
A14: it still folds there, measured at 0.985 times fp16, a near-zero
bandwidth gain, and it first streams from the A15 generation. On the M5,
of the H17s generation, all four forms stream, at a measured 1.6 to 1.8
times fp16 on bandwidth-bound layers.

The streaming-versus-folding split is a hardware-abstraction-layer
decision. Every weight-bearing operation is legal on every chip, so the
boundary is in a set of feature bytes the compiler reads from the
per-chip table. In that table one master byte enables weight streaming
at all and a cluster of per-format bytes admits each compressed form by
generation. \Cref{tbl:c7-hal-bytes} gives those feature bytes and the
generation each switches on.

\begin{longtable}[]{@{}
  >{\raggedright\arraybackslash}p{(\linewidth - 10\tabcolsep) * \real{0.1364}}
  >{\raggedright\arraybackslash}p{(\linewidth - 10\tabcolsep) * \real{0.1364}}
  >{\raggedleft\arraybackslash}p{(\linewidth - 10\tabcolsep) * \real{0.1818}}
  >{\raggedleft\arraybackslash}p{(\linewidth - 10\tabcolsep) * \real{0.1818}}
  >{\raggedleft\arraybackslash}p{(\linewidth - 10\tabcolsep) * \real{0.1818}}
  >{\raggedleft\arraybackslash}p{(\linewidth - 10\tabcolsep) * \real{0.1818}}@{}}
\caption{The hardware-abstraction-layer feature bytes that gate
compressed-weight streaming, by chip
generation.}\label{tbl:c7-hal-bytes}\tabularnewline
\toprule\noalign{}
\begin{minipage}[b]{\linewidth}\raggedright
\textbf{Hardware-abstraction-layer byte}
\end{minipage} & \begin{minipage}[b]{\linewidth}\raggedright
\textbf{Role}
\end{minipage} & \begin{minipage}[b]{\linewidth}\raggedleft
\textbf{M1/H13}
\end{minipage} & \begin{minipage}[b]{\linewidth}\raggedleft
\textbf{A14/M2}
\end{minipage} & \begin{minipage}[b]{\linewidth}\raggedleft
\textbf{A15/M3}
\end{minipage} & \begin{minipage}[b]{\linewidth}\raggedleft
\textbf{A16, A17, M5}
\end{minipage} \\
\midrule\noalign{}
\endfirsthead
\toprule\noalign{}
\begin{minipage}[b]{\linewidth}\raggedright
\textbf{Hardware-abstraction-layer byte}
\end{minipage} & \begin{minipage}[b]{\linewidth}\raggedright
\textbf{Role}
\end{minipage} & \begin{minipage}[b]{\linewidth}\raggedleft
\textbf{M1/H13}
\end{minipage} & \begin{minipage}[b]{\linewidth}\raggedleft
\textbf{A14/M2}
\end{minipage} & \begin{minipage}[b]{\linewidth}\raggedleft
\textbf{A15/M3}
\end{minipage} & \begin{minipage}[b]{\linewidth}\raggedleft
\textbf{A16, A17, M5}
\end{minipage} \\
\midrule\noalign{}
\endhead
\bottomrule\noalign{}
\endlastfoot
\texttt{+0x48f} & kernel-streaming master & 1 & 1 & 1 & 1 \\
\rowcolor{zebra}\texttt{+0x528}, \texttt{+0x532}, \texttt{+0x537} &
per-format gates that switch on at A14 & 0 & 1 & 1 & 1 \\
\texttt{+0x520}, \texttt{+0x523}, \texttt{+0x533}, \texttt{+0x539} &
per-format gates that switch on at A15 & 0 & 0 & 1 & 1 \\
\rowcolor{zebra}\texttt{+0x529} & palette and stride gate & 1 & 1 & 1 &
1 \\
\end{longtable}

The master byte is set from the A13 generation, which is why the M1
streams anything at all, and the palette gate is set on the M1, which is
why the int4 lookup-table form streams there. The per-format gates for
the affine int8 and blockwise forms are clear on the M1 and switch on at
the A14 and A15 generations, which is why those two forms fold on the M1
and stream on the newer parts. Structured sparsity streams on the M1 by
a separate route: it is held as a mask-and-values weight operand under
the master byte rather than as a palettized kernel coefficient, so it
streams under the master gate independent of the per-format cluster. The
int8 floor at A14 is confirmed on the M2 silicon, and the blockwise
floor at A15 is read from that gate pattern and not yet confirmed on the
intermediate silicon.

On the M1 the int8 fold is a stored-size saving only. The weight is half
the size on disk, but it is expanded to a dense fp16 constant in DRAM
before the data-movement step, so a weight-streaming-bound matmul moves
full-width fp16 bytes and runs at the fp16 latency, with no bandwidth
gain. The int8 weight first streams as int8 on the A14 and M2
generation, where it is dispatched as int8 and dequantized at the
multiplier input rather than materialized to a dense fp16 constant in
DRAM. A matmul-path measurement on the A14 and M2 makes this concrete:
at a weight matrix wide enough to leave the dispatch floor, the int8 and
fp16 weight streams reach the same effective bandwidth against their
stored bytes, while the int8 form moves half the bytes
(\cref{tbl:c7-int8-matmul}).

\begin{longtable}[]{@{}lrrr@{}}
\caption{The int8 matmul-path weight stream against fp16 on the A14 and
M2, the latency ratio approaching one half as the weight
grows.}\label{tbl:c7-int8-matmul}\tabularnewline
\toprule\noalign{}
\textbf{Weight width \(K=N\)} & \textbf{fp16 latency} & \textbf{int8
latency} & \textbf{int8 / fp16} \\
\midrule\noalign{}
\endfirsthead
\toprule\noalign{}
\textbf{Weight width \(K=N\)} & \textbf{fp16 latency} & \textbf{int8
latency} & \textbf{int8 / fp16} \\
\midrule\noalign{}
\endhead
\bottomrule\noalign{}
\endlastfoot
2048 & 0.290 ms & 0.253 ms & 0.87 \\
\rowcolor{zebra}4096 & 0.740 ms & 0.447 ms & 0.60 \\
8192 & 2.610 ms & 1.351 ms & 0.52 \\
\end{longtable}

\section{Choosing a form}\label{choosing-a-form}

What streams on the target drives the choice of compression form, chip
by chip. On the M1, prefer structured sparsity for a weight that is half
zeros or more, since it streams and is lossless apart from fp16
rounding, and prefer the int4 lookup table otherwise, since it is the
densest form that streams there. Reserve int8 on the M1 for the case
where sixteen levels are too coarse, taking it for its halved stored
size. On the M5 and the newer generations, where every form streams,
choose by accuracy per byte rather than by what streams.

The wide accumulator of chapter \hyperref[numerics]{3} is what makes
streamed low-precision weights safe on the layers that tolerate them. A
streamed weight is reconstructed to fp16 and then enters the same fp16
multiply and wide reduction as a dense weight, so the quantization of
the weight itself is the only precision lost. On convolution, matrix
multiply, and normalization, whose partial sums stay in range, the
reduction holds and the compressed weight keeps the layer's accuracy. A
cancellation-heavy step has no fp16-safe form, compressed or not, for
the reason given in chapter \hyperref[numerics]{3}.

\section{Patching weights in a compiled
program}\label{patching-weights-in-a-compiled-program}

A host can patch a compiled program's weights in place without
recompiling. Each weight tensor occupies a decoded region of the program
image with a known tiling. A convolution weight is in a
\texttt{0xC0}-stride layout and a matrix-multiply weight in a
\texttt{0x40}-stride layout, and editing the weight values leaves the
program descriptor unchanged. A host can thus swap new weights into an
already-compiled program, which makes a weight-only update inexpensive.

The driver has a per-patch-mutable-buffer accounting path that confirms
the route, \texttt{ANEScheduler::pendingRequestsPerPatchMutableBuffer}.
The descriptor names the operations and binds the buffers, and a
weight-value edit touches neither, so the same compiled program runs
with the new coefficients.

\section{Automating the choice}\label{automating-the-choice}

The estimate classifies a layer as compute bound or bandwidth bound so
the procedure chooses a form only where the layer is bandwidth bound and
a stream would help. The procedure keeps the smallest form that streams
natively on the target and clears an accuracy tolerance against an fp32
reference. Which forms stream depends on the target: the M1 streams int4
and sparse while int8 and blockwise fold, and the A14 and later stream
int8 as well. Sparsity applies only when at least half the weight is
zero, and the candidates are tried smallest-bytes-first, int4 before
sparse before int8. Here \texttt{accuracy\_error} round-trips the weight
through the candidate form and compares the layer output against the
fp32 reference.

\begin{Shaded}
\begin{Highlighting}[]
\NormalTok{tolerance }\OperatorTok{=} \FloatTok{0.01}    \CommentTok{\# max relative error of the layer vs an fp32 reference}

\KeywordTok{def}\NormalTok{ native\_streams(chip):}
    \ControlFlowTok{if}\NormalTok{ chip }\OperatorTok{==}\NormalTok{ H13:   }\ControlFlowTok{return}\NormalTok{ [int4, sparse]       }\CommentTok{\# M1: int8 and blockwise fold}
    \ControlFlowTok{if}\NormalTok{ chip }\OperatorTok{\textgreater{}=}\NormalTok{ H14:   }\ControlFlowTok{return}\NormalTok{ [int4, sparse, int8]}

\KeywordTok{def}\NormalTok{ choose\_weight\_form(layer, weights, chip):}
    \ControlFlowTok{if} \KeywordTok{not}\NormalTok{ is\_bandwidth\_bound(layer, chip):}
        \ControlFlowTok{return}\NormalTok{ fp16}
\NormalTok{    candidates }\OperatorTok{=}\NormalTok{ native\_streams(chip)}
    \ControlFlowTok{if}\NormalTok{ fraction\_zero(weights) }\OperatorTok{\textless{}} \FloatTok{0.5}\NormalTok{:}
\NormalTok{        candidates.remove(sparse)}
    \ControlFlowTok{for}\NormalTok{ form }\KeywordTok{in} \BuiltInTok{sorted}\NormalTok{(candidates, key}\OperatorTok{=}\NormalTok{bytes\_per\_weight):}
        \ControlFlowTok{if}\NormalTok{ accuracy\_error(form, weights, layer) }\OperatorTok{\textless{}=}\NormalTok{ tolerance:}
            \ControlFlowTok{return}\NormalTok{ form}
    \ControlFlowTok{return}\NormalTok{ fp16}

\NormalTok{form }\OperatorTok{=}\NormalTok{ choose\_weight\_form(layer, W, chip}\OperatorTok{=}\NormalTok{H13)     }\CommentTok{\# M1 {-}\textgreater{} int4 (streams, 2.37x)}
\NormalTok{program }\OperatorTok{=} \BuiltInTok{compile}\NormalTok{(graph\_with(W, form), chip}\OperatorTok{=}\NormalTok{H13)}
\end{Highlighting}
\end{Shaded}

On the M1 the procedure chooses a folding form, int8 or blockwise, only
where the halved stored size or a finer scale pays on the matmul path,
not for a stream gain.

\section{Reference: per-form streaming and measured
cost}\label{reference-per-form-streaming-and-measured-cost}

The figures behind this chapter's streaming claims are collected in
\cref{tbl:c7-reference}.

\begin{longtable}[]{@{}
  >{\raggedright\arraybackslash}p{(\linewidth - 4\tabcolsep) * \real{0.3000}}
  >{\raggedright\arraybackslash}p{(\linewidth - 4\tabcolsep) * \real{0.3000}}
  >{\raggedleft\arraybackslash}p{(\linewidth - 4\tabcolsep) * \real{0.4000}}@{}}
\caption{The per-form streaming behavior and measured cost, by chip
generation.}\label{tbl:c7-reference}\tabularnewline
\toprule\noalign{}
\begin{minipage}[b]{\linewidth}\raggedright
\textbf{Constant}
\end{minipage} & \begin{minipage}[b]{\linewidth}\raggedright
\textbf{Generation}
\end{minipage} & \begin{minipage}[b]{\linewidth}\raggedleft
\textbf{Value}
\end{minipage} \\
\midrule\noalign{}
\endfirsthead
\toprule\noalign{}
\begin{minipage}[b]{\linewidth}\raggedright
\textbf{Constant}
\end{minipage} & \begin{minipage}[b]{\linewidth}\raggedright
\textbf{Generation}
\end{minipage} & \begin{minipage}[b]{\linewidth}\raggedleft
\textbf{Value}
\end{minipage} \\
\midrule\noalign{}
\endhead
\bottomrule\noalign{}
\endlastfoot
int4 lookup-table stream speedup & M1/H13 & 2.37x fp16 \\
\rowcolor{zebra}Structured sparsity stream speedup & M1/H13 & 1.55-1.64x
fp16 \\
Structured sparsity stored bytes & M1/H13 & 0.43x dense \\
\rowcolor{zebra}int8 fold on M1 (stored-size saving, no stream gain) &
M1/H13 & 1.0x fp16 latency \\
int8 stored size on disk & M1/H13 & 0.5x fp16 \\
\rowcolor{zebra}int8 accuracy versus fp32 reference & M1/H13 & cosine
near 1.0, relative error near 1\% \\
int8 matmul latency, 2k-wide weight & A14/M2 & 0.85x fp16 \\
\rowcolor{zebra}int8 matmul latency, 8k-wide weight & A14/M2 & 0.52x
fp16 \\
\end{longtable}

The compressed weight element types are a subset of the engine
element-type catalog the kernel descriptor records. The forms that bear
on weight encoding are the integer lanes, fp16 dense and codebook
entries, and packed palette indices. \Cref{tbl:c7-element-types} gives
those entries of the catalog.

\begin{longtable}[]{@{}lrl@{}}
\caption{The weight-relevant entries of the engine element-type catalog,
with the width each has.}\label{tbl:c7-element-types}\tabularnewline
\toprule\noalign{}
\textbf{Element type} & \textbf{Width} & \textbf{Role in weight
encoding} \\
\midrule\noalign{}
\endfirsthead
\toprule\noalign{}
\textbf{Element type} & \textbf{Width} & \textbf{Role in weight
encoding} \\
\midrule\noalign{}
\endhead
\bottomrule\noalign{}
\endlastfoot
int8 & 1 byte & affine int8 lane \\
\rowcolor{zebra}uint8 & 1 byte & affine uint8 lane \\
float16 & 2 bytes & dense weight, codebook entry, scale, and bias \\
\rowcolor{zebra}uint4 & 4 bits & 4-bit palette index \\
int4 & 4 bits & 4-bit signed, palette-only on the M1 \\
\rowcolor{zebra}e4m3 & 1 byte & fp8, gated off on the M1 \\
\end{longtable}

There is no int4 arithmetic lane in the datapath, so the four-bit value
is always a palette index into the sixteen-entry fp16 codebook, which is
why the element-type table marks int4 palette-only on the M1.

\chapter{Entitlement boundary}\label{entitlement-boundary}

\begin{summarybox}

The direct route reaches the engine's useful compute. Four features are
behind the framework loader or an entitlement: three-dimensional
convolution, native stateful types, bf16 program input and output, and
flexible or symbolic shapes. Each passes an earlier layer and fails at
the one that runs the work. None of the four runs on the engine on the
M1 under either route. One hard limit is below all of this: a hand-built
or self-compiled program is rejected at load with error
\texttt{0xe00002e2}. The reachable surface is therefore everything the
system daemon's compiler will accept.

\end{summarybox}

The direct route reaches the engine through the execution runtime, below
the model framework, with no placement planner in the path. The
sanctioned route reaches it through the model framework, which adds a
segmenter, a model container, and a set of loader features the direct
route does not have. The direct route reaches the engine's useful
compute; four features are behind the framework loader or behind an
entitlement. This boundary is about technical reachability alone;
whether a path is supported by Apple or permitted for distribution is a
separate and stricter matter, treated in the front-matter status note.

\section{What the direct route
reaches}\label{what-the-direct-route-reaches}

The operations that on-device perception and numerics networks are built
from compile and run on the direct route without any entitlement.
Two-dimensional convolution and its transpose, matrix multiply, fused
attention, the normalizations, the activation tables, pooling,
elementwise arithmetic, and the data-movement operations all run,
bounded only by the per-chip limits of chapter
\hyperref[capability-surface]{4}. Compressed weights stream where the
family supports it and reconstruct to fp16 where it does not. The
compiler accepts these operations and the runtime dispatches them, and
an entitlement gates none of them. An entitlement gates a separate set
of loader-tier features instead.

\section{Features that are gated}\label{features-that-are-gated}

Four features are attested in the hardware capability tables or
recognized by the compiler frontend, yet do not run on the direct route.
Each is gated, and the gate is at a different layer in each case
(\cref{tbl:c8-gated}).

\begin{longtable}[]{@{}
  >{\raggedright\arraybackslash}p{(\linewidth - 6\tabcolsep) * \real{0.2500}}
  >{\raggedright\arraybackslash}p{(\linewidth - 6\tabcolsep) * \real{0.2500}}
  >{\raggedright\arraybackslash}p{(\linewidth - 6\tabcolsep) * \real{0.2500}}
  >{\raggedright\arraybackslash}p{(\linewidth - 6\tabcolsep) * \real{0.2500}}@{}}
\caption{Features blocked on the direct path, what gates each, and how
the direct and framework routes
behave.}\label{tbl:c8-gated}\tabularnewline
\toprule\noalign{}
\begin{minipage}[b]{\linewidth}\raggedright
\textbf{Feature}
\end{minipage} & \begin{minipage}[b]{\linewidth}\raggedright
\textbf{Gated by}
\end{minipage} & \begin{minipage}[b]{\linewidth}\raggedright
\textbf{Direct path}
\end{minipage} & \begin{minipage}[b]{\linewidth}\raggedright
\textbf{Entitled/framework path}
\end{minipage} \\
\midrule\noalign{}
\endfirsthead
\toprule\noalign{}
\begin{minipage}[b]{\linewidth}\raggedright
\textbf{Feature}
\end{minipage} & \begin{minipage}[b]{\linewidth}\raggedright
\textbf{Gated by}
\end{minipage} & \begin{minipage}[b]{\linewidth}\raggedright
\textbf{Direct path}
\end{minipage} & \begin{minipage}[b]{\linewidth}\raggedright
\textbf{Entitled/framework path}
\end{minipage} \\
\midrule\noalign{}
\endhead
\bottomrule\noalign{}
\endlastfoot
Three-dimensional convolution & Missing backend lowering on every device
mask & Frontend recognizes it, lowering fails with a not-implemented
rejection & Segmenter places it on the central or graphics processor,
not the engine \\
\rowcolor{zebra}Native stateful types (state, ring buffer) & Absent
counter-and-event direct-memory-access engine in the M1 descriptor &
Compile fails; reserved operation classes do not lower & Model container
wires the stateful types; engine still lacks the primitive on the M1 \\
bf16 program input and output & Serialization layer and datapath; bf16
is not among the eleven program-I/O dtype codes & Compile fails with an
unsupported-dtype rejection & Framework accepts the bf16 array, then
casts to fp16 before the engine segment \\
\rowcolor{zebra}Flexible and symbolic shapes & Runtime path, not the
compiler; the symbolic-shape gate is on for the M1 & A symbolic
dimension parses, then fails to lower & Compiles a small set of fixed
shapes ahead of time and dispatches the nearest \\
\end{longtable}

\Cref{lst:c8-attempts} shows where each is admitted and where it stops.

\begin{listing}[H]
\caption{Where each gated feature is admitted at one layer and rejected at the next, for three-dimensional convolution, a bf16 program output, and a symbolic shape.}\label{lst:c8-attempts}

\begin{Shaded}
\begin{Highlighting}[]
\CommentTok{\# 3D convolution: the frontend recognizes a static{-}weight 3D conv (it enforces}
\CommentTok{\# "3D Convolution does not support dynamic weights"); lowering then fails on every mask.}
\NormalTok{conv(}\BuiltInTok{input}\NormalTok{, weight, kernel}\OperatorTok{=}\NormalTok{[d, h, w])}
  \OperatorTok{{-}\textgreater{}}\NormalTok{ frontend: accepted}
  \OperatorTok{{-}\textgreater{}}\NormalTok{ lowering: }\StringTok{"Not implemented"}\NormalTok{: Some ops are }\KeywordTok{not}\NormalTok{ supported on }\BuiltInTok{any}\NormalTok{ of the specified backends}

\CommentTok{\# bf16 program input/output: not among the eleven program{-}I/O dtype codes.}
\CommentTok{\# ANECIRDataType = \{0 int4, 1 uint8, 2 int8, 3 fp16, 4 fp32, 5 int16, 6 uint16,}
\CommentTok{\#                   7 int32, 8 uint32, 9 int64, 10 uint64\}   \# bf16 absent}
\NormalTok{function\_output dtype }\OperatorTok{=}\NormalTok{ bf16}
  \OperatorTok{{-}\textgreater{}} \BuiltInTok{compile}\NormalTok{: }\StringTok{"Unsupported function output dtype bf16"}   \CommentTok{\# no code to serialize it}

\CommentTok{\# symbolic shape: a "?" dimension parses, then will not lower on the direct path.}
\NormalTok{tensor}\OperatorTok{\textless{}}\NormalTok{fp16, [}\DecValTok{1}\NormalTok{, ?, }\DecValTok{64}\NormalTok{]}\OperatorTok{\textgreater{}}
  \OperatorTok{{-}\textgreater{}}\NormalTok{ parse: accepted (the symbolic variable s0 binds)}
  \OperatorTok{{-}\textgreater{}}\NormalTok{ lowering: }\StringTok{"Not implemented"}\NormalTok{: Some ops are }\KeywordTok{not}\NormalTok{ supported on }\BuiltInTok{any}\NormalTok{ of the specified backends}
\end{Highlighting}
\end{Shaded}

\end{listing}

The hardware capability tables advertise a kernel-depth dimension, and
the frontend recognizes a three-dimensional convolution with a static
weight, but lowering then fails on every device mask with a
not-implemented rejection. No backend on the M1 has a code-generation
path for the three-dimensional convolution operation class. The
sanctioned route does not run it on the engine either: its segmenter
places the operation on the central processor or the graphics processor
instead, so on this silicon there is no on-engine path to it for any
caller.

Native stateful types need a hardware primitive the silicon does not
have. The model framework documents the stateful type and flexible input
shapes at developer.apple.com/documentation/coreml, and this account
corrects the impression that those documented capabilities run on the
engine directly \hyperref[ref-applecoreml]{{[}AppleCoreML{]}}. The
compiler reserves the operation classes for resident state and ring
buffers, the operation table even naming a ring-buffer writer and reader
and the counter-and-event direct-memory-access opcodes, but the in-place
update needs that counter-and-event engine, and it is stubbed out of the
M1 hardware descriptor. Every register setter for it, the source and
destination base addresses, shape and stride setters, atomic-operation
and counter-mode setters, and wait-event address and value setters, has
a body that asserts the engine is unsupported on this architecture. A
program that declares native resident state fails the compile, and this
gate is stronger than an entitlement: the silicon path does not contain
the engine, so no entitlement, property, or compiler option synthesizes
it. A resident cache on the M1 is built instead from a shared buffer
kept live across dispatches, the documented zero-copy path for this
generation. The counter-and-event engine arrives on a later generation,
so native resident state is a per-family feature of the generations
after the M1.

The serialization layer and the datapath both block the bf16 program
input and output type. The dtype enumeration that declares a program's
input and output types has eleven codes, covering fp16, fp32, and the
integer widths, and bf16 is not among them, so a program that declares a
bf16 input or output fails the compile with an unsupported-dtype
rejection. The framework accepts a bf16 array at its boundary, but it
casts to fp16 before the engine segment, so the engine runs no bf16
datapath under the sanctioned route either. The fp16 datapath of chapter
\hyperref[numerics]{3} already gives the accumulation precision a bf16
declaration would otherwise imply.

Flexible and symbolic shapes stop at the runtime path. The compiler has
a full symbolic-shape system, with its master gate on for the M1, yet
the direct runtime path will not drive it: a program with a symbolic
dimension parses and then fails to lower. The sanctioned route reaches
flexible shapes by compiling a small set of fixed shapes ahead of time
and dispatching the nearest, with a dynamic remainder on the central
processor. That is bucketed fixed-shape specialization, not one symbolic
program on the engine, and the direct route matches it by padding to a
fixed maximum or compiling a small set of length buckets, with the
compile cache making a repeated shape free.

\section{Image-input boundary}\label{image-input-boundary}

A fifth feature is at the boundary in a sharper form: direct
image-format input, where a camera or video surface is sent to the
engine in its native four-character-code pixel format with no host-side
conversion to fp16. The operation that performs this on-engine
conversion, \texttt{pixel\_buffer\_to\_tensor}, is present in the
intermediate-representation parser on the direct route and parses
without an unknown-operation rejection. Its input is a distinct surface
type,
\texttt{pixel\_buffer\textless{}format,\ shape,\ bytes\_per\_row\textgreater{}},
and the format token is a \texttt{FMT\_*} enumeration that maps onto the
per-chip interchange-format table, so the grammar, format enumeration,
and type rules are all reachable and solvable on the direct route.

The boundary is at lowering. On the direct route the intermediate
program parses and type-checks and reaches the engine backend, but it
does not lower there, so the conversion cannot be compiled on the direct
route. The entitled framework route supplies the image-input descriptor
and the surface setup that the lowering needs, so direct
four-character-code input is a framework-route feature on the M1. The
terminal direct-route form is an on-engine integer-to-fp16
dequantization of a plain byte input, which avoids the unsupported
operation and still removes the host-side conversion.

\section{What the boundary means
concretely}\label{what-the-boundary-means-concretely}

The choice between the routes is a choice of loader and convenience, not
of reachable compute. Vision, encoders, on-device numerics, and training
of the supported operation set are built from the operations the direct
route compiles and dispatches, none of which the gated set touches. The
entitled path adds a bounded list: a segmenter that places arbitrary
models across three devices with fallback, a model container with
flexible-shape and stateful wiring, and a loader-tier feature set. It
does not add a different engine: the four gated features are absent on
the M1 under both routes, or are matched by a direct-route construction
that does the same work.

\section{Load-time signature check}\label{load-time-signature-check}

One hard limit is below the compute and defines the whole access model:
a hand-built or self-compiled program cannot be loaded onto the engine.
The kernel driver verifies every submitted program before it reaches the
firmware, by a corecrypto signature check over the program bytes and a
trustcache check on the backing file's vnode, and rejects a program that
fails either check at load with error \texttt{0xe00002e2}. The only
program the kernel will load is one the system daemon compiled and
signed in place, so a caller cannot author a network binary by hand and
submit it. This is why the direct route does not build a loadable
binary: it authors a network in the intermediate representation and
hands it to the daemon, which compiles and signs it on the caller's
behalf, and the caller then drives the resulting signed program. The
reachable surface is thus everything the daemon's compiler will accept,
not everything the engine could in principle run.

\section{Unentitled dispatch path}\label{unentitled-dispatch-path}

The kernel driver rejects a binary the client signs itself at load with
\texttt{0xe00002e2}, so the daemon's compile-and-sign step is not
optional. The client authors the network as intermediate language, since
a self-authored binary never passes the load check. It hands that
intermediate language to the system daemon, the one process able to
sign, and the daemon compiles and signs it in place, so the returned
program carries the signature and trustcache trust the kernel load check
requires. The client then drives that signed program over the kernel
interface directly, where the corecrypto signature and trustcache checks
pass because it is daemon-signed, and the program loads onto the engine.

\section{Reference: the boundary
constants}\label{reference-the-boundary-constants}

\Cref{tbl:c8-constants} collects the constants of the entitlement
boundary on the M1.

\begin{longtable}[]{@{}
  >{\raggedright\arraybackslash}p{(\linewidth - 2\tabcolsep) * \real{0.5000}}
  >{\raggedright\arraybackslash}p{(\linewidth - 2\tabcolsep) * \real{0.5000}}@{}}
\caption{The constants of the entitlement boundary,
M1/H13.}\label{tbl:c8-constants}\tabularnewline
\toprule\noalign{}
\begin{minipage}[b]{\linewidth}\raggedright
\textbf{Constant}
\end{minipage} & \begin{minipage}[b]{\linewidth}\raggedright
\textbf{Value}
\end{minipage} \\
\midrule\noalign{}
\endfirsthead
\toprule\noalign{}
\begin{minipage}[b]{\linewidth}\raggedright
\textbf{Constant}
\end{minipage} & \begin{minipage}[b]{\linewidth}\raggedright
\textbf{Value}
\end{minipage} \\
\midrule\noalign{}
\endhead
\bottomrule\noalign{}
\endlastfoot
Program-load rejection code & \texttt{0xe00002e2} \\
\rowcolor{zebra}Program-I/O dtype codes (\texttt{ANECIRDataType}) & 11
codes: 0 int4, 1 uint8, 2 int8, 3 fp16, 4 fp32, 5 int16, 6 uint16, 7
int32, 8 uint32, 9 int64, 10 uint64 \\
bf16 program input or output & absent from the 11 codes; compile
rejected \\
\rowcolor{zebra}Three-dimensional convolution & no backend lowering on
any device mask \\
Native stateful types & counter-and-event direct-memory-access engine
stubbed out of the M1 descriptor \\
\rowcolor{zebra}Flexible and symbolic shapes & parse and bind, then fail
to lower on the direct runtime path \\
Load-time signature checks & corecrypto signature over program bytes
plus trustcache vnode-trust check \\
\rowcolor{zebra}Entitlement-rejection return code & \texttt{0xe00002c7}
(\texttt{kIOReturnUnsupported}) \\
Native counter-and-event engine & every register setter asserts
unsupported on this architecture \\
\rowcolor{zebra}Direct image-format input &
\texttt{pixel\_buffer\_to\_tensor} parses, then does not lower on the
direct route \\
\end{longtable}

The entitlement check the direct route never trips returns a distinct
code from the program-load rejection. The program-load check returns
\texttt{0xe00002e2} from the signature and trustcache check, while the
entitlement gate that guards the higher inference-tier features returns
\texttt{0xe00002c7}, \texttt{kIOReturnUnsupported}, the code a client
gets when a gated feature's entitlement is absent. The four gated
features above fail upstream of either kernel code, inside the compiler
or the firmware descriptor, which is why a host-side entitlement moves
none of them.

\anepartopen{Part III}{Performance and Fit}
\anechap{09}{Roofline}{Two ceilings, the ridge point, the 2 MB working set, and the dispatch floor.}
\anechap{10}{Power and efficiency}{Engine draw, energy per FLOP, and the efficiency margin over the GPU.}
\anechap{11}{ANE, GPU, and CPU}{Which processor leads on each workload class, by speed and by energy.}
\anechap{12}{Across the chip family}{How the roofline scales from the M1 to the M5, predicted then measured.}
\anepartclose

\chapter{Roofline}\label{roofline}

\begin{summarybox}

A layer is compute-bound only above 141 FLOP per byte; below it the
engine runs on the memory slope. Keep every operation's working set
under 2 MB or it streams its full activation from DRAM. Every dispatch
pays a 0.23 ms floor, so batch or fuse until the compute clears it.

\end{summarybox}

The Apple Neural Engine is a roofline machine
\hyperref[ref-williams2009]{{[}Williams2009{]}}. Two ceilings bound its
performance on any layer: the multiply array sets a peak compute rate,
and how fast operands stream from DRAM sets a peak bandwidth. A layer
runs at whichever ceiling its arithmetic intensity reaches first. The
constants below are the M1 instance, the measured silicon. The roofline
has the same form on every chip in the family; only its constants
change, scaling with core count and clock, which chapter
\hyperref[across-the-chip-family]{12} covers and the M5 confirms.

\section{Roofline ceilings and ridge
point}\label{roofline-ceilings-and-ridge-point}

The compute ceiling depends on how it is measured, and two figures
bracket it. The overhead-isolated matmul slope is about 12 fp16 TFLOP/s:
a fused matmul chain runs as one dispatch, so deepening it grows the
compute at fixed input and output cost. The per-layer slope cancels that
fixed cost, leaving the marginal rate. A direct absolute measurement of
one large matmul is lower, about 4.8 fp16 TFLOP/s on a 4096-dimension
matrix multiply whose call time is far above the dispatch floor. That is
about 87 percent of the roughly 5.5 TFLOP/s theoretical peak. A single
conv dispatch is lower still, about 1.8 TFLOP/s, because it pays the
fixed cost the slope subtracts away.

The M1 fp16 peak is best read as the sequence of consistent numbers
listed in \cref{tbl:roofline}, from the 1.8 TFLOP/s conservative
headline through the 3.25 TFLOP/s effective peak to the 4.8 TFLOP/s
saturating ceiling. Apple markets the engine in int8 operations per
second, the M1 at about 11 TOPS \hyperref[ref-appleane]{{[}AppleANE{]}},
and the fp16 rate is by convention about half of an int8 figure, which
puts the fp16 theoretical peak near 5.5 TFLOP/s. The 4.8 TFLOP/s figure
is the rate one large matrix multiply holds after it crosses the 2 MB
working-set threshold, while a fused chain whose activations stay under
the threshold sustains more. That is why chapter
\hyperref[power-and-efficiency]{10} gives a fused steady state near 8.1
TFLOP/s. The 5.5 TFLOP/s figure is a convention derived from the
marketed int8 TOPS rather than a measured hardware ceiling.

The int8 path is faster than fp16. A measured int8 convolution runs
about 1.4 times faster than the fp16 form and trends toward 2 times at
scale. The multiply array has a double-int8 mode that packs two int8
products into one multiply-accumulate, so int8 arithmetic runs at about
twice the fp16 rate on the same array. This is distinct from the
compressed-weight path of chapter \hyperref[weights-and-compression]{7},
where an int8 weight is dequantized to fp16 and then enters an ordinary
fp16 multiply: there the saving is in stored bytes, here the saving is
in compute cycles. The int8 path thus cuts compute cycles as well as
storage.

The bandwidth ceiling is the rate at which operands cross the DRAM
boundary. The engine sustains about 85 GB/s of DRAM bandwidth measured
at the memory controller, and the roofline below uses that figure as its
B. A direct wall-clock measurement of weight streaming runs lower: a
large matrix multiply with one output row reads each weight once and
saturates at about 51 GB/s, roughly 60 percent of that 85 GB/s ceiling.
That figure matches the compiler's own internal bandwidth constant of 50
GB/s. A small elementwise op reaches less than either. A single relu
stream saturates near 10 GB/s effective, because the per-dispatch
overhead caps it well below the DRAM rate.

The attainable rate is the lower of the two ceilings at a layer's
arithmetic intensity \(I\), measured in FLOP per byte.

\[R(I) = \min(P,\ I\,B), \qquad P \approx 12\ \text{fp16 TFLOP/s}, \qquad B \approx 85\ \text{GB/s}\]

A layer with intensity \(I\) is compute-bound when \(I\,B \ge P\) and
bandwidth-bound otherwise. The ridge point is where the two ceilings
cross, near 141 FLOP per byte, the compute roof over the bandwidth roof,
\(12 \times 10^{12}\) over \(85 \times 10^9\).

\[I^{*} = \frac{P}{B} = \frac{12 \times 10^{12}}{85 \times 10^9} \approx 141\ \text{FLOP/byte}\]

\begin{figure}[H]\centering
\includegraphics[width=4.917in]{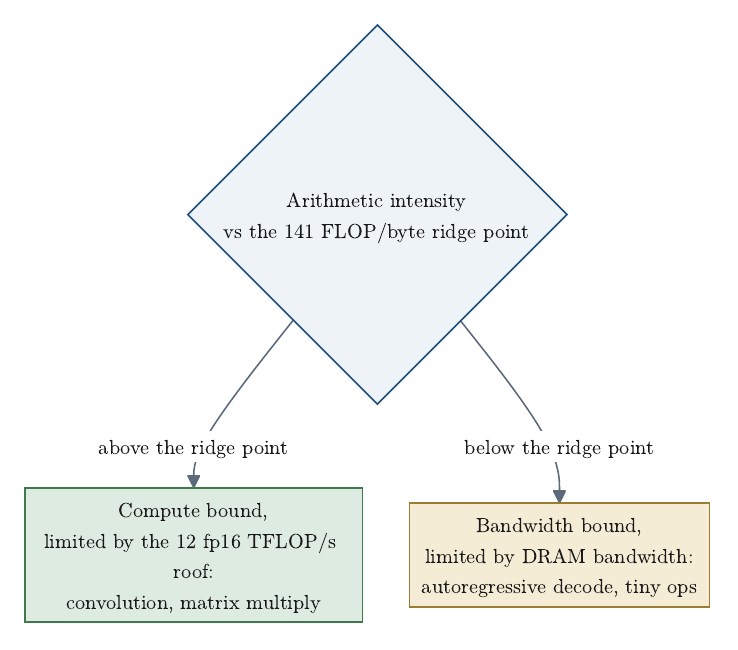}
\caption{How a kernel's arithmetic intensity locates it against the roofline.}
\label{fig:c9-ridge}
\end{figure}

Comparing a kernel's arithmetic intensity against that 141 FLOP-per-byte
crossing routes it to its bound (\cref{fig:c9-ridge}).

Convolutions are far to the right of the ridge point: a 3x3 conv at 256
channels reaches 466 FLOP per byte, so the engine rarely touches its
DRAM ceiling on conv work.

\section{On-chip working-set
threshold}\label{on-chip-working-set-threshold}

The primary design limit is not the 141 FLOP-per-byte crossing but the
on-chip working set, 2 MB on the M1. While a layer's activation fits in
the 2 MB on-chip memory, its intermediate values stay on chip and do not
cross DRAM. Once the activation exceeds 2 MB, every layer streams its
full activation in and out of DRAM and arithmetic intensity collapses.
The hardware counters show the threshold directly: at a batch where the
activation is exactly 2 MB, a 96-layer matmul chain moved 426 MB of DRAM
per dispatch, and arithmetic intensity fell to 60 FLOP per byte. The
same matmul that reached 12 TFLOP/s with a 1 MB activation dropped to
about 4.8 TFLOP/s. Crossing 2 MB moves a workload from compute-bound to
bandwidth-bound on one step, so the working set is the first thing to
tune. The working-set limit scales with on-chip memory: it is near 2 MB
on the M1 and 4.72 MB on the M5, and chapter
\hyperref[across-the-chip-family]{12} gives the per-generation values.

\section{Dispatch floor}\label{dispatch-floor}

Every dispatch pays a fixed minimum cost regardless of the work it
contains. On the M1 that floor is about 0.23 ms per evaluation. A relu,
sigmoid, average pool, and small convolution are all at 0.23 to 0.26 ms,
and a 64-element linear is 0.23 ms: below the floor, neither the
operation nor its size matters. Host dispatch and operand transfer set
the floor, so a small op spends almost all of its wall time outside the
multiply array. A live decomposition of a tiny model puts the full-call
floor near 0.19 ms, of which about 98 percent is dispatch overhead and
about 0.13 ms is the firmware round trip; chapter
\hyperref[execution-model]{2} breaks the budget down stage by stage.

The wall time of a dispatch adds the fixed floor \(t_0\) to the time the
work takes at the attainable rate \(R\).

\[t \approx t_0 + \frac{\text{work}}{R}, \qquad t_0 \approx 0.23\ \text{ms}\]

When the work is small the second term vanishes and \(t \approx t_0\),
so small operations are dispatch-bound: a layer whose compute time is
under 0.23 ms gains nothing from the engine being faster. The same
convolution runs at 63 GFLOP/s at a 16x16 spatial size, where the floor
dominates, and at 1247 GFLOP/s at 256x256, where compute amortizes the
floor, a 20x span from one shape change. Batching and fusion grow the
work term until it dominates the fixed \(t_0\), amortizing the floor
across more useful FLOP.

\section{Fusion economics}\label{fusion-economics}

Fusing a chain of operations into one program removes the per-dispatch
floor from every operation but the first and removes the intermediate
round-trips between them. A network run as N separate dispatches pays
the 0.23 ms floor N times and copies each intermediate back to the host
and forward again. The same network fused into one program pays the
floor once and keeps the intermediates resident in the engine's working
set. The slope measurement isolated that fixed per-dispatch cost at
roughly 0.76 ms for a program with 1 MB of operand input and output, and
fusion removes it from every dispatch it eliminates at close to no
compute cost.

The amortization is measured directly, and depth and batch are the two
controls. A stack of conv-relu layers fused into one program keeps its
per-call latency flat near 0.19 ms from one layer to thirty-two. A
thirty-two-layer model thus pays one firmware round trip just as a
one-layer model does, while the cost charged to each operation falls
from about 222 microseconds at one layer to about 6.3 microseconds at
thirty-two. Batching the same program to 512 samples amortizes that one
round trip across the batch, dropping the per-sample cost from about 196
microseconds to about 1.5 microseconds, a 127-fold reduction. A deep
model on a large batch thus pays a single round trip for all of it,
which is why packing work into one fused program is the main way to cut
latency below the compute roof.

\section{Cross-device roofline}\label{cross-device-roofline}

The M1 ceilings above locate a layer against one engine. Taken on three
devices, the same two measurements locate a layer against the whole
machine, and the difference between the three rooflines is the device
map. Each device's compute roof comes from a saturation sweep and its
bandwidth roof from a streaming sweep, and the ratio of the two is the
device's ridge point, the arithmetic intensity at which a layer crosses
from bandwidth-bound to compute-bound.

The three ridge points are far apart (\cref{tbl:c9-perdevice}).

\begin{longtable}[]{@{}
  >{\raggedright\arraybackslash}p{(\linewidth - 6\tabcolsep) * \real{0.2000}}
  >{\raggedleft\arraybackslash}p{(\linewidth - 6\tabcolsep) * \real{0.2667}}
  >{\raggedleft\arraybackslash}p{(\linewidth - 6\tabcolsep) * \real{0.2667}}
  >{\raggedleft\arraybackslash}p{(\linewidth - 6\tabcolsep) * \real{0.2667}}@{}}
\caption{Per-device compute roof, bandwidth roof, and ridge point,
M5/H17s measured; the engine bandwidth is its standalone-activation
path.}\label{tbl:c9-perdevice}\tabularnewline
\toprule\noalign{}
\begin{minipage}[b]{\linewidth}\raggedright
\textbf{Device}
\end{minipage} & \begin{minipage}[b]{\linewidth}\raggedleft
\textbf{Compute roof (GFLOP/s)}
\end{minipage} & \begin{minipage}[b]{\linewidth}\raggedleft
\textbf{Bandwidth roof (GB/s)}
\end{minipage} & \begin{minipage}[b]{\linewidth}\raggedleft
\textbf{Ridge (FLOP/byte)}
\end{minipage} \\
\midrule\noalign{}
\endfirsthead
\toprule\noalign{}
\begin{minipage}[b]{\linewidth}\raggedright
\textbf{Device}
\end{minipage} & \begin{minipage}[b]{\linewidth}\raggedleft
\textbf{Compute roof (GFLOP/s)}
\end{minipage} & \begin{minipage}[b]{\linewidth}\raggedleft
\textbf{Bandwidth roof (GB/s)}
\end{minipage} & \begin{minipage}[b]{\linewidth}\raggedleft
\textbf{Ridge (FLOP/byte)}
\end{minipage} \\
\midrule\noalign{}
\endhead
\bottomrule\noalign{}
\endlastfoot
Engine & 10191 (matmul) / 18771 (conv) & 24.1 (standalone) & 424 \\
\rowcolor{zebra}GPU & 30862 & 229.7 & 134 \\
CPU & 1898 & 130.4 & 15 \\
\end{longtable}

The engine ridge near 424 FLOP per byte is about three times the GPU
ridge near 134 and about twenty-nine times the CPU ridge near 15. A
layer must reach roughly three times the arithmetic intensity to be
compute-bound on the engine that it would need on the GPU. The high
engine ridge is a direct consequence of the standalone bandwidth: the
engine still delivers about 10 TFLOP/s of compute while a standalone
elementwise stream moves only about 24 GB/s, so the crossover is far to
the right.

The engine has two distinct effective bandwidths, and the gap between
them sets which regime each workload runs in. A standalone elementwise
or memory-bound operation streams at roughly 24 GB/s, the activation
path measured on a relu or a reduction in isolation. A compiled matmul
streams its weights faster than the standalone path, though the rate
depends on the measurement. A decode GEMV at arithmetic intensity near 1
reaches about 112 GFLOP/s, which implies a weight-direct-memory-access
path near 112 GB/s. A direct wall-clock measurement of one large
single-row matrix multiply saturates lower, at the 51 GB/s established
by the M1 bandwidth ceiling above. Either way the weight path is far
faster than the throttled standalone activation path.

A standalone bandwidth-bound operation, a large softmax or a large
elementwise op, reaches only the 24 GB/s activation path on the engine,
and the GPU reaches roughly 230 GB/s on the same stream. Such an op is
thus dispatched to the GPU, not the engine: against the engine's own
bandwidth roof a standalone layer\_norm reaches 18 percent and a
standalone softmax 63 percent, so dispatching them one at a time leaves
the engine mostly idle.

\section{Fusion as a control}\label{fusion-as-a-control}

Fusion moves the engine's operating point to the right of its ridge, and
the measurements below fix how far. Fusing operations into one program
keeps intermediate activations on chip and streams each weight once for
reuse across the whole graph, raising the program's effective arithmetic
intensity above that of any single operation in it. A real
three-convolution stack fused into one program reaches an effective rate
of 20718 GFLOP/s. That rate is above the single-convolution saturation
peak of 18771 GFLOP/s, at 110 percent of the convolution roof. Fusion
lifts the block's effective arithmetic intensity from about 1076 for one
standalone convolution to about 2854 for the fused stack, moving the
operating point past the ridge into the compute-bound region. The
over-roof GEMV at 466 percent of the standalone roof is the same effect
for a single compiled weight stream. A fused convolution or attention
graph runs in the compute-bound region where the engine is efficient,
and the identical operations dispatched standalone do not.

\section{Locating a layer before it is
built}\label{locating-a-layer-before-it-is-built}

Arithmetic intensity and working-set size together locate any layer. A
layer above 141 FLOP per byte is compute-bound against the 12 TFLOP/s
roof and one below it is bandwidth-bound against the 85 GB/s roof, but
an activation over 2 MB is forced onto the memory slope whatever its
nominal intensity, so the working set must shrink before any other
tuning helps. The 0.23 ms dispatch floor sets a latency a small op
cannot beat, so the developer batches or fuses small ops until their
compute clears the floor.

The estimate below is static: it takes a layer's shape and returns the
bound and the control that moves it.

\begin{Shaded}
\begin{Highlighting}[]
\CommentTok{\# Locate a layer against the roofline BEFORE building it, from its shape alone,}
\CommentTok{\# with no hardware in hand. Three numbers decide where it falls and which}
\CommentTok{\# control moves it.}

\CommentTok{\# M1 roofline constants (see the constants table below):}
\NormalTok{P            }\OperatorTok{=} \FloatTok{12e12}    \CommentTok{\# compute roof, fp16 FLOP per second}
\NormalTok{B            }\OperatorTok{=} \FloatTok{85e9}     \CommentTok{\# bandwidth roof, bytes per second}
\NormalTok{ridge\_point  }\OperatorTok{=} \DecValTok{141}      \CommentTok{\# P / B, in FLOP per byte: the compute{-}vs{-}bandwidth crossover}
\NormalTok{working\_cap  }\OperatorTok{=} \FloatTok{2e6}      \CommentTok{\# on{-}chip working set, in bytes: above this it streams from DRAM}
\NormalTok{t0           }\OperatorTok{=} \FloatTok{0.00023}  \CommentTok{\# per{-}dispatch floor, in seconds (\textasciitilde{}0.23 ms), paid no matter the work}

\CommentTok{\# Count the layer\textquotesingle{}s work and the bytes it must move, from its shape.}
\NormalTok{function arithmetic\_intensity(layer):}
    \ControlFlowTok{return}\NormalTok{ total\_flops(layer) }\OperatorTok{/}\NormalTok{ dram\_bytes(layer)        }\CommentTok{\# FLOP per byte}

\CommentTok{\# Classify the layer and name the control it needs.}
\NormalTok{function place\_layer(layer):}
\NormalTok{    intensity    }\OperatorTok{=}\NormalTok{ arithmetic\_intensity(layer)}
\NormalTok{    working\_set  }\OperatorTok{=}\NormalTok{ activation\_bytes(layer)               }\CommentTok{\# live intermediate size}
\NormalTok{    work\_seconds }\OperatorTok{=}\NormalTok{ total\_flops(layer) }\OperatorTok{/}\NormalTok{ attainable\_rate(intensity)}
\NormalTok{    latency      }\OperatorTok{=}\NormalTok{ t0 }\OperatorTok{+}\NormalTok{ work\_seconds                     }\CommentTok{\# t \textasciitilde{}= t0 + work / R(I)}

    \CommentTok{\# The working{-}set threshold dominates: above it, intensity collapses regardless.}
    \ControlFlowTok{if}\NormalTok{ working\_set }\OperatorTok{\textgreater{}}\NormalTok{ working\_cap:}
        \ControlFlowTok{return}\NormalTok{ (}\StringTok{"bandwidth"}\NormalTok{, }\StringTok{"shrink the working set: reshape or split below 2 MB"}\NormalTok{)}

    \CommentTok{\# Below the dispatch floor the op size does not matter at all.}
    \ControlFlowTok{if}\NormalTok{ work\_seconds }\OperatorTok{\textless{}}\NormalTok{ t0:}
        \ControlFlowTok{return}\NormalTok{ (}\StringTok{"dispatch"}\NormalTok{, }\StringTok{"batch or fuse until the compute clears the 0.23 ms floor"}\NormalTok{)}

    \CommentTok{\# Otherwise the ridge point decides which roof binds.}
    \ControlFlowTok{if}\NormalTok{ intensity }\OperatorTok{\textgreater{}=}\NormalTok{ ridge\_point:}
        \ControlFlowTok{return}\NormalTok{ (}\StringTok{"compute"}\NormalTok{, }\StringTok{"near the 12 TFLOP/s roof: no tuning needed"}\NormalTok{)}
    \ControlFlowTok{else}\NormalTok{:}
        \ControlFlowTok{return}\NormalTok{ (}\StringTok{"bandwidth"}\NormalTok{, }\StringTok{"compress the weights or fuse to raise the intensity"}\NormalTok{)}

\CommentTok{\# attainable\_rate is the lower of the two roofs at this intensity.}
\NormalTok{function attainable\_rate(intensity):}
    \ControlFlowTok{return} \BuiltInTok{min}\NormalTok{(P, intensity }\OperatorTok{*}\NormalTok{ B)                         }\CommentTok{\# R(I) = min(P, I*B)}

\CommentTok{\# Worked placement of a 3x3 convolution, 256 channels:}
\NormalTok{(bound, lever) }\OperatorTok{=}\NormalTok{ place\_layer(conv\_3x3\_256ch)}
\CommentTok{\# intensity \textasciitilde{}= 466 FLOP/byte  {-}\textgreater{} well right of the 141 ridge point}
\CommentTok{\# working\_set \textless{} 2 MB          {-}\textgreater{} not on the memory slope}
\CommentTok{\# bound == "compute", lever  == "no tuning needed"}

\CommentTok{\# A layer that comes back "bandwidth" or with a working set above 2 MB is}
\CommentTok{\# reshaped, batched, or fused before any other tuning, because no clock{-}rate}
\CommentTok{\# advantage helps a layer pinned to the memory slope.}
\end{Highlighting}
\end{Shaded}

\section{Reference: the M1 roofline
constants}\label{reference-the-m1-roofline-constants}

\begin{longtable}[]{@{}llr@{}}
\caption{The M1 roofline constants, with the symbol each has in this
chapter.}\label{tbl:roofline}\tabularnewline
\toprule\noalign{}
\textbf{Constant} & \textbf{Symbol} & \textbf{M1/H13 value} \\
\midrule\noalign{}
\endfirsthead
\toprule\noalign{}
\textbf{Constant} & \textbf{Symbol} & \textbf{M1/H13 value} \\
\midrule\noalign{}
\endhead
\bottomrule\noalign{}
\endlastfoot
Compute roof, overhead-isolated matmul slope & \(P\) & 12 fp16
TFLOP/s \\
\rowcolor{zebra}Compute roof, saturating large matmul & & 4.8 fp16
TFLOP/s \\
Effective peak the analytic model is fit to & & 3.25 fp16 TFLOP/s \\
\rowcolor{zebra}Convolution end-to-end ceiling & & 1.8 TFLOP/s \\
int8 over fp16 compute rate & & 1.4 to 2 times \\
\rowcolor{zebra}Bandwidth roof, DRAM ceiling & \(B\) & 85 GB/s \\
Bandwidth roof, saturating weight-stream wall clock & & 51 GB/s \\
\rowcolor{zebra}Standalone activation-stream rate & & 24 GB/s \\
Single-relu effective stream & & 10 GB/s \\
\rowcolor{zebra}Roofline ridge point & \(I^{*}\) & 141 FLOP/byte \\
On-chip working-set threshold & & 2 MB \\
\rowcolor{zebra}Per-dispatch floor, slope method & \(t_0\) & 0.23 ms \\
Per-dispatch floor, measured tiny model & & 0.19 ms \\
\rowcolor{zebra}Fused per-dispatch cost, 1 MB operands & & 0.76 ms \\
\end{longtable}

\chapter{Power and efficiency}\label{power-and-efficiency}

\begin{summarybox}

The engine draws less power than the GPU on every workload class
measured, including the ones where the GPU is faster. On the M1 it
delivers 2 to 14 times the GPU energy efficiency. The advantage holds
across the M1, M2, and M5, so it is a property of the fixed-function
fp16 datapath, not of one chip.

\end{summarybox}

That 2 to 14 times envelope is throughput per watt of idle-subtracted
total-package power. A stack of sixteen 3x3 convolutions sets the high
mark: the engine reaches 2063 GFLOP/s per watt against the GPU's 142, a
14.5 times efficiency advantage, while also running about 2 times
faster. The M5 has the same shape at 2289 GFLOP/s per watt on the engine
against 175 on the GPU, a 13 times advantage.

The efficiency lead survives even where the engine loses on latency. On
a large square matrix multiply with inner dimension 4096 the M1 GPU is
about 2 times faster in raw throughput, yet the engine still leads on
energy. It computes that workload at 4.4 watts where the GPU draws 32.5
watts, a 4.0 times efficiency edge. The engine's draw stays in the
single digits of watts across these classes; the GPU climbs into the
tens.

\section{Watt-complete map}\label{watt-complete-map}

The map covers the workload classes a perception or encoder deployment
is built from. Convolution appears as a single 3x3 layer and as a
sixteen-deep resnet-style stack. There are three matrix-multiply sizes,
a dispatch-bound floor, a bandwidth-bound mid size, and a compute-bound
4096 square, and two attention lengths, a short vision-transformer
sequence and a longer 512 sequence. Layer norm, rms norm, and group norm
stand in for the normalization family. The scientific kernels are a
discrete Fourier transform as a matrix multiply, a five-point stencil,
and a fixed-iteration linear solve.

Only the efficiency standouts of that set are tabulated below, with the
ratio on each chip generation.

\begin{longtable}[]{@{}lrrrl@{}}
\caption{Engine versus GPU energy efficiency across workloads and chip
generations, in GFLOP/s/W or absolute
watts.}\label{tbl:c10-headline}\tabularnewline
\toprule\noalign{}
\textbf{Workload} & \textbf{Engine GFLOP/s/W} & \textbf{GPU GFLOP/s/W} &
\textbf{Ratio} & \textbf{Silicon} \\
\midrule\noalign{}
\endfirsthead
\toprule\noalign{}
\textbf{Workload} & \textbf{Engine GFLOP/s/W} & \textbf{GPU GFLOP/s/W} &
\textbf{Ratio} & \textbf{Silicon} \\
\midrule\noalign{}
\endhead
\bottomrule\noalign{}
\endlastfoot
Conv-resnet stack (16x 3x3) & 2063 & 142 & 14.5x & M1/H13 \\
\rowcolor{zebra}Conv-resnet stack (16x 3x3) & 2289 & 175 & 13x &
M5/H17s \\
Large square GEMM (inner 4096) & 4.4 W & 32.5 W & 4.0x & M1/H13 \\
\rowcolor{zebra}Five-point stencil (fused) & n/a & n/a & 49x &
M5/H17s \\
Workload-class envelope & n/a & n/a & 2 to 14x & M1/H13 \\
\end{longtable}

The large square GEMM row of \cref{tbl:c10-headline} reports absolute
package power rather than throughput per watt, because the GPU is faster
on that workload on raw throughput while the engine still draws fewer
watts to compute it. The efficiency lead is largest on the convolution
stack and on the fused multi-step stencil, where the engine does
sustained compute at low power and the GPU spends proportionally more
power to keep its units supplied. It narrows on the small normalization
reductions, where both devices finish in well under a millisecond and
the per-watt numbers are close to a tie, and again on the large 4096
square matrix multiply, which favors the GPU on speed and on fp16
accuracy though the engine keeps its lower absolute draw there. Only at
the trivial dispatch-bound region, the 256-inner matrix multiply, does
the engine lose the efficiency comparison outright, where the FLOP count
is too small to matter and the lowest call overhead is the better
choice.

\section{Two-generation result}\label{two-generation-result}

The same sixteen-class harness was run on an M1 and on an M5, with
package power taken from hardware instrumentation on both. Both silicons
return the same verdict: the engine is the efficiency device on every
substantial workload class. On the convolution stack the advantage holds
at 14.5 times and 13 times across the generation gap. A third run on the
M2, an A14 part, fills the middle and reproduces the same shape, with
the convolution stack at 2234.6 GFLOP/s per watt on the engine against
172.8 on the GPU, a 12.9 times advantage. The engine rail never exceeds
about 6 watts where the GPU pulls 13 to 21 watts on the same classes.

The per-class efficiency rows on the M2 trace the same split the M1 and
M5 maps draw (\cref{tbl:c10-m2}).

\begin{longtable}[]{@{}
  >{\raggedright\arraybackslash}p{(\linewidth - 6\tabcolsep) * \real{0.2000}}
  >{\raggedleft\arraybackslash}p{(\linewidth - 6\tabcolsep) * \real{0.2667}}
  >{\raggedleft\arraybackslash}p{(\linewidth - 6\tabcolsep) * \real{0.2667}}
  >{\raggedleft\arraybackslash}p{(\linewidth - 6\tabcolsep) * \real{0.2667}}@{}}
\caption{Per-class engine versus GPU efficiency on the M2, an A14 part,
with the total-package energy ratio.}\label{tbl:c10-m2}\tabularnewline
\toprule\noalign{}
\begin{minipage}[b]{\linewidth}\raggedright
\textbf{Workload}
\end{minipage} & \begin{minipage}[b]{\linewidth}\raggedleft
\textbf{Engine GFLOP/s/W}
\end{minipage} & \begin{minipage}[b]{\linewidth}\raggedleft
\textbf{GPU GFLOP/s/W}
\end{minipage} & \begin{minipage}[b]{\linewidth}\raggedleft
\textbf{Energy ratio (GPU over engine)}
\end{minipage} \\
\midrule\noalign{}
\endfirsthead
\toprule\noalign{}
\begin{minipage}[b]{\linewidth}\raggedright
\textbf{Workload}
\end{minipage} & \begin{minipage}[b]{\linewidth}\raggedleft
\textbf{Engine GFLOP/s/W}
\end{minipage} & \begin{minipage}[b]{\linewidth}\raggedleft
\textbf{GPU GFLOP/s/W}
\end{minipage} & \begin{minipage}[b]{\linewidth}\raggedleft
\textbf{Energy ratio (GPU over engine)}
\end{minipage} \\
\midrule\noalign{}
\endhead
\bottomrule\noalign{}
\endlastfoot
GEMM floor (M=64, K=256, N=256) & 11.4 & 16.6 & 0.7x \\
\rowcolor{zebra}GEMM bandwidth (M=128, K=1024, N=1024) & 298.2 & 129.9 &
2.3x \\
GEMM compute (M=256, K=4096, N=4096) & 773.3 & 213.7 & 3.6x \\
\rowcolor{zebra}Conv single (C=64, 32x32, k=3) & 25.1 & 31.4 & 0.8x \\
Conv resnet stack (C=256, 32x32, k=3, d=16) & 2234.6 & 172.8 & 12.9x \\
\rowcolor{zebra}Attention ViT (S=197) & 480.8 & 120.2 & 4.0x \\
Attention long sequence (S=512) & 567.4 & 150.1 & 3.8x \\
\rowcolor{zebra}Five-point stencil (256x256, 32 steps) & 37.6 & 0.8 &
47.8x \\
\end{longtable}

The real-model rows on the M2 are the engine's widest margin: a full
twelve-layer ViT-B/16 forward runs at 67.9 millijoules per inference
against the GPU's 714.3, a 10.5 times energy advantage, and a ResNet-18
forward at 2.6 against 22.3 millijoules per inference, an 8.5 times
advantage. Both also run from 1.7 to 4 times faster. On the M1 the
engine draws under 2.3 watts at its compute roof and spends roughly 0.5
picojoules per FLOP in its fixed-function datapath.

The advantage is wider on the older M1 because that chip is more
bandwidth-limited. The M1 streams at roughly 10 GB/s effective against
the M5's roughly 57 GB/s, and the smaller memory bandwidth makes the GPU
spend proportionally more power to lead on raw throughput there. On the
M1 the engine sustains a 256-channel 3x3 convolution at 643 GFLOP/s per
watt, drawing 1.78 watts, with the rail reading about zero at idle. The
1.78 watt figure is the canonical short-run draw for this workload, and
a longer 176 second run of the same convolution in chapter
\hyperref[power-and-thermal]{31} settles slightly lower at about 1.66
watts. The two figures are thus the same workload measured over
different durations rather than a disagreement.

\section{Rail-off idle floor}\label{rail-off-idle-floor}

The engine reads about zero milliwatts at idle, and that figure is a
property of the power-state machine, not a measurement artifact. The
firmware keeps the engine in a fully gated state,
\texttt{ANE\_POWER\_STATE\_ALL\_OFF}, with every domain off, until a job
arrives. Power-up is lazy and job-triggered: the engine is at
\texttt{ALL\_OFF} and brings up its base domain only on the first call,
transitioning through \texttt{ANE\_POWER\_STATE\_BASE\_PS\_ON\_WAIT}
before it gates in the compute sets. There is no clock-gated but powered
idle floor; the idle state is rail-off.

The M1 engine has five independently gated power domains: a base domain
for the control and front-end fabric, brought up first, and four compute
sets, one per ANE cluster. Dynamic power-gating, on by default,
collapses the compute sets back toward \texttt{ALL\_OFF} between jobs,
so idle work pays no rail power. Disabling it keeps the sets powered
between jobs, trading idle power for lower per-job latency. Thermal
management and voltage are off-engine: the firmware image has no
thermal, temperature, or throttle path, so those decisions are in the
system-on-chip power manager. The rail-off idle is why the efficiency
map subtracts a flat zero for the engine baseline while the GPU and CPU
baselines subtract a nonzero idle draw.

\section{Power scales with
utilization}\label{power-scales-with-utilization}

The idle rail reads zero and the dispatch floor draws about 0.9 watts,
and from there the draw climbs with how much of the multiply array a
workload keeps active. A twenty-point sweep on an M1 Max, read at the
root power sampler, traces the curve: a dispatch-floor matrix multiply
draws about 0.9 watts, a compute-bound 1024 by 4096 fp16 matrix multiply
about 4.3 watts, and the int8 form about 5.8 watts. That is a roughly
fivefold range set by utilization alone. The draw is regime-dependent as
well: a bandwidth-bound matrix multiply that mostly streams a large
weight draws about 1.4 watts, a convolution about 1.3 watts because its
lowering leaves part of the array idle, and an elementwise operation
almost nothing on the engine rail. Efficiency peaks at the fp16 compute
optimum near 2.68 trillion operations per watt, about 0.37 picojoules
per FLOP, and falls to about 22 picojoules per FLOP at the dispatch
floor, which is a further reason to amortize work past the floor. The
engine exposes only the power reading and a binary on-or-off state even
to the root sampler, with no frequency or voltage telemetry, so the draw
is observable but the sequence of operating points behind it is not.

\section{Sustained load holds one clock
state}\label{sustained-load-holds-one-clock-state}

The efficiency figures above are warmup measurements, so they leave open
whether the engine stays in its power state under minutes of continuous
load. A compute-bound probe answers it directly: a chain of eight
1024-square fp16 matrix multiplies run as one program in a low-overhead
zero-copy loop, at about 2.11 milliseconds per call, roughly ten times
the dispatch floor, so the loop is compute-bound. Run continuously for
210 seconds and 98,092 calls on the M1, the throughput curve is flat.
The whole-run median is 2109.6 microseconds, the start bucket reads 2153
microseconds and the end bucket 205 seconds later reads 2143. The
per-five-second bucket medians vary by about 4 percent in a
non-monotonic band that is sampling noise, not thermal decay. The engine
rail, sampled read-only at five-second cadence, pins at about 5.5 watts
for the whole run, with start, middle, and end thirds reading 5530,
5522, and 5391 milliwatts, a drift under 3 percent. Thermal pressure
reads Nominal on every sample, and the engine never leaves its single
clock state, with the steady draw set by the workload's utilization as
the previous section describes. The M1 engine does not throttle on this
workload over 3.5 minutes: it runs at one clock state and stays there.

What moves first-call latency is at the boundaries of a burst
(\cref{tbl:c10-phases}).

\begin{longtable}[]{@{}
  >{\raggedright\arraybackslash}p{(\linewidth - 2\tabcolsep) * \real{0.5000}}
  >{\raggedright\arraybackslash}p{(\linewidth - 2\tabcolsep) * \real{0.5000}}@{}}
\caption{Sustained-load phases on the M1, with the warmup and idle
re-wake costs that the flat steady state
hides.}\label{tbl:c10-phases}\tabularnewline
\toprule\noalign{}
\begin{minipage}[b]{\linewidth}\raggedright
\textbf{Phase}
\end{minipage} & \begin{minipage}[b]{\linewidth}\raggedright
\textbf{Behavior on the M1}
\end{minipage} \\
\midrule\noalign{}
\endfirsthead
\toprule\noalign{}
\begin{minipage}[b]{\linewidth}\raggedright
\textbf{Phase}
\end{minipage} & \begin{minipage}[b]{\linewidth}\raggedright
\textbf{Behavior on the M1}
\end{minipage} \\
\midrule\noalign{}
\endhead
\bottomrule\noalign{}
\endlastfoot
Warmup & About a three-call ramp: call 1 about 7.6 ms (3.6x steady),
call 2 about 4.6 ms, call 3 within 13 percent of steady, then flat at
about 2.15 ms. \\
\rowcolor{zebra}Steady state & Flat at about 8.1 TFLOP/s and about 5.5 W
for 3.5 minutes, p50 2110 microseconds, p99 only 36 percent over p50. \\
Sub-second idle & A modest first-call penalty of 1.2x to 1.7x, then the
next call is back to steady. \\
\rowcolor{zebra}Multi-second idle & The first call after a 5-second gap
costs about 260 ms, roughly 123x the steady p50, then the very next call
returns to about 2.1 ms. \\
\end{longtable}

The idle re-wake penalty is the consequence of the rail-off idle
described above. Once idle crosses a few seconds the compute sets gate
fully off, and the first call after the gap pays a one-time cold re-wake
of tens to hundreds of milliseconds before the engine returns to its
steady state on the following call. The cost falls on the first call of
each active burst. Latency-sensitive code that must answer immediately
after an idle gap should thus keep the engine warm with a sub-second
dispatch cadence or a low-cost keep-alive call.

The engine rail sampled directly on the M1 confirms that average power
tracks utilization. Idle draws about 0 milliwatts. A sustained
batch-of-512 hot loop draws about 1.48 watts. A single-call loop draws
about 755 milliwatts, about half the hot-loop figure, because a single
small call leaves the engine idle for most of the dispatch window.

\section{Compression as an energy
control}\label{compression-as-an-energy-control}

Streaming a weight in a narrower format cuts energy per inference even
though instantaneous power stays roughly flat. On a bandwidth-bound
4096-inner matrix multiply on the M2, the engine draws between 2.06 and
2.59 watts across fp16, int8, int4, and sparse, with the sparse stream
drawing the most watts. The energy per inference still falls with the
narrower format, because latency falls faster than power rises
(\cref{tbl:c10-formats}).

\begin{longtable}[]{@{}
  >{\raggedright\arraybackslash}p{(\linewidth - 8\tabcolsep) * \real{0.1579}}
  >{\raggedleft\arraybackslash}p{(\linewidth - 8\tabcolsep) * \real{0.2105}}
  >{\raggedleft\arraybackslash}p{(\linewidth - 8\tabcolsep) * \real{0.2105}}
  >{\raggedleft\arraybackslash}p{(\linewidth - 8\tabcolsep) * \real{0.2105}}
  >{\raggedleft\arraybackslash}p{(\linewidth - 8\tabcolsep) * \real{0.2105}}@{}}
\caption{Power, latency, and energy per inference across weight formats
on a bandwidth-bound matrix multiply,
M2.}\label{tbl:c10-formats}\tabularnewline
\toprule\noalign{}
\begin{minipage}[b]{\linewidth}\raggedright
\textbf{Weight format}
\end{minipage} & \begin{minipage}[b]{\linewidth}\raggedleft
\textbf{Engine power}
\end{minipage} & \begin{minipage}[b]{\linewidth}\raggedleft
\textbf{Latency}
\end{minipage} & \begin{minipage}[b]{\linewidth}\raggedleft
\textbf{Energy per inference}
\end{minipage} & \begin{minipage}[b]{\linewidth}\raggedleft
\textbf{Versus fp16 energy}
\end{minipage} \\
\midrule\noalign{}
\endfirsthead
\toprule\noalign{}
\begin{minipage}[b]{\linewidth}\raggedright
\textbf{Weight format}
\end{minipage} & \begin{minipage}[b]{\linewidth}\raggedleft
\textbf{Engine power}
\end{minipage} & \begin{minipage}[b]{\linewidth}\raggedleft
\textbf{Latency}
\end{minipage} & \begin{minipage}[b]{\linewidth}\raggedleft
\textbf{Energy per inference}
\end{minipage} & \begin{minipage}[b]{\linewidth}\raggedleft
\textbf{Versus fp16 energy}
\end{minipage} \\
\midrule\noalign{}
\endhead
\bottomrule\noalign{}
\endlastfoot
fp16 & 2.18 W & 0.690 ms & 1.50 mJ & 1.00x \\
\rowcolor{zebra}int8 & 2.06 W & 0.431 ms & 0.89 mJ & 0.59x \\
int4 & 2.29 W & 0.270 ms & 0.62 mJ & 0.41x \\
\rowcolor{zebra}Sparse (about 32 percent dense) & 2.59 W & 0.330 ms &
0.86 mJ & 0.57x \\
\end{longtable}

An int4 weight stream reaches the same answer at 0.41 times the fp16
energy on this part, a 2.4 times energy reduction at equal work, with no
efficiency-versus-latency tradeoff to weigh: the narrower format is
faster and more efficient.

\section{Where the lead comes from}\label{where-the-lead-comes-from}

The fp16 datapath is why the lead holds even where the GPU is faster on
latency: a narrow multiply and a fixed-function pipeline move and
compute far fewer bits per result than a general-purpose vector unit.
The engine thus spends less energy reaching the same answer wherever the
math stays in its precision range.

\section{Reading a layer's energy off its
shape}\label{reading-a-layers-energy-off-its-shape}

A workload's efficiency follows from where the roofline locates it. A
layer that is compute-bound on the engine spends its energy in the
fixed-function multiply array, which is the regime where the engine
outruns the GPU on throughput per watt by the wide margins in the map
above. A layer pinned to the memory slope, or below the dispatch floor,
spends its energy moving bytes or paying call overhead, where the
efficiency lead narrows toward a tie.

So the energy regime can be read off a shape at design time: count the
work and the bytes, take the bound, and look up which region of the map
the layer falls in.

\begin{Shaded}
\begin{Highlighting}[]
\CommentTok{\# Estimate the energy of one layer before any hardware is in hand.}
\CommentTok{\# A 3x3 stride{-}1 convolution, 256 channels, on a 56 by 56 feature map.}

\NormalTok{given layer L }\OperatorTok{=}\NormalTok{ conv\_3x3(}\BuiltInTok{input} \OperatorTok{=}\NormalTok{ [}\DecValTok{1}\NormalTok{, }\DecValTok{256}\NormalTok{, }\DecValTok{56}\NormalTok{, }\DecValTok{56}\NormalTok{], stride }\OperatorTok{=} \DecValTok{1}\NormalTok{)}
\NormalTok{given target chip }\OperatorTok{=}\NormalTok{ H13            }\CommentTok{\# M1; peak and bandwidth come from its roofline}

\CommentTok{\# 1. Count the work the layer does and the bytes it must move.}
\NormalTok{flops }\OperatorTok{=}\NormalTok{ total multiply\_adds }\KeywordTok{in}\NormalTok{ L           }\CommentTok{\# arithmetic operations}
\BuiltInTok{bytes} \OperatorTok{=}\NormalTok{ weight\_bytes(L) }\OperatorTok{+}\NormalTok{ input\_bytes(L) }\OperatorTok{+}\NormalTok{ output\_bytes(L)}

\CommentTok{\# 2. Locate the layer on the roofline of the chosen chip.}
\NormalTok{peak      }\OperatorTok{=}\NormalTok{ compute\_peak(chip)             }\CommentTok{\# operations per second the array can sustain}
\NormalTok{bandwidth }\OperatorTok{=}\NormalTok{ memory\_bandwidth(chip)         }\CommentTok{\# bytes per second from on{-}chip and DRAM}
\NormalTok{floor     }\OperatorTok{=}\NormalTok{ dispatch\_floor(chip)           }\CommentTok{\# fixed per{-}call overhead, about 0.23 ms on H13}

\NormalTok{compute\_time }\OperatorTok{=}\NormalTok{ flops }\OperatorTok{/}\NormalTok{ peak                }\CommentTok{\# time if the array is the limit}
\NormalTok{memory\_time  }\OperatorTok{=} \BuiltInTok{bytes} \OperatorTok{/}\NormalTok{ bandwidth           }\CommentTok{\# time if moving bytes is the limit}
\NormalTok{latency      }\OperatorTok{=} \BuiltInTok{max}\NormalTok{(compute\_time, memory\_time) }\OperatorTok{+}\NormalTok{ floor}

\ControlFlowTok{if}\NormalTok{ compute\_time }\OperatorTok{\textgreater{}=}\NormalTok{ memory\_time:  regime }\OperatorTok{=} \StringTok{"compute"}     \CommentTok{\# array busy, most efficient regime}
\ControlFlowTok{else}\NormalTok{:                            regime }\OperatorTok{=} \StringTok{"bandwidth"}    \CommentTok{\# moving bytes, efficiency lead narrows}
\ControlFlowTok{if}\NormalTok{ latency }\KeywordTok{is}\NormalTok{ dominated by floor:  regime }\OperatorTok{=} \StringTok{"dispatch"}  \CommentTok{\# call overhead, leads converge}

\CommentTok{\# 3. Energy = power drawn in that regime, times how long the layer runs.}
\NormalTok{power  }\OperatorTok{=}\NormalTok{ sustained\_power(chip, regime)     }\CommentTok{\# e.g. about 1.78 W compute{-}bound on H13}
\NormalTok{energy }\OperatorTok{=}\NormalTok{ power }\OperatorTok{*}\NormalTok{ latency}

\CommentTok{\# 4. Amortized versus per{-}call: a layer called once pays the full floor,}
\CommentTok{\#    a layer called many times spreads that fixed floor across the calls.}
\NormalTok{energy\_per\_call\_standalone }\OperatorTok{=}\NormalTok{ power }\OperatorTok{*}\NormalTok{ (}\BuiltInTok{max}\NormalTok{(compute\_time, memory\_time) }\OperatorTok{+}\NormalTok{ floor)}
\NormalTok{energy\_per\_call\_amortized  }\OperatorTok{=}\NormalTok{ power }\OperatorTok{*}\NormalTok{ (}\BuiltInTok{max}\NormalTok{(compute\_time, memory\_time) }\OperatorTok{+}\NormalTok{ floor }\OperatorTok{/}\NormalTok{ number\_of\_calls)}

\ControlFlowTok{return}\NormalTok{ regime, latency, energy\_per\_call\_standalone, energy\_per\_call\_amortized}
\end{Highlighting}
\end{Shaded}

A layer that prints \texttt{compute} runs in the region where the
convolution-stack figures hold, up to 2063 GFLOP/s per watt on the M1; a
layer that prints \texttt{dispatch} or \texttt{bandwidth} falls toward
the tie at the small-reduction and trivial-matmul regions.

\section{Reference: engine versus GPU efficiency
constants}\label{reference-engine-versus-gpu-efficiency-constants}

\Cref{tbl:c10-constants} lists every figure quoted in this chapter.

\begin{longtable}[]{@{}
  >{\raggedright\arraybackslash}p{(\linewidth - 4\tabcolsep) * \real{0.3000}}
  >{\raggedleft\arraybackslash}p{(\linewidth - 4\tabcolsep) * \real{0.4000}}
  >{\raggedright\arraybackslash}p{(\linewidth - 4\tabcolsep) * \real{0.3000}}@{}}
\caption{The power and efficiency constants of this chapter, with the
silicon each was measured on.}\label{tbl:c10-constants}\tabularnewline
\toprule\noalign{}
\begin{minipage}[b]{\linewidth}\raggedright
\textbf{Quantity}
\end{minipage} & \begin{minipage}[b]{\linewidth}\raggedleft
\textbf{Value}
\end{minipage} & \begin{minipage}[b]{\linewidth}\raggedright
\textbf{Silicon}
\end{minipage} \\
\midrule\noalign{}
\endfirsthead
\toprule\noalign{}
\begin{minipage}[b]{\linewidth}\raggedright
\textbf{Quantity}
\end{minipage} & \begin{minipage}[b]{\linewidth}\raggedleft
\textbf{Value}
\end{minipage} & \begin{minipage}[b]{\linewidth}\raggedright
\textbf{Silicon}
\end{minipage} \\
\midrule\noalign{}
\endhead
\bottomrule\noalign{}
\endlastfoot
Convolution-stack efficiency, engine & 2063 GFLOP/s/W & M1/H13 \\
\rowcolor{zebra}Convolution-stack efficiency, GPU & 142 GFLOP/s/W &
M1/H13 \\
Convolution-stack efficiency ratio & 14.5x & M1/H13 \\
\rowcolor{zebra}Convolution-stack efficiency, engine & 2289 GFLOP/s/W &
M5/H17s \\
Convolution-stack efficiency, GPU & 175 GFLOP/s/W & M5/H17s \\
\rowcolor{zebra}Convolution-stack efficiency ratio & 13x & M5/H17s \\
Convolution-stack efficiency, engine & 2234.6 GFLOP/s/W & M2/H14 \\
\rowcolor{zebra}Convolution-stack efficiency, GPU & 172.8 GFLOP/s/W &
M2/H14 \\
\end{longtable}

\chapter{ANE, GPU, and CPU}\label{ane-gpu-and-cpu}

\begin{summarybox}

The engine leads on convolution, vision, short-sequence attention, and
mid-size matrix multiply. A sixteen-deep convolution stack runs about
4.2 times faster than on the GPU, at about 13 times the energy
efficiency on the M5. The GPU is about 2.8 times faster on large square
matrix multiply, long-sequence attention, and bandwidth-bound decode.
The CPU is fastest on the trivial operations that fall below the 0.23 ms
dispatch floor. Serving moves the line: the GPU overtakes the engine on
a batched encoder block near a batch of 23 and on self-attention near a
batch of 6.

\end{summarybox}

A workload on an Apple system on chip has no single fastest processor.
It has three candidates, and the right one depends on the form of the
work. This chapter maps which of the engine, GPU, and CPU leads on which
class of work, on raw speed and on energy per result, from a single
harness across sixteen workload classes on the same silicon. The map is
two-dimensional: a processor can lead on latency, on energy, on both, or
on neither.

\section{Where the engine leads}\label{where-the-engine-leads}

The engine is faster and more efficient on convolution, convolution
stacks, stencil-like fixed-iteration numerics, short-sequence attention,
and mid-size matrix multiply. The convolution stack leads the set: a
sixteen-deep stack of 3x3 convolutions at 256 channels runs about 4.2
times faster than the GPU and at about 13 times its energy efficiency on
the M5 generation. On the M1 generation it runs about 2 times faster and
at 14.5 times the GPU's efficiency, the widest power gap in the set. A
five-point stencil iterated thirty-two times as one fused graph runs
about 10 times faster and at about 49 times the GPU's efficiency on the
M5, because the fixed-iteration graph amortizes the per-dispatch floor
across every step. Short-sequence attention, a transformer block at
sequence length 197, is fastest, most efficient, and most accurate in
fp16 on the engine.

The real models track the primitives: a ResNet-18 forward runs about 6.1
times faster than the GPU reference and at about 11 times the energy
efficiency per inference, and a twelve-layer encoder forward about 1.5
times faster and at about 18 times the energy efficiency per inference.
A single-sentence encoder runs about 4.4 times faster. The energy
advantage is wider than the speed advantage and persists where the speed
advantage does not. On the M1 the engine draws 4.4 W on a large
compute-bound matrix multiply where the GPU draws 32.5 W, a 4.0 times
efficiency edge even where the GPU is about 2 times faster.

\section{Where the GPU leads}\label{where-the-gpu-leads}

Large square matrix multiply, long-sequence attention, and
bandwidth-bound autoregressive decode all run faster on the GPU. A 256
by 4096 by 4096 matrix multiply runs about 2.8 times faster on the GPU
than on the engine, and the gap grows to 3.0 times at saturation, about
30.9 fp16 TFLOP/s against the engine's 10.2 TFLOP/s peak. The engine's
single fused matrix multiply stalls on weight streaming once a square
operand passes its on-chip working set near N of 2048. Long-sequence
attention shifts to the GPU as the sequence grows: at sequence length
512 the GPU is about 2.1 times faster, though the engine keeps a 1.8
times energy edge.

On very large reductions the GPU also leads on fp16 accuracy. On the
large square matrix multiply the GPU keeps an fp16 relative error of
\(3.6 \times 10^{-4}\) flat across size, while the engine's error grows
with the contraction dimension, from \(1.2 \times 10^{-2}\) at N of 2048
to \(6.2 \times 10^{-2}\) at N of 8192. That workload is thus disfavored
on the engine on both axes.

\section{Where the CPU leads}\label{where-the-cpu-leads}

The CPU is the better choice for the trivial cases, where call overhead
decides the result. On the floor matrix multiply, 64 by 256 by 256, the
CPU completes in about 0.026 ms, below the per-dispatch floor of either
accelerator. The engine pays a fixed per-eval overhead of about 0.23 ms
on the M1, so any single small operation is overhead-bound and cheaper
on the CPU. On any larger class the CPU is not competitive, reaching
about 1.9 fp32 TFLOP/s on a saturated matrix multiply.

\section{M1 per-class measurement}\label{m1-per-class-measurement}

The verdicts above are taken from one harness across sixteen workload
classes, and the M1 rows make the per-class split concrete. The engine
leads on energy on every class but the dispatch-bound floor, and it is
fastest on the convolution-heavy and mid-bandwidth classes
(\cref{tbl:c11-m1class}).

\begin{longtable}[]{@{}
  >{\raggedright\arraybackslash}p{(\linewidth - 10\tabcolsep) * \real{0.1304}}
  >{\raggedleft\arraybackslash}p{(\linewidth - 10\tabcolsep) * \real{0.1739}}
  >{\raggedleft\arraybackslash}p{(\linewidth - 10\tabcolsep) * \real{0.1739}}
  >{\raggedleft\arraybackslash}p{(\linewidth - 10\tabcolsep) * \real{0.1739}}
  >{\raggedleft\arraybackslash}p{(\linewidth - 10\tabcolsep) * \real{0.1739}}
  >{\raggedleft\arraybackslash}p{(\linewidth - 10\tabcolsep) * \real{0.1739}}@{}}
\caption{The M1 per-class speed and efficiency comparison, engine versus
GPU, idle-subtracted package
power.}\label{tbl:c11-m1class}\tabularnewline
\toprule\noalign{}
\begin{minipage}[b]{\linewidth}\raggedright
\textbf{Workload}
\end{minipage} & \begin{minipage}[b]{\linewidth}\raggedleft
\textbf{Engine GFLOP/s}
\end{minipage} & \begin{minipage}[b]{\linewidth}\raggedleft
\textbf{Engine GFLOP/s/W}
\end{minipage} & \begin{minipage}[b]{\linewidth}\raggedleft
\textbf{GPU GFLOP/s}
\end{minipage} & \begin{minipage}[b]{\linewidth}\raggedleft
\textbf{GPU GFLOP/s/W}
\end{minipage} & \begin{minipage}[b]{\linewidth}\raggedleft
\textbf{Engine efficiency advantage}
\end{minipage} \\
\midrule\noalign{}
\endfirsthead
\toprule\noalign{}
\begin{minipage}[b]{\linewidth}\raggedright
\textbf{Workload}
\end{minipage} & \begin{minipage}[b]{\linewidth}\raggedleft
\textbf{Engine GFLOP/s}
\end{minipage} & \begin{minipage}[b]{\linewidth}\raggedleft
\textbf{Engine GFLOP/s/W}
\end{minipage} & \begin{minipage}[b]{\linewidth}\raggedleft
\textbf{GPU GFLOP/s}
\end{minipage} & \begin{minipage}[b]{\linewidth}\raggedleft
\textbf{GPU GFLOP/s/W}
\end{minipage} & \begin{minipage}[b]{\linewidth}\raggedleft
\textbf{Engine efficiency advantage}
\end{minipage} \\
\midrule\noalign{}
\endhead
\bottomrule\noalign{}
\endlastfoot
GEMM floor (K=256) & 51 & 17 & 46 & 27 & 0.6x \\
\rowcolor{zebra}GEMM bandwidth (K=1024) & 662 & 470 & 1109 & 207 &
2.3x \\
GEMM compute (K=4096) & 3588 & 775 & 7286 & 192 & 4.0x \\
\rowcolor{zebra}Conv single (C=64) & 99 & 23 & 89 & 18 & 1.3x \\
Conv resnet (d=16) & 9602 & 2063 & 4671 & 142 & 14.5x \\
\rowcolor{zebra}Attention ViT (S=197) & 1711 & 370 & 1831 & 187 &
2.0x \\
Attention long sequence (S=512) & 1775 & 544 & 3768 & 217 & 2.5x \\
\end{longtable}

fp16 accuracy tracks the same boundary the M5 shows: the engine matches
or beats GPU fp16 where the arithmetic is well-conditioned and degrades
on the large-K reduction, reaching \(2.7 \times 10^{-2}\) relative error
at K of 4096.

\section{Dispatch floor and conv
throughput}\label{dispatch-floor-and-conv-throughput}

A single small operation is overhead-bound on the engine: it costs about
0.23 milliseconds regardless of the operation or its size. A relu,
sigmoid, average pool, and small convolution are all between 0.24 and
0.26 milliseconds, with a 64-element linear at 0.23 milliseconds. That
floor is host dispatch and operand transfer, not engine compute, so it
sets a latency a small operation cannot beat.

A convolution amortizes the floor as its spatial size grows, and its
throughput climbs roughly twenty-fold from the floor to the saturated
shape (\cref{tbl:c11-convramp}).

\begin{longtable}[]{@{}lrr@{}}
\caption{A 3x3 convolution at 64 channels on the M1, throughput climbing
as spatial size amortizes the dispatch
floor.}\label{tbl:c11-convramp}\tabularnewline
\toprule\noalign{}
\textbf{Spatial size} & \textbf{Latency} & \textbf{GFLOP/s} \\
\midrule\noalign{}
\endfirsthead
\toprule\noalign{}
\textbf{Spatial size} & \textbf{Latency} & \textbf{GFLOP/s} \\
\midrule\noalign{}
\endhead
\bottomrule\noalign{}
\endlastfoot
16x16 & 0.231 ms & 63 \\
\rowcolor{zebra}32x32 & 0.248 ms & 267 \\
64x64 & 0.395 ms & 718 \\
\rowcolor{zebra}128x128 & 1.062 ms & 1102 \\
256x256 & 3.814 ms & 1247 \\
\end{longtable}

Raising channels rather than spatial size reaches the higher M1
end-to-end conv peak, about 2212 GFLOP/s at 256 channels. A matrix
multiply with a single output row stays overhead-bound regardless of
inner size, reaching only about 5.9 GFLOP/s at an inner dimension of
1024, which is why a decode-shaped projection does not supply the array.

\section{Batch threshold for serving}\label{batch-threshold-for-serving}

The single-stream map is the device choice for one request, but serving
batches requests, and the choice moves with batch size. On a
true-batched encoder block the GPU overtakes the engine on throughput
near a batch of 23, and on a self-attention block near a batch of 6, as
the GPU scales with batch while the engine saturates near a batch of 1.
The energy crossover is at a larger batch than the throughput crossover,
and on three of four serving workloads it never appears. Vision
convolution serving never crosses on either axis, leading throughput by
3.6 to 5.7 times and energy by 6 to 10 times at every batch from 1 to
256. Only bare large-batch matrix multiply converges, to an energy tie
by a batch of 16.

\Cref{tbl:c11-serving} collects the per-workload serving crossovers.

\begin{longtable}[]{@{}
  >{\raggedright\arraybackslash}p{(\linewidth - 4\tabcolsep) * \real{0.3333}}
  >{\raggedright\arraybackslash}p{(\linewidth - 4\tabcolsep) * \real{0.3333}}
  >{\raggedright\arraybackslash}p{(\linewidth - 4\tabcolsep) * \real{0.3333}}@{}}
\caption{The batch at which the GPU overtakes the engine for each
serving workload, on throughput and on
energy.}\label{tbl:c11-serving}\tabularnewline
\toprule\noalign{}
\begin{minipage}[b]{\linewidth}\raggedright
\textbf{Serving workload}
\end{minipage} & \begin{minipage}[b]{\linewidth}\raggedright
\textbf{Throughput crossover to GPU}
\end{minipage} & \begin{minipage}[b]{\linewidth}\raggedright
\textbf{Energy crossover to GPU}
\end{minipage} \\
\midrule\noalign{}
\endfirsthead
\toprule\noalign{}
\begin{minipage}[b]{\linewidth}\raggedright
\textbf{Serving workload}
\end{minipage} & \begin{minipage}[b]{\linewidth}\raggedright
\textbf{Throughput crossover to GPU}
\end{minipage} & \begin{minipage}[b]{\linewidth}\raggedright
\textbf{Energy crossover to GPU}
\end{minipage} \\
\midrule\noalign{}
\endhead
\bottomrule\noalign{}
\endlastfoot
Encoder block, true-batched & near batch 23 & none from batch 1 to
256 \\
\rowcolor{zebra}Self-attention block & near batch 6 & none from batch 1
to 256 \\
Vision convolution & none; engine leads 3.6 to 5.7x & none; engine leads
6 to 10x \\
\rowcolor{zebra}Large-batch matrix multiply & GPU regime throughout &
energy tie near batch 16 \\
\end{longtable}

\section{Verdict table}\label{verdict-table}

\begin{figure}[H]\centering
\includegraphics[width=5.210in]{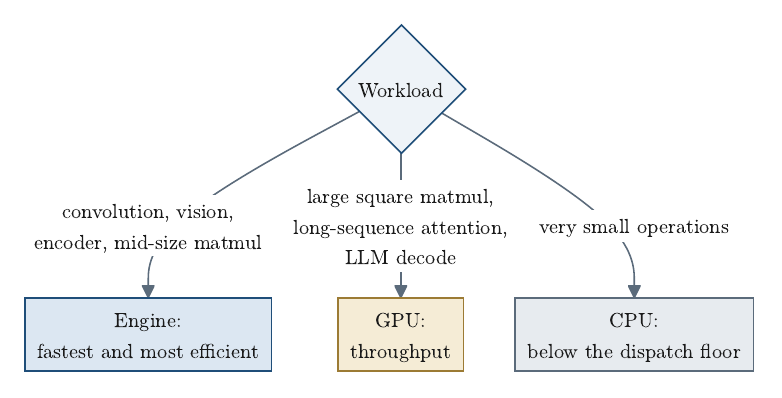}
\caption{Which processor suits which workload.}
\label{fig:c11-verdict}
\end{figure}

Taken together, the verdicts above route a workload to one of the three
processors (\cref{fig:c11-verdict}).

The per-class verdicts collapse into one table (\cref{tbl:c11-verdict}),
which pairs each class with its fastest processor, its most efficient
one on energy per result, and the headline figure behind the verdict.

\begin{longtable}[]{@{}
  >{\raggedright\arraybackslash}p{(\linewidth - 6\tabcolsep) * \real{0.2500}}
  >{\raggedright\arraybackslash}p{(\linewidth - 6\tabcolsep) * \real{0.2500}}
  >{\raggedright\arraybackslash}p{(\linewidth - 6\tabcolsep) * \real{0.2500}}
  >{\raggedright\arraybackslash}p{(\linewidth - 6\tabcolsep) * \real{0.2500}}@{}}
\caption{The fastest and most efficient device among engine, GPU, and
CPU for each workload class, with a standout
figure.}\label{tbl:c11-verdict}\tabularnewline
\toprule\noalign{}
\begin{minipage}[b]{\linewidth}\raggedright
\textbf{Workload}
\end{minipage} & \begin{minipage}[b]{\linewidth}\raggedright
\textbf{Fastest}
\end{minipage} & \begin{minipage}[b]{\linewidth}\raggedright
\textbf{Most efficient}
\end{minipage} & \begin{minipage}[b]{\linewidth}\raggedright
\textbf{Standout number}
\end{minipage} \\
\midrule\noalign{}
\endfirsthead
\toprule\noalign{}
\begin{minipage}[b]{\linewidth}\raggedright
\textbf{Workload}
\end{minipage} & \begin{minipage}[b]{\linewidth}\raggedright
\textbf{Fastest}
\end{minipage} & \begin{minipage}[b]{\linewidth}\raggedright
\textbf{Most efficient}
\end{minipage} & \begin{minipage}[b]{\linewidth}\raggedright
\textbf{Standout number}
\end{minipage} \\
\midrule\noalign{}
\endhead
\bottomrule\noalign{}
\endlastfoot
Convolution and vision & Engine & Engine & Convolution stack about 4.2
times faster and about 13 times the GPU efficiency on M5; about 2 times
faster and 14.5 times on M1 \\
\rowcolor{zebra}Encoder and embedding serving, low to moderate batch &
Engine & Engine & Single-sentence encoder about 4.4 times faster;
throughput crossover to GPU near a batch of 23 \\
Mid-size matrix multiply & Engine & Engine & Engine is faster and more
efficient below the square-operand working set near N of 2048 \\
\rowcolor{zebra}Short-sequence attention & Engine & Engine & Transformer
block at sequence length 197 is fastest, most efficient, and most
accurate in fp16 on the engine \\
Large square matrix multiply, large K & GPU & GPU & About 2.8 times
faster, growing to 3.0 times at saturation: 30.9 against 10.2 fp16
TFLOP/s \\
\rowcolor{zebra}Long-sequence attention & GPU & Engine & At sequence
length 512 the GPU is about 2.1 times faster; the engine keeps a 1.8
times energy edge \\
Autoregressive decode & GPU & GPU & Bandwidth-bound, the GPU's
throughput regime \\
\rowcolor{zebra}Tiny operations & CPU & CPU & Floor matrix multiply 64
by 256 by 256 in about 0.026 ms, below the per-eval floor of about 0.23
ms on M1 \\
\end{longtable}

\section{Picking the processor for a
workload}\label{picking-the-processor-for-a-workload}

The device choice is the workload's regime read against the thresholds
above. Estimate the layer, then route it: compute-bound and resident to
the engine, large-square or long-sequence to the GPU, below the floor to
the CPU.

\begin{Shaded}
\begin{Highlighting}[]
\CommentTok{\# Pick the processor for a workload from its regime, before choosing a device.}
\CommentTok{\# Example workload: a 256 by 4096 times 4096 by 4096 matrix multiply.}

\NormalTok{given workload W }\OperatorTok{=}\NormalTok{ matmul([}\DecValTok{256}\NormalTok{, }\DecValTok{4096}\NormalTok{], [}\DecValTok{4096}\NormalTok{, }\DecValTok{4096}\NormalTok{])}
\NormalTok{given target chip }\OperatorTok{=}\NormalTok{ H13            }\CommentTok{\# M1; thresholds below are read from its roofline}

\CommentTok{\# 1. Estimate the workload statically on the chip\textquotesingle{}s roofline.}
\NormalTok{flops        }\OperatorTok{=}\NormalTok{ total multiply\_adds }\KeywordTok{in}\NormalTok{ W}
\BuiltInTok{bytes}        \OperatorTok{=}\NormalTok{ weight\_bytes(W) }\OperatorTok{+}\NormalTok{ input\_bytes(W) }\OperatorTok{+}\NormalTok{ output\_bytes(W)}
\NormalTok{working\_set  }\OperatorTok{=} \BuiltInTok{bytes}\NormalTok{ that must stay resident at once   }\CommentTok{\# in MB}
\NormalTok{compute\_time }\OperatorTok{=}\NormalTok{ flops }\OperatorTok{/}\NormalTok{ compute\_peak(chip)}
\NormalTok{memory\_time  }\OperatorTok{=} \BuiltInTok{bytes} \OperatorTok{/}\NormalTok{ memory\_bandwidth(chip)}
\NormalTok{floor        }\OperatorTok{=}\NormalTok{ dispatch\_floor(chip)        }\CommentTok{\# about 0.23 ms on H13}

\ControlFlowTok{if} \BuiltInTok{max}\NormalTok{(compute\_time, memory\_time) }\OperatorTok{\textless{}}\NormalTok{ floor:  bound }\OperatorTok{=} \StringTok{"dispatch"}
\ControlFlowTok{else} \ControlFlowTok{if}\NormalTok{ compute\_time }\OperatorTok{\textgreater{}=}\NormalTok{ memory\_time:        bound }\OperatorTok{=} \StringTok{"compute"}
\ControlFlowTok{else}\NormalTok{:                                        bound }\OperatorTok{=} \StringTok{"bandwidth"}

\CommentTok{\# 2. Classify the workload by shape and bound.}
\NormalTok{is\_engine\_shape }\OperatorTok{=}\NormalTok{ (bound }\OperatorTok{==} \StringTok{"compute"}\NormalTok{) }\KeywordTok{and}\NormalTok{ (working\_set }\OperatorTok{\textless{}=} \FloatTok{2.0}\NormalTok{)}
                  \CommentTok{\# convolution, encoder block, or mid{-}size GEMM that stays on chip}
\NormalTok{is\_gpu\_shape    }\OperatorTok{=}\NormalTok{ (working\_set }\OperatorTok{\textgreater{}} \FloatTok{2.0}\NormalTok{)                  }\CommentTok{\# large square GEMM, streams from DRAM}
                  \KeywordTok{or}\NormalTok{ (sequence\_length(W) }\KeywordTok{is} \BuiltInTok{long}\NormalTok{)      }\CommentTok{\# long{-}sequence attention}
\NormalTok{is\_cpu\_shape    }\OperatorTok{=}\NormalTok{ (bound }\OperatorTok{==} \StringTok{"dispatch"}\NormalTok{)                }\CommentTok{\# below the floor, too small to dispatch}

\CommentTok{\# 3. Apply the regime rules to pick the processor.}
\ControlFlowTok{if}\NormalTok{ is\_cpu\_shape:        choose CPU       }\CommentTok{\# call overhead dominates, keep it local}
\ControlFlowTok{else} \ControlFlowTok{if}\NormalTok{ is\_gpu\_shape:   choose GPU       }\CommentTok{\# GPU saturates wide work and holds fp16 better here}
\ControlFlowTok{else} \ControlFlowTok{if}\NormalTok{ is\_engine\_shape: choose ENGINE   }\CommentTok{\# where the engine is fastest and most efficient}
\ControlFlowTok{else}\NormalTok{:                    choose ENGINE   }\CommentTok{\# default for compute{-}bound resident work}

\CommentTok{\# The 256 by 4096 by 4096 multiply has a working set past the 2 MB on{-}chip limit,}
\CommentTok{\# so the rule routes it to the GPU, which the verdict table records as about 2.8x faster.}
\ControlFlowTok{return}\NormalTok{ chosen\_processor, bound, working\_set}
\end{Highlighting}
\end{Shaded}

\chapter{Across the chip family}\label{across-the-chip-family}

\begin{summarybox}

An M(n) chip has the H(n+12) ANE architecture, so M1 is H13 and M5 is
H17. A network that compiles and runs on one generation does so on the
others. One compiler binary builds every target, and only a per-target
data table changes. Only fp16 numerics vary from chip to chip, and only
at a unit in the last place. Target the generation a network needs by
its operation set. Then verify per chip only the cancellation-sensitive
reductions and the width-axis slices.

\end{summarybox}

\section{Naming rule}\label{naming-rule}

The M-series engine and the contemporaneous A-series engine are the same
architecture under two product names, offset by a fixed amount. An M(n)
chip has the H(n+12) ANE architecture, the compact relation
\(M(n) \rightarrow H(n+12)\). M1 is H13, M2 is H14, M3 is H15, M4 is
H16, and M5 is H17. The A-series anchor is one generation over: the A13
and the M1 share H13, and the A17 and the M5 share H17.

Core count and clock are what scale across the generations
(\cref{tbl:c12-family}).

\begin{longtable}[]{@{}
  >{\raggedright\arraybackslash}p{(\linewidth - 8\tabcolsep) * \real{0.2000}}
  >{\raggedright\arraybackslash}p{(\linewidth - 8\tabcolsep) * \real{0.2000}}
  >{\raggedright\arraybackslash}p{(\linewidth - 8\tabcolsep) * \real{0.2000}}
  >{\raggedright\arraybackslash}p{(\linewidth - 8\tabcolsep) * \real{0.2000}}
  >{\raggedright\arraybackslash}p{(\linewidth - 8\tabcolsep) * \real{0.2000}}@{}}
\caption{The M-series chips with their A-series anchor, engine
architecture string, core count, and
clock.}\label{tbl:c12-family}\tabularnewline
\toprule\noalign{}
\begin{minipage}[b]{\linewidth}\raggedright
\textbf{Chip}
\end{minipage} & \begin{minipage}[b]{\linewidth}\raggedright
\textbf{A-series anchor}
\end{minipage} & \begin{minipage}[b]{\linewidth}\raggedright
\textbf{ANE architecture}
\end{minipage} & \begin{minipage}[b]{\linewidth}\raggedright
\textbf{NE cores}
\end{minipage} & \begin{minipage}[b]{\linewidth}\raggedright
\textbf{Clock}
\end{minipage} \\
\midrule\noalign{}
\endfirsthead
\toprule\noalign{}
\begin{minipage}[b]{\linewidth}\raggedright
\textbf{Chip}
\end{minipage} & \begin{minipage}[b]{\linewidth}\raggedright
\textbf{A-series anchor}
\end{minipage} & \begin{minipage}[b]{\linewidth}\raggedright
\textbf{ANE architecture}
\end{minipage} & \begin{minipage}[b]{\linewidth}\raggedright
\textbf{NE cores}
\end{minipage} & \begin{minipage}[b]{\linewidth}\raggedright
\textbf{Clock}
\end{minipage} \\
\midrule\noalign{}
\endhead
\bottomrule\noalign{}
\endlastfoot
M1 & A13 & H13 & 4 (base), 8 (Pro and Max) & \textasciitilde1.14 GHz \\
\rowcolor{zebra}M2 & A14 & H14g & 4 (base), 8 (Pro and Max), 32
(Max-class) & measured \\
M3 & A15 & H15 & 4 (base), predicted & predicted \\
\rowcolor{zebra}M4 & A16 & H16 & 4 (base), predicted & predicted \\
M5 & A17 & H17s & 16 & \textasciitilde1.89 GHz \\
\end{longtable}

Three rows are measured on physical silicon: M1 (H13), M2 (a Pro
reporting H14g), and M5 (H17s); the M3 and M4 rows are decompile-derived
from the per-family device tables and not individually measured, with
the A15/M3 generation the one rail that remains unmeasured. The M2
measurement closes the middle of the sequence: a seeded classifier
trains to the M1 number to the digit, the four fp16 axes match the M1
bit for bit, and a watt-complete device map reproduces the M1 and M5
shape. The A14 is thus the A13's numerical twin on every axis measured.
The fuller silicon-to-target table, the board-type sequence, and the
full set of 28 compiler targets are the subject of a later chapter
\hyperref[ref-appleane]{{[}AppleANE{]}}.

The rule reads directly off the device tables and is confirmed on the
measured parts. The live M1 reports the architecture string h13g, the
Pro and Max variant of H13, and the M5 compiles to the H17 target and
resolves the H17s variant on disk. The system frameworks corroborate the
sequence: the on-device video upscaler includes exactly the five targets
H13 through H17, which are the five Mac generations M1 through M5. The
target a network needs can thus be named by generation, and the mapping
holds without probing the silicon.

\section{What holds across the
family}\label{what-holds-across-the-family}

A network that compiles and runs on one generation compiles and runs on
the others. The compiler is a single binary that constructs any target
on demand, so the program format, operation legality, and datapath are
shared, and only a per-target data table changes underneath them.

The operation limits are properties of the family, not of one chip. The
operations with no hardware path on any current part, such as the
product reduction, scatter family, and recurrent cells, are absent on
every generation from the M1 through the M5. Gated operations arrive at
a known generation and stay: the texture-engine sampler operations turn
on at the A14, and native sin and cos turn on at the A15. A network that
avoids them runs everywhere, and a network that uses them runs from its
unlock generation forward. No operation the M1 can compute is missing on
the M5, because each newer engine is an operation superset of the one
before it.

Each generation adds one capability over the one before it, then stops
(\cref{tbl:c12-caps}).

\begin{longtable}[]{@{}
  >{\raggedright\arraybackslash}p{(\linewidth - 2\tabcolsep) * \real{0.5000}}
  >{\raggedright\arraybackslash}p{(\linewidth - 2\tabcolsep) * \real{0.5000}}@{}}
\caption{The single capability each engine generation adds, read off the
per-target capability bytes.}\label{tbl:c12-caps}\tabularnewline
\toprule\noalign{}
\begin{minipage}[b]{\linewidth}\raggedright
\textbf{Generation}
\end{minipage} & \begin{minipage}[b]{\linewidth}\raggedright
\textbf{Capability added}
\end{minipage} \\
\midrule\noalign{}
\endfirsthead
\toprule\noalign{}
\begin{minipage}[b]{\linewidth}\raggedright
\textbf{Generation}
\end{minipage} & \begin{minipage}[b]{\linewidth}\raggedright
\textbf{Capability added}
\end{minipage} \\
\midrule\noalign{}
\endhead
\bottomrule\noalign{}
\endlastfoot
A13 (M1) & three-dimensional convolution, the sixteen-deep kernel,
native softmax, layer norm, all reductions, and fused attention \\
\rowcolor{zebra}A14 (M2) & the texture-engine samplers (resize,
crop-resize, resample, affine, hardware gather) and cross-die
addressing \\
A15 (M3) & native sin and cos, the dropout and random path, and global
argument-min and argument-max \\
\rowcolor{zebra}A16 (M4) & the tensor dimension limit rises from 16384
to 65536, and the fp16 kernel-width ceiling rises from 13 to 15 \\
A17 (M5) & no operation over A16; the NE-core count scales \\
\end{longtable}

The A17 and A18 add no operation over the A16: identical dimension
limits, the same texture engine, the same legal operation set, differing
only in NE-core count, which scales throughput. The dimension limit is
not a single number per chip: on the M2 the spatial and contraction
extents cap at 16384 while the channel axis caps at 65536, exactly four
times the spatial cap. The limit thus belongs to the axis an operation
uses rather than to the tensor.

The newer parts scale the core count and the clock; they do not change
the programming model. The core count runs 4 on the M1, 8 on its Pro and
Max variant, and 16 on the M5, and the operating clock rises from
roughly 1.14 GHz to roughly 1.89 GHz across that span. The fp16
datapath, the wide accumulator, and the form of the roofline hold across
the family unchanged. The measured M5 confirms the scaling: about 19.6
fp16 TFLOP/s on the matmul slope and about 14.3 fp16 TFLOP/s on the
convolution peak, both from the fused-chain probe on the same network
with no source change, against the roofline saturation peak of 18.8
TFLOP/s in Chapter \hyperref[roofline]{9}. A single large matmul above
the dispatch floor runs at about 9.5 fp16 TFLOP/s, the M5 analogue of
the M1's 4.8, and the engine streams weights at about 145 GB/s over two
DRAM read channels, near three times the M1's 51 GB/s. Those peaks are
set by the larger core count and the higher clock; the datapath, the
metric on which the generations compare, is unchanged. The working-set
threshold moves with the silicon, from near 2 MB on the M1 to a measured
4.72 MB on the M5, scaling with the larger 16-core on-chip memory.

\section{One thing that varies}\label{one-thing-that-varies}

The single property that differs from chip to chip is fp16 numerics. The
accumulator width is uniform across every engine and the compiler text
is identical, so a cross-chip value difference can only come from a
data-selected codegen route that changes the order in which fp16
operations combine. That surface is limited: most route changes are a
numerical no-op, because the wide accumulator absorbs the reordering,
and the rest are at a unit in the last place, set by tiling-boundary
alignment.

The cross-generation measurement fixes the scale. The same seeded
convolutional classifier trained on the M1, M2, and M5 reaches 0.9080,
0.9080, and 0.9070 test accuracy, each deterministic across repeated
runs, a difference of one test sample in a thousand between the ends.
The M2 is exactly on the M1 number, which puts the entire fp16 training
drift at the A16 generation rather than spread across the family: the M1
and the M2 are numerical twins, and the gap opens only at the M5. That
gap is the drift of sub-unit-in-the-last-place fp16 differences
compounding over a few hundred training steps, real and negligible. The
cross-silicon predictions extracted from the device tables were
confirmed on the M5: all ten, covering throughput, the working-set
threshold, operation limits, texture engine, and fp16 slice behavior,
held on the real part. The one finite-to-infinity axis, a slice
saturation that occurs on the M1, takes the non-saturating route on the
M5, as predicted.

\section{fp16 divergence axes}\label{fp16-divergence-axes}

That data-selected codegen surface reduces to four axes. Three of them
are at most a unit in the last place, and one is a finite-to-infinity
saturation that occurs on the older parts (\cref{tbl:c12-axes}).

\begin{longtable}[]{@{}
  >{\raggedright\arraybackslash}p{(\linewidth - 6\tabcolsep) * \real{0.2500}}
  >{\raggedright\arraybackslash}p{(\linewidth - 6\tabcolsep) * \real{0.2500}}
  >{\raggedright\arraybackslash}p{(\linewidth - 6\tabcolsep) * \real{0.2500}}
  >{\raggedright\arraybackslash}p{(\linewidth - 6\tabcolsep) * \real{0.2500}}@{}}
\caption{The four fp16 cross-chip divergence axes, the codegen route
each selects, and the bounded magnitude of
each.}\label{tbl:c12-axes}\tabularnewline
\toprule\noalign{}
\begin{minipage}[b]{\linewidth}\raggedright
\textbf{Axis}
\end{minipage} & \begin{minipage}[b]{\linewidth}\raggedright
\textbf{Mechanism}
\end{minipage} & \begin{minipage}[b]{\linewidth}\raggedright
\textbf{Effect}
\end{minipage} & \begin{minipage}[b]{\linewidth}\raggedright
\textbf{Magnitude}
\end{minipage} \\
\midrule\noalign{}
\endfirsthead
\toprule\noalign{}
\begin{minipage}[b]{\linewidth}\raggedright
\textbf{Axis}
\end{minipage} & \begin{minipage}[b]{\linewidth}\raggedright
\textbf{Mechanism}
\end{minipage} & \begin{minipage}[b]{\linewidth}\raggedright
\textbf{Effect}
\end{minipage} & \begin{minipage}[b]{\linewidth}\raggedright
\textbf{Magnitude}
\end{minipage} \\
\midrule\noalign{}
\endhead
\bottomrule\noalign{}
\endlastfoot
Slice saturation & A width-axis slice with a nonzero offset routes
through a fixed-point crop that multiplies by sixteen & A source value
above 4094 saturates to plus or minus infinity, since 4094 times sixteen
is the 65504 fp16 ceiling & finite to infinity \\
\rowcolor{zebra}Reduction then square fusion & A reduction immediately
followed by a square or multiply can fuse, removing one intermediate
rounding step & Drops one fp16 rounding step & at most one unit in the
last place \\
Reduction route & A reduction selects a transpose route or a reshape
route by an extent threshold, 192 on the older parts and 384 from the M3
& Reorders partial sums & numerical no-op, the wide accumulator absorbs
it \\
\rowcolor{zebra}Tiling granularity & The partial-sum tile alignment is
set by the patch and core-count fields, granularity near 128 & A sum off
a tile boundary loses one rounding increment & at most one unit in the
last place \\
\end{longtable}

The saturation axis is the only one that changes a finite value into an
infinity, and it is magnitude-gated: it triggers only when a
width-offset slice contains a value above 4094. Measured on the M1 the
threshold is exact: a width-offset slice is finite at 4094 and goes to
plus or minus infinity at 4100, while the zero-offset control stays
finite even at 60000. The M2 saturates bit-identically to the M1, which
corrects the earlier reading that the non-saturating route arrives at
the A14: the saturation persists through the A14 and the M5 is the part
that takes the non-saturating route. The reduction-then-square fusion
never manifests on silicon: the M1, M2, and M5 all measure the unfused
result, so that axis is uniform across the family. A divergence is thus
predictable from a small set of fields without running every chip: a
reduction or normalization denominator can differ by at most one unit in
the last place where the tiling or route fields differ. A width-offset
slice has saturation risk only when its values can exceed 4094.

\section{Targeting and per-chip
verification}\label{targeting-and-per-chip-verification}

Choose the generation a network requires from its operation set, then
let the program run across every part at and above that generation.
Verify per chip only the numerics that can move: the
cancellation-sensitive reductions, the variance and normalization
denominators, and any width-axis slice whose values can exceed the
saturation bound. Everything else is portable by construction: the same
source, the same operation legality, faster on the newer silicon by the
core and clock scaling.

\section{Compiling for a target
generation}\label{compiling-for-a-target-generation}

The compiler constructs any target from its per-chip table, so a
developer compiles a program for the oldest generation it must support,
then runs it unchanged on every part at and above that generation. A
static estimate against a target reads back that target's core-and-clock
scaling, no silicon required.

Set the floor generation in the compiler options, then estimate against
the newer targets to read the core-and-clock speedup.

\begin{Shaded}
\begin{Highlighting}[]
\CommentTok{/* The target generation is held in the compiler options as a TargetArchitecture string, */}
\CommentTok{/* so the same source compiles for whatever floor the network must support.            */}
\NormalTok{e5rt\_e5\_compiler\_options\_create}\OperatorTok{(\&}\NormalTok{options}\OperatorTok{);}
\NormalTok{e5rt\_e5\_compiler\_options\_set\_custom\_ane\_compiler\_options}\OperatorTok{(}\NormalTok{options}\OperatorTok{,} \StringTok{"TargetArchitecture=h13"}\OperatorTok{);}
\NormalTok{e5rt\_e5\_compiler\_compile}\OperatorTok{(}\NormalTok{compiler}\OperatorTok{,}\NormalTok{ model\_path}\OperatorTok{,}\NormalTok{ options}\OperatorTok{,} \OperatorTok{\&}\NormalTok{library}\OperatorTok{);}  \CommentTok{/* M1 floor, runs M1..M5 */}

\CommentTok{/* Then the same drive: retain the function, build the op, and dispatch on a stream. */}
\NormalTok{e5rt\_program\_library\_retain\_program\_function}\OperatorTok{(}\NormalTok{library}\OperatorTok{,}\NormalTok{ fn\_name}\OperatorTok{,} \OperatorTok{\&}\NormalTok{function}\OperatorTok{);}
\NormalTok{e5rt\_precompiled\_compute\_op\_create\_options\_create\_with\_program\_function}\OperatorTok{(\&}\NormalTok{op\_opts}\OperatorTok{,}\NormalTok{ function}\OperatorTok{);}
\NormalTok{e5rt\_execution\_stream\_operation\_create\_precompiled\_compute\_operation\_with\_options}\OperatorTok{(\&}\NormalTok{op}\OperatorTok{,}\NormalTok{ op\_opts}\OperatorTok{);}
\NormalTok{e5rt\_execution\_stream\_encode\_operation}\OperatorTok{(}\NormalTok{stream}\OperatorTok{,}\NormalTok{ op}\OperatorTok{);}
\NormalTok{e5rt\_execution\_stream\_execute\_sync}\OperatorTok{(}\NormalTok{stream}\OperatorTok{);}
\end{Highlighting}
\end{Shaded}

A network compiled to \texttt{h13} runs on every generation from the M1
through the M5; estimating the same graph against \texttt{h17s} reads
back the M5 scaling without the M5 in hand.

\section{Reference: per-family scaling
constants}\label{reference-per-family-scaling-constants}

Everything this chapter pins down is gathered in
\cref{tbl:c12-constants}.

\begin{longtable}[]{@{}
  >{\raggedright\arraybackslash}p{(\linewidth - 4\tabcolsep) * \real{0.3333}}
  >{\raggedright\arraybackslash}p{(\linewidth - 4\tabcolsep) * \real{0.3333}}
  >{\raggedright\arraybackslash}p{(\linewidth - 4\tabcolsep) * \real{0.3333}}@{}}
\caption{The per-family scaling constants and the figures that fix
cross-generation behavior, with the silicon each was measured
on.}\label{tbl:c12-constants}\tabularnewline
\toprule\noalign{}
\begin{minipage}[b]{\linewidth}\raggedright
\textbf{Quantity}
\end{minipage} & \begin{minipage}[b]{\linewidth}\raggedright
\textbf{Value}
\end{minipage} & \begin{minipage}[b]{\linewidth}\raggedright
\textbf{Silicon}
\end{minipage} \\
\midrule\noalign{}
\endfirsthead
\toprule\noalign{}
\begin{minipage}[b]{\linewidth}\raggedright
\textbf{Quantity}
\end{minipage} & \begin{minipage}[b]{\linewidth}\raggedright
\textbf{Value}
\end{minipage} & \begin{minipage}[b]{\linewidth}\raggedright
\textbf{Silicon}
\end{minipage} \\
\midrule\noalign{}
\endhead
\bottomrule\noalign{}
\endlastfoot
Naming rule & \(M(n) \rightarrow H(n+12)\) & family-wide \\
\rowcolor{zebra}NE cores & 4 (base), 8 (Pro and Max) & M1/H13 \\
NE cores & 16 & M5/H17s \\
\rowcolor{zebra}Operating clock & \textasciitilde1.14 GHz & M1/H13 \\
Operating clock & \textasciitilde1.89 GHz & M5/H17s \\
\rowcolor{zebra}Matmul-slope peak & about 19.6 fp16 TFLOP/s & M5/H17s \\
Single-program matmul peak & about 9.5 fp16 TFLOP/s & M5/H17s \\
\rowcolor{zebra}Convolution peak & about 14.3 fp16 TFLOP/s & M5/H17s \\
\end{longtable}

Part IV turns from the engine in isolation to the workloads that run on
it.

\anepartopen{Part IV}{Workloads}
\anechap{13}{Vision, convolution, and encoders}{The engine's strongest classes, fitted as one fused program.}
\anechap{14}{LLM case study}{Why decode belongs on the GPU, and the hybrid placement that remains.}
\anechap{15}{Training on the engine}{Gradients as graph inputs, resident optimizer state, and cross-generation parity.}
\anechap{16}{Numerical and scientific computing}{Stencils, Fourier transforms, and other non-network work cast as matmuls.}
\anepartclose

\chapter{Vision, convolution, and encoders}\label{vision-convolution-and-encoders}

\begin{summarybox}

A 256-channel 3x3 convolution runs about 3.8 times faster than the GPU,
at about 9 times the energy efficiency. The engine also draws less
absolute power on every workload class measured. Serve encoders and
embeddings on the engine below a batch of about 23, and a self-attention
block below a batch of about 6. Serve vision convolution on it at every
batch. On-engine image preprocessing arrives on the A14 generation and
later; the M1 has no texture engine. On the M1 and the A14, keep slice
and crop source magnitudes below 4094 on the width axis or they saturate
to infinity.

\end{summarybox}

Convolution and vision are the work the engine is built for. This
chapter gives the measured economics of that work against the GPU, and
the batch threshold below which encoder serving stays on the engine. It
also covers the on-engine image preprocessing path and the family that
gates it, and the model-construction rules that keep a vision or encoder
model on the engine end to end.

\section{Convolution datapath}\label{convolution-datapath}

A 3x3 convolution at 256 channels runs about 3.8 times faster on the
engine than the same kernel on the GPU, at about 9 times the energy
efficiency per result \hyperref[ref-applecoreml]{{[}AppleCoreML{]}}. The
advantage widens with depth. Stack sixteen 3x3 convolutions at 256
channels and the engine is about 2 times faster than the GPU at about
14.5 times its energy efficiency on the M1, the largest power difference
in the M1 workload set. On the M5 it runs about 4.2 times faster at
about 13 times the efficiency. A ResNet-18 forward pass runs about 6.1
times faster than the GPU reference and at about 11 times the energy
efficiency per inference.

Two properties of the datapath produce this. Winograd lowering is one.
Any dense 3x3 stride-1 convolution with enough channels lowers to the
F(2x2, 3x3) transform, which replaces 36 direct multiplies per 2x2
output tile with 16, a factor of about 2.25 reduction in multiplies. The
compiler selects this path automatically for dense non-unicast 3x3
stride-1 layers and falls back to direct convolution otherwise, so a
model written with 3x3 stride-1 layers gets the reduction without any
annotation. A layer takes Winograd only with enough channels to amortize
the transform, and a float kernel requires more than a non-float one;
the full eligibility test, two tile sizes, and work-threshold derivation
are in chapter \hyperref[datapath-and-mac-geometry]{20}. There is no
accumulator widening tied to the transform; its precision safety rests
entirely on that higher float work threshold. The engine's lower power
draw is the other. On a large compute-bound matrix multiply the engine
draws 4.4 W against the GPU's 32.5 W, a 4.0 times efficiency advantage
that holds even on the matrix-multiply class where the GPU runs faster.
Across the M1 workload set the engine runs at 2 to 14.5 times the GPU's
energy efficiency, drawing less absolute power on every class measured.

The end-to-end convolution ceiling on the M1 is about 1.8 fp16 TFLOP/s
at about 1.78 W, or about 643 GFLOP/s per watt under load. A 3x3
convolution at 256 channels and a 28 by 28 feature map reaches that
peak: it runs in about 0.51 ms at about 1823 GFLOP/s. Throughput rises
with spatial size as the fixed per-eval overhead amortizes, and the same
3x3 64-channel kernel runs at about 63 GFLOP/s at 16 by 16 and about
1102 GFLOP/s at 128 by 128.

\section{MAC array and its tiling}\label{mac-array-and-its-tiling}

The engine is a fixed-geometry multiply array supplied by DMA engines
that re-base per tile. On the M1 the array is four NE cores, read from
the per-chip parameter table as a core count of 4, with each core a
two-dimensional multiply tile backed by an accumulator file of 8
work-units. The compiler assigns output channels across the cores by a
strided round-robin: channel \(c\) is on core \(c \bmod 4\), so the four
cores are independent parallel slices and the scheduling granule is one
output channel.

The parallelism is measured directly. Driving a single heavy convolution
at each output-channel count in a back-to-back dispatch loop, where the
per-dispatch overhead pins the dispatch rate constant, the sustained
multiply rate rises in exact integer multiples with the number of cores
active, as \cref{tbl:c13-core-scaling} records at one through four
active cores.

\begin{longtable}[]{@{}
  >{\raggedleft\arraybackslash}p{(\linewidth - 8\tabcolsep) * \real{0.1905}}
  >{\centering\arraybackslash}p{(\linewidth - 8\tabcolsep) * \real{0.2381}}
  >{\raggedleft\arraybackslash}p{(\linewidth - 8\tabcolsep) * \real{0.1905}}
  >{\raggedleft\arraybackslash}p{(\linewidth - 8\tabcolsep) * \real{0.1905}}
  >{\raggedleft\arraybackslash}p{(\linewidth - 8\tabcolsep) * \real{0.1905}}@{}}
\caption{Per-core throughput and power scaling on the M1, measured at a
fixed dispatch rate.}\label{tbl:c13-core-scaling}\tabularnewline
\toprule\noalign{}
\begin{minipage}[b]{\linewidth}\raggedleft
\textbf{Output channels}
\end{minipage} & \begin{minipage}[b]{\linewidth}\centering
\textbf{Cores active (\(c \bmod 4\))}
\end{minipage} & \begin{minipage}[b]{\linewidth}\raggedleft
\textbf{Engine power, net}
\end{minipage} & \begin{minipage}[b]{\linewidth}\raggedleft
\textbf{GMAC/s}
\end{minipage} & \begin{minipage}[b]{\linewidth}\raggedleft
\textbf{Ratio to one core}
\end{minipage} \\
\midrule\noalign{}
\endfirsthead
\toprule\noalign{}
\begin{minipage}[b]{\linewidth}\raggedleft
\textbf{Output channels}
\end{minipage} & \begin{minipage}[b]{\linewidth}\centering
\textbf{Cores active (\(c \bmod 4\))}
\end{minipage} & \begin{minipage}[b]{\linewidth}\raggedleft
\textbf{Engine power, net}
\end{minipage} & \begin{minipage}[b]{\linewidth}\raggedleft
\textbf{GMAC/s}
\end{minipage} & \begin{minipage}[b]{\linewidth}\raggedleft
\textbf{Ratio to one core}
\end{minipage} \\
\midrule\noalign{}
\endhead
\bottomrule\noalign{}
\endlastfoot
1 & 1 & 811 mW & 3.8 & 1.00x \\
\rowcolor{zebra}2 & 2 & 822 mW & 7.6 & 2.00x \\
3 & 3 & 831 mW & 11.4 & 3.00x \\
\rowcolor{zebra}4 & 4 & 843 mW & 15.4 & 4.05x \\
\end{longtable}

The power rail steps about 10 to 11 mW per added core over an always-on
floor near 800 mW, the floor being the base plus dispatch domain and
each step one of the four independently power-gated compute domains
turning on. Above four output channels the array fills more lanes inside
the same four cores, and the rate keeps rising linearly while the
dispatch rate stays floor-bound.

The compiler tiles output channels into output-channel groups sized to
the accumulator file. The group size is about the 8-accumulator budget
divided by the kernel-element count \(k_w k_h k_d\), rounded down to a
power of two and capped by a per-element byte budget of 32, 16, or 8
bytes depending on the weight format. A 3x3 convolution thus has about 9
times the per-channel accumulator pressure of a 1x1, so its group is
about 9 times smaller and it needs more passes. The measured signature
is a super-linear cost step: a 1x1 fp16 convolution doubles its
per-layer cost once the output channel count crosses the accumulator
file between about 192 and 256 channels. A 3x3 convolution reaches that
pass-doubling threshold at fewer channels for the same reason. This is
the mechanism that makes Winograd worth selecting for 3x3: cutting the
effective kernel-element count enlarges the group and relieves the
accumulator pressure the threshold makes visible.

\section{Convolution variants and their
lowering}\label{convolution-variants-and-their-lowering}

The convolution variants all map onto the same multiply array, and what
differs is how the compiler tiles the weights and sets up the channel
grouping, which \cref{tbl:c13-conv-variants} gives variant by variant.

\begin{longtable}[]{@{}
  >{\raggedright\arraybackslash}p{(\linewidth - 2\tabcolsep) * \real{0.5000}}
  >{\raggedright\arraybackslash}p{(\linewidth - 2\tabcolsep) * \real{0.5000}}@{}}
\caption{How each convolution variant lowers onto the multiply array on
the M1.}\label{tbl:c13-conv-variants}\tabularnewline
\toprule\noalign{}
\begin{minipage}[b]{\linewidth}\raggedright
\textbf{Variant}
\end{minipage} & \begin{minipage}[b]{\linewidth}\raggedright
\textbf{Lowering onto the datapath}
\end{minipage} \\
\midrule\noalign{}
\endfirsthead
\toprule\noalign{}
\begin{minipage}[b]{\linewidth}\raggedright
\textbf{Variant}
\end{minipage} & \begin{minipage}[b]{\linewidth}\raggedright
\textbf{Lowering onto the datapath}
\end{minipage} \\
\midrule\noalign{}
\endhead
\bottomrule\noalign{}
\endlastfoot
Standard 2D convolution & native; weights tiled into output-channel
groups, channels assigned across the cores by a strided round-robin \\
\rowcolor{zebra}Transpose or deconvolution & native; runs as a
fractionally-strided forward convolution on the same array, and serves
the convolution data gradient \\
Dilated & lowered by a space-to-batch decomposition with a factor list
of 2, 3, 4, or 8, folding the dilation into a strided input gather so
the kernel itself stays dense \\
\rowcolor{zebra}Depthwise & native; each channel is its own group with
no cross-channel reduction, one input channel per output channel \\
Grouped & native; the group count partitions the channel-to-core
assignment, and runs best when the group count divides the channel count
and the core count \\
\rowcolor{zebra}3D convolution & capability present in the hardware
parameter table but not reachable on the direct path; the compiler
reports the operation as not implemented on every backend \\
\end{longtable}

The speed and efficiency advantage holds across the workload classes the
engine is built for (\cref{tbl:c13-engine-vs-gpu}).

\begin{longtable}[]{@{}
  >{\raggedright\arraybackslash}p{(\linewidth - 4\tabcolsep) * \real{0.3333}}
  >{\raggedright\arraybackslash}p{(\linewidth - 4\tabcolsep) * \real{0.3333}}
  >{\raggedright\arraybackslash}p{(\linewidth - 4\tabcolsep) * \real{0.3333}}@{}}
\caption{Engine versus GPU speed and efficiency across convolution,
ResNet, matrix multiply, and encoder
workloads.}\label{tbl:c13-engine-vs-gpu}\tabularnewline
\toprule\noalign{}
\begin{minipage}[b]{\linewidth}\raggedright
\textbf{Workload}
\end{minipage} & \begin{minipage}[b]{\linewidth}\raggedright
\textbf{Engine vs GPU speed}
\end{minipage} & \begin{minipage}[b]{\linewidth}\raggedright
\textbf{Engine vs GPU efficiency}
\end{minipage} \\
\midrule\noalign{}
\endfirsthead
\toprule\noalign{}
\begin{minipage}[b]{\linewidth}\raggedright
\textbf{Workload}
\end{minipage} & \begin{minipage}[b]{\linewidth}\raggedright
\textbf{Engine vs GPU speed}
\end{minipage} & \begin{minipage}[b]{\linewidth}\raggedright
\textbf{Engine vs GPU efficiency}
\end{minipage} \\
\midrule\noalign{}
\endhead
\bottomrule\noalign{}
\endlastfoot
3x3 convolution (256 channels) & 3.8x faster & 9x more efficient \\
\rowcolor{zebra}Convolution stack, 16 deep at 256 channels (M1) & 2x
faster & 14.5x more efficient \\
Convolution stack, 16 deep at 256 channels (M5) & 4.2x faster & 13x more
efficient \\
\rowcolor{zebra}ResNet-18 forward & 6.1x faster & 11x more efficient \\
Batched matrix multiply & faster below N of 2048 & more efficient at
every batch size \\
\rowcolor{zebra}Single-sentence encoder & 4.4x faster & faster at low to
moderate batch \\
\end{longtable}

\section{Encoders and embeddings}\label{encoders-and-embeddings}

An encoder forward pass favors the engine at low to moderate batch on
both latency and energy. A single-sentence encoder runs about 4.4 times
faster than the GPU. A twelve-layer encoder forward runs about 1.5 times
faster and at about 18 times the energy efficiency per inference.
Short-sequence attention, a transformer block at sequence length 197, is
at once the fastest, most efficient, and most accurate in fp16 on the
engine, because its partial sums stay in range and the wide accumulator
retains their precision.

Batch size moves the choice, because serving batches requests. The
engine saturates near a batch of 1 while the GPU scales with batch,
which produces a throughput crossover. On a true-batched encoder block
the GPU overtakes the engine on throughput near a batch of 23, and on a
self-attention block near a batch of 6. The energy crossover is at a
larger batch than the throughput crossover. On vision convolution
serving the energy crossover never appears: the engine leads throughput
by 3.6 to 5.7 times and energy by 6 to 10 times at every batch from 1 to
256. Above the encoder and self-attention crossovers the GPU is the
throughput device.

\section{On-engine image
preprocessing}\label{on-engine-image-preprocessing}

Resize, crop-and-resize, grid sample, affine warp, and reflective or
symmetric padding run on a single hardware sampling datapath, the
texture engine, on the A14 generation and later
\hyperref[ref-applevision]{{[}AppleVision{]}}. The datapath is a
DMA-side sampler fused onto a layer: it reads a coordinate or box tensor
through a fixed interleave and applies bilinear or nearest-neighbor
interpolation in line with the convolution that consumes its output. The
payoff is the elimination of a host preprocessing stage. Image
dequantization and resize execute on the engine in the same program as
the model, so the resampled tensor never makes a round trip to the host
between the camera frame and the first convolution.

One datapath backs a fixed set of front-end operations, each gated by
the same parameter-table bool (\cref{tbl:c13-texture-ops}).

\begin{longtable}[]{@{}
  >{\raggedright\arraybackslash}p{(\linewidth - 4\tabcolsep) * \real{0.3000}}
  >{\raggedleft\arraybackslash}p{(\linewidth - 4\tabcolsep) * \real{0.4000}}
  >{\raggedright\arraybackslash}p{(\linewidth - 4\tabcolsep) * \real{0.3000}}@{}}
\caption{The texture-engine operation set and how each lowers on the A14
generation and later.}\label{tbl:c13-texture-ops}\tabularnewline
\toprule\noalign{}
\begin{minipage}[b]{\linewidth}\raggedright
\textbf{Operation}
\end{minipage} & \begin{minipage}[b]{\linewidth}\raggedleft
\textbf{Bottoms}
\end{minipage} & \begin{minipage}[b]{\linewidth}\raggedright
\textbf{Lowering on the A14 and later}
\end{minipage} \\
\midrule\noalign{}
\endfirsthead
\toprule\noalign{}
\begin{minipage}[b]{\linewidth}\raggedright
\textbf{Operation}
\end{minipage} & \begin{minipage}[b]{\linewidth}\raggedleft
\textbf{Bottoms}
\end{minipage} & \begin{minipage}[b]{\linewidth}\raggedright
\textbf{Lowering on the A14 and later}
\end{minipage} \\
\midrule\noalign{}
\endhead
\bottomrule\noalign{}
\endlastfoot
Resize, upsample & 1 & native texture-engine resize unit \\
\rowcolor{zebra}Crop-and-resize, ROI-align & 2 & native; a
denormalization scale-and-bias pair precedes the index input, box tensor
in fp16 \\
Resample, grid-sample, warp & 2 & native; coordinate tensor of 1 or 2
channels read through the index interleave \\
\rowcolor{zebra}Affine transform & 2 & decomposes to resample with a
computed coordinate grid; matrix fp16, six coefficients \\
Resize-as & 2 & decomposes to resize \\
\rowcolor{zebra}Reflective or symmetric padding & 1 & uses the
texture-engine pad mode \\
Hardware gather & 2 & uses the texture-engine index interleave path \\
\end{longtable}

The sampler reads its coordinate or box tensor through an interleave of
1, 2, 3, 4, or 8, set by the box-coordinate layout. A two-corner box
such as \texttt{Y0X0Y1X1} selects interleave 4, an origin-and-size box
selects 4, a two-coordinate point selects 2, and a full four-coordinate
batched box selects the default 8. The sampling method is linear, that
is bilinear, or nearest-neighbor, and the interpolation weights and the
crop scale program two register arrays at the DMA-side sampler. On the
M1 the whole set decomposes or rejects. Resize becomes a channelwise
deconvolution for integer upsample or a transpose followed by a
convolution otherwise, affine warp rejects outright with a
not-supported-on-this-architecture diagnostic, and symmetric padding
rejects. Gather routes through a limited software envelope that requires
a gather-axis size of 3 and a batch and depth of 1.

A single bool in the per-chip parameter table gates the capability, and
that bool is zero on the M1. The gate is one bit with no finer
per-operation granularity: crop-and-resize, resample, affine, native
resize, the hardware gather index path, and reflective or symmetric
padding all turn on together at the A14.

The same gate has a precision caveat on the M1 and the A14. The crop and
slice path applies a width-axis gain of 16, so a source value above 4094
on the width axis saturates to plus or minus infinity, while height,
channel, and batch offsets stay free of the saturation. Chapter
\hyperref[numerics]{3} derives the bound and chapter
\hyperref[pitfalls-and-limits]{19} gives the build-time guard. So keep
slice and crop source magnitudes below 4094 on the width axis on the M1
and the A14. The A15 generation and later sampler avoids this
saturation.

\section{Keeping the model on the
engine}\label{keeping-the-model-on-the-engine}

The datapath runs best with a small set of model-construction choices.

\begin{itemize}
\tightlist
\item
  Prefer 3x3 stride-1 dense convolutions over 5x5 or strided forms where
  the model allows, so the Winograd path is taken and each layer gets
  the factor of about 2.25 multiply reduction.
\item
  Keep the largest single per-layer operand at or below the 2 MB on-chip
  working set. A convolution whose live tiles exceed 2 MB is tiled and
  streamed from DRAM, which adds transfer traffic and moves the layer
  off the compute ceiling.
\item
  Fuse the whole graph into one program so the about 0.23 ms per-eval
  dispatch floor on the M1 is paid once rather than per layer. A
  fixed-iteration stencil fused as one graph amortizes that floor across
  every step.
\item
  Fold normalization and activation into the convolution.
  Per-output-channel scale and bias and a fused nonlinearity run on the
  convolution output at no additional cost, so a batch-normalization or
  a clamped activation after a convolution adds no separate dispatch.
\item
  Serve encoders below the batch threshold of the previous section, and
  keep preprocessing on the engine only on the A14 and later, where the
  texture engine is present.
\end{itemize}

The native convolution backend operation records its geometry in a fixed
set of attributes (\cref{lst:c13-conv-backend}).

\begin{listing}[H]
\caption{The native convolution backend operation and the geometry attributes the compiler reads to tile a layer onto the multiply array.}\label{lst:c13-conv-backend}

\begin{Shaded}
\begin{Highlighting}[]
\FunctionTok{anec.convolution}\NormalTok{(}\VariableTok{\%input}\NormalTok{, }\VariableTok{\%weights}\NormalTok{, }\VariableTok{\%bias}\NormalTok{) \{}
\NormalTok{  strides         = [sh, sw],          }\CommentTok{// 1 keeps the Winograd path for a 3x3 kernel}
\NormalTok{  explicit\_padding = [pt, pb, pl, pr], }\CommentTok{// or padding\_style}
\NormalTok{  groups          = g,                 }\CommentTok{// 1 standard; g == Cin == Cout is depthwise}
\NormalTok{  dilation\_rates  = [dh, dw],          }\CommentTok{// folded to a strided input gather, up to 8}
\NormalTok{  kernel\_sizes    = [kh, kw],          }\CommentTok{// fp16 kernel width bounded at 13 on the M1}
\NormalTok{  weights\_layout                        }\CommentTok{// weights are output{-}channel{-}major, [Cout, D, Cin, H, W]}
\NormalTok{\}  }\CommentTok{// filter and input rank 4 or 5}
\end{Highlighting}
\end{Shaded}

\end{listing}

The encoder paths and the convolution weight gradient both reduce to the
matrix-multiply backend operation (\cref{lst:c13-matmul-backend}).

\begin{listing}[H]
\caption{The matrix-multiply backend operation, which holds its contraction direction in two transpose attributes rather than a fixed operand order.}\label{lst:c13-matmul-backend}

\begin{Shaded}
\begin{Highlighting}[]
\FunctionTok{anec.matmul}\NormalTok{(}\VariableTok{\%lhs}\NormalTok{, }\VariableTok{\%rhs}\NormalTok{) \{ transpose\_lhs, transpose\_rhs \}  }\CommentTok{// depth D must be 1 on both operands}
\end{Highlighting}
\end{Shaded}

\end{listing}

A trainable convolution lowers its forward and its data gradient onto
the native convolution and deconvolution operations, and its weight
gradient through an image-to-column expansion into this matrix multiply,
because the hardware cross-correlation operation is single-channel only
and cannot serve the multi-channel weight gradient directly.

\section{Fitting a convolution or encoder block on the
engine}\label{fitting-a-convolution-or-encoder-block-on-the-engine}

A vision or encoder block stays on the engine when its layers take the
Winograd path, its working set fits on chip, and the whole graph
compiles as one program. \Cref{lst:c13-fit-block} applies those
conditions in order to a two-stage 3x3 block.

\begin{listing}[H]
\caption{Fitting a convolution or encoder block on the engine as one fused program.}\label{lst:c13-fit-block}

\begin{Shaded}
\begin{Highlighting}[]
\CommentTok{\# Fit a convolution or encoder block on the engine as one fused program.}
\CommentTok{\# Each stage is convolution, then normalization, then activation.}

\NormalTok{build graph G:}
    \BuiltInTok{input}\NormalTok{ x : [}\DecValTok{1}\NormalTok{, }\DecValTok{256}\NormalTok{, }\DecValTok{28}\NormalTok{, }\DecValTok{28}\NormalTok{] fp16}

    \CommentTok{\# Stage 1: keep the convolution 3x3 stride{-}1 so it takes the Winograd path.}
\NormalTok{    h }\OperatorTok{=}\NormalTok{ conv(x, weights }\OperatorTok{=}\NormalTok{ w1, kernel }\OperatorTok{=}\NormalTok{ [}\DecValTok{3}\NormalTok{, }\DecValTok{3}\NormalTok{], stride }\OperatorTok{=} \DecValTok{1}\NormalTok{)}
\NormalTok{    h }\OperatorTok{=}\NormalTok{ normalize(h)                       }\CommentTok{\# batch or layer norm, folded into the conv output}
\NormalTok{    h }\OperatorTok{=}\NormalTok{ activation(h)                      }\CommentTok{\# e.g. relu or gelu, on the same output pass}

    \CommentTok{\# Stage 2: another 3x3 stride{-}1 stage, same fused shape.}
\NormalTok{    h }\OperatorTok{=}\NormalTok{ conv(h, weights }\OperatorTok{=}\NormalTok{ w2, kernel }\OperatorTok{=}\NormalTok{ [}\DecValTok{3}\NormalTok{, }\DecValTok{3}\NormalTok{], stride }\OperatorTok{=} \DecValTok{1}\NormalTok{)}
\NormalTok{    h }\OperatorTok{=}\NormalTok{ normalize(h)}
\NormalTok{    h }\OperatorTok{=}\NormalTok{ activation(h)}

\NormalTok{    output h}

\CommentTok{\# 1. Check the working set stays on chip before committing to the engine.}
\NormalTok{working\_set }\OperatorTok{=} \BuiltInTok{max}\NormalTok{ over stages of resident\_bytes(stage)   }\CommentTok{\# in MB}
\ControlFlowTok{if}\NormalTok{ working\_set }\OperatorTok{\textgreater{}} \FloatTok{2.0}\NormalTok{:}
    \CommentTok{\# Too large: it would tile and stream from DRAM. Shrink the operand}
    \CommentTok{\# (fewer channels or smaller tile) and re{-}check before any other tuning.}
\NormalTok{    reshape operands until working\_set }\OperatorTok{\textless{}=} \FloatTok{2.0}

\CommentTok{\# 2. Fuse the whole block into ONE program so the per{-}call floor is paid once.}
\NormalTok{program }\OperatorTok{=}\NormalTok{ fuse\_and\_compile(G, target }\OperatorTok{=}\NormalTok{ H13)   }\CommentTok{\# one dispatch, the 0.23 ms floor paid a single time}

\CommentTok{\# 3. Run the fused program on an input image.}
\NormalTok{output }\OperatorTok{=}\NormalTok{ run(program, x }\OperatorTok{=}\NormalTok{ image)}

\CommentTok{\# Note: input preprocessing can join the same graph only on A14 and later,}
\CommentTok{\# where the texture engine is present.}
\ControlFlowTok{return}\NormalTok{ output}
\end{Highlighting}
\end{Shaded}

\end{listing}

\section{Reference: convolution and encoder economics on the
M1}\label{reference-convolution-and-encoder-economics-on-the-m1}

The constants below cover the GPU speed ratios, the datapath geometry,
and the batch thresholds (\cref{tbl:c13-reference}).

\begin{longtable}[]{@{}
  >{\raggedright\arraybackslash}p{(\linewidth - 2\tabcolsep) * \real{0.4286}}
  >{\raggedleft\arraybackslash}p{(\linewidth - 2\tabcolsep) * \real{0.5714}}@{}}
\caption{The M1 convolution and encoder constants for vision and encoder
workloads.}\label{tbl:c13-reference}\tabularnewline
\toprule\noalign{}
\begin{minipage}[b]{\linewidth}\raggedright
\textbf{Constant}
\end{minipage} & \begin{minipage}[b]{\linewidth}\raggedleft
\textbf{M1/H13 value}
\end{minipage} \\
\midrule\noalign{}
\endfirsthead
\toprule\noalign{}
\begin{minipage}[b]{\linewidth}\raggedright
\textbf{Constant}
\end{minipage} & \begin{minipage}[b]{\linewidth}\raggedleft
\textbf{M1/H13 value}
\end{minipage} \\
\midrule\noalign{}
\endhead
\bottomrule\noalign{}
\endlastfoot
3x3 convolution (256 channels) speed versus GPU & 3.8x faster \\
\rowcolor{zebra}3x3 convolution (256 channels) efficiency versus GPU &
9x more efficient \\
Convolution-stack speed, 16 deep at 256 channels & 2x faster \\
\rowcolor{zebra}Convolution-stack efficiency, 16 deep at 256 channels &
14.5x more efficient \\
Matrix-multiply power, engine versus GPU & 4.4 W versus 32.5 W \\
\rowcolor{zebra}Energy-efficiency range across the workload set & 2x to
14.5x \\
End-to-end convolution ceiling & 1.8 fp16 TFLOP/s at 1.78 W \\
\rowcolor{zebra}Convolution efficiency under load & 643 GFLOP/s per
watt \\
\end{longtable}

\chapter{LLM case study}\label{llm-case-study}

\begin{summarybox}

Autoregressive decode is bandwidth-bound and dispatch-bound, the two
regimes the engine loses, so it belongs on the GPU at every batch size.
At a batch of 16 the GPU runs decode about 2.7 times faster and about
4.6 times more energy efficient than the engine. Int8 weights halve the
weight traffic but leave the hybrid decoder at about 0.99 times the fp16
rate. The step is dispatch-bound, not weight-bandwidth-bound. Send the
prefill, encoder, and vision front end to the engine, and send the
autoregressive decode to the GPU.

\end{summarybox}

Autoregressive decode is the one major workload class where the engine
is not the fastest. The single-token decode step runs against the
roofline of chapter \hyperref[roofline]{9}, batching is the only serving
control that moves throughput, and the engine's niche is on the other
side of the model from the decoder.

\section{Why decode is not a compute
problem}\label{why-decode-is-not-a-compute-problem}

A decoder step generates one token. It reads every weight in the model
once, multiplies each against a single activation row, and produces one
new row. The arithmetic per byte of weight read is thus near one
multiply-add per weight, which puts the step far below the 141
FLOP-per-byte ridge point of chapter \hyperref[roofline]{9}. Decode is
bandwidth-bound: the step spends its time streaming weights from DRAM.

A second cost adds on top of the bandwidth cost. A single decoder layer
is a chain of projections, a normalization, an attention block, and a
feed-forward pair, and on the direct path each of them is a separate
dispatch. A token through a transformer layer stack issues on the order
of forty to fifty small dispatches, every one paying the per-eval floor
of about 0.23 ms measured on the M1 in chapter
\hyperref[ane-gpu-and-cpu]{11}. Decode is thus dispatch-bound as well as
bandwidth-bound, and both regimes are exactly the ones the engine loses,
per the verdict table of chapter \hyperref[ane-gpu-and-cpu]{11}.

The wide accumulator does not help the decoder either. A transformer
down-projection loses precision in fp16 from the per-product rounding of
its inputs under cancellation, the hazard named in chapter
\hyperref[numerics]{3}, not from the accumulator. The decoder thus has
an accuracy penalty on the engine on top of the speed penalty.

\section{Hybrid decoder placement}\label{hybrid-decoder-placement}

A decoder runs as a hybrid: the bandwidth-heavy, well-conditioned
projections execute on the engine in fp16, and two precision-critical
sites stay in wider precision off the engine. One is the residual
stream. Over twenty-two layers an fp16 residual add drops the small
per-layer update, because adding a tiny delta to a large running sum
loses the low bits. The following normalization rescales but cannot
recover them, so the stream must accumulate in wider precision off the
engine. The other is the down-projection, the cancellation-heavy step of
chapter \hyperref[numerics]{3}, whose near-cancelled output over a
contraction of 5632 takes about 3 percent error in fp16 on the engine,
enough to flip the greedy argmax.

Per-operation placement follows \cref{tbl:c14-placement}. The query,
key, gate, up, and output-embedding projections fit the engine in fp16;
the value projection, output projection, and down-projection take a
wider-precision path off the engine.

\begin{longtable}[]{@{}lll@{}}
\caption{Per-projection placement for a hybrid decoder: the engine holds
the fp16 projections, while three are held off-engine in wider
precision.}\label{tbl:c14-placement}\tabularnewline
\toprule\noalign{}
\textbf{Projection} & \textbf{End-to-end fp16} & \textbf{Placement} \\
\midrule\noalign{}
\endfirsthead
\toprule\noalign{}
\textbf{Projection} & \textbf{End-to-end fp16} & \textbf{Placement} \\
\midrule\noalign{}
\endhead
\bottomrule\noalign{}
\endlastfoot
Query, key & Survives & Engine, fp16 \\
\rowcolor{zebra}Gate, up & Survives & Engine, fp16 \\
Output embedding & Survives & Engine, fp16 \\
\rowcolor{zebra}Value & Fails & Wider precision, off-engine \\
Output projection & Fails & Wider precision, off-engine \\
\rowcolor{zebra}Down-projection & Fails & Wider precision, off-engine \\
\end{longtable}

Placement follows position, not per-operation error. The gate and up
projections survive on the engine despite per-operation error comparable
to the value and output projections. An operation survives on the engine
when its result passes through a wider-precision step downstream. The
gate and up projections send the feed-forward delta through the
off-engine down-projection, which insulates them. The value and output
projections send to the attention output and the residual directly, so
their error compounds through the key-value cache across the layer
stack.

\section{Single stream against many
streams}\label{single-stream-against-many-streams}

A single decode stream is serial by construction. Token \(t+1\) cannot
start until token \(t\) is produced, because it consumes that token, so
there is nothing to overlap within one stream. The host issues a
dispatch, waits for the result, and issues the next, and the engine is
idle between the dispatch floor and the next submission.

Concurrent independent streams behave differently. Executing one stream
releases the host thread, so several streams interleave their host-side
work against each other's engine-side work and lift the aggregate token
rate even though each individual stream stays serial. This lifts
aggregate throughput for many requests; it does not cut the latency of
any one of them. The three modes separate cleanly on what each improves
(\cref{tbl:c14-serving-modes}).

\begin{longtable}[]{@{}
  >{\raggedright\arraybackslash}p{(\linewidth - 4\tabcolsep) * \real{0.3333}}
  >{\raggedright\arraybackslash}p{(\linewidth - 4\tabcolsep) * \real{0.3333}}
  >{\raggedright\arraybackslash}p{(\linewidth - 4\tabcolsep) * \real{0.3333}}@{}}
\caption{Decode serving modes, what each improves, and how its
throughput compares against the
GPU.}\label{tbl:c14-serving-modes}\tabularnewline
\toprule\noalign{}
\begin{minipage}[b]{\linewidth}\raggedright
\textbf{Serving mode}
\end{minipage} & \begin{minipage}[b]{\linewidth}\raggedright
\textbf{What it improves}
\end{minipage} & \begin{minipage}[b]{\linewidth}\raggedright
\textbf{Throughput against the GPU}
\end{minipage} \\
\midrule\noalign{}
\endfirsthead
\toprule\noalign{}
\begin{minipage}[b]{\linewidth}\raggedright
\textbf{Serving mode}
\end{minipage} & \begin{minipage}[b]{\linewidth}\raggedright
\textbf{What it improves}
\end{minipage} & \begin{minipage}[b]{\linewidth}\raggedright
\textbf{Throughput against the GPU}
\end{minipage} \\
\midrule\noalign{}
\endhead
\bottomrule\noalign{}
\endlastfoot
Single-stream decode & Latency for one request & Below the GPU; serial,
no overlap \\
\rowcolor{zebra}Multi-stream concurrent decode & Aggregate tokens per
second & Higher aggregate than single stream, still below the GPU \\
GPU batched decode & Aggregate tokens per second & Fastest and most
energy efficient at batch \\
\end{longtable}

The multi-stream gain is real but bounded. A fair batched comparison at
a batch of 16 puts the GPU about 2.7 times faster and about 4.6 times
more energy efficient than the engine on decode, because the engine path
stays host-dominated while the GPU scales with batch.

\section{What int8 weights do and do not
buy}\label{what-int8-weights-do-and-do-not-buy}

Native int8 weights stream at half the bytes of fp16 and cut the
weight-read traffic of the bandwidth-bound step. That is a reduction in
the primary cost of decode, and it is reachable on the direct path.

It does not make the engine faster on decode. The int8 projections come
back numerically correct, but the hybrid decoder runs at about 0.99
times the fp16 version, because once the model is small or the dispatch
count is high the step is dispatch-bound and host-bound. Halving the
weight traffic of a step whose wall time is set by forty-plus dispatch
floors and host marshalling does not move the wall time. An int8 gain
needs a fused, engine-dominated decoder where the weight stream is the
bottleneck, which the per-layer dispatch structure of a general decoder
does not provide. Batching, not quantization, is the control that moves
serving throughput.

\section{Two serving controls and two hard
caps}\label{two-serving-controls-and-two-hard-caps}

Two controls make the hybrid practical for serving. Speculative decoding
drafts several tokens with an inexpensive proposer and verifies all of
them in one batched forward. Because the weights are read once per
forward regardless of the number of drafted tokens, the verification
adds almost no cost. On a TinyLlama decoder this lifts the rate from 35
tokens per second to as high as 128 on repetitive text and 42 to 56 on
factual prose, with output identical to plain greedy decoding. Batched
prefill processes the prompt in chunks of several tokens at once instead
of token by token, and because the engine matrix multiplies are flat in
batch width, prefill latency collapses by about 2.3 to 5.9 times,
bit-equal to serial prefill.

A multi-shape decoder also reaches two hard caps enforced by the device
daemon. The in-flight cap is 127 outstanding requests per program, set
by a dispatch semaphore of 127 inside the in-memory model. The
loaded-program cap is near 128 programs per process: the next load fails
with \texttt{GetANEFModel:\ must\ re-compile} and forces a recompile.
For multi-shape serving, where each sequence length or batch width is a
distinct compiled program, the loaded-program cap bounds how many
resident shapes a process can hold at once.

\section{Cache that stays resident}\label{cache-that-stays-resident}

The one part of the decoder that fits the engine's execution model is
the key-value cache. The cache stores the keys and values of every prior
token and grows by one row per step. Re-streaming the whole cache
through the host each token would add a copy proportional to the
sequence length on top of the weight stream.

Instead, the cache stays resident on the engine across steps, as chapter
\hyperref[execution-model]{2} describes for resident state. A program
declares the cache as both an input and an output, and after compilation
the runtime aliases the output buffer onto the input buffer, so the
updated cache produced by one step is the cache consumed by the next
without leaving the device. A masked update writes the new key and value
into the cache at the current position, with the position supplied as a
small one-hot vector each step, as \cref{lst:c14-cache-schematic}
sketches.

\begin{listing}[H]
\caption{One resident-cache decode step, where aliasing the cache output buffer onto its input buffer keeps the key-value cache on the engine across steps.}\label{lst:c14-cache-schematic}

\begin{Shaded}
\begin{Highlighting}[]
\CommentTok{\# Schematic of one resident{-}cache decode step.}
\CommentTok{\# The cache output buffer is aliased onto the cache input buffer,}
\CommentTok{\# so the cache never round{-}trips through the host.}

\BuiltInTok{compile}\NormalTok{(graph }\ControlFlowTok{with}\NormalTok{ inputs [x, slot\_onehot, cache\_in]}
\NormalTok{                outputs [logits, cache\_out])}

\NormalTok{alias\_output\_to\_input(}\StringTok{"cache\_out"} \OperatorTok{{-}\textgreater{}} \StringTok{"cache\_in"}\NormalTok{)   }\CommentTok{\# output buffer IS input buffer}

\CommentTok{\# per token, the host sends only the new row and the slot, not the cache:}
\NormalTok{cache\_out }\OperatorTok{=}\NormalTok{ cache\_in }\OperatorTok{*}\NormalTok{ (}\DecValTok{1} \OperatorTok{{-}}\NormalTok{ slot\_onehot) }\OperatorTok{+}\NormalTok{ new\_kv }\OperatorTok{*}\NormalTok{ slot\_onehot}
\end{Highlighting}
\end{Shaded}

\end{listing}

With the cache resident the host sends only the new token and the
position one-hot each step, a few bytes. This removes the per-token
cache copy and is the form in which a decoder is most efficiently
expressed on the engine. A resident cache removes one copy from a step
whose wall time is still set by weight bandwidth and dispatch count, so
it makes engine decode more efficient without making it as fast as the
GPU.

\section{Two residency mechanisms}\label{two-residency-mechanisms}

The engine has a native persistent-state type and a pair of read-state
and write-state operations, the obvious mechanism for a resident cache.
On the unentitled runtime path that type parses but does not reach the
engine. The front end recognizes the state operations, but the backend
rejects them during the conversion to the engine program, because the
read-state has no state object to bind to: the runtime registers the
cache operand as a plain input rather than as a persistent state, so the
binding step fails before code generation. The native route thus stays
closed pending a compiler option that moves the state operations off the
engine subgraph.

The buffer-aliasing route is open and verified. It compiles a program
whose cache is both an input and an output, then aliases the output
buffer onto the input buffer after compilation, so the tensor persists
on the engine across dispatches with no host re-supply. The aliasing
primitive is the same one that keeps optimizer state resident for
on-engine training in chapter \hyperref[training-on-the-engine]{15}. Two
measurements confirm it on the M1. A resident accumulator that adds one
each step, aliased output onto input with no re-supply, returns 1, 2, 3,
4 over four dispatches, accumulating in place. A resident cache of six
slots, written by a masked positional update of \((t+1) \times 10\) at
slot \(t\), returns 10, 20, 30, 40, 50, 0 over five steps, each token
written to its slot with the cache never leaving the device. The
coordination behind residency is a hardware event signal-and-wait
primitive that sequences the producer and consumer of the cache buffer,
which the compiler emits on its own.

\section{A resident-cache decode
step}\label{a-resident-cache-decode-step}

Compile the decoder once with the cache as a paired input and output,
alias the two buffers, and thereafter send only the new token and its
slot (\cref{lst:c14-cache-step}).

\begin{listing}[H]
\caption{The runtime calls for a resident-cache decode step, binding one cache buffer to both ports so the cache updates in place each token.}\label{lst:c14-cache-step}

\begin{Shaded}
\begin{Highlighting}[]
\CommentTok{/* Allocate one cache buffer object and bind it to BOTH the cache\_in and cache\_out ports, */}
\CommentTok{/* so the engine writes the updated cache back into the same resident buffer in place.    */}
\NormalTok{e5rt\_buffer\_object\_alloc}\OperatorTok{(\&}\NormalTok{cache\_buf}\OperatorTok{,}\NormalTok{ cache\_nbytes}\OperatorTok{,} \CommentTok{/*type=*/}\DecValTok{0}\OperatorTok{);}
\NormalTok{e5rt\_execution\_stream\_operation\_retain\_input\_port}\OperatorTok{(}\NormalTok{op}\OperatorTok{,} \StringTok{"cache\_in"}\OperatorTok{,} \OperatorTok{\&}\NormalTok{cache\_in\_port}\OperatorTok{);}
\NormalTok{e5rt\_execution\_stream\_operation\_retain\_output\_port}\OperatorTok{(}\NormalTok{op}\OperatorTok{,} \StringTok{"cache\_out"}\OperatorTok{,} \OperatorTok{\&}\NormalTok{cache\_out\_port}\OperatorTok{);}
\NormalTok{e5rt\_io\_port\_bind\_buffer\_object}\OperatorTok{(}\NormalTok{cache\_in\_port}\OperatorTok{,}\NormalTok{ cache\_buf}\OperatorTok{);}
\NormalTok{e5rt\_io\_port\_bind\_buffer\_object}\OperatorTok{(}\NormalTok{cache\_out\_port}\OperatorTok{,}\NormalTok{ cache\_buf}\OperatorTok{);}  \CommentTok{/* alias: cache stays resident */}

\ControlFlowTok{for} \OperatorTok{(}\DataTypeTok{int}\NormalTok{ t }\OperatorTok{=} \DecValTok{0}\OperatorTok{;}\NormalTok{ t }\OperatorTok{\textless{}}\NormalTok{ max\_tokens}\OperatorTok{;}\NormalTok{ t}\OperatorTok{++)} \OperatorTok{\{}      \CommentTok{/* one encode + execute per token */}
    \CommentTok{/* host marshals only the new row and slot; the cache is NOT re{-}sent */}
\NormalTok{    e5rt\_execution\_stream\_operation\_prepare\_op\_for\_encode}\OperatorTok{(}\NormalTok{op}\OperatorTok{);}
\NormalTok{    e5rt\_execution\_stream\_encode\_operation}\OperatorTok{(}\NormalTok{stream}\OperatorTok{,}\NormalTok{ op}\OperatorTok{);}
\NormalTok{    e5rt\_execution\_stream\_execute\_sync}\OperatorTok{(}\NormalTok{stream}\OperatorTok{);}
\NormalTok{    e5rt\_execution\_stream\_reset}\OperatorTok{(}\NormalTok{stream}\OperatorTok{);}
\OperatorTok{\}}
\end{Highlighting}
\end{Shaded}

\end{listing}

\section{Reference: decode placement and the serving
caps}\label{reference-decode-placement-and-the-serving-caps}

The chapter's figures are in \cref{tbl:c14-reference}.

\begin{longtable}[]{@{}
  >{\raggedright\arraybackslash}p{(\linewidth - 2\tabcolsep) * \real{0.4286}}
  >{\raggedleft\arraybackslash}p{(\linewidth - 2\tabcolsep) * \real{0.5714}}@{}}
\caption{The decode placement figures and the two device-daemon serving
caps.}\label{tbl:c14-reference}\tabularnewline
\toprule\noalign{}
\begin{minipage}[b]{\linewidth}\raggedright
\textbf{Quantity}
\end{minipage} & \begin{minipage}[b]{\linewidth}\raggedleft
\textbf{Value}
\end{minipage} \\
\midrule\noalign{}
\endfirsthead
\toprule\noalign{}
\begin{minipage}[b]{\linewidth}\raggedright
\textbf{Quantity}
\end{minipage} & \begin{minipage}[b]{\linewidth}\raggedleft
\textbf{Value}
\end{minipage} \\
\midrule\noalign{}
\endhead
\bottomrule\noalign{}
\endlastfoot
Per-eval dispatch floor (M1) & 0.23 ms \\
\rowcolor{zebra}Dispatches per transformer layer stack & 40 to 50 \\
Down-projection fp16 error over contraction 5632 & about 3 percent \\
\rowcolor{zebra}Int8-hybrid decode rate versus fp16 & 0.99x \\
GPU versus engine decode speed at batch 16 & 2.7x faster \\
\rowcolor{zebra}GPU versus engine decode energy at batch 16 & 4.6x more
efficient \\
Speculative-decoding rate, TinyLlama & 35 to 128 tokens per second \\
\rowcolor{zebra}Speculative-decoding rate, factual prose & 42 to 56
tokens per second \\
Batched-prefill latency reduction & 2.3x to 5.9x \\
\rowcolor{zebra}In-flight cap per program & 127 outstanding requests \\
Loaded-program cap per process & near 128
(\texttt{GetANEFModel:\ must\ re-compile}) \\
\end{longtable}

\section{Verdict: encoders, not
decoders}\label{verdict-encoders-not-decoders}

The engine's niche is the other side of the same model. Encoders and
embedding models process a whole sequence in one forward pass of
compute-bound matrix multiplies and normalizations, the engine's strong
regime. Chapter \hyperref[ane-gpu-and-cpu]{11} measures a
single-sentence encoder about 4.4 times faster than the GPU at low
batch, with the crossover to the GPU only near a batch of 23. Vision and
convolution hold the same verdict more strongly.

\chapter{Training on the engine}\label{training-on-the-engine}

\begin{summarybox}

The engine has no backward operation. A full forward, backward, and
optimizer loop still runs as ordinary inference-style graph operations,
with the optimizer state resident across steps. The registered gradient
set matches the closed form to a cosine of \(1.0000\). Transformers,
normalization-based convolutional networks, and gated linear networks
all train end to end. A small convolutional network trains to a final
test accuracy of \(0.9080\) on the M1 and \(0.9070\) on the M5, at a
loss scale of 1024. That is a difference of one test sample in one
thousand. The conv weight-gradient saturates above 4094 on the width
axis on the M1 and M2, so a preflight check guards that one numeric
edge.

\end{summarybox}

There is no backward operation, no gradient layer, and no adjoint
primitive in the compiler's operation catalog. A network still trains
end to end anyway, optimizer step included, because every stage of the
loop is expressible in the operations the engine already runs. This
chapter states how that expression works and which gradients are
numerically trustworthy, then isolates the one path that diverges in
fp16. The result is a capability, not a speed claim.

\section{A backward pass built from forward
operations}\label{a-backward-pass-built-from-forward-operations}

A search of the compiler binary for an engine-native gradient layer
returns nothing. Every gradient operation held in the shared compiler
belongs to the graphics-processor dialect, is costed by the
graphics-processor cost model, and executes there. The convolution
data-gradient and weight-gradient, the max-pool and average-pool
gradients, normalization, recurrent, top-k, pad, strided-slice, and
resize gradients are all graphics-processor operations. At the operation
level the engine is inference-only.

The engine-native convolution family is the forward convolution and a
single-channel cross-correlation. The forward convolution has a
transpose mode, so the data-gradient, which is a transposed convolution,
has a native operation. The weight-gradient has no native operation,
because the only cross-correlation is single-channel and cannot serve
the multi-channel correlation a weight-gradient requires. The
weight-gradient thus lowers through a patch-extraction expansion into a
matrix multiply, which is why a trainable convolution has its weight as
a graph input and why its backward cost grows with the minibatch.

The public on-device update path meets the same boundary
\hyperref[ref-applecoreml]{{[}AppleCoreML{]}}. When the system framework
fine-tunes a model on device, the forward pass may run on the engine,
but the backward pass runs on the graphics processor or the host through
those gradient operations.

Training the whole loop on the engine thus means not calling any
backward operation. The forward pass is a graph of engine operations.
The backward pass is a second graph, built by the host from the same
forward operations and their analytic gradients, that computes
cotangents through the network. The optimizer update is a third graph
that combines the gradients with the parameter and optimizer state. All
three are inference-style graphs of operations the compiler already
accepts, so all three compile and dispatch on the direct route from
chapter \hyperref[dispatching-without-core-ml]{6}.

A trainable weight is a graph input, not an embedded constant. The
forward graph reads the weight from a bound buffer, so the update graph
can write a new value into that buffer between steps without
recompiling. This is what lets the optimizer step run on the engine: the
weight is an operand, and an operand can be both read by the forward
pass and written by the update.

\section{Gradient vocabulary and its
correctness}\label{gradient-vocabulary-and-its-correctness}

A gradient is built by composing each forward operation with its
vector-Jacobian product. The registered set was checked against
closed-form derivatives by forming a linear loss
\(L = \sum_i (\mathrm{op}(x)_i \cdot w_i)\), whose exact gradient is
\(w \cdot \mathrm{op}'(x)\), and comparing the engine-computed gradient
to that reference.

The core set matches to a cosine of \(1.0000\) against the closed form,
at the fp16 error level. That set covers the activations \texttt{relu},
\texttt{sigmoid}, \texttt{tanh}, and \texttt{gelu}, the elementwise
\texttt{mul}, \texttt{add}, \texttt{sub}, \texttt{square}, the linear
\texttt{matmul} and batched \texttt{matmul}, \texttt{softmax},
\texttt{conv}, \texttt{avg\_pool}, \texttt{max\_pool}, the reductions
\texttt{reduce\_sum} and \texttt{reduce\_mean}, and the shape operations
\texttt{transpose}, \texttt{reshape}, \texttt{flatten}, \texttt{slice},
and \texttt{concat}. The four normalization layers \texttt{layer\_norm},
\texttt{rms\_norm}, \texttt{group\_norm}, and \texttt{l2\_norm}, the
\texttt{silu} activation, and a set of unary math operations including
\texttt{exp}, \texttt{sqrt}, \texttt{rsqrt}, \texttt{log}, \texttt{erf},
and \texttt{cos} were added later and verified against a
finite-difference reference at the same cosine. The normalization
gradients re-inject the per-channel scale as a supplied value-input,
since the engine has no constant-tensor operation to hold it. The
registered operations group by kind (\cref{tbl:c15-gradients}).

\begin{longtable}[]{@{}
  >{\raggedright\arraybackslash}p{(\linewidth - 2\tabcolsep) * \real{0.5000}}
  >{\raggedright\arraybackslash}p{(\linewidth - 2\tabcolsep) * \real{0.5000}}@{}}
\caption{The registered gradient vocabulary, each verified to a cosine
of \(1.0000\) against the closed form or a finite-difference reference
at the fp16 error level.}\label{tbl:c15-gradients}\tabularnewline
\toprule\noalign{}
\begin{minipage}[b]{\linewidth}\raggedright
\textbf{Group}
\end{minipage} & \begin{minipage}[b]{\linewidth}\raggedright
\textbf{Operations with a registered gradient}
\end{minipage} \\
\midrule\noalign{}
\endfirsthead
\toprule\noalign{}
\begin{minipage}[b]{\linewidth}\raggedright
\textbf{Group}
\end{minipage} & \begin{minipage}[b]{\linewidth}\raggedright
\textbf{Operations with a registered gradient}
\end{minipage} \\
\midrule\noalign{}
\endhead
\bottomrule\noalign{}
\endlastfoot
Activations & \texttt{relu}, \texttt{sigmoid}, \texttt{tanh},
\texttt{gelu}, \texttt{silu} \\
\rowcolor{zebra}Elementwise & \texttt{mul}, \texttt{add}, \texttt{sub},
\texttt{square} \\
Linear & \texttt{matmul}, batched \texttt{matmul}, \texttt{softmax},
\texttt{conv}, \texttt{avg\_pool}, \texttt{max\_pool} \\
\rowcolor{zebra}Reductions & \texttt{reduce\_sum},
\texttt{reduce\_mean} \\
Shape & \texttt{transpose}, \texttt{reshape}, \texttt{flatten},
\texttt{slice}, \texttt{concat} \\
\rowcolor{zebra}Normalization & \texttt{layer\_norm},
\texttt{rms\_norm}, \texttt{group\_norm}, \texttt{l2\_norm} \\
Unary math & \texttt{exp}, \texttt{sqrt}, \texttt{rsqrt}, \texttt{log},
\texttt{erf}, \texttt{cos} \\
\end{longtable}

With the normalization gradients present, a transformer block,
normalization-based convolutional network, and gated linear network all
train end to end on the engine, where previously only a plain multilayer
network did. Some operations have a forward but no registered gradient,
among them \texttt{cumsum}, \texttt{amax}, \texttt{amin}, and several
parametric activations. A model that uses one of those compiles and runs
its forward pass, then the backward construction raises an explicit
error rather than producing a silently wrong gradient. The gap there is
coverage, not correctness.

\section{A resident-state training
step}\label{a-resident-state-training-step}

The optimizer state stays resident on the engine across steps, through
the buffer-aliasing mechanism of chapter \hyperref[execution-model]{2}.
The first and second moments of an adaptive optimizer, along with the
weights themselves, persist as buffers in the engine working set from
one dispatch to the next. The host sends only the per-step minibatch and
the scalar learning rate, and reads the weight buffers back at a
checkpoint. The large held tensors never cross the host boundary on
every step.

One dispatch advances the network by one optimizer step, as
\cref{lst:c15-step} shows: forward, backward, and update in a single
submitted graph.

\begin{listing}[H]
\caption{A resident-state training step run as one engine graph, with weights and optimizer moments resident across steps.}\label{lst:c15-step}

\begin{Shaded}
\begin{Highlighting}[]
\CommentTok{\# Resident buffers, allocated once and kept in the engine working set:}
\CommentTok{\#   W           trainable weights        (read by forward, written by update)}
\CommentTok{\#   M, V        optimizer moments        (read and written by update)}
\CommentTok{\# Per{-}step host inputs: minibatch (x, y), learning rate lr\_t}

\ControlFlowTok{for}\NormalTok{ step }\KeywordTok{in} \BuiltInTok{range}\NormalTok{(num\_steps):}
    \CommentTok{\# all three stages are one engine graph, one dispatch}
\NormalTok{    logits }\OperatorTok{=}\NormalTok{ forward(W, x)                 }\CommentTok{\# inference{-}style ops}
\NormalTok{    g      }\OperatorTok{=}\NormalTok{ backward(W, x, y, logits)     }\CommentTok{\# vjp graph, no native backward op}
\NormalTok{    M      }\OperatorTok{=}\NormalTok{ beta1 }\OperatorTok{*}\NormalTok{ M }\OperatorTok{+}\NormalTok{ (}\DecValTok{1} \OperatorTok{{-}}\NormalTok{ beta1) }\OperatorTok{*}\NormalTok{ g           }\CommentTok{\# update, on engine}
\NormalTok{    V      }\OperatorTok{=}\NormalTok{ beta2 }\OperatorTok{*}\NormalTok{ V }\OperatorTok{+}\NormalTok{ (}\DecValTok{1} \OperatorTok{{-}}\NormalTok{ beta2) }\OperatorTok{*}\NormalTok{ (g }\OperatorTok{*}\NormalTok{ g)}
\NormalTok{    W      }\OperatorTok{=}\NormalTok{ W }\OperatorTok{{-}}\NormalTok{ lr\_t }\OperatorTok{*}\NormalTok{ M }\OperatorTok{/}\NormalTok{ (sqrt(V) }\OperatorTok{+}\NormalTok{ eps)        }\CommentTok{\# writes resident W in place}
    \CommentTok{\# M, V, W remain resident; host sends only (x, y, lr\_t) next step}
\end{Highlighting}
\end{Shaded}

\end{listing}

The adaptive update is

\[W_{t+1} = W_t - \eta_t \frac{M_t}{\sqrt{V_t} + \epsilon}, \quad M_t = \beta_1 M_{t-1} + (1 - \beta_1) g_t, \quad V_t = \beta_2 V_{t-1} + (1 - \beta_2) g_t^2\]

where \(g_t\) is the gradient from the backward graph. The host-supplied
learning rate \(\eta_t\) already absorbs any loss scaling, because the
scale factor cancels in the ratio \(M_t / \sqrt{V_t}\) and must not be
divided out a second time.

A small convolutional network trains this way to a final test accuracy
of \(0.9080\). On the M1 generation, the seeded handwritten-digit
network reaches that accuracy after 300 steps, deterministic and
reproducible to the digit across runs.

\section{Conv weight-gradient
divergence}\label{conv-weight-gradient-divergence}

One gradient path diverges in fp16 on the M1 and the M2 generations. The
convolution is built from a patch-extraction step, a set of width-offset
slices, followed by a matmul. A nonzero-offset width slice on these
generations saturates above 4094, the bound derived in chapter
\hyperref[numerics]{3} and guarded in chapter
\hyperref[pitfalls-and-limits]{19}.

The weight-gradient runs the backward activations back through those
same width-offset slices. When the loss-scaled backward activations
exceed about 4094, a few weight-gradient elements saturate to infinity
and the gradient is corrupted. The break is magnitude-gated and
finite-to-infinity, not a small rounding error: at a fixed shape the
path is exact at loss scale 384 and produces its first infinity at loss
scale 512. A larger input magnitude crosses the same threshold at a
lower loss scale.

The input magnitude and the loss scale cross the threshold
independently, as \cref{tbl:c15-saturation} records across a sweep of
both.

\begin{longtable}[]{@{}rrlr@{}}
\caption{Conv weight-gradient saturation at a fixed shape,
finite-to-infinity and
magnitude-gated.}\label{tbl:c15-saturation}\tabularnewline
\toprule\noalign{}
\textbf{Input scale} & \textbf{Loss scale} & \textbf{Result} &
\textbf{Infinities} \\
\midrule\noalign{}
\endfirsthead
\toprule\noalign{}
\textbf{Input scale} & \textbf{Loss scale} & \textbf{Result} &
\textbf{Infinities} \\
\midrule\noalign{}
\endhead
\bottomrule\noalign{}
\endlastfoot
1 & 256 & free & 0 \\
\rowcolor{zebra}1 & 384 & free & 0 \\
1 & 512 & saturates & 1 \\
\rowcolor{zebra}1 & 768 & saturates & 3 \\
1 & 1024 & saturates & 8 \\
\rowcolor{zebra}4 & 256 & saturates & 36 \\
\end{longtable}

A 1x1 convolution does not touch the slice path, because it has no
width-offset patch slice. The hazard is limited. A convolution-first
network sends the normalization-bounded input through the width-offset
slices on the forward side, where values stay well below 4094 regardless
of loss scale. Its conv input-gradient, the path that would take the
loss-scaled gradient through the slice, is discarded because no layer
precedes the first convolution. The handwritten-digit network trains
correctly at loss scales of 128, 1024, and 65536, with overlapping
curves and final accuracies of \(0.9070\), \(0.9080\), and \(0.9100\).
The residual risk is a network that pushes width-offset-slice values in
the weight-gradient path past 4094, which a preflight check can flag on
the M1 and M2 generations. The M5 generation takes a different slice
route and has no such saturation; the non-saturating route arrives on
the A15 generation and later.

\section{Cross-generation parity}\label{cross-generation-parity}

The same seeded network trains deterministically on both generations,
with the data order and initialization fixed. The two accuracy curves
track to three decimals through all 300 steps and part by a single
borderline prediction at the end (\cref{tbl:c15-parity}).

\begin{longtable}[]{@{}rrr@{}}
\caption{The seeded handwritten-digit network trained identically on
both generations, deterministic and
reproducible.}\label{tbl:c15-parity}\tabularnewline
\toprule\noalign{}
\textbf{Step} & \textbf{M1 test accuracy} & \textbf{M5 test accuracy} \\
\midrule\noalign{}
\endfirsthead
\toprule\noalign{}
\textbf{Step} & \textbf{M1 test accuracy} & \textbf{M5 test accuracy} \\
\midrule\noalign{}
\endhead
\bottomrule\noalign{}
\endlastfoot
0 & 0.0850 & 0.0850 \\
\rowcolor{zebra}50 & 0.8790 & 0.8810 \\
100 & 0.9020 & 0.9020 \\
\rowcolor{zebra}150 & 0.9050 & 0.9030 \\
200 & 0.9070 & 0.9070 \\
\rowcolor{zebra}250 & 0.9080 & 0.9070 \\
300 & 0.9080 & 0.9070 \\
\end{longtable}

That one-sample gap is the end-to-end signature of the per-operation
cross-generation fp16 difference, at most one unit in the last place per
operation, accumulating across 300 steps until it flips one
near-threshold logit. The difference does not vanish, but it is far too
small to affect the trained model, so training is portable across the
two generations.

\section{A resident-state training
loop}\label{a-resident-state-training-loop}

The weights are declared as trainable graph inputs and the optimizer
moments allocated once, both outside the step loop
(\cref{lst:c15-loop}).

\begin{listing}[H]
\caption{A resident-state training loop, keeping the weights and optimizer moments resident across all 300 steps.}\label{lst:c15-loop}

\begin{Shaded}
\begin{Highlighting}[]
\NormalTok{graph G:}
    \BuiltInTok{input}\NormalTok{   x : [B, }\DecValTok{1}\NormalTok{, }\DecValTok{28}\NormalTok{, }\DecValTok{28}\NormalTok{] fp16          }\CommentTok{\# one minibatch of images}
    \BuiltInTok{input}\NormalTok{   y : [B]                           }\CommentTok{\# one minibatch of labels}
\NormalTok{    weights W                                 }\CommentTok{\# trainable, an input not an embedded constant}
\NormalTok{    logits }\OperatorTok{=}\NormalTok{ forward(W, x)                    }\CommentTok{\# inference{-}style ops only}
\NormalTok{    loss   }\OperatorTok{=}\NormalTok{ cross\_entropy(logits, y)}
\NormalTok{    output  loss, grad(loss, W)               }\CommentTok{\# autodiff appends the backward pass}

\NormalTok{program P }\OperatorTok{=} \BuiltInTok{compile}\NormalTok{(G, target }\OperatorTok{=}\NormalTok{ H13)   }\CommentTok{\# forward + backward in one program, one dispatch}
\NormalTok{M, V }\OperatorTok{:=} \DecValTok{0}                                     \CommentTok{\# adaptive optimizer moments, resident on the engine}
\ControlFlowTok{for}\NormalTok{ step }\KeywordTok{in} \FloatTok{1..300}\NormalTok{:}
\NormalTok{    grads }\OperatorTok{=}\NormalTok{ dispatch(P, x\_batch, y\_batch)     }\CommentTok{\# forward and backward in one submitted graph}
\NormalTok{    M }\OperatorTok{=}\NormalTok{ beta1 }\OperatorTok{*}\NormalTok{ M }\OperatorTok{+}\NormalTok{ (}\DecValTok{1} \OperatorTok{{-}}\NormalTok{ beta1) }\OperatorTok{*}\NormalTok{ grads               }\CommentTok{\# update, on{-}device}
\NormalTok{    V }\OperatorTok{=}\NormalTok{ beta2 }\OperatorTok{*}\NormalTok{ V }\OperatorTok{+}\NormalTok{ (}\DecValTok{1} \OperatorTok{{-}}\NormalTok{ beta2) }\OperatorTok{*}\NormalTok{ (grads }\OperatorTok{*}\NormalTok{ grads)}
\NormalTok{    W }\OperatorTok{=}\NormalTok{ W }\OperatorTok{{-}}\NormalTok{ lr\_t }\OperatorTok{*}\NormalTok{ M }\OperatorTok{/}\NormalTok{ (sqrt(V) }\OperatorTok{+}\NormalTok{ eps)        }\CommentTok{\# optimizer step writes resident W in place}
    \CommentTok{\# M, V, W stay resident; host sends only (x\_batch, y\_batch, lr\_t) next step}
    \CommentTok{\# lr\_t already absorbs loss scaling: it cancels in M / sqrt(V), do not divide out again}
\NormalTok{W\_final }\OperatorTok{=}\NormalTok{ read(W)                             }\CommentTok{\# read the weights back at a checkpoint}
\end{Highlighting}
\end{Shaded}

\end{listing}

\section{Reference: training correctness and the numeric
edge}\label{reference-training-correctness-and-the-numeric-edge}

Everything this chapter establishes about correctness reduces to a
handful of figures and one guarded numeric edge
(\cref{tbl:c15-reference}).

\begin{longtable}[]{@{}
  >{\raggedright\arraybackslash}p{(\linewidth - 2\tabcolsep) * \real{0.4286}}
  >{\raggedleft\arraybackslash}p{(\linewidth - 2\tabcolsep) * \real{0.5714}}@{}}
\caption{The training correctness figures and the single guarded numeric
edge.}\label{tbl:c15-reference}\tabularnewline
\toprule\noalign{}
\begin{minipage}[b]{\linewidth}\raggedright
\textbf{Quantity}
\end{minipage} & \begin{minipage}[b]{\linewidth}\raggedleft
\textbf{Value}
\end{minipage} \\
\midrule\noalign{}
\endfirsthead
\toprule\noalign{}
\begin{minipage}[b]{\linewidth}\raggedright
\textbf{Quantity}
\end{minipage} & \begin{minipage}[b]{\linewidth}\raggedleft
\textbf{Value}
\end{minipage} \\
\midrule\noalign{}
\endhead
\bottomrule\noalign{}
\endlastfoot
Registered-gradient cosine versus closed form & \(1.0000\) \\
\rowcolor{zebra}Final test accuracy, M1 & \(0.9080\) after 300 steps \\
Final test accuracy, M5 & \(0.9070\) \\
\rowcolor{zebra}Cross-generation accuracy gap & one test sample in one
thousand \\
Per-operation cross-generation fp16 difference & at most one unit in the
last place \\
\rowcolor{zebra}Loss-scale accuracies, M1 (128, 1024, 65536) &
\(0.9070\), \(0.9080\), \(0.9100\) \\
Conv weight-gradient saturation, width axis & 4094, where
\(65504 / 16 = 4094\) \\
\rowcolor{zebra}First infinity at fixed shape & loss scale 512 (exact at
loss scale 384) \\
Generations with the saturation & M1 and M2; non-saturating on A15 and
later \\
\end{longtable}

\section{Capability and its scale}\label{capability-and-its-scale}

Training reaches the engine for the supported operation set, with the
optimizer state resident across steps and the conv weight-gradient as
the one guarded numeric edge. The loop is dispatch-bound: the per-step
work is small relative to the dispatch cost, so at this scale the engine
is no faster than the host or the graphics processor.

\chapter{Numerical and scientific computing}\label{numerical-and-scientific-computing}

\begin{summarybox}

Dense linear algebra and fixed-iteration numerics fit the engine. The
wide accumulator keeps precision to a condition number of roughly
\(10^4\), after which fp16 input rounding sets the error floor. Dense
factorizations and full spectral decompositions run by unrolling into
one static graph. They reach about \(n = 32\) for the factorizations and
about \(n = 20\) for the spectral decompositions. The discrete Fourier
transform runs as a matrix multiply, fp16-clean to about \(N = 2048\).
The single architectural limit is a runtime-sized output, not pivoting.
A data-dependent pivot is reached by cutting the graph at the pivot
selection.

\end{summarybox}

The Apple Neural Engine is a dense fp16 matrix engine with a wide
accumulator. That hardware structure decides which numerical and
scientific kernels run on it without a precision penalty and which the
architecture bars. Dense linear algebra and fixed-iteration numerics
fit; the dense factorizations and full spectral decompositions run by
unrolling into one static graph; the discrete Fourier transform fits as
a matrix multiply while the fast-transform butterfly fits only as a
static unroll.

\section{What the dense engine computes
well}\label{what-the-dense-engine-computes-well}

The three levels of dense linear algebra map onto the multiply array
directly \hyperref[ref-appleaccelerate]{{[}AppleAccelerate{]}}. A
level-1 operation is a vector reduction or scaling: a dot product is a
single contraction, and an axpy or a norm is a scaled add or a
reduction. A level-2 operation is a matrix-vector product, one matmul
with a unit output dimension. A level-3 operation is a matrix-matrix
product, the native shape of the array. The wide accumulator (chapter 3)
stores the reduction in a register of fp32 class, so a representable sum
comes back near exact and the only quantization is the fp16 rounding of
the inputs and the output.

Iterative kernels inherit the same precision behavior because each step
is a matmul or a reduction. Power iteration for a dominant eigenpair,
conjugate-gradient sweep for a symmetric positive-definite solve, and
fixed-cycle generalized-minimal-residual solve for a general system all
fold the matrix as a constant weight and run as one program. Because the
matrix is a fixed weight, the program size does not grow with \(n\): a
symmetric positive-definite solve runs unchanged from a small system up
to \(n = 512\) at the same relative error. Across these solvers the wide
accumulator keeps the precision to a condition number of roughly
\(10^4\), above which the fp16 rounding of the inputs sets the error
floor.

The measured envelope for the on-engine iterative solvers is a relative
error near \(10^{-3}\) at a condition number at or below \(10^2\). A
dominant eigenpair from power iteration returns to about \(10^{-15}\),
limited only by the spectral gap, because the iteration is a repeated
matmul whose fixed point is exact in the representable range.

The accumulator that retains the precision is measured, not assumed: a
reduction accumulates in a register of fp32 class and rounds to fp16
only at the output port, with the silicon-probe worked examples and the
radix-4 fan-in derived in chapter \hyperref[numerics]{3}. A
representable sum thus comes back near exact, while a cancellation-heavy
reduction has its input lanes round in radix-4 groups first, which makes
the matrix-multiply contraction path the more accurate route, exact to
fp32 and rounded once at the output. The compiler emits a precision
warning steering a narrow reduction toward it.

\section{Factorizations and spectral decompositions by static
unrolling}\label{factorizations-and-spectral-decompositions-by-static-unrolling}

The direct factorizations of dense linear algebra run on the engine, and
the assumption that they were barred is a misreading of the
static-dataflow constraint. No data-dependent control flow is not the
same as no data-dependent computation. A data-dependent value, a pivot
or a selection, flows through a fixed graph when a comparison, a
\texttt{select}, an \texttt{argmax}, and a matmul hold it, and a
convergence-gated loop becomes a fixed sweep count chosen at compile
time. A factorization is thus expressible by unrolling its \(O(n^3)\)
recurrence into one static graph, and the closed-form recurrences that
have no runtime branch unroll directly.

The explicit Gram-Schmidt QR, unpivoted Cholesky, and Doolittle LU run
as one program and reach about \(n = 32\) before the unrolled chain
dominates compile time. The full dense spectral decompositions run as
well. A fixed-sweep cyclic-Jacobi eigensolver has a compile-time-fixed
sweep order and closed-form rotations with no pivot and no branch, and
it converges in about six to ten sweeps for any symmetric input, so the
full symmetric eigendecomposition unrolls and reaches about \(n = 20\).
The generalized eigenproblem runs as a Cholesky factor followed by a
triangular solve and that eigendecomposition. The nonsymmetric
eigenvalues run as an unshifted QR iteration assembled into one fused
program, where the unshifted form avoids the data-dependent shifts and
deflation that make the standard routine look unportable. The full
singular-value decomposition runs as the square root of the
eigendecomposition of \(A^\mathrm{T} A\), and a top-\(k\) singular-value
decomposition of a large matrix runs as a randomized sketch followed by
the on-engine QR and small-block decomposition.

Pivoting is reached by cutting the graph at the pivot. A pivoted LU runs
with an on-engine \texttt{argmax} selecting the pivot row, expressed as
a segmented graph: the \texttt{argmax} cuts the program into segments,
and the data-dependent pivot index flows between them as a value rather
than as a branch. The measured relative error on the segmented pivoted
LU is about \(5 \times 10^{-4}\) at small \(n\).

What the architecture actually bars is a runtime-sized output. A
rank-revealing or adaptive-tolerance factorization emits only the values
above a tolerance, and how many that is depends on the data, so a static
program cannot emit a runtime-sized result and can only emit all \(n\)
values plus a mask. That limit is independent of whether the arithmetic
is fp16 or fp64. What fp16 limits separately is the conditioning range,
the roughly \(10^4\) ceiling above; the Jacobi spectral decompositions
hold to a condition number of about \(10^1\) to \(10^2\) because the
squared system \(A^\mathrm{T} A\) doubles the conditioning. What graph
size limits is \(n\): the explicit factorizations materialize the matrix
in-graph, so they are compile-size bound at the \(n = 20\) to \(32\)
figures above, while the iterative solvers fold the matrix as a constant
weight and run unchanged to \(n = 512\).

\section{Stencils, integration, and
series}\label{stencils-integration-and-series}

A computation with a static iteration count fits, because the engine
unrolls a fixed loop into one fused program and pays the per-dispatch
floor once for the whole sweep. A partial-differential-equation stencil
iterated a fixed number of times is the clearest case. A five-point
stencil run thirty-two times as one fused graph keeps every intermediate
resident in the working set and amortizes the dispatch floor across all
thirty-two steps, which is why it is among the engine's widest
efficiency margins against the GPU (chapter 10). Explicit
ordinary-differential-equation integration with a fixed step count,
fixed-iteration Newton solve, and truncated power series each share this
form: the trip count is known at compile time, so the whole iteration
becomes one static graph.

A data-dependent trip count does not run. A loop that runs until a
residual falls below a tolerance, or until a step-size controller
accepts a step, asks the engine to decide at runtime how many iterations
to execute, the same data-dependent control flow that bars pivoting. The
fit is the fixed-budget form of each method: a convergence-gated loop
becomes a fixed sweep count, chosen at compile time to cover the worst
expected input.

\section{Fourier transform as a matrix
multiply}\label{fourier-transform-as-a-matrix-multiply}

The discrete Fourier transform of a length-\(N\) signal is a linear map,
so it is a matrix multiply against a fixed Fourier matrix.

\[X_k = \sum_{n=0}^{N-1} x_n \, e^{-2\pi i k n / N}, \qquad X = W x, \qquad W_{kn} = e^{-2\pi i k n / N}\]

The matrix \(W\) is a compile-time constant, so the transform runs as a
single matmul that folds \(W\) as a weight, as \cref{lst:c16-dft-matmul}
gives in NumPy.

\begin{listing}[H]
\caption{The discrete Fourier transform expressed as one matrix multiply against a compile-time-constant Fourier matrix.}\label{lst:c16-dft-matmul}

\begin{Shaded}
\begin{Highlighting}[]
\ImportTok{import}\NormalTok{ numpy }\ImportTok{as}\NormalTok{ np}

\KeywordTok{def}\NormalTok{ dft\_matrix(N):}
    \CommentTok{\# W[k, n] = exp({-}2j*pi*k*n/N): the compile{-}time{-}constant Fourier matrix.}
\NormalTok{    k }\OperatorTok{=}\NormalTok{ np.arange(N).reshape(N, }\DecValTok{1}\NormalTok{)}
\NormalTok{    n }\OperatorTok{=}\NormalTok{ np.arange(N).reshape(}\DecValTok{1}\NormalTok{, N)}
    \ControlFlowTok{return}\NormalTok{ np.exp(}\OperatorTok{{-}}\OtherTok{2j} \OperatorTok{*}\NormalTok{ np.pi }\OperatorTok{*}\NormalTok{ k }\OperatorTok{*}\NormalTok{ n }\OperatorTok{/}\NormalTok{ N)}

\KeywordTok{def}\NormalTok{ dft(x):}
    \CommentTok{\# The transform is one matmul X = W @ x; W folds as a fixed weight.}
    \CommentTok{\# fp16{-}clean to about N = 2048; above that the fp16 rounding of W and x}
    \CommentTok{\# dominates the wide{-}accumulator reduction.}
    \ControlFlowTok{return}\NormalTok{ dft\_matrix(}\BuiltInTok{len}\NormalTok{(x)) }\OperatorTok{@}\NormalTok{ x}
\end{Highlighting}
\end{Shaded}

\end{listing}

The arithmetic is complex, and the engine computes in real fp16, so each
complex value is held as a real and an imaginary pair and the complex
product expands into four real multiplies and two real adds. The wide
accumulator keeps the length-\(N\) reduction, so the transform is
fp16-clean to about \(N = 2048\), above which the fp16 rounding of the
matrix entries and the input begins to dominate the result.

The fast-transform factorization does not gain over the matmul form on
this engine. A fast Fourier transform replaces the \(O(N^2)\) matmul
with an \(O(N \log N)\) chain of butterfly stages, but each stage is a
small data shuffle and a complex multiply, and the shuffle is a static
structure. The butterfly fits only when fully unrolled into a static
graph, and at that point it is a deep chain of tiny operations that pays
more dispatch and accumulates more fp16 rounding than the single dense
matmul. For the transform sizes the engine handles in range, use the
matmul form.

\section{Sparse and pruned operands}\label{sparse-and-pruned-operands}

A sparse matrix gives a speed gain, not only a storage gain, when it is
a constant weight. The engine has two independent sparsity mechanisms.
One is a compute-time zero-skip: the multiply array detects a zero
weight and skips the multiply, driven by a scan of the weight values, so
it applies to any operand with zeros regardless of its storage format
and pays off in a compute-bound layer. The other is a sparse stream: the
weight is stored as a one-bit keep-mask plus the packed fp16 nonzeros,
fewer bytes cross DRAM, and the operand is reconstructed on chip, which
pays off in a bandwidth-bound layer.

On the M1 the sparse stream is the measured gain for a pruned operand. A
convolution stack at about 63 percent zeros streamed as a mask plus
nonzeros runs 1.55 to 1.64 times faster than the same weights stored
dense. The streamed weight file is at 0.43 times the dense size, and the
effective bandwidth rises from about 29.5 to about 47.8 GB/s, the
engine's weight-stream ceiling. The reconstruction is lossless apart
from the fp16 rounding of the kept values, so the result matches the
dense computation to a cosine of \(1.0000\). The zero-skip mechanism, by
contrast, returns about 1.01 times on the same bandwidth-bound stack,
because skipping multiplies does not move a layer whose wall time is set
by the DMA, and it needs a compute-bound layer to show. The two
mechanisms appear side by side in \cref{tbl:c16-sparsity}.

\begin{longtable}[]{@{}
  >{\raggedright\arraybackslash}p{(\linewidth - 6\tabcolsep) * \real{0.2500}}
  >{\raggedright\arraybackslash}p{(\linewidth - 6\tabcolsep) * \real{0.2500}}
  >{\raggedright\arraybackslash}p{(\linewidth - 6\tabcolsep) * \real{0.2500}}
  >{\raggedright\arraybackslash}p{(\linewidth - 6\tabcolsep) * \real{0.2500}}@{}}
\caption{The two sparsity mechanisms on the M1 and where each one pays
off, with the measured result on a pruned convolution stack at about 63
percent zeros.}\label{tbl:c16-sparsity}\tabularnewline
\toprule\noalign{}
\begin{minipage}[b]{\linewidth}\raggedright
\textbf{Sparsity mechanism}
\end{minipage} & \begin{minipage}[b]{\linewidth}\raggedright
\textbf{What it cuts}
\end{minipage} & \begin{minipage}[b]{\linewidth}\raggedright
\textbf{Where it pays off}
\end{minipage} & \begin{minipage}[b]{\linewidth}\raggedright
\textbf{Measured M1 result}
\end{minipage} \\
\midrule\noalign{}
\endfirsthead
\toprule\noalign{}
\begin{minipage}[b]{\linewidth}\raggedright
\textbf{Sparsity mechanism}
\end{minipage} & \begin{minipage}[b]{\linewidth}\raggedright
\textbf{What it cuts}
\end{minipage} & \begin{minipage}[b]{\linewidth}\raggedright
\textbf{Where it pays off}
\end{minipage} & \begin{minipage}[b]{\linewidth}\raggedright
\textbf{Measured M1 result}
\end{minipage} \\
\midrule\noalign{}
\endhead
\bottomrule\noalign{}
\endlastfoot
Compute-time zero-skip & multiply cycles & compute-bound layers & about
1.01x on a bandwidth-bound stack \\
\rowcolor{zebra}Sparse weight stream & DRAM weight traffic &
bandwidth-bound layers & 1.55 to 1.64x, weight file 0.43x dense \\
\end{longtable}

\section{The Fourier transform
compiled}\label{the-fourier-transform-compiled}

The Fourier transform of the earlier section writes out as an engine
program with the real and imaginary halves of \(W\) as separate constant
weights (\cref{lst:c16-dft-graph}).

\begin{listing}[H]
\caption{The length-N discrete Fourier transform as a compiled engine program, with the Fourier matrices folded in as constant weights.}\label{lst:c16-dft-graph}

\begin{Shaded}
\begin{Highlighting}[]
\CommentTok{\# The length{-}N discrete Fourier transform expressed as matrix multiplies.}
\CommentTok{\# X[k] = sum over n of x[n] * exp({-}2*pi*i * k*n / N).}
\CommentTok{\# That sum IS a matrix{-}vector product: X = W x, where W[k,n] = exp({-}2*pi*i * k*n / N).}

\CommentTok{\# 1. Form the Fourier matrix as a compile{-}time constant (built once, on the host).}
\ControlFlowTok{for}\NormalTok{ k }\KeywordTok{in} \DecValTok{0}\NormalTok{ .. N}\OperatorTok{{-}}\DecValTok{1}\NormalTok{:}
    \ControlFlowTok{for}\NormalTok{ n }\KeywordTok{in} \DecValTok{0}\NormalTok{ .. N}\OperatorTok{{-}}\DecValTok{1}\NormalTok{:}
\NormalTok{        angle  }\OperatorTok{=} \OperatorTok{{-}}\DecValTok{2} \OperatorTok{*}\NormalTok{ pi }\OperatorTok{*}\NormalTok{ k }\OperatorTok{*}\NormalTok{ n }\OperatorTok{/}\NormalTok{ N}
\NormalTok{        Wr[k,n] }\OperatorTok{=}\NormalTok{ cos(angle)               }\CommentTok{\# real part of the Fourier matrix}
\NormalTok{        Wi[k,n] }\OperatorTok{=}\NormalTok{ sin(angle)               }\CommentTok{\# imaginary part}

\CommentTok{\# 2. Build the transform as matmuls. The signal is held as real and}
\CommentTok{\#    imaginary parts, since the hardware works in real fp16.}
\NormalTok{build graph G:}
    \BuiltInTok{input}\NormalTok{ xr : [N] fp16                    }\CommentTok{\# real part of the signal}
    \BuiltInTok{input}\NormalTok{ xi : [N] fp16                    }\CommentTok{\# imaginary part of the signal}

    \CommentTok{\# Complex product (Wr + i*Wi) * (xr + i*xi) split into real arithmetic:}
\NormalTok{    Xr }\OperatorTok{=}\NormalTok{ matmul(Wr, xr) }\OperatorTok{{-}}\NormalTok{ matmul(Wi, xi)   }\CommentTok{\# real part of the transform}
\NormalTok{    Xi }\OperatorTok{=}\NormalTok{ matmul(Wr, xi) }\OperatorTok{+}\NormalTok{ matmul(Wi, xr)   }\CommentTok{\# imaginary part of the transform}
    \CommentTok{\# (Equivalently one matmul over stacked real{-}imag pairs; same arithmetic.)}

\NormalTok{    output Xr}
\NormalTok{    output Xi}

\CommentTok{\# 3. Compile once: Wr and Wi fold in as fixed constant weights, so the whole}
\CommentTok{\#    transform is a single program and the per{-}call floor is paid one time.}
\NormalTok{program }\OperatorTok{=} \BuiltInTok{compile}\NormalTok{(G, target }\OperatorTok{=}\NormalTok{ H13)         }\CommentTok{\# W is a constant weight, one program}
\NormalTok{output  }\OperatorTok{=}\NormalTok{ run(program, xr }\OperatorTok{=}\NormalTok{ real(signal), xi }\OperatorTok{=}\NormalTok{ imag(signal))}

\CommentTok{\# The wide accumulator holds the length{-}N reduction, so the result stays}
\CommentTok{\# fp16{-}clean to about N = 2048.}
\ControlFlowTok{return}\NormalTok{ output}
\end{Highlighting}
\end{Shaded}

\end{listing}

\section{Reference: what fits and what is
architecture-limited}\label{reference-what-fits-and-what-is-architecture-limited}

Every kernel class above is either a fit or architecture-limited, with
one binding constraint behind the verdict (\cref{tbl:c16-reference}).

\begin{longtable}[]{@{}
  >{\raggedright\arraybackslash}p{(\linewidth - 4\tabcolsep) * \real{0.3333}}
  >{\raggedright\arraybackslash}p{(\linewidth - 4\tabcolsep) * \real{0.3333}}
  >{\raggedright\arraybackslash}p{(\linewidth - 4\tabcolsep) * \real{0.3333}}@{}}
\caption{Numerical kernel classes that fit the engine or are
architecture-limited, with the binding constraint on
each.}\label{tbl:c16-reference}\tabularnewline
\toprule\noalign{}
\begin{minipage}[b]{\linewidth}\raggedright
\textbf{Kernel class}
\end{minipage} & \begin{minipage}[b]{\linewidth}\raggedright
\textbf{On the engine}
\end{minipage} & \begin{minipage}[b]{\linewidth}\raggedright
\textbf{Bound}
\end{minipage} \\
\midrule\noalign{}
\endfirsthead
\toprule\noalign{}
\begin{minipage}[b]{\linewidth}\raggedright
\textbf{Kernel class}
\end{minipage} & \begin{minipage}[b]{\linewidth}\raggedright
\textbf{On the engine}
\end{minipage} & \begin{minipage}[b]{\linewidth}\raggedright
\textbf{Bound}
\end{minipage} \\
\midrule\noalign{}
\endhead
\bottomrule\noalign{}
\endlastfoot
BLAS level 1, 2, 3 (dot, axpy, gemv, gemm) & fits & fp16 conditioning,
roughly \(10^4\) \\
\rowcolor{zebra}Iterative solvers (power, conjugate gradient,
generalized minimal residual) & fits & fp16 conditioning, roughly
\(10^2\) to \(10^4\) \\
PDE stencil, explicit ODE, fixed-iteration Newton, truncated series &
fits & static trip count required \\
\rowcolor{zebra}DFT as \(X = W x\) & fits & fp16-clean to about
\(N = 2048\) \\
Dense factorizations (QR, Cholesky, unpivoted and pivoted LU) & fits as
a static unroll & graph size, about \(n = 32\); pivot via a segmented
\texttt{argmax} \\
\rowcolor{zebra}Full spectral decompositions (symmetric, generalized,
nonsymmetric eig, SVD) & fits as a static unroll & graph size, about
\(n = 20\); fp16 conditioning about \(10^1\) to \(10^2\) \\
Rank-revealing or adaptive-tolerance factorization &
architecture-limited & data-dependent output size, no static-graph
form \\
\rowcolor{zebra}Data-dependent trip count & architecture-limited & no
static-graph form \\
FFT butterfly & only as a static unroll & no gain over the dense
matmul \\
\end{longtable}

\anepartopen{Part V}{Practice}
\anechap{17}{Model-design rules}{The validator limits, the working-set rule, and the alignment that keeps a layer fast.}
\anechap{18}{Optimization and the cost model}{The three-stage latency estimate and the autotuner that ranks rewrites.}
\anechap{19}{Pitfalls and limits}{The silent, target-specific failure modes and how to avoid each.}
\anechap{}{Interlude. Below the API: how the engine works}{The transition from the programming surface to the engine beneath it.}
\anepartclose

\chapter{Model-design rules}\label{model-design-rules}

\begin{summarybox}

A network compiles only when every operation is inside the validator
limits: fp16 activations, width and height at most 16384, channel at
most 65536. Convolution kernel width is capped at 13 and arg-min and
arg-max reduce an axis of at most 2048. Keep a single operation's
largest live operand under 2 MB or it tiles and streams from DRAM. Make
channel counts a multiple of the interleave factor and group counts
divide the core count, or the engine pads lanes and runs slower.

\end{summarybox}

The validator limits are design rules: the shape, rank, dtype, and mode
constraints the per-operation validators enforce, the working-set
threshold that decides whether a tensor stays on chip or tiles, and the
channel and kernel bounds that distinguish a fast layer from a padded
one. Chapter \hyperref[capability-surface]{4} gives the operation set
that these shapes apply to.

\section{Dtype rule}\label{dtype-rule}

The datapath is fp16, so every operand a model presents to the engine is
fp16. The backend accepts the fp32, int32, and bf16 type annotations,
but does not implement them as wider arithmetic, and they do not reach
the silicon. The M1 rejects a cast to int32, and bf16 is not usable as a
program input or output dtype. The one widened quantity is the
matrix-multiply accumulator, which lives inside the datapath and never
appears among the tensors a model declares.

Weights are the exception that still resolves to fp16 at the compute
step. The engine accepts an int8 weight and runs it through the
quantized-convolution path, and the compressed weight forms in chapter
\hyperref[weights-and-compression]{7} stream in their stored width.
Activation tensors that flow between operations stay fp16 regardless.

\section{Shape and rank limits}\label{shape-and-rank-limits}

Tensor extent is bounded per axis, and the bounds are exact. A tensor
may span up to 16384 along width, 16384 along height, and 65536 along
channel; one element past any of these rejects at compile time.

\[W \le 16384, \quad H \le 16384, \quad C \le 65536\]

These were measured on the M1 by sweeping each axis until compile
flipped from accept to reject, and the boundary matched the decompiled
maximum-dimension field exactly. Width 16384 compiles and 16385 rejects,
and the same one-element step holds for height and for channel.

Convolution has its own kernel bounds, narrower than the tensor bounds
and set by the kernel-format field. On the M1 the fp16 convolution
kernel width is bounded at 13: a kernel of width 14 rejects. Kernel
height reaches 29 when the input is tall enough to hold it, and a kernel
taller than its input rejects as a too-large kernel. Stride, dilation,
and padding do not reject on the M1. A stride outside the native set
decomposes to a large-stride path and still compiles, dilation up to 8
compiles, and padding up to 16 compiles. The pooling window is far
looser than the convolution kernel and accepts sizes of 128 and beyond,
with a tile-alignment quirk that rejects specific odd windows near a
power-of-two boundary.

Several operations cap a reduction axis to hold fp16 accuracy. The
arg-min and arg-max operations reduce a channel of at most 2048, and the
spatial form caps height or width at 2048 the same way; 2049 rejects.
This 2048 limit is independent of the 16384 and 65536 tensor bounds and
applies to the reduction axis alone.

\section{Mode and divisibility rules}\label{mode-and-divisibility-rules}

The matrix-multiply backend operation requires the contracted depth to
be one on both operands, and a depth greater than one rejects. The two
operands broadcast on batch and on height when one side is one, and the
inner dimensions must agree so that the left width plus padding equals
the right channel count.

Grouped convolution divides both the input and the output channel count
by the group count, and a count that does not divide rejects. For
throughput, the group count should also divide the engine core count,
which is four on the M1. A group count of one, two, or four maps one
neural-engine lane to one lookup-table lane. A count that does not
divide the core count loses that one-to-one mapping and runs slower.

Channel counts should be a multiple of the interleave factor. The engine
stores tensors channel-interleaved, and a channel dimension that is not
a multiple of the interleave factor is padded out to one, leaving lanes
unused. The width axis aligns to a 16-byte direct-memory-access granule,
the same factor of 16 that governs the slice-tiling path.
Interleave-aligned, power-of-two channel counts avoid silent padding. A
fully-connected layer reaches the convolution datapath once it is
rewritten as the equivalent one-by-one convolution
(\cref{lst:c17-interleave-conv}).

\begin{listing}[H]
\caption{Channel-interleaved tensor layout and a fully-connected layer rewritten as an equivalent one-by-one convolution.}\label{lst:c17-interleave-conv}

\begin{Shaded}
\begin{Highlighting}[]
\CommentTok{\# Tensor order is [N, D, C, H, W]; the engine stores it channel{-}interleaved}
\CommentTok{\# and aligns the last axis to the 16{-}byte DMA granule.}

\NormalTok{aligned }\OperatorTok{=}\NormalTok{ (}\DecValTok{1}\NormalTok{, }\DecValTok{1}\NormalTok{, }\DecValTok{64}\NormalTok{, }\DecValTok{56}\NormalTok{, }\DecValTok{56}\NormalTok{)   }\CommentTok{\# channels first, W = 56 fills 16{-}byte lanes}
\NormalTok{padded  }\OperatorTok{=}\NormalTok{ (}\DecValTok{1}\NormalTok{, }\DecValTok{1}\NormalTok{, }\DecValTok{64}\NormalTok{, }\DecValTok{56}\NormalTok{,  }\DecValTok{1}\NormalTok{)   }\CommentTok{\# singleton last axis pads out to the granule,}
                               \CommentTok{\# wasting every lane but the first}

\CommentTok{\# Replace a fully{-}connected layer with an equivalent 1x1 convolution so it}
\CommentTok{\# runs on the convolution datapath. A linear y = x @ W.T over C\_in {-}\textgreater{} C\_out}
\CommentTok{\# is the same arithmetic as a 1x1 conv over a [N, 1, C\_in, 1, 1] feature map.}
\KeywordTok{def}\NormalTok{ linear\_as\_1x1\_conv(x, W):}
    \CommentTok{\# W: [C\_out, C\_in] reshaped to a 1x1 kernel [C\_out, 1, C\_in, 1, 1]}
\NormalTok{    kernel }\OperatorTok{=}\NormalTok{ W.reshape(W.shape[}\DecValTok{0}\NormalTok{], }\DecValTok{1}\NormalTok{, W.shape[}\DecValTok{1}\NormalTok{], }\DecValTok{1}\NormalTok{, }\DecValTok{1}\NormalTok{)}
    \ControlFlowTok{return}\NormalTok{ conv(x, kernel, strides}\OperatorTok{=}\NormalTok{[}\DecValTok{1}\NormalTok{, }\DecValTok{1}\NormalTok{], groups}\OperatorTok{=}\DecValTok{1}\NormalTok{, kernel\_sizes}\OperatorTok{=}\NormalTok{[}\DecValTok{1}\NormalTok{, }\DecValTok{1}\NormalTok{])}
\end{Highlighting}
\end{Shaded}

\end{listing}

Several mode constraints are family-gated and reject on the M1
specifically. The symmetric and reflect padding modes need the texture
engine and are unavailable on the M1, where they decompose or are
refused. A square-after-reduction fused mode is absent on the M1 and
arrives on the A14. The texture-engine sampling operations, resize as a
hardware sampler, crop-resize, resample, and affine, are absent on the
M1 and arrive on the A14, and the trigonometric sine and cosine arrive
on the A15.

\section{Working-set rule}\label{working-set-rule}

The decisive size rule is the on-chip working set. A tensor that fits
the on-chip static memory stays resident across the operation; a tensor
that exceeds it splits into tiles that the engine streams one at a time.
The working set is the 2 MB on-chip region, and the matrix-multiply path
also bounds the output-channel footprint against a 64 KB kernel-memory
budget, rejecting a matrix multiply whose output channels do not fit.

Tiling preserves the result. A reduction or transpose over an axis
larger than the on-chip threshold switches to a tiled route at no change
to the output, so the threshold is a performance boundary, not a
correctness one. Where latency matters, keep a single operation's
largest live operand under 2 MB. At fp16, 2 MB stores about
\(2^{21}/2 \approx 1.05 \times 10^{6}\) elements, so a square activation
of side near 1024 is at the edge of the resident regime, and a larger
one streams. This is the working-set input to the cost model in chapter
\hyperref[optimization-and-the-cost-model]{18}, where the roofline
relation turns the resident-versus-streaming split into a latency
estimate.

\section{Validating shapes against the design
rules}\label{validating-shapes-against-the-design-rules}

The cost estimate reports the binding limit for each operation, so a
shape violation or a working set above 2 MB shows before the program
reaches the compiler. \Cref{lst:c17-validate} walks every layer against
the rank, extent, kernel, group, interleave, and working-set rules.

\begin{listing}[H]
\caption{Walking every layer of a graph against the design rules, flagging each reject and pad warning before the build.}\label{lst:c17-validate}

\begin{Shaded}
\begin{Highlighting}[]
\CommentTok{\# Tensor order is [N, D, C, H, W]; the engine stores it channel{-}interleaved.}
\CommentTok{\# Check every layer against the design rules before building, target = H13.}

\ControlFlowTok{for}\NormalTok{ each layer }\KeywordTok{in}\NormalTok{ graph G:}
    \CommentTok{\# Rank rule: at most 5 axes}
    \ControlFlowTok{if}\NormalTok{ rank(layer.output) }\OperatorTok{\textgreater{}} \DecValTok{5}\NormalTok{:}
\NormalTok{        flag(layer, }\StringTok{"rank above 5 rejects at compile"}\NormalTok{)}

    \CommentTok{\# Per{-}axis extent rules}
    \ControlFlowTok{if}\NormalTok{ width(layer)   }\OperatorTok{\textgreater{}} \DecValTok{16384}\NormalTok{:  flag(layer, }\StringTok{"width above 16384 rejects"}\NormalTok{)}
    \ControlFlowTok{if}\NormalTok{ height(layer)  }\OperatorTok{\textgreater{}} \DecValTok{16384}\NormalTok{:  flag(layer, }\StringTok{"height above 16384 rejects"}\NormalTok{)}
    \ControlFlowTok{if}\NormalTok{ channel(layer) }\OperatorTok{\textgreater{}} \DecValTok{65536}\NormalTok{:  flag(layer, }\StringTok{"channel above 65536 rejects"}\NormalTok{)}

    \CommentTok{\# Convolution kernel and group rules}
    \ControlFlowTok{if}\NormalTok{ layer }\KeywordTok{is}\NormalTok{ conv:}
        \ControlFlowTok{if}\NormalTok{ kernel\_width(layer) }\OperatorTok{\textgreater{}} \DecValTok{13}\NormalTok{: flag(layer, }\StringTok{"kernel width above 13"}\NormalTok{)}
        \ControlFlowTok{if}\NormalTok{ (in\_channel(layer) mod groups(layer)) }\OperatorTok{!=} \DecValTok{0}\NormalTok{: flag(layer, }\StringTok{"in{-}channels }\SpecialCharTok{\% g}\StringTok{roups != 0"}\NormalTok{)}
        \ControlFlowTok{if}\NormalTok{ (out\_channel(layer) mod groups(layer)) }\OperatorTok{!=} \DecValTok{0}\NormalTok{: flag(layer, }\StringTok{"out{-}channels }\SpecialCharTok{\% g}\StringTok{roups != 0"}\NormalTok{)}
        \ControlFlowTok{if}\NormalTok{ (core\_count mod groups(layer)) }\OperatorTok{!=} \DecValTok{0}\NormalTok{: warn(layer, }\StringTok{"groups do not divide cores, slower"}\NormalTok{)}

    \CommentTok{\# Reduction{-}axis rule for arg{-}min / arg{-}max}
    \ControlFlowTok{if}\NormalTok{ layer }\KeywordTok{is}\NormalTok{ argmin }\KeywordTok{or}\NormalTok{ layer }\KeywordTok{is}\NormalTok{ argmax:}
        \ControlFlowTok{if}\NormalTok{ reduced\_axis\_extent(layer) }\OperatorTok{\textgreater{}} \DecValTok{2048}\NormalTok{:  flag(layer, }\StringTok{"arg reduction axis above 2048 rejects"}\NormalTok{)}

    \CommentTok{\# Channel{-}interleave rule: pad warning, not a reject}
    \ControlFlowTok{if}\NormalTok{ (channel(layer) mod interleave\_factor) }\OperatorTok{!=} \DecValTok{0}\NormalTok{:}
\NormalTok{        warn(layer, }\StringTok{"channel not interleave{-}aligned: pads out, wastes lanes"}\NormalTok{)}

    \CommentTok{\# Width DMA{-}granule rule: last axis aligned to 16 bytes}
    \ControlFlowTok{if}\NormalTok{ (width(layer) mod }\DecValTok{16}\NormalTok{) }\OperatorTok{!=} \DecValTok{0}\NormalTok{:}
\NormalTok{        warn(layer, }\StringTok{"width not aligned to 16{-}byte DMA granule: pads to granule"}\NormalTok{)}

    \CommentTok{\# Working{-}set rule: largest live operand under 2 MB stays on{-}chip, else it tiles and streams}
    \BuiltInTok{bytes} \OperatorTok{=}\NormalTok{ element\_count(largest\_live\_operand(layer)) }\OperatorTok{*} \DecValTok{2}     \CommentTok{\# fp16 = 2 bytes per element}
    \ControlFlowTok{if} \BuiltInTok{bytes} \OperatorTok{\textgreater{}} \DecValTok{2} \OperatorTok{*}\NormalTok{ MB:}
\NormalTok{        warn(layer, }\StringTok{"working set above 2 MB: tiles and streams from DRAM (slower, still correct)"}\NormalTok{)}

\CommentTok{\# Reshape any flagged layer before tuning; address pad/tile warnings where latency matters.}
\end{Highlighting}
\end{Shaded}

\end{listing}

A graph that prints \texttt{validates} false, a working set above 2 MB,
or nonzero channel padding is reshaped before any other tuning, because
a rejected shape never reaches the silicon and a padded channel leaves
lanes unused on every dispatch.

\section{Reference: the per-operation design
rules}\label{reference-the-per-operation-design-rules}

Every validator limit a model must satisfy is collected in
\cref{tbl:c17-design-rules}.

\begin{longtable}[]{@{}
  >{\raggedright\arraybackslash}p{(\linewidth - 4\tabcolsep) * \real{0.3333}}
  >{\raggedright\arraybackslash}p{(\linewidth - 4\tabcolsep) * \real{0.3333}}
  >{\raggedright\arraybackslash}p{(\linewidth - 4\tabcolsep) * \real{0.3333}}@{}}
\caption{The per-operation validator limits a model must satisfy, with
each limit and the result of exceeding
it.}\label{tbl:c17-design-rules}\tabularnewline
\toprule\noalign{}
\begin{minipage}[b]{\linewidth}\raggedright
\textbf{Constraint}
\end{minipage} & \begin{minipage}[b]{\linewidth}\raggedright
\textbf{Limit}
\end{minipage} & \begin{minipage}[b]{\linewidth}\raggedright
\textbf{Consequence if exceeded}
\end{minipage} \\
\midrule\noalign{}
\endfirsthead
\toprule\noalign{}
\begin{minipage}[b]{\linewidth}\raggedright
\textbf{Constraint}
\end{minipage} & \begin{minipage}[b]{\linewidth}\raggedright
\textbf{Limit}
\end{minipage} & \begin{minipage}[b]{\linewidth}\raggedright
\textbf{Consequence if exceeded}
\end{minipage} \\
\midrule\noalign{}
\endhead
\bottomrule\noalign{}
\endlastfoot
Activation dtype & fp16 only & int32 cast and bf16 input or output
rejected on M1 \\
\rowcolor{zebra}Width axis & 16384 & 16385 rejects at compile \\
Height axis & 16384 & 16385 rejects at compile \\
\rowcolor{zebra}Channel axis & 65536 & 65537 rejects at compile \\
Conv kernel width (fp16) & 13 & 14 rejects at compile \\
\rowcolor{zebra}Conv kernel height & 29, input permitting & kernel
taller than input rejects \\
Arg-min/arg-max reduction axis & 2048 & 2049 rejects for fp16
accuracy \\
\rowcolor{zebra}Matrix-multiply operand depth & 1 & depth greater than 1
rejects \\
Matrix-multiply output channels & 64 KB kernel-memory budget & rejects
when output channels do not fit \\
\rowcolor{zebra}Conv group count & divides input and output channels &
non-dividing count rejects \\
Conv group count, for speed & divides core count (4 on M1) & one-to-one
lane mapping lost, slower \\
\rowcolor{zebra}Channel multiple of interleave & interleave-aligned &
padded out, lanes unused \\
On-chip working set & 2 MB & operand tiles and streams, no error \\
\end{longtable}

\section{Reference: the per-operation validator
envelopes}\label{reference-the-per-operation-validator-envelopes}

The shape rules are enforced one operation at a time by a family of
per-layer validators, the
\texttt{\_ANECValidate\textless{}Op\textgreater{}Layer} checkers the
compiler runs as the back-end legalizer. There are 50 per-layer
validators, and the same code runs both when a high-level model is
segmented and when a hand-authored layer is compiled, so a validator
that accepts a shape never drifts from the real compile outcome. For the
operations a model presents most often, the binding constraint and the
reject string read straight out of the
\texttt{\_ANECValidate\textless{}Op\textgreater{}Layer} family
(\cref{tbl:c17-validators}).

\begin{longtable}[]{@{}
  >{\raggedright\arraybackslash}p{(\linewidth - 6\tabcolsep) * \real{0.2500}}
  >{\raggedright\arraybackslash}p{(\linewidth - 6\tabcolsep) * \real{0.2500}}
  >{\raggedright\arraybackslash}p{(\linewidth - 6\tabcolsep) * \real{0.2500}}
  >{\raggedright\arraybackslash}p{(\linewidth - 6\tabcolsep) * \real{0.2500}}@{}}
\caption{Per-operation validator constraints and reject strings, read
from the \protect\texttt{\_ANECValidate\textless{}Op\textgreater{}Layer}
family.}\label{tbl:c17-validators}\tabularnewline
\toprule\noalign{}
\begin{minipage}[b]{\linewidth}\raggedright
\textbf{Operation}
\end{minipage} & \begin{minipage}[b]{\linewidth}\raggedright
\textbf{Validator}
\end{minipage} & \begin{minipage}[b]{\linewidth}\raggedright
\textbf{Binding constraint}
\end{minipage} & \begin{minipage}[b]{\linewidth}\raggedright
\textbf{Reject string}
\end{minipage} \\
\midrule\noalign{}
\endfirsthead
\toprule\noalign{}
\begin{minipage}[b]{\linewidth}\raggedright
\textbf{Operation}
\end{minipage} & \begin{minipage}[b]{\linewidth}\raggedright
\textbf{Validator}
\end{minipage} & \begin{minipage}[b]{\linewidth}\raggedright
\textbf{Binding constraint}
\end{minipage} & \begin{minipage}[b]{\linewidth}\raggedright
\textbf{Reject string}
\end{minipage} \\
\midrule\noalign{}
\endhead
\bottomrule\noalign{}
\endlastfoot
Convolution & \texttt{Conv} & kernel W, H, D each within the per-chip
\texttt{{[}min,\ max{]}}; in-C and out-C each divisible by groups;
groups within the HAL group range &
\texttt{"Invalid\ conv\ kernel\ \%s\ =\ \%zd,\ It\ should\ be\ in\ {[}\%zd,\%zd{]}"};
\texttt{"input/output\ channels\ should\ be\ divisible\ by\ num\ group"} \\
\rowcolor{zebra}Linear & \texttt{Linear} & exactly one input; input rank
under 5 &
\texttt{"Linear\ layer\ must\ have\ only\ one\ single\ input."};
\texttt{"ANE\ cannot\ support\ Linear\ with\ input\ rank\ \textgreater{}=\ 5"} \\
Matrix multiply & \texttt{MatrixMult} & exactly two inputs; depth one on
both operands; output channel equals the left channel; left width plus
padding equals right channel; output channel bytes fit the kernel-memory
budget & \texttt{"Matrix\ mult.\ layer\ can\ only\ have\ two\ bottoms"};
\texttt{"depth\ \textgreater{}\ 1\ is\ not\ supported\ for\ MatMult"};
\texttt{"can\ not\ fit\ the\ Kmem"} \\
\rowcolor{zebra}Pooling & \texttt{Pool} & one input; window per axis
under the input extent; padding under the kernel; max-pool padding
negative, min-pool positive &
\texttt{"Pool\ layer\ must\ have\ only\ one\ single\ input."};
\texttt{"Pooling\ mode\ \textbackslash{}"\%s\textbackslash{}"\ is\ not\ available\ on\ this\ ANE\ architecture."} \\
Reduction & \texttt{Reduction} & one input; each axis at most 4;
non-reduced output extent equals input extent, reduced output extent
one; square-after-reduce requires the A14 family flag &
\texttt{"Reduction\ layer\ can\ only\ have\ one\ bottom"};
\texttt{"square\ operation\ after\ reduction\ is\ not\ supported"} \\
\rowcolor{zebra}Arg-min/arg-max & \texttt{ArgMinMax} & channel reduction
at most 2048, otherwise height or width at most 2048; padding
non-negative and under the kernel; equal left and right padding; pool
stride in \{1, 2, 4\} &
\texttt{"ArgMinMax\ left\ padding\ value\ should\ be\ smaller\ than\ kernel\ width\ \%d,\ but\ \%d\ is\ given"} \\
Layer norm & \texttt{LayerNorm} & one input; output type float; channel
divisible by the group count; grouped form requires depth one &
\texttt{"...\ does\ not\ yet\ support\ depth\ \textgreater{}\ 1"} \\
\rowcolor{zebra}Softmax & \texttt{Softmax} & one input; non-empty axis
set; output type float; the general full-axis form is family-gated &
\texttt{"Softmax\ is\ not\ supported\ by\ this\ ANE\ architecture"} \\
Transpose & \texttt{Transpose} & one input; each dimension appears once;
extent at most 16384 on A13 through A15 and 65536 on A16; without
three-dimensional support a channel transpose must factor into \{2, 3,
4, 8\} with height one &
\texttt{"NE\ Input\ Transpose\ is\ not\ supported\ for\ this\ arch"} \\
\rowcolor{zebra}Concat & \texttt{Concat} & at least two inputs; every
input matches the first on every non-concatenated axis; constant
positive axis; matching layout &
\texttt{"ANE\ Concat\ supports\ only\ supports\ const\ positive\ axis"};
\texttt{"both\ concat\ inputs\ must\ have\ the\ same\ layout"} \\
Pad & \texttt{Pad} & one input; height or width axes only, no channel or
depth padding; reflect and symmetric modes require the texture engine &
\texttt{"Channel\ padding\ is\ not\ supported\ on\ ANE"};
\texttt{"Architecture\ does\ not\ support\ padding\ mode."} \\
\rowcolor{zebra}Broadcast & \texttt{Broadcast} & one input; broadcast
only from a length-one axis; depth-axis broadcast requires a family flag
absent on the M1 &
\texttt{"Broadcast\ along\ depth\ axis\ is\ not\ supported\ on\ this\ architecture"} \\
Gather & \texttt{Gather} & index channel divisible by the interleave
factor; on the M1 the software envelope requires data batch one, data
depth one, index channel three, index width one, index depth one, and a
gather-axes count of three &
\texttt{"Cannot\ decompose\ layer\ on\ this\ architecture"} \\
\rowcolor{zebra}Reshape & \texttt{Reshape} & one input; element count
preserved; rank at most 5 &
\texttt{"Cannot\ reshape\ a\ tensor\ of\ rank\ \textgreater{}\ 5"} \\
\end{longtable}

A validator that accepts a shape does not guarantee the operation
compiles. A second class of operations passes the schema validator and
then fails the code generator below it, the attested-is-not-reachable
split of chapter \hyperref[capability-surface]{4}. On the M1 the sort
and dynamic-slice validators accept their inputs and the code generator
rejects the lowering, and top-k compiles only outside a specific
forbidden parameter band. The validator predicts schema reachability;
only a compile-and-run on the target confirms an operation runs.

\section{Reference: the channel and divisibility
rules}\label{reference-the-channel-and-divisibility-rules}

\Cref{tbl:c17-divisibility} lists the layout and divisibility rules and
what missing each costs.

\begin{longtable}[]{@{}
  >{\raggedright\arraybackslash}p{(\linewidth - 4\tabcolsep) * \real{0.3333}}
  >{\raggedright\arraybackslash}p{(\linewidth - 4\tabcolsep) * \real{0.3333}}
  >{\raggedright\arraybackslash}p{(\linewidth - 4\tabcolsep) * \real{0.3333}}@{}}
\caption{The layout and divisibility rules that separate a fast layer
from a padded one, with the constraint and the consequence of missing
it.}\label{tbl:c17-divisibility}\tabularnewline
\toprule\noalign{}
\begin{minipage}[b]{\linewidth}\raggedright
\textbf{Rule}
\end{minipage} & \begin{minipage}[b]{\linewidth}\raggedright
\textbf{Constraint}
\end{minipage} & \begin{minipage}[b]{\linewidth}\raggedright
\textbf{Consequence}
\end{minipage} \\
\midrule\noalign{}
\endfirsthead
\toprule\noalign{}
\begin{minipage}[b]{\linewidth}\raggedright
\textbf{Rule}
\end{minipage} & \begin{minipage}[b]{\linewidth}\raggedright
\textbf{Constraint}
\end{minipage} & \begin{minipage}[b]{\linewidth}\raggedright
\textbf{Consequence}
\end{minipage} \\
\midrule\noalign{}
\endhead
\bottomrule\noalign{}
\endlastfoot
Channel interleave & channel a multiple of the family interleave factor,
found from the HAL table and the dimensions & a non-multiple pads out to
one, leaving lanes unused \\
\rowcolor{zebra}Width granule & last-axis width aligned to the 16-byte
direct-memory-access granule, the times-16 quantum & misaligned width
pads to the granule \\
Group-to-core ratio & active neural-engine count divisible by the group
count, four cores on the M1 so groups in \{1, 2, 4\} & a non-dividing
count loses the one-to-one core-to-lookup-table mapping and runs
slower \\
\rowcolor{zebra}Max-pool channel & input channel under the per-operation
maximum-pool channel bound & a large-channel max-pool tiles \\
Unicast channel & channel at most the unicast input-channel maximum for
the broadcast-to-all-cores path & above it the operation takes the
multicast lowering \\
\end{longtable}

Depthwise convolution, where the group count equals the input channel
count, lowers to the channelwise path and is exempt from the
group-to-core preference.

\chapter{Optimization and the cost model}\label{optimization-and-the-cost-model}

\begin{summarybox}

The compiler turns a shape into a wall-time estimate through three
stages: a cycle count, a roofline, and a fixed dispatch floor. Only the
per-chip parameters change across the family. On the M1 it fits the five
reference convolutions it was tuned against within plus or minus 17
percent. Those fits assume a peak of 3.25 fp16 TFLOP/s, bandwidth of 9.0
GB/s, and a 0.23 ms floor. Across a broader sweep the median error is
about 31 percent, so the estimate is an ordinal placement tool rather
than an absolute-latency oracle. The same model estimates latency for
any of the 28 targets without the chip in hand. Bandwidth scales by core
count, the floor by clock. The autotuner ranks equivalent rewrites by
this estimate and preserves accuracy by default. On attention blocks it
gains 3.7 to 5.3 times at cosine similarity 1.0.

\end{summarybox}

The compiler estimates how long a layer will take before it runs. The
same model drives an ahead-of-dispatch latency estimate for any chip in
the family and a deterministic autotuner that ranks equivalent graph
rewrites.

\section{Three stages of the model}\label{three-stages-of-the-model}

Compute cycles come first, estimated from the operation and its
dimensions. For a convolution the cycle count is the input-channel
passes times the kernel volume times the output spatial extent, divided
by the active compute units, and sparsity scales it down by the fraction
of zero weights. Pooling and elementwise operations use their own cycle
variants of the same form.

Next comes a roofline. Compute time is the cycle count divided by the
throughput rate, memory time is the operand bytes divided by the
bandwidth, and the layer latency is the larger of the two. A layer whose
memory time exceeds its compute time is bandwidth-bound; the reverse is
compute-bound.

The last stage converts to wall time, a latency in microseconds. The
model runs at a clock near 0.8 times the maximum frequency, scales the
rates by a per-chip efficiency curve, and adds a fixed overhead for the
per-input transfer setup plus a constant dispatch cost.

\[t \approx \max\left(\frac{\mathrm{cycles}}{f},\ \frac{\mathrm{bytes}}{B}\right) + t_0\]

The throughput is set by the compute-unit geometry and the efficiency
curve,
\(f = \mathrm{cores} \times 4 \times \mathrm{eff}(f) \times \mathrm{clock}\),
and the bandwidth \(B\) is the DMA rate the chip sustains across the
operand stream. On the M1 the model fits the measured convolution
latencies with a peak of 3.25 fp16 TFLOP/s, bandwidth of 9.0 GB/s, and
dispatch overhead \(t_0\) of 0.23 ms, all five reference convolutions
within plus or minus 17 percent.

\section{Per-chip parameters}\label{per-chip-parameters}

Only the parameter values change from chip to chip; the three-stage
structure is identical across the family. The compute-unit count,
per-cycle divisor, clock range, and efficiency curve are read from the
hardware abstraction table for each target, the table chapter
\hyperref[hal-and-capability-gates]{24} decodes. The fitted M1 and M5
values appear side by side in \cref{tbl:c18-chip-params}.

\begin{longtable}[]{@{}llll@{}}
\caption{The cost-model parameters that change from chip to chip, with
their source and the fitted values for the M1 and
M5.}\label{tbl:c18-chip-params}\tabularnewline
\toprule\noalign{}
\textbf{Quantity} & \textbf{Source} & \textbf{M1/H13} &
\textbf{M5/H17s} \\
\midrule\noalign{}
\endfirsthead
\toprule\noalign{}
\textbf{Quantity} & \textbf{Source} & \textbf{M1/H13} &
\textbf{M5/H17s} \\
\midrule\noalign{}
\endhead
\bottomrule\noalign{}
\endlastfoot
Compute units & core-count field & 4 & 16 \\
\rowcolor{zebra}Cycle divisor & divisor field & 64 & 64 \\
Clock range & DVFS table & 0.40 to 1.43 GHz & 0.74 to 2.36 GHz \\
\rowcolor{zebra}Efficiency & frequency curve & 1.0 & about 0.84 \\
Fitted peak & silicon anchor & 3.25 fp16 TFLOP/s & 8.9 fp16 TFLOP/s \\
\rowcolor{zebra}Fitted bandwidth & silicon anchor & 9.0 GB/s & 57
GB/s \\
Dispatch floor & silicon anchor & 0.23 ms & 0.11 ms \\
\end{longtable}

The M1 model takes efficiency as 1.0, meaning it sustains peak at any
clock; later generations derate to about 0.84 because the array cannot
hold the peak multiply rate at the higher clock ceiling. The compute
units scale 4 to 16 to 32 to 64 across the die variants, and the clock
ceiling rose from 1.43 GHz to 2.36 GHz, which together put the M5
theoretical peak near 5.5 times the M1. The fitted peaks in the table
scale more conservatively, about 2.7 times, from 3.25 to 8.9 fp16
TFLOP/s, since the cost model anchors to the rate the array sustains
rather than the theoretical product.

Bandwidth scales with the compute-unit count rather than the clock. The
M5 streams the same bandwidth-bound model at 57 GB/s against the M1's
9.0 GB/s, close to the 16-to-4 core ratio, so the per-chip anchor sets
bandwidth by core count and the dispatch floor by clock. The 9.0 GB/s
here is the cost model's jointly calibrated effective fit, and it is not
the 51 GB/s single-row saturating weight-stream of chapter
\hyperref[roofline]{9} nor the roughly 40 GB/s broad-shape effective
rate, which measure different things. A model anchored to the M1 alone
over-predicts M5 latency by a mean of 99 percent: clock-scaling the M1
bandwidth yields about 15 GB/s, which the true 57 GB/s rate exceeds by
about 3.8 times. Substituting the M5 measured bandwidth, floor, and peak
brings four of the five reference convolutions within 13 percent
(\cref{lst:c18-estimate-latency}).

\begin{listing}[H]
\caption{The three-stage cost model evaluated against per-chip anchors, taking the larger of the compute and memory terms plus the dispatch floor.}\label{lst:c18-estimate-latency}

\begin{Shaded}
\begin{Highlighting}[]
\KeywordTok{def}\NormalTok{ estimate\_latency(cycles, op\_bytes, chip):}
    \CommentTok{\# Stage 1 input: cycles = input{-}channel passes * kernel volume *}
    \CommentTok{\#                output spatial extent / active compute units.}
    \CommentTok{\# Stage 2: roofline picks the binding term.}
\NormalTok{    compute\_time }\OperatorTok{=}\NormalTok{ cycles }\OperatorTok{/}\NormalTok{ chip[}\StringTok{"peak\_flops"}\NormalTok{]   }\CommentTok{\# f: cores * 4 * eff * clock}
\NormalTok{    memory\_time  }\OperatorTok{=}\NormalTok{ op\_bytes }\OperatorTok{/}\NormalTok{ chip[}\StringTok{"bandwidth"}\NormalTok{]  }\CommentTok{\# B: DMA rate the chip sustains}
    \CommentTok{\# Stage 3: add the fixed per{-}eval dispatch floor.}
    \ControlFlowTok{return} \BuiltInTok{max}\NormalTok{(compute\_time, memory\_time) }\OperatorTok{+}\NormalTok{ chip[}\StringTok{"dispatch\_floor"}\NormalTok{]}

\CommentTok{\# Fitted M1/H13 anchors: peak 3.25 fp16 TFLOP/s, B 9.0 GB/s, floor 0.23 ms.}
\NormalTok{m1 }\OperatorTok{=}\NormalTok{ \{}\StringTok{"peak\_flops"}\NormalTok{: }\FloatTok{3.25e12}\NormalTok{, }\StringTok{"bandwidth"}\NormalTok{: }\FloatTok{9.0e9}\NormalTok{, }\StringTok{"dispatch\_floor"}\NormalTok{: }\FloatTok{0.23e{-}3}\NormalTok{\}}
\CommentTok{\# A small operation collapses to the dispatch floor; a large one tracks}
\CommentTok{\# whichever of compute\_time and memory\_time binds.}
\end{Highlighting}
\end{Shaded}

\end{listing}

\section{Reading the model before
dispatch}\label{reading-the-model-before-dispatch}

The model returns a per-chip latency estimate for an output graph
without running it, which locates a layer against the roofline of
chapter \hyperref[roofline]{9} ahead of time. The estimate is the same
three-stage number described above, evaluated for a named target. For
chips that cannot be run locally it is the only available latency
figure, since the structure is identical across all 28 targets and only
the per-chip parameters differ.

Two readings of the estimate matter most. A small operation collapses to
the dispatch floor, and the floor is the optimization target: an
operation whose compute term is under 0.23 ms on the M1 gains nothing
from a faster array, so the work must grow to clear the floor. A large
operation tracks the binding roofline term, and the question is whether
the layer is compute-bound or bandwidth-bound, which decides whether
shape or streaming is the control.

Fusion removes floors and intermediate round-trips. A network run as
separate dispatches pays the floor on every one and copies each
intermediate back to the host and forward again; the same network fused
into one program pays the floor once and keeps the intermediates
resident. The model accounts for this: fusing operations removes their
separate \(t_0\) terms and the operand bytes of the eliminated
round-trips, which is why a network is compiled as one program.

\section{What the optimizer does}\label{what-the-optimizer-does}

The optimizer ranks equivalent rewrites of a graph using an op-agnostic
cost estimate and a deterministic autotuner. The cost estimate is the
analytic roofline-plus-floor number, applied uniformly to every
operation. The autotuner enumerates rewrites that compute the same
result, estimates each, caches the measured outcomes, and selects the
lowest deterministically, so the same graph yields the same choice on
every run.

The rewrites preserve accuracy by default. A route rewrite that
decomposes attention into a fused form is selected only when it computes
the same result to within tolerance, and a rewrite that trades
precision, such as an integer-quantized weight stream, is taken only
under an explicit tolerance and a speedup margin. The default never
changes the numerical result of the graph.

The reported gains follow the roofline. A fused attention route rewrite
is 3.7 to 5.3 times faster on attention blocks and on a full vision
transformer at cosine similarity 1.0, because it removes per-dispatch
floors and intermediate round-trips. The optimizer's gate is the test
corpus, which contains the cross-chip latency table as a regression so a
change to the model cannot silently move an estimate.

\section{Estimating latency and tuning a
graph}\label{estimating-latency-and-tuning-a-graph}

The estimate locates a graph against any target's roofline before
dispatch, and the autotuner selects the lowest-latency equivalent
rewrite under an accuracy bound (\cref{lst:c18-estimate-tune}).

\begin{listing}[H]
\caption{Estimating a graph’s latency for a named target with no hardware, then ranking accuracy-preserving rewrites by the same estimate.}\label{lst:c18-estimate-tune}

\begin{Shaded}
\begin{Highlighting}[]
\CommentTok{\# Estimate latency for a named target with no hardware in hand, then tune.}
\CommentTok{\# The three{-}stage estimate: compute\_time = cycles / f, memory\_time = bytes / B,}
\CommentTok{\# latency = max(compute\_time, memory\_time) + dispatch\_floor (t0).}

\NormalTok{target }\OperatorTok{=}\NormalTok{ H17s                                 }\CommentTok{\# per{-}chip f, B, t0 read from the hardware table}

\NormalTok{function estimate(graph, target):}
\NormalTok{    cycles }\OperatorTok{=} \BuiltInTok{sum}\NormalTok{ over ops of compute\_cycles(op)        }\CommentTok{\# stage 1: per{-}op cycle count}
    \BuiltInTok{bytes}  \OperatorTok{=} \BuiltInTok{sum}\NormalTok{ over ops of operand\_bytes(op)}
\NormalTok{    compute\_time }\OperatorTok{=}\NormalTok{ cycles }\OperatorTok{/}\NormalTok{ f(target)                  }\CommentTok{\# stage 2: roofline}
\NormalTok{    memory\_time  }\OperatorTok{=} \BuiltInTok{bytes}  \OperatorTok{/}\NormalTok{ B(target)}
\NormalTok{    latency }\OperatorTok{=} \BuiltInTok{max}\NormalTok{(compute\_time, memory\_time) }\OperatorTok{+}\NormalTok{ t0(target)   }\CommentTok{\# stage 3: add dispatch floor}
    \ControlFlowTok{if}\NormalTok{   latency near t0(target):       bound }\OperatorTok{=} \StringTok{"dispatch"}
    \ControlFlowTok{elif}\NormalTok{ memory\_time }\OperatorTok{\textgreater{}}\NormalTok{ compute\_time:    bound }\OperatorTok{=} \StringTok{"bandwidth"}
    \ControlFlowTok{else}\NormalTok{:                               bound }\OperatorTok{=} \StringTok{"compute"}
    \ControlFlowTok{return}\NormalTok{ latency, bound}

\NormalTok{latency, bound }\OperatorTok{=}\NormalTok{ estimate(G, target)}
\CommentTok{\# If bound == "dispatch": batch or fuse until the compute term clears the floor;}
\CommentTok{\# no faster array helps a layer pinned to the dispatch floor.}

\CommentTok{\# Tune: enumerate equivalent rewrites, estimate each, keep the lowest that holds the tolerance.}
\NormalTok{best }\OperatorTok{:=}\NormalTok{ G                                      }\CommentTok{\# the unmodified graph}
\NormalTok{best\_latency }\OperatorTok{:=}\NormalTok{ latency}
\ControlFlowTok{for}\NormalTok{ each rewrite R that computes the same result }\ImportTok{as}\NormalTok{ G:}
    \ControlFlowTok{if}\NormalTok{ max\_abs\_difference(R, G) }\OperatorTok{\textless{}=}\NormalTok{ tolerance:  }\CommentTok{\# accuracy{-}preserving only (tolerance = 0 by default)}
\NormalTok{        latency\_R, \_ }\OperatorTok{=}\NormalTok{ estimate(R, target)}
        \ControlFlowTok{if}\NormalTok{ latency\_R }\OperatorTok{\textless{}}\NormalTok{ best\_latency:}
\NormalTok{            best }\OperatorTok{:=}\NormalTok{ R}
\NormalTok{            best\_latency }\OperatorTok{:=}\NormalTok{ latency\_R}
\CommentTok{\# A fused{-}attention route rewrite is 3.7 to 5.3 times faster on attention blocks at cosine 1.0.}
\CommentTok{\# Note: attention has no analytic cost form, so its estimate is absent and must be timed on device.}
\end{Highlighting}
\end{Shaded}

\end{listing}

No faster array helps a layer pinned to the 0.23 ms floor on the M1, so
a graph whose estimate prints \texttt{dispatch} is batched or fused
until the compute term clears it.

\section{Two cost models and their
coefficients}\label{two-cost-models-and-their-coefficients}

There are two cost models in the toolchain. The model above is the
intra-engine analytic model the compiler uses, the
cycles-to-roofline-to-wall-time chain, fit to silicon for a per-chip
latency estimate. A second model decides backend placement: whether each
operation runs on the engine, central processor, or graphics processor,
and it is in the segmenter rather than the engine compiler. The
placement model costs an operation as the same roofline form,
\texttt{max(flops/peak,\ bytes/bandwidth)\ +\ launch}, but with coarse
abstract anchors rather than silicon-fit ones.

A live read of the compiler shows how the intra-engine model scores one
layer. A layer's cycles are the larger of the core compute time and the
direct-memory-access and L2 transfer time, plus the dependency stalls,
computed in \texttt{ZinEnginePerf::ComputeRunTime}. The layer total is
the execute cycles plus the overhead cycles. The tiling choice is the
argument that minimizes the split cost against the unsplit cost,
searched once and cached. \texttt{CalculateExeCycles} reads a
precomputed double at layer offset \texttt{+0x1e0} and saturates at
65535, the width of the 16-bit task-descriptor field.

The two answer different questions at different layers: the per-layer
model sizes and tiles work already bound for the engine, while the
placement model orders the three backends.

The live measurement validates the analytic roofline. The measured
full-call floor of about 190 microseconds is about 16 percent below the
frontend's analytic per-chip anchor of 220 microseconds, inside the
model's stated plus-or-minus 17 percent fit.

The placement-model anchors are per-backend, not per-chip, and each
backend is charged against an abstract roofline peak and bandwidth read
from the segmenter cost functions (\cref{tbl:c18-backend-anchors}).

\begin{longtable}[]{@{}lrr@{}}
\caption{The abstract per-backend roofline anchors the placement model
uses to order the three backends, read from the segmenter cost
functions.}\label{tbl:c18-backend-anchors}\tabularnewline
\toprule\noalign{}
\textbf{Backend} & \textbf{Peak (GFLOP/s)} & \textbf{Bandwidth
(GB/s)} \\
\midrule\noalign{}
\endfirsthead
\toprule\noalign{}
\textbf{Backend} & \textbf{Peak (GFLOP/s)} & \textbf{Bandwidth
(GB/s)} \\
\midrule\noalign{}
\endhead
\bottomrule\noalign{}
\endlastfoot
Engine & 800 & 50 \\
\rowcolor{zebra}Graphics processor & 120 & 40 \\
Central processor & 20 & 10 \\
\end{longtable}

These anchors are deliberately coarse. The 800 GFLOP/s engine peak is
about four times under the M1 silicon fit of 3.25 fp16 TFLOP/s, and the
50 GB/s engine bandwidth is above the 9 GB/s effective fit, so the two
errors run in opposite directions and the net placement order survives.
The job of these numbers is to separate the engine from the graphics
processor by 6.7 times and from the central processor by 40 times, a
separation that holds under any single miscalibration. Absolute latency
accuracy is the job of the learned per-operation layer above them.

The placement model also charges a launch cost on every segment and a
transfer cost on every backend crossing, which is why the segmenter
prefers long single-backend runs (\cref{tbl:c18-penalties}, in the
model's relative units).

\begin{longtable}[]{@{}lrr@{}}
\caption{The fixed launch and transfer penalties the placement model
charges, in the model's relative
units.}\label{tbl:c18-penalties}\tabularnewline
\toprule\noalign{}
\textbf{Penalty} & \textbf{To the engine} & \textbf{To another
backend} \\
\midrule\noalign{}
\endfirsthead
\toprule\noalign{}
\textbf{Penalty} & \textbf{To the engine} & \textbf{To another
backend} \\
\midrule\noalign{}
\endhead
\bottomrule\noalign{}
\endlastfoot
Launch cost per segment & 0.05 & 0.10 \\
\rowcolor{zebra}Transfer cost across backends & 0.09 & 0.23 \\
\end{longtable}

A transfer between two operations on the same backend costs zero, so a
long run kept on the engine pays neither the per-segment launch nor the
per-crossing transfer.

\section{Learned per-operation
leaves}\label{learned-per-operation-leaves}

Above the coarse roofline the placement model has a learned
per-operation layer: 322 regression trees, one set per platform class
and compute path, that map the coarse roofline ratio onto a calibrated
cost in nanoseconds. The trees split on a normalized feature built from
the operation flops and bytes, and the leaf is the predicted cost. The
engine trees are the small roofline-form ones, and the cost they return
is monotone in operation cost: elementwise below pooling below linear
below convolution below transposed convolution.
\Cref{tbl:c18-tree-leaves} gives representative engine cost-tree leaves,
spanning a constant-leaf elementwise operation to the largest
transposed-convolution leaf.

\begin{longtable}[]{@{}
  >{\raggedright\arraybackslash}p{(\linewidth - 4\tabcolsep) * \real{0.3000}}
  >{\raggedright\arraybackslash}p{(\linewidth - 4\tabcolsep) * \real{0.3000}}
  >{\raggedleft\arraybackslash}p{(\linewidth - 4\tabcolsep) * \real{0.4000}}@{}}
\caption{Representative learned engine cost-tree leaves, in the model's
nanosecond output.}\label{tbl:c18-tree-leaves}\tabularnewline
\toprule\noalign{}
\begin{minipage}[b]{\linewidth}\raggedright
\textbf{Operation (engine)}
\end{minipage} & \begin{minipage}[b]{\linewidth}\raggedright
\textbf{Tree form}
\end{minipage} & \begin{minipage}[b]{\linewidth}\raggedleft
\textbf{Leaf cost}
\end{minipage} \\
\midrule\noalign{}
\endfirsthead
\toprule\noalign{}
\begin{minipage}[b]{\linewidth}\raggedright
\textbf{Operation (engine)}
\end{minipage} & \begin{minipage}[b]{\linewidth}\raggedright
\textbf{Tree form}
\end{minipage} & \begin{minipage}[b]{\linewidth}\raggedleft
\textbf{Leaf cost}
\end{minipage} \\
\midrule\noalign{}
\endhead
\bottomrule\noalign{}
\endlastfoot
\texttt{relu} & constant leaf & 50 ns \\
\rowcolor{zebra}\texttt{add} & constant leaf & 26.5 to 36 ns \\
\texttt{mul} & constant leaf & 40 to 80 ns \\
\rowcolor{zebra}\texttt{reduce\_sum} & constant leaf & 22 ns \\
\texttt{max\_pool} & single-threshold leaf & 26 to 29 ns \\
\rowcolor{zebra}\texttt{conv} (main tree) & seven-split learned tree &
417.7 ns to 8609.6 ns \\
\texttt{matmul} & three-threshold tree & 116.9 ns to 66408.3 ns \\
\rowcolor{zebra}\texttt{linear} & three-threshold tree & 67 ns to 14.7
microseconds \\
\texttt{conv\_transpose} & three-threshold tree & 3.5 to 37.8
microseconds \\
\end{longtable}

Three operations have no engine tree at all, because the placement model
never puts them on the engine: gather, the recurrent cell, and the
argument-maximum reduction are central-processor or graphics-processor
placements only.

The learned leaves and the silicon-fit analytic anchors reach calibrated
accuracy by different routes. The analytic model above replaces the
coarse anchors with silicon-fit ones and brings all five reference
convolutions within plus or minus 17 percent. The placement model keeps
the coarse anchors as a binning feature and bolts a learned
per-operation tree on top. The convolution case shows both against
silicon. Four convolutions set the coarse placement roofline and the
silicon-anchored analytic estimate against the measured M1 latency
(\cref{tbl:c18-conv-compare}).

\begin{longtable}[]{@{}
  >{\raggedright\arraybackslash}p{(\linewidth - 6\tabcolsep) * \real{0.2000}}
  >{\raggedleft\arraybackslash}p{(\linewidth - 6\tabcolsep) * \real{0.2667}}
  >{\raggedleft\arraybackslash}p{(\linewidth - 6\tabcolsep) * \real{0.2667}}
  >{\raggedleft\arraybackslash}p{(\linewidth - 6\tabcolsep) * \real{0.2667}}@{}}
\caption{Coarse roofline, anchored estimate, and measured M1 latency for
four convolutions.}\label{tbl:c18-conv-compare}\tabularnewline
\toprule\noalign{}
\begin{minipage}[b]{\linewidth}\raggedright
\textbf{Convolution}
\end{minipage} & \begin{minipage}[b]{\linewidth}\raggedleft
\textbf{Coarse roofline only}
\end{minipage} & \begin{minipage}[b]{\linewidth}\raggedleft
\textbf{Analytic, h13 anchors}
\end{minipage} & \begin{minipage}[b]{\linewidth}\raggedleft
\textbf{Measured, M1}
\end{minipage} \\
\midrule\noalign{}
\endfirsthead
\toprule\noalign{}
\begin{minipage}[b]{\linewidth}\raggedright
\textbf{Convolution}
\end{minipage} & \begin{minipage}[b]{\linewidth}\raggedleft
\textbf{Coarse roofline only}
\end{minipage} & \begin{minipage}[b]{\linewidth}\raggedleft
\textbf{Analytic, h13 anchors}
\end{minipage} & \begin{minipage}[b]{\linewidth}\raggedleft
\textbf{Measured, M1}
\end{minipage} \\
\midrule\noalign{}
\endhead
\bottomrule\noalign{}
\endlastfoot
3x3 C256 to 256 at 28 & 1156 microseconds & 505 microseconds & 507
microseconds \\
\rowcolor{zebra}1x1 C512 to 512 at 32 & 671 microseconds & 511
microseconds & 444 microseconds \\
1x1 C1024 to 1024 at 16 & 671 microseconds & 569 microseconds & 686
microseconds \\
\rowcolor{zebra}1x1 C2048 to 2048 at 8 & 671 microseconds & 1210
microseconds & 1047 microseconds \\
\end{longtable}

The coarse roofline cannot even order these four: it ties the three
one-by-one convolutions and misclassifies the memory-bound deep-narrow
C2048 at 8 as compute-bound, predicting it faster than it runs. The
anchored analytic estimate reproduces the non-obvious silicon result
that the deep-narrow C2048 at 8, with about a twentieth the
multiply-accumulate count, runs about twice as slowly as the
compute-bound 3x3 C256 at 28. This is because the former is
bandwidth-bound at arithmetic intensity 60 and the latter is
compute-bound at 466.

\section{Fidelity the estimate actually
has}\label{fidelity-the-estimate-actually-has}

The plus-or-minus 17 percent fit is a property of the five reference
convolutions the M1 anchors were tuned against, not of the estimate at
large. A broad sweep of 68 graphs across eight operation families, each
estimated by the cost model for the \texttt{h13} target and timed
against the on-device per-call latency, locates the median absolute
error at about 31 percent. Only 11 of the 68 shapes are inside plus or
minus 17 percent. The estimate is sound as an ordinal placement tool: it
identifies the binding roofline term and orders shapes correctly, so the
backend choice and the relative ranking the optimizer needs survive the
error. The error is directional and concentrated in the bandwidth-bound
regime, where the 9.0 GB/s anchor undershoots the roughly 40 GB/s
effective rate the engine sustains and the estimate over-predicts the
large-weight shapes by several times.

Attention has no estimate at all. The cost model returns no estimate for
any scaled-dot-product-attention graph, because attention compiles to a
segmented plan whose native sub-programs the analytic model has no cost
form for. This is a coverage gap: the estimate is absent for attention,
so such a graph must be timed on device.

\chapter{Pitfalls and limits}\label{pitfalls-and-limits}

\begin{summarybox}

Five direct-path failure modes are silent or target-specific. An M1
last-axis-offset slice saturates above 4094. Dynamic-weight convolution
fails to compile above batch one. The saturating slice kernel is keyed
to the M1 target. Repeated failed compiles in quick succession can stall
the shared compile service. A true four-character-code image input is
not reachable on the direct path. Each is a property of the private
compiler or its data tables, so the defense is a build-time rule rather
than a runtime check. Pace compiles after a failure, roughly 15 seconds
apart, so a run of failed compiles does not stall the service.

\end{summarybox}

A few of the engine's failure modes are silent, target-specific, or
affect the shared compile service. The ones below are what a developer
building on the direct path runs into.

\section{Slice saturation hazard on the
M1}\label{slice-saturation-hazard-on-the-m1}

A last-axis slice with a nonzero start offset is not a free descriptor
edit on the M1. On that generation the offset copy routes through a
crop-DMA that stores values in a fixed-point format with four fractional
bits, an implied scale of 16. The stored value is the input times 16,
and the storage port clamps at the fp16 maximum, so any element whose
magnitude exceeds the fp16 maximum divided by 16 overflows to infinity.

The threshold is exact and given by

\[V_{\max} = \frac{65504}{16} = 4094\]

A slice element at 4094 survives, an element at 4100 becomes infinity,
and a control value of 60000 that never enters the offset copy stays
finite. The hazard is limited and easy to miss: it occurs only on the M1
family, only on a last-axis slice, only with a nonzero start offset, and
only when a value on that axis can exceed 4094. On this one axis a
cross-chip route change turns a finite number into an infinity. The A14
generation saturates on the same slice; the A15 generation and every
part above it take a plain fp16 route and do not saturate.

The consequence shows up in training. A convolution weight gradient
scaled up by a large loss-scale factor can push values past 4094 on the
M1 and silently produce infinities. Keep the values on a width-offset
slice under the bound, which for the training case means capping the
loss-scale on M1 targets, and prefer a zero start offset where the
layout allows, since a zero last-axis begin avoids the offset-DMA path
entirely. That 4094 bound on the M1 is worth asserting before the
compile (\cref{lst:c19-slice-guard}).

\begin{listing}[H]
\caption{A build-time check that a nonzero width-offset slice on the M1 stays under the 4094 saturation bound of the times-16 crop datapath.}\label{lst:c19-slice-guard}

\begin{Shaded}
\begin{Highlighting}[]
\NormalTok{V\_MAX }\OperatorTok{=} \DecValTok{65504} \OperatorTok{/} \DecValTok{16}   \CommentTok{\# = 4094.0; the fp16 maximum divided by the Q.4 gain of 16}

\KeywordTok{def}\NormalTok{ slice\_offset\_is\_safe(x, begin\_w, on\_m1):}
    \CommentTok{\# A nonzero width{-}offset slice on the M1 routes through the times{-}16 crop{-}DMA.}
    \CommentTok{\# A zero begin avoids the offset{-}DMA path entirely; on A15 and above the}
    \CommentTok{\# route is plain fp16 and never saturates.}
    \ControlFlowTok{if} \KeywordTok{not}\NormalTok{ on\_m1 }\KeywordTok{or}\NormalTok{ begin\_w }\OperatorTok{==} \DecValTok{0}\NormalTok{:}
        \ControlFlowTok{return} \VariableTok{True}
    \ControlFlowTok{return} \BuiltInTok{float}\NormalTok{(}\BuiltInTok{abs}\NormalTok{(x).}\BuiltInTok{max}\NormalTok{()) }\OperatorTok{\textless{}=}\NormalTok{ V\_MAX   }\CommentTok{\# 4094 survives, 4100 {-}\textgreater{} inf}
\end{Highlighting}
\end{Shaded}

\end{listing}

\section{Dynamic-weight convolution above batch
one}\label{dynamic-weight-convolution-above-batch-one}

A convolution whose weight is a supplied runtime tensor rather than an
embedded constant is an engine capability. At a batch of one it compiles
and runs and matches a reference convolution to a cosine of 1.0000. At a
batch of two or more the same program crashes the compiler service: the
helper process dies mid-compile and the error reads as a lost helper
application rather than a plain rejection of an unsupported operation.

The crash is specific to the dynamic-kernel path combined with a batch
above one. A constant-weight convolution at the same batch compiles
without issue, and batches of 64 are routine on that path. The
reproduction is stable across fresh restarts of the compile daemon:
batch one passes, batch two fails, every time.

Single-image dynamic-weight convolution is usable for hypernetwork and
per-sample-kernel inference where the batch is one. Batched trainable
convolution must take a different route, since the dynamic-kernel path
is closed above batch one. A build that needs the batched form should
reject a dynamic-weight convolution at batch two or more before the
program ever reaches the compiler service
(\cref{lst:c19-dynweight-guard}).

\begin{listing}[H]
\caption{A build-time guard that rejects a dynamic-weight convolution at batch two or more before it reaches the compiler service that the path crashes.}\label{lst:c19-dynweight-guard}

\begin{Shaded}
\begin{Highlighting}[]
\KeywordTok{def}\NormalTok{ guard\_dynamic\_weight\_conv(weight\_is\_runtime\_tensor, batch\_n):}
    \CommentTok{\# A const{-}weight conv shares one kernel across the batch; a dynamic kernel}
    \CommentTok{\# must be re{-}split per batch element, and that path is broken on the M1.}
    \CommentTok{\# Reject at build time so the program never reaches the compiler service.}
    \ControlFlowTok{if}\NormalTok{ weight\_is\_runtime\_tensor }\KeywordTok{and}\NormalTok{ batch\_n }\OperatorTok{\textgreater{}=} \DecValTok{2}\NormalTok{:}
        \ControlFlowTok{raise} \PreprocessorTok{ValueError}\NormalTok{(}\StringTok{"dynamic{-}weight convolution is closed above batch one"}\NormalTok{)}
\end{Highlighting}
\end{Shaded}

\end{listing}

\section{Saturating slice kernel is
target-keyed}\label{saturating-slice-kernel-is-target-keyed}

The saturating slice kernel is keyed to the target, not to the compiler
build. The same compiler binary emits a saturating slice kernel on the
M1 target and a non-saturating one on the M5 target, because the slice
lowering is specialized per hardware family.

Inside the single binary the slice conversion is a C++ template
instantiated once per family, so the same intermediate-language slice
lowers through a different compiled converter for the M1 than for the
M5. Per-target hardware-abstraction fields then drive the DMA source
path: a width granule used as both a divisor and a stride multiplier,
and a set of patch-width clamps that resolve to a power-of-two granule.
On the M1 the family converter and the format selection pick the
times-16 fixed-point DMA format for a last-axis-offset copy; on the M5
the same copy stays in plain fp16. The trigger is exactly the M1 family
plus a last-axis slice with a nonzero start offset, which is the
saturation hazard again, so the defense belongs in the per-target build
path.

\section{Pacing compiles after a
failure}\label{pacing-compiles-after-a-failure}

A failed compile is not free of side effects on the shared compile
service. A compile that fails restarts the service, which takes a few
seconds to come back, and failures that keep arriving faster than the
service can restart between them keep it from making progress, so
unrelated compiles slow down until the failures stop. The effect is a
function of how fast failures arrive, not how many occur: failures
spaced out past the restart interval cause no degradation at all. On
detecting a failed compile, wait at least one restart interval, roughly
15 seconds, before the next compile, so a burst of failures cannot
accumulate. No hard failure-count cap is needed.

\section{A single slow compile is a separate failure
mode}\label{a-single-slow-compile-is-a-separate-failure-mode}

A slow compile is its own failure mode, and the back-off rule that
handles repeated failures does nothing for it. A trainable convolution
compiled over a large batch has a compile cost that scales with the
batch, because the image-to-column tensors grow with the batch and the
tiling and partition cost grows with them. On the M1 the same backward
convolution graph compiles in about 1.9 seconds at a batch of 4, about
35 seconds at a batch of 64, about 79 seconds at a batch of 128, and
never finishes at a batch of 1000. Shrinking the spatial size does not
help, since the cost tracks the batch. The large-batch
trainable-convolution stall is the firmware-overload throttle of chapter
\hyperref[firmware]{29} in action: a submission that overruns the
firmware command queue is throttled, so the fix is to cap the batch or
split the program, not to wait it out. Real training is mini-batched at
a modest batch per step, and the same graph compiles without issue on
the M5.

A second slow-compile path is a combinatorial subgraph search. The
compiler clusters a graph by a memoized search over cut points that
collapses linear chains safely but is exponential in the width of
parallel branches that reconverge into one consumer. The search has no
iteration cap, but an internal time budget abandons it and falls back to
a cheaper partition, which on the M1 plateaus the worst measured fan-in
at about 9.2 seconds. A single cluster with roughly twelve or more
independent branches reconverging into one consumer is worth flagging at
build time, since the 9-second per-cluster cost is itself worth avoiding
and the budget bail is not guaranteed across compiler versions.

\section{Direct four-character-code image
input}\label{direct-four-character-code-image-input}

A direct image input that declares a true four-character-code
interchange format, so the engine reads a camera or video surface with
no host-side conversion, is a no-go on the unentitled direct path. The
capability is real and the syntax is fully reachable: the pixel-buffer
input type, format enum, grammar, and type rules all parse and
type-check, and the program reaches the backend. The failure is at
backend lowering: the program does not compile on the direct route.

The lowering needs setup that only the entitled model-input route
supplies, the four-character-code format descriptor backed by a surface,
which the unentitled path lacks. The supported and terminal form on the
unentitled path is the uint8 image input, which dequantizes in the graph
and saves the host-side conversion at the input. It produces output
byte-identical to a host conversion, saves roughly two milliseconds per
frame at 1080p, and avoids the unsupported operation entirely.

\section{A build-time pitfall check}\label{a-build-time-pitfall-check}

The pitfalls share one defense: gate the graph before it reaches the
compiler, keyed on the target family and the offending shape. The
estimate records the per-target hazard flags, so a saturating slice,
batched dynamic-weight convolution, or four-character-code input is
rejected at build time (\cref{lst:c19-pitfall-scan}).

\begin{listing}[H]
\caption{Scanning a graph for the five direct-path pitfalls and stopping the build before a flagged graph reaches the compiler.}\label{lst:c19-pitfall-scan}

\begin{Shaded}
\begin{Highlighting}[]
\CommentTok{\# Scan the graph for the known pitfalls and reject or rewrite before compiling, target = H13.}
\CommentTok{\# Each hazard is keyed on the target family plus an offending shape.}

\NormalTok{target }\OperatorTok{=}\NormalTok{ H13}
\NormalTok{on\_m1  }\OperatorTok{=}\NormalTok{ (family(target) }\KeywordTok{is}\NormalTok{ M1 }\KeywordTok{or}\NormalTok{ family(target) }\KeywordTok{is}\NormalTok{ M2)   }\CommentTok{\# the saturating{-}slice generations}
\NormalTok{V\_MAX  }\OperatorTok{=} \DecValTok{65504} \OperatorTok{/} \DecValTok{16}                                        \CommentTok{\# = 4094: fp16 max / 16 (Q.4 gain)}

\ControlFlowTok{for}\NormalTok{ each op }\KeywordTok{in}\NormalTok{ graph G:}

    \CommentTok{\# Pitfall 1: last{-}axis slice saturation on the M1 (finite turns to infinity above 4094)}
    \ControlFlowTok{if}\NormalTok{ op }\KeywordTok{is} \BuiltInTok{slice} \KeywordTok{and}\NormalTok{ last\_axis\_begin\_offset(op) }\OperatorTok{!=} \DecValTok{0} \KeywordTok{and}\NormalTok{ on\_m1:}
        \ControlFlowTok{if}\NormalTok{ max\_abs\_value\_on\_axis(op) }\OperatorTok{\textgreater{}}\NormalTok{ V\_MAX:}
\NormalTok{            reject(op, }\StringTok{"width{-}offset slice can saturate above 4094 on M1"}\NormalTok{)}
        \CommentTok{\# rewrite where possible: use a zero start offset to avoid the offset{-}DMA path entirely}

    \CommentTok{\# Pitfall 2: dynamic{-}weight convolution above batch one (fails to compile)}
    \ControlFlowTok{if}\NormalTok{ op }\KeywordTok{is}\NormalTok{ conv }\KeywordTok{and}\NormalTok{ weight\_is\_runtime\_tensor(op) }\KeywordTok{and}\NormalTok{ batch(op) }\OperatorTok{\textgreater{}=} \DecValTok{2}\NormalTok{:}
\NormalTok{        reject(op, }\StringTok{"dynamic{-}weight conv needs batch one; use batch one or another route"}\NormalTok{)}

    \CommentTok{\# Pitfall 3: an unsupported op (validator may accept it, code generator rejects on this target)}
    \ControlFlowTok{if} \KeywordTok{not}\NormalTok{ runs\_on\_target(op, target):}
\NormalTok{        reject(op, }\StringTok{"op does not lower on this target; decompose or choose a later family"}\NormalTok{)}

    \CommentTok{\# Pitfall 4: true four{-}char{-}code image input (does not lower on the direct path, unentitled)}
    \ControlFlowTok{if}\NormalTok{ op }\KeywordTok{is}\NormalTok{ image\_input }\KeywordTok{and} \BuiltInTok{format}\NormalTok{(op) }\KeywordTok{is}\NormalTok{ fourcc:}
\NormalTok{        reject(op, }\StringTok{"true 4CC input faults at lowering; use the uint8 in{-}graph image input instead"}\NormalTok{)}

\ControlFlowTok{if} \BuiltInTok{any}\NormalTok{ op was rejected:}
\NormalTok{    stop before compiling                                  }\CommentTok{\# never let a flagged graph compile}

\NormalTok{program P }\OperatorTok{=} \BuiltInTok{compile}\NormalTok{(G, target)                             }\CommentTok{\# reached only when the graph is clean}

\CommentTok{\# Pitfall 5 is a rate effect, not a graph shape: after any compile FAILURE, back off at least one}
\CommentTok{\# restart interval (\textasciitilde{}15 s) before the next compile, so a burst does not stall the service.}
\NormalTok{on compile\_failure:}
\NormalTok{    wait(restart\_interval)                                 }\CommentTok{\# about 15 seconds}
\end{Highlighting}
\end{Shaded}

\end{listing}

\section{Symbolic shapes are not reachable on the direct
path}\label{symbolic-shapes-are-not-reachable-on-the-direct-path}

The engine compiler has a full symbolic-shape system: it accepts a
dimension written as an unknown, type-checks two dynamic operands for
matching symbolic expressions, and enforces affine constraints on
symbolic dimensions. The silicon can thus run a program that accepts a
variable sequence length on one compile in principle. The direct compile
path does not reach it. A relu over a tensor with an unknown dimension
parses, but the compile fails with an unsupported-operation error,
because the symbolic-shape machinery is behind the entitled model-input
route that lowers enumerated or range shapes onto symbolic engine
programs.

The direct path thus compiles one program per concrete shape, and there
is no one-compile-many-shapes form. For a variable-length workload the
options are to pad to a fixed maximum length and compile once, to bucket
a small set of lengths and dispatch the nearest with the compile cache
making repeats free, or to recompile per length under the per-compile
cost. The same limit closes the native dynamic-slice and dynamic-offset
key-value-cache primitive on the direct path, so the host manages cache
state with fixed-shape windows.

\section{Reference: the five direct-path
pitfalls}\label{reference-the-five-direct-path-pitfalls}

All five are set out together in \cref{tbl:c19-pitfalls}.

\begin{longtable}[]{@{}
  >{\raggedright\arraybackslash}p{(\linewidth - 6\tabcolsep) * \real{0.2500}}
  >{\raggedright\arraybackslash}p{(\linewidth - 6\tabcolsep) * \real{0.2500}}
  >{\raggedright\arraybackslash}p{(\linewidth - 6\tabcolsep) * \real{0.2500}}
  >{\raggedright\arraybackslash}p{(\linewidth - 6\tabcolsep) * \real{0.2500}}@{}}
\caption{The five direct-path pitfalls a developer encounters, each with
its symptom, its cause, and the
workaround.}\label{tbl:c19-pitfalls}\tabularnewline
\toprule\noalign{}
\begin{minipage}[b]{\linewidth}\raggedright
\textbf{Pitfall}
\end{minipage} & \begin{minipage}[b]{\linewidth}\raggedright
\textbf{Symptom}
\end{minipage} & \begin{minipage}[b]{\linewidth}\raggedright
\textbf{Cause}
\end{minipage} & \begin{minipage}[b]{\linewidth}\raggedright
\textbf{Workaround}
\end{minipage} \\
\midrule\noalign{}
\endfirsthead
\toprule\noalign{}
\begin{minipage}[b]{\linewidth}\raggedright
\textbf{Pitfall}
\end{minipage} & \begin{minipage}[b]{\linewidth}\raggedright
\textbf{Symptom}
\end{minipage} & \begin{minipage}[b]{\linewidth}\raggedright
\textbf{Cause}
\end{minipage} & \begin{minipage}[b]{\linewidth}\raggedright
\textbf{Workaround}
\end{minipage} \\
\midrule\noalign{}
\endhead
\bottomrule\noalign{}
\endlastfoot
Slice saturation (M1) & Last-axis-offset slice silently produces
infinities above \(4094\) & Offset copy routes through a Q.4 times-16
DMA that clamps at the fp16 maximum & Keep width-offset values under
4094; cap loss-scale on M1; prefer a zero start offset \\
\rowcolor{zebra}Dynamic-weight conv at batch \(\geq 2\) & Fails to
compile & Dynamic-kernel path is closed above batch one & Use it only at
batch one; reject batch \(\geq 2\) at build time; take a different
batched route \\
Target-keyed saturating slice kernel & Same source saturates on M1,
non-saturating on M5 & Slice lowering is template-specialized per
family; M1 picks the times-16 DMA format & Treat the slice route as
M1-specific in the per-target build path \\
\rowcolor{zebra}Failed compiles in quick succession & Unrelated compiles
slow while failures keep arriving & Failures arriving faster than the
shared service can restart between them & Pace compiles after a failure
by at least one restart interval (\textasciitilde15 s) \\
Direct four-character-code image input & Does not lower on the direct
path & The entitled surface descriptor the lowering needs is unavailable
unentitled & Use the supported uint8 in-graph image input, the terminal
unentitled form \\
\end{longtable}

\section{Reference: the compile-time and capability
limits}\label{reference-the-compile-time-and-capability-limits}

Beyond the five silent or service-level pitfalls, three further limits
reject or stall a build rather than corrupt a result
(\cref{tbl:c19-compile-limits}).

\begin{longtable}[]{@{}
  >{\raggedright\arraybackslash}p{(\linewidth - 6\tabcolsep) * \real{0.2500}}
  >{\raggedright\arraybackslash}p{(\linewidth - 6\tabcolsep) * \real{0.2500}}
  >{\raggedright\arraybackslash}p{(\linewidth - 6\tabcolsep) * \real{0.2500}}
  >{\raggedright\arraybackslash}p{(\linewidth - 6\tabcolsep) * \real{0.2500}}@{}}
\caption{The compile-time and capability limits that reject or stall a
build, distinct from the silent direct-path pitfalls
above.}\label{tbl:c19-compile-limits}\tabularnewline
\toprule\noalign{}
\begin{minipage}[b]{\linewidth}\raggedright
\textbf{Limit}
\end{minipage} & \begin{minipage}[b]{\linewidth}\raggedright
\textbf{Symptom}
\end{minipage} & \begin{minipage}[b]{\linewidth}\raggedright
\textbf{Cause}
\end{minipage} & \begin{minipage}[b]{\linewidth}\raggedright
\textbf{Workaround}
\end{minipage} \\
\midrule\noalign{}
\endfirsthead
\toprule\noalign{}
\begin{minipage}[b]{\linewidth}\raggedright
\textbf{Limit}
\end{minipage} & \begin{minipage}[b]{\linewidth}\raggedright
\textbf{Symptom}
\end{minipage} & \begin{minipage}[b]{\linewidth}\raggedright
\textbf{Cause}
\end{minipage} & \begin{minipage}[b]{\linewidth}\raggedright
\textbf{Workaround}
\end{minipage} \\
\midrule\noalign{}
\endhead
\bottomrule\noalign{}
\endlastfoot
Large-batch trainable convolution & Single compile runs for minutes and
hangs on the M1 & Image-to-column tiling and partition cost scales with
the batch & Mini-batch at a modest batch per step; do not compile one
full-batch graph \\
\rowcolor{zebra}Wide reconvergent fan-in & A single cluster compile
climbs super-linearly and plateaus near 9 seconds & Memoized subgraph
cut-search is exponential in parallel-branch width; an internal time
budget bails & Flag a cluster with roughly twelve or more branches
reconverging into one consumer \\
Symbolic shape on the direct path & A tensor with an unknown dimension
parses but fails to compile & The symbolic-shape system is reachable
only through the entitled model-input route & Compile per concrete
shape; pad, bucket, or recompile per length \\
\end{longtable}

\section{Reference: the firmware fault
surface}\label{reference-the-firmware-fault-surface}

When a fault does reach the silicon rather than the compiler, the
firmware classifies it into one of a few responses, and the response
decides whether a workload is rejected, silently dropped, or recovered
through a reset (\cref{tbl:c19-fault-surface}).

\begin{longtable}[]{@{}
  >{\raggedright\arraybackslash}p{(\linewidth - 4\tabcolsep) * \real{0.3333}}
  >{\raggedright\arraybackslash}p{(\linewidth - 4\tabcolsep) * \real{0.3333}}
  >{\raggedright\arraybackslash}p{(\linewidth - 4\tabcolsep) * \real{0.3333}}@{}}
\caption{The firmware fault surface, with each fault class, its
detection mechanism, and whether the firmware rejects, drops, or
recovers through a reset.}\label{tbl:c19-fault-surface}\tabularnewline
\toprule\noalign{}
\begin{minipage}[b]{\linewidth}\raggedright
\textbf{Fault class}
\end{minipage} & \begin{minipage}[b]{\linewidth}\raggedright
\textbf{Mechanism}
\end{minipage} & \begin{minipage}[b]{\linewidth}\raggedright
\textbf{Firmware response}
\end{minipage} \\
\midrule\noalign{}
\endfirsthead
\toprule\noalign{}
\begin{minipage}[b]{\linewidth}\raggedright
\textbf{Fault class}
\end{minipage} & \begin{minipage}[b]{\linewidth}\raggedright
\textbf{Mechanism}
\end{minipage} & \begin{minipage}[b]{\linewidth}\raggedright
\textbf{Firmware response}
\end{minipage} \\
\midrule\noalign{}
\endhead
\bottomrule\noalign{}
\endlastfoot
Command integrity & Magic-word, checksum, padding, and 32-bit-address
asserts on each command & Reject the command and return an error status
to the host \\
\rowcolor{zebra}Program section & Section bounds and overlap audit at
load & Refuse to load with a sanity-check failure \\
Runtime mis-target & A call to a torn-down process or a wrong cache
state & Drop and count, no error returned \\
\rowcolor{zebra}Assertion or exception & A runtime-assert or a processor
exception & Register dump, coredump, and reset \\
Uncorrectable cache error & The L2 controller error block and its
overflow state & Coredump and reset, not a soft retry \\
\rowcolor{zebra}Task-queue watchdog & A queue that will not quiesce
within two seconds & Abort the queue \\
Host-side timeout & No completion within the deadline & Cancel
outstanding commands and re-initialize the firmware \\
\end{longtable}

The reject-versus-drop split is explicit. Integrity and section and
argument validation reject a command and return an error, so the failure
is visible to the host. The firmware instead silently drops runtime
mis-targeting, a call to a process that has been torn down or a trigger
on a cache handle in the wrong state, and counts it in a soft-failure
telemetry counter. A misbehaving client thus shows up in the drop
counters well before it faults, and a host-visible restart counter
increments each time the firmware re-initializes after a fault.

\chapter*{Interlude. Below the API: how the engine
works}\label{interlude.-below-the-api-how-the-engine-works}
\addcontentsline{toc}{chapter}{Interlude. Below the API: how the engine
works}

\begin{summarybox}

A reader who only dispatches work to the engine can stop at the end of
Part V. The back half is the mechanism beneath the programming surface.
Read it to reason about a numeric result, to predict an unmeasured chip,
or to debug a compile failure.

\end{summarybox}

Parts I through V describe the engine as a developer uses it: what it
computes, how to reach it, how it performs, what it suits, and how to
fit work to it. That account stops at the programming surface. The rest
of the guide goes beneath that surface, to the chip and the system
around it.

The back half is written at a lower level than the front half, and shows
the structures, registers, and command protocol directly.

Four back-half parts and the back matter divide the work
(\cref{tbl:handoff-backhalf}).

\renewcommand{\thetable}{I.\arabic{table}}\setcounter{table}{0}

\begin{longtable}[]{@{}
  >{\raggedright\arraybackslash}p{(\linewidth - 4\tabcolsep) * \real{0.3333}}
  >{\raggedright\arraybackslash}p{(\linewidth - 4\tabcolsep) * \real{0.3333}}
  >{\raggedright\arraybackslash}p{(\linewidth - 4\tabcolsep) * \real{0.3333}}@{}}
\caption{The back-half parts and the back matter, each with its subject
and the reader it serves.}\label{tbl:handoff-backhalf}\tabularnewline
\toprule\noalign{}
\begin{minipage}[b]{\linewidth}\raggedright
\textbf{Part}
\end{minipage} & \begin{minipage}[b]{\linewidth}\raggedright
\textbf{Subject}
\end{minipage} & \begin{minipage}[b]{\linewidth}\raggedright
\textbf{For the reader who needs}
\end{minipage} \\
\midrule\noalign{}
\endfirsthead
\toprule\noalign{}
\begin{minipage}[b]{\linewidth}\raggedright
\textbf{Part}
\end{minipage} & \begin{minipage}[b]{\linewidth}\raggedright
\textbf{Subject}
\end{minipage} & \begin{minipage}[b]{\linewidth}\raggedright
\textbf{For the reader who needs}
\end{minipage} \\
\midrule\noalign{}
\endhead
\bottomrule\noalign{}
\endlastfoot
VI. The Silicon & The datapath and MAC geometry, and the memory
hierarchy & The mechanism behind the numerics and the roofline \\
\rowcolor{zebra}VII. The Toolchain and Encoding & The compiler, the
program and container format, the HAL and its gates, the compression
pipeline, and direct netplist authoring & The mechanism behind
compilation, the program format, capability gating, and compression \\
VIII. System Internals & The kernel driver and its ABI, the
address-translation unit, the firmware, the host-to-firmware command
protocol, power and thermal, security and isolation, and telemetry & The
path from a host call to the silicon, and what is and is not
observable \\
\rowcolor{zebra}IX. Cross-Silicon Reference & The full target set,
per-family code generation, and the upper tier & To compile for and
reason about chips other than the one in hand \\
Back matter & The methodology, the open questions, and the reference
appendices & To trust the numbers and to find a specific value \\
\end{longtable}

\renewcommand{\thetable}{\thechapter.\arabic{table}}

\anepartopen{Part VI}{The Silicon}
\anechap{20}{Datapath and MAC geometry}{The multiply array, the output-channel groups, and the per-core tiling.}
\anechap{21}{Memory hierarchy}{The on-chip working set, the 2 MB threshold, and the streaming boundary.}
\anepartclose

\chapter{Datapath and MAC geometry}\label{datapath-and-mac-geometry}

\begin{summarybox}

The multiply-accumulate array is four cores on the M1, each with an
eight-deep accumulator file, reading every geometry constant from a
per-chip hardware-abstraction table. The lane width emits eight output
channels per cycle in the int8 fast path and four in the default fp16
path. Output channels are the dimension that splits across cores and
across accumulator passes. Each core reduces through a radix-4 tree of
fp16-rounded tiles into one wide running accumulator of fp32 class,
rounding to fp16 only at the output port. The firmware has no
convolution, Winograd, or systolic code: it is the real-time kernel and
the tile, kernel, and output direct-memory-access plumbing, and nothing
else.

\end{summarybox}

The array is a set of cores, accumulators, and lanes whose dimensions
are constants in the compiler and in a per-chip hardware-abstraction
table. The compiler supplies it by re-basing direct-memory-access
engines per tile. Those dimensions and the roofline that follows from
them are in \cref{tbl:c20-geometry}.

\section{Geometry constants}\label{geometry-constants}

\begin{longtable}[]{@{}llll@{}}
\caption{The multiply-accumulate array geometry constants, with
hardware-abstraction-table offset, M1 value, and the scope over which
each holds.}\label{tbl:c20-geometry}\tabularnewline
\toprule\noalign{}
\textbf{quantity} & \textbf{table offset} & \textbf{M1 value} &
\textbf{scope} \\
\midrule\noalign{}
\endfirsthead
\toprule\noalign{}
\textbf{quantity} & \textbf{table offset} & \textbf{M1 value} &
\textbf{scope} \\
\midrule\noalign{}
\endhead
\bottomrule\noalign{}
\endlastfoot
core count & \texttt{0x238} & 4 & per die variant \\
\rowcolor{zebra}accumulator budget per core & \texttt{0x0c} & 8 & all
chips \\
accumulator-split granule & \texttt{0x230} & 64 & all chips \\
\rowcolor{zebra}performance cycle divisor & \texttt{0x228} & 64 & A12
and later \\
output channels per cycle, fp16 & accessor table & 4 & architectural \\
\rowcolor{zebra}output channels per cycle, int8 & accessor table & 8 &
architectural \\
patch-width floor, log2 & \texttt{0x400} & 4 & M1 \\
\rowcolor{zebra}patch-width cap, log2 & \texttt{0x410} & 9 & M1 \\
working-set cap & \texttt{0x1b8} & 2 MB & M1 \\
\end{longtable}

The compiler's performance model reads the array geometry from these
fields in the per-chip hardware-abstraction table. The core count is
keyed to the die rather than to the generation. The decoded four-core
figure on the M1 counts physical compute sets, which the per-core
power-rail step confirms below. Apple's published per-chip figure, the
16-core count the I/O registry also reports, is a different quantity
\hyperref[ref-appleane]{{[}AppleANE{]}}. The M1 has four cores; the M5
generation has sixteen; small reference variants have one. The
accumulator budget of eight, uniform across every chip, is the depth of
the accumulator file per core in work units. An operation fits
single-buffered when its accumulator demand is at most eight, and
double-buffered when twice its demand is at most eight, so
double-buffering holds up to a demand of four.

\section{Output channels per cycle and the lane
width}\label{output-channels-per-cycle-and-the-lane-width}

Two accessors of the performance model set the lane width
(\cref{tbl:c20-lane-accessors}).

\begin{longtable}[]{@{}
  >{\raggedright\arraybackslash}p{(\linewidth - 4\tabcolsep) * \real{0.3333}}
  >{\raggedright\arraybackslash}p{(\linewidth - 4\tabcolsep) * \real{0.3333}}
  >{\raggedright\arraybackslash}p{(\linewidth - 4\tabcolsep) * \real{0.3333}}@{}}
\caption{The two lane-width accessors of the performance model, with
what each returns and the input that selects the
value.}\label{tbl:c20-lane-accessors}\tabularnewline
\toprule\noalign{}
\begin{minipage}[b]{\linewidth}\raggedright
\textbf{accessor}
\end{minipage} & \begin{minipage}[b]{\linewidth}\raggedright
\textbf{returns}
\end{minipage} & \begin{minipage}[b]{\linewidth}\raggedright
\textbf{selected by}
\end{minipage} \\
\midrule\noalign{}
\endfirsthead
\toprule\noalign{}
\begin{minipage}[b]{\linewidth}\raggedright
\textbf{accessor}
\end{minipage} & \begin{minipage}[b]{\linewidth}\raggedright
\textbf{returns}
\end{minipage} & \begin{minipage}[b]{\linewidth}\raggedright
\textbf{selected by}
\end{minipage} \\
\midrule\noalign{}
\endhead
\bottomrule\noalign{}
\endlastfoot
\texttt{GetNumOutputChannelsPerCycle} & 8 (int8 fast path), 4 (fp16
default), 2 or 1 (narrow modes) & data type and mode \\
\rowcolor{zebra}\texttt{GetNumOutputChannelsPerAccumulator} & 8 / 4 / 2
/ 1 (table below) & source-patch size \\
\end{longtable}

The second accessor selects among four per-accumulator channel counts by
source-patch size, each held at its own hardware-abstraction-table
offset (\cref{tbl:c20-oc-per-acc}).

\begin{longtable}[]{@{}lll@{}}
\caption{Output channels per physical accumulator slot, the
hardware-abstraction-table offset holding each value, and the
source-patch size that selects
it.}\label{tbl:c20-oc-per-acc}\tabularnewline
\toprule\noalign{}
\textbf{HAL offset} & \textbf{output channels per accumulator} &
\textbf{selected when} \\
\midrule\noalign{}
\endfirsthead
\toprule\noalign{}
\textbf{HAL offset} & \textbf{output channels per accumulator} &
\textbf{selected when} \\
\midrule\noalign{}
\endhead
\bottomrule\noalign{}
\endlastfoot
\texttt{0x3a8} & 8 & tiny-source mode (small-source-mode field reads
2) \\
\rowcolor{zebra}\texttt{0x3b0} & 4 & small-source mode
(small-source-mode field reads 1) \\
\texttt{0x3b8} & 2 & default, format 3 / fp16-packed \\
\rowcolor{zebra}\texttt{0x3c0} & 1 & default, non-format-3 and not
half-work-unit \\
\end{longtable}

Each of the four values is its own field in the hardware-abstraction
table, at offsets \texttt{0x3a8}, \texttt{0x3b0}, \texttt{0x3b8}, and
\texttt{0x3c0}, and \texttt{GetNumOutputChannelsPerAccumulator} returns
the one the operation's source-patch mode selects. The values are
uniform across every chip in the set, so the multiplexing law is
architectural, not per-die. The source-patch mode is set by
\texttt{ComputeSmallSourceMode}, which classifies an operation as
default, small, or tiny from the output tensor dimensions against the
kernel: tiny mode requires a source of at least 2 that fits the field at
\texttt{0x338}, which is 4 on the M1.

The lane width is the column dimension of the systolic tile.
\texttt{GetNumOutputChannelsPerCycle} sets the per-cycle count: eight in
the int8 fast path, four in the default fp16 path, and two or one for
narrow modes. One core streams an input patch and emits up to that many
output channels per cycle, the eight-wide case being double-int8 mode.
\texttt{GetNumOutputChannelsPerAccumulator} sets how many output
channels time-share one physical accumulator slot. A tiny input patch
underuses the array spatially, so the compiler packs eight output
channels onto one accumulator to keep the multipliers busy; with a large
patch each output channel keeps its own accumulator.

\section{Radix-4 reduction and wide
accumulator}\label{radix-4-reduction-and-wide-accumulator}

The reduction that supplies one accumulator runs in two stages. The
first stage groups input lanes into tiles of four, each rounded to fp16.
Those fp16-rounded tiles then feed a single wide running accumulator of
fp32 class, with the rounding back to fp16 deferred to the output port.

The fan-in of four comes from measurement rather than from the table. A
reduction of \texttt{{[}+B,\ -B,\ +1{]}} triples, repeated sixteen times
so that the true sum is sixteen, returns sixteen survivors below a
partial magnitude of 4096 and saturates to exactly four survivors for
every \texttt{B} at or above 4096. The saturation count of four is the
radix-4 tile signature: once a tile has a partial at or above 4096, the
unit increments sharing that tile fall below half of the fp16 spacing
and vanish, and the three-element triple period beats against the
four-lane tile to leave four unaffected lanes. The threshold is at 4096
because that is where the fp16 spacing first reaches four. The result is
layout-independent: the \texttt{{[}+B,\ -B,\ +1{]}} and
\texttt{{[}+B,\ +1,\ -B{]}} orderings return byte-identical values, so
the hardware reduction order is a fixed lattice over lane index.

The cross-tile accumulator is wider than fp16. A reduction of one value
of 4096 followed by 1024 ones returns 5116, between the naive-fp16
result 4096 and the exact 5120, and sixteen unit increments all survive
next to a partial of 8192 where an fp16 accumulator would drop most of
them. The radix-4 first stage is consistent with the eight-work-unit
accumulator file, since four input lanes plus margin fit the per-core
budget.

\section{Output-channel-group tiling}\label{output-channel-group-tiling}

The compiler tiles output channels into output-channel groups sized to
the accumulator file. \texttt{ComputeMaxOcgSize} derives the group size
from the accumulator budget, the kernel-element count, and a per-format
byte cap.

\[\mathrm{OCG} = \min\!\left( \mathrm{floor\_pow2}\!\left( \frac{8}{k_W \, k_H \, k_D} \right), \; \mathrm{byte\_cap} \right),\]

where the byte cap is 32, 16, or 8 bytes per kernel element depending on
the weight format, read from the hardware-abstraction table at offset
\texttt{0x388}, \texttt{0x390}, or \texttt{0x398}. In compiler terms,
\texttt{ComputeMaxOcgSize} reads the accumulator budget at field
\texttt{0x0c}, divides it by the kernel-element count, rounds down to a
power of two, and clamps to the per-format byte cap, as
\cref{lst:c20-ocg-size} gives.

\begin{listing}[H]
\caption{How the compiler computes the output-channel-group size for one pass from the accumulator budget, kernel-element count, and per-format byte cap.}\label{lst:c20-ocg-size}

\begin{Shaded}
\begin{Highlighting}[]
\CommentTok{\# ComputeMaxOcgSize: output{-}channel{-}group size for one pass}
\NormalTok{acc\_per\_oc }\OperatorTok{=}\NormalTok{ GetNumOutputChannelsPerAccumulator(mode)   }\CommentTok{\# 1, 2, 4, or 8}
\NormalTok{budget     }\OperatorTok{=}\NormalTok{ floor\_pow2( HAL[}\BaseNTok{0x0c}\NormalTok{] }\OperatorTok{/}\NormalTok{ (kW }\OperatorTok{*}\NormalTok{ kH }\OperatorTok{*}\NormalTok{ kD) )    }\CommentTok{\# HAL[0x0c] = 8 accumulators}
\NormalTok{byte\_cap   }\OperatorTok{=}\NormalTok{ HAL[ocg\_cap\_offset(}\BuiltInTok{format}\NormalTok{)] }\OperatorTok{/}\NormalTok{ (kW }\OperatorTok{*}\NormalTok{ kH }\OperatorTok{*}\NormalTok{ kD)   }\CommentTok{\# 0x388/0x390/0x398 = 32/16/8 B/elem}
\NormalTok{OCG        }\OperatorTok{=} \BuiltInTok{min}\NormalTok{( ComputeMaxOcg(budget, acc\_per\_oc), byte\_cap )}
\end{Highlighting}
\end{Shaded}

\end{listing}

A 1-by-1 convolution has a kernel-element count of one, so it admits a
large group. At a kernel-element count of nine, a 3-by-3 convolution
gets a group roughly nine times smaller and needs more passes over the
input. The compiler relieves that pressure by selecting Winograd for
dense 3-by-3 stride-1 convolutions, since the transform cuts the
effective kernel-element count.

Once the output-channel count exceeds one output-channel group, the
array re-streams the input for a second pass. The number of passes is

\[\mathrm{OCG\ passes} = \left\lceil \frac{C_{out}}{\mathrm{OCG}} \right\rceil .\]

For a 1-by-1 fp16 convolution at a 32-by-32 spatial size, the per-layer
slope shows a distinct super-linear jump in cost between an
output-channel count of 192 and 256: a step from 20 microseconds to 61
microseconds per layer, a roughly threefold cost step for a 1.33-fold
increase in arithmetic. That step is the pass count incrementing from
one to two as the group cap is reached. A 3-by-3 convolution reaches the
same threshold at fewer output channels per pass, matching the nine-fold
accumulator pressure.

\section{Winograd selection gate}\label{winograd-selection-gate}

The eligibility gate is \texttt{CanUseWinogradMode}, a fixed conjunction
of conditions that all must hold (\cref{tbl:c20-winograd-gate}).

\begin{longtable}[]{@{}
  >{\raggedright\arraybackslash}p{(\linewidth - 2\tabcolsep) * \real{0.5000}}
  >{\raggedright\arraybackslash}p{(\linewidth - 2\tabcolsep) * \real{0.5000}}@{}}
\caption{The conditions of the Winograd eligibility gate, each of which
must hold for the mode to be
considered.}\label{tbl:c20-winograd-gate}\tabularnewline
\toprule\noalign{}
\begin{minipage}[b]{\linewidth}\raggedright
\textbf{condition}
\end{minipage} & \begin{minipage}[b]{\linewidth}\raggedright
\textbf{meaning}
\end{minipage} \\
\midrule\noalign{}
\endfirsthead
\toprule\noalign{}
\begin{minipage}[b]{\linewidth}\raggedright
\textbf{condition}
\end{minipage} & \begin{minipage}[b]{\linewidth}\raggedright
\textbf{meaning}
\end{minipage} \\
\midrule\noalign{}
\endhead
\bottomrule\noalign{}
\endlastfoot
\texttt{HAL{[}0x680{]}} bit 0 set & the per-chip Winograd-enable bit,
present on the M1 \\
\rowcolor{zebra}kernel underlying type not 5, not 2, not unity & the
format is Winograd-eligible and the weights are not 1-by-1 or
identity \\
tensor format not 12 & the input format class is eligible \\
\rowcolor{zebra}one kernel axis equals 3 & the transformed axis is fixed
at 3, so 3-by-3 only \\
the orthogonal kernel axis is below 6 & the input-tile width of the
larger tile is 6 \\
\rowcolor{zebra}no third (depth) kernel extent & two-dimensional
convolution \\
unit stride & input and output extents are consistent with stride 1 \\
\rowcolor{zebra}\texttt{OCG\ x\ kH\ x\ kW\ x\ kD\ x\ 2} clears the work
threshold & enough work to amortize the transform \\
\end{longtable}

The axis-equals-3 plus orthogonal-axis-below-6 pair pins the supported
tile set to two forms: \(F(2 \times 2, 3 \times 3)\), with an output
tile of 2 and an input tile of 4, and \(F(4 \times 4, 3 \times 3)\),
with an output tile of 4 and an input tile of 6. The work threshold on
the term \(\mathrm{OCG} \times k_H \times k_W \times k_D \times 2\) is
precision-dependent: it is 32 for work-unit modes 1 and 2, 8 for a
non-float kernel, and 16 for a float kernel. The higher float threshold
is the compiler's guard against the precision cost of the
\(F(4 \times 4, 3 \times 3)\) transform, since the float convolution
must clear a larger work bar before the transform is taken. No
accumulator widening is tied to Winograd; the precision safety comes
from the eligibility and threshold gating instead.

Eligibility is not selection. Even when the gate passes, Winograd is
taken only when it beats both direct convolution and the
accumulator-double-buffering path, decided by a cost-model comparison
that reads the same accumulator and lane-width accessors. Winograd is
rejected when the convolution uses unicast, when the weight is sparse
(sparse weights keep their own datapath and skip zeros), or when
double-buffering the accumulator file already saturates the array. The
transform matrices themselves live in the engine: the compiler sets a
single \texttt{winograd\_mode} config field, and the input, filter, and
output transforms run inside the engine datapath. The textbook
\(F(2 \times 2, 3 \times 3)\) and \(F(4 \times 4, 3 \times 3)\) forms
describe the behavior, but the coefficients are resident in hardware and
not recoverable from the program.

\section{Four-core granule}\label{four-core-granule}

Output channels are the dimension the compiler splits across cores.
\texttt{ZinMirNECoreAssignment} builds a strided round-robin map: core
\(k\) owns output channels \(\{k, k + N, k + 2N, \ldots\}\) for a core
count \(N\). On the M1 with four cores, channel \(c\) is on core

\[\mathrm{core}(c) = c \bmod 4 .\]

The active-core count is not something the developer sets. The compiler
sets it from the output-channel count and the group size, rounds it up
to the next power of two, and writes it into the task descriptor.
\texttt{GetNumNeededNEsNextPow2} sets the count
(\cref{lst:c20-core-map}).

\begin{listing}[H]
\caption{The strided round-robin channel-to-core map alongside the pass count and active-core count, both driven by the output-channel count and the group size.}\label{lst:c20-core-map}

\begin{Shaded}
\begin{Highlighting}[]
\CommentTok{\# ZinMirNECoreAssignment: strided round{-}robin channel{-}to{-}core map (M1: num\_nes = 4)}
\KeywordTok{def}\NormalTok{ core(c, num\_nes):}
    \ControlFlowTok{return}\NormalTok{ c }\OperatorTok{\%}\NormalTok{ num\_nes                       }\CommentTok{\# channel c {-}\textgreater{} core c mod 4 on the M1}

\CommentTok{\# OCG passes and active{-}core count, both driven by C\_out and the OCG size}
\KeywordTok{def}\NormalTok{ ocg\_passes(c\_out, ocg):}
    \ControlFlowTok{return}\NormalTok{ ceil(c\_out }\OperatorTok{/}\NormalTok{ ocg)                 }\CommentTok{\# input re{-}streamed once per pass}

\KeywordTok{def}\NormalTok{ cores\_needed(c\_out, ocg, num\_nes):       }\CommentTok{\# num\_nes = HAL[0x238] = 4 on the M1}
    \ControlFlowTok{return} \BuiltInTok{min}\NormalTok{(num\_nes, next\_pow2(ceil(c\_out }\OperatorTok{/}\NormalTok{ ocg)))}
\end{Highlighting}
\end{Shaded}

\end{listing}

\Cref{tbl:c20-core-scaling} reports the measured throughput as the
active-core count rises from one to four, showing near-linear integer
scaling of the single-core rate.

\begin{longtable}[]{@{}llll@{}}
\caption{Measured throughput as the active-core count rises from one to
four.}\label{tbl:c20-core-scaling}\tabularnewline
\toprule\noalign{}
\textbf{output channels} & \textbf{cores lit} & \textbf{throughput,
GMAC/s} & \textbf{ratio to one core} \\
\midrule\noalign{}
\endfirsthead
\toprule\noalign{}
\textbf{output channels} & \textbf{cores lit} & \textbf{throughput,
GMAC/s} & \textbf{ratio to one core} \\
\midrule\noalign{}
\endhead
\bottomrule\noalign{}
\endlastfoot
1 & 1 & 3.8 & 1.00 \\
\rowcolor{zebra}2 & 2 & 7.6 & 2.00 \\
3 & 3 & 11.4 & 3.00 \\
\rowcolor{zebra}4 & 4 & 15.4 & 4.05 \\
\end{longtable}

Driving a single 1-by-1 convolution at each output-channel count below
the core count, in a dispatch loop pinned at a fixed rate by
per-dispatch overhead, the sustained throughput rises in exact integer
multiples of the single-core rate. Each added core does one more output
channel's worth of multiply-accumulates in the same wall time. The power
rail confirms the count independently: the engine draws a step of about
10 milliwatts per added core over an always-on floor of about 800
milliwatts, matching the firmware's one always-on base domain plus four
independently power-gated compute sets, one per core.

A live capture of the compiler during a real M1 compile pins the
geometry it costs against (\cref{tbl:c20-live-geometry}).

\begin{longtable}[]{@{}ll@{}}
\caption{Costed-geometry constants captured live during an M1 compile,
with the active-engine count and the candidate split
geometries.}\label{tbl:c20-live-geometry}\tabularnewline
\toprule\noalign{}
\textbf{quantity} & \textbf{live M1 value} \\
\midrule\noalign{}
\endfirsthead
\toprule\noalign{}
\textbf{quantity} & \textbf{live M1 value} \\
\midrule\noalign{}
\endhead
\bottomrule\noalign{}
\endlastfoot
clusters costed & 1 \\
\rowcolor{zebra}active engines costed
(\texttt{GetTotalNumberOfActiveNEs}) & 8 \\
candidate split geometries speculated & 1, 4, and 16 engines \\
\rowcolor{zebra}kernel-memory budget & 64 units \\
fixed per-layer overhead & 44 cycles \\
\end{longtable}

The 8 active engines the compiler costs against differ from the 16 cores
the I/O registry reports for the M1: the costed geometry is 8 active
engines on 1 cluster, a model-internal figure rather than the registry
core count or the four physical compute sets that the power rail
confirms above. The model also sets a 64-unit kernel-memory budget that
weight residency must fit under, and a 44-cycle fixed per-layer overhead
that is added to every layer's execute cycles.

\begin{figure}[H]\centering
\includegraphics[width=4.627in]{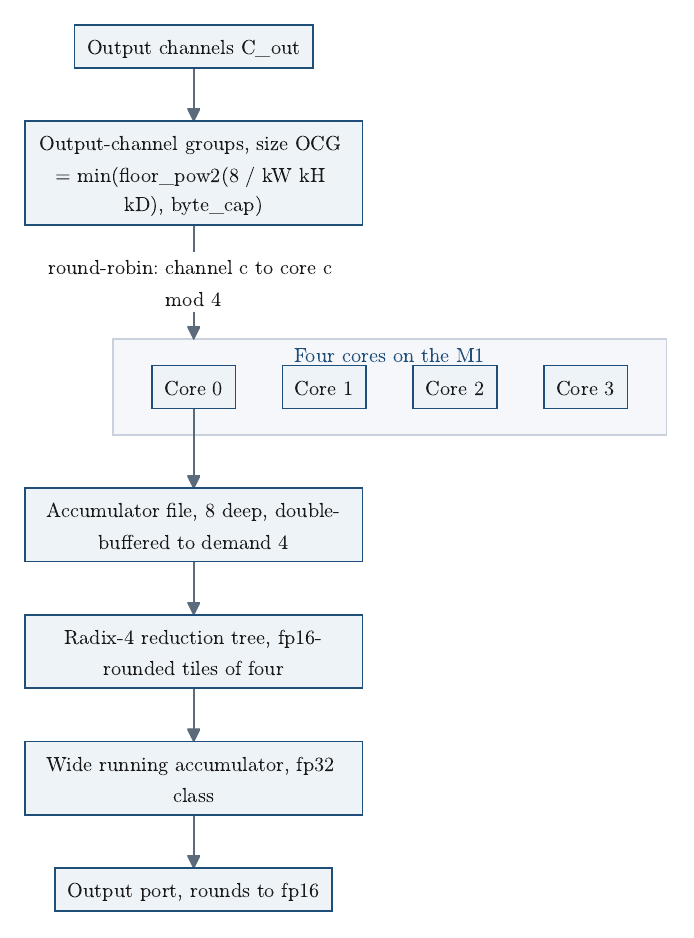}
\caption{The multiply-accumulate datapath, with output-channel groups tiled across four cores, each core holding a wide accumulator fed by a radix-4 reduction tree.}
\label{fig:c20-datapath}
\end{figure}

The full datapath reads top to bottom from the output-channel groups
down through the four cores to each core's accumulator and its radix-4
reduction, as \cref{fig:c20-datapath} draws it.

\section{Roofline from the geometry}\label{roofline-from-the-geometry}

Peak fp16 throughput is the product of the core count, lane width, and
clock. Writing the lane width as output channels per cycle, the peak
rate of multiply-accumulate operations is

\[P_{\mathrm{MAC}} = \mathrm{cores} \times \mathrm{lanes} \times f,\]

and the floating-point rate is twice that, since one multiply-accumulate
is two floating-point operations. On the M1 the best fp16 mode is four
cores times four output channels per cycle, and the int8 fast path
doubles the lane width to eight. The measured all-four-cores-saturated
rate for a large 1-by-1 convolution is about 3.48 trillion
multiply-accumulates per second, about 7 fp16 TFLOP/s at the engine
output: above the 4.8 fp16 TFLOP/s large-matmul saturating ceiling of
chapter \hyperref[roofline]{9}, because this convolution stays under the
2 MB working-set threshold and remains compute-bound, and below the
overhead-subtracted roofline anchor of 12 fp16 TFLOP/s. The
full-saturation power at that rate is about 4.7 watts, which is on the
convolution power anchor and within 0.3 watts of the 4.4-watt
large-matrix-multiply anchor, so the rail readings are calibrated.

The frontend does not reach that doubled lane. The int8 compile flag
quantizes the weights and leaves the multiply-accumulate in fp16, so it
halves the streamed weight bytes and leaves the eight-channel int8
compute path unused. A 3-by-3 convolution at 256 input and output
channels and a square matmul both run within a few percent of fp16. The
int8 path reaches about 1.5 times faster only where the weight is large
enough to stream from main memory, near a 4096-by-4096 weight at a batch
of 256 or more.

The working-set cap closes the roofline at the memory edge. The largest
single operand that stays on chip is 2 MB. Past that, the operand is
tiled and streamed from main memory, which adds traffic and moves the
workload onto the bandwidth side of the roofline.

\section{Task descriptor that holds the
geometry}\label{task-descriptor-that-holds-the-geometry}

The compiler reaches the array through a per-partition task descriptor,
a flat register image organized into seven register groups, and
serializes each group into the address-value pairs the firmware writes.
On the M1 the descriptor is the version-10 layout, one of fourteen
versioned descriptor structs the compiler builds, selected by chip
family. \Cref{tbl:c20-descriptor-groups} lists the seven register groups
of that layout.

\begin{longtable}[]{@{}
  >{\raggedright\arraybackslash}p{(\linewidth - 6\tabcolsep) * \real{0.2500}}
  >{\raggedright\arraybackslash}p{(\linewidth - 6\tabcolsep) * \real{0.2500}}
  >{\raggedright\arraybackslash}p{(\linewidth - 6\tabcolsep) * \real{0.2500}}
  >{\raggedright\arraybackslash}p{(\linewidth - 6\tabcolsep) * \real{0.2500}}@{}}
\caption{The seven register groups of the version-10 task descriptor,
each with its register count, address base, and
contents.}\label{tbl:c20-descriptor-groups}\tabularnewline
\toprule\noalign{}
\begin{minipage}[b]{\linewidth}\raggedright
\textbf{group}
\end{minipage} & \begin{minipage}[b]{\linewidth}\raggedright
\textbf{register count}
\end{minipage} & \begin{minipage}[b]{\linewidth}\raggedright
\textbf{address base}
\end{minipage} & \begin{minipage}[b]{\linewidth}\raggedright
\textbf{contents}
\end{minipage} \\
\midrule\noalign{}
\endfirsthead
\toprule\noalign{}
\begin{minipage}[b]{\linewidth}\raggedright
\textbf{group}
\end{minipage} & \begin{minipage}[b]{\linewidth}\raggedright
\textbf{register count}
\end{minipage} & \begin{minipage}[b]{\linewidth}\raggedright
\textbf{address base}
\end{minipage} & \begin{minipage}[b]{\linewidth}\raggedright
\textbf{contents}
\end{minipage} \\
\midrule\noalign{}
\endhead
\bottomrule\noalign{}
\endlastfoot
kernel and common & 34 & \texttt{0x5500} & kernel-DMA enable, format,
stride, common config, task type, output transpose, network id \\
\rowcolor{zebra}dimensions & 19 & none & input and output width, height,
depth, channels, group count, broadcast, transpose, interleave,
output-channel-group size \\
tile DMA & 69 & \texttt{0x4d00} & the three tile-DMA engines: enable,
cache hints, base-address halves, row, channel, depth, group strides,
format, wrap \\
\rowcolor{zebra}elementwise, planar engine, padding & 30 &
\texttt{0x4100} & elementwise and planar-engine config, padding mode,
planar-engine index \\
L2 and texture & 14 & \texttt{0x4500} & L2 source and result config,
texture mode, source dimensions \\
\rowcolor{zebra}kernel format and op mode & 11 & \texttt{0x4900} & op
mode, kernel alignment, sparse and palette flags, padding constant, bias
and post-scale enables \\
L2 result & 21 & \texttt{0x5100} & L2-result base, strides, wrap, result
format \\
\end{longtable}

The dimension group encodes the per-axis caps directly in its field
widths. Input width, height, and depth are 15-bit fields masked
\texttt{0x7fff}, while the channel fields are 17-bit, masked
\texttt{0x1ffff}, so a channel axis reaches 131071 where a spatial axis
stops at 32767. The output-channel-group size is a 3-bit field, and the
active-core count is the \texttt{ActiveNE} field, a 3-bit value that
records how many cores the task runs on, set from the shape by
\texttt{GetNumNeededNEsNextPow2}.

A fused convolution folds its per-channel affine and its activation into
four separate kernel-coefficient streams, each a distinct sub-buffer
with its own base offset and a relocation slot the loader patches at
load. Each of the four streams has its own relocation register
(\cref{tbl:c20-coeff-streams}).

\begin{longtable}[]{@{}
  >{\raggedright\arraybackslash}p{(\linewidth - 4\tabcolsep) * \real{0.3333}}
  >{\raggedright\arraybackslash}p{(\linewidth - 4\tabcolsep) * \real{0.3333}}
  >{\raggedright\arraybackslash}p{(\linewidth - 4\tabcolsep) * \real{0.3333}}@{}}
\caption{The four kernel-coefficient streams a fused convolution emits,
each with the relocation register the loader patches and the role it
holds.}\label{tbl:c20-coeff-streams}\tabularnewline
\toprule\noalign{}
\begin{minipage}[b]{\linewidth}\raggedright
\textbf{stream}
\end{minipage} & \begin{minipage}[b]{\linewidth}\raggedright
\textbf{relocation register}
\end{minipage} & \begin{minipage}[b]{\linewidth}\raggedright
\textbf{role}
\end{minipage} \\
\midrule\noalign{}
\endfirsthead
\toprule\noalign{}
\begin{minipage}[b]{\linewidth}\raggedright
\textbf{stream}
\end{minipage} & \begin{minipage}[b]{\linewidth}\raggedright
\textbf{relocation register}
\end{minipage} & \begin{minipage}[b]{\linewidth}\raggedright
\textbf{role}
\end{minipage} \\
\midrule\noalign{}
\endhead
\bottomrule\noalign{}
\endlastfoot
bias & \texttt{0x1554} & the per-channel additive offset \\
\rowcolor{zebra}post-scale & \texttt{0x1558} & the per-channel output
multiply, where a dequantize scale folds \\
palette lookup & \texttt{0x155c} & the palettized-weight lookup table \\
\rowcolor{zebra}activation lookup & \texttt{0x1560} & the 33-segment
piecewise-linear activation table \\
\end{longtable}

A convolution with batch normalization and a following activation thus
stays one operation: the scale and bias fold into the post-scale and
bias streams, the activation runs as the activation-lookup stream, and
one fused operation streams all four coefficient banks alongside its
weights.

\section{Cross-source corroboration}\label{cross-source-corroboration}

The accumulator and output-channel-group geometry decoded in this
chapter from the compiler is independently described in Apple's
neural-engine patents, analyzed in Maynard Handley's AArch64-Explore
vol.~7 (ANE). The patents describe a 256-by-8-by-32-byte accumulator
with eight storage channels per multiply-accumulator, splittable into
two four-channel halves, with a per-pass channel interleave; this
matches the eight-deep accumulator file, the demand-four
double-buffering, and the output-channel-group tiling derived here. The
patents' rasterizer loop manager is the kernel-rasterizer unit, and
their account of descriptors held in a configuration queue and switched
in order is the register-write task descriptor of chapter
\hyperref[program-and-container-format]{23}. The two derivations are
independent: one read from the shipped compiler and the live engine, the
other from the patent record.

The patents describe a configurable design spanning many parts, so the
absolute counts there (a 256-wide multiply-accumulator array, an
eight-core organization) are family ceilings rather than M1 figures; the
per-chip core count read from the hardware-abstraction table, four on
the M1, is the value that holds for a given part.

\chapter{Memory hierarchy}\label{memory-hierarchy}

\begin{summarybox}

The engine keeps its working data in one on-chip pool. Its 2 MB size, at
hardware-abstraction-table field \texttt{0x1b8} on the M1, is the
dominant size limit. A layer whose largest single operand fits stays
on-chip; one that exceeds it is tiled and streamed from DRAM. The pool
is interleaved across 64 banks at a 16-byte granule, with the bank index
\texttt{floor(addr\ /\ 16)\ mod\ 64}. A compile-time stride optimizer
spreads accesses to avoid conflicts. The pool is a compiler-managed
scratchpad, not a demand-filled cache. Residency and stride are decided
at compile time, so re-reference order does not change which operands
are resident.

\end{summarybox}

The numbers below come from the engine's hardware-abstraction table,
indexed by field offset (for example \texttt{0x1b8}); quoted values are
the M1/H13 entries.

\section{2 MB working-set threshold}\label{mb-working-set-threshold}

The bound shows up first in the matmul lowering, which rejects a
resident right-hand operand and streams a tiled copy instead
(\cref{lst:c21-matmul-lower}).

\begin{listing}[H]
\caption{Matmul lowering: reject the resident RHS and stream a tiled copy once the operand reaches the working-set bound.}\label{lst:c21-matmul-lower}

\begin{Shaded}
\begin{Highlighting}[]
\CommentTok{/* ZinMirMatMul::LowerNEMatMulToNEConv */}
\ControlFlowTok{if} \OperatorTok{(}\NormalTok{GetTensorSizeInBytes}\OperatorTok{(}\NormalTok{rhs}\OperatorTok{)} \OperatorTok{\textgreater{}=}\NormalTok{ HAL}\OperatorTok{[}\BaseNTok{0x1b8}\OperatorTok{])} \OperatorTok{\{}  \CommentTok{/* HAL[0x1b8] = 2 MB on M1 */}
    \CommentTok{/* reject resident copy{-}cast; tile and stream the RHS from DRAM */}
\OperatorTok{\}}
\end{Highlighting}
\end{Shaded}

\end{listing}

The largest single operand that stays in on-chip SRAM on the M1 is 2 MB,
the value at field \texttt{0x1b8}, the maximum operand bytes for the
SRAM working set. The same field is 1 MB on the efficiency-class engine
of the M9 part: the value is per-chip. The compiler names this field
\texttt{MemCacheSize}, also \texttt{L2Size}, and exposes overrides for
it through \texttt{fl2-size} and the related cache-mode and allocator
options.

A layer runs entirely from the on-chip pool when its largest single
operand, the larger of the input activation, weight, or output, fits
within this bound. For a half-precision tensor the operand byte count is
the element count times two. A \texttt{{[}1,\ 512,\ 32,\ 32{]}}
activation is exactly 1 MB and stays resident. A
\texttt{{[}1,\ 512,\ 64,\ 64{]}} activation is 4 MB and a
\texttt{{[}4096,\ 4096{]}} linear weight is 32 MB, and both exceed the
bound.

When the largest operand exceeds \texttt{0x1b8}, the compiler tiles it
and streams the pieces from DRAM. Only one tensor's size enters the
comparison; the operands are not summed. Past the bound, the streamed
tile traffic adds DMA bytes that did not exist below it, so the
arithmetic intensity of the layer, the ratio of multiply-accumulate work
to bytes moved, falls. The design rule that follows is to keep the
largest per-layer operand at or under 2 MB by tiling the batch or the
spatial extent or by shrinking the channel count, so the layer stays
on-chip.

The M5 softens this threshold. Its compiler tiles over the batch and the
output dimensions together, so neither operand need be fully resident
and the crossing is a smooth throughput ramp where the M1 shows a sharp
step. The working-set cost then appears in DRAM energy per operation,
which bottoms near a 2 MB operand and rises beyond about 4 to 5 MB,
around the M5 bound of 4.72 MB.

The measured throughput threshold is slightly above the table value,
near 2.28 to 2.34 MB, because the tiler uses roughly 0.3 MB of
double-buffer and alignment margin before it splits the weight. Sweeping
a batched matrix multiply whose weight grows from 1.25 to 3.1 MB
resolves the threshold directly, with a smooth plateau below the bound
and a sharp step just above it (\cref{tbl:c21-threshold-sweep}).

\begin{longtable}[]{@{}ll@{}}
\caption{Measured throughput of a batched matrix multiply as the weight
crosses the 2 MB operand
bound.}\label{tbl:c21-threshold-sweep}\tabularnewline
\toprule\noalign{}
\textbf{weight size} & \textbf{throughput, GFLOP/s} \\
\midrule\noalign{}
\endfirsthead
\toprule\noalign{}
\textbf{weight size} & \textbf{throughput, GFLOP/s} \\
\midrule\noalign{}
\endhead
\bottomrule\noalign{}
\endlastfoot
1.25 to 2.25 MB & smooth plateau, 700 to 850, gently rising \\
\rowcolor{zebra}2.28 MB & 992, the first step up \\
2.34 MB & 1038, a sharp threshold, 185 higher in one 0.03 MB step \\
\rowcolor{zebra}2.50 MB & 1120 \\
2.75 MB & 1180 \\
\rowcolor{zebra}3.00 MB & 1226 \\
3.06 MB & 1368 \\
\end{longtable}

Throughput is higher above the threshold than below it. Below the bound
the whole weight is one resident operand that must fill before compute
starts, a serialized stream into the array. Above the bound the weight
is tiled and double-buffered, so the fill of tile \(n + 1\) overlaps
compute on tile \(n\) and throughput climbs. The threshold is at the
same 2.31 MB sweeping the grid up or down, so the boundary has no memory
state.

\section{64-bank pool and bank
function}\label{bank-pool-and-bank-function}

The engine interleaves the on-chip pool across 64 banks at a 16-byte
granule. The bank count is field \texttt{0x1c8} and equals 64; the
interleave granule is field \texttt{0x1c0} and equals 16 bytes. An
address maps to a bank by

\[\mathrm{bank} = \left\lfloor \frac{\mathrm{addr}}{16} \right\rfloor \bmod 64.\]

In compiler terms the two operands are the two table fields directly
(\cref{lst:c21-bank-index}).

\begin{listing}[H]
\caption{The on-chip bank index computed from a byte address using the interleave granule and the bank count read straight from the table.}\label{lst:c21-bank-index}

\begin{Shaded}
\begin{Highlighting}[]
\CommentTok{/* SRAM/L2 bank index from a byte address */}
\DataTypeTok{unsigned}\NormalTok{ bank}\OperatorTok{(}\DataTypeTok{unsigned} \DataTypeTok{long}\NormalTok{ addr}\OperatorTok{)} \OperatorTok{\{}
    \ControlFlowTok{return} \OperatorTok{(}\NormalTok{addr }\OperatorTok{/}\NormalTok{ HAL}\OperatorTok{[}\BaseNTok{0x1c0}\OperatorTok{])} \OperatorTok{\%}\NormalTok{ HAL}\OperatorTok{[}\BaseNTok{0x1c8}\OperatorTok{];}   \CommentTok{/* (addr / 16) mod 64 */}
\OperatorTok{\}}
\end{Highlighting}
\end{Shaded}

\end{listing}

The map is direct and modular: there is no second-level grouping above
the 64 banks. The bank-conflict optimizer reads the same two table
fields and chooses a row stride that spreads accesses across banks. Over
candidate strides bounded by the maximum stride field \texttt{0x1f8}, it
selects the stride whose per-bank cost is least
(\cref{lst:c21-bank-optimizer}).

\begin{listing}[H]
\caption{The bank-conflict optimizer searching candidate row strides and selecting the one with the least per-bank cost.}\label{lst:c21-bank-optimizer}

\begin{Shaded}
\begin{Highlighting}[]
\CommentTok{/* ZinMirBankConflictOptimizer::OptimizeL2StrideMinCost */}
\NormalTok{uVar1 }\OperatorTok{=}\NormalTok{ HAL}\OperatorTok{[}\BaseNTok{0x1c0}\OperatorTok{];}                       \CommentTok{/* granule    = 16   */}
\NormalTok{uVar2 }\OperatorTok{=}\NormalTok{ HAL}\OperatorTok{[}\BaseNTok{0x1c8}\OperatorTok{];}                       \CommentTok{/* bank count = 64   */}
\NormalTok{uVar4 }\OperatorTok{=}\NormalTok{ param\_1 }\OperatorTok{/}\NormalTok{ uVar1}\OperatorTok{;}                  \CommentTok{/* stride in 16{-}byte granules */}
\CommentTok{/* loop over candidate strides, bounded by HAL[0x1f8] (= 2 MB) */}
\NormalTok{    uVar5 }\OperatorTok{=} \OperatorTok{(}\NormalTok{uVar4 }\OperatorTok{+}\NormalTok{ uVar9}\OperatorTok{)} \OperatorTok{/}\NormalTok{ uVar2}\OperatorTok{;}            \CommentTok{/* granules / 64 */}
\NormalTok{    index }\OperatorTok{=} \OperatorTok{(}\NormalTok{uVar4 }\OperatorTok{+}\NormalTok{ uVar9}\OperatorTok{)} \OperatorTok{{-}}\NormalTok{ uVar5 }\OperatorTok{*}\NormalTok{ uVar2}\OperatorTok{;}    \CommentTok{/* granules mod 64 = bank index */}
\NormalTok{    fVar10 }\OperatorTok{=}\NormalTok{ cost\_vector}\OperatorTok{[}\NormalTok{index}\OperatorTok{];}                \CommentTok{/* per{-}bank conflict cost */}
    \ControlFlowTok{if} \OperatorTok{(}\NormalTok{fVar10 }\OperatorTok{\textless{}}\NormalTok{ best}\OperatorTok{)} \OperatorTok{\{}\NormalTok{ best }\OperatorTok{=}\NormalTok{ fVar10}\OperatorTok{;}\NormalTok{ chosen }\OperatorTok{=}\NormalTok{ uVar8}\OperatorTok{;} \OperatorTok{\}}
\end{Highlighting}
\end{Shaded}

\end{listing}

The cost vector has exactly 64 float entries, one per bank, which fixes
the modulus at 64. The modelled conflict depth for a row stride \(s\)
over the 64 banks is

\[\mathrm{requests} = \left\lceil \frac{\mathrm{rows}}{64 / g} \right\rceil, \qquad g = \gcd\!\left(\frac{s}{16},\ 64\right).\]

The conflict is worst when the granule stride is a high power of two,
which drives \(g\) up, and is conflict-free when \(g = 1\), which gives
one request per bank. Every accepted stride is a multiple of the 16-byte
granule, and the search never proposes a stride above the 2 MB ceiling
in field \texttt{0x1f8}. The on-device throughput period that this
structure produces is 64 bytes, which is 32 half-precision elements and
four 16-byte granules; that period is the granule-quantized echo of the
64-bank interleave after the optimizer has set the stride.

There is no runtime replacement policy on the operand path. The compiler
decides residency and stride at compile time and writes them into the
program, which is why the order in which weights are re-referenced does
not change which ones are resident.

\section{Resident vs shared output
buffers}\label{resident-vs-shared-output-buffers}

Whether a convolution gets a dedicated resident buffer or the shared
default turns on the residency-threshold field
(\cref{lst:c21-resident-buffer}).

\begin{listing}[H]
\caption{The output-buffer allocator deciding between a dedicated resident buffer and the shared default, gated by the residency threshold.}\label{lst:c21-resident-buffer}

\begin{Shaded}
\begin{Highlighting}[]
\CommentTok{/* ZinIrLocalRegAlloc::AllocateOutputDMADefaultBuffer */}
\ControlFlowTok{if} \OperatorTok{(}\NormalTok{HAL}\OperatorTok{[}\BaseNTok{0x491}\OperatorTok{]} \OperatorTok{==} \DecValTok{1} \OperatorTok{\&\&}
    \OperatorTok{(}\NormalTok{footprint }\OperatorTok{=}\NormalTok{ HAL}\OperatorTok{[}\BaseNTok{0x1c0}\OperatorTok{]} \OperatorTok{*}\NormalTok{ dst\_buf\_count}\OperatorTok{,}        \CommentTok{/* 16 * count, in bytes */}
\NormalTok{     HAL}\OperatorTok{[}\BaseNTok{0x1f0}\OperatorTok{]} \OperatorTok{\textless{}=}\NormalTok{ footprint }\OperatorTok{\&\&}\NormalTok{ footprint }\OperatorTok{!=}\NormalTok{ HAL}\OperatorTok{[}\BaseNTok{0x1f0}\OperatorTok{]))} \OperatorTok{\{}
\NormalTok{    use\_default\_buffer}\OperatorTok{;}                  \CommentTok{/* no dedicated conv\_res\_l2\_buf */}
\OperatorTok{\}} \ControlFlowTok{else} \OperatorTok{\{}
\NormalTok{    allocate }\StringTok{"\textless{}layer\textgreater{}/conv\_res\_l2\_buf"}\OperatorTok{;}  \CommentTok{/* dedicated resident buffer */}
\OperatorTok{\}}
\end{Highlighting}
\end{Shaded}

\end{listing}

A convolution can be given a dedicated on-chip buffer that keeps its
output residency across the convolution so the result is not
re-streamed. Field \texttt{0x1f0}, the resident-buffer threshold,
decides whether that buffer is allocated by comparison against the
convolution's output DMA footprint. On the M1 the field is 0, and the
allocator skips the dedicated buffer when the footprint is strictly
greater than the threshold.

With \texttt{0x1f0} equal to 0 on the M1, any buffer of at least one
destination slot has a footprint greater than 0, so the condition is
always true and the M1 always takes the default-buffer branch. The
dedicated residency buffer is thus never allocated on this chip. On the
A16-class and M5 parts the same field is 262144, which is 256 KB, so
only a convolution whose output DMA footprint exceeds 256 KB receives
the dedicated buffer; on the A15 part it is 32768, which is 32 KB. The
field rises with the larger on-chip pools of the newer parts.

The dedicated convolution buffer lives in the one on-chip pool. It is
carved from the same memory as the 2 MB operand working set and the
other DMA buffers, aligned to the same 16-byte granule, and its stride
runs through the same 64-bank optimizer. Field \texttt{0x1f0} decides
only whether a convolution's output residency gets its own named
sub-allocation in that one pool. The kernel-coefficient store is a
distinct on-chip budget: 64 KB on the M1, separate from the 2 MB operand
working set.

The on-chip granules are finer than the page granule at which buffers
are mapped into the engine's address space. Three alignment scales
coexist on the M1: the 16-byte DMA-width granule of field
\texttt{0x1c0}, a 256-byte segment alignment, and the 16 KB page at
which buffers map to the engine.

\section{Two-stage address
translation}\label{two-stage-address-translation}

A host buffer reaches the array through two stacked translations: the
kernel driver pins the pages to physical memory, then maps them through
a device input-output memory-management unit to a device virtual address
the engine's DMA engines read. The translation unit that serves the
engine is the t6000-generation controller bound to the device node
\texttt{dart-ane0}. \Cref{tbl:c21-translation} gives the page size,
aperture base, and aperture size.

\begin{longtable}[]{@{}
  >{\raggedright\arraybackslash}p{(\linewidth - 4\tabcolsep) * \real{0.3333}}
  >{\raggedright\arraybackslash}p{(\linewidth - 4\tabcolsep) * \real{0.3333}}
  >{\raggedright\arraybackslash}p{(\linewidth - 4\tabcolsep) * \real{0.3333}}@{}}
\caption{The device-translation parameters for the engine, read from the
live device node.}\label{tbl:c21-translation}\tabularnewline
\toprule\noalign{}
\begin{minipage}[b]{\linewidth}\raggedright
\textbf{property}
\end{minipage} & \begin{minipage}[b]{\linewidth}\raggedright
\textbf{value}
\end{minipage} & \begin{minipage}[b]{\linewidth}\raggedright
\textbf{meaning}
\end{minipage} \\
\midrule\noalign{}
\endfirsthead
\toprule\noalign{}
\begin{minipage}[b]{\linewidth}\raggedright
\textbf{property}
\end{minipage} & \begin{minipage}[b]{\linewidth}\raggedright
\textbf{value}
\end{minipage} & \begin{minipage}[b]{\linewidth}\raggedright
\textbf{meaning}
\end{minipage} \\
\midrule\noalign{}
\endhead
\bottomrule\noalign{}
\endlastfoot
page size & \texttt{0x4000} = 16 KB & the device virtual page; sub-page
buffers still consume a full 16 KB page \\
\rowcolor{zebra}aperture base & \texttt{0x0} & the device-virtual window
starts at zero \\
aperture size & \texttt{0xE0000000} = 3.5 GiB & the managed
device-virtual window \\
\end{longtable}

The aperture is 3.5 GiB, but the firmware clamps every address it
programs into a DMA engine to the bottom 4 GiB by asserting that the
high 32 bits of every buffer address are zero, so the engine is confined
to the bottom of its device-virtual space. Each page maps to physical
memory through a single 64-bit leaf descriptor: for every live engine
data page the descriptor is the physical frame address with bit 63 set,
the valid bit, and no other field. The descriptor reads
\(\mathrm{phys} \mathbin{|} \mathtt{0x8000000000000000}\) for an input,
weight, intermediate, or output page alike, since the physical frames
are 16 KB-aligned and leave the high bits free; unmap clears the
descriptor to zero, dropping the valid bit. Read-only protection of the
inputs is outside the leaf descriptor on the M1: a configuration bit
collapses the would-be read-only class into read-write at the
translation unit, so input protection is enforced upstream.

Each mapped buffer has a usage code that records its role
(\cref{tbl:c21-usage-codes}).

\begin{longtable}[]{@{}ll@{}}
\caption{The buffer-role usage code each mapped buffer holds, recovered
from the mapping call sites.}\label{tbl:c21-usage-codes}\tabularnewline
\toprule\noalign{}
\textbf{usage code} & \textbf{role} \\
\midrule\noalign{}
\endfirsthead
\toprule\noalign{}
\textbf{usage code} & \textbf{role} \\
\midrule\noalign{}
\endhead
\bottomrule\noalign{}
\endlastfoot
1 & client input tensor \\
\rowcolor{zebra}2 & client output tensor \\
7 & intermediate buffer \\
\rowcolor{zebra}8 & kernel and weights \\
9 & program text, task descriptor, and working set \\
\rowcolor{zebra}11 & program constant and scratch \\
\texttt{0x1019}, \texttt{0x101a} & firmware shared surface and resident
heap \\
\end{longtable}

The address-translation registers and the full fault path are the
subject of chapter \hyperref[address-translation-and-the-dart]{28}.

\section{Hierarchy on the M1}\label{hierarchy-on-the-m1}

The on-chip levels and their M1/H13 values are gathered in
\cref{tbl:c21-hierarchy}.

\begin{longtable}[]{@{}
  >{\raggedright\arraybackslash}p{(\linewidth - 6\tabcolsep) * \real{0.2500}}
  >{\raggedright\arraybackslash}p{(\linewidth - 6\tabcolsep) * \real{0.2500}}
  >{\raggedright\arraybackslash}p{(\linewidth - 6\tabcolsep) * \real{0.2500}}
  >{\raggedright\arraybackslash}p{(\linewidth - 6\tabcolsep) * \real{0.2500}}@{}}
\caption{The on-chip memory levels on the M1, each with its value,
hardware-abstraction-table field, and
role.}\label{tbl:c21-hierarchy}\tabularnewline
\toprule\noalign{}
\begin{minipage}[b]{\linewidth}\raggedright
\textbf{Level}
\end{minipage} & \begin{minipage}[b]{\linewidth}\raggedright
\textbf{M1/H13 value}
\end{minipage} & \begin{minipage}[b]{\linewidth}\raggedright
\textbf{Table field}
\end{minipage} & \begin{minipage}[b]{\linewidth}\raggedright
\textbf{Role}
\end{minipage} \\
\midrule\noalign{}
\endfirsthead
\toprule\noalign{}
\begin{minipage}[b]{\linewidth}\raggedright
\textbf{Level}
\end{minipage} & \begin{minipage}[b]{\linewidth}\raggedright
\textbf{M1/H13 value}
\end{minipage} & \begin{minipage}[b]{\linewidth}\raggedright
\textbf{Table field}
\end{minipage} & \begin{minipage}[b]{\linewidth}\raggedright
\textbf{Role}
\end{minipage} \\
\midrule\noalign{}
\endhead
\bottomrule\noalign{}
\endlastfoot
SRAM operand working set & 2 MB & \texttt{0x1b8} & largest single
operand that stays on-chip; over this, tile and stream \\
\rowcolor{zebra}Operand threshold (measured) & 2.28 to 2.34 MB & derived
from \texttt{0x1b8} & observed throughput threshold, 2 MB plus tiler
double-buffer margin \\
Bank count & 64 & \texttt{0x1c8} & banks the pool is interleaved
across \\
\rowcolor{zebra}Interleave granule & 16 B & \texttt{0x1c0} & DMA-width
granule and bank-stride unit \\
Maximum L2 destination stride & 2 MB & \texttt{0x1f8} & upper bound of
the bank-conflict stride search \\
\rowcolor{zebra}Residency-buffer threshold & 0 & \texttt{0x1f0} & over
this footprint, skip the dedicated buffer; 0 means M1 never allocates
one \\
Kernel-coefficient store & 64 KB & \texttt{0x200} & on-chip
weight-coefficient budget at \texttt{0x200}/\texttt{0x210}, with the 64
KB-or-16 MB mode select at \texttt{0x288}, distinct from the operand
working set \\
\end{longtable}

The residency-buffer threshold is the field that varies most across
parts (\cref{tbl:c21-residency-gens}).

\begin{longtable}[]{@{}
  >{\raggedright\arraybackslash}p{(\linewidth - 4\tabcolsep) * \real{0.3333}}
  >{\raggedright\arraybackslash}p{(\linewidth - 4\tabcolsep) * \real{0.3333}}
  >{\raggedright\arraybackslash}p{(\linewidth - 4\tabcolsep) * \real{0.3333}}@{}}
\caption{The residency-buffer threshold across generations and its
effect on whether a dedicated convolution output buffer is
allocated.}\label{tbl:c21-residency-gens}\tabularnewline
\toprule\noalign{}
\begin{minipage}[b]{\linewidth}\raggedright
\textbf{Part}
\end{minipage} & \begin{minipage}[b]{\linewidth}\raggedright
\textbf{\texttt{0x1f0} value}
\end{minipage} & \begin{minipage}[b]{\linewidth}\raggedright
\textbf{Effect}
\end{minipage} \\
\midrule\noalign{}
\endfirsthead
\toprule\noalign{}
\begin{minipage}[b]{\linewidth}\raggedright
\textbf{Part}
\end{minipage} & \begin{minipage}[b]{\linewidth}\raggedright
\textbf{\texttt{0x1f0} value}
\end{minipage} & \begin{minipage}[b]{\linewidth}\raggedright
\textbf{Effect}
\end{minipage} \\
\midrule\noalign{}
\endhead
\bottomrule\noalign{}
\endlastfoot
M1/H13 & 0 & dedicated residency buffer never allocated \\
\rowcolor{zebra}A15-class & 32 KB & dedicated buffer when convolution
output footprint exceeds 32 KB \\
A16-class and M5 & 256 KB & dedicated buffer when convolution output
footprint exceeds 256 KB \\
\end{longtable}

\begin{figure}[H]\centering
\includegraphics[width=4.917in]{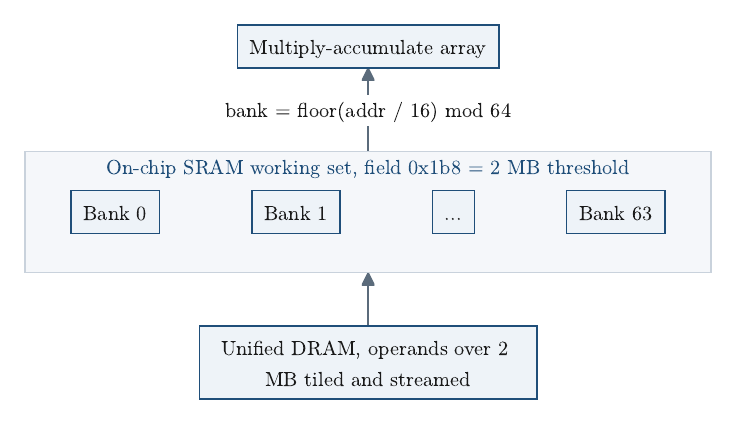}
\caption{The memory hierarchy, with unified DRAM at the bottom, the on-chip SRAM working set and its 2 MB threshold, and the 64-bank interleave feeding the multiply-accumulate array.}
\label{fig:c21-hierarchy}
\end{figure}

The levels stack from unified DRAM at the bottom, up through the on-chip
working set and its 64 banks, into the multiply-accumulate array, as
\cref{fig:c21-hierarchy} draws them.

\anepartopen{Part VII}{The Toolchain and Encoding}
\anechap{22}{Compiler}{The frontend, the anec backend dialect, lowering, and the validators.}
\anechap{23}{Program and container format}{The on-disk program, its descriptors, and the resolved layout sidecar.}
\anechap{24}{HAL and capability gates}{The per-chip parameter table and the capability bytes that gate each operation.}
\anechap{25}{Compression internals}{The reconstruction codecs and the wire layout of each compressed weight form.}
\anechap{26}{Hidden layers and direct netplist authoring}{Native layers reachable only by authoring the netplist directly.}
\anepartclose

\chapter{Compiler}\label{compiler}

\begin{summarybox}

The compiler turns a network description into a loadable program through
four phases. They are fusion to one compute operation, legalization to
the hardware envelope, schedule and task-descriptor partition under the
on-chip working set, and memory and direct-memory-access optimization. A
frontend operation lowers to one or more backend \texttt{anec.*}
operations, most of them one to one. A convolution absorbs its bias,
activation, padding, and dequantize into one fused operation. An
exported \texttt{\_ANECValidate\{Op\}Layer} entry point checks each
operation first, one of 50 per-layer validators among 55
\texttt{\_ANECValidate*} exports. A callable validator signals that an
operation is reachable on the direct path. A validator that accepts does
not guarantee code generation: top-k, sort, dynamic-slice, and
three-dimensional convolution pass validation and fail backend lowering
on the M1.

\end{summarybox}

\section{Pipeline}\label{pipeline}

The compiler runs four phases in execution order, each listed in
\Cref{tbl:c22-phases} with its job and the passes that represent it.

\begin{longtable}[]{@{}
  >{\raggedright\arraybackslash}p{(\linewidth - 4\tabcolsep) * \real{0.3333}}
  >{\raggedright\arraybackslash}p{(\linewidth - 4\tabcolsep) * \real{0.3333}}
  >{\raggedright\arraybackslash}p{(\linewidth - 4\tabcolsep) * \real{0.3333}}@{}}
\caption{The four compiler phases in execution order, each with its job
and representative passes.}\label{tbl:c22-phases}\tabularnewline
\toprule\noalign{}
\begin{minipage}[b]{\linewidth}\raggedright
\textbf{phase}
\end{minipage} & \begin{minipage}[b]{\linewidth}\raggedright
\textbf{job}
\end{minipage} & \begin{minipage}[b]{\linewidth}\raggedright
\textbf{representative passes}
\end{minipage} \\
\midrule\noalign{}
\endfirsthead
\toprule\noalign{}
\begin{minipage}[b]{\linewidth}\raggedright
\textbf{phase}
\end{minipage} & \begin{minipage}[b]{\linewidth}\raggedright
\textbf{job}
\end{minipage} & \begin{minipage}[b]{\linewidth}\raggedright
\textbf{representative passes}
\end{minipage} \\
\midrule\noalign{}
\endhead
\bottomrule\noalign{}
\endlastfoot
1. Fusion & collapse the graph to one fused compute operation &
transpose fusion, bias and activation hoisting into conv and matmul,
elementwise-copy elimination \\
\rowcolor{zebra}2. Legalization & make the fused graph hardware-legal &
dimension and rank legalization (rank \texttt{\textbackslash{}le\ 5}),
kernel-memory split, multi-segment graph cuts \\
3. Schedule and task-descriptor partition & linearize the graph and
carve it into on-engine partitions & parallel-execution discovery, the
greedy list scheduler, partition emission \\
\rowcolor{zebra}4. Memory and direct-memory-access optimization & lay
out buffers and streams & bank-conflict optimization, pad optimization,
width-concat, weight packing \\
\end{longtable}

\begin{figure}[H]\centering
\includegraphics[width=1.960in]{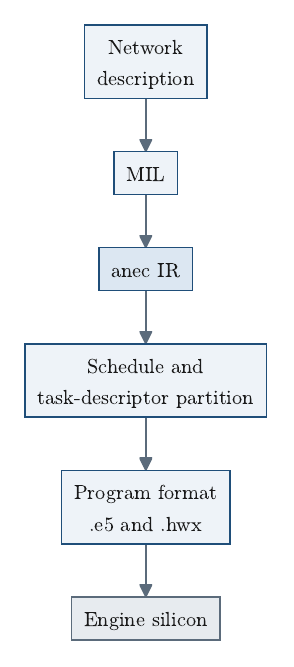}
\caption{The compile pipeline from a network description to the loadable program.}
\label{fig:c22-pipeline}
\end{figure}

Traced as a dataflow, the same pipeline runs from the network
description through the intermediate forms to the loadable program on
the engine (\cref{fig:c22-pipeline}).

Phase 1 collapses the whole graph to a single compute operation.
Transposes are free: they fold into the adjacent operation or route to
the dedicated transpose engine. A bias add and an activation that follow
a convolution or a matrix multiply hoist into that operation, so
\texttt{conv\ +\ bias\ +\ relu} and \texttt{matmul\ +\ bias\ +\ gelu}
each become one fused operation rather than three.

Phase 2 enforces the hardware envelope. Tensor rank is capped at five,
per-dimension maxima apply, and an operation whose coefficients exceed
the per-buffer kernel-memory cap is tiled or split. The compiler cuts a
graph that cannot run as one segment into several at bridge operations.

Phase 3 turns a fused operation into real on-engine work. A greedy
priority-queue list scheduler linearizes the operation-layer graph into
one execution order, then chunks that order into task-descriptor
partitions under the on-chip working-set budget. The scheduler pops the
highest-priority ready node each step, ordered by a fixed tie-break
sequence: ring-buffer membership first, then branch-sibling score, then
merge-sibling score, then the working-set-aware branch test, then
topological order, then raw node id. The partition boundary is a memory
test: the scheduler tentatively schedules a layer, queries the peak L2
pressure over its live range, and closes the current partition when

\[\mathrm{peak\ L2\ pressure} > \mathrm{HAL}[\mathtt{0x1b8}] \times f,\]

where \(\mathrm{HAL}[\mathtt{0x1b8}]\) is the 2 MB on-chip static-memory
ceiling on the M1 and \(f\) is a margin fraction below one. A partition
grows until the next layer's live range would push the peak past that
scaled budget. The branch decision that keeps the live set on chip reads
the same 2 MB field: when a diamond's endpoints fit, the scheduler runs
the larger-footprint branch first while its peak still fits in 2 MB,
draining it before the working set grows.

Phase 4 lays out the buffers the partitions read and write: it
re-strides the destination buffers to avoid bank conflicts on the main
engine, folds padding, and packs the weight stream.

The allocator runs alongside phases 3 and 4. It tags every root tensor
with an allocation type from a nine-state set that records whether the
tensor is resident on chip, streamed from dynamic memory, chained
directly into the next operation in the on-chip cache, or held as a ring
buffer (\Cref{tbl:c22-alloc-states}).

\begin{longtable}[]{@{}
  >{\raggedright\arraybackslash}p{(\linewidth - 2\tabcolsep) * \real{0.5000}}
  >{\raggedright\arraybackslash}p{(\linewidth - 2\tabcolsep) * \real{0.5000}}@{}}
\caption{The nine allocation-type states the allocator assigns to each
root tensor, decoded from the bit-test predicates that read
them.}\label{tbl:c22-alloc-states}\tabularnewline
\toprule\noalign{}
\begin{minipage}[b]{\linewidth}\raggedright
\textbf{value}
\end{minipage} & \begin{minipage}[b]{\linewidth}\raggedright
\textbf{meaning}
\end{minipage} \\
\midrule\noalign{}
\endfirsthead
\toprule\noalign{}
\begin{minipage}[b]{\linewidth}\raggedright
\textbf{value}
\end{minipage} & \begin{minipage}[b]{\linewidth}\raggedright
\textbf{meaning}
\end{minipage} \\
\midrule\noalign{}
\endhead
\bottomrule\noalign{}
\endlastfoot
0 & plain on-chip resident operand \\
\rowcolor{zebra}1 & streamed from dynamic memory \\
2 & L2-chained, supplies the next operation directly in L2: the
double-buffer state \\
\rowcolor{zebra}3 & depends on an L2 producer \\
4 & in-place rewrite in L2 \\
\rowcolor{zebra}5 & in-place rewrite, L2-dependent \\
6 & ring-buffer resident \\
\rowcolor{zebra}7 & ring-buffer resident, L2-dependent \\
8 & in-place in dynamic memory \\
\end{longtable}

A single-fanout producer immediately followed by a same-engine consumer
with matching tile geometry is marked chainable, allocation type 2, the
structural form of double-buffering: the producer output stays on chip
and supplies the consumer in place instead of round-tripping through
dynamic memory. Chainability is a structural pattern, not a size
threshold. The decision is a conjunction. The per-chip chaining-enable
bit must be set, the producer must have exactly one outgoing layer, and
both ends must be the same engine class on the same engine. The
producer's emitted tile geometry must match the consumer's declared
input geometry, the consumer must be scheduled immediately after the
producer with no other layer in the gap, and the producer's emitted rows
must cover the consumer's first tile including overlap. The ring-buffer
state, allocation type 6, is the state the scheduler ranks first when it
orders ready operations.

Phase 3 also discovers operations that can co-issue on the two engines.
The rule is restrictive: two operations run concurrently only when one
is a multiply-accumulate operation and the other is a planar-engine
operation, since two operations on one engine cannot overlap. The
discovery disqualifies a pair when either is unscheduled, when both are
the same engine class, when they are on different engines, or when they
stand in a producer-consumer relation. A pair is also disqualified when
either input is chained or chainable, or when both outputs stream from
dynamic memory and would serialize on the shared channel. A matrix
multiply's resident weight is exempt from the last test, so it does not
block its partner. The attention block is the operation this filter
selects: a softmax on the planar engine co-issued with a value matrix
multiply on the multiply-accumulate array. Chaining and co-issue are
mutually exclusive: an operation is either pipelined into its neighbor
or co-issued with an independent operation on the other engine, never
both.

\section{Lowering one operation}\label{lowering-one-operation}

A frontend operation lowers to one or more backend \texttt{anec.*}
operations through the operation-converter pass, most of them one to
one. A matrix multiply lowers to \texttt{anec.matmul}, which records the
contraction as two attributes (\cref{lst:c22-matmul-lower}).

\begin{listing}[H]
\caption{The matrix-multiply lowering, with its operands, the transpose attributes that set the contraction direction, and the pre-lowering constraints.}\label{lst:c22-matmul-lower}

\begin{Shaded}
\begin{Highlighting}[]
\NormalTok{mps.MatMulOp  {-}\textgreater{}  anec.matmul}
\NormalTok{  bottoms:  2   (A, B)}
\NormalTok{  attrs:    transpose\_lhs : bool}
\NormalTok{            transpose\_rhs : bool}
\NormalTok{  constraints (checked before lowering):}
\NormalTok{    depth(A) == 1 and depth(B) == 1}
\NormalTok{    out\_channels == channels(A)}
\NormalTok{    width(A) + pad == channels(B)}
\NormalTok{    out{-}channel bytes of A must fit kernel memory}
\end{Highlighting}
\end{Shaded}

\end{listing}

The contraction direction is parametric: \texttt{transpose\_lhs} and
\texttt{transpose\_rhs} select which operands are transposed before the
product, so one backend operation covers every transposed and
non-transposed matrix-multiply form. The compiler rewrites a matrix
multiply whose right-hand weight fits the on-chip working set into a
resident convolution, a single fused operation; a weight above the 2 MB
ceiling stays a tiled matrix multiply.

A convolution lowers the same way and absorbs more of its neighborhood
(\cref{lst:c22-conv-lower}).

\begin{listing}[H]
\caption{The convolution lowering, with its folded weight and bias, its attributes, and the constraints checked before lowering.}\label{lst:c22-conv-lower}

\begin{Shaded}
\begin{Highlighting}[]
\NormalTok{mps.Conv2DOp  {-}\textgreater{}  anec.convolution}
\NormalTok{  bottoms:  1   (+ weight and bias folded in)}
\NormalTok{  attrs:    strides}
\NormalTok{            dilation\_rates}
\NormalTok{            groups}
\NormalTok{            explicit\_padding / padding\_style}
\NormalTok{            kernel\_sizes}
\NormalTok{            weights\_layout}
\NormalTok{  constraints (checked before lowering):}
\NormalTok{    filter and input rank 4 or 5}
\NormalTok{    kernel within the per{-}chip [min, max]}
\NormalTok{    in{-}channels and out{-}channels divisible by groups}
\NormalTok{    padded input \textgreater{}= kernel}
\end{Highlighting}
\end{Shaded}

\end{listing}

The bias becomes the convolution's gain-offset control, and a following
activation becomes a post-operation lookup table on the same operation.
A padded, biased, activated convolution thus emits as one backend
operation with the pad folded, bias as a per-channel affine, and
activation as a 33-segment piecewise-linear table. A quantization scale
folds into the per-channel output multiply that precedes that table, so
a dequantize between the convolution and its activation does not become
its own operation either.

\section{Fusion rules}\label{fusion-rules}

Fusion does not happen in the MIL the host hands over. The MIL is the
pre-fusion handoff: a biased, activated convolution arrives as three
separate operations, \texttt{conv} then
\texttt{add(x=conv,\ y=bias\_const)} then \texttt{relu}, and the
collapse into one engine layer happens entirely inside ANECompiler's
\texttt{ZinIr}/\texttt{ZinMir} layer-graph stage. The fused hardware
unit is the gain-offset control, \texttt{anec.gain\_offset\_control}, a
per-channel or singular affine
\(y = \mathrm{gain} \cdot x + \mathrm{offset}\) applied at engine
output. The \texttt{ZinNEBypassLayer} container wraps a fused engine
operation in a fixed seven-slot epilogue chain whose constructor
argument order, given in \cref{lst:c22-nebypass}, fixes the seven slots.

\begin{listing}[H]
\caption{The ZinNEBypassLayer constructor, whose argument order fixes the seven epilogue slots a single engine layer can absorb.}\label{lst:c22-nebypass}

\begin{Shaded}
\begin{Highlighting}[]
\NormalTok{ZinNEBypassLayer( engine\_op,}
\NormalTok{    ZinTextureLayer,      }\CommentTok{// in{-}place spatial/texture remap}
\NormalTok{    ZinBroadcastLayer,    }\CommentTok{// broadcast of a fused operand}
\NormalTok{    ZinActivationLayer,   }\CommentTok{// pre{-}GOC activation}
\NormalTok{    ZinGOCLayer,          }\CommentTok{// the gain/offset (scale + bias) unit}
\NormalTok{    ZinActivationLayer,   }\CommentTok{// post{-}GOC activation}
\NormalTok{    ZinTransposeLayer,    }\CommentTok{// output transpose / layout}
\NormalTok{    ZinQuantLayer )       }\CommentTok{// output (re)quantization}
\end{Highlighting}
\end{Shaded}

\end{listing}

A convolution, matrix multiply, pool, or elementwise operation absorbs,
in slot order, a texture remap, broadcast, pre-GOC activation, the
gain-offset affine, a post-GOC activation, output transpose, and output
requantization. The bias-add fills the GOC offset, the scale or
batch-norm gain fills the GOC gain, and the activation fills one of the
two activation slots. The pass that forms the offset is
\texttt{ScaledEWOrEWWithConstInToGOC}, and an internal guard reads
\texttt{"Must\ have\ 2\ inputs\ when\ convert\ EW\ to\ GOC"}: the
elementwise-to-GOC rewrite needs exactly one live operand and one
constant operand. That guard is the dividing line between a fusable
epilogue and a fusion barrier (\Cref{tbl:c22-fusion}).

\begin{longtable}[]{@{}
  >{\raggedright\arraybackslash}p{(\linewidth - 4\tabcolsep) * \real{0.3333}}
  >{\raggedright\arraybackslash}p{(\linewidth - 4\tabcolsep) * \real{0.3333}}
  >{\raggedright\arraybackslash}p{(\linewidth - 4\tabcolsep) * \real{0.3333}}@{}}
\caption{The fusable epilogues that collapse into the producing engine
layer and the barriers that keep operations as separate
layers.}\label{tbl:c22-fusion}\tabularnewline
\toprule\noalign{}
\begin{minipage}[b]{\linewidth}\raggedright
\textbf{pattern}
\end{minipage} & \begin{minipage}[b]{\linewidth}\raggedright
\textbf{fuses as}
\end{minipage} & \begin{minipage}[b]{\linewidth}\raggedright
\textbf{gating pass}
\end{minipage} \\
\midrule\noalign{}
\endfirsthead
\toprule\noalign{}
\begin{minipage}[b]{\linewidth}\raggedright
\textbf{pattern}
\end{minipage} & \begin{minipage}[b]{\linewidth}\raggedright
\textbf{fuses as}
\end{minipage} & \begin{minipage}[b]{\linewidth}\raggedright
\textbf{gating pass}
\end{minipage} \\
\midrule\noalign{}
\endhead
\bottomrule\noalign{}
\endlastfoot
bias-add (\texttt{add} with one constant) & GOC offset &
\texttt{ScaledEWOrEWWithConstInToGOC},
\texttt{ZinMirHoistGOCorActivationForConvFusion} \\
\rowcolor{zebra}scale or batch-norm (\texttt{mul} with one constant) &
GOC gain, or folded into the weights & \texttt{FoldScale},
\texttt{FoldWeightsWithScale} \\
activation (relu, leaky, clamped, swish, sigmoid, tanh, gelu) & a pre-
or post-GOC activation slot; transcendentals become a palette lookup
table & \texttt{MergeFusableActivationPairs},
\texttt{SimpleActivation} \\
\rowcolor{zebra}transpose and reshape chains & the NEBypass transpose
slot; symmetric pairs cancel & \texttt{CollapseTranspose},
\texttt{ZinMirNETransposeFusion} \\
dequantize & folded into the convolution weight path as
\texttt{kernel\_scale} or \texttt{kernel\_palettized\_LUT} &
\texttt{CollapseQuantDequant},
\texttt{DeduplicateStandaloneDequantGOCsAcrossConcat} \\
\rowcolor{zebra}matrix multiply or linear & rewritten to a convolution,
then takes the convolution epilogue rules &
\texttt{ReplaceMatmulWithConv} \\
two-live-input elementwise (\texttt{add(conv\_a,\ conv\_b)}) & does not
fuse: stays a real \texttt{anec.add} engine layer (fails the
one-constant guard) & barrier \\
\rowcolor{zebra}concat & a fusion boundary unless
\texttt{IsFusableConcat} qualifies it & barrier \\
scaled-dot-product attention & a hard segment cut between programs &
barrier \\
\end{longtable}

A two-live-input elementwise operation is the one barrier that follows
from the GOC definition itself. The GOC is an affine of one tensor whose
gain and offset are constants or per-channel vectors, so
\texttt{add(conv\_a,\ conv\_b)} cannot be a GOC and remains a real
\texttt{anec.add} engine layer that separates the two convolutions. A
trailing activation can still fill the add's own post-GOC activation
slot, but the add itself does not vanish. Concat is a boundary unless it
qualifies as fusable, scaled-dot-product attention is a hard segment cut
that matches the host's own segmentation, and the \texttt{ZinMir*Split}
capacity passes can re-cut a fused layer that exceeds the tile or engine
limit, so a fusion formed earlier can be undone downstream.

\section{Validator surface}\label{validator-surface}

Before an operation lowers, a per-layer validator checks it. The
validators are a family of exported \texttt{\_ANECValidate\{Op\}Layer}
entry points, one per operation class, with an umbrella
\texttt{\_ANECValidateNetworkCreate} that iterates a network's layers
and invokes the per-layer validator for each. The export table has 55
symbols whose name starts \texttt{\_ANECValidate}, which split into 50
per-layer \texttt{…Layer} validators and 5 non-layer exports, whose
roles are in \Cref{tbl:c22-nonlayer}.

\begin{longtable}[]{@{}
  >{\raggedright\arraybackslash}p{(\linewidth - 2\tabcolsep) * \real{0.5000}}
  >{\raggedright\arraybackslash}p{(\linewidth - 2\tabcolsep) * \real{0.5000}}@{}}
\caption{The five non-layer \protect\texttt{\_ANECValidate*} exports and
the role of each, alongside the 50 per-layer
validators.}\label{tbl:c22-nonlayer}\tabularnewline
\toprule\noalign{}
\begin{minipage}[b]{\linewidth}\raggedright
\textbf{non-layer export}
\end{minipage} & \begin{minipage}[b]{\linewidth}\raggedright
\textbf{role}
\end{minipage} \\
\midrule\noalign{}
\endfirsthead
\toprule\noalign{}
\begin{minipage}[b]{\linewidth}\raggedright
\textbf{non-layer export}
\end{minipage} & \begin{minipage}[b]{\linewidth}\raggedright
\textbf{role}
\end{minipage} \\
\midrule\noalign{}
\endhead
\bottomrule\noalign{}
\endlastfoot
\texttt{\_ANECValidateNetworkCreate} & the umbrella oracle: a dry-run of
network creation that iterates the layers and invokes each per-layer
validator \\
\rowcolor{zebra}\texttt{\_ANECValidate} & the top-level wrapper for a
single operation or descriptor \\
\texttt{\_ANECValidateMPSModule}, \texttt{\_ANECValidateMPSModuleCreate}
& validate an intermediate-language module before its conversion to the
backend form \\
\rowcolor{zebra}\texttt{\_ANECValidateMutableProcedureInfo} & validate a
weight-editing procedure descriptor \\
\end{longtable}

Five per-layer validators have no matching descriptor constructor, since
they validate computed or umbrella operations rather than directly
authored layers: argmin-max, global argmin-max, broadcast,
cross-product, and elementwise. The same per-layer code runs in two
places: the placement dry-run that decides whether an operation is
eligible for the engine, and the backend legalizer during a real
compile, so the eligibility prediction matches the compile outcome.

Each validator checks the operand count and the shape, type, and
attribute envelope for its operation, and emits a fixed reject string
when a constraint fails. The matrix-multiply validator requires exactly
two inputs, depth one on both, the output channel count equal to the
first input's channel count, and the first input's output-channel bytes
to fit kernel memory. It rejects with
\texttt{"Matrix\ mult.\ layer\ can\ only\ have\ two\ bottoms"},
\texttt{"depth\ \textgreater{}\ 1\ is\ not\ supported\ for\ MatMult"},
and \texttt{"can\ not\ fit\ the\ Kmem"}. The convolution validator
checks the kernel against the per-chip bounds and the channel-group
divisibility, rejecting with
\texttt{"Invalid\ conv\ kernel\ \%s\ =\ \%zd,\ It\ should\ be\ in\ {[}\%zd,\%zd{]}"}
and
\texttt{"input/output\ channels\ should\ be\ divisible\ by\ num\ group"}.

The validators that gate a feature on the chip read a per-chip support
byte, so the same operation can validate on one generation and reject on
another (\Cref{tbl:c22-feature-gates}).

\begin{longtable}[]{@{}
  >{\raggedright\arraybackslash}p{(\linewidth - 4\tabcolsep) * \real{0.3333}}
  >{\raggedright\arraybackslash}p{(\linewidth - 4\tabcolsep) * \real{0.3333}}
  >{\raggedright\arraybackslash}p{(\linewidth - 4\tabcolsep) * \real{0.3333}}@{}}
\caption{Representative feature-gated validators, the per-chip support
byte each reads, and the generation it gates
on.}\label{tbl:c22-feature-gates}\tabularnewline
\toprule\noalign{}
\begin{minipage}[b]{\linewidth}\raggedright
\textbf{operation}
\end{minipage} & \begin{minipage}[b]{\linewidth}\raggedright
\textbf{support byte}
\end{minipage} & \begin{minipage}[b]{\linewidth}\raggedright
\textbf{gated on}
\end{minipage} \\
\midrule\noalign{}
\endfirsthead
\toprule\noalign{}
\begin{minipage}[b]{\linewidth}\raggedright
\textbf{operation}
\end{minipage} & \begin{minipage}[b]{\linewidth}\raggedright
\textbf{support byte}
\end{minipage} & \begin{minipage}[b]{\linewidth}\raggedright
\textbf{gated on}
\end{minipage} \\
\midrule\noalign{}
\endhead
\bottomrule\noalign{}
\endlastfoot
softmax & \texttt{0x815} & rejected on the older targets \\
\rowcolor{zebra}instance norm & \texttt{0x816} & per-chip \\
dropout, random & \texttt{0x4a9} & the A15 and later generations only \\
\rowcolor{zebra}affine transform, resize, resample, padding mode &
\texttt{0x81d} & the texture engine, absent on the M1 \\
dynamic gain-offset control & \texttt{0x814} & absent only on the
smallest legacy parts \\
\rowcolor{zebra}three-dimensional convolution & dimension and
kernel-depth fields & has the capability byte yet fails backend
lowering \\
\end{longtable}

The reject strings name the constraint plainly: a softmax on a target
without \texttt{0x815} returns
\texttt{"Softmax\ is\ not\ supported\ by\ this\ ANE\ architecture"}, a
dropout without \texttt{0x4a9} returns
\texttt{"Dropout\ layer\ is\ not\ supported\ on\ this\ architecture."},
and an affine transform without the texture engine returns
\texttt{"affine\ transform\ is\ not\ supported\ on\ this\ architecture"}.
The network-level umbrella adds its own checks, rejecting a module newer
than the current schema, a bonded network the target cannot run, a
unit-name collision, or a missing conversion record. The reject strings
include
\texttt{"Bonded\ networks\ are\ not\ supported\ on\ the\ target"} and
\texttt{"ANE\ internal\ error:\ Unit\ name\ \textbackslash{}"\%s\textbackslash{}"\ collision\ during\ MIL\ to\ ANEC\ IR\ conversion."}.

Every per-layer validator is an exported, user-callable symbol: an
operation whose validator accepts the schema is reachable by direct
network authoring. This is the attested-is-not-reachable rule from
chapter \hyperref[capability-surface]{4} in its compiler form: an
operation can pass its \texttt{\_ANECValidate*Layer} schema check and
still fail at backend lowering. On the M1 the top-k, sort, and
dynamic-slice validators accept their inputs and then reject at code
generation, and the unflatten validator passes yet fails lowering for
every variant. Three-dimensional convolution has its capability byte yet
fails backend lowering on every device mask. One opcode-surface gap
exists below the validator layer: the contrast-adaptive-sharpening and
reverse operations have an internal validation routine but no exported
per-layer validator, so they are not reachable by direct authoring at
all. The reachable surface here is confirmed against an on-device sweep.

Validation happens in three distinct layers. The exported
\texttt{\_ANECValidate*} catalog, the per-layer and umbrella validators,
is the outward face. Behind it the in-binary
\texttt{ValidateSemantics\_Impl} family has the \texttt{MatchParams} and
\texttt{MatchStatus} pattern grammar that the exported validators are
generated from. Below both, an in-kernel
\texttt{ZinComputeProgramValidate*} family validates the compiled
program again at load, across single-plane, multi-plane, uncompressed,
and tiled-compressed matrix forms.

\section{Native layers and netplist
authoring}\label{native-layers-and-netplist-authoring}

The 50 per-layer validators pair with 45 native hardware layer
descriptors, the operation classes the engine silicon implements. A few
of these never reach the engine through the conversion-tool front door,
since that path always decomposes them, yet they are reachable by
authoring the network description directly. The fused
scaled-dot-product-attention layer is one such: its descriptor takes
four or five inputs, the query, key, value, scale, and an optional
additive mask, and its validator rejects with
\texttt{"4\ or\ 5\ bottoms\ must\ be\ present\ for\ SDPA"} and
\texttt{"Mask\ format\ must\ be\ same\ as\ Q,\ K\ and\ V"}. Its one
parsed parameter is the max-subtraction flag, which defaults to off and
must be set on for a numerically correct softmax, and the causal-decode
form is not a separate descriptor but the additive-mask variant.

The network description is a property-list document, the netplist,
schema version 1.0.10, whose keys the compiler reads verbatim and
\Cref{tbl:c22-netplist} names with the role of each.

\begin{longtable}[]{@{}
  >{\raggedright\arraybackslash}p{(\linewidth - 2\tabcolsep) * \real{0.5000}}
  >{\raggedright\arraybackslash}p{(\linewidth - 2\tabcolsep) * \real{0.5000}}@{}}
\caption{The netplist dictionary keys the compiler reads, with the role
of each.}\label{tbl:c22-netplist}\tabularnewline
\toprule\noalign{}
\begin{minipage}[b]{\linewidth}\raggedright
\textbf{key}
\end{minipage} & \begin{minipage}[b]{\linewidth}\raggedright
\textbf{role}
\end{minipage} \\
\midrule\noalign{}
\endfirsthead
\toprule\noalign{}
\begin{minipage}[b]{\linewidth}\raggedright
\textbf{key}
\end{minipage} & \begin{minipage}[b]{\linewidth}\raggedright
\textbf{role}
\end{minipage} \\
\midrule\noalign{}
\endhead
\bottomrule\noalign{}
\endlastfoot
\texttt{Networks}, \texttt{ProcedureList} & the network names and the
callable entry points \\
\rowcolor{zebra}\texttt{Units}, \texttt{Weights} & the ordered layer
list and the weight-blob files \\
\texttt{InputList}, \texttt{OutputList}, \texttt{OperationList} & the
external ports and the operation order \\
\rowcolor{zebra}\texttt{Bottom}, \texttt{Type}, \texttt{Params} &
per-unit wiring, layer-type tag, and typed attributes \\
\texttt{BatchSize}, \texttt{InputChannels}, \texttt{InputDepth},
\texttt{InputHeight}, \texttt{InputWidth}, \texttt{InputInterleave} &
the port shape five-tuple and the channel-interleave packing \\
\end{longtable}

A layer is a unit with a \texttt{Type} tag, a \texttt{Bottom} wiring
list, and a \texttt{Params} sub-dictionary of typed attributes, and the
wiring is by symbol name. The compiled program that comes out is a
container with the magic \texttt{0xbeefface}, whose text section
contains the task-descriptor register-write stream and whose
\texttt{\_\_KERN\_N} sections store the weight coefficients, the same
image the kernel loads.

\chapter{Program and container format}\label{program-and-container-format}

\begin{summarybox}

A compiled network reaches the engine as two layered files. The upper
one is a dispatch descriptor, a FlatBuffer that names the operations and
binds the buffers. Beneath it the hardware container is a Mach-O-shaped
file with the magic \texttt{0xbeefface}, holding the register-write
program and the weight banks. The descriptor tracks dispatch count, not
operation count. A fused graph of any depth reduces to one inference
operation bracketed by input and output casts, and a bridge cut adds one
per segment. Between them is the hardware task descriptor
\texttt{ZinAneTdHw\_v10}, a register image in seven groups. It
serializes sparsely as 44-byte records whose buffer bases are relocation
slots resolved at load.

\end{summarybox}

\section{Two layers}\label{two-layers}

A model passes through six on-disk and in-memory forms
(\Cref{tbl:c23-forms}).

\begin{longtable}[]{@{}
  >{\raggedright\arraybackslash}p{(\linewidth - 6\tabcolsep) * \real{0.2500}}
  >{\raggedright\arraybackslash}p{(\linewidth - 6\tabcolsep) * \real{0.2500}}
  >{\raggedright\arraybackslash}p{(\linewidth - 6\tabcolsep) * \real{0.2500}}
  >{\raggedright\arraybackslash}p{(\linewidth - 6\tabcolsep) * \real{0.2500}}@{}}
\caption{The six on-disk and in-memory forms a model passes through,
each with its format, identifying tag, and
role.}\label{tbl:c23-forms}\tabularnewline
\toprule\noalign{}
\begin{minipage}[b]{\linewidth}\raggedright
\textbf{Stage}
\end{minipage} & \begin{minipage}[b]{\linewidth}\raggedright
\textbf{Format}
\end{minipage} & \begin{minipage}[b]{\linewidth}\raggedright
\textbf{Tag}
\end{minipage} & \begin{minipage}[b]{\linewidth}\raggedright
\textbf{Role}
\end{minipage} \\
\midrule\noalign{}
\endfirsthead
\toprule\noalign{}
\begin{minipage}[b]{\linewidth}\raggedright
\textbf{Stage}
\end{minipage} & \begin{minipage}[b]{\linewidth}\raggedright
\textbf{Format}
\end{minipage} & \begin{minipage}[b]{\linewidth}\raggedright
\textbf{Tag}
\end{minipage} & \begin{minipage}[b]{\linewidth}\raggedright
\textbf{Role}
\end{minipage} \\
\midrule\noalign{}
\endhead
\bottomrule\noalign{}
\endlastfoot
Intermediate-language text & text & versioned \texttt{3520.4.1} & the
compiler's input \\
\rowcolor{zebra}Network description & property list & version
\texttt{1.0.10} & the layer and wiring graph \\
Hardware container & Mach-O shape & \texttt{0xbeefface} & register-write
program plus weights \\
\rowcolor{zebra}Dispatch descriptor & FlatBuffer & vtable
\texttt{0c00\ 1400\ 0400\ 0800} & the operation chain the runtime
submits \\
Loaded program image & Mach-O shape, signed & \texttt{0xbeefface} & the
same container, virtualized per load \\
\rowcolor{zebra}Firmware container & sectioned blob & \texttt{ANEH} /
\texttt{ANEP} / \texttt{ANES} & the on-package load form \\
\end{longtable}

The compiler produces a dispatch descriptor and a hardware container for
one network, cached together. The dispatch descriptor is above the
operation level of the intermediate language: it gives the operation
chain as a parametric program structure, with each operation reduced to
an argument frame and a kernel attribute blob. The hardware container is
below that level: it stores the register-write stream, weight
coefficients laid out for the streaming datapath, and typed input and
output layout, all in a Mach-O-shaped file with the magic
\texttt{0xbeefface}. The hardware container and the loaded program image
share one byte format: the loaded copy is the on-disk container after
the address relocations are resolved.

\begin{figure}[H]\centering
\includegraphics[width=5.153in]{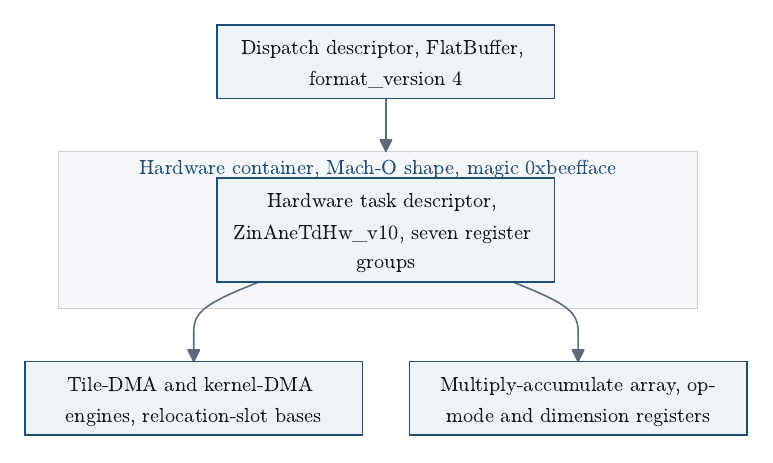}
\caption{The program format, with the dispatch descriptor over the hardware container over the task-descriptor register groups that control the DMA engines and the multiply array.}
\label{fig:c23-stack}
\end{figure}

The \cref{fig:c23-stack} shows the layering, with the dispatch
descriptor over the hardware container over the task-descriptor register
groups that drive the DMA engines and the multiply array.

\section{Dispatch descriptor}\label{dispatch-descriptor}

The dispatch descriptor is a FlatBuffer whose root table has four
fields. The serializer method set yields the root layout, validated byte
for byte against a real descriptor (\Cref{tbl:c23-root}).

\begin{longtable}[]{@{}
  >{\raggedright\arraybackslash}p{(\linewidth - 6\tabcolsep) * \real{0.2500}}
  >{\raggedright\arraybackslash}p{(\linewidth - 6\tabcolsep) * \real{0.2500}}
  >{\raggedright\arraybackslash}p{(\linewidth - 6\tabcolsep) * \real{0.2500}}
  >{\raggedright\arraybackslash}p{(\linewidth - 6\tabcolsep) * \real{0.2500}}@{}}
\caption{The four fields of the dispatch-descriptor root table, with
their vtable offsets, contents, and
types.}\label{tbl:c23-root}\tabularnewline
\toprule\noalign{}
\begin{minipage}[b]{\linewidth}\raggedright
\textbf{Field}
\end{minipage} & \begin{minipage}[b]{\linewidth}\raggedright
\textbf{Voffset}
\end{minipage} & \begin{minipage}[b]{\linewidth}\raggedright
\textbf{Contents}
\end{minipage} & \begin{minipage}[b]{\linewidth}\raggedright
\textbf{Type}
\end{minipage} \\
\midrule\noalign{}
\endfirsthead
\toprule\noalign{}
\begin{minipage}[b]{\linewidth}\raggedright
\textbf{Field}
\end{minipage} & \begin{minipage}[b]{\linewidth}\raggedright
\textbf{Voffset}
\end{minipage} & \begin{minipage}[b]{\linewidth}\raggedright
\textbf{Contents}
\end{minipage} & \begin{minipage}[b]{\linewidth}\raggedright
\textbf{Type}
\end{minipage} \\
\midrule\noalign{}
\endhead
\bottomrule\noalign{}
\endlastfoot
0 & 4 & the symbol and section name vector &
\texttt{symbol\_names:\ {[}string{]}} \\
\rowcolor{zebra}1 & 8 & build-info key and value pairs &
\texttt{build\_info:\ BuildInfo} \\
2 & 12 & the section descriptor vector &
\texttt{sections:\ {[}Section{]}} \\
\rowcolor{zebra}3 & 16 & inline scalar, value \texttt{4} &
\texttt{format\_version:\ int} \\
\end{longtable}

The operation set the descriptor can contain is fixed. The serializer
instantiates one attribute-serialization template per operation kind,
and the twelve instantiations enumerate the complete set that
\cref{lst:c23-schema} recovers as the root table and the operation-type
enumeration.

\begin{listing}[H]
\caption{The recovered FlatBuffer schema for the dispatch-descriptor root table and the twelve-member operation-type enumeration.}\label{lst:c23-schema}

\begin{Shaded}
\begin{Highlighting}[]
\CommentTok{// E5Program root table, recovered from the E5Serializer symbol family}
\CommentTok{// and validated against a real H13C dispatch descriptor (3816 B).}
\KeywordTok{table}\NormalTok{ E5Program \{}
\NormalTok{  symbol\_names: [}\DataTypeTok{string}\NormalTok{];     }\CommentTok{// operation, tensor, and section names}
\NormalTok{  build\_info: BuildInfo;      }\CommentTok{// compiler version and source path k/v}
\NormalTok{  sections: [Section];        }\CommentTok{// per{-}operation arg\_frame + op\_attrs refs}
\NormalTok{  format\_version: }\DataTypeTok{int}\NormalTok{;        }\CommentTok{// observed == 4}
\NormalTok{\}}

\KeywordTok{enum}\NormalTok{ OpType : }\DataTypeTok{ubyte}\NormalTok{ \{}
\NormalTok{  Cast, AneInference, EirInference, CpuInference, BnnsCpuInference,}
\NormalTok{  MlcCpuInference, MpsGraphInference, E5MinimalCpu, Quant, Dequant,}
\NormalTok{  Barrier, JitCall            }\CommentTok{// ordinals inferred; membership measured}
\NormalTok{\}}
\end{Highlighting}
\end{Shaded}

\end{listing}

The schema was recovered from the serializer symbol family rather than
from embedded reflection, since the runtime strips the binary schema and
the schema source. The deciding evidence is the \texttt{E5Serializer}
method set, one serializer method per schema table, with the
attribute-serialization template instantiated once per operation kind,
so the table list and the operation-type enumeration fall straight out
of the symbols. \Cref{tbl:c23-serializers} pairs each demangled
serializer method with the schema table or operation-type union it
proves.

\begin{longtable}[]{@{}
  >{\raggedright\arraybackslash}p{(\linewidth - 2\tabcolsep) * \real{0.5000}}
  >{\raggedright\arraybackslash}p{(\linewidth - 2\tabcolsep) * \real{0.5000}}@{}}
\caption{The dispatch-descriptor serializer methods, each proving one
schema table or the operation-type
union.}\label{tbl:c23-serializers}\tabularnewline
\toprule\noalign{}
\begin{minipage}[b]{\linewidth}\raggedright
\textbf{Demangled serializer method}
\end{minipage} & \begin{minipage}[b]{\linewidth}\raggedright
\textbf{Schema element it proves}
\end{minipage} \\
\midrule\noalign{}
\endfirsthead
\toprule\noalign{}
\begin{minipage}[b]{\linewidth}\raggedright
\textbf{Demangled serializer method}
\end{minipage} & \begin{minipage}[b]{\linewidth}\raggedright
\textbf{Schema element it proves}
\end{minipage} \\
\midrule\noalign{}
\endhead
\bottomrule\noalign{}
\endlastfoot
\texttt{SerializeFunction} & \texttt{table\ Function} \\
\rowcolor{zebra}\texttt{SerializeBlock} & \texttt{table\ Block} \\
\texttt{SerializeOperation} & \texttt{table\ Operation} \\
\rowcolor{zebra}\texttt{SerializeOpArgFrame} & the
\texttt{\_\_arg\_frame} section of an operation \\
\texttt{SerializeOperand} & \texttt{table\ Operand} \\
\rowcolor{zebra}\texttt{SerializeIOPort} & \texttt{table\ IOPort} \\
\texttt{SerializeAliasSymbol(string,\ uint,\ uint)} &
\texttt{table\ AliasSymbol\{name,\ symbol\_index,\ addr\_offset\}} \\
\rowcolor{zebra}\texttt{SerializeBuildInfo} &
\texttt{table\ BuildInfo} \\
\texttt{SerializeOpAttrs\textless{}CastOpT\textgreater{}} and 11 more &
\texttt{union\ OpAttrs} and \texttt{enum\ OpType} \\
\end{longtable}

The tensor and surface descriptor each have a fixed field list, lifted
verbatim from one runtime type-encoding string and reproduced in
\cref{lst:c23-tensordesc}.

\begin{listing}[H]
\caption{The tensor and surface descriptor field list, recovered verbatim from a runtime type-encoding string.}\label{lst:c23-tensordesc}

\begin{Shaded}
\begin{Highlighting}[]
\CommentTok{/* TensorDescriptor, the per{-}operand layout record, from the runtime type encoding.}
\CommentTok{   Q = u64, i = i32; the two leading pointers are runtime{-}only and not serialized. */}
\KeywordTok{struct}\NormalTok{ TensorDescriptor }\OperatorTok{\{}
  \DataTypeTok{void}    \OperatorTok{*}\NormalTok{data}\OperatorTok{,} \OperatorTok{*}\NormalTok{reserved}\OperatorTok{;}     \CommentTok{/* runtime pointers, not on the wire */}
  \DataTypeTok{uint64\_t}\NormalTok{ dim}\OperatorTok{[}\DecValTok{4}\OperatorTok{],}\NormalTok{ stride}\OperatorTok{[}\DecValTok{4}\OperatorTok{];}
  \DataTypeTok{uint64\_t}\NormalTok{ width}\OperatorTok{,}\NormalTok{ height}\OperatorTok{,}\NormalTok{ channels}\OperatorTok{,}\NormalTok{ batch\_number}\OperatorTok{,}\NormalTok{ sequence\_length}\OperatorTok{;}
  \DataTypeTok{uint64\_t}\NormalTok{ stride\_width}\OperatorTok{,}\NormalTok{ stride\_height}\OperatorTok{,}\NormalTok{ stride\_channels}\OperatorTok{;}
  \DataTypeTok{uint64\_t}\NormalTok{ stride\_batch\_number}\OperatorTok{,}\NormalTok{ stride\_sequence\_length}\OperatorTok{;}
  \DataTypeTok{int32\_t}\NormalTok{  storage\_type}\OperatorTok{;}        \CommentTok{/* the element{-}type code */}
\OperatorTok{\};}
\end{Highlighting}
\end{Shaded}

\end{listing}

The canonical program shape is a fused compute operation bracketed by
input and output format casts. A network that fuses to one engine
dispatch decodes as \texttt{Op0\_Cast} then \texttt{Op1\_AneInference}
then \texttt{Op2\_Cast}. The cast operations convert the host
half-precision layout to the engine-internal interleaved layout on input
and back on output; the single inference operation is the entire fused
graph.

Each operation has two sections: an argument frame that names the
operand binding and the per-axis tensor layout, and an attribute blob
that sets the kernel parameters. The attribute blob is a nested
FlatBuffer sub-descriptor, not expanded microcode: its byte diversity is
0.19, the low value of a structured and sparse record, and the
descriptor size does not change with the compute shape. A contraction
over an inner dimension of 64 and one over an inner dimension of 256
both produce the same descriptor size, so the compute shape is a
parameter the descriptor records.

Two properties follow from this structure. The descriptor is
depth-invariant: a one-operation graph and a six-operation fused graph
both reduce to a single inference operation of the same size. It tracks
the dispatch count: a bridge operation that cuts the graph into three
segments produces three inference operations and a larger descriptor.

\section{Hardware task descriptor}\label{hardware-task-descriptor}

The hardware task descriptor is the register-level program the engine
runs: each group below is a block of register writes that configures one
engine unit.

\begin{longtable}[]{@{}
  >{\raggedright\arraybackslash}p{(\linewidth - 8\tabcolsep) * \real{0.2000}}
  >{\raggedright\arraybackslash}p{(\linewidth - 8\tabcolsep) * \real{0.2000}}
  >{\raggedright\arraybackslash}p{(\linewidth - 8\tabcolsep) * \real{0.2000}}
  >{\raggedright\arraybackslash}p{(\linewidth - 8\tabcolsep) * \real{0.2000}}
  >{\raggedright\arraybackslash}p{(\linewidth - 8\tabcolsep) * \real{0.2000}}@{}}
\caption{The seven register groups of the hardware task descriptor, each
with its struct base offset, register count, addressable aperture base,
and contents.}\label{tbl:c23-td-groups}\tabularnewline
\toprule\noalign{}
\begin{minipage}[b]{\linewidth}\raggedright
\textbf{Group}
\end{minipage} & \begin{minipage}[b]{\linewidth}\raggedright
\textbf{Struct base}
\end{minipage} & \begin{minipage}[b]{\linewidth}\raggedright
\textbf{Registers}
\end{minipage} & \begin{minipage}[b]{\linewidth}\raggedright
\textbf{Register base}
\end{minipage} & \begin{minipage}[b]{\linewidth}\raggedright
\textbf{Contents}
\end{minipage} \\
\midrule\noalign{}
\endfirsthead
\toprule\noalign{}
\begin{minipage}[b]{\linewidth}\raggedright
\textbf{Group}
\end{minipage} & \begin{minipage}[b]{\linewidth}\raggedright
\textbf{Struct base}
\end{minipage} & \begin{minipage}[b]{\linewidth}\raggedright
\textbf{Registers}
\end{minipage} & \begin{minipage}[b]{\linewidth}\raggedright
\textbf{Register base}
\end{minipage} & \begin{minipage}[b]{\linewidth}\raggedright
\textbf{Contents}
\end{minipage} \\
\midrule\noalign{}
\endhead
\bottomrule\noalign{}
\endlastfoot
Kernel and common & \texttt{+0x2c} & 34 & \texttt{0x5500} & kernel-DMA
enable, format, stride, task type, network id \\
\rowcolor{zebra}Dimensions & \texttt{+0xfc} & 19 & none & input and
output width, height, depth, channels, groups, transpose \\
Tile DMA & \texttt{+0x150} & 69 & \texttt{0x4d00} & the three tile-DMA
engines: enable, cache hints, base address, strides \\
\rowcolor{zebra}Element-wise and planar & \texttt{+0x26c} & 30 &
\texttt{0x4100} & element-wise and planar-engine config, padding mode \\
L2 and texture & \texttt{+0x2ec} & 14 & \texttt{0x4500} & L2 source and
result config, texture mode \\
\rowcolor{zebra}Kernel format and op mode & \texttt{+0x32c} & 11 &
\texttt{0x4900} & op mode, kernel alignment, sparse and palette format,
bias enables \\
L2 result & \texttt{+0x360} & 21 & \texttt{0x5100} & L2-result base,
strides, wrap, result format \\
\end{longtable}

Between the parametric attribute blob and the register-write records the
engine executes is the hardware task descriptor. On the M1 generation it
is the versioned descriptor \texttt{ZinAneTdHw\_v10}, a flat register
image the compiler fills field by field and then serializes into address
and value pairs, partitioned into the seven groups
\Cref{tbl:c23-td-groups} lists. The version family is per generation:
the same descriptor is instantiated under a different version number for
each silicon family, and the byte offsets, register addresses, and field
widths move between versions.

The dimension fields show the hardware encoding behind the per-axis size
limits. Spatial dimensions are 15-bit and channel dimensions are 17-bit,
read from the register getters and confirmed against the vendor's own
field symbols, as \cref{lst:c23-td-fields} gives with each field's
descriptor offset and bit mask.

\begin{listing}[H]
\caption{The dimension and direct-memory-access fields of the hardware task descriptor, with their descriptor offsets and bit masks.}\label{lst:c23-td-fields}

\begin{Shaded}
\begin{Highlighting}[]
\CommentTok{/* ZinAneTdHw\_v10 dimension and DMA fields, offsets on the descriptor base.}
\CommentTok{   Field names and widths confirmed against the vendor TD symbol table. */}
\NormalTok{Win        @}\BaseNTok{0x0f4}\NormalTok{  mask }\BaseNTok{0x07fff}   \CommentTok{/* input width,  15{-}bit */}
\NormalTok{Hin        @}\BaseNTok{0x0f6}\NormalTok{  mask }\BaseNTok{0x07fff}   \CommentTok{/* input height, 15{-}bit */}
\NormalTok{Cin        @}\BaseNTok{0x100}\NormalTok{  mask }\BaseNTok{0x1ffff}   \CommentTok{/* input channels,  17{-}bit (up to 131071) */}
\NormalTok{Cout       @}\BaseNTok{0x104}\NormalTok{  mask }\BaseNTok{0x1ffff}   \CommentTok{/* output channels, 17{-}bit */}
\NormalTok{OCGSize    @}\BaseNTok{0x118}\NormalTok{  mask }\BaseNTok{0x07}      \CommentTok{/* output{-}channel{-}group size, 3{-}bit */}
\NormalTok{numGroups  @}\BaseNTok{0x11c}\NormalTok{  mask }\BaseNTok{0x1fff}    \CommentTok{/* conv groups, 13{-}bit */}

\CommentTok{/* The four DMA engines. Base addresses are relocation slots, resolved}
\CommentTok{   at load; strides and formats are written in{-}image. */}
\NormalTok{TileDMASrc1 base   reloc reg }\BaseNTok{0x1344}   \CommentTok{/* input tile read  */}
\NormalTok{TileDMASrc2 base   reloc reg }\BaseNTok{0x134a}   \CommentTok{/* second operand   */}
\NormalTok{TileDMADst  base   reloc reg }\BaseNTok{0x1442}   \CommentTok{/* output tile write */}
\NormalTok{KernelDMASrc       group G0 @}\BaseNTok{0x2c}     \CommentTok{/* weight banks, up to 16 sub{-}buffers */}
\end{Highlighting}
\end{Shaded}

\end{listing}

The compiler does not write buffer base addresses into the image at
compile time. It leaves each as a named relocation slot keyed by a
register address, and the loader patches the device address in. The full
v10 address-register relocation table is the input tile, second operand,
output tile, and four kernel coefficient streams, which
\Cref{tbl:c23-reloc} maps slot by slot.

\begin{longtable}[]{@{}ll@{}}
\caption{The v10 relocation-register map, the address slots left
symbolic at compile time and patched to device addresses at
load.}\label{tbl:c23-reloc}\tabularnewline
\toprule\noalign{}
\textbf{Relocation register} & \textbf{Slot} \\
\midrule\noalign{}
\endfirsthead
\toprule\noalign{}
\textbf{Relocation register} & \textbf{Slot} \\
\midrule\noalign{}
\endhead
\bottomrule\noalign{}
\endlastfoot
\texttt{0x1344} & input tile read base \\
\rowcolor{zebra}\texttt{0x134a} & second-operand read base \\
\texttt{0x1442} & output tile write base \\
\rowcolor{zebra}\texttt{0x1554} & kernel bias stream \\
\texttt{0x1558} & kernel post-scale stream \\
\rowcolor{zebra}\texttt{0x155c} & kernel palette-lookup stream \\
\texttt{0x1560} & kernel activation-lookup stream \\
\end{longtable}

The four weight sub-streams, bias, post-scale, palette lookup, and
activation lookup, each get their own relocation slot, so a convolution
with a folded batch normalization lowers to four independent weight
streams. Each direct-memory-access stride is a 26-bit signed field
located at bit 6 of its register word, range-checked against a per-chip
bound table. A tensor's element strides become these register words: a
row stride, channel stride, depth stride, and group stride per engine.

The loader patches more than buffer bases. Each relocation is a custom
bar that writes a resolved value, a device address, an on-chip-memory
data-set identifier, or a buffer offset, into an exact descriptor
bit-field named by its register, bit offset, and bit width.
Runtime-variable shapes and strides bind separately as live-in
parameters, each tied to a named procedure input and range-checked
against a start, stop, and step, so one compiled program serves a range
of input shapes without recompiling.

The descriptor layout is versioned per silicon generation: the M1 builds
the v10 image, and other generations build a different version under the
same family of accessors, with the dimension block, relocation-register
addresses, and field widths all moving between versions. The M1 v10
image omits the compute-cache direct-memory-access engine entirely:
every setter for the on-chip atomic, counter, and wait-event primitives
asserts that it is unsupported on this architecture. This is the
silicon-level reason a resident in-place state buffer cannot be
expressed on the M1 and falls back to a shared buffer. Any tool that
reads task-descriptor bytes must dispatch on the version number.

The serializer emits the image sparsely: a register reaches the stream
only when its value differs from the architectural default, which is why
a small contraction yields only six to eight records. Each surviving
record is a fixed 44 bytes: a count marker, the register address in one
IOMMU aperture, and one or two device addresses in a second aperture
that are written to the DMA base-address registers. The register
addresses are per-load device addresses, virtualized by the IOMMU, so
the same logical register resolves to different addresses across loads.

\section{A decoded program}\label{a-decoded-program}

A 64-to-64 half-precision linear layer with an identity weight matrix is
the smallest real program on the M1 host that includes a resolved layout
sidecar. Its container is a Mach-O-shaped file with the magic
\texttt{0xbeefface}, a pseudo-architecture marker, and the executable
file type. The resolved frame runs \texttt{{[}1,\ 1,\ 1,\ 64{]}}
half-precision input to \texttt{{[}1,\ 1,\ 1,\ 64{]}} half-precision
output, all strides 128 bytes.

The container maps each external tensor and the weight bank to its own
page-aligned aperture in the engine virtual address space, base
\texttt{0x30000000}, decoded segment by segment in
\Cref{tbl:c23-segmap}.

\begin{longtable}[]{@{}
  >{\raggedright\arraybackslash}p{(\linewidth - 10\tabcolsep) * \real{0.1667}}
  >{\raggedright\arraybackslash}p{(\linewidth - 10\tabcolsep) * \real{0.1667}}
  >{\raggedright\arraybackslash}p{(\linewidth - 10\tabcolsep) * \real{0.1667}}
  >{\raggedright\arraybackslash}p{(\linewidth - 10\tabcolsep) * \real{0.1667}}
  >{\raggedright\arraybackslash}p{(\linewidth - 10\tabcolsep) * \real{0.1667}}
  >{\raggedright\arraybackslash}p{(\linewidth - 10\tabcolsep) * \real{0.1667}}@{}}
\caption{The segment map of the decoded identity linear layer container,
with each segment's virtual address, size, protection, and
role.}\label{tbl:c23-segmap}\tabularnewline
\toprule\noalign{}
\begin{minipage}[b]{\linewidth}\raggedright
\textbf{Segment}
\end{minipage} & \begin{minipage}[b]{\linewidth}\raggedright
\textbf{Section}
\end{minipage} & \begin{minipage}[b]{\linewidth}\raggedright
\textbf{Virtual address}
\end{minipage} & \begin{minipage}[b]{\linewidth}\raggedright
\textbf{Size}
\end{minipage} & \begin{minipage}[b]{\linewidth}\raggedright
\textbf{Protection}
\end{minipage} & \begin{minipage}[b]{\linewidth}\raggedright
\textbf{Role}
\end{minipage} \\
\midrule\noalign{}
\endfirsthead
\toprule\noalign{}
\begin{minipage}[b]{\linewidth}\raggedright
\textbf{Segment}
\end{minipage} & \begin{minipage}[b]{\linewidth}\raggedright
\textbf{Section}
\end{minipage} & \begin{minipage}[b]{\linewidth}\raggedright
\textbf{Virtual address}
\end{minipage} & \begin{minipage}[b]{\linewidth}\raggedright
\textbf{Size}
\end{minipage} & \begin{minipage}[b]{\linewidth}\raggedright
\textbf{Protection}
\end{minipage} & \begin{minipage}[b]{\linewidth}\raggedright
\textbf{Role}
\end{minipage} \\
\midrule\noalign{}
\endhead
\bottomrule\noalign{}
\endlastfoot
guard & none & \texttt{0x00000000} & \texttt{0x4000} & none & guard
page \\
\rowcolor{zebra}input & const & \texttt{0x30008000} & \texttt{0x80} &
read & input aperture \\
output & data & \texttt{0x3000c000} & \texttt{0x80} & write & output
aperture \\
\rowcolor{zebra}text & text & \texttt{0x30010000} & \texttt{0x474} &
read and execute & register-write program \\
weights & kernel & \texttt{0x30018000} & \texttt{0x2000} & read & weight
coefficients \\
\end{longtable}

The segment protection encodes the hardware read and write direction:
read for the input and the weights, write for the output. Two named port
descriptors form the external binding table, each a 24-byte body of
\texttt{\{byteSize,\ apertureVA,\ nameRef\}}, binding \texttt{t0} to the
input aperture and \texttt{t2} to the output aperture.

The register-write program in the text section is 0x474 bytes: a 12-byte
header then a sparse packed register image holding exactly two
task-descriptor blocks, an input layout-convert and the fused
matrix-multiply, matching the two operations the symbol table names.
\Cref{tbl:c23-regprog} decodes the program record by record, each with
its offset, bytes, and meaning.

\begin{longtable}[]{@{}
  >{\raggedright\arraybackslash}p{(\linewidth - 4\tabcolsep) * \real{0.3333}}
  >{\raggedright\arraybackslash}p{(\linewidth - 4\tabcolsep) * \real{0.3333}}
  >{\raggedright\arraybackslash}p{(\linewidth - 4\tabcolsep) * \real{0.3333}}@{}}
\caption{The register-write program decoded record by record, with each
record's offset, bytes, and
meaning.}\label{tbl:c23-regprog}\tabularnewline
\toprule\noalign{}
\begin{minipage}[b]{\linewidth}\raggedright
\textbf{Offset}
\end{minipage} & \begin{minipage}[b]{\linewidth}\raggedright
\textbf{Bytes}
\end{minipage} & \begin{minipage}[b]{\linewidth}\raggedright
\textbf{Meaning}
\end{minipage} \\
\midrule\noalign{}
\endfirsthead
\toprule\noalign{}
\begin{minipage}[b]{\linewidth}\raggedright
\textbf{Offset}
\end{minipage} & \begin{minipage}[b]{\linewidth}\raggedright
\textbf{Bytes}
\end{minipage} & \begin{minipage}[b]{\linewidth}\raggedright
\textbf{Meaning}
\end{minipage} \\
\midrule\noalign{}
\endhead
\bottomrule\noalign{}
\endlastfoot
header & \texttt{00\ 00\ 00\ 00\ 00\ 00\ 9c\ 00\ 08\ 04} & size and
version words, register-count base \texttt{0x408} \\
\rowcolor{zebra}TD-0 +0x010 & \texttt{28\ 00\ 00\ 00\ ...} & record
marker \texttt{0x28}, a tile-DMA base relocation slot \\
TD-0 +0x034 & 16 times \texttt{80\ 00\ 00\ 00} & per-group value
\texttt{0x80} for the 16 output-channel-group partitions \\
\rowcolor{zebra}TD-0 +0x0c8 & \texttt{21\ a0\ 00\ 50\ 41\ 20\ 00\ 00} &
register \texttt{0x5021} set to \texttt{0x2041}, the kernel-common
record \\
TD-0 +0x128 & \texttt{31\ 20\ 00\ 01} & op-config descriptor
\texttt{0x2031}, starts the operation \\
\rowcolor{zebra}TD-1 +0x234 & 16 times \texttt{81\ 00\ 00\ 00} & the
second descriptor's 16 partitions \\
TD-1 +0x274 & \texttt{00\ 02\ 04\ ...\ 1e} & the 16 output-channel-group
offset indices \\
\rowcolor{zebra}TD-1 +0x470 & \texttt{31\ 20\ 30\ 01} & op-config
descriptor variant, terminates the program \\
\end{longtable}

The 16-wide register runs are the binary form of the 16
output-channel-group partitions, one per weight sub-kernel. The
kernel-common record \texttt{0x5021} set to \texttt{0x2041} is identical
across both descriptor blocks and matches the firmware's own
program-manager log byte for byte.

A live M1 program image has the same shape in the task-descriptor stream
itself. Its program-image text section contains three chained
\texttt{ZinAneTd\textless{}7u\textgreater{}} records linked by a
next-record pointer at offset \texttt{+0x1c} and followed to a null
terminator. Each operation descriptor has a 32-bit opcode word, and the
three operations of this program decode as the opcode words
\Cref{tbl:c23-opcodes} lists.

\begin{longtable}[]{@{}ll@{}}
\caption{The version-7 codegen opcode words of three operations decoded
from a live M1 task-descriptor
stream.}\label{tbl:c23-opcodes}\tabularnewline
\toprule\noalign{}
\textbf{Operation} & \textbf{Opcode word} \\
\midrule\noalign{}
\endfirsthead
\toprule\noalign{}
\textbf{Operation} & \textbf{Opcode word} \\
\midrule\noalign{}
\endhead
\bottomrule\noalign{}
\endlastfoot
Convolution & \texttt{0x5042a063} \\
\rowcolor{zebra}Reduce-mean & \texttt{0x5000a021} \\
Matrix multiply & \texttt{0x5000b021} \\
\end{longtable}

These are the version-7 H13 codegen opcodes. The high half-word,
\texttt{0x5042} or \texttt{0x5000}, is shared across operations, and the
low 16 bits distinguish the operation. The full operation identity is
spread across the lookup-table and configuration words as well as the
opcode word. The container closes with an \texttt{LC\_THREAD} trailer
holding the procedure and tensor names \texttt{main\_ane},
\texttt{t0\_ane}, and \texttt{t5\_ane@output}.

The container header is a 32-byte Mach-O-shaped block
(\Cref{tbl:c23-header}), and its load-command census names the binding
and configuration records the loader walks.

\begin{longtable}[]{@{}llll@{}}
\caption{The hardware-container header fields of the decoded identity
linear layer, with each field's value and
meaning.}\label{tbl:c23-header}\tabularnewline
\toprule\noalign{}
\textbf{Offset} & \textbf{Field} & \textbf{Value} & \textbf{Meaning} \\
\midrule\noalign{}
\endfirsthead
\toprule\noalign{}
\textbf{Offset} & \textbf{Field} & \textbf{Value} & \textbf{Meaning} \\
\midrule\noalign{}
\endhead
\bottomrule\noalign{}
\endlastfoot
\texttt{0x00} & magic & \texttt{0xbeefface} & engine compute
executable \\
\rowcolor{zebra}\texttt{0x04} & cputype & \texttt{0x80} & engine
pseudo-architecture \\
\texttt{0x08} & cpusubtype & \texttt{0x4} & the H13 codegen revision \\
\rowcolor{zebra}\texttt{0x0c} & filetype & \texttt{0x2} & executable \\
\texttt{0x10} & ncmds & 15 & load-command count \\
\rowcolor{zebra}\texttt{0x14} & sizeofcmds & \texttt{0x32d8} &
load-command bytes \\
\texttt{0x18} & flags & \texttt{0x200000} & \\
\end{longtable}

The 15 load commands of this container break down as seven segment
commands, two input and output port descriptors, three operation
descriptors, one build-info banner, one source-note, and one symbol
table. The two-port, three-operation-descriptor signature is the form of
a single-input, single-output graph: one header descriptor that lists
the aperture bases plus one descriptor per engine pass. The header
operation descriptor has an aperture-address table that pre-resolves the
segment bases to the engine virtual map, re-virtualized per load by the
input-output memory-management unit.

The build-info banner is the compiler invocation that produced the
container, recorded verbatim. It names the toolchain
(\texttt{zin\_ane\_compiler\ v9.509.0}, the intermediate-language
component \texttt{3520.4.1}) and the target plus the live flag set, of
which the relevant entries \cref{lst:c23-buildinfo} reproduces are the
target selector and the streaming and cache policy.

\begin{listing}[H]
\caption{The entries of the build-info banner, the target selector plus the streaming, cache, and fp8 policy flags.}\label{lst:c23-buildinfo}

\begin{Shaded}
\begin{Highlighting}[]
\NormalTok{{-}t h13g                              target H13, the M1 engine}
\NormalTok{{-}{-}fl2{-}cache{-}mode=resident            keep the L2 working set resident}
\NormalTok{{-}{-}fkernel{-}rewind=enabled             kernel{-}stream rewind}
\NormalTok{{-}{-}split{-}kernel{-}section=true          split the weight section}
\NormalTok{{-}{-}max{-}kernel{-}section{-}size=134217728  128 MB weight{-}section ceiling}
\NormalTok{{-}{-}e4m3{-}overflow{-}setting=Saturate     fp8 overflow policy: clamp, not NaN}
\NormalTok{{-}{-}memcache{-}size=4194304  {-}{-}bss{-}limit=3221225472}
\NormalTok{{-}{-}foptimize{-}ne{-}utilization=true  {-}{-}enable{-}global{-}cw{-}optimization=true}
\NormalTok{{-}i  .../model.mil   {-}o  .../model.hwx.tmp}
\end{Highlighting}
\end{Shaded}

\end{listing}

The container is self-describing. Its symbol table lists the 24-entry
element-type catalog with range encodings, half-precision as type 5, the
8-bit floating form as type 16, and 4-bit integer as type 23, alongside
the per-axis stride frame for each tensor as a typed array, both
reproduced in \cref{lst:c23-typecat}.

\begin{listing}[H]
\caption{The container’s element-type catalog and per-axis stride frame.}\label{lst:c23-typecat}

\begin{Shaded}
\begin{Highlighting}[]
\DataTypeTok{void}\OperatorTok{:}\NormalTok{t1  int8}\OperatorTok{:}\NormalTok{t2}\OperatorTok{=}\NormalTok{r2}\OperatorTok{;}\DecValTok{0}\OperatorTok{;}\DecValTok{127}\NormalTok{  uint8}\OperatorTok{:}\NormalTok{t3  int16}\OperatorTok{:}\NormalTok{t4  float16}\OperatorTok{:}\NormalTok{t5}\OperatorTok{=}\NormalTok{r1}\OperatorTok{;}\DecValTok{2}\OperatorTok{;}\DecValTok{0}  \DataTypeTok{float}\OperatorTok{:}\NormalTok{t6}\OperatorTok{=}\NormalTok{r1}\OperatorTok{;}\DecValTok{4}\OperatorTok{;}\DecValTok{0}
\OperatorTok{...}\NormalTok{ e4m3}\OperatorTok{:}\NormalTok{t16}\OperatorTok{=}\NormalTok{r1}\OperatorTok{;}\DecValTok{1}\OperatorTok{;}\DecValTok{0} \OperatorTok{...}\NormalTok{ int4}\OperatorTok{:}\NormalTok{t23}\OperatorTok{=}\NormalTok{r1}\OperatorTok{;{-}}\DecValTok{8}\OperatorTok{;}\DecValTok{7}
\NormalTok{t0}\OperatorTok{:}\NormalTok{t44}\OperatorTok{=}\NormalTok{ar1}\OperatorTok{;}\DecValTok{0}\OperatorTok{;}\DecValTok{1}\OperatorTok{;}\NormalTok{ s128n}\OperatorTok{:}\NormalTok{ s128c}\OperatorTok{:}\NormalTok{ s128h}\OperatorTok{:}\NormalTok{ s2w}\OperatorTok{:}\DecValTok{5}   \CommentTok{// batch, channel, height, width strides}
\end{Highlighting}
\end{Shaded}

\end{listing}

The container stores each weight tensor at a decoded symbol region with
a fixed tiling. A convolution weight is in a \texttt{0xC0}-stride layout
and a matrix-multiply weight in a \texttt{0x40}-stride layout, the two
strides \Cref{tbl:c23-wtstride} gives, and a weight-value edit leaves
the program descriptor unchanged, so weights are patchable in place.

\begin{longtable}[]{@{}lr@{}}
\caption{The decoded weight-region tiling stride per weight kind, the
fixed layout that lets a host patch weight values without
recompiling.}\label{tbl:c23-wtstride}\tabularnewline
\toprule\noalign{}
\textbf{Weight kind} & \textbf{Tiling stride} \\
\midrule\noalign{}
\endfirsthead
\toprule\noalign{}
\textbf{Weight kind} & \textbf{Tiling stride} \\
\midrule\noalign{}
\endhead
\bottomrule\noalign{}
\endlastfoot
Convolution weight & \texttt{0xC0} \\
\rowcolor{zebra}Matrix-multiply weight & \texttt{0x40} \\
\end{longtable}

The weight bank has 16 sub-kernels at a 512-byte stride. Decoding the
bytes proves the model: only 64 of the 8192 bytes are nonzero, each the
half-precision value \texttt{0x3c00} equal to 1.0, laid one per
partition across the 16 sub-kernels, the diagonal of a 64-by-64 identity
matrix. The container is the complete compiled form of an identity
linear layer that copies its input through the multiply-accumulate
array.

The dispatch descriptor and the container correspond to one network at a
time. A stateful recurrent model with one input and five held states
decodes as six input casts, one fused inference operation, and six
output casts, with the state buffers laid at 16-kilobyte strides. The
whole fused graph is still a single inference operation and a single
dispatch.

\section{On-disk container, decoded segment by
segment}\label{on-disk-container-decoded-segment-by-segment}

A second M1 sample, an \texttt{h13g} container of 64 kibibytes, confirms
the same format and contributes the deltas below; its segment map,
\texttt{+0x1c} next-record chain, \texttt{0xC0} and \texttt{0x40} weight
strides, build-info banner, and element-type catalog all match the
identity-layer decode above. The delta that proves the format is a
Mach-O variant rather than a Mach-O proper is the header magic:
\texttt{0xbeefface} is little-endian bytes \texttt{CE\ FA\ EF\ BE} where
a real 64-bit Mach-O would have \texttt{0xfeedfacf}. Patching those four
bytes lets \texttt{otool} and \texttt{rabin2} parse the file, while the
in-kernel loader accepts the engine magic directly.

The segment commands name \texttt{\_\_FVMLIB}, \texttt{\_\_TEXT}, and
\texttt{\_\_KERN\_0} regions. The two \texttt{\_\_FVMLIB} windows
declare the input and output tensors with \texttt{fileoff} and
\texttt{filesize} both zero: they have no on-disk bytes and exist only
as virtual-memory declarations that the kernel maps into the
address-translation aperture at submit time. The window sections align
to \texttt{0x4000}, the 16-kibibyte page granule of the
address-translation unit, while the task-descriptor and weight sections
align to \texttt{0x40}, the 64-byte engine tile granule. A pair of
window-binding load commands, each a
\texttt{\{u32\ size,\ u32\ pad,\ u64\ vmaddr,\ char\ name{[}8{]}\}}
record, binds the symbolic names \texttt{t0\_ane} and \texttt{t5\_ane}
to the input and output virtual addresses.

\subsection{Task-descriptor linked
list}\label{task-descriptor-linked-list}

This sample has three \texttt{ZinAneTd} records linked through the
\texttt{+0x1c} next-offset field at \texttt{0x000} to \texttt{0x300} to
\texttt{0x500} to terminator. Each descriptor header decodes into a
small set of fields, among them the next-offset field that chains the
list (\Cref{tbl:c23-td-header}).

\begin{longtable}[]{@{}
  >{\raggedright\arraybackslash}p{(\linewidth - 4\tabcolsep) * \real{0.3333}}
  >{\raggedright\arraybackslash}p{(\linewidth - 4\tabcolsep) * \real{0.3333}}
  >{\raggedright\arraybackslash}p{(\linewidth - 4\tabcolsep) * \real{0.3333}}@{}}
\caption{The decoded header fields of a \protect\texttt{ZinAneTd} task
descriptor, with the next-offset field at \protect\texttt{+0x1c} that
chains the list.}\label{tbl:c23-td-header}\tabularnewline
\toprule\noalign{}
\begin{minipage}[b]{\linewidth}\raggedright
\textbf{Offset}
\end{minipage} & \begin{minipage}[b]{\linewidth}\raggedright
\textbf{Field}
\end{minipage} & \begin{minipage}[b]{\linewidth}\raggedright
\textbf{Meaning}
\end{minipage} \\
\midrule\noalign{}
\endfirsthead
\toprule\noalign{}
\begin{minipage}[b]{\linewidth}\raggedright
\textbf{Offset}
\end{minipage} & \begin{minipage}[b]{\linewidth}\raggedright
\textbf{Field}
\end{minipage} & \begin{minipage}[b]{\linewidth}\raggedright
\textbf{Meaning}
\end{minipage} \\
\midrule\noalign{}
\endhead
\bottomrule\noalign{}
\endlastfoot
\texttt{+0x00} & index and flags & a 16-bit index plus a flags high
byte, where \texttt{0x03} marks the last or barrier record \\
\rowcolor{zebra}\texttt{+0x04} & size & a per-descriptor size or latency
field \\
\texttt{+0x08} & op and engine config & an operation-kind word whose low
byte distinguishes the engine pass \\
\rowcolor{zebra}\texttt{+0x10} & config mask & an engine configuration
mask sharing a \texttt{0x00fff8xx} prefix across records \\
\texttt{+0x18} & DMA word & the direct-memory-access or compute
descriptor word \\
\rowcolor{zebra}\texttt{+0x1c} & next offset & the byte offset of the
next descriptor, zero at the tail \\
\end{longtable}

The weight bases each descriptor consumes are not in the header but in
an inline array of relocations against \texttt{\_\_KERN\_0}. The twelve
relocations on \texttt{\_\_text} are the tile base offsets: the first
weighted descriptor reads eight tiles, the third reads four, and the
middle descriptor has no kernel relocation and so is a non-weighted pass
such as an activation, reshape, or pool.

\subsection{Weight tiling and the per-lane
naming}\label{weight-tiling-and-the-per-lane-naming}

The \texttt{\_\_kern\_0} section stores the half-precision weight
coefficients split into 64-byte-aligned tiles, one tile per engine
compute lane, named in the symbol table as
\texttt{K\textless{}sha256\textgreater{}\_ne\_\textless{}i\textgreater{}}
where the index \texttt{i} selects the lane. The first weight in this
sample, \texttt{K596A4B73…}, splits into eight tiles \texttt{\_ne\_0}
through \texttt{\_ne\_7} at the \texttt{0xc0}-byte convolution stride of
\texttt{3\ ×\ 64}. The second weight, \texttt{KE125552B…}, splits into
four tiles \texttt{\_ne\_0} through \texttt{\_ne\_3} at the
\texttt{0x40}-byte matrix-multiply stride, each tile a small weight
padded up to the 64-byte tile granule. The symbol table also lists tiles
\texttt{\_ne\_4} through \texttt{\_ne\_15} for that second weight, all
pointing at the section end as empty sentinel tiles for the unused
lanes, distinguished by a \texttt{desc} field of zero against
\texttt{0x0002} on the live tiles. The tiling rule this sample exhibits
is that a weight splits into \texttt{min(8,\ lanes)} tiles, each padded
to a multiple of 64 bytes and laid out contiguously, with each
descriptor referencing tile \texttt{i} by its 64-byte-aligned offset
through a relocation.

\subsection{Stabs-style shape
descriptors}\label{stabs-style-shape-descriptors}

The input and output tensors are shaped through stabs-style type
descriptors in the symbol table, the entries with symbol type
\texttt{0x20}, encoded as
\texttt{s\textless{}stride\textgreater{}\textless{}axis\textgreater{}}
byte-stride lists over the four axes \texttt{n} for batch, \texttt{c}
for channel, \texttt{h} for height, and \texttt{w} for width, as
\cref{lst:c23-stabs} gives for the bound input and output.

\begin{listing}[H]
\caption{The stabs-style shape descriptors for the bound input and output tensors, encoding the per-axis byte stride.}\label{lst:c23-stabs}

\begin{Shaded}
\begin{Highlighting}[]
\NormalTok{t0\_ane  }\OperatorTok{(}\NormalTok{input}\OperatorTok{,}\NormalTok{  vm }\BaseNTok{0x30000000}\OperatorTok{):}\NormalTok{  s8192n }\OperatorTok{/}\NormalTok{ s1024c }\OperatorTok{/}\NormalTok{ s64h }\OperatorTok{/}\NormalTok{ s2w}
\NormalTok{t5\_ane  }\OperatorTok{(}\NormalTok{output}\OperatorTok{,}\NormalTok{ vm }\BaseNTok{0x30004000}\OperatorTok{):}\NormalTok{  s64n   }\OperatorTok{/}\NormalTok{ s64c   }\OperatorTok{/}\NormalTok{ s64h }\OperatorTok{/}\NormalTok{ s2w}
\end{Highlighting}
\end{Shaded}

\end{listing}

The input strides describe an eight-channel tensor packed channel-major
into the 64-byte tile and 16-kibibyte page tiling. The width stride of 2
bytes is one half-precision element, the height stride of 64 bytes is
one tile row, and the channel stride of 1024 bytes is sixteen height
rows. The same symbol table enumerates the full hardware element-type
catalog, the wider set decoded under the element-type enumeration
section below.

\subsection{Build manifest}\label{build-manifest}

The build manifest matches the banner decoded above and also names the
bundle \texttt{com.apple.ANECompilerFramework}. Three
\texttt{LC\_THREAD} commands store the per-engine register and thread
state, the context-restore blobs the kernel pushes before it launches
the descriptor chain.

\section{Element-type and operation
enumerations}\label{element-type-and-operation-enumerations}

The serialization element-type enumeration the descriptor uses for an
operand is a closed set of eleven codes, which \Cref{tbl:c23-eltcodes}
gives.

\begin{longtable}[]{@{}llll@{}}
\caption{The eleven serialization element-type codes the dispatch
descriptor uses for an operand.}\label{tbl:c23-eltcodes}\tabularnewline
\toprule\noalign{}
\textbf{Code} & \textbf{Type} & \textbf{Code} & \textbf{Type} \\
\midrule\noalign{}
\endfirsthead
\toprule\noalign{}
\textbf{Code} & \textbf{Type} & \textbf{Code} & \textbf{Type} \\
\midrule\noalign{}
\endhead
\bottomrule\noalign{}
\endlastfoot
0 & int4 & 6 & uint16 \\
\rowcolor{zebra}1 & uint8 & 7 & int32 \\
2 & int8 & 8 & uint32 \\
\rowcolor{zebra}3 & float16 & 9 & int64 \\
4 & float32 & 10 & uint64 \\
\rowcolor{zebra}5 & int16 & & \\
\end{longtable}

The 8-bit floating forms and the palettized index types are a separate
hardware element-type space, not part of this serialization enumeration:
they are gated by the hardware-abstraction table and held as typed
fields. The container's symbol-table catalog runs to 24 entries because
it lists that wider hardware set.

The single inference operation the descriptor names is a fused graph of
backend operations, each one of a fixed micro-operation opcode space and
one operation-class selector. The operation-class enumeration, the unit
the descriptor calls the operation mode, has 79 members; the
compute-unit entries a developer reaches appear in
\Cref{tbl:c23-opclass}.

\begin{longtable}[]{@{}llll@{}}
\caption{Selected entries of the 79-member operation-class enumeration,
the selector the descriptor calls the operation
mode.}\label{tbl:c23-opclass}\tabularnewline
\toprule\noalign{}
\textbf{Class} & \textbf{Code} & \textbf{Class} & \textbf{Code} \\
\midrule\noalign{}
\endfirsthead
\toprule\noalign{}
\textbf{Class} & \textbf{Code} & \textbf{Class} & \textbf{Code} \\
\midrule\noalign{}
\endhead
\bottomrule\noalign{}
\endlastfoot
Conv & 1 & MatrixMultiplication & 18 \\
\rowcolor{zebra}Pooling & 2 & Reduction & 20 \\
Concat & 3 & Linear & 60 \\
\rowcolor{zebra}ElementWise & 4 & NEConv & 68 \\
ScaledElementWise & 5 & NEMatMul & 69 \\
\rowcolor{zebra}Neuron & 6 & NEPool & 70 \\
GOC & 8 & SDPA & 77 \\
\rowcolor{zebra}Softmax & 24 & AllReduce & 78 \\
\end{longtable}

There is no transposed-convolution class: a transposed convolution
lowers to a convolution, cross-correlation, or kernel rasterizer. The
full 79-member set and the 126-member micro-operation opcode space are
decoded in full in Appendix
\hyperref[appendix-c.-decoded-reference-tables]{C}.

\section{Submission ABI}\label{submission-abi}

The runtime hands the dispatch descriptor and the bound buffers to the
driver through one external-method interface keyed by a selector and a
structure size, and the kernel disambiguates an overloaded selector by
the structure size, method by method in \Cref{tbl:c23-abi}.

\begin{longtable}[]{@{}llll@{}}
\caption{The external-method selectors of the submission interface, with
the structure size that disambiguates each overloaded
selector.}\label{tbl:c23-abi}\tabularnewline
\toprule\noalign{}
\textbf{Selector} & \textbf{Method} & \textbf{In bytes} & \textbf{Out
bytes} \\
\midrule\noalign{}
\endfirsthead
\toprule\noalign{}
\textbf{Selector} & \textbf{Method} & \textbf{In bytes} & \textbf{Out
bytes} \\
\midrule\noalign{}
\endhead
\bottomrule\noalign{}
\endlastfoot
0 & device open & 104 & 104 \\
\rowcolor{zebra}2 & program send request & 2376 & 40 async \\
3 & program create & 32 & 0 \\
\rowcolor{zebra}3 & program output set enqueue & 40 & 0 \\
4 & program prepare & 56 & pointer \\
\rowcolor{zebra}4 & program inputs ready & 3104 & 0 \\
5 & program memory map request & 1 scalar & 2080 \\
\rowcolor{zebra}6 & program destroy & 16 & 0 \\
8 & program create instance & 32 & 0 \\
\rowcolor{zebra}8 & program chaining set active procedure & 32 & 0 \\
9 & program chaining prepare & 16 & pointer \\
\end{longtable}

The submit call is the asynchronous selector 2, whose 2376-byte argument
structure contains the program token, a sequence number, the
quality-of-service pair, and the array of input, output, and
intermediate surface identifiers. The 40-byte asynchronous reply gives
the sequence result and the echoed token, and completion arrives on a
wake port. The chaining-prepare and chaining-set-active selectors build
the firmware chaining cache, the structure a multi-segment descriptor
and the resident-state path use to wire one inference segment to the
next.

\chapter{HAL and capability gates}\label{hal-and-capability-gates}

\begin{summarybox}

The compiler is one binary that builds any chip in the line from a
per-chip data table, the hardware abstraction layer, read at compile
time. The table has scalar fields indexed by byte offset for the numeric
limits. A dense capability-byte region runs from offset \texttt{0x48f}
through \texttt{0x8cc}, each byte read as
\texttt{hal{[}offset{]}\ \&\ 1} to gate one operation or format. A
\texttt{MinimumFamily\textless{}N\textgreater{}} trait on the operation
declares its legality: native only when the target family index is
\texttt{N} or greater, decomposed below that floor. No compute operation
floors above A15. A capability in the table attests support at the layer
that reads it; it does not prove the operation runs. Three-dimensional
convolution has its kernel-depth attestation at \texttt{0x70} and fails
backend lowering on every device mask.

\end{summarybox}

What separates one target from the next is a per-chip data table, read
at compile time, that records every size limit and every per-operation
switch for that silicon.

\section{HAL property table}\label{hal-property-table}

The hardware-abstraction-layer table is the compiler's profile of a
target, one packed structure that contains both the numeric limits and
the feature gates for a chip, with representative scalar fields and
their M1 values in \Cref{tbl:c24-hal-scalars}.

\begin{longtable}[]{@{}
  >{\raggedright\arraybackslash}p{(\linewidth - 8\tabcolsep) * \real{0.2000}}
  >{\raggedright\arraybackslash}p{(\linewidth - 8\tabcolsep) * \real{0.2000}}
  >{\raggedright\arraybackslash}p{(\linewidth - 8\tabcolsep) * \real{0.2000}}
  >{\raggedright\arraybackslash}p{(\linewidth - 8\tabcolsep) * \real{0.2000}}
  >{\raggedright\arraybackslash}p{(\linewidth - 8\tabcolsep) * \real{0.2000}}@{}}
\caption{Representative scalar fields of the hardware-abstraction table;
the full decoded register map is in Appendix
\hyperref[appendix-c.-decoded-reference-tables]{C}.}\label{tbl:c24-hal-scalars}\tabularnewline
\toprule\noalign{}
\begin{minipage}[b]{\linewidth}\raggedright
\textbf{Offset}
\end{minipage} & \begin{minipage}[b]{\linewidth}\raggedright
\textbf{Field}
\end{minipage} & \begin{minipage}[b]{\linewidth}\raggedright
\textbf{Meaning}
\end{minipage} & \begin{minipage}[b]{\linewidth}\raggedright
\textbf{M1 (H13)}
\end{minipage} & \begin{minipage}[b]{\linewidth}\raggedright
\textbf{Changes at}
\end{minipage} \\
\midrule\noalign{}
\endfirsthead
\toprule\noalign{}
\begin{minipage}[b]{\linewidth}\raggedright
\textbf{Offset}
\end{minipage} & \begin{minipage}[b]{\linewidth}\raggedright
\textbf{Field}
\end{minipage} & \begin{minipage}[b]{\linewidth}\raggedright
\textbf{Meaning}
\end{minipage} & \begin{minipage}[b]{\linewidth}\raggedright
\textbf{M1 (H13)}
\end{minipage} & \begin{minipage}[b]{\linewidth}\raggedright
\textbf{Changes at}
\end{minipage} \\
\midrule\noalign{}
\endhead
\bottomrule\noalign{}
\endlastfoot
\texttt{0x1b8} & \texttt{max\_operand\_bytes} & on-chip SRAM working set
& 2 MB & constant across the line (1 MB on M9) \\
\rowcolor{zebra}\texttt{0x1c0} & \texttt{dram\_alignment} & DMA width
granule in bytes & 16 & constant (1 only on the small profiles) \\
\texttt{0x1c8} & \texttt{l2\_bank\_align} & DMA bank-conflict modulo &
64 & constant \\
\rowcolor{zebra}\texttt{0x1f0} & L2-resident buffer threshold &
dedicated-buffer trip & 0 & 32768 at A15, 262144 at A16 \\
\texttt{0x200} & dense kernel-memory cap & non-streamed weight ceiling &
64 KB & the fold-path budget \\
\rowcolor{zebra}\texttt{0x210} & streamed kernel-memory cap & streamed
weight ceiling & 16 MB & the stream-path budget \\
\texttt{0x218} & instruction or segment alignment & record packing
granule & 256 & 16 at A14 \\
\rowcolor{zebra}\texttt{0x228} & \texttt{ne\_perf\_cycle\_divisor} &
cost-model per-cycle divisor & 64 & 32 on H11, 16 on M9 \\
\end{longtable}

The hardware abstraction layer is a single packed structure the compiler
constructs for its target, holding two kinds of entry. The first is a
block of scalar fields, indexed by byte offset, that record numeric
limits: maximum kernel sizes, maximum tensor dimensions per axis, the
on-chip working-set size, data-movement alignment granule, and
cost-model curve. The second is a dense region of single-byte boolean
flags, the capability bytes, each gating one operation or one format on
or off for the target. A family of constructors, one per architecture,
builds the structure per target inside the compiler. The scalar region
runs from offset \texttt{0x18} to roughly \texttt{0x348} as plain data,
with non-scalar members such as the format map and the cost-model curve
extending past it. The capability-byte region occupies offsets
\texttt{0x48f} through \texttt{0x8cc} and has on the order of 165
single-byte flags, of which 24 have recovered field names and the rest
are enumerated by offset and classified by the family that enables them.
The compiler reads a scalar as a value at its offset and a capability
byte as \texttt{hal{[}offset{]}\ \&\ 1}, so every limit and every gate
is one indexed read into this one table (\Cref{lst:c24-hal-struct}). The
same structure stores the cost model: a policy-name string at
\texttt{0x580}, frequency-to-efficiency curve at \texttt{0x7a8}, and
per-cycle divisor at \texttt{0x228}, which the roofline of chapter
\hyperref[optimization-and-the-cost-model]{18} reads from this table.

\begin{listing}[H]
\caption{A partial C view of the per-target hardware-abstraction structure, with selected scalar fields and the capability-byte region at their recovered offsets.}\label{lst:c24-hal-struct}

\begin{Shaded}
\begin{Highlighting}[]
\CommentTok{/* ZinIrHalParameters, selected fields at their byte offsets (M1/H13 values) */}
\KeywordTok{struct}\NormalTok{ ZinIrHalParameters }\OperatorTok{\{}
    \CommentTok{/* ... */}
    \DataTypeTok{uint64\_t}\NormalTok{ max\_large\_conv\_kernel\_dim\_z}\OperatorTok{;}  \CommentTok{/* 0x70:  3D{-}conv kernel depth   = 16  */}
    \CommentTok{/* ... */}
    \DataTypeTok{uint64\_t}\NormalTok{ max\_tensor\_width}\OperatorTok{;}             \CommentTok{/* 0x138: max tensor width       = 16384 */}
    \CommentTok{/* ... */}
    \DataTypeTok{uint64\_t}\NormalTok{ max\_operand\_bytes}\OperatorTok{;}            \CommentTok{/* 0x1b8: SRAM working set       = 2 MB  */}
    \DataTypeTok{uint64\_t}\NormalTok{ dram\_alignment}\OperatorTok{;}               \CommentTok{/* 0x1c0: DMA width granule      = 16    */}
    \DataTypeTok{uint64\_t}\NormalTok{ l2\_bank\_align}\OperatorTok{;}                \CommentTok{/* 0x1c8: DMA/L2 bank count      = 64    */}
    \CommentTok{/* ... */}
    \DataTypeTok{uint64\_t}\NormalTok{ num\_nes}\OperatorTok{;}                      \CommentTok{/* 0x238: NE{-}core count          = 4     */}
    \CommentTok{/* ... */}
    \DataTypeTok{uint8\_t}\NormalTok{  cap\_bytes}\OperatorTok{[}\BaseNTok{0x8cd} \OperatorTok{{-}} \BaseNTok{0x48f}\OperatorTok{];}     \CommentTok{/* 0x48f..0x8cc: per{-}op capability flags */}
\OperatorTok{\};}
\end{Highlighting}
\end{Shaded}

\end{listing}

The capability bytes are read one at a time as
\texttt{hal{[}offset{]}\ \&\ 1}, for example the texture engine at
\texttt{0x81d} and the kernel-streaming master at \texttt{0x48f}.

\section{Operation gate}\label{operation-gate}

Operation legality is declared not in the HAL table but on the operation
itself, as a trait the compiler attaches to every backend operation. The
trait is a minimum-family index: the operation
\texttt{MinimumFamily\textless{}N\textgreater{}} is natively legal only
inside a compilation whose family index is \texttt{N} or greater, and
below that floor the compiler decomposes it into legal operations. The
family index orders the generations: \texttt{A11Legacy} is 0,
\texttt{A12} is 1, \texttt{A13} is 2, \texttt{A14} is 3, \texttt{A15} is
4, and so on, with the M1 at \texttt{A13} and the M5 at \texttt{A17}.

The check the compiler runs on each backend operation is the trait floor
against the target family index (\cref{lst:c24-family-gate}).

\begin{listing}[H]
\caption{The minimum-family gate, where an operation is emitted natively when the target family meets its floor and decomposed otherwise.}\label{lst:c24-family-gate}

\begin{Shaded}
\begin{Highlighting}[]
\CommentTok{\# mlir::OpTrait::anec::MinimumFamily\textless{}N\textgreater{}: native iff target family \textgreater{}= N}
\KeywordTok{def}\NormalTok{ op\_is\_native(op, target\_family):}
    \ControlFlowTok{return}\NormalTok{ target\_family }\OperatorTok{\textgreater{}=}\NormalTok{ op.minimum\_family   }\CommentTok{\# e.g. softmax N=2 (A13), sin N=4 (A15)}

\KeywordTok{def}\NormalTok{ lower\_op(op, target\_family):}
    \ControlFlowTok{if}\NormalTok{ op\_is\_native(op, target\_family):}
\NormalTok{        emit\_native(op)                          }\CommentTok{\# one anec op}
    \ControlFlowTok{else}\NormalTok{:}
\NormalTok{        decompose(op)                            }\CommentTok{\# rewrite into ops legal below the floor}
\end{Highlighting}
\end{Shaded}

\end{listing}

The M1 has family index two and the M5 has family index six, so an
operation with floor four, such as sin, is native on the M5 and
decomposed on the M1.

The floors fall into a small number of tiers. The base tier, family 0,
lists the operations every engine runs: convolution, matrix multiply,
pooling, the elementwise and activation set, reshape, transpose, and
concat. At A13 come softmax, the normalizations, the reductions, fused
attention, and the square-root and error functions. A14 brings the
texture-engine samplers, crop-resize, and resample; A15 brings native
sin and cos. No compute operation floors above A15, so the newest
generations add core count and clock.

The two gate mechanisms work together. A capability byte read as
\texttt{hal{[}offset{]}\ \&\ 1} decides a route inside a single
operation, for example whether the texture engine at byte \texttt{0x81d}
is present, which on the M1 reads 0 and forces resize to a
decomposition. The minimum-family trait decides whether the operation is
native at all. When either gate is closed, the compiler either emits a
decomposition into legal operations or rejects the operation with a
message naming the architecture, depending on whether a legal
decomposition exists.

The floors a developer reaches, and the families each is native on, are
in \Cref{tbl:c24-floors}.

\begin{longtable}[]{@{}
  >{\raggedright\arraybackslash}p{(\linewidth - 4\tabcolsep) * \real{0.3333}}
  >{\raggedright\arraybackslash}p{(\linewidth - 4\tabcolsep) * \real{0.3333}}
  >{\raggedright\arraybackslash}p{(\linewidth - 4\tabcolsep) * \real{0.3333}}@{}}
\caption{The minimum-family floors a developer reaches, with the
families each is native on and representative
operations.}\label{tbl:c24-floors}\tabularnewline
\toprule\noalign{}
\begin{minipage}[b]{\linewidth}\raggedright
\textbf{Floor}
\end{minipage} & \begin{minipage}[b]{\linewidth}\raggedright
\textbf{Native on}
\end{minipage} & \begin{minipage}[b]{\linewidth}\raggedright
\textbf{Representative operations}
\end{minipage} \\
\midrule\noalign{}
\endfirsthead
\toprule\noalign{}
\begin{minipage}[b]{\linewidth}\raggedright
\textbf{Floor}
\end{minipage} & \begin{minipage}[b]{\linewidth}\raggedright
\textbf{Native on}
\end{minipage} & \begin{minipage}[b]{\linewidth}\raggedright
\textbf{Representative operations}
\end{minipage} \\
\midrule\noalign{}
\endhead
\bottomrule\noalign{}
\endlastfoot
F0 & all families & convolution, matmul, pooling, elementwise, reshape,
transpose, concat \\
\rowcolor{zebra}F2 (A13+) & A13 onward & softmax, layer and instance and
batch norm, reductions, attention, erf, sqrt \\
F3 (A14+) & A14 onward & crop-resize, resample \\
\rowcolor{zebra}F4 (A15+) & A15 onward & sin, cos, global argmin and
argmax \\
\end{longtable}

The compiler text that rewrites an operation is identical across the
family, and the chip selects a different limit, gate, or decomposition
strategy from the table beneath it.

\section{Capability-byte gates across the
line}\label{capability-byte-gates-across-the-line}

A capability byte is a single-byte switch in that table that turns one
operation or feature on or off for a target. The named capability bytes,
each with its gate and its value across the M1 and the later
generations, are in \Cref{tbl:c24-capbytes}.

\begin{longtable}[]{@{}
  >{\raggedright\arraybackslash}p{(\linewidth - 12\tabcolsep) * \real{0.1429}}
  >{\raggedright\arraybackslash}p{(\linewidth - 12\tabcolsep) * \real{0.1429}}
  >{\raggedright\arraybackslash}p{(\linewidth - 12\tabcolsep) * \real{0.1429}}
  >{\raggedright\arraybackslash}p{(\linewidth - 12\tabcolsep) * \real{0.1429}}
  >{\raggedright\arraybackslash}p{(\linewidth - 12\tabcolsep) * \real{0.1429}}
  >{\raggedright\arraybackslash}p{(\linewidth - 12\tabcolsep) * \real{0.1429}}
  >{\raggedright\arraybackslash}p{(\linewidth - 12\tabcolsep) * \real{0.1429}}@{}}
\caption{The named capability
bytes.}\label{tbl:c24-capbytes}\tabularnewline
\toprule\noalign{}
\begin{minipage}[b]{\linewidth}\raggedright
\textbf{Byte}
\end{minipage} & \begin{minipage}[b]{\linewidth}\raggedright
\textbf{Gate}
\end{minipage} & \begin{minipage}[b]{\linewidth}\raggedright
\textbf{M1 (H13)}
\end{minipage} & \begin{minipage}[b]{\linewidth}\raggedright
\textbf{A14}
\end{minipage} & \begin{minipage}[b]{\linewidth}\raggedright
\textbf{A15}
\end{minipage} & \begin{minipage}[b]{\linewidth}\raggedright
\textbf{A16}
\end{minipage} & \begin{minipage}[b]{\linewidth}\raggedright
\textbf{A18}
\end{minipage} \\
\midrule\noalign{}
\endfirsthead
\toprule\noalign{}
\begin{minipage}[b]{\linewidth}\raggedright
\textbf{Byte}
\end{minipage} & \begin{minipage}[b]{\linewidth}\raggedright
\textbf{Gate}
\end{minipage} & \begin{minipage}[b]{\linewidth}\raggedright
\textbf{M1 (H13)}
\end{minipage} & \begin{minipage}[b]{\linewidth}\raggedright
\textbf{A14}
\end{minipage} & \begin{minipage}[b]{\linewidth}\raggedright
\textbf{A15}
\end{minipage} & \begin{minipage}[b]{\linewidth}\raggedright
\textbf{A16}
\end{minipage} & \begin{minipage}[b]{\linewidth}\raggedright
\textbf{A18}
\end{minipage} \\
\midrule\noalign{}
\endhead
\bottomrule\noalign{}
\endlastfoot
\texttt{0x48f} & kernel-streaming master, the 64 KB to 16 MB select & 1
& 1 & 1 & 1 & 1 \\
\rowcolor{zebra}\texttt{0x494} & square-after-reduction fusion & 0 & 1 &
1 & 1 & 1 \\
\texttt{0x4a9} & dropout and random & 0 & 0 & 1 & 1 & 1 \\
\rowcolor{zebra}\texttt{0x4f2} & global argmin and argmax & 1 & 1 & 1 &
1 & 1 \\
\texttt{0x529} & per-format kernel-stride enable, the palette stream & 1
& 1 & 1 & 1 & 1 \\
\rowcolor{zebra}\texttt{0x52d} & fp8 E4M3 kernel format & 0 & 0 & 0 & 0
& 1 \\
\texttt{0x563} & FIFO-mode direct memory access & 0 & 0 & 0 & 0 & 1 \\
\rowcolor{zebra}\texttt{0x815} & softmax, native & 1 & 1 & 1 & 1 & 1 \\
\end{longtable}

The texture engine at byte \texttt{0x81d} is the largest M1 functional
gap: it reads 0 on the M1 and 1 from A14 onward. It gates resize,
crop-resize, resample, affine transform, hardware gather, and symmetric
padding all together, so each of those routes through a software
decomposition on the M1. The fp8 byte \texttt{0x52d} is set on the A18
generation alone of the 28 targets, so the M5, an A17 part, does not
have it. The streaming master at byte \texttt{0x48f} and the
palette-stream byte \texttt{0x529} both read 1 on the M1, which is why
the int4 palette and the sparse form stream on the M1, while int8 and
blockwise fold, a mechanism chapter \hyperref[compression-internals]{25}
develops. The compiler builds the table for a target by calling that
target's constructor, so a single host recovers the table for every chip
in the line whether or not it is the chip that is running.

\section{Per-family scalar matrix}\label{per-family-scalar-matrix}

The scalar parameters show the same pattern across the generation
anchors: a value holds for a span of generations and then steps once
(\Cref{tbl:c24-scalar-matrix}).

\begin{longtable}[]{@{}llllll@{}}
\caption{The scalar parameters at the generation
anchors.}\label{tbl:c24-scalar-matrix}\tabularnewline
\toprule\noalign{}
\textbf{Field (offset)} & \textbf{M1 (H13)} & \textbf{A14} &
\textbf{A15} & \textbf{A16 (M4)} & \textbf{A17 (M5)} \\
\midrule\noalign{}
\endfirsthead
\toprule\noalign{}
\textbf{Field (offset)} & \textbf{M1 (H13)} & \textbf{A14} &
\textbf{A15} & \textbf{A16 (M4)} & \textbf{A17 (M5)} \\
\midrule\noalign{}
\endhead
\bottomrule\noalign{}
\endlastfoot
\texttt{num\_nes} (\texttt{0x238}) & 4 & 4 & 4 & 4 & 16 \\
\rowcolor{zebra}\texttt{max\_operand\_bytes} (\texttt{0x1b8}) & 2 MB & 2
MB & 2 MB & 2 MB & 2 MB \\
\texttt{max\_tensor\_width} (\texttt{0x138}) & 16384 & 16384 & 16384 &
65536 & 65536 \\
\rowcolor{zebra}\texttt{max\_tensor\_depth} (\texttt{0x158}) & 16384 &
16384 & 16384 & 65536 & 65536 \\
\texttt{max\_large\_conv\_kernel\_dim\_z} (\texttt{0x70}) & 16 & 16 & 16
& 16 & 16 \\
\rowcolor{zebra}L2-resident threshold (\texttt{0x1f0}) & 0 & 0 & 32768 &
262144 & 262144 \\
instruction alignment (\texttt{0x218}) & 256 & 16 & 16 & 16 & 16 \\
\rowcolor{zebra}reduction-transpose extent (\texttt{0x3f0}) & 192 & 192
& 384 & 384 & 384 \\
interchange-format count (\texttt{0x668}) & 3 & 13 & 16 & 14 & 14 \\
\end{longtable}

The base-name M5 reads \texttt{num\_nes} of 16 because the column is the
16-core Pro-class profile, while the base A17 profile has 4. The per-die
sequence runs 4 for the base name, 8 for the \texttt{g} suffix, 16 for
\texttt{s} and the legacy 16-core profile, 32 for \texttt{c}, and 64 for
the \texttt{d} Ultra-class die.

\section{Kernel-memory split}\label{kernel-memory-split}

The streaming master byte does more than gate the compressed-weight
stream: it selects which of two kernel-memory caps a layer's weights are
sized against. The legalization check is two lines of logic, reading a
streamable flag and the master byte to pick the offset of the cap, then
comparing the demand against it, as \cref{lst:c24-kmem-split} gives.

\begin{listing}[H]
\caption{The kernel-memory split, where a streamable weight under the streaming master is sized against the 16 MB cap and a dense weight against the 64 KB cap.}\label{lst:c24-kmem-split}

\begin{Shaded}
\begin{Highlighting}[]
\CommentTok{\# ExceedKmemSizeLimit: split{-}legalize a layer\textquotesingle{}s weights when they exceed the cap}
\KeywordTok{def}\NormalTok{ exceeds\_kmem(hal, demand, is\_streamable):}
\NormalTok{    cap }\OperatorTok{=}\NormalTok{ hal[}\BaseNTok{0x210}\NormalTok{] }\ControlFlowTok{if}\NormalTok{ (is\_streamable }\KeywordTok{and}\NormalTok{ hal[}\BaseNTok{0x48f}\NormalTok{]) }\ControlFlowTok{else}\NormalTok{ hal[}\BaseNTok{0x200}\NormalTok{]}
    \ControlFlowTok{return}\NormalTok{ cap }\OperatorTok{\textless{}}\NormalTok{ demand   }\CommentTok{\# 0x200 = 64 KB dense, 0x210 = 16 MB streamed}
\end{Highlighting}
\end{Shaded}

\end{listing}

An ordinary non-streamed weight over 64 KB, or any weight over 16 MB, is
thus split into multiple sub-layers on the M1, which raises the dispatch
count and the compile time. A streamed compressed weight is sized
against the 16 MB cap and has far more weight per layer. This is the
weight path; it does not bound the activations, which stay within the
maximum-tensor-dimension caps, so a layer with a tiny weight and a large
activation hits the tiling cost in the partition passes before it hits
this discrete limit.

\section{Dead and family-gated
fields}\label{dead-and-family-gated-fields}

Not every per-target value in the table is a live gate. A byte-granular
re-diff of all 28 target blobs leaves zero undecoded scalar fields, but
several offsets that vary by family are populated by the per-chip
builder and never read back through the table pointer, so their value is
a write-only mirror. Five scalar offsets are dead as table fields in
this fashion: the global element cap at \texttt{0x18}, kernel-depth
constant at \texttt{0x80}, legacy tiling granule at \texttt{0x260},
offset at \texttt{0x320}, and die-class flag at \texttt{0x29c}. Each
varies meaningfully by family, but the value a reader consumes is read
off a different object that shares the byte displacement, a
tensor-dimensions, compiler-parameters, or memory-pools structure, not
the table. The distinguishing test is whether the base register at the
access stores the table pointer, since the same displacement aliases
dozens of other by-reference structures, so a raw displacement match
inside a table-typed function is not proof of a table read.

The one offset that looks dead on the M1 but is not is the FIFO-mode
byte at \texttt{0x563}: it reads 0 on the M1 and is read through the
table pointer under a branch that is taken only when the byte is set,
which happens on the A18 generation. A per-family value pattern alone
does not establish a live gate; only a traced reader off the table base
does.

\section{Naming the remaining capability
flags}\label{naming-the-remaining-capability-flags}

The capability bytes and the scalar limits are both fields of one
struct, \texttt{ZinIrHalParameters}, the per-family blob the compiler
builds for its target. The struct has no per-field getter method, so the
compiler reads a field through an inlined \texttt{ldrb} or \texttt{ldr}
off the table pointer at a fixed offset, which is why a first pass
recovers offsets and values but not names. The names survive in one
place only. The compiler retains full mangled C++ symbols, and a reader
function whose signature has \texttt{ZinIrHalParameters\ const\&} reads
each field, so the reader's name labels the field it reads. A read is
attributed to the table only when the load's base register is the
function's \texttt{ZinIrHalParameters\ const\&} argument, since the same
byte displacement aliases dozens of other by-reference structures.

Cross-referencing the unnamed offsets against these reader functions
names 95 more of them, of which roughly 30 resolve to a precise
individual meaning with the base register verified against the table
argument (\Cref{tbl:c24-named-flags}).

\begin{longtable}[]{@{}
  >{\raggedright\arraybackslash}p{(\linewidth - 4\tabcolsep) * \real{0.3333}}
  >{\raggedright\arraybackslash}p{(\linewidth - 4\tabcolsep) * \real{0.3333}}
  >{\raggedright\arraybackslash}p{(\linewidth - 4\tabcolsep) * \real{0.3333}}@{}}
\caption{Precise capability-flag names recovered this round, each
attributed to its reader
function.}\label{tbl:c24-named-flags}\tabularnewline
\toprule\noalign{}
\begin{minipage}[b]{\linewidth}\raggedright
\textbf{Offset}
\end{minipage} & \begin{minipage}[b]{\linewidth}\raggedright
\textbf{Field}
\end{minipage} & \begin{minipage}[b]{\linewidth}\raggedright
\textbf{Reader function}
\end{minipage} \\
\midrule\noalign{}
\endfirsthead
\toprule\noalign{}
\begin{minipage}[b]{\linewidth}\raggedright
\textbf{Offset}
\end{minipage} & \begin{minipage}[b]{\linewidth}\raggedright
\textbf{Field}
\end{minipage} & \begin{minipage}[b]{\linewidth}\raggedright
\textbf{Reader function}
\end{minipage} \\
\midrule\noalign{}
\endhead
\bottomrule\noalign{}
\endlastfoot
\texttt{0x4a8} & PE work-unit-shape supported &
\texttt{PERasterization::ComputeWUShape} \\
\rowcolor{zebra}\texttt{0x4ac} & small-source-mode compression supported
& \texttt{ZinANELayer::AllowCompressionBasedOnSmallSourceMode} \\
\texttt{0x4b0} & non-power-of-2 work-unit width supported &
\texttt{NERasterization::CanUseNonPowerOf2WUs} \\
\rowcolor{zebra}\texttt{0x4f0} & preferred kernel layout format &
\texttt{ZinIrKernel::GetPreferredKernelLayoutFormat} \\
\texttt{0x500} & transpose and multicast configuration &
\texttt{ZinNELayer::FindValidMirInfoForTransposeCore} \\
\rowcolor{zebra}\texttt{0x520} & secure-mode cache-hint DSID gate &
\texttt{GetDSIDFromPriorityHalAndSecureMode} \\
\texttt{0x52c} & tensor-format support flag, pairs with the named
\texttt{0x52d} fp8 byte &
\texttt{ZinLayerValidationUtils::ValidateFormat} \\
\rowcolor{zebra}\texttt{0x54c} & cache-prefetch kernel-task-interval
limit &
\texttt{ZinValidateTd\textless{}17\textgreater{}::ValidateCachePrefetchKernelTaskInterval} \\
\end{longtable}

The remaining 95-minus-30 additions are class-named: the reader
identifies the subsystem the field gates without the exact semantics.
Examples are the per-axis DMA range bounds read by
\texttt{ZinValidateTd\textless{}N\textgreater{}::CheckInRangeDmaAccess}
and the texture-engine plane-equation coefficients in the \texttt{0x820}
to \texttt{0x8f8} block gated by the named \texttt{0x81d} texture-engine
byte on A14 and later. Two candidates were rejected as table fields
despite matching a displacement inside a table-typed function:
\texttt{0xcf8} loads off an \texttt{adrp}-formed read-only constant, and
\texttt{0x678} loads off a nested object two pointers deep. The same
base-register test that found the five dead fields above also rules
these out.

With this round the silicon-capability subset of the packed bitfield,
the part the compiler reads to gate a feature per family, is fully
named. The struct is \texttt{0x938} bytes, and the few entries inside it
that are not capability flags are the cost-model coefficient block at
offsets \texttt{0x580} through \texttt{0x7f0}. This block contains the
frequency-to-efficiency curve, rate indices, performance multiplier that
the roofline of chapter \hyperref[optimization-and-the-cost-model]{18}
reads, and about two soft fp64 coefficients. These are all performance
coefficients, with no legality cap among them. A small number of
capability fields are also true holdouts, read only off aliased bases.
The offsets past \texttt{0x938} have no table: an earlier reading that
located an A12 operation-emulation catalog at \texttt{0xa30} through
\texttt{0xe84} was a read into the adjacent zeroed memory beyond the
struct, not a real field.

\section{Attested is not reachable}\label{attested-is-not-reachable-1}

A capability recorded in the HAL table attests support at the layer that
reads the table. It does not by itself prove that the operation lowers
to a task descriptor and runs on the silicon. These are distinct layers,
and a capability present at the first can fail at the second.

The case that fixes the rule is three-dimensional convolution. The HAL
scalar at offset \texttt{0x70} records a 3D-conv kernel depth of 16 on
the M1, attesting that the kernel geometry is permitted, and the
compiler frontend recognizes the operation. It still fails backend
lowering on every device mask, returning the message that it is not
supported on any backend. The capability is in the table and the
operation does not run.

The gap appears in the other direction as well, where a checker accepts
an operation the code generator rejects: on the M1 the top-k, sort, and
dynamic-slice validators are all callable and all three are refused at
code generation. A bit in the table, a frontend that recognizes an
operation, or a validator that passes are each a claim about one layer;
only a compile-and-run on the target confirms the operation at the layer
that executes it. This is why the reachable surface of chapter
\hyperref[capability-surface]{4} is smaller than the surface the table
advertises, and why each native entry there was compiled and run on the
M1 rather than inferred from a capability byte.

\chapter{Compression internals}\label{compression-internals}

\begin{summarybox}

Every weight blob has a format enum, an integer 1 through 31, keying
five parallel helper tables that control all packing. The palette range
is 7 through 27; the unity range is 28 through 30. The engine
reconstructs four compression forms: per-tensor and per-channel int8
affine, int4 lookup-table, structured sparsity, and blockwise affine.
Each is a distinct dequantization chain the converter decodes into a
kernel format. Two gates decide whether a compressed weight streams or
folds to a dense half-precision constant. They are the kernel-streaming
master at offset \texttt{0x48f} and the per-format palette and affine
cluster at \texttt{0x520} through \texttt{0x539}. On the M1 only the
int4 lookup-table and structured-sparsity forms stream; int8 and
blockwise fold to dense through the dequantize-to-dense path.

\end{summarybox}

Chapter \hyperref[weights-and-compression]{7} stated which compressed
weight forms reach the engine and which save bandwidth. The mechanism
beneath that account follows: how a weight tensor becomes the compiled
weight blob, and how the compiler decides whether the compressed bytes
stream to the engine or fold to a dense half-precision constant first.
The path runs from a high-level dequantization operation chain down to
the per-output-channel-group records the firmware re-bases into on every
dispatch. Bit-layout and address detail are M1/H13, table-descriptor
codegen version five (family two, A13), unless another version or family
is named.

\section{Kernel-format enum and its
tables}\label{kernel-format-enum-and-its-tables}

Every weight blob has a format enum that names its storage encoding, an
integer in the range 1 through 31. All packing is controlled by five
parallel tables keyed on that enum (\Cref{tbl:c25-helpers}), and two
structural ranges fall out of their guards and recur throughout the
pipeline.

\begin{longtable}[]{@{}
  >{\raggedright\arraybackslash}p{(\linewidth - 2\tabcolsep) * \real{0.5000}}
  >{\raggedright\arraybackslash}p{(\linewidth - 2\tabcolsep) * \real{0.5000}}@{}}
\caption{The kernel-format helper functions keyed on the format enum and
what each one returns.}\label{tbl:c25-helpers}\tabularnewline
\toprule\noalign{}
\begin{minipage}[b]{\linewidth}\raggedright
\textbf{helper}
\end{minipage} & \begin{minipage}[b]{\linewidth}\raggedright
\textbf{returns}
\end{minipage} \\
\midrule\noalign{}
\endfirsthead
\toprule\noalign{}
\begin{minipage}[b]{\linewidth}\raggedright
\textbf{helper}
\end{minipage} & \begin{minipage}[b]{\linewidth}\raggedright
\textbf{returns}
\end{minipage} \\
\midrule\noalign{}
\endhead
\bottomrule\noalign{}
\endlastfoot
\texttt{\_ZinKernelFormatGetBitDepth(f)} & the index bit-width per
element \\
\rowcolor{zebra}\texttt{\_ZinKernelFormatGetPaletteFormat(f)} & the
palette sub-format, valid for \texttt{f} in 7 through 27 \\
\texttt{\_ZinKernelFormatGetUnderlyingType(f)} & the underlying scalar
type, int8, uint8, e4m3, or half \\
\rowcolor{zebra}\texttt{\_ZinKernelGetPaletteLUTSize(f,\ n)} & the
per-codebook byte stride times the codebook count \\
\texttt{ZinKernelFormatGetTypeno(f)} & the type number written into the
descriptor \\
\end{longtable}

The palette formats are the enum range 7 through 27, selected by the
guard \texttt{f\ -\ 7\ \textless{}\ 0x15}. The unity formats, the
identity and scale-only weights that skip the multiply array, are the
enum range 28 through 30, selected by
\texttt{f\ -\ 0x1c\ \textless{}\ 3}.

A compiled weight unit is a fixed set of sections, and the descriptor
that names them is the authoritative field list. The descriptor stores
an activation lookup table, palette lookup table, output-channel-group
table, per-output-channel scale vector, per-output-channel bias vector,
and the packed weight or index blob (\cref{lst:c25-kernel-unit}).

\begin{listing}[H]
\caption{The per-output-channel-group kernel-unit descriptor, the authoritative six-section field list emitted into the compiled weight section.}\label{lst:c25-kernel-unit}

\begin{Shaded}
\begin{Highlighting}[]
\NormalTok{t\%u = s\%lu}
\NormalTok{  lut:     ; activation LUT     (typeno, bitdepth, lut\_bytes)}
\NormalTok{  pal:     ; palette LUT        (palette\_format, bitdepth, num\_luts, lut\_bytes)}
\NormalTok{  ocgs:    ; output{-}channel{-}group table  (the per{-}OCG records)}
\NormalTok{  scale:   ; per{-}output{-}channel scale    (half or single precision)}
\NormalTok{  bias:    ; per{-}output{-}channel bias}
\NormalTok{  weights: ; the packed weight or index blob}
\end{Highlighting}
\end{Shaded}

\end{listing}

The palette table tuple gives the palette format, index bit-depth,
codebook count, and codebook byte size; a codebook count above one marks
the vector-palettized case. The activation table and the palette table
are peers in one descriptor.

The element types the descriptor writes for a weight section are a
subset of the 24-entry catalog. They are int8 as type 2, uint8 as type
3, half-precision as type 5, the lookup-table tag as type 8, 4-bit
unsigned as type 9, the 8-bit floating form as type 16, and 4-bit signed
as type 23. The half-precision type stores the codebook entries, scale,
and bias, so a palette codebook supplies the multiply array with no
further conversion.

\section{Four codecs}\label{four-codecs}

Each of the four codecs has its own dequantization relation, on-device
representation, native-stream family, and selecting compiler gate
(\Cref{tbl:c25-codecs}).

\begin{longtable}[]{@{}
  >{\raggedright\arraybackslash}p{(\linewidth - 8\tabcolsep) * \real{0.2000}}
  >{\raggedright\arraybackslash}p{(\linewidth - 8\tabcolsep) * \real{0.2000}}
  >{\raggedright\arraybackslash}p{(\linewidth - 8\tabcolsep) * \real{0.2000}}
  >{\raggedright\arraybackslash}p{(\linewidth - 8\tabcolsep) * \real{0.2000}}
  >{\raggedright\arraybackslash}p{(\linewidth - 8\tabcolsep) * \real{0.2000}}@{}}
\caption{The four weight-compression codecs, each with its
dequantization relation, native-stream family, and selecting compiler
gate.}\label{tbl:c25-codecs}\tabularnewline
\toprule\noalign{}
\begin{minipage}[b]{\linewidth}\raggedright
\textbf{form}
\end{minipage} & \begin{minipage}[b]{\linewidth}\raggedright
\textbf{dequant relation}
\end{minipage} & \begin{minipage}[b]{\linewidth}\raggedright
\textbf{representation}
\end{minipage} & \begin{minipage}[b]{\linewidth}\raggedright
\textbf{streams natively on}
\end{minipage} & \begin{minipage}[b]{\linewidth}\raggedright
\textbf{gate}
\end{minipage} \\
\midrule\noalign{}
\endfirsthead
\toprule\noalign{}
\begin{minipage}[b]{\linewidth}\raggedright
\textbf{form}
\end{minipage} & \begin{minipage}[b]{\linewidth}\raggedright
\textbf{dequant relation}
\end{minipage} & \begin{minipage}[b]{\linewidth}\raggedright
\textbf{representation}
\end{minipage} & \begin{minipage}[b]{\linewidth}\raggedright
\textbf{streams natively on}
\end{minipage} & \begin{minipage}[b]{\linewidth}\raggedright
\textbf{gate}
\end{minipage} \\
\midrule\noalign{}
\endhead
\bottomrule\noalign{}
\endlastfoot
int8 per-tensor / per-channel affine & \(w = s(q - z)\), \(z = 0\) on M1
& int8 byte stream plus per-channel half-precision scale & A14 and later
& folds on M1 via the dequantize-to-dense path \\
\rowcolor{zebra}int4 lookup-table &
\(w = \mathrm{LUT}[g/v][k][c \bmod v]\) & four-bit index stream plus
sixteen-entry half-precision codebook & M1/H13 and later & streaming
master plus the kernel-stride enable \\
structured sparsity & scatter of packed values into the mask & one-bit
mask plus packed half-precision nonzeros & M1/H13 and later & streaming
master, as a separate mask-and-values operand \\
\rowcolor{zebra}blockwise affine & \(w = s_b\,q\) per block & int8 byte
stream plus per-block half-precision scale & A15 and later & folds on M1
and M2 via the dequantize-to-dense path \\
\end{longtable}

Per-tensor and per-channel int8 use the affine form. The stored byte is
dequantized as

\[w = s\,(q - z)\]

with quantized byte \(q\), half-precision scale \(s\), and zero point
\(z\), the encode being \(q = \mathrm{round}(x / s) + z\). The scale is
scalar or one value per output channel. The M1 forces the zero point to
zero, so the form is symmetric and reduces to \(w = s\,q\): the
asymmetric setter hard-asserts on the version-five table descriptor, and
the firmware invariant requires the asymmetric-quantization
configuration bit clear.

The per-output-channel scale and bias do not cost a runtime operation:
the compiler folds them into the weight coefficients at compile time and
keeps them in the descriptor's scale and bias arrays as
per-output-channel vectors. The engine applies them at the
gain-offset-control stage as
\(y = s_{\mathrm{out}}\,(\sum_k w'_k x_k) + b_{\mathrm{out}}\) where the
coefficient already absorbs the dequantization scale. This is why a
symmetric quantized convolution on the M1 costs nothing extra for the
affine: the compiler folds the scale and the zero point into the stored
weights and the per-output-channel arrays before the dispatch.

The int4 lookup-table form stores a 4-bit index per element into a
16-entry half-precision codebook. Reconstruction is a table lookup with
no arithmetic, since the codebook entries are already half-precision and
supply the multiply array directly. The reconstructed weight is

\[w = \mathrm{LUT}\big[g / v\big]\big[\,k\,\big]\big[\,c \bmod v\,\big]\]

with output-channel-group index \(g\), palette vector size \(v\), the
four-bit index \(k\) for the element, and channel position \(c\). For
the common per-tensor case the vector size is one and a single codebook
has all 16 entries, so the relation flattens to one table lookup. A
4-bit weight has no affine path at all: there is no 4-bit underlying
scalar type in the type table, so it can only be a palette index. That
is why the lookup-table form is structurally a palette and is the one
form that streams on the M1.

The palette index width is 2, 4, 6, or 8 bits, but the M1-practical
widths are 4-bit, a 16-entry codebook, and 8-bit, a 256-entry codebook.
The 1-bit and 2-bit widths are rejected on the general path, and the
3-bit and 6-bit widths are version-gated to a later table descriptor,
while the quantized-palette combination is capped below 256 entries. A
legality mask constrains the vector size, the count of consecutive
output channels that share one codebook, to the set \(\{1, 2, 4, 5\}\).
The format selector tests \(2^{v}\) against the value \texttt{0x36},
which is \texttt{0b110110}, so vector sizes 0, 3, and any value of 6 or
more are illegal. A 4-bit codebook is 32 bytes and an 8-bit codebook is
512 bytes per lookup table, and the palette lookup table and the
activation lookup table share one on-chip-memory budget bounded by the
per-target palette-lookup-table size field.

Structured sparsity stores a one-bit nonzero mask plus the packed
half-precision values of the surviving nonzeros. The mask costs one bit
per element and the values cost two bytes per survivor, so a weight that
is half zeros or more stores well below its dense size. Reconstruction
walks the mask, consuming one packed value for each set bit and emitting
a zero for each clear bit, exact apart from the half-precision rounding
of the kept values (\Cref{lst:c25-recon}). Blockwise affine assigns a
separate scale to each contiguous block of elements, finer than a
per-channel scale and so lower in quantization error.

\begin{listing}[H]
\caption{The three reconstruction codecs in pseudocode, the affine dequantization, lookup-table reconstruction, and structured-sparsity scatter.}\label{lst:c25-recon}

\begin{Shaded}
\begin{Highlighting}[]
\CommentTok{\# int8 / uint8 affine dequant (constexpr\_affine\_dequantize)}
\KeywordTok{def}\NormalTok{ dequant\_affine(q, scale, zero\_point):     }\CommentTok{\# zero\_point forced to 0 on the M1}
    \ControlFlowTok{return}\NormalTok{ scale }\OperatorTok{*}\NormalTok{ (q }\OperatorTok{{-}}\NormalTok{ zero\_point)           }\CommentTok{\# symmetric M1 form: scale * q}

\CommentTok{\# int4 lookup{-}table reconstruction (constexpr\_lut\_to\_dense)}
\KeywordTok{def}\NormalTok{ dequant\_lut(index, codebook, g, vector\_size):}
    \ControlFlowTok{return}\NormalTok{ codebook[g }\OperatorTok{//}\NormalTok{ vector\_size][index][g }\OperatorTok{\%}\NormalTok{ vector\_size]   }\CommentTok{\# already fp16, no arithmetic}
    \CommentTok{\# per{-}tensor case: vector\_size == 1, one 16{-}entry codebook {-}\textgreater{} codebook[0][index][0]}

\CommentTok{\# structured{-}sparsity scatter (constexpr\_sparse\_to\_dense): 1{-}bit mask + packed fp16 nonzeros}
\KeywordTok{def}\NormalTok{ densify\_sparse(mask\_bits, nonzeros, n):}
\NormalTok{    out }\OperatorTok{=}\NormalTok{ [}\FloatTok{0.0}\NormalTok{] }\OperatorTok{*}\NormalTok{ n                           }\CommentTok{\# mask bit 1 = keep (nonzero)}
\NormalTok{    j }\OperatorTok{=} \DecValTok{0}
    \ControlFlowTok{for}\NormalTok{ i }\KeywordTok{in} \BuiltInTok{range}\NormalTok{(n):}
        \ControlFlowTok{if}\NormalTok{ mask\_bits[i]:                      }\CommentTok{\# LSB{-}first packed, one bit per element}
\NormalTok{            out[i] }\OperatorTok{=}\NormalTok{ nonzeros[j]              }\CommentTok{\# consume the next packed fp16 value}
\NormalTok{            j }\OperatorTok{+=} \DecValTok{1}
    \ControlFlowTok{return}\NormalTok{ out}
\end{Highlighting}
\end{Shaded}

\end{listing}

The sparse layout on the wire is the bitmask blob followed by the packed
values blob (\cref{lst:c25-sparse-layout}).

\begin{listing}[H]
\caption{The on-the-wire operand layout of the structured-sparsity form, the bitmask blob followed by the packed half-precision nonzeros.}\label{lst:c25-sparse-layout}

\begin{Shaded}
\begin{Highlighting}[]
\NormalTok{constexpr\_sparse\_to\_dense operand layout}
\NormalTok{  mask     : ceil(n }\OperatorTok{/} \DecValTok{8}\NormalTok{) }\BuiltInTok{bytes}   \CommentTok{\# 1 bit per element, LSB{-}first, dtype code UINT1 (9)}
\NormalTok{  nonzeros : }\DecValTok{2} \OperatorTok{*}\NormalTok{ popcount(mask) }\BuiltInTok{bytes}   \CommentTok{\# fp16 survivors in scan order, dtype code FP16 (1)}
  \CommentTok{\# streamed size \textasciitilde{} (1/16 + density) x the dense fp16 size}
\end{Highlighting}
\end{Shaded}

\end{listing}

\section{Stream-versus-fold decision}\label{stream-versus-fold-decision}

Two cooperating gates decide whether a compressed weight reaches the
engine in its compressed bytes or is materialized to a dense
half-precision constant before the dispatch. The master gate is the
kernel-streaming check. It returns false unless the
hardware-abstraction-layer streaming master bit at offset \texttt{0x48f}
is set, which holds from the A13 generation onward. It then requires the
format be primary or convertible to half-precision by the
direct-memory-access engine, plus unit stride on all axes, no dilation,
no tile overlap, and an immutable weight: a mutable weight, a
trained-in-place parameter, cannot stream.

A second gate guards the palette path. The native lookup-table form
streams only when the weight is vector-palettized and the per-format
kernel-stride-enable bit at offset \texttt{0x529} is set. When both
hold, the compiler sets the palette-enable and stream flags and zeroes
the fold sub-fields. On the M1 the offset \texttt{0x529} bit is set, so
the palette stream is live while the other formats fold.

Structured sparsity reaches the engine through the master gate as a
separate operand. The sparse weight lowers to a mask producer and a
nonzero-values producer held as their own blobs. Those stream under the
master at offset \texttt{0x48f}, which is set on the M1, not under the
per-format palette and affine cluster at offsets \texttt{0x520} through
\texttt{0x539}, which is clear on the M1. A live byte comparison
confirms the form: the sparse weight blob stores the one-bit mask and
the half-precision nonzeros at about 0.43 times the dense size, with no
dense half-precision constant present, whereas a fold would show a
single full-width half-precision blob.

The int8 and blockwise forms have no streamed encoding on the M1. The
converter builds the native quantized kernel, but the lowering routes it
through the dequantize-to-dense path, which reconstructs the weight to a
dense half-precision constant before the direct-memory-access transfer.
The stored bytes are then plain half-precision and move at full width,
so a weight-streaming-bound layer gets no bandwidth gain from these
forms on the M1.

A budget relaxation accompanies the stream decision. A streamed weight
is sized against the relaxed on-chip-memory cap at offset
\texttt{0x210}, while a folded or dense weight is sized against the
dense cap at offset \texttt{0x200}, which is sixty-four kilobytes on the
M1. The compressed path thus has far more weight per layer.

\section{Per-output-channel-group
packing}\label{per-output-channel-group-packing}

The weight section is a sequence of per-output-channel-group records,
each a fixed 14-word block holding a computed offset and size and ten
unrelocated address slots (\Cref{tbl:c25-ocg-record}).

\begin{longtable}[]{@{}
  >{\raggedright\arraybackslash}p{(\linewidth - 4\tabcolsep) * \real{0.3333}}
  >{\raggedright\arraybackslash}p{(\linewidth - 4\tabcolsep) * \real{0.3333}}
  >{\raggedright\arraybackslash}p{(\linewidth - 4\tabcolsep) * \real{0.3333}}@{}}
\caption{The per-output-channel-group record, the 14-word block with its
computed offset and size and its ten unrelocated address
slots.}\label{tbl:c25-ocg-record}\tabularnewline
\toprule\noalign{}
\begin{minipage}[b]{\linewidth}\raggedright
\textbf{Word}
\end{minipage} & \begin{minipage}[b]{\linewidth}\raggedright
\textbf{Field}
\end{minipage} & \begin{minipage}[b]{\linewidth}\raggedright
\textbf{Initial value}
\end{minipage} \\
\midrule\noalign{}
\endfirsthead
\toprule\noalign{}
\begin{minipage}[b]{\linewidth}\raggedright
\textbf{Word}
\end{minipage} & \begin{minipage}[b]{\linewidth}\raggedright
\textbf{Field}
\end{minipage} & \begin{minipage}[b]{\linewidth}\raggedright
\textbf{Initial value}
\end{minipage} \\
\midrule\noalign{}
\endhead
\bottomrule\noalign{}
\endlastfoot
\texttt{{[}0{]}} & output-channel-group byte offset into the section &
computed \\
\rowcolor{zebra}\texttt{{[}1{]}} & output-channel-group size, the
channel count for this engine & computed \\
\texttt{{[}2{]}} through \texttt{{[}0xb{]}} & ten address relocation
slots & \texttt{-1} \\
\rowcolor{zebra}\texttt{{[}0xc{]}}, \texttt{{[}0xd{]}} & reserved flags
& 0 \\
\end{longtable}

The ten slots left at \texttt{-1} are the per-record device addresses
the loader patches in: the input and output tile bases, per-buffer
coefficient sub-buffer bases, and four coefficient-stream bases for
bias, post-scale, palette lookup, and activation lookup. Those four
streams map one-to-one to four kernel direct-memory-access sub-channels
in the kernel-and-common register group, each with the enable,
base-offset, and relocation register in \Cref{tbl:c25-dma-streams}.

\begin{longtable}[]{@{}
  >{\raggedright\arraybackslash}p{(\linewidth - 6\tabcolsep) * \real{0.2500}}
  >{\raggedright\arraybackslash}p{(\linewidth - 6\tabcolsep) * \real{0.2500}}
  >{\raggedright\arraybackslash}p{(\linewidth - 6\tabcolsep) * \real{0.2500}}
  >{\raggedright\arraybackslash}p{(\linewidth - 6\tabcolsep) * \real{0.2500}}@{}}
\caption{The four kernel direct-memory-access coefficient streams, each
with its enable, base-offset, and relocation
register.}\label{tbl:c25-dma-streams}\tabularnewline
\toprule\noalign{}
\begin{minipage}[b]{\linewidth}\raggedright
\textbf{Descriptor slot}
\end{minipage} & \begin{minipage}[b]{\linewidth}\raggedright
\textbf{Enable register}
\end{minipage} & \begin{minipage}[b]{\linewidth}\raggedright
\textbf{Base-offset register}
\end{minipage} & \begin{minipage}[b]{\linewidth}\raggedright
\textbf{Relocation register}
\end{minipage} \\
\midrule\noalign{}
\endfirsthead
\toprule\noalign{}
\begin{minipage}[b]{\linewidth}\raggedright
\textbf{Descriptor slot}
\end{minipage} & \begin{minipage}[b]{\linewidth}\raggedright
\textbf{Enable register}
\end{minipage} & \begin{minipage}[b]{\linewidth}\raggedright
\textbf{Base-offset register}
\end{minipage} & \begin{minipage}[b]{\linewidth}\raggedright
\textbf{Relocation register}
\end{minipage} \\
\midrule\noalign{}
\endhead
\bottomrule\noalign{}
\endlastfoot
bias & \texttt{0x5548} & \texttt{0x554c} & \texttt{0x1554} \\
\rowcolor{zebra}post-scale & \texttt{0x5558} & \texttt{0x555c} &
\texttt{0x1558} \\
palette lookup & \texttt{0x5568} & \texttt{0x556c} & \texttt{0x155c} \\
\rowcolor{zebra}activation lookup & \texttt{0x5578} & \texttt{0x557c} &
\texttt{0x1560} \\
\end{longtable}

There is no zero-point stream, since the M1 symmetric form folds the
zero point to zero; the per-output-channel scale and bias streams supply
the dequantization scale instead. The M1 replicates the kernel
coefficients per engine core, one copy per core, because it lowers to
the per-core layout where the A14 generation and later use a shared
kernel-memory layout. A four-core M1 thus has four per-core records each
with its own output-channel-group vector.

The activation lookup table shares the descriptor with the palette
lookup table. It is a 43-entry half-precision record: two input-clamp
values, a mode flag, a scale, the 33 knot values uniform in the input
domain, four tail extrapolation coefficients, and a packed mode word. At
runtime the input maps affinely onto the 32 segments between the 33
knots, the integer part selects a segment, and the fraction controls a
half-precision linear interpolation, with the output clamping to the
end-knot beyond the input-clamp domain. The activation lookup table and
the palette lookup table are co-located in the weight section and
located by one offset cursor, which is why they draw on one shared
budget.

\section{Sparsity datapath}\label{sparsity-datapath}

The engine has two independent sparsity mechanisms. The first is
compute-time zero-skip. The table descriptor has a detect-zeros bit that
is implemented on the M1, and when it is set the multiply array skips a
multiply-accumulate whose weight is zero. The cost model scans the
weight values for their zero density once, caches the ratio, and reduces
the convolution cycle estimate by it, adding the implicit zeros that
strided and padded convolutions contribute. This mechanism is
format-independent: it fires on any kernel with zeros regardless of
storage encoding, including weights that fold to dense for storage. On a
bandwidth-bound stack the zero-skip alone moves the limit by about one
percent, because skipping multiply-accumulates in the array does not
help a layer whose limit is the weight stream.

The second is the sparse-binary store with on-chip decompress, which is
the source of the bandwidth gain. The mask and the packed nonzeros cross
the direct-memory-access stream, and the engine decompresses them on
chip into the dense tile. The on-device kernel-configuration register
has a single packed sparse-format field beside the palette-enable and
palette-bits fields in the same word. A six-by convolution stack at
sixty-three percent zeros streams the sparse form at about 0.43 times
the dense weight bytes and runs 1.55 to 1.64 times faster than the same
weights stored dense. Effective bandwidth rises from about 29 to about
48 gigabytes per second, which is the M1 weight-stream ceiling. The
compiled program is byte-identical between the dense and sparse bundles
at 2416 bytes, so the difference is entirely in the streamed weight
payload and its descriptor, which is the signature of a native stream
rather than a fold or a changed program.

The version-five table descriptor asserts that sparse-binary mode is not
supported, and this assert reconciles with the live stream because it
governs a different path. That assert bounds the sparse packing of the
palette-index plane inside the compiled weight blob, reachable only
through the vector-palette flag, so a non-palettized weight never
reaches it. The streamed sparse weight is the separate mask-and-values
operand described above, which does not pass through that
table-descriptor bit.

\section{fp8 datapath}\label{fp8-datapath}

The compiler has a complete fp8 datapath that no M1 generation can
reach. The 8-bit floating form is two distinct things in the binary: the
E4M3 form, a gated hardware weight and activation format with element
code \texttt{0xc}, and the E5M2 form, a conversion format with element
code \texttt{0xd}. E4M3 is the gated one: its kernel-format validity
admits it as a native weight element type, but the encoder that writes
its element code exists only from the A17 generation. The runtime
capability byte at \texttt{0x52d} that enables the format is set on the
A18 generation alone. On the M1 the format register is two bits wide
with no E4M3 codepoint at all, so the format never reaches a runtime
refusal; it triggers a compile-time assert. The accumulator stays at the
wide half-precision-class width on every family, so the 8-bit form is an
input and storage width supplying the same multiply array, not an
accumulate width. Its throughput runs on the same double-multiply path
the int8 form uses. The full fp8 datapath, format-register delta,
E5M2-to-half fold asymmetry, and conversion functions are decoded in
chapter \hyperref[predicted-upper-tier]{36}.

\section{Codegen version matrix}\label{codegen-version-matrix}

The same kernel format produces different weight bytes per
table-descriptor version, which is why the M1 and the M5 differ at the
bit level, as \Cref{tbl:c25-version-matrix} records capability by
capability.

\begin{longtable}[]{@{}
  >{\raggedright\arraybackslash}p{(\linewidth - 4\tabcolsep) * \real{0.3333}}
  >{\raggedright\arraybackslash}p{(\linewidth - 4\tabcolsep) * \real{0.3333}}
  >{\raggedright\arraybackslash}p{(\linewidth - 4\tabcolsep) * \real{0.3333}}@{}}
\caption{Codegen capability per table-descriptor version: the M1
version-five descriptor lacks the sparse-binary, multi-codebook, and
asymmetric encoders.}\label{tbl:c25-version-matrix}\tabularnewline
\toprule\noalign{}
\begin{minipage}[b]{\linewidth}\raggedright
\textbf{Capability}
\end{minipage} & \begin{minipage}[b]{\linewidth}\raggedright
\textbf{M1 version-five descriptor}
\end{minipage} & \begin{minipage}[b]{\linewidth}\raggedright
\textbf{Later descriptor}
\end{minipage} \\
\midrule\noalign{}
\endfirsthead
\toprule\noalign{}
\begin{minipage}[b]{\linewidth}\raggedright
\textbf{Capability}
\end{minipage} & \begin{minipage}[b]{\linewidth}\raggedright
\textbf{M1 version-five descriptor}
\end{minipage} & \begin{minipage}[b]{\linewidth}\raggedright
\textbf{Later descriptor}
\end{minipage} \\
\midrule\noalign{}
\endhead
\bottomrule\noalign{}
\endlastfoot
sparse-binary index packing & asserts unsupported & a packed flag at
descriptor offset \texttt{0x424} \\
\rowcolor{zebra}multi-codebook palette & asserts unsupported &
implemented \\
asymmetric quantization & asserts unsupported & implemented \\
\rowcolor{zebra}zero-skip detect & implemented & implemented \\
palette enable and bit width & implemented, 4 and 8 bit & implemented,
plus 3 and 6 bit \\
\end{longtable}

The M1 legal kernel-format space is thus the intersection of the format
enumeration and the version-five descriptor: dense half-precision,
palettized 4-bit and 8-bit including the single-codebook vector form,
symmetric int8 that folds to half-precision, and the unity forms. The
sparse-binary index packing, multi-codebook palette, asymmetric
quantization, fp8 form, and 3-bit and 6-bit palette widths are absent
from the version-five codegen entirely, a missing encoder rather than a
runtime refusal.

\section{Winograd and the compressed-weight
interaction}\label{winograd-and-the-compressed-weight-interaction}

The full Winograd eligibility gate is derived in chapter
\hyperref[datapath-and-mac-geometry]{20}: the enable bit, kernel-axis
and tile-size conditions, and the
\(\mathrm{OCG} \times K_y \times K_x \times K_d \times 2\) work
threshold against its non-float, float, and packed floors. The
compression-relevant point is the weight-format interaction: eligibility
requires a non-unity, non-sparse weight, so a sparse or unity-format
weight is excluded from the Winograd path and keeps its own datapath.
The transform matrices \(G\), \(B^\top\), and \(A^\top\) are resident in
the array and appear in no shippable weight blob, so no codec contains
them.

\chapter{Hidden layers and direct netplist authoring}\label{hidden-layers-and-direct-netplist-authoring}

\begin{summarybox}

The compiler validates and lowers a native catalog of 45 hardware layer
descriptors. The model converter surfaces only the subset its public
operation set covers. Authoring the network description directly reaches
the rest, including fused attention, ranking, spatial rearrangement, and
geometry layers, each cutting its own dispatch segment. Authoring does
not bypass a family gate. The texture-engine samplers are accepted from
the A14 and the whole-tensor arg-min and arg-max from the A15, both
rejected on the M1. A layer is confirmed only by a compile-and-run on
the target.

\end{summarybox}

The engine implements more layer kinds than the high-level conversion
path emits. Authoring the network description directly reaches the rest,
including several layers a later chip accepts and the M1 rejects.

The back half decodes this machinery in full: the \texttt{.espresso.net}
program format in chapter \hyperref[program-and-container-format]{23},
the compiler's layer parsers and validators in chapter
\hyperref[compiler]{22}, and the per-family capability gates in chapter
\hyperref[hal-and-capability-gates]{24}. Here it is enough that the
format is hand-authorable and that each layer has a validator and a
family gate; Appendix \hyperref[appendix-b.-hidden-layer-catalog]{B} is
the full per-layer schema.

\section{Framework and native
catalogs}\label{framework-and-native-catalogs}

The compiler has two parallel layer catalogs. The framework-level set is
the engine-agnostic kernel abstraction, about 190 layer types, of which
some run only on the host or the GPU. Below it is the native hardware
set: 45 descriptor structs, each with a constructor named
\texttt{\_ANEC\textless{}Name\textgreater{}LayerDescInitialize} and a
checker named \texttt{\_ANECValidate\textless{}Name\textgreater{}Layer},
and this set is the list of operations the engine silicon runs.

The conversion path from a trained model walks the documented
intermediate-language operation set
\hyperref[ref-applecoremltools]{{[}AppleCoreMLTools{]}} and emits only
the native layers that set maps onto. Several native descriptors have no
operation in that set, so the converter never produces them, and a model
passed through it decomposes the work into the operations it does know.
The converter always splits scaled dot-product attention into a matrix
multiply, scale, softmax, and second matrix multiply, because its
operation set has no atom that lowers to the single fused descriptor.
The descriptor and its validator are present in the compiler the whole
time. Reaching them means handing the compiler a network description
that names the layer directly.

\section{Method}\label{method}

A network expressed for the compiler is a property list, the
\texttt{.espresso.net} representation the compiler accepts alongside the
intermediate language. Its layers are dictionary entries the compiler
calls Units, each holding a \texttt{Type} tag, a \texttt{Bottom} wiring
list, an output type, and a \texttt{Params} sub-dictionary of typed
attributes, all under a \texttt{ProcedureList} of callable entry points
that name the \texttt{InputList}, \texttt{OperationList}, and
\texttt{OutputList}. The representation is hand-authorable, which is
what makes the hidden layers reachable.

The reusable method has four steps. Author one Unit whose \texttt{Type}
is the native layer name and whose \texttt{Params} contain the
descriptor attributes that layer's parser expects. Supply the wiring and
any constant weight blobs the layer needs. Compile the description
through the runtime, which lowers the Unit to its
\texttt{ANEC\textless{}Name\textgreater{}LayerDesc}, runs the matching
validator, and assigns it to an engine. Then load, bind buffers to the
named ports, and dispatch, exactly as chapter
\hyperref[dispatching-without-core-ml]{6} gives for any compiled
program. A raw native descriptor enters the network without
re-expressing it as a framework kernel through the tunneled-unit path:
the path passes the descriptor through the framework layer untouched,
with its float16 and integer operands intact, so a \texttt{Type} of
\texttt{SDPA} or \texttt{CostVolume} reaches the silicon directly.

The descriptor attributes are read out of the compiler's per-layer
parsers, the \texttt{ZinParse\textless{}Name\textgreater{}Unit}
routines, and cross-referenced constant-string tables. A required key
the parser does not find raises a parse error such as
\texttt{InvalidParamSyntax}, so an authored layer either has the exact
attribute set the parser expects or fails at compile time rather than at
dispatch.

\section{Classes reached this way}\label{classes-reached-this-way}

The reachable native layer kinds fall into several groups. Fused
attention is a single descriptor, scaled dot-product attention over four
or five operands of query, key, value, and constant scale, where the
optional fifth operand is an additive mask that records the causal and
decode cases as data. Ranking and selection cover the top-k, sort, and
argument-minimum-and-maximum descriptors, including the whole-tensor
argument form; their integer index outputs come back float16-encoded and
exact for the index ranges these layers produce. The spatial rearrangers
come as pairs, pixel-shuffle with pixel-unshuffle, the channel-and-space
pair, and the space-and-batch pair, each parameterized by per-axis
integer factors, and the two members of a pair differ by
channel-ordering convention, so neither is an alias of the other. Three
further normalizations appear: the range normalization that maps a
tensor to its minimum-to-maximum span, local response normalization, and
per-channel affine gain-offset control in static and runtime-tensor
forms. Geometry and point-cloud work draws on the template
cross-correlation, the three-vector cross product, furthest-point
sampling, radius neighborhood search, and the stereo cost volume. What
is left is data movement: the re-strided input view, runtime-offset
dynamic slice, and tile and concatenate descriptors. The full catalog of
these layers, with each layer's \texttt{Type} tag, its descriptor, and
its \texttt{Params} schema, is Appendix
\hyperref[appendix-b.-hidden-layer-catalog]{B}.

\section{A native layer descriptor}\label{a-native-layer-descriptor}

Take the fused-attention descriptor: the high-level path never emits it,
and its validator pins the shape exactly, down to the four-or-five
operand contract and the one attribute its parser reads
(\Cref{lst:c26-sdpa-unit}).

\begin{listing}[H]
\caption{A hand-authored fused-attention Unit naming the native scaled dot-product attention descriptor in the network description.}\label{lst:c26-sdpa-unit}

\begin{Shaded}
\begin{Highlighting}[]
\NormalTok{Unit }\StringTok{"attn"}\NormalTok{ \{}
\NormalTok{  Type    }\OperatorTok{=} \StringTok{"SDPA"}
\NormalTok{  Bottom  }\OperatorTok{=}\NormalTok{ [ }\StringTok{"q"}\NormalTok{, }\StringTok{"k"}\NormalTok{, }\StringTok{"v"}\NormalTok{, }\StringTok{"scale"}\NormalTok{ ]   }\CommentTok{\# optional 5th: additive mask}
\NormalTok{  Params  }\OperatorTok{=}\NormalTok{ \{ SubtractMax }\OperatorTok{=}\NormalTok{ true \}        }\CommentTok{\# ANECSDPALayerDesc byte 0x00}
\NormalTok{  OutputType }\OperatorTok{=} \StringTok{"Float16"}
\NormalTok{\}}
\end{Highlighting}
\end{Shaded}

\end{listing}

The validator enforces the operand count with
\texttt{4\ or\ 5\ bottoms\ must\ be\ present\ for\ SDPA}, requires the
key and value to share a shape, and checks that the query times the
transposed key contracts against the value. The \texttt{SubtractMax}
attribute is the single key the attention parser reads, and it defaults
to false in the descriptor constructor, which is numerically wrong for
softmax, so an authored attention Unit must set it true. The optional
fifth operand is an additive float16 mask broadcast over heads, zero on
and under the diagonal and a large negative bias above it for the causal
case.

At the backend-dialect level the same fused operation is one atom with a
parametric contraction, the same form the matrix multiply uses
(\cref{lst:c26-sdpa-atom}).

\begin{listing}[H]
\caption{The fused attention atom at the backend-dialect level, a single parametric contraction that covers the matrix-multiply, softmax, and transpose path.}\label{lst:c26-sdpa-atom}

\begin{Shaded}
\begin{Highlighting}[]
\FunctionTok{anec.sdpa}\NormalTok{(}\VariableTok{\%q}\NormalTok{, }\VariableTok{\%k}\NormalTok{, }\VariableTok{\%v}\NormalTok{, }\VariableTok{\%scale}\NormalTok{)   }\CommentTok{// 4 or 5 bottoms; covers matmul + softmax + transpose}
\FunctionTok{anec.matmul}\NormalTok{(}\VariableTok{\%lhs}\NormalTok{, }\VariableTok{\%rhs}\NormalTok{) \{ transpose\_lhs, transpose\_rhs \}   }\CommentTok{// depth D must be 1 on both operands}
\end{Highlighting}
\end{Shaded}

\end{listing}

The attention atom covers the matrix-multiply, softmax, and transpose
path and is not gated behind the texture engine, which is why it runs on
every family from the M1 onward.

\section{Authoring a hidden layer}\label{authoring-a-hidden-layer}

A native layer reaches the silicon by naming its descriptor in the graph
directly, then compiling and dispatching it as any other program, with
the target chosen as the first family that runs the layer. The compile
step runs the layer's validator and assigns the layer to an engine,
failing at compile time where the family gate rejects it, as
\cref{lst:c26-author} walks step by step.

\begin{listing}[H]
\caption{Authoring a hidden layer through the bridge route, then compiling and dispatching to confirm it lowers and runs.}\label{lst:c26-author}

\begin{Shaded}
\begin{Highlighting}[]
\CommentTok{\# Author a hidden layer via the bridge route: name the native layer kind and its parameters,}
\CommentTok{\# cut it into the graph as a bridge node, then compile and confirm it lowers.}

\NormalTok{graph G:}
    \BuiltInTok{input}\NormalTok{ q : [}\DecValTok{1}\NormalTok{, }\DecValTok{8}\NormalTok{, }\DecValTok{197}\NormalTok{, }\DecValTok{64}\NormalTok{] fp16}
    \BuiltInTok{input}\NormalTok{ k : [}\DecValTok{1}\NormalTok{, }\DecValTok{8}\NormalTok{, }\DecValTok{197}\NormalTok{, }\DecValTok{64}\NormalTok{] fp16}
    \BuiltInTok{input}\NormalTok{ v : [}\DecValTok{1}\NormalTok{, }\DecValTok{8}\NormalTok{, }\DecValTok{197}\NormalTok{, }\DecValTok{64}\NormalTok{] fp16}
\NormalTok{    const scale}

    \CommentTok{\# Describe the native layer directly as a bridge node, passed through untouched to the silicon.}
\NormalTok{    bridge\_node attn:}
\NormalTok{        kind    }\OperatorTok{=} \StringTok{"SDPA"}                      \CommentTok{\# the native descriptor name, not a decomposition}
\NormalTok{        inputs  }\OperatorTok{=}\NormalTok{ [ q, k, v, scale ]          }\CommentTok{\# 4 operands; optional 5th mask is the causal case}
\NormalTok{        params  }\OperatorTok{=}\NormalTok{ \{ subtract\_max }\OperatorTok{=}\NormalTok{ true \}     }\CommentTok{\# required: the default is false, wrong for softmax}

\NormalTok{    output attn}

\NormalTok{program P }\OperatorTok{=} \BuiltInTok{compile}\NormalTok{(G, target }\OperatorTok{=}\NormalTok{ H13)   }\CommentTok{\# runs the validator and assigns it to an engine}
\CommentTok{\# Fused attention runs from the M1 onward (not texture{-}gated). A family{-}gated layer would fail here,}
\CommentTok{\# at compile time, below its minimum family.}
\NormalTok{output }\OperatorTok{=}\NormalTok{ dispatch(P, q\_data, k\_data, v\_data)  }\CommentTok{\# confirm it lowers and runs on the target}
\end{Highlighting}
\end{Shaded}

\end{listing}

An authored layer is confirmed only by this compile-and-run on the
target, not by the presence of its descriptor, since a gated layer fails
the code generator below its minimum family.

\section{Arch-gated negatives}\label{arch-gated-negatives}

The same family gates as chapter \hyperref[across-the-chip-family]{12}
accept some authored layers on a later chip and reject them on the M1. A
native descriptor exists in the compiler binary on every target, but its
validator has a minimum-family trait, and below that family the code
generator rejects it.

The texture-engine samplers are the largest group: resize as a hardware
sampler, crop-and-resize, grid resample, and the affine spatial
transform are accepted from the A14 generation and rejected on the M1,
where the compiler reports that the affine transform is not supported on
this architecture. The whole-tensor argument-minimum-and-maximum layer
is gated to the A15 generation and rejected on the M1. Range
normalization is arch-gated and rejected on the M1. The native circular
state layers behind an on-device key-value cache are hard-gated on the
M1 and reached there only through a shared resident buffer. Authoring
the Unit does not bypass the gate: the layer compiles where its family
allows and fails at compile time where it does not, so the same network
description targets the generation that first runs the layer and every
generation above it.

A second class of rejection is not about family but about a layer that
passes an earlier check and fails the code generator, the
attested-is-not-reachable rule from chapter
\hyperref[capability-surface]{4}. On the M1 the top-k, sort, and
dynamic-slice validators are all callable, yet the code generator
rejects sort and dynamic-slice and accepts top-k only outside a small
forbidden parameter band.

\section{Two compile routes and their
gates}\label{two-compile-routes-and-their-gates}

The same operation can have different availability on the same chip
depending on which route reaches it. The conversion route from a
high-level model is gated by a minimum-family trait on each operation,
the floor the public conversion path checks. The direct-authoring route
is gated instead by the per-chip hardware-abstraction feature bytes the
layer validators read. The whole-tensor argument-minimum-and-maximum
layer splits along exactly that line: the conversion route floors it
above the M1, yet the direct-authoring route gates it on a feature byte
that is set from the A13 onward, so it is rejected through conversion on
the M1 and runs through direct authoring on the same chip. Trigonometric
sine and cosine have no direct-authoring bridge, so they stay
conversion-only and reject on the M1; sort and top-k have a bridge but
the code generator rejects it on the M1.

\section{Reference: the native layer classes reached by direct
authoring}\label{reference-the-native-layer-classes-reached-by-direct-authoring}

The native catalog is the set of layer kinds the engine silicon runs,
each with a constructor and a validator in the compiler. Each class has
a binding validator gate and a first family that runs it
(\cref{tbl:c26-classes}).

\begin{longtable}[]{@{}
  >{\raggedright\arraybackslash}p{(\linewidth - 6\tabcolsep) * \real{0.2500}}
  >{\raggedright\arraybackslash}p{(\linewidth - 6\tabcolsep) * \real{0.2500}}
  >{\raggedright\arraybackslash}p{(\linewidth - 6\tabcolsep) * \real{0.2500}}
  >{\raggedright\arraybackslash}p{(\linewidth - 6\tabcolsep) * \real{0.2500}}@{}}
\caption{The native layer classes direct authoring reaches; the full
per-layer descriptor and parameter schema is in Appendix
\hyperref[appendix-b.-hidden-layer-catalog]{B}.}\label{tbl:c26-classes}\tabularnewline
\toprule\noalign{}
\begin{minipage}[b]{\linewidth}\raggedright
\textbf{Class}
\end{minipage} & \begin{minipage}[b]{\linewidth}\raggedright
\textbf{Native layers}
\end{minipage} & \begin{minipage}[b]{\linewidth}\raggedright
\textbf{Binding gate}
\end{minipage} & \begin{minipage}[b]{\linewidth}\raggedright
\textbf{First family}
\end{minipage} \\
\midrule\noalign{}
\endfirsthead
\toprule\noalign{}
\begin{minipage}[b]{\linewidth}\raggedright
\textbf{Class}
\end{minipage} & \begin{minipage}[b]{\linewidth}\raggedright
\textbf{Native layers}
\end{minipage} & \begin{minipage}[b]{\linewidth}\raggedright
\textbf{Binding gate}
\end{minipage} & \begin{minipage}[b]{\linewidth}\raggedright
\textbf{First family}
\end{minipage} \\
\midrule\noalign{}
\endhead
\bottomrule\noalign{}
\endlastfoot
Fused attention & scaled dot-product attention & four or five operands;
key and value share a shape; the subtract-maximum attribute must be set
& M1, not texture-gated \\
\rowcolor{zebra}Ranking and selection & top-k, sort,
argument-minimum-and-maximum, whole-tensor argument form & index outputs
returned float16-encoded; sort and top-k pass the validator and fail the
M1 code generator & whole-tensor form A15; top-k and sort code-generated
above the M1 \\
Spatial rearrangement & pixel-shuffle, pixel-unshuffle,
space-and-channel, space-and-batch & per-axis integer factors that
factor into \{2, 3, 4, 8\}; depth factor one & M1 \\
\rowcolor{zebra}Normalization & range normalization, local response
normalization, per-channel gain-offset control & range normalization
arch-gated; gain-offset in static and runtime forms & range
normalization above the M1; gain-offset M1 \\
Geometry and point cloud & template cross-correlation, three-vector
cross product, furthest-point sampling, radius neighborhood, stereo cost
volume & cross product requires interleave one and float16 operands;
cross-correlation template depth one & M1 for the cross and correlation
forms \\
\rowcolor{zebra}Texture samplers & resize as a hardware sampler,
crop-and-resize, grid resample, affine transform & the texture-engine
feature byte & A14; rejected on the M1 \\
Data movement & re-strided input view, runtime-offset dynamic slice,
tile, concatenate & input view must follow a reshape; dynamic slice
passes the validator and fails the M1 code generator & M1 for the static
movers \\
\rowcolor{zebra}Streaming state & live state, ring-buffer reader and
writer, tensor-to-buffer movers & the ring-buffer writer must connect to
a live-state buffer; circular mode arch-gated & above the M1; reached on
the M1 only through a shared resident buffer \\
\end{longtable}

\section{Reference: the descriptor lowering of a fused
operation}\label{reference-the-descriptor-lowering-of-a-fused-operation}

Each atom's operand contract and the binding constraint its validator
enforces are in \Cref{tbl:c26-lowering}.

\begin{longtable}[]{@{}
  >{\raggedright\arraybackslash}p{(\linewidth - 4\tabcolsep) * \real{0.3333}}
  >{\raggedright\arraybackslash}p{(\linewidth - 4\tabcolsep) * \real{0.3333}}
  >{\raggedright\arraybackslash}p{(\linewidth - 4\tabcolsep) * \real{0.3333}}@{}}
\caption{The descriptor lowering of four representative atoms, with the
operand contract and the binding constraint each validator
enforces.}\label{tbl:c26-lowering}\tabularnewline
\toprule\noalign{}
\begin{minipage}[b]{\linewidth}\raggedright
\textbf{Atom}
\end{minipage} & \begin{minipage}[b]{\linewidth}\raggedright
\textbf{Operands}
\end{minipage} & \begin{minipage}[b]{\linewidth}\raggedright
\textbf{Binding constraint}
\end{minipage} \\
\midrule\noalign{}
\endfirsthead
\toprule\noalign{}
\begin{minipage}[b]{\linewidth}\raggedright
\textbf{Atom}
\end{minipage} & \begin{minipage}[b]{\linewidth}\raggedright
\textbf{Operands}
\end{minipage} & \begin{minipage}[b]{\linewidth}\raggedright
\textbf{Binding constraint}
\end{minipage} \\
\midrule\noalign{}
\endhead
\bottomrule\noalign{}
\endlastfoot
\texttt{sdpa} & four or five: query, key, value, scale, optional mask &
key and value share a shape; query times the transposed key contracts
against the value; softmax uses the family path \\
\rowcolor{zebra}\texttt{matmul} & two & depth one on both operands;
output channel equals the left channel; contraction parametric through
the transpose flags \\
\texttt{gain\_offset\_control} & one, plus scale and bias & per-channel
affine, height one; the fold target for a bias or activation \\
\rowcolor{zebra}\texttt{layer\_norm} & one, with gamma and beta folded &
channel divisible by the group count; output type float; the grouped
form is the same atom with a group count above one \\
\end{longtable}

\section{Programmable activation LUT}\label{programmable-activation-lut}

The named activations are not distinct hardware. The compiler
synthesizes each one, sigmoid, tanh, gelu, swish, and the rest, into a
33-knot piecewise-linear table over a fixed domain, and the engine
evaluates that table. The format is recovered byte for byte, 33 knot
values at a fixed step and 32 inter-knot deltas behind a short header,
and the operation set includes a custom-table opcode,
\texttt{kZinIrNonLinearCustomLUT}, that runs an arbitrary pointwise
function the same way.

The raw custom-table path is not reachable from a netplist. The unit
parser requires a saturation set and a version-specific set together,
and a consistency check in the same routine then rejects their
coexistence, so no authored table satisfies both and every attempt fails
to compile. The capability is real but sits below the user artifact: the
firmware synthesizes the table from the program at load.

An arbitrary pointwise function still runs on the engine by composition.
A piecewise-linear curve over chosen knots is a \texttt{linear},
\texttt{relu}, \texttt{linear} chain, the same form the hardware table
evaluates, so a small rectifier basis reproduces any knot table to fp16
exactly, demonstrated on a Gaussian bump that no named activation
provides. The engine's exposed model is thus a linear map, a pointwise
nonlinearity, and a linear map: custom weights and any scalar function,
but not a new arithmetic primitive, a custom reduction, or
data-dependent control flow.

\anepartopen{Part VIII}{System Internals}
\anechap{27}{Kernel driver and IOKit ABI}{The user clients, the selectors, and the IOKit ABI of the coprocessor endpoint.}
\anechap{28}{Address translation and the DART}{The engine IOMMU, the leaf page-table entry, and the firmware rebase.}
\anechap{29}{Firmware}{The real-time controller, its task model, and the dispatch loop.}
\anechap{30}{Host-to-firmware command protocol}{The ninety-three-command mailbox protocol across the host boundary.}
\anechap{31}{Power and thermal}{Idle and active draw, the credit sequence, and sustained thermal behavior.}
\anechap{32}{Security and isolation}{The signed-load chain, secure mode, and the exclave path on the M1.}
\anechap{33}{Telemetry and hardware counters}{The counter block, the stats-mask gate, and the free-running timestamp.}
\anepartclose

\chapter{Kernel driver and IOKit ABI}\label{kernel-driver-and-iokit-abi}

\begin{summarybox}

The engine is reached through a single kernel driver and a flat
user-client selector ABI. There are 17 control-client selectors and 9
direct-path selectors, each pinned to one exact size tuple. Opening
either user client is gated on the kernel entitlement
\texttt{com.apple.ane.iokit-user-access}, which exactly two system
binaries have. Every application therefore reaches the engine through a
privileged broker daemon. Selector 2 alone reaches a hardware doorbell
from user space, through a four-layer call path ending in an MMIO
mailbox write. The size tuples match across the M1 and M2-class kernel
cache, so the ABI at this layer is family-invariant.

\end{summarybox}

\section{Driver stack and class
hierarchy}\label{driver-stack-and-class-hierarchy}

The three cooperating kexts, their bundles, principal classes, and
provider matches appear in \Cref{tbl:c27-kext-stack}.

\begin{longtable}[]{@{}
  >{\raggedright\arraybackslash}p{(\linewidth - 6\tabcolsep) * \real{0.2500}}
  >{\raggedright\arraybackslash}p{(\linewidth - 6\tabcolsep) * \real{0.2500}}
  >{\raggedright\arraybackslash}p{(\linewidth - 6\tabcolsep) * \real{0.2500}}
  >{\raggedright\arraybackslash}p{(\linewidth - 6\tabcolsep) * \real{0.2500}}@{}}
\caption{Bundle, principal class, and provider match for each of the
three kernel extensions.}\label{tbl:c27-kext-stack}\tabularnewline
\toprule\noalign{}
\begin{minipage}[b]{\linewidth}\raggedright
\textbf{Kext role}
\end{minipage} & \begin{minipage}[b]{\linewidth}\raggedright
\textbf{Bundle}
\end{minipage} & \begin{minipage}[b]{\linewidth}\raggedright
\textbf{Principal class}
\end{minipage} & \begin{minipage}[b]{\linewidth}\raggedright
\textbf{Provider match}
\end{minipage} \\
\midrule\noalign{}
\endfirsthead
\toprule\noalign{}
\begin{minipage}[b]{\linewidth}\raggedright
\textbf{Kext role}
\end{minipage} & \begin{minipage}[b]{\linewidth}\raggedright
\textbf{Bundle}
\end{minipage} & \begin{minipage}[b]{\linewidth}\raggedright
\textbf{Principal class}
\end{minipage} & \begin{minipage}[b]{\linewidth}\raggedright
\textbf{Provider match}
\end{minipage} \\
\midrule\noalign{}
\endhead
\bottomrule\noalign{}
\endlastfoot
interface and hardware driver & \texttt{AppleH11ANEInterface} &
\texttt{ANEHWDevice} (registered as \texttt{H11ANEIn}) &
\texttt{RTBuddyService}, role \texttt{ANE} \\
\rowcolor{zebra}per-die abstraction layer & \texttt{AppleT8132ANEHAL} &
\texttt{AppleT8132ANEHAL} & \texttt{IOResources} \\
multi-engine arbiter & \texttt{AppleANELoadBalancer} &
\texttt{ANEDriver} (\texttt{H1xANELoadBalancer}) & \texttt{IOResources},
\texttt{IOKit} \\
\end{longtable}

All three are version-locked, at build \texttt{9.511.3} on the M1
generation. The interface and hardware driver registers the device,
vends the user clients, and rings the firmware. A per-die hardware
abstraction layer supplies the clock, power, and topology constants. The
load balancer owns the program-to-engine residency map and arbitrates
across physical engines on parts that have more than one. The engine
attaches as an Apple RTKit coprocessor endpoint, so the interface driver
is the host side of a real-time-operating-system mailbox client. The
work is spread over three kernel families: the address-translation
family that drives the device IOMMU, the surface family that backs
zero-copy tensor buffers, and the real-time mailbox family that queues
firmware commands.

\texttt{ANEHWDevice::newUserClient(task*,\ void*,\ uint\ type,\ IOUserClient**)}
vends two distinct user clients plus a hint-only client
(\Cref{lst:c27-userclients}).

\begin{listing}[H]
\caption{The three user-client types, by requested type.}\label{lst:c27-userclients}

\begin{Shaded}
\begin{Highlighting}[]
\CommentTok{/* ANEHWDevice::newUserClient vends, by requested type: */}
\NormalTok{H11ANEInUserClient        }\CommentTok{/* control client: program lifecycle + the inference hot path */}
\NormalTok{H11ANEInDirectPathClient  }\CommentTok{/* direct path: enqueue, memory{-}map, session{-}hint            */}
\NormalTok{ANEClientHints            }\CommentTok{/* scheduling{-}hint client (setClientHint)                    */}
\end{Highlighting}
\end{Shaded}

\end{listing}

The two functional clients use separate connections and separate
selector spaces that both start at zero, so one small selector integer
names different methods on each. Each stores its connection at object
offset \texttt{+0x40}, and every selector call loads that offset before
issuing the kernel call.

On the user-space side an \texttt{IOServiceOpen} whose return type is
\texttt{0xe00002c5} opens the device, and the connection object is at
offset \texttt{0x40} in the device handle. That type is opened by two
front-ends: \texttt{ANEServicesDevice} for inference and
\texttt{ANEHWDevice} for administration. The user-space selector
immediates read on the M1 are selector 0 for the device open, selector 2
for the send-request, selector 3 for create, selector 4 for prepare,
selector 6 for destroy, selector 10 for the version query, and selector
16 for the firmware load. The lifecycle indices differ from the
kernel-registered table because the kernel has two user-client classes
with independent selector tables. The two selectors agree across both
sides, selector 0 opening the device at 104 bytes and selector 2 sending
a request with a 2376-byte input and a 40-byte output.

Every selector shim reads its arguments out of the framework
\texttt{IOExternalMethodArguments} block at the fixed set of offsets in
\Cref{lst:c27-extmethodargs}, identical across all 26 selectors.

\begin{listing}[H]
\caption{The IOExternalMethodArguments field offsets every selector shim reads.}\label{lst:c27-extmethodargs}

\begin{Shaded}
\begin{Highlighting}[]
\CommentTok{/* IOExternalMethodArguments field offsets, used by every selector shim: */}
\OperatorTok{+}\BaseNTok{0x08}\NormalTok{  asyncWakePort       }\CommentTok{/* mach completion port (async selectors only)        */}
\OperatorTok{+}\BaseNTok{0x10}\NormalTok{  scalarInput}\OperatorTok{[]}       \CommentTok{/* also holds the 0x20{-}byte asyncReference block       */}
\OperatorTok{+}\BaseNTok{0x20}\NormalTok{  scalarInputCount}
\OperatorTok{+}\BaseNTok{0x30}\NormalTok{  structureInput      }\CommentTok{/* pointer to the typed argument struct                */}
\OperatorTok{+}\BaseNTok{0x38}\NormalTok{  structureInputSize}
\OperatorTok{+}\BaseNTok{0x48}\NormalTok{  scalarOutput}\OperatorTok{[]}
\OperatorTok{+}\BaseNTok{0x50}\NormalTok{  scalarOutputCount}
\OperatorTok{+}\BaseNTok{0x58}\NormalTok{  structureOutput}
\OperatorTok{+}\BaseNTok{0x60}\NormalTok{  structureOutputSize}
\end{Highlighting}
\end{Shaded}

\end{listing}

The shims share three return constants: \texttt{0xe00002c2} is
\texttt{kIOReturnBadArgument}, returned on a size or null-pointer check
failure, \texttt{0xe00002c7} is \texttt{kIOReturnUnsupported}, returned
on a disabled or stub path, and \texttt{0xe00002c5} is the
closed-or-closing client state.

\section{User-client dispatch-array
format}\label{user-client-dispatch-array-format}

Each user client routes a selector through the standard IOKit 2022
dispatch pattern. \texttt{H11ANEInUserClient::externalMethod} loads the
dispatch-array pointer and the method count, then tail-calls the
framework dispatcher, which validates the declared scalar and structure
sizes against the array entry before it calls the named handler. Both
dispatch arrays were read byte for byte out of the read-only data
section of the kernel cache (\Cref{lst:c27-dispatch-disasm}).

\begin{listing}[H]
\caption{The disassembled selector dispatch for each user client, with its dispatch-array pointer and selector count.}\label{lst:c27-dispatch-disasm}

\begin{Shaded}
\begin{Highlighting}[]
\CommentTok{/* H11ANEInUserClient::externalMethod, disassembled (M1, T6000): */}
\NormalTok{add x3}\OperatorTok{,}\NormalTok{ x3}\OperatorTok{,}\NormalTok{ \#}\BaseNTok{0xc08}    \CommentTok{/* x3 = \&\_sANEDriverClientMethods           */}
\NormalTok{mov w4}\OperatorTok{,}\NormalTok{ \#}\BaseNTok{0x11}         \CommentTok{/* count = 17 selectors (0..16)             */}
\NormalTok{bl  IOUserClient2022}\OperatorTok{::}\NormalTok{dispatchExternalMethod}

\CommentTok{/* H11ANEInDirectPathClient::externalMethod: */}
\NormalTok{add x3}\OperatorTok{,}\NormalTok{ x3}\OperatorTok{,}\NormalTok{ \#}\BaseNTok{0xeb0}    \CommentTok{/* x3 = \&\_sANEDriverDirectPathClientMethods */}
\NormalTok{mov w4}\OperatorTok{,}\NormalTok{ \#}\BaseNTok{0x9}          \CommentTok{/* count = 9 selectors (0..8)               */}
\NormalTok{bl  IOUserClient2022}\OperatorTok{::}\NormalTok{dispatchExternalMethod}
\end{Highlighting}
\end{Shaded}

\end{listing}

Each array element is one \texttt{IOExternalMethodDispatch2022} record
at a 40-byte stride, packing the authenticated handler pointer and the
four size checks the dispatcher enforces, shown in
\Cref{lst:c27-dispatch-elem}.

\begin{listing}[H]
\caption{The dispatch-array element layout, packing the handler pointer and the four size checks the dispatcher enforces.}\label{lst:c27-dispatch-elem}

\begin{Shaded}
\begin{Highlighting}[]
\CommentTok{/* IOExternalMethodDispatch2022 element, 40{-}byte stride, field offsets: */}
\KeywordTok{struct}\NormalTok{ IOExternalMethodDispatch2022 }\OperatorTok{\{}
  \DataTypeTok{void} \OperatorTok{*}\NormalTok{function}\OperatorTok{;}                  \CommentTok{/* +0x00  pointer{-}authenticated handler   */}
  \DataTypeTok{uint32\_t}\NormalTok{ checkScalarInputCount}\OperatorTok{;}  \CommentTok{/* +0x08  exact scalar{-}input count        */}
  \DataTypeTok{uint32\_t}\NormalTok{ checkStructureInputSize}\OperatorTok{;}\CommentTok{/* +0x0c  exact struct{-}input bytes        */}
  \DataTypeTok{uint32\_t}\NormalTok{ checkScalarOutputCount}\OperatorTok{;} \CommentTok{/* +0x10  exact scalar{-}output count       */}
  \DataTypeTok{uint32\_t}\NormalTok{ checkStructureOutputSize}\OperatorTok{;} \CommentTok{/* +0x14 exact struct{-}output bytes      */}
  \DataTypeTok{uint8\_t}\NormalTok{  reserved}\OperatorTok{[}\BaseNTok{0x10}\OperatorTok{];}         \CommentTok{/* +0x18  reserved (debug{-}WP group flag)  */}
\OperatorTok{\};}
\end{Highlighting}
\end{Shaded}

\end{listing}

No entry on either client uses the sentinel \texttt{0xffffffff} that
means ``do not check''. Every selector pins an exact scalar count and an
exact struct size in exactly one record, and the dispatcher rejects any
other size tuple with \texttt{kIOReturnBadArgument}.

\section{Control-client selector
table}\label{control-client-selector-table}

The control client has 17 selectors covering the open handshake, the
program lifecycle, status and version reads, and the firmware-driven
debug work-processor channel, with each handler and its
kernel-authoritative sizes read out of the dispatch array into
\Cref{tbl:c27-control-selectors}.

\begin{longtable}[]{@{}rlrrrr@{}}
\caption{The seventeen control-client selectors, with handler name and
kernel-authoritative scalar and structure
sizes.}\label{tbl:c27-control-selectors}\tabularnewline
\toprule\noalign{}
\textbf{Sel} & \textbf{Handler} & \textbf{Scalar in} & \textbf{Struct
in} & \textbf{Scalar out} & \textbf{Struct out} \\
\midrule\noalign{}
\endfirsthead
\toprule\noalign{}
\textbf{Sel} & \textbf{Handler} & \textbf{Scalar in} & \textbf{Struct
in} & \textbf{Scalar out} & \textbf{Struct out} \\
\midrule\noalign{}
\endhead
\bottomrule\noalign{}
\endlastfoot
0 & \texttt{ANE\_DeviceOpen} & 0 & 104 & 0 & 104 \\
\rowcolor{zebra}1 & \texttt{ANE\_DeviceClose} & 0 & 0 & 0 & 0 \\
2 & \texttt{ANE\_ProgramSendRequest} & 1 & 2376 & 0 & 40 \\
\rowcolor{zebra}3 & \texttt{ANE\_ProgramCreate} & 0 & 32 & 0 & 0 \\
4 & \texttt{ANE\_ProgramPrepare} & 0 & 56 & 0 & 56 \\
\rowcolor{zebra}5 & \texttt{ANE\_ProgramUnprepare} & 0 & 56 & 0 & 0 \\
6 & \texttt{ANE\_ProgramDestroy} & 0 & 16 & 0 & 0 \\
\rowcolor{zebra}7 & \texttt{ANE\_GetStatus} & 0 & 0 & 0 & 32 \\
8 & \texttt{ANE\_ProgramCreateInstance} & 0 & 32 & 0 & 0 \\
\rowcolor{zebra}9 & \texttt{ANE\_ProgramChainingPrepare} & 0 & 16 & 0 &
24 \\
10 & \texttt{ANE\_GetVersion} & 0 & 0 & 1 & 0 \\
\rowcolor{zebra}11 & \texttt{ANE\_RegisterDebugWorkProcessor} & 0 & 24 &
0 & 0 \\
12 & \texttt{ANE\_UnregisterDebugWorkProcessor} & 0 & 0 & 0 & 0 \\
\rowcolor{zebra}13 & \texttt{ANE\_GetDebugWorkProcessorItem} & 2 & 0 & 0
& 0 \\
14 & \texttt{ANE\_CompleteDebugWorkProcessorItem} & 2 & 0 & 0 & 0 \\
\rowcolor{zebra}15 & \texttt{ANE\_ReleaseDebugWorkProcessorBuffers} & 0
& 0 & 0 & 0 \\
16 & \texttt{ANE\_LoadFirmware} & 3 & 0 & 0 & 0 \\
\end{longtable}

Selector 0 is the open handshake. It passes a 104-byte device-info
structure as both the input and the output buffer, echoes the caller
header back, fills the output half with the device descriptor, and
returns the session token at offset \texttt{+0x00} that every later
request stores, with the handshake fields given in
\Cref{tbl:c27-deviceinfo}.

\begin{longtable}[]{@{}
  >{\raggedright\arraybackslash}p{(\linewidth - 4\tabcolsep) * \real{0.3333}}
  >{\raggedright\arraybackslash}p{(\linewidth - 4\tabcolsep) * \real{0.3333}}
  >{\raggedright\arraybackslash}p{(\linewidth - 4\tabcolsep) * \real{0.3333}}@{}}
\caption{The \protect\texttt{ANEDeviceInfo} handshake structure passed
in and echoed back by selector
0.}\label{tbl:c27-deviceinfo}\tabularnewline
\toprule\noalign{}
\begin{minipage}[b]{\linewidth}\raggedright
\textbf{Offset}
\end{minipage} & \begin{minipage}[b]{\linewidth}\raggedright
\textbf{Input (client to kernel)}
\end{minipage} & \begin{minipage}[b]{\linewidth}\raggedright
\textbf{Output (kernel to client)}
\end{minipage} \\
\midrule\noalign{}
\endfirsthead
\toprule\noalign{}
\begin{minipage}[b]{\linewidth}\raggedright
\textbf{Offset}
\end{minipage} & \begin{minipage}[b]{\linewidth}\raggedright
\textbf{Input (client to kernel)}
\end{minipage} & \begin{minipage}[b]{\linewidth}\raggedright
\textbf{Output (kernel to client)}
\end{minipage} \\
\midrule\noalign{}
\endhead
\bottomrule\noalign{}
\endlastfoot
\texttt{+0x00} & usage type byte (1 standard, 2 unsupported) & program /
session token (u64) \\
\rowcolor{zebra}\texttt{+0x08} & callback function pointer; board id
\texttt{0x1111222233334444} & echoed \\
\texttt{+0x10} & receiver context pointer & echoed \\
\rowcolor{zebra}\texttt{+0x18} & timeout \texttt{0x2710} = 10000 &
echoed \\
\texttt{+0x48} & (output only) & ANE version \texttt{0x20} = 32, 256 \\
\rowcolor{zebra}\texttt{+0x50} & (output only) & number of engines =
1 \\
\texttt{+0x60} & (output only) & CPU subtype = 4 \\
\end{longtable}

The usage-type byte selects the standard client profile: usage
\texttt{1} opens, usage \texttt{2} returns the unsupported code
\texttt{24}. Selector 16 is inactive on this build: its shim returns
\texttt{kIOReturnUnsupported} unconditionally, and the real firmware
load runs internally at driver start.

A compiled program reaches the kernel in one of two representations:
\texttt{ANEProgramLegacyResource}, a loader for the program-image
executable, and \texttt{ANEProgramRTResource}, a runtime op-graph
variant.

\section{Direct-path selector table}\label{direct-path-selector-table}

The direct-path client has 9 selectors, listed in
\Cref{tbl:c27-directpath-selectors} with their handlers and sizes.
Selectors 0, 1, and 2 reuse the control client's handler functions; the
remaining six are the enqueue, memory-map, and session-hint methods of
the low-latency submission model.

\begin{longtable}[]{@{}rlrrrr@{}}
\caption{The nine direct-path selectors with their handler names and
scalar and structure sizes; the full reference is in Appendix
\hyperref[appendix-c.-decoded-reference-tables]{C}.}\label{tbl:c27-directpath-selectors}\tabularnewline
\toprule\noalign{}
\textbf{Sel} & \textbf{Handler} & \textbf{Scalar in} & \textbf{Struct
in} & \textbf{Scalar out} & \textbf{Struct out} \\
\midrule\noalign{}
\endfirsthead
\toprule\noalign{}
\textbf{Sel} & \textbf{Handler} & \textbf{Scalar in} & \textbf{Struct
in} & \textbf{Scalar out} & \textbf{Struct out} \\
\midrule\noalign{}
\endhead
\bottomrule\noalign{}
\endlastfoot
0 & \texttt{ANE\_DeviceOpen} & 0 & 104 & 0 & 104 \\
\rowcolor{zebra}1 & \texttt{ANE\_DeviceClose} & 0 & 0 & 0 & 0 \\
2 & \texttt{ANE\_ProgramSendRequest} & 1 & 2376 & 0 & 40 \\
\rowcolor{zebra}3 & \texttt{ANE\_ProgramOutputSetEnqueue} & 0 & 40 & 0 &
0 \\
4 & \texttt{ANE\_ProgramInputsReady} & 0 & 3104 & 0 & 0 \\
\rowcolor{zebra}5 & \texttt{ANE\_MemoryMapRequest} & 1 & 2080 & 1 & 0 \\
6 & \texttt{ANE\_MemoryUnMapRequest} & 0 & 2080 & 0 & 0 \\
\rowcolor{zebra}7 & \texttt{ANE\_SessionHintRequest} & 0 & 16 & 0 &
24 \\
8 & \texttt{ANE\_ProgramChainingSetActiveProcedure} & 0 & 32 & 0 & 0 \\
\end{longtable}

Selector 5 is the device-IOMMU map. Its 2080-byte parameter structure
describes a host buffer, and on success the handler writes the resulting
engine-visible device address back into the single scalar output slot.
Selectors 3 and 4 are the pre-post and trigger of the resident
submission model: an output buffer set is enqueued, the inputs-ready
signal fires, and the same doorbell path as selector 2 rings the engine.

\section{Register and exclave method
catalog}\label{register-and-exclave-method-catalog}

Beyond the nine kernel selectors, the direct-path client exports a wider
register, power, firmware, and secure-world method surface. These are
not distinct kernel selector indices: each routes through one of the
nine kernel selectors or through a separate entry point, and
\Cref{tbl:c27-method-catalog} names the surface by role.

\begin{longtable}[]{@{}
  >{\raggedright\arraybackslash}p{(\linewidth - 2\tabcolsep) * \real{0.5000}}
  >{\raggedright\arraybackslash}p{(\linewidth - 2\tabcolsep) * \real{0.5000}}@{}}
\caption{The register, power, firmware, and exclave method surface
exported by the direct-path client beyond its nine kernel
selectors.}\label{tbl:c27-method-catalog}\tabularnewline
\toprule\noalign{}
\begin{minipage}[b]{\linewidth}\raggedright
\textbf{Method}
\end{minipage} & \begin{minipage}[b]{\linewidth}\raggedright
\textbf{Role}
\end{minipage} \\
\midrule\noalign{}
\endfirsthead
\toprule\noalign{}
\begin{minipage}[b]{\linewidth}\raggedright
\textbf{Method}
\end{minipage} & \begin{minipage}[b]{\linewidth}\raggedright
\textbf{Role}
\end{minipage} \\
\midrule\noalign{}
\endhead
\bottomrule\noalign{}
\endlastfoot
\texttt{ANE\_PowerOn} / \texttt{ANE\_PowerOff} / \texttt{ANE\_IsPowered}
& power-domain control \\
\rowcolor{zebra}\texttt{ANE\_LoadFirmware} /
\texttt{ANE\_ForgetFirmware} & firmware image lifecycle \\
\texttt{ANE\_SendCommand} & raw firmware command injection \\
\rowcolor{zebra}\texttt{ANE\_SetPowerManagement} /
\texttt{ANE\_SetDynamicPowerGating} /
\texttt{ANE\_SetPowerGatingHysteresisTime} & power policy \\
\texttt{ANE\_SetThrottlingPercentage} & thermal throttle \\
\rowcolor{zebra}\texttt{ANE\_SetDARTCacheTTL} /
\texttt{ANE\_FlushInactiveDARTMappings} / \texttt{ANE\_UnmapDartBuffers}
& address-translation controls \\
\texttt{ANE\_ReadANERegister} / \texttt{ANE\_WriteANERegister} & raw
memory-mapped register read and write \\
\rowcolor{zebra}\texttt{ANE\_FWSharedEventDoorbellRing} & ring the
firmware shared-event doorbell \\
\end{longtable}

A second access check beyond the kernel entitlement gates the raw
register read and write, command injection, and exclave methods: a
privileged-virtual-machine-access property probed at client open,
distinct from the device-open entitlement.

\section{From a user-space call to the
doorbell}\label{from-a-user-space-call-to-the-doorbell}

A submit on selector 2 crosses four layers from the user-space call down
to the hardware doorbell write, traced in \Cref{lst:c27-doorbell-path}.

\begin{listing}[H]
\caption{The four-layer call path from a user-space submit selector down to the hardware doorbell write.}\label{lst:c27-doorbell-path}

\begin{Shaded}
\begin{Highlighting}[]
\CommentTok{/* The submit path for selector 2 / direct{-}path selector 4: */}
\NormalTok{H11ANEInUserClient}\OperatorTok{::}\NormalTok{externalMethod}\OperatorTok{(}\NormalTok{sel}\OperatorTok{=}\DecValTok{2}\OperatorTok{,}\NormalTok{ args}\OperatorTok{)}
  \OperatorTok{{-}\textgreater{}}\NormalTok{ dispatchExternalMethod                 }\CommentTok{/* validates 2376{-}in / 40{-}out */}
  \OperatorTok{{-}\textgreater{}}\NormalTok{ ANE\_ProgramSendRequest}\OperatorTok{(}\NormalTok{client}\OperatorTok{,}\NormalTok{ ref}\OperatorTok{,}\NormalTok{ args}\OperatorTok{)}            \CommentTok{/* arg shim     */}
  \OperatorTok{{-}\textgreater{}}\NormalTok{ ANEClientDevice}\OperatorTok{::}\NormalTok{programSendRequest}\OperatorTok{(}\NormalTok{ANEProgramRequestArgs}\OperatorTok{*,} \OperatorTok{...)}
  \OperatorTok{{-}\textgreater{}}\NormalTok{ ANEDriver}\OperatorTok{::}\NormalTok{ANE\_ProgramSendRequest}\OperatorTok{(...)}               \CommentTok{/* gated        */}
  \OperatorTok{{-}\textgreater{}}\NormalTok{ ANEHWDevice}\OperatorTok{::}\NormalTok{doorBellRing}\OperatorTok{(}\NormalTok{db}\OperatorTok{)}
  \OperatorTok{{-}\textgreater{}}\NormalTok{ ANERegisterControl}\OperatorTok{::}\NormalTok{write32}\OperatorTok{(}\NormalTok{reg}\OperatorTok{,} \DecValTok{1} \OperatorTok{\textless{}\textless{}}\NormalTok{ idx}\OperatorTok{)}           \CommentTok{/* MMIO mailbox */}
\end{Highlighting}
\end{Shaded}

\end{listing}

Below the dispatcher, the thin shim re-checks the argument sizes and
unmarshals the typed argument structure, the client object method builds
the memory descriptors and retains the shared-event fences, and the
gated driver method runs on the command-gate workloop. The 2376-byte
request structure contains the program handle minted at create time, a
sequence number, the quality-of-service and execution-priority pair, and
the array of surface identifiers for the input, output, and intermediate
buffers, with the measured field layout in
\Cref{tbl:c27-request-struct}.

\begin{longtable}[]{@{}
  >{\raggedright\arraybackslash}p{(\linewidth - 4\tabcolsep) * \real{0.3333}}
  >{\raggedright\arraybackslash}p{(\linewidth - 4\tabcolsep) * \real{0.3333}}
  >{\raggedright\arraybackslash}p{(\linewidth - 4\tabcolsep) * \real{0.3333}}@{}}
\caption{The measured field layout of the 2376-byte request structure
submitted on selector 2.}\label{tbl:c27-request-struct}\tabularnewline
\toprule\noalign{}
\begin{minipage}[b]{\linewidth}\raggedright
\textbf{Offset}
\end{minipage} & \begin{minipage}[b]{\linewidth}\raggedright
\textbf{Field}
\end{minipage} & \begin{minipage}[b]{\linewidth}\raggedright
\textbf{Observed}
\end{minipage} \\
\midrule\noalign{}
\endfirsthead
\toprule\noalign{}
\begin{minipage}[b]{\linewidth}\raggedright
\textbf{Offset}
\end{minipage} & \begin{minipage}[b]{\linewidth}\raggedright
\textbf{Field}
\end{minipage} & \begin{minipage}[b]{\linewidth}\raggedright
\textbf{Observed}
\end{minipage} \\
\midrule\noalign{}
\endhead
\bottomrule\noalign{}
\endlastfoot
\texttt{+0x000} & program / instance token (u64) & the handle minted by
program-create \\
\rowcolor{zebra}\texttt{+0x008} & sequence number & 0, then 1 on the
next submit \\
\texttt{+0x010} & priority / quality-of-service pair &
\texttt{(5,\ 21)}, qos class and execution priority \\
\rowcolor{zebra}\texttt{+0x01c} & io category & 2 \\
\texttt{+0x020} & surface identifier array & input, output, intermediate
surfaces \\
\end{longtable}

The 40-byte output returns the sequence and the echoed token:
\texttt{+0x00} is the sequence and result, \texttt{+0x08} is the echoed
token, and \texttt{+0x20} is a status flag. Selector 2 alone uses the
asynchronous machinery, and it is the only path that reaches a hardware
doorbell from user space. Completion arrives at a mach wake port as a
callback, not by a shared-memory poll. The doorbell write itself reads
the doorbell index from the request, requires it below 32, computes the
mask \texttt{1\ \textless{}\textless{}\ index}, and stores that mask
into the engine register aperture, the mailbox signal that triggers the
firmware.

The reverse signal, the engine telling the host a job is complete,
travels the same windowed store mechanism in firmware. The engine rings
a host interrupt with an interrupt-atomic memory-mapped store to the
host-supplied target register, bracketed by clearing and then setting
bit 39 of the implementation-defined AArch64 system register
\texttt{S3\_3\_C15\_C8\_0}. Clearing bit 39 opens the posted-write
window, the firmware stores the doorbell value into the host aperture,
and a barrier separates the store from the status sample. The firmware
then reads back the uncorrectable-cache-error bit (bit 1) and the
transaction-reject bit (bit 7) to confirm the store has committed before
setting bit 39 to close the window. The whole sequence runs with
interrupts disabled so a nested handler cannot corrupt the status read.

\section{Entitlement gate and broker
model}\label{entitlement-gate-and-broker-model}

The client open checks two kernel entitlements, the hard device-open
gate and the resident data-chaining gate (\Cref{lst:c27-entitlements}).

\begin{listing}[H]
\caption{The two kernel entitlements checked when a user client is opened.}\label{lst:c27-entitlements}

\begin{Shaded}
\begin{Highlighting}[]
\CommentTok{/* checked at H11ANEInUserClient::init via copyClientEntitlement: */}
\StringTok{"com.apple.ane.iokit{-}user{-}access"}            \CommentTok{/* the hard device{-}open gate   */}
\StringTok{"com.apple.ane.allow{-}dataChaining{-}access"}    \CommentTok{/* resident data{-}chaining gate */}
\end{Highlighting}
\end{Shaded}

\end{listing}

A single kernel entitlement gates opening either user client. The check
runs once at client construction and is a boolean on the client object.
Across the whole system, exactly two binaries have
\texttt{com.apple.ane.iokit-user-access}: the system broker daemon and
its per-user sibling. No application process opens the device. Every
other consumer reaches the engine through the broker over a
cross-process call, where the credential is an entitlement from the
broker's own private family and never the kernel gate.

The driver stamps the capability at client creation in
\texttt{ANEClientInfo::create}, which reads each entitlement through
\texttt{copyClientEntitlement} and records \texttt{isPrivileged} and
\texttt{allowDataChaining} as bits on the client. Beyond the two open
gates, the driver enforces six further \texttt{com.apple.ane} and
\texttt{com.apple.private.ane} entitlements covering scheduling
priority, memory and data access, and client and coalition hints, as
\Cref{tbl:c27-driver-entitlements} lists.

\begin{longtable}[]{@{}
  >{\raggedright\arraybackslash}p{(\linewidth - 2\tabcolsep) * \real{0.5000}}
  >{\raggedright\arraybackslash}p{(\linewidth - 2\tabcolsep) * \real{0.5000}}@{}}
\caption{The kernel-driver entitlements beyond the two open gates, with
the capability each
grants.}\label{tbl:c27-driver-entitlements}\tabularnewline
\toprule\noalign{}
\begin{minipage}[b]{\linewidth}\raggedright
\textbf{Entitlement}
\end{minipage} & \begin{minipage}[b]{\linewidth}\raggedright
\textbf{Capability}
\end{minipage} \\
\midrule\noalign{}
\endfirsthead
\toprule\noalign{}
\begin{minipage}[b]{\linewidth}\raggedright
\textbf{Entitlement}
\end{minipage} & \begin{minipage}[b]{\linewidth}\raggedright
\textbf{Capability}
\end{minipage} \\
\midrule\noalign{}
\endhead
\bottomrule\noalign{}
\endlastfoot
\texttt{com.apple.ane.realtime-priority-client} & the real-time-priority
client grant \\
\rowcolor{zebra}\texttt{com.apple.ane.allow-system-reserved-priorities}
& use of the system-reserved scheduling priorities \\
\texttt{com.apple.ane.memory} & a memory-access grant \\
\rowcolor{zebra}\texttt{com.apple.ane.allow-data} & a data-access
grant \\
\texttt{com.apple.private.ane.allow-set-client-hints} & set per-client
hints \\
\rowcolor{zebra}\texttt{com.apple.private.ane.allow-share-coalition-hints}
& share hints across a coalition \\
\end{longtable}

The broker keys are a private entitlement family, checked per connection
and per method over a cross-process call, with each key, its capability,
and its holder count given in \Cref{tbl:c27-entitlement-family}.

\begin{longtable}[]{@{}
  >{\raggedright\arraybackslash}p{(\linewidth - 4\tabcolsep) * \real{0.3333}}
  >{\raggedright\arraybackslash}p{(\linewidth - 4\tabcolsep) * \real{0.3333}}
  >{\raggedright\arraybackslash}p{(\linewidth - 4\tabcolsep) * \real{0.3333}}@{}}
\caption{The entitlement family that gates the engine, from a static
scan of the M1 system binaries, independent of boot-security
state.}\label{tbl:c27-entitlement-family}\tabularnewline
\toprule\noalign{}
\begin{minipage}[b]{\linewidth}\raggedright
\textbf{Entitlement}
\end{minipage} & \begin{minipage}[b]{\linewidth}\raggedright
\textbf{Capability}
\end{minipage} & \begin{minipage}[b]{\linewidth}\raggedright
\textbf{Holders}
\end{minipage} \\
\midrule\noalign{}
\endfirsthead
\toprule\noalign{}
\begin{minipage}[b]{\linewidth}\raggedright
\textbf{Entitlement}
\end{minipage} & \begin{minipage}[b]{\linewidth}\raggedright
\textbf{Capability}
\end{minipage} & \begin{minipage}[b]{\linewidth}\raggedright
\textbf{Holders}
\end{minipage} \\
\midrule\noalign{}
\endhead
\bottomrule\noalign{}
\endlastfoot
\texttt{com.apple.ane.iokit-user-access} & the hard kernel gate: open
the user client, privileged device open & 2: the broker and its per-user
sibling \\
\rowcolor{zebra}\texttt{com.apple.ane.allow-dataChaining-access} &
resident data-chaining on the direct-path client & kernel-checked at
client init \\
\texttt{com.apple.aned.private.allow} & baseline: compile, load,
instantiate through the broker & 18 \\
\rowcolor{zebra}\texttt{com.apple.aned.private.ANEAccess.allow} &
inference-client access-grant variant & 14 \\
\texttt{com.apple.aned.private.adapterWeight.allow} & stream adapter
weights onto a shared resident base model & 5 \\
\rowcolor{zebra}\texttt{com.apple.aned.private.processModelShare.allow}
& share one resident model across processes & 4 \\
\texttt{com.apple.aned.private.secondaryANECompilerServiceAccess.allow}
& the longer-duration compiler service for large models & 1 \\
\rowcolor{zebra}\texttt{com.apple.aned.private.aggressivePowerSaving} &
the aggressive-power-saving execution mode & gate helper only \\
\texttt{com.apple.aned.private.modelPurgeInAllPartitions} & purge models
across all cache partitions & gate helper only \\
\rowcolor{zebra}\texttt{com.apple.security.temporary-exception.iokit-user-client-class}
& open the direct-path user client for own submission & 27 \\
\texttt{com.apple.security.ts.ane-client} & trust-cache blessed-client
slot for latency-critical consumers & 5 \\
\end{longtable}

The broker is a listener that enforces per-connection and per-method
entitlement checks. It sorts admitted clients into a restricted tier,
unrestricted tier, and per-user tier, and it threads a
quality-of-service argument through every compile, load, and instantiate
method. The restricted tier admits the adapter-weight, model-share, and
aggressive-power-saving requests through a per-method admission helper,
and the per-user tier serves the per-user broker.

Swappable model weights without recompilation rest on the adapter-weight
path. A base model is loaded once, and each adapter is a new instance
bound to a named base-model identifier holding only its per-adapter
weight files, through a create-instance-with-weights method that names
the base-model identifier and the weight-file count. Residency and power
are explicit per-instance arguments on the create-instance method: an
enable-power-saving flag, a more-aggressive variant gated by the
restricted tier, an opt-out-of-model-memory-unwiring flag that keeps a
hot client's weights resident at the cost of footprint, and a
queue-depth that the broker down-adjusts under contention. A queue-index
function and a program-priority function map the quality-of-service
argument to hardware scheduling. A privileged subset of system daemons
also has a sandbox exception that opens the direct-path user client and
drives per-inference submission on its own connection, skipping the
per-call round trip through the broker. That exception is a latency
optimization, not a capability grant: the privileged device open still
happens in the broker, which hands the client a program handle and an
intermediate-buffer handle.

The kernel binds residency to code-signing identity. A second client may
attach to an already-resident program only when its team identifier and
code-directory hash match the owner, so a shared resident model or
key-value cache cannot leak across tenants. The kernel resolves the
caller's team identifier and code-directory hash, and the attach path
tests them against the resident owner before a sibling instance reuses
the shared intermediate-buffer handle. With one physical engine on this
generation, cross-client arbitration is time-division multiplexing on a
single gated request queue, biased by the per-stream quality of service
the clients declare.

\section{Live device properties}\label{live-device-properties}

The driver publishes its topology and version constants into the
registry, read live on the M1 host and decoded in
\Cref{tbl:c27-device-props}.

\begin{longtable}[]{@{}
  >{\raggedright\arraybackslash}p{(\linewidth - 4\tabcolsep) * \real{0.3333}}
  >{\raggedright\arraybackslash}p{(\linewidth - 4\tabcolsep) * \real{0.3333}}
  >{\raggedright\arraybackslash}p{(\linewidth - 4\tabcolsep) * \real{0.3333}}@{}}
\caption{The device properties the driver publishes into the registry on
the M1 host.}\label{tbl:c27-device-props}\tabularnewline
\toprule\noalign{}
\begin{minipage}[b]{\linewidth}\raggedright
\textbf{Property}
\end{minipage} & \begin{minipage}[b]{\linewidth}\raggedright
\textbf{Value}
\end{minipage} & \begin{minipage}[b]{\linewidth}\raggedright
\textbf{Meaning}
\end{minipage} \\
\midrule\noalign{}
\endfirsthead
\toprule\noalign{}
\begin{minipage}[b]{\linewidth}\raggedright
\textbf{Property}
\end{minipage} & \begin{minipage}[b]{\linewidth}\raggedright
\textbf{Value}
\end{minipage} & \begin{minipage}[b]{\linewidth}\raggedright
\textbf{Meaning}
\end{minipage} \\
\midrule\noalign{}
\endhead
\bottomrule\noalign{}
\endlastfoot
architecture type string & \texttt{h13g} & microarchitecture family,
drives per-family codegen \\
\rowcolor{zebra}version & 96 = \texttt{0x60} & major hardware version \\
minor version & 17 & minor revision \\
\rowcolor{zebra}board type & 96 & board and system-on-chip type id \\
board subtype & 0 & board sub-variant \\
\rowcolor{zebra}number of cores & 16 & compute cores in this engine \\
number of engines & 1 & distinct engine units, load balancer is a
pass-through \\
\rowcolor{zebra}CPU subtype & 4 & program-ABI gate \\
internal build & No & release build, gates the debug surfaces \\
\end{longtable}

The number-of-cores value is a topology count and not the
throughput-relevant multiply-array width, so a floating-point rate is
taken from the measured cost-model anchor, never inferred from the core
count. At rest the registry shows the device, load-balancer instance,
and standing hints client present, with zero control clients and zero
direct-path clients open, confirming the brokered, lazily-opened model.
The driver opens the user clients on demand per active client and tears
them down when idle.

\chapter{Address translation and the DART}\label{address-translation-and-the-dart}

\begin{summarybox}

The engine reads and writes DRAM through its own IOMMU, the DART. It
translates a device address into a physical page at a 16 KB granule
across a 3.5 GiB aperture. A normal engine page is one 64-bit leaf word,
the physical frame located unshifted with bit 63 set:
\(\mathrm{leaf} = \mathrm{phys} \mathbin{|} \mathtt{0x8000000000000000}\).
Program-create re-bases each host-mapped buffer into a firmware aperture
with high half \texttt{0x1bc4}, a second translation with no constant
offset. The fault-capture registers cannot be read on the M1, because
every function on the fault path ends in a kernel panic.

\end{summarybox}

Chapter \hyperref[memory-hierarchy]{21} covered the on-chip working set:
the pool the engine reads and writes once data is resident. The off-chip
step that puts data there is the subject here. The engine reads its
inputs and writes its outputs directly from DRAM through an input-output
memory management unit, the DART, which translates a device-visible
address into a physical DRAM page before any DMA engine touches memory.

\section{DART and its address space}\label{dart-and-its-address-space}

The page granule, aperture base and size, and active stream set come
straight off the live device tree (\Cref{tbl:c28-address-space}).

\begin{longtable}[]{@{}
  >{\raggedright\arraybackslash}p{(\linewidth - 4\tabcolsep) * \real{0.3333}}
  >{\raggedright\arraybackslash}p{(\linewidth - 4\tabcolsep) * \real{0.3333}}
  >{\raggedright\arraybackslash}p{(\linewidth - 4\tabcolsep) * \real{0.3333}}@{}}
\caption{The page granule, aperture base and size, and active stream
set, read from the live device
tree.}\label{tbl:c28-address-space}\tabularnewline
\toprule\noalign{}
\begin{minipage}[b]{\linewidth}\raggedright
\textbf{Property}
\end{minipage} & \begin{minipage}[b]{\linewidth}\raggedright
\textbf{Value}
\end{minipage} & \begin{minipage}[b]{\linewidth}\raggedright
\textbf{Meaning}
\end{minipage} \\
\midrule\noalign{}
\endfirsthead
\toprule\noalign{}
\begin{minipage}[b]{\linewidth}\raggedright
\textbf{Property}
\end{minipage} & \begin{minipage}[b]{\linewidth}\raggedright
\textbf{Value}
\end{minipage} & \begin{minipage}[b]{\linewidth}\raggedright
\textbf{Meaning}
\end{minipage} \\
\midrule\noalign{}
\endhead
\bottomrule\noalign{}
\endlastfoot
page granule & \texttt{0x4000} = 16384 = 16 KB & the IOVA page; every
buffer maps at this granularity \\
\rowcolor{zebra}aperture base & \texttt{0x0} & IOVA window starts at
zero \\
aperture size & \texttt{0xE0000000} = 3.5 GiB & the managed IOVA span,
\texttt{0x0} to \texttt{0xE0000000} \\
\rowcolor{zebra}active streams & \texttt{\{0,\ 1,\ 2\}} & the engine
streams sharing one translation-table base; client isolation is
separate, per-client contexts \texttt{mapper-ane0-iso1} through
\texttt{iso7} \\
\end{longtable}

A host tensor buffer never reaches the engine as a host virtual address.
The DART maps it into the engine's own address space, the device virtual
address or IOVA, and the DMA engines issue reads and writes against
IOVAs. The DART contains the page tables that translate each IOVA back
to a physical DRAM page, so a buffer physically scattered across DRAM
appears IOVA-contiguous to the engine. The DART instance serving the
engine on the M1 is a single controller bound to the engine stream set,
and its managed address window and granule come from the live device
tree. The controller is the device-tree node \texttt{dart-ane0} at
physical base \texttt{0x85800000}, bound to the t6000-generation driver
class, not the t8020 class. It has four 16 KB register windows at
\texttt{0x85800000}, \texttt{0x85810000}, \texttt{0x85820000}, and
\texttt{0x85804000}. \Cref{tbl:c28-dart-props} gives the live
\texttt{dart-ane0} device-tree properties read from the controller node.

\begin{longtable}[]{@{}lll@{}}
\caption{The live \protect\texttt{dart-ane0} device-tree properties read
from the controller node.}\label{tbl:c28-dart-props}\tabularnewline
\toprule\noalign{}
\textbf{Property} & \textbf{Raw} & \textbf{Decoded} \\
\midrule\noalign{}
\endfirsthead
\toprule\noalign{}
\textbf{Property} & \textbf{Raw} & \textbf{Decoded} \\
\midrule\noalign{}
\endhead
\bottomrule\noalign{}
\endlastfoot
compatible & \texttt{dart,t6000} & the t6000-generation controller \\
\rowcolor{zebra}page size & \texttt{0x00004000} & 16 KB IOVA page \\
stream-ID enable bitmap & \texttt{0x0000a001} & bits 0, 13, 15 \\
\rowcolor{zebra}bypass bitmap & \texttt{0x0000a000} & bits 13, 15 \\
stream count & \texttt{0x10} & 16 stream slots \\
\rowcolor{zebra}options & \texttt{0x25} & low byte of the live config
word \texttt{0x80000025} \\
\end{longtable}

The host-side translation object is a three-level mapper chain: the
controller driver owns a mapper nub, which owns the translation mapper
the engine driver is handed. That mapper has the highest retain count of
any mapper on the system, consistent with its role as the live
tensor-mapping object. The page granule is 16 KB, not the 4 KB of the
host page tables. The firmware validates the same value independently
and rejects a wrong one with \texttt{MMU\ invalid\ page\ size:\ \%x}, so
the host maps and wires every engine DMA buffer at 16 KB granularity. A
sub-page buffer still consumes a full 16 KB IOVA page. This 16 KB page
is the coarsest of the three alignment scales on the M1: the 16-byte
DMA-width granule of chapter \hyperref[memory-hierarchy]{21} is below a
256-byte segment alignment, which is below the 16 KB DART page.

A two-level page-table walk, the L2 and L3 tables under the per-stream
translation-table base, over the 16 KB page covers the 3.5 GiB span. The
per-stream translation-table base is in the controller's
translation-table base register, measured live at \texttt{0x90022320}
for the active streams: bit 31 marks the base valid, and the remaining
field shifts left by 12 to the physical table base
\texttt{0x10022320000}. Streams 0, 1, and 2 share one table base, a
single table for the active engine streams. That base is in the same
DRAM band as the measured leaf physical frames, so it points at the
controller's own page-table memory.

Client isolation is separate from this engine-stream table. The DART
gives each client its own isolation context, the eight
address-translation mappers \texttt{mapper-ane0-iso1} through
\texttt{iso7} plus a base in the live IORegistry, so each client's
buffers map into its own translation domain. The secure exclave receives
these contexts as the capabilities \texttt{ANEIsoID1} through
\texttt{ID7}, confining each client's DMA to its own domain, the
address-translation half of the exclave capability model in chapter
\hyperref[security-and-isolation]{32}.

The controller register layout is recovered from its capture routine as
byte offsets into a stream's 16 KB register window
(\Cref{tbl:c28-reg-offsets}).

\begin{longtable}[]{@{}
  >{\raggedright\arraybackslash}p{(\linewidth - 2\tabcolsep) * \real{0.5000}}
  >{\raggedright\arraybackslash}p{(\linewidth - 2\tabcolsep) * \real{0.5000}}@{}}
\caption{The controller register-offset map, recovered from the
register-capture routine.}\label{tbl:c28-reg-offsets}\tabularnewline
\toprule\noalign{}
\begin{minipage}[b]{\linewidth}\raggedright
\textbf{Offset}
\end{minipage} & \begin{minipage}[b]{\linewidth}\raggedright
\textbf{Register}
\end{minipage} \\
\midrule\noalign{}
\endfirsthead
\toprule\noalign{}
\begin{minipage}[b]{\linewidth}\raggedright
\textbf{Offset}
\end{minipage} & \begin{minipage}[b]{\linewidth}\raggedright
\textbf{Register}
\end{minipage} \\
\midrule\noalign{}
\endhead
\bottomrule\noalign{}
\endlastfoot
\texttt{+0x40} & error status: fault flag at bit 31, plus stream id and
fault code \\
\rowcolor{zebra}\texttt{+0x50}, \texttt{+0x54} & error address low and
high, the faulting device address \\
\texttt{+0xfc} & enabled-streams global bitmap \\
\rowcolor{zebra}\texttt{+0x100} + 4·sid & per-stream translation-control
array \\
\texttt{+0x200} + 4·idx & per-stream translation-table base array \\
\rowcolor{zebra}\texttt{+0x1000}, \texttt{+0x100c} & translation-buffer
control and status \\
\texttt{+0x1020}, \texttt{+0x1028} & translation-buffer error
registers \\
\end{longtable}

\section{Leaf page-table entry}\label{leaf-page-table-entry}

The leaf entry that the DART stores per page is one 64-bit word. For a
normal engine data page it is the physical frame with the valid bit set,
given by

\[\mathrm{leaf} = \mathrm{phys} \mathbin{|} \mathtt{0x8000000000000000}.\]

Its bit fields are in \Cref{tbl:c28-leaf-entry}.

\begin{longtable}[]{@{}
  >{\raggedright\arraybackslash}p{(\linewidth - 6\tabcolsep) * \real{0.2500}}
  >{\raggedright\arraybackslash}p{(\linewidth - 6\tabcolsep) * \real{0.2500}}
  >{\raggedright\arraybackslash}p{(\linewidth - 6\tabcolsep) * \real{0.2500}}
  >{\raggedright\arraybackslash}p{(\linewidth - 6\tabcolsep) * \real{0.2500}}@{}}
\caption{The bit-field layout of the 64-bit leaf page-table entry the
DART stores per page.}\label{tbl:c28-leaf-entry}\tabularnewline
\toprule\noalign{}
\begin{minipage}[b]{\linewidth}\raggedright
\textbf{bits}
\end{minipage} & \begin{minipage}[b]{\linewidth}\raggedright
\textbf{field}
\end{minipage} & \begin{minipage}[b]{\linewidth}\raggedright
\textbf{value}
\end{minipage} & \begin{minipage}[b]{\linewidth}\raggedright
\textbf{note}
\end{minipage} \\
\midrule\noalign{}
\endfirsthead
\toprule\noalign{}
\begin{minipage}[b]{\linewidth}\raggedright
\textbf{bits}
\end{minipage} & \begin{minipage}[b]{\linewidth}\raggedright
\textbf{field}
\end{minipage} & \begin{minipage}[b]{\linewidth}\raggedright
\textbf{value}
\end{minipage} & \begin{minipage}[b]{\linewidth}\raggedright
\textbf{note}
\end{minipage} \\
\midrule\noalign{}
\endhead
\bottomrule\noalign{}
\endlastfoot
63 & valid / active & \texttt{1} on map, \texttt{0} on unmap & the only
flag set for an engine page \\
\rowcolor{zebra}62 & aux protection class & \texttt{0} & set only when
prot bit 3 is set; never reached on the engine path \\
60 & aux protection class & \texttt{0} & set only when prot bit 4 is
set; never reached \\
\rowcolor{zebra}59 & aux protection class & \texttt{0} & set only when
prot bit 5 is set; never reached \\
46:14 & physical frame & full physical address, unshifted & low 14 bits
zero at the 16 KB granule \\
\rowcolor{zebra}13:0 & within-page offset & \texttt{0} & always zero at
page granularity \\
\end{longtable}

The physical frame is the full 16 KB-aligned physical address located
unshifted, so its low 14 bits are zero and the frame and the valid bit
do not overlap.

The word keeps the same shape for every live engine usage type. The
mapping software names two software protection classes, read-write for
inputs, weights, and intermediates, and a device-write class for
outputs, yet both collapse to the identical leaf template of bit 63
alone. The DART encodes access permission in the per-stream
translation-control configuration. Bits 62, 60, and 59 are
aux-protection classes that the engine driver never sets, because its
direction-to-protection mapper produces only the values \(\{1, 2, 3\}\),
whose bits 3, 4, and 5 are always zero. Across 26178 measured leaf-map
events none of those three bits was ever set.

The mapping path that produces this word stacks two translations. The
host pins the physical pages, then fills the leaf table by accumulating
per-page segments and flushing them through the page-protection-layer
write. That write uses a 40-byte per-page segment structure, from which
the leaf word is assembled per page (\Cref{lst:c28-ppl-seg}).

\begin{listing}[H]
\caption{The per-page segment structure passed into the page-protection-layer write and how the leaf word is assembled from it.}\label{lst:c28-ppl-seg}

\begin{Shaded}
\begin{Highlighting}[]
\KeywordTok{struct}\NormalTok{ ppl\_iommu\_seg }\OperatorTok{\{}   \CommentTok{/* 0x28 bytes, measured layout */}
    \DataTypeTok{uint64\_t}\NormalTok{ iova}\OperatorTok{;}       \CommentTok{/* +0x00  device virtual address      */}
    \DataTypeTok{uint64\_t}\NormalTok{ phys}\OperatorTok{;}       \CommentTok{/* +0x08  physical DRAM page           */}
    \DataTypeTok{uint64\_t}\NormalTok{ size}\OperatorTok{;}       \CommentTok{/* +0x10  0x4000 (16 KB granule)       */}
    \DataTypeTok{uint64\_t}\NormalTok{ prot}\OperatorTok{;}       \CommentTok{/* +0x18  3 = RW, 1 = device{-}write     */}
    \DataTypeTok{uint64\_t}\NormalTok{ reserved}\OperatorTok{;}   \CommentTok{/* +0x20  0                            */}
\OperatorTok{\};}
\CommentTok{/* assembled leaf word, per page i:  leaf[i] = seg[i].phys | template */}
\CommentTok{/* template = 0x8000000000000000 (bit 63) on map, 0x0 on unmap        */}
\end{Highlighting}
\end{Shaded}

\end{listing}

The leaf table itself is page-protection-layer memory. The word above is
the value the kernel hands to that layer to store, captured at the store
register on a live serialized dispatch, not a read-back of the stored
page.

The host maps each buffer under a usage code that names its role, held
in the segment structure and recovered from the map call sites, with the
codes and their protection classes in \Cref{tbl:c28-usage-codes}.

\begin{longtable}[]{@{}lll@{}}
\caption{The buffer-role usage codes and the protection class each maps
under.}\label{tbl:c28-usage-codes}\tabularnewline
\toprule\noalign{}
\textbf{Code} & \textbf{Role} & \textbf{Protection class} \\
\midrule\noalign{}
\endfirsthead
\toprule\noalign{}
\textbf{Code} & \textbf{Role} & \textbf{Protection class} \\
\midrule\noalign{}
\endhead
\bottomrule\noalign{}
\endlastfoot
1 & client input tensor & read-write (3) on the M1 \\
\rowcolor{zebra}2 & client output tensor & device-write (1) \\
7 & intermediate buffer & read-write (3) \\
\rowcolor{zebra}8 & kernel and weights & read-write (3) \\
9 & program text, task descriptors, working set & read-write (3) \\
\rowcolor{zebra}11 & program constants and scratch & read-write (3) \\
\texttt{0x1019} = 4121 & firmware power-on shared surface & firmware
class \\
\rowcolor{zebra}\texttt{0x101a} = 4122 & firmware resident heap &
firmware class \\
\end{longtable}

A direction-to-protection mapper produces only the values 1, 2, and 3: a
device-write output is class 1, a read-only input would be class 2, and
read-write or no-direction is class 3. On the M1 a configuration bit in
the controller collapses the would-be read-only class 2 into read-write
3, so the controller maps read-only inputs read-write and enforces input
protection upstream of the leaf word. The high \texttt{0x1000} bit on
the firmware codes flags the firmware-owned shared class, distinct from
the per-program client and kernel buffers.

\section{Host-to-firmware rebase
boundary}\label{host-to-firmware-rebase-boundary}

The host-side translation resolves to a physical address for every
buffer. For a single matmul-with-activation load, every host-programmed
buffer resolves to its named buffer role: input tensor, output tensor,
weights, program text, constants, intermediate, working set, and two
firmware shared surfaces. \Cref{tbl:c28-nine-buffers} shows the nine
buffers of that load, each with its usage code.

\begin{longtable}[]{@{}ll@{}}
\caption{The nine buffers of one matmul-with-activation load, each shown
with its usage code.}\label{tbl:c28-nine-buffers}\tabularnewline
\toprule\noalign{}
\textbf{Named buffer} & \textbf{Usage} \\
\midrule\noalign{}
\endfirsthead
\toprule\noalign{}
\textbf{Named buffer} & \textbf{Usage} \\
\midrule\noalign{}
\endhead
\bottomrule\noalign{}
\endlastfoot
input tensor & 1 \\
\rowcolor{zebra}output tensor & 2 \\
weights & 8 \\
\rowcolor{zebra}program text & 9 \\
constants & 11 \\
\rowcolor{zebra}intermediate & 7 \\
working set & 9 \\
\rowcolor{zebra}firmware shared surface & 4121 \\
firmware resident heap & 4122 \\
\end{longtable}

An in-place operation maps two distinct device addresses onto one
physical page, a single shared memory descriptor that the runtime reads
as input and overwrites as output. The physical pages are scattered
across the DRAM band and are not physically contiguous: the page table
is what makes each buffer appear contiguous to the engine.

One further translation is past the host boundary. The values the engine
reads from its instruction stream are not host IOVAs. At program-create
the firmware re-bases each host-mapped buffer into its own resident
aperture and patches that rebased address into the engine registers, so
the engine-register operand values are in a firmware window with high
half \texttt{0x1bc4}, shown in \Cref{lst:c28-rebase}.

\begin{listing}[H]
\caption{The firmware rebase from the host device-address band into the resident firmware aperture.}\label{lst:c28-rebase}

\begin{Shaded}
\begin{Highlighting}[]
\NormalTok{host IOVA band     {-}{-}\textgreater{}    firmware DRAM{-}tile aperture     (no constant host{-}side offset)}
\end{Highlighting}
\end{Shaded}

\end{listing}

The host IOVA and the firmware-aperture address are unequal and have no
constant offset between them, so the rebase is a second translation. The
bridge that establishes it runs once at program-create and is cached. On
the cached dispatch path the host never re-emits the host-to-firmware
pair, so the firmware rebase is the only translation below the host
boundary there.

One load spans three address spaces, recovered from the loaded program
text and the page-table fill and named in \Cref{tbl:c28-address-spaces}.

\begin{longtable}[]{@{}lll@{}}
\caption{The three coexisting address spaces for one load: the host
device addresses, engine-register aperture, and firmware DRAM-tile
aperture.}\label{tbl:c28-address-spaces}\tabularnewline
\toprule\noalign{}
\textbf{Space} & \textbf{Address space} & \textbf{What it holds} \\
\midrule\noalign{}
\endfirsthead
\toprule\noalign{}
\textbf{Space} & \textbf{Address space} & \textbf{What it holds} \\
\midrule\noalign{}
\endhead
\bottomrule\noalign{}
\endlastfoot
A & Host IOVA & host page table, fully resolved to physical \\
\rowcolor{zebra}B & Engine-register & where the firmware writes the
DMA-engine bases \\
C & DRAM-tile aperture & what streams: the weight and data tiles \\
\end{longtable}

The program text is a list of 44-byte register-write records, each
pairing an engine-register address in space B with two DRAM-tile operand
values in space C. The engine-register address selects which
data-movement engine to re-base: the weight-streaming sources,
input-tile reader, and output-tile writer. The operand values are the
high half \texttt{0x1bc4} of space C, the firmware-resident rebase of
the host buffers, which is why a naive comparison of raw register
addresses across loads fails: the aperture base moves with each program
load.

\subsection{Firmware rebase
arithmetic}\label{firmware-rebase-arithmetic}

The rebase that produces space C from a host IOVA is a pure linear
translation recovered from the firmware itself. The firmware
validated-translate routine reads the descriptor IOVA, range-checks it
against the mapped region, then computes the runtime address as

\[\mathrm{runtime} = \mathrm{mappedBase} + (\mathrm{IOVA} - \mathrm{dvaBase}).\]

There is one subtract of the IOVA-region base and one add of the
firmware aperture base. There is no shift, no mask, and no page
rounding, which is why the host IOVA and the firmware-aperture address
have no constant offset: the offset is the difference of two independent
region bases that each move per load.

The three runtime quantities the formula needs are fields of the
firmware aperture-config object at \texttt{engine+0xa000}
(\Cref{tbl:c28-aperture-config}).

\begin{longtable}[]{@{}
  >{\raggedright\arraybackslash}p{(\linewidth - 4\tabcolsep) * \real{0.3333}}
  >{\raggedright\arraybackslash}p{(\linewidth - 4\tabcolsep) * \real{0.3333}}
  >{\raggedright\arraybackslash}p{(\linewidth - 4\tabcolsep) * \real{0.3333}}@{}}
\caption{The firmware aperture-config fields at
\protect\texttt{engine+0xa000} that drive the rebase, populated at
buffer-map time and absent from the firmware
image.}\label{tbl:c28-aperture-config}\tabularnewline
\toprule\noalign{}
\begin{minipage}[b]{\linewidth}\raggedright
\textbf{Offset}
\end{minipage} & \begin{minipage}[b]{\linewidth}\raggedright
\textbf{Field}
\end{minipage} & \begin{minipage}[b]{\linewidth}\raggedright
\textbf{Meaning}
\end{minipage} \\
\midrule\noalign{}
\endfirsthead
\toprule\noalign{}
\begin{minipage}[b]{\linewidth}\raggedright
\textbf{Offset}
\end{minipage} & \begin{minipage}[b]{\linewidth}\raggedright
\textbf{Field}
\end{minipage} & \begin{minipage}[b]{\linewidth}\raggedright
\textbf{Meaning}
\end{minipage} \\
\midrule\noalign{}
\endhead
\bottomrule\noalign{}
\endlastfoot
\texttt{+0xb98} & \texttt{dvaBase} & host / DART IOVA region base \\
\rowcolor{zebra}\texttt{+0xba0} & \texttt{regionSize} & mapped region
size \\
\texttt{+0xba8} & \texttt{mappedBase} & firmware runtime aperture base,
the rebase target base \\
\end{longtable}

These three are firmware-runtime state set at buffer-map time; the image
has only the arithmetic and the field offsets. The firmware then writes
the rebased value into the engine DMA bar registers at
\texttt{0x285c25020\ +\ engine*0x148\ +\ barId*4}, where
\texttt{engine*0x148} is the per-engine MMIO stride and \texttt{barId*4}
selects the word-indexed bar register. Only the low 32 bits of the
rebased address reach this MMIO path; a per-bar config flag gates wider
bars onto a separate software descriptor table. Both the rebase tail and
the register-target prologue were emulated under unicorn and reproduced
the formula and the \texttt{0x285c25020\ +\ engine*0x148} register
address exactly.

The firmware also clamps every address it programs into the DMA engines
to a 32-bit ceiling, below the 3.5 GiB aperture. Each device address the
engine touches, text, weights, descriptors, intermediate, output, and
chained buffers, must satisfy
\texttt{addr\ \textgreater{}\textgreater{}\ 32\ ==\ 0}, asserted per
buffer in the firmware. The host allocator thus hands out only sub-4-GiB
IOVAs, and the engine operates in the bottom 3.5 GiB of its address
space.

An engine buffer is governed by three distinct alignment scales, each
coarser than the last, collected in \Cref{tbl:c28-alignment}.

\begin{longtable}[]{@{}
  >{\raggedright\arraybackslash}p{(\linewidth - 4\tabcolsep) * \real{0.3333}}
  >{\raggedright\arraybackslash}p{(\linewidth - 4\tabcolsep) * \real{0.3333}}
  >{\raggedright\arraybackslash}p{(\linewidth - 4\tabcolsep) * \real{0.3333}}@{}}
\caption{The three alignment scales on the M1, from the finest DMA width
granule to the coarsest page
granule.}\label{tbl:c28-alignment}\tabularnewline
\toprule\noalign{}
\begin{minipage}[b]{\linewidth}\raggedright
\textbf{Scale}
\end{minipage} & \begin{minipage}[b]{\linewidth}\raggedright
\textbf{Value}
\end{minipage} & \begin{minipage}[b]{\linewidth}\raggedright
\textbf{What it governs}
\end{minipage} \\
\midrule\noalign{}
\endfirsthead
\toprule\noalign{}
\begin{minipage}[b]{\linewidth}\raggedright
\textbf{Scale}
\end{minipage} & \begin{minipage}[b]{\linewidth}\raggedright
\textbf{Value}
\end{minipage} & \begin{minipage}[b]{\linewidth}\raggedright
\textbf{What it governs}
\end{minipage} \\
\midrule\noalign{}
\endhead
\bottomrule\noalign{}
\endlastfoot
DMA width granule & 16 B & the data-movement quantum \\
\rowcolor{zebra}segment alignment & 256 B & program-text and segment
packing \\
page granule & 16 KB & the device-address page, allocation and wiring
unit \\
\end{longtable}

The host may pre-map a buffer before the inference that uses it, through
a pre-map command that establishes its device-address mapping ahead of
the hot path. The firmware keeps explicit buffer pools, tagging each
mapped buffer with a pool identifier, and runs a buffer-recycle state
machine that reuses output buffers across chained calls. A per-process
pool tracks outstanding requests with an in-flight count capped at 127
per request.

Under device-address pressure the kernel applies a least-recently-used
eviction policy over its mappings, scoring each with
\texttt{getDartBufferFreeUpScore} and freeing through the
\texttt{FreeUpDart*} family. A long-running process that maps more than
the address window has room for thus loses its coldest mappings; its
maps do not fail.

\section{Fault-capture registers}\label{fault-capture-registers}

The DART captures a translation or protection fault into a single-shot
register block: an error-status word, the faulting device address split
low and high, and a translation-buffer status word
(\Cref{tbl:c28-fault-registers}).

\begin{longtable}[]{@{}
  >{\raggedright\arraybackslash}p{(\linewidth - 4\tabcolsep) * \real{0.3333}}
  >{\raggedright\arraybackslash}p{(\linewidth - 4\tabcolsep) * \real{0.3333}}
  >{\raggedright\arraybackslash}p{(\linewidth - 4\tabcolsep) * \real{0.3333}}@{}}
\caption{The DART fault-capture register block, giving each register
device offset and the field it
holds.}\label{tbl:c28-fault-registers}\tabularnewline
\toprule\noalign{}
\begin{minipage}[b]{\linewidth}\raggedright
\textbf{register}
\end{minipage} & \begin{minipage}[b]{\linewidth}\raggedright
\textbf{device offset}
\end{minipage} & \begin{minipage}[b]{\linewidth}\raggedright
\textbf{field}
\end{minipage} \\
\midrule\noalign{}
\endfirsthead
\toprule\noalign{}
\begin{minipage}[b]{\linewidth}\raggedright
\textbf{register}
\end{minipage} & \begin{minipage}[b]{\linewidth}\raggedright
\textbf{device offset}
\end{minipage} & \begin{minipage}[b]{\linewidth}\raggedright
\textbf{field}
\end{minipage} \\
\midrule\noalign{}
\endhead
\bottomrule\noalign{}
\endlastfoot
error / status & \texttt{+0x40} & fault flag at bit 31, plus stream id
and fault code \\
\rowcolor{zebra}error address low & \texttt{+0x50} & faulting device
address, low half \\
error address high & \texttt{+0x54} & faulting device address, high
half \\
\rowcolor{zebra}translation-buffer status & \texttt{+0x100c} & busy /
error status \\
\end{longtable}

These registers cannot be read safely on the M1. The capture routine
writes the block into a DRAM snapshot only on the fault path, and on
this controller that path is unconditionally a kernel panic. Every
function on the fault path ends in a direct call to the kernel panic
routine, and control flow reaches it: the driver implements a DART fault
as a \texttt{REQUIRE(...)} assertion, which panics the machine rather
than returning the captured registers. A read-only probe that reads the
snapshot at the function boundary is built and validated against the
disassembly, but it cannot fire without a fault, and any fault panics
the box. The faulting-address and status values are thus stated here as
the structure of the block, not as measured values, since obtaining them
on this hardware would require panicking the machine. The register
layout above is recovered from the capture routine; the field decode of
the words a contained fault would return follows the published
controller field layout.

\subsection{\texorpdfstring{\texttt{IODARTErrorInfo} fault
descriptor}{IODARTErrorInfo fault descriptor}}\label{iodarterrorinfo-fault-descriptor}

Above the hardware DART-side capture is a software fault descriptor,
\texttt{IODARTErrorInfo}, that the kernel \texttt{t6000dart} core
constructs from the DART fault MMIO and hands to each registered
consumer. This descriptor is the structure the driver fault callback
reads and logs, and it is a kernel-wide ABI: the sibling DART consumers
\texttt{AppleAVD} and \texttt{AVE\_DART} read the identical layout at
the identical offsets, which confirms it is not an engine-private
struct. Most fields are object pointers whose stringifier the callback
invokes; \Cref{tbl:c28-errorinfo} gives the byte offsets into the
descriptor.

\begin{longtable}[]{@{}
  >{\raggedright\arraybackslash}p{(\linewidth - 6\tabcolsep) * \real{0.2500}}
  >{\raggedright\arraybackslash}p{(\linewidth - 6\tabcolsep) * \real{0.2500}}
  >{\raggedright\arraybackslash}p{(\linewidth - 6\tabcolsep) * \real{0.2500}}
  >{\raggedright\arraybackslash}p{(\linewidth - 6\tabcolsep) * \real{0.2500}}@{}}
\caption{The \protect\texttt{IODARTErrorInfo} software fault descriptor,
the shared kernel ABI the \protect\texttt{t6000dart} core hands to its
registered fault consumers.}\label{tbl:c28-errorinfo}\tabularnewline
\toprule\noalign{}
\begin{minipage}[b]{\linewidth}\raggedright
\textbf{Offset}
\end{minipage} & \begin{minipage}[b]{\linewidth}\raggedright
\textbf{Field}
\end{minipage} & \begin{minipage}[b]{\linewidth}\raggedright
\textbf{Type}
\end{minipage} & \begin{minipage}[b]{\linewidth}\raggedright
\textbf{Meaning}
\end{minipage} \\
\midrule\noalign{}
\endfirsthead
\toprule\noalign{}
\begin{minipage}[b]{\linewidth}\raggedright
\textbf{Offset}
\end{minipage} & \begin{minipage}[b]{\linewidth}\raggedright
\textbf{Field}
\end{minipage} & \begin{minipage}[b]{\linewidth}\raggedright
\textbf{Type}
\end{minipage} & \begin{minipage}[b]{\linewidth}\raggedright
\textbf{Meaning}
\end{minipage} \\
\midrule\noalign{}
\endhead
\bottomrule\noalign{}
\endlastfoot
\texttt{+0x00} & \texttt{Type} & string & fault type, also the event-id
header \\
\rowcolor{zebra}\texttt{+0x08} & \texttt{HwClass} & string & hardware
fault class \\
\texttt{+0x10} & \texttt{HwError} & string & hardware error code \\
\rowcolor{zebra}\texttt{+0x18} & \texttt{HwStatus} & \texttt{u32} & raw
DART fault-status word \\
\texttt{+0x20} & \texttt{IsWrite} & bool & \texttt{0} read fault,
\texttt{1} write fault \\
\rowcolor{zebra}\texttt{+0x28} & \texttt{SID} & \texttt{u32} & DART
stream id of the faulting agent \\
\texttt{+0x30} & \texttt{Address} & \texttt{u32} & faulting IOVA \\
\rowcolor{zebra}\texttt{+0x40} & \texttt{TTBRIndex} & \texttt{u32} &
translation-table base index \\
\end{longtable}

The four slots at \texttt{+0x90} are the \texttt{AXI\_ID{[}0..3{]}}
array, the four fault descriptors a prior decode reported. What
localizes a fault is the walk-index triple, \texttt{TTBRIndex},
\texttt{L2Index}, and \texttt{L3Index}, which names the failing
page-table entry; the remaining scalars are the taxonomy, the faulting
page and stream, and the read-write bit.

One register the callback reads is not part of the descriptor: it reads
an engine status word directly off its own device object at
\texttt{{[}engine+0xe028{]}}. The callback treats the fault as benign
and returns early when \texttt{(status\ \textbar{}\ 0x80)\ ==\ 0xa0},
that is when the low seven bits equal \texttt{0x20} and bit 7 is a
don't-care. Any other value is a real fault. The same word gates the
engine clock elsewhere in the driver, where \texttt{0xa0} is the
powered-idle state.

The ANE kext does not raise the panic. The kext fault path returns and,
on a real fault, sets a sticky latch, dumps the shared-memory allocation
table and firmware debug state, then parks for 250 ms awaiting external
recovery. The machine-halting panic is in the kernel \texttt{t6000dart}
core, where the mapping and page-table-walk fault detection is a
\texttt{REQUIRE} assertion that calls \texttt{panic()} directly. That
substrate panic fires before, or instead of, the kext recovery park.

\chapter{Firmware}\label{firmware}

\begin{summarybox}

The engine controller runs a C++ application over a real-time kernel,
distributed as an unencrypted preload executable. Every subsystem is a
long-lived message-pumped task, and all kernel objects sit in fixed
build-time pools. The execution loop accepts four command classes on the
shared channel, three procedure-call variants plus the cache-request
trigger, and rejects anything else as unsupported. The scheduler is
fixed-priority with eight levels, no hardware watchdog, and a software
deadline of about 2 seconds that aborts a stuck queue. Recovery is
serialized behind a single command gate and rate-limited, so crashes
arriving faster than the recovery cycle keep the channel not-ready.

\end{summarybox}

The engine's controller runs a small real-time operating system, not
bare metal. A real-time kernel hosts an application that loads compiled
programs, schedules their task descriptors, drives the multiply array
and the data-movement engines, and reports faults back to the host. That
controller is CHINOOK, an embedded ARM core with eleven hardware thread
contexts (\texttt{CHINOOK\_CPU\_IMPL\_THID0} through \texttt{THID10}),
its own level-two cache, and pipeline error-capture and power-down-save
registers, so the firmware runs on a real multithreaded processor.

\section{Substrate}\label{substrate}

The firmware is a C++ application over a real-time kernel, distributed
on the package as an unencrypted preload Mach-O wrapped in an Image4
container with payload tag \texttt{anef}. The kernel layer provides
tasks, a priority scheduler, semaphores, mutexes, message queues, two
heap arenas, a fault handler, and a generic finite-state-machine
framework. The application layer above it provides the program loader,
execution loop, task-descriptor driver, tensor-mover driver,
address-translation manager, and power-control service.

Every subsystem is a long-lived task with its own stack, supplied by a
message queue. A task blocks on its queue, wakes on a posted message,
runs that message to completion, and loops. Each task has two queues: a
task-context input queue and a separate interrupt-context queue. An
interrupt handler never runs subsystem logic in interrupt context. It
pushes a record into an interrupt buffer, posts the target task's
interrupt queue, and returns, deferring the work to task context. The
named tasks include an idle task, task terminator, execution-loop pump,
address-translation manager, tunable-register loader, load monitor,
interrupt manager, call manager, server, and host remote-procedure-call
daemon.

Each task is in one of five scheduler states, recovered from the
task-list dumper (\Cref{tbl:c29-task-states}).

\begin{longtable}[]{@{}ll@{}}
\caption{The five scheduler task
states.}\label{tbl:c29-task-states}\tabularnewline
\toprule\noalign{}
\textbf{State} & \textbf{Meaning} \\
\midrule\noalign{}
\endfirsthead
\toprule\noalign{}
\textbf{State} & \textbf{Meaning} \\
\midrule\noalign{}
\endhead
\bottomrule\noalign{}
\endlastfoot
\texttt{RUNNABLE} & ready or running \\
\rowcolor{zebra}\texttt{SUSPENDED} & explicitly parked \\
\texttt{WAITING} & blocked on a generic wait, such as a message queue \\
\rowcolor{zebra}\texttt{SEMWAIT} & blocked specifically on a
semaphore \\
\texttt{BLOCKED} & blocked \\
\end{longtable}

The distinct semaphore-wait state, separate from the generic wait, is
the heritage of the real-time executive model the primitives follow. The
message queue each task blocks on is a bounded-buffer control block
holding a maximum message width, a ring depth, read and write pointers,
a lock, and two counting semaphores. The send side blocks when full and
the receive side blocks when empty. The kernel keeps all its objects in
fixed pools sized at build time: tasks, semaphores, mailboxes, queues,
signals, timers, and mutexes each occupy a slot in a pre-sized table.
There is no unbounded dynamic task creation at run time, and the
capacity of each pool is fixed and reportable as a total-and-available
count. A hard boot invariant ties the heap to the system reservation:
the initial heap must exceed twice the reserved framework allocation,
and the kernel refuses to boot if the arena is too small.

\section{Bring-up}\label{bring-up}

A three-level call chain separates device attach from firmware bring-up
(\Cref{lst:c29-bringup}).

\begin{listing}[H]
\caption{The three-level kernel bring-up call chain.}\label{lst:c29-bringup}

\begin{Shaded}
\begin{Highlighting}[]
\NormalTok{ANEHWDevice}\OperatorTok{::}\NormalTok{start}\OperatorTok{()}\NormalTok{              attach }\OperatorTok{+}\NormalTok{ provider }\OperatorTok{+}\NormalTok{ RTBuddy client }\OperatorTok{+} \DecValTok{7}\NormalTok{ MMIO banks }\OperatorTok{+}\NormalTok{ clocks}
\NormalTok{  ANEHWDevice}\OperatorTok{::}\NormalTok{power\_on\_hardware}\OperatorTok{()}\NormalTok{    clocks}\OperatorTok{/}\NormalTok{power }\OperatorTok{+}\NormalTok{ MPM }\OperatorTok{+}\NormalTok{ resume}
\NormalTok{    ANEHWDevice}\OperatorTok{::}\NormalTok{ANE\_Init}\OperatorTok{()}\NormalTok{             RTBuddy endpoints }\OperatorTok{+}\NormalTok{ scratch handshakes }\OperatorTok{+}\NormalTok{ first commands}
\end{Highlighting}
\end{Shaded}

\end{listing}

Bring-up splits into two phases on distinct entry points.
\texttt{ANEHWDevice::start} attaches the device, wires the RTBuddy
client, maps the register apertures, and enables clocks and power.
\texttt{ANEHWDevice::ANE\_Init}, reached from
\texttt{ANEHWDevice::power\_on\_hardware} on first power-on and on every
wake, drives the firmware handshake. This separation keeps device attach
independent of the firmware boot that runs again after each sleep.

The kernel maps seven named register apertures, not six. The six-bank
count belongs to a different system-on-chip family and does not hold on
the M1. The apertures are constructed in order, each with a device-tree
\texttt{reg} aperture index, and \Cref{tbl:c29-apertures} names the
seven by role.

\begin{longtable}[]{@{}
  >{\raggedright\arraybackslash}p{(\linewidth - 4\tabcolsep) * \real{0.3333}}
  >{\raggedright\arraybackslash}p{(\linewidth - 4\tabcolsep) * \real{0.3333}}
  >{\raggedright\arraybackslash}p{(\linewidth - 4\tabcolsep) * \real{0.3333}}@{}}
\caption{The seven memory-mapped register apertures the kernel maps
during \protect\texttt{ANEHWDevice::start} on the
M1.}\label{tbl:c29-apertures}\tabularnewline
\toprule\noalign{}
\begin{minipage}[b]{\linewidth}\raggedright
\textbf{Order}
\end{minipage} & \begin{minipage}[b]{\linewidth}\raggedright
\textbf{Bank}
\end{minipage} & \begin{minipage}[b]{\linewidth}\raggedright
\textbf{Role}
\end{minipage} \\
\midrule\noalign{}
\endfirsthead
\toprule\noalign{}
\begin{minipage}[b]{\linewidth}\raggedright
\textbf{Order}
\end{minipage} & \begin{minipage}[b]{\linewidth}\raggedright
\textbf{Bank}
\end{minipage} & \begin{minipage}[b]{\linewidth}\raggedright
\textbf{Role}
\end{minipage} \\
\midrule\noalign{}
\endhead
\bottomrule\noalign{}
\endlastfoot
1 & \texttt{ANE} & main control aperture, holding the scratch registers
and the reset-vector register \\
\rowcolor{zebra}2 & \texttt{PS} & power-state \\
3 & \texttt{PWGATE} & power-gate \\
\rowcolor{zebra}4 & \texttt{PTD} & page-table and translation-domain
control \\
5 & \texttt{ANEHAL1} & hardware-abstraction register aperture \\
\rowcolor{zebra}6 & \texttt{ANEHAL2} & hardware-abstraction register
aperture \\
7 & \texttt{ANEHAL3} & hardware-abstraction register aperture \\
\end{longtable}

The first four apertures take a contiguous index base and three
successive offsets. The three \texttt{ANEHAL} aperture indices are read
from per-family device fields, which is why the aperture count and
indices differ across chips.

The kernel allocates the RTBuddy client during attach and stores it on
the device object, with a name object retrieved from the \texttt{ANE}
property and a separate mailbox object. The endpoint itself comes up
inside \texttt{ANE\_Init} through \texttt{EnableRTBuddyEndpoints}, which
uses the endpoint identifier literal \texttt{0xe400}. The path first
tries \texttt{ANE1Endpoint1}, and on failure cleans up and retries
\texttt{ANEEndpoint1}. The endpoint lookup resolves the named service,
retrieves the endpoint object, and registers the inbound message
handler, with the doorbell paths wired alongside.

\texttt{ANE\_Init} then runs the firmware handshake in program order. It
quiesces pending work, brings up the \texttt{0xe400} endpoint,
initializes the scratch mailbox registers, and selects a warm or cold
boot by writing \texttt{1} or \texttt{0} to scratch register 7. It reads
the firmware boot address and writes it to the reset-vector register
\texttt{rANE\_H11\_CHINOOK\_IO\_RVBAR}, then releases reset and polls.
Three scratch handshakes follow, in the order the kernel observes them
(\Cref{tbl:c29-handshakes}).

\begin{longtable}[]{@{}
  >{\raggedright\arraybackslash}p{(\linewidth - 4\tabcolsep) * \real{0.3333}}
  >{\raggedright\arraybackslash}p{(\linewidth - 4\tabcolsep) * \real{0.3333}}
  >{\raggedright\arraybackslash}p{(\linewidth - 4\tabcolsep) * \real{0.3333}}@{}}
\caption{The three scratch-register handshakes the kernel waits on
during firmware bring-up, in
order.}\label{tbl:c29-handshakes}\tabularnewline
\toprule\noalign{}
\begin{minipage}[b]{\linewidth}\raggedright
\textbf{Handshake}
\end{minipage} & \begin{minipage}[b]{\linewidth}\raggedright
\textbf{Register}
\end{minipage} & \begin{minipage}[b]{\linewidth}\raggedright
\textbf{Meaning}
\end{minipage} \\
\midrule\noalign{}
\endfirsthead
\toprule\noalign{}
\begin{minipage}[b]{\linewidth}\raggedright
\textbf{Handshake}
\end{minipage} & \begin{minipage}[b]{\linewidth}\raggedright
\textbf{Register}
\end{minipage} & \begin{minipage}[b]{\linewidth}\raggedright
\textbf{Meaning}
\end{minipage} \\
\midrule\noalign{}
\endhead
\bottomrule\noalign{}
\endlastfoot
1 & scratch 7, first wake & the engine controller is alive after the
reset-vector handoff \\
\rowcolor{zebra}2 & scratch 7, second wake & the channel-description
table the firmware published is ready \\
3 & scratch 3, plus a magic-number-1 check & final acknowledgement
before interrupts are enabled \\
\end{longtable}

Between the second and third handshakes the kernel walks the
channel-description table the firmware published and resolves the named
inter-process channels: \texttt{SHAREDMALLOC}, \texttt{TERMINAL},
\texttt{BUF\_H2T}, \texttt{BUF\_T2H}, \texttt{IO}, \texttt{IO\_T2H},
\texttt{DEBUG}, and \texttt{DATA\_CHAIN\_H2T}. If the second wake never
arrives, the kernel powers the hardware off and on in a bounded retry
loop and increments a failure count.

On the firmware side the reset-vector entry brings up the real-time
kernel: it sets up the memory-management unit and the exception and
interrupt stacks, then the kernel heap, subject to the boot invariant
that the initial heap exceeds twice the framework reservation. The
firmware then checks its boot arguments and chip revision against the
values the kernel passed, exchanges an inter-process-communication
protocol version, and validates each ring buffer in the channel table
against a ring-buffer version constant. The firmware marks each control
channel ready in turn and rejects commands addressed to a channel before
it is ready.

Bring-up is ready to accept the first command when all of the following
hold.

\begin{itemize}
\tightlist
\item
  The kernel has enabled the \texttt{0xe400} endpoint, mapped the seven
  apertures, and enabled clocks and power.
\item
  The firmware has booted from the reset vector, brought up its heap,
  passed the chip-revision and protocol-version checks, published the
  channel table, and marked every control channel ready.
\item
  The three scratch handshakes have completed in order.
\item
  The kernel has logged that the engine controller is ready, enabled
  interrupts, and round-tripped the start command.
\end{itemize}

Only then does the kernel issue its first commands over the channel, in
order: print-enable, start, the performance-monitoring-unit base set, a
host-to-engine time synchronization, channel-property write,
resource-information query, default-setting write, memory-cache
power-on, and the shared-event-information initialization. The
resource-information query returns the engine count, cache-request
limits, and maximum procedure count, and the default-setting write sets
the context-switch latency threshold.

\section{Execution loop}\label{execution-loop}

The execution loop accepts four command classes on the shared channel
(\Cref{lst:c29-exeloop-cmds}).

\begin{listing}[H]
\caption{The four execution-loop command classes accepted on the channel.}\label{lst:c29-exeloop-cmds}

\begin{Shaded}
\begin{Highlighting}[]
\KeywordTok{union}\NormalTok{ uCExeLoopSupportedCmd          }\OperatorTok{;}\NormalTok{ insize }\OperatorTok{\textless{}=} \KeywordTok{sizeof}\OperatorTok{(}\KeywordTok{union}\NormalTok{ uCExeLoopSupportedCmd}\OperatorTok{)}

\OperatorTok{[}\NormalTok{PROC CALL}\OperatorTok{]}\NormalTok{              ProgId}\OperatorTok{=\%}\NormalTok{d  ProcId}\OperatorTok{=\%}\NormalTok{d  Proc}\OperatorTok{=\%}\NormalTok{d  Pri}\OperatorTok{=\%}\NormalTok{d}
\OperatorTok{[}\NormalTok{PROC CALL WITH BARS}\OperatorTok{]}    \OperatorTok{+}\NormalTok{ nbrOfCustomBars }\OperatorTok{\textless{}=} \DecValTok{32}        \OperatorTok{;}\NormalTok{ custom buffer}\OperatorTok{{-}}\NormalTok{access}\OperatorTok{{-}}\DataTypeTok{register}\NormalTok{ overrides}
\OperatorTok{[}\NormalTok{PROC CALL WITH EVENTS}\OperatorTok{]}  \OperatorTok{+}\NormalTok{ nbrOfSignalEvents in }\FloatTok{1.}\ErrorTok{.16}   \OperatorTok{;}\NormalTok{ wait}\OperatorTok{/}\NormalTok{signal events}
                         \OperatorTok{+} \OperatorTok{(}\NormalTok{nbrOfWaitEvents }\OperatorTok{+}\NormalTok{ nbrOfSignalEvents}\OperatorTok{)} \OperatorTok{\textgreater{}} \DecValTok{0}
\OperatorTok{[}\NormalTok{CR TRIGGER}\OperatorTok{]}\NormalTok{             cacheHandler }\DecValTok{0}\ErrorTok{x}\OperatorTok{\%}\NormalTok{llx            }\OperatorTok{;}\NormalTok{ data}\OperatorTok{{-}}\NormalTok{chaining cache}\OperatorTok{{-}}\NormalTok{request trigger}
\end{Highlighting}
\end{Shaded}

\end{listing}

The execution loop is the hot path that turns a host procedure-call
command into work for the task-descriptor driver. A host writes a
command into the shared ring buffer. The loop receives that command
holding a program identifier, procedure identifier, procedure index, and
priority. It builds the request list and pushes the program's
task-descriptor partitions onto a task queue; the task-descriptor driver
runs them on the cores while the tensor-mover streams the operands. The
command is a fixed-layout record on the shared channel, and the loop
logs each class as it accepts it. A bounds check rejects any record
larger than the supported-command union before the loop reads its
fields. The bar field is a buffer-access register, the buffer base and
size binding a task descriptor references by index, and it logs as
\texttt{bar{[}\%d{]}:\ type=\%d\ cfg=\%d\ barId=\%d\ value=0x\%llx\ bufSize=\%lld}.
The loop rejects an unknown opcode with
\texttt{Cmd\ 0x\%x\ is\ not\ supported\ through\ ExeLoop\ cmd\ channel}
and drops the record.

The main command classes on this channel are a plain procedure call,
procedure call holding buffer-base overrides, procedure call holding
wait-and-signal events, and data-chaining cache-request trigger. The
dispatcher checks each command identifier on every event against a fixed
accepted set and drops anything else as unsupported: the procedure-call
family (\texttt{0x204}, \texttt{0x20c}, \texttt{0x211}, and
\texttt{0x212}), the cache-request trigger (\texttt{0x209} with its path
variant \texttt{0x20a}), the channel data-file load \texttt{0x2d}, and
the back-channel and process-id controls (\texttt{0x404},
\texttt{0xff00}). The complete decoded command set is in chapter
\hyperref[host-to-firmware-command-protocol]{30}. The firmware drops and
counts a procedure call addressed to a process that is not running.

The procedure-call command structures list the buffer-access registers
and optional event and execute-order arrays under fixed field limits,
recovered from the firmware assertion strings
(\Cref{tbl:c29-field-limits}).

\begin{longtable}[]{@{}
  >{\raggedright\arraybackslash}p{(\linewidth - 2\tabcolsep) * \real{0.5000}}
  >{\raggedright\arraybackslash}p{(\linewidth - 2\tabcolsep) * \real{0.5000}}@{}}
\caption{The field limits on a procedure-call command, recovered from
the firmware assertion
strings.}\label{tbl:c29-field-limits}\tabularnewline
\toprule\noalign{}
\begin{minipage}[b]{\linewidth}\raggedright
\textbf{Field}
\end{minipage} & \begin{minipage}[b]{\linewidth}\raggedright
\textbf{Limit}
\end{minipage} \\
\midrule\noalign{}
\endfirsthead
\toprule\noalign{}
\begin{minipage}[b]{\linewidth}\raggedright
\textbf{Field}
\end{minipage} & \begin{minipage}[b]{\linewidth}\raggedright
\textbf{Limit}
\end{minipage} \\
\midrule\noalign{}
\endhead
\bottomrule\noalign{}
\endlastfoot
custom buffer-access registers & at most 32 \\
\rowcolor{zebra}signal events & greater than 0 and at most 16 \\
wait events plus signal events & greater than 0 \\
\rowcolor{zebra}task-descriptor partitions & at least 1 and below the
per-procedure maximum \\
custom execute-order entries & at most 128 \\
\rowcolor{zebra}output buffer sets & exactly 1 \\
event masks & exactly 1 \\
\end{longtable}

Each buffer-access register binds a buffer base and size that a task
descriptor references by index, with at most 32 hardware register slots
addressing the distinct buffer regions for one operation, and at most 16
input buffers per trigger.

Resident state across dispatches comes from the data-chaining trigger. A
cache request chains one execution's output set onto the next
execution's input set, so a value produced by one dispatch is available
to the next without a copy back to the host. The resident key-and-value
cache and the resident optimizer state described in chapter
\hyperref[execution-model]{2} use this native chaining path. The
firmware can chain procedures, running several in sequence on the engine
without returning to the host between them.

\section{Doorbell emit}\label{doorbell-emit}

The host-notify store is bracketed by the window-gate bit clear and set
(\Cref{lst:c29-doorbell-emit}).

\begin{listing}[H]
\caption{The guarded doorbell-emit sequence, bracketing the host-notify store with the window-gate bit clear and set.}\label{lst:c29-doorbell-emit}

\begin{Shaded}
\begin{Highlighting}[]
\NormalTok{ring\_doorbell(DoorBellReg, DoorBellBit):     ; host{-}notify site @0x4c890}
\NormalTok{  w0 = disable\_irq()                         ; DAIFSet \#0x2, make the quad atomic}
\NormalTok{  x8 = mrs S3\_3\_C15\_C8\_0}
\NormalTok{  x8 = x8 \& \textasciitilde{}(1 \textless{}\textless{} 39)                        ; CLEAR bit 39: arm the doorbell window}
\NormalTok{       msr S3\_3\_C15\_C8\_0, x8}
\NormalTok{  str DoorBellBit {-}\textgreater{} [DoorBellReg]            ; *** the doorbell store (the interrupt) ***}
\NormalTok{  dsb sy ; isb                                ; force the store + status update to commit}
\NormalTok{  x8 = mrs S3\_3\_C15\_C8\_0}
\NormalTok{  if x8 \& (1 \textless{}\textless{} 1): fatal "Uncorrectable L2C error overflow"}
\NormalTok{  if x8 \& (1 \textless{}\textless{} 7): rc = 1, clear sticky \{1,7\} ; transaction rejected, retry}
\NormalTok{  x8 = mrs S3\_3\_C15\_C8\_0}
\NormalTok{  x8 = x8 | (1 \textless{}\textless{} 39)                         ; SET bit 39: close the window}
\NormalTok{       msr S3\_3\_C15\_C8\_0, x8}
\NormalTok{  restore\_irq(w0)}
\NormalTok{  return rc}
\end{Highlighting}
\end{Shaded}

\end{listing}

After staging a task-descriptor list the driver notifies the host by
ringing a doorbell. The engine-to-host interrupt is a single 32-bit
memory-mapped store to a host-supplied register, bracketed by the
windowed-store gate of chapter
\hyperref[kernel-driver-and-iokit-abi]{27}. Bit 39 of the
implementation-defined system register \texttt{S3\_3\_C15\_C8\_0} is
cleared to arm the doorbell window, and the store is forwarded as a
fabric-coherent transaction. Bit 39 is set again to close the window.
The firmware masks interrupts across the four steps so the arm, store,
sample, and disarm cannot be interrupted, and the two status bits
sampled after the store record the fabric response. The same guarded
sequence drives the engine-to-graphics-processor synchronization
doorbell, which writes the value \texttt{1} to the fixed memory-mapped
target \texttt{0x2\_0646\_8000} between the identical bit-39 clear and
set.

The system register is not a single-bit gate. At least six bit positions
have distinct meaning, recovered from the read, write, and sample
patterns at the 28 access sites and decoded in
\Cref{tbl:c29-fabric-reg}.

\begin{longtable}[]{@{}
  >{\raggedright\arraybackslash}p{(\linewidth - 2\tabcolsep) * \real{0.5000}}
  >{\raggedright\arraybackslash}p{(\linewidth - 2\tabcolsep) * \real{0.5000}}@{}}
\caption{The decoded bit map of the implementation-defined fabric system
register
\protect\texttt{S3\_3\_C15\_C8\_0}.}\label{tbl:c29-fabric-reg}\tabularnewline
\toprule\noalign{}
\begin{minipage}[b]{\linewidth}\raggedright
\textbf{Bit}
\end{minipage} & \begin{minipage}[b]{\linewidth}\raggedright
\textbf{Role}
\end{minipage} \\
\midrule\noalign{}
\endfirsthead
\toprule\noalign{}
\begin{minipage}[b]{\linewidth}\raggedright
\textbf{Bit}
\end{minipage} & \begin{minipage}[b]{\linewidth}\raggedright
\textbf{Role}
\end{minipage} \\
\midrule\noalign{}
\endhead
\bottomrule\noalign{}
\endlastfoot
0 & fabric-path enable, cleared with bits 2 and 4 on the power-gate
path \\
\rowcolor{zebra}1 & uncorrectable last-level-cache error, read-only and
sticky \\
2 & fabric-link status, must be 1 for the power-gate condition \\
\rowcolor{zebra}4 & fabric-link secondary status, must be 0 for the
power-gate condition \\
7 & transaction-reject status, clearable on a rejected doorbell \\
\rowcolor{zebra}39 & doorbell-window arm and disarm: clear opens the
store window, set closes it \\
\end{longtable}

Clearing bit 39 opens a window in which a store to a fabric doorbell
aperture is forwarded as a coherent doorbell transaction, and bits 1 and
7 are the fabric response status latched by the barrier after the store.
The power-gate path reads the register to sample bits 2 and 4 as
fabric-link status before it clock-gates the engine, requiring bit 2 set
and bit 4 clear, then clears bits 0, 2, and 4.

A completion event can fan out to up to three host doorbells. The host
doorbell target is a register-and-bit pair the host supplies in an
endpoint descriptor and the engine stores resident, in three slots, so
the shared-event path can notify the inference client, a cross-agent
waiter, and a telemetry sink from one completion.

\section{Scheduler}\label{scheduler}

\Cref{tbl:c29-scheduler} collects the firmware scheduler properties,
covering its priority model, bands, preemption rules, watchdog, and
software deadline.

\begin{longtable}[]{@{}ll@{}}
\caption{The firmware scheduler
properties.}\label{tbl:c29-scheduler}\tabularnewline
\toprule\noalign{}
\textbf{Scheduler property} & \textbf{Value} \\
\midrule\noalign{}
\endfirsthead
\toprule\noalign{}
\textbf{Scheduler property} & \textbf{Value} \\
\midrule\noalign{}
\endhead
\bottomrule\noalign{}
\endlastfoot
Priority model & fixed-priority, eight levels (0 through 7) \\
\rowcolor{zebra}System band & levels 0 through 1 \\
Application band & levels 2 through 7 \\
\rowcolor{zebra}Preemption within a task & none, run to completion \\
Preemption between tasks & by kernel thread priority \\
\rowcolor{zebra}Hardware watchdog & none \\
Software deadline & about 2 seconds, queue-idle wait then abort \\
\rowcolor{zebra}Task-queue identifiers & 1 through 255, identifier 0
reserved \\
\end{longtable}

The scheduler is fixed-priority with eight levels, numbered zero through
seven. The levels split into two bands: levels zero and one are reserved
for the system, and levels two through seven are application priorities.
Preemption within a task does not occur: each message pump runs to
completion, and preemption happens only between pump threads by kernel
priority. Priority exists on two axes: the kernel thread priority orders
the worker tasks, and the application job priority orders queued
inference jobs onto the per-priority task queues. They are distinct
numbers on distinct objects. No hardware watchdog panics the engine. The
firmware self-polices with a software deadline: a task queue that will
not quiesce within about two seconds is force-aborted.

The submit path pushes work onto numbered task queues keyed by a network
identifier in the range 1 through 255, with identifier 0 reserved. The
submit path guards the queue identifier below 8, polls the queue for
vacancy, enables the queue, copies a 35-word task-descriptor register
block into the queue's register file, and rings a submit doorbell. Each
queue occupies a fixed stride of \texttt{0x148} bytes in the engine
register aperture, with the per-queue registers the submit path writes
given in \Cref{tbl:c29-queue-regs}.

\begin{longtable}[]{@{}ll@{}}
\caption{The per-queue register map, at the queue base plus the queue
identifier times \protect\texttt{0x148}, that the submit path writes to
start hardware work.}\label{tbl:c29-queue-regs}\tabularnewline
\toprule\noalign{}
\textbf{Offset within queue} & \textbf{Register} \\
\midrule\noalign{}
\endfirsthead
\toprule\noalign{}
\textbf{Offset within queue} & \textbf{Register} \\
\midrule\noalign{}
\endhead
\bottomrule\noalign{}
\endlastfoot
\texttt{+0x24000} & submit doorbell \\
\rowcolor{zebra}\texttt{+0x24004} & size and count \\
\texttt{+0x24008} & priority and network identifier \\
\rowcolor{zebra}\texttt{+0x25000} & queue enable \\
\texttt{+0x2500c} & queue status and vacancy \\
\rowcolor{zebra}\texttt{+0x250a0} & buffer-access-register file \\
\end{longtable}

The eight application priorities map to a fixed in-firmware table of
queue weights, \texttt{\{1,\ 2,\ 3,\ 4,\ 5,\ 6,\ 30,\ 31\}}, and
scheduling is strict fixed-priority with per-priority credit and no
preemption. The scheduler bounds how many requests are in flight on the
firmware at once and applies backpressure when it saturates, through
\texttt{ANEScheduler::wakePendingRequestsQueueWithFWOverload} and
\texttt{ANEScheduler::pendingRequestsWithFirmwareCount}, with a
dedicated unwire thread that releases resources behind completed work.
The firmware-overload throttle is the mechanism behind the large-batch
trainable-convolution stall described in chapter
\hyperref[pitfalls-and-limits]{19}: a submission that overruns the
firmware queue is throttled.

\section{Control state machine}\label{control-state-machine}

A finite-state machine drives the execution loop
(\Cref{tbl:c29-control-fsm}).

\begin{longtable}[]{@{}
  >{\raggedright\arraybackslash}p{(\linewidth - 2\tabcolsep) * \real{0.5000}}
  >{\raggedright\arraybackslash}p{(\linewidth - 2\tabcolsep) * \real{0.5000}}@{}}
\caption{The execution-loop control state machine, with what each state
does.}\label{tbl:c29-control-fsm}\tabularnewline
\toprule\noalign{}
\begin{minipage}[b]{\linewidth}\raggedright
\textbf{State}
\end{minipage} & \begin{minipage}[b]{\linewidth}\raggedright
\textbf{What it does}
\end{minipage} \\
\midrule\noalign{}
\endfirsthead
\toprule\noalign{}
\begin{minipage}[b]{\linewidth}\raggedright
\textbf{State}
\end{minipage} & \begin{minipage}[b]{\linewidth}\raggedright
\textbf{What it does}
\end{minipage} \\
\midrule\noalign{}
\endhead
\bottomrule\noalign{}
\endlastfoot
\texttt{INIT} (0) & the start node, after the loop is constructed but
before it is armed; no hardware work dispatches \\
\rowcolor{zebra}\texttt{RUN} (1) & non-secure and runnable, the engine
is powered and owned but nothing is in flight \\
\texttt{EXEC} (2) & a task-descriptor partition is in flight,
non-secure; the state in which a handler is permitted to touch
hardware \\
\rowcolor{zebra}\texttt{EXEC-SECURE} (3) & in flight while the secure
phase is asserted, the engine owned by the secure tenant \\
\texttt{PAUSE} & fully quiesced for a secure-mode boundary, rejects all
events \\
\end{longtable}

The framework stores no name for each state, so the meaning of each is
reconstructed from the handlers that read the current state and from the
events that drive it. Four states have framework identifiers zero
through three, and a fifth is a named transient crossed during a
secure-mode switch. Posted events drive the machine. The event alphabet
is four entries: a command-dispatch event when a procedure call or
trigger is accepted, drain event toward idle, enter-secure event, and
exit-secure event. A command accepted on the channel posts the dispatch
event, which moves the machine from \texttt{RUN} to \texttt{EXEC} and
pushes the task-descriptor list. A request finishing returns the machine
from \texttt{EXEC} to \texttt{RUN}.

Roughly seven independent handlers gate hardware work on the same test:
the current state must be \texttt{EXEC} or \texttt{EXEC-SECURE} before
the handler writes the memory-mapped registers or rings a doorbell. A
switch into or out of the secure phase quiesces the engine first, with
the task queues disabled, and the machine passes through \texttt{PAUSE}
between the two phases. Any event delivered while in \texttt{PAUSE} is a
firmware invariant violation and asserts.

Alongside the execution loop sit three smaller state machines, each with
explicit named states. A two-state process machine marks each process
slot idle or running. A three-state cache-request machine moves a
resident chain through available, in-use, and invalidating. A two-state
output-set machine marks a chained output buffer available or
in-execution.

\section{Per-run statistics buffer}\label{per-run-statistics-buffer}

The path runs from the host size query and map through to the firmware
completion write (\Cref{lst:c29-stats-path}).

\begin{listing}[H]
\caption{The per-run statistics-buffer path, from the host size query and map through the firmware completion write.}\label{lst:c29-stats-path}

\begin{Shaded}
\begin{Highlighting}[]
\NormalTok{host: stats{-}buffer size{-}get            {-}\textgreater{} \{hdr 0x28, evDesc 0x14, dbg 0x10, perEngine 0x20\}}
\NormalTok{host: PreMap stats buffer \{addr,size\}  {-}\textgreater{} DART{-}map into the firmware aperture}
\NormalTok{        if addr == 0 or size == 0:  cbz skips the map; later writes have no destination}
\NormalTok{host: procedure call (inference)}
\NormalTok{firmware completion (CAneProgramManagerH11):}
\NormalTok{        per engine: copy the 16{-}byte counter quad from engine\_block+0x30 into a descriptor}
\NormalTok{        header: magic 0x0101, per{-}call timing, per{-}call counter, descriptor{-}area size}
\NormalTok{host: reads validCount, logEvents, statEvents back out of the mapped buffer}
\end{Highlighting}
\end{Shaded}

\end{listing}

The firmware reports per-run timing and event counts into a statistics
buffer the host supplies. \texttt{CAneProgramManagerH11} packs the
records on inference completion, but it never allocates the destination.
The host first queries the layout sizes, then maps a buffer through the
address-translation manager, and the firmware writes into that mapping
at completion.

The record begins with a \texttt{0x28}-byte header
(\Cref{tbl:c29-stats-header}). The first two bytes are the magic value
\texttt{0x0101}, a version stamp.

\begin{longtable}[]{@{}
  >{\raggedright\arraybackslash}p{(\linewidth - 4\tabcolsep) * \real{0.3333}}
  >{\raggedright\arraybackslash}p{(\linewidth - 4\tabcolsep) * \real{0.3333}}
  >{\raggedright\arraybackslash}p{(\linewidth - 4\tabcolsep) * \real{0.3333}}@{}}
\caption{The \protect\texttt{0x28}-byte statistics-buffer header
\protect\texttt{CAneProgramManagerH11} writes on inference
completion.}\label{tbl:c29-stats-header}\tabularnewline
\toprule\noalign{}
\begin{minipage}[b]{\linewidth}\raggedright
\textbf{Offset}
\end{minipage} & \begin{minipage}[b]{\linewidth}\raggedright
\textbf{Width}
\end{minipage} & \begin{minipage}[b]{\linewidth}\raggedright
\textbf{Field}
\end{minipage} \\
\midrule\noalign{}
\endfirsthead
\toprule\noalign{}
\begin{minipage}[b]{\linewidth}\raggedright
\textbf{Offset}
\end{minipage} & \begin{minipage}[b]{\linewidth}\raggedright
\textbf{Width}
\end{minipage} & \begin{minipage}[b]{\linewidth}\raggedright
\textbf{Field}
\end{minipage} \\
\midrule\noalign{}
\endhead
\bottomrule\noalign{}
\endlastfoot
\texttt{+0x00} & 2 & magic \texttt{0x0101} \\
\rowcolor{zebra}\texttt{+0x04} & 8 & per-call timing value, from the
real-time-kernel timebase \\
\texttt{+0x0c} & 4 & word from the call object \\
\rowcolor{zebra}\texttt{+0x10} & 8 & reserved, cleared \\
\texttt{+0x18} & 4 & word from the call object \\
\rowcolor{zebra}\texttt{+0x1c} & 8 & per-call counter value \\
\texttt{+0x24} & 4 & descriptor-area byte length, the total size minus
\texttt{0x28} \\
\end{longtable}

After the header the firmware copies one descriptor per active engine,
looping over up to 32 engines at a stride of \texttt{0x48} bytes in the
engine table. Each descriptor contains the engine identifier and type,
call index, and engine index, followed by a 16-byte counter quad read
from \texttt{engine\_block+0x30}. The size query returns four constants
the host uses to size the buffer: a \texttt{0x28}-byte header, a
\texttt{0x14}-byte per-event descriptor, a \texttt{0x10}-byte
per-debug-event record, and a \texttt{0x20}-byte per-engine descriptor.
The per-engine values are software-tracked latency and event
accumulators maintained by the firmware profiler, not raw register
reads. The buffer also stores running totals the host reads back: a
valid-record count, log-event total, and stat-event total.

The read is gated by a host-side null check. The map handler loads the
host buffer address and size from the command, and a \texttt{cbz} on
each skips the address-translation map when either is null, as
\Cref{lst:c29-stats-nullcheck} shows.

\begin{listing}[H]
\caption{The null check that skips the statistics-buffer map when the host supplies no address or size.}\label{lst:c29-stats-nullcheck}

\begin{Shaded}
\begin{Highlighting}[]
\NormalTok{PreMap stats buffer:}
\NormalTok{  x22 = host stats{-}buffer address     ; from the command}
\NormalTok{  cbz x22, skip                       ; address == 0 {-}\textgreater{} no map}
\NormalTok{  x23 = host stats{-}buffer size}
\NormalTok{  cbz x23, skip                       ; size == 0    {-}\textgreater{} no map}
\NormalTok{  ... DART{-}map \{x22, x23\} into the firmware aperture ...}
\end{Highlighting}
\end{Shaded}

\end{listing}

When the host never maps a buffer, the mapped pointer stays null, the
completion writer has no destination, and every counter value the host
observes is null. A separate mechanism reads the
performance-monitoring-unit aperture at the fixed base
\texttt{0x2\_8e08\_c000}. Those reads are 32-bit power-status registers
at a stride of eight bytes, up to five entries, whose low byte is a
per-engine power state. They drive power and frequency-scaling decisions
and are never packed into the statistics buffer. The kernel hardware
performance-counter block is a third mechanism again, armed and drained
by the kernel, and the firmware does not touch it.

\section{Faults and recovery}\label{faults-and-recovery}

A failed invariant routes into the kernel fault handler, which produces
a post-mortem dump in a fixed order. The dump records the exception
class, fault-address and exception-syndrome registers, full
general-purpose register file, cache-controller error registers, a
restart count that persists across restarts, and a replay of the
recent-command ring newest first. The firmware keeps that command ring
so a crash dump shows what the engine was running when it died. The
platform exception handler stringifies four exception classes for the
dump: non-maskable interrupt, interrupt, fast interrupt, and synchronous
abort. The dump names the faulting task by name, marks whether the fault
occurred in interrupt context or in a task, gives the processor
identifier, and prints a call stack.

The fault model below a true exception is abort-and-recover per task
queue. The firmware aborts a stuck queue through the engine driver's
hardware-abort path, and an abort that does not clear within the
two-second deadline escalates to a panic. A dedicated timeout interrupt
fires on a global stuck flag, logs an event record, and drives the
abort. When a per-queue abort cannot clear the fault, the firmware
escalates to a host-coordinated reset: it takes a reset mutex and sends
a reset notification holding the task-queue range the reset spans,
telling the host to tear down and re-initialize.

The engine cannot write a filesystem, so it depends on the host to drain
the dump. After staging the dump the firmware blocks while the host
pulls it section by section, then the host resets the channel and
re-initializes the firmware to the ready handshake.

Firmware status is not a flat numeric enum. A namespace of notification
commands sent to the host records it, alongside inline return values at
the point of failure. A generic per-channel error notification has an
opaque error index, a reset notification gives the task-queue range the
reset spans, and a tile-sync error notification reports a data-movement
error.

Faults divide into four dispositions. A version-minor mismatch warns and
continues. A command addressed to a torn-down process or a wrong-state
cache handle is dropped and counted. A command failing integrity,
section, or argument validation is rejected without running. A failed
invariant, a processor exception, or an uncorrectable cache error faults
into the dump-and-reset path.

The recovery sequence is serialized behind a single command gate. A
fault stages a firmware dump, cancels outstanding commands, and re-boots
the firmware to the ready handshake before the gate frees.

\chapter{Host-to-firmware command protocol}\label{host-to-firmware-command-protocol}

\begin{summarybox}

The host drives the engine with one message protocol over a shared
mailbox. A fixed \texttt{sCSneControllerCmdHdr} header names a program,
process, and procedure, and a command body follows. The dispatched
command space on the M1 is 93 identifiers, \texttt{0x00} through
\texttt{0x5c}, indexed by \texttt{eCSneCmdId}, with \texttt{0xFFFFFFFF}
reserved as the invalid sentinel. A submission posts to a bound ring
endpoint. The controller validates it, queues it by priority, runs it,
and returns completion as a firmware-to-host notification in the same
identifier space.

\end{summarybox}

The in-package controller is a real-time-kernel firmware image, and the
host communicates with it through a ring-buffer endpoint with a single
command vocabulary, \texttt{CSNE\_CMD\_*}, the controller-side
neural-engine command set. This chapter covers the wire protocol between
the host-side execution model of chapter \hyperref[execution-model]{2}
and the program file layers of chapter
\hyperref[program-and-container-format]{23}.

\section{Command header}\label{command-header}

\texttt{sCSneControllerCmdHdr} prefixes every ring message with six
fields that name the command and bind it to a loaded program
(\Cref{tbl:c30-cmd-header}).

\begin{longtable}[]{@{}
  >{\raggedright\arraybackslash}p{(\linewidth - 6\tabcolsep) * \real{0.2500}}
  >{\raggedright\arraybackslash}p{(\linewidth - 6\tabcolsep) * \real{0.2500}}
  >{\raggedright\arraybackslash}p{(\linewidth - 6\tabcolsep) * \real{0.2500}}
  >{\raggedright\arraybackslash}p{(\linewidth - 6\tabcolsep) * \real{0.2500}}@{}}
\caption{The six fields of the command
header.}\label{tbl:c30-cmd-header}\tabularnewline
\toprule\noalign{}
\begin{minipage}[b]{\linewidth}\raggedright
\textbf{Field}
\end{minipage} & \begin{minipage}[b]{\linewidth}\raggedright
\textbf{Offset}
\end{minipage} & \begin{minipage}[b]{\linewidth}\raggedright
\textbf{Width}
\end{minipage} & \begin{minipage}[b]{\linewidth}\raggedright
\textbf{Meaning}
\end{minipage} \\
\midrule\noalign{}
\endfirsthead
\toprule\noalign{}
\begin{minipage}[b]{\linewidth}\raggedright
\textbf{Field}
\end{minipage} & \begin{minipage}[b]{\linewidth}\raggedright
\textbf{Offset}
\end{minipage} & \begin{minipage}[b]{\linewidth}\raggedright
\textbf{Width}
\end{minipage} & \begin{minipage}[b]{\linewidth}\raggedright
\textbf{Meaning}
\end{minipage} \\
\midrule\noalign{}
\endhead
\bottomrule\noalign{}
\endlastfoot
\texttt{id} & \texttt{0x00} & \texttt{u32} & the \texttt{eCSneCmdId}
selector, logged as \texttt{\%\#04x} \\
\rowcolor{zebra}\texttt{size} & \texttt{0x04} & \texttt{u32} & byte
length of the command body that follows \\
\texttt{priority} & \texttt{0x08} & \texttt{u32} & scheduling band,
\texttt{0..7} \\
\rowcolor{zebra}\texttt{programId} & \texttt{0x0c} & \texttt{i32} &
loaded-program slot, \texttt{-1} when invalid \\
\texttt{processId} & \texttt{0x10} & \texttt{i32} & per-program process
instance, \texttt{-1} when none \\
\rowcolor{zebra}\texttt{procedureId} & \texttt{0x14} & \texttt{u32} &
index into the program's procedure table \\
\end{longtable}

The dispatcher validates the full message length before reading the
body:
\texttt{maxCmdBufSize\ \textgreater{}=\ sizeof(*pMsg)\ +\ pCmdHdr-\textgreater{}size}.
The byte offsets follow the field order under a packed 32-bit layout,
and the field presence and widths are read from the firmware's assert
and log format strings. The \texttt{priority} field has two scheduling
classes: values \texttt{0} and \texttt{1} are the privileged realtime
band and values \texttt{2} through \texttt{7} are the normal queue.
These are encoded by the firmware checks
\texttt{(priority\ \textgreater{}=\ 0\ \&\&\ priority\ \textless{}=\ 1)\ \textbar{}\textbar{}\ (priority\ \textgreater{}=\ 2\ \&\&\ priority\ \textless{}=\ 7)}
and the normal-only
\texttt{(priority\ \textgreater{}=\ 2\ \&\&\ priority\ \textless{}=\ 7)},
under the full-range bound \texttt{fwPriority\ \textless{}\ 8}. The
signed \texttt{programId} and \texttt{processId} use \texttt{-1}, which
reads as \texttt{0xFFFFFFFF}, as the unbound-slot marker. The firmware
resolves the pair against \texttt{progProcInfo{[}builtinProgramId{]}}
with the fields \texttt{.valid}, \texttt{.progId}, and \texttt{.procId},
under \texttt{builtinProgramId\ \textless{}\ maxProgramServerSupported}
and \texttt{id\ \textless{}\ CAneBuiltInNetworkId::ANE\_NET\_TOT}. The
resident cache-request commands prepend one further field, an opaque
64-bit \texttt{cacheHandler} logged as \texttt{cacheHandler:\ 0x\%llx},
naming a long-lived request object.

The scheduler maps the eight priority levels onto two admitted classes
(\Cref{tbl:c30-priority}).

\begin{longtable}[]{@{}
  >{\raggedright\arraybackslash}p{(\linewidth - 4\tabcolsep) * \real{0.3333}}
  >{\raggedright\arraybackslash}p{(\linewidth - 4\tabcolsep) * \real{0.3333}}
  >{\raggedright\arraybackslash}p{(\linewidth - 4\tabcolsep) * \real{0.3333}}@{}}
\caption{The eight priority levels and the two scheduling classes they
map onto.}\label{tbl:c30-priority}\tabularnewline
\toprule\noalign{}
\begin{minipage}[b]{\linewidth}\raggedright
\textbf{Level}
\end{minipage} & \begin{minipage}[b]{\linewidth}\raggedright
\textbf{Class}
\end{minipage} & \begin{minipage}[b]{\linewidth}\raggedright
\textbf{Firmware gate}
\end{minipage} \\
\midrule\noalign{}
\endfirsthead
\toprule\noalign{}
\begin{minipage}[b]{\linewidth}\raggedright
\textbf{Level}
\end{minipage} & \begin{minipage}[b]{\linewidth}\raggedright
\textbf{Class}
\end{minipage} & \begin{minipage}[b]{\linewidth}\raggedright
\textbf{Firmware gate}
\end{minipage} \\
\midrule\noalign{}
\endhead
\bottomrule\noalign{}
\endlastfoot
\texttt{0}, \texttt{1} & high, realtime &
\texttt{(priority\ \textgreater{}=\ 0\ \&\&\ priority\ \textless{}=\ 1)},
separately admitted \\
\rowcolor{zebra}\texttt{2} through \texttt{7} & normal queue &
\texttt{(priority\ \textgreater{}=\ 2\ \&\&\ priority\ \textless{}=\ 7)} \\
\end{longtable}

The full-range bound is \texttt{fwPriority\ \textless{}\ 8}, with
\texttt{maxPriorityNbr\ \textless{}=\ 8}, and a submission is queued
under the scheduler key
\texttt{(userId,\ jobId,\ fwPriority,\ hwqId,\ tqId,\ queueId)}, where
\texttt{queueId\ \textless{}\ priorityQCount} selects the schedule-info
slot.

\section{Command set}\label{command-set}

The dispatched space is 93 entries, identifiers \texttt{0x00} through
\texttt{0x5c}. The firmware enumerates them in one contiguous, ordered
string table, \texttt{CAneControllerStringsH11}, which begins at
firmware file offset \texttt{0xa83b4} and runs sequentially. The command
logger indexes that table by identifier to render each
\texttt{CMD\ =\ \%\#04x\ {[}\%s{]}} log line. The dispatched identifiers
group into subsystems. Lifecycle and power bring the controller and
engine up and down, property reads and writes config and registers, and
stats drives printing, profiling, and tracing. The rest bind and feed
the engine: ipc binds endpoints, buffer configures and recycles the
channel pool, and the program, execution, cache, and secure families
carry the inference path itself. The complete set in numeric-identifier
order, with each command's direction, subsystem, and recovered request
struct, is in \Cref{tbl:apc-command-table}.

Most identifiers are host-to-firmware requests; fourteen in the same
space are firmware-to-host notifications, and one,
\texttt{BACK\_CHANNEL\_RPC} at \texttt{0x5b}, flows in both directions.
The sentinel \texttt{ECSneCmdId\_Invalid}, value \texttt{0xFFFFFFFF}, is
the no-command marker and has no table entry. Two further strings the
firmware contains are not dispatched identifiers:
\texttt{CSNE\_CMD\_START} is a standalone log string with no table slot,
and \texttt{CSNE\_CMD\_IPC\_ENDPOINT\_TYPE\_DATA\_CHAINING} is an
endpoint-type enum value rather than a command.

\section{Command lifecycle}\label{command-lifecycle}

\begin{figure}[H]\centering
\includegraphics[width=6.373in]{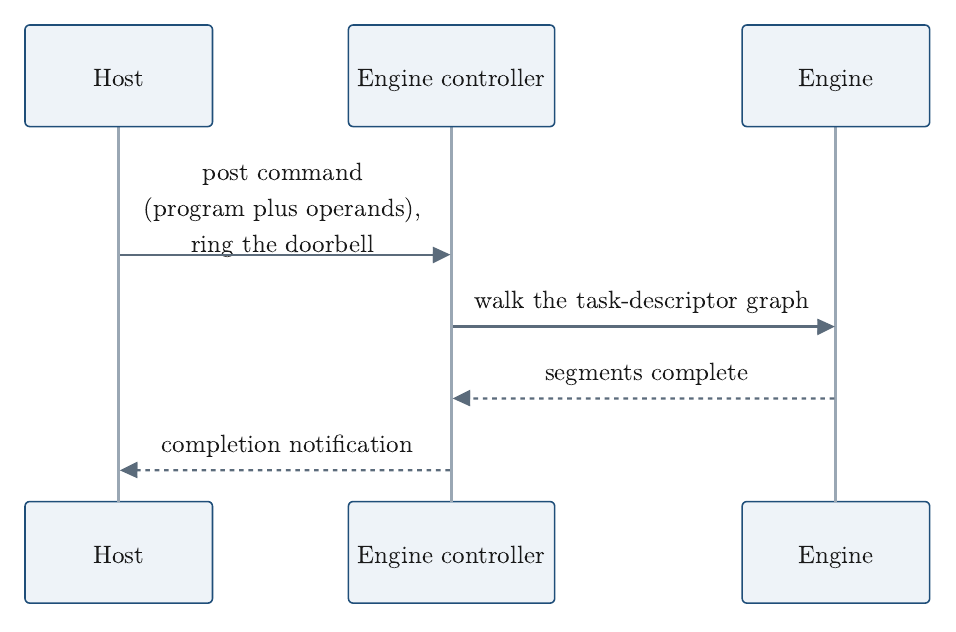}
\caption{A submission moving between the host, controller, and engine.}
\label{fig:c30-lifecycle}
\end{figure}

A submission moves through four stages: the host posts, the controller
picks it up, executes it, and notifies completion
(\Cref{fig:c30-lifecycle}).

The host posts by writing the header and body into the ring-buffer slot
of a bound endpoint and signalling the controller. The host must bind
the endpoint first with \texttt{CSNE\_CMD\_IPC\_ENDPOINT\_SET}, and the
execution-loop endpoint is required to be the data-chaining type: the
firmware asserts
\texttt{endPointId\ ==\ CSNE\_CMD\_IPC\_ENDPOINT\_TYPE\_DATA\_CHAINING}.
The flag \texttt{ipcEndpointSetDone} guards endpoint state, false before
the bind and true after, so the firmware rejects a submission on an
unbound endpoint before any work begins.

The controller picks the message off the ring and dispatches on
\texttt{id}. For a procedure call it validates the named procedure
against the loaded program,
\texttt{procedureId\ \textless{}\ pProg-\textgreater{}pProgram-\textgreater{}procNbr},
and resolves the content type through
\texttt{getProcedureCallType(programId,\ procedureId)}, which must fall
in the admitted set or the firmware rejects the call with
\texttt{getProcedureCallType():\ invalid\ procedureId=\%d,\ tot=\%d}.
The validated call enters the priority queue keyed by
\texttt{(userId,\ jobId,\ fwPriority,\ hwqId,\ tqId,\ queueId)}, and a
full queue is reported by
\texttt{{[}isHWReady{]}\ priority\ queue\ is\ full!}. The controller
drops a submission whose target program is not running, logged for the
inference path as
\texttt{CSNE\_CMD\_INFERENCE\_CALL\ dropped.\ prog\ \%d\ process\ \%d\ not\ running.}

Completion and asynchronous status return as firmware-to-host
notifications that reuse the command identifier space: a per-program
event as \texttt{CSNE\_CMD\_PROGRAM\_EVENT}, a data-chaining stage
completion as \texttt{CSNE\_CMD\_DATA\_CHAINING\_EVENT}, a prefetch
completion as \texttt{CSNE\_CMD\_PREFETCH\_DSID\_EVENT} logged
\texttt{Dsid\ (\%d)\ Event\ (0x\%x)}, and an error as
\texttt{CSNE\_CMD\_CH\_ERROR\_NOTIFICATION}. Performance signposts
captured during a run originate in the firmware as
\texttt{CSNE\_CMD\_CH\_SIGNPOST*} notifications, not as a host-side
instrumentation artifact.

\section{Procedure-call body}\label{procedure-call-body}

The body of a procedure call is \texttt{sCSneCmdProcedureCall}, a shared
container whose trailing variable-length arrays spill into an
over-allocated \texttt{sCSneCmdProcedureCallCopyContainer}. The counted
fields after the header describe the call's shape, each bounded by a
firmware assert (\Cref{tbl:c30-proccall-body}).

\begin{longtable}[]{@{}
  >{\raggedright\arraybackslash}p{(\linewidth - 4\tabcolsep) * \real{0.3333}}
  >{\raggedright\arraybackslash}p{(\linewidth - 4\tabcolsep) * \real{0.3333}}
  >{\raggedright\arraybackslash}p{(\linewidth - 4\tabcolsep) * \real{0.3333}}@{}}
\caption{The counted fields of the procedure-call
body.}\label{tbl:c30-proccall-body}\tabularnewline
\toprule\noalign{}
\begin{minipage}[b]{\linewidth}\raggedright
\textbf{Field}
\end{minipage} & \begin{minipage}[b]{\linewidth}\raggedright
\textbf{Type}
\end{minipage} & \begin{minipage}[b]{\linewidth}\raggedright
\textbf{Bound}
\end{minipage} \\
\midrule\noalign{}
\endfirsthead
\toprule\noalign{}
\begin{minipage}[b]{\linewidth}\raggedright
\textbf{Field}
\end{minipage} & \begin{minipage}[b]{\linewidth}\raggedright
\textbf{Type}
\end{minipage} & \begin{minipage}[b]{\linewidth}\raggedright
\textbf{Bound}
\end{minipage} \\
\midrule\noalign{}
\endhead
\bottomrule\noalign{}
\endlastfoot
\texttt{nbrOfOutputBufferSets} & \texttt{u32} & \texttt{==\ 1} on the M1
generation \\
\rowcolor{zebra}\texttt{nbrOfWaitEvents} & \texttt{u32} &
\texttt{nbrOfWaitEvents\ +\ nbrOfSignalEvents\ \textgreater{}\ 0} \\
\texttt{nbrOfSignalEvents} & \texttt{u32} & \texttt{\textless{}=\ 16} \\
\rowcolor{zebra}\texttt{nbrOfCustomBars} & \texttt{u32} &
\texttt{\textless{}=\ 32} on the wire \\
\texttt{stats.buffer} & \texttt{u64} &
\texttt{\textgreater{}=\ pProc-\textgreater{}pStatArrayBaseOrig} \\
\rowcolor{zebra}\texttt{stats.size} & \texttt{u64} & covers the three
stats records \\
\end{longtable}

The firmware logs the decoded shape on one line:
\texttt{Bufs\ =\ \%d\ Priority\ =\ \%d\ CustomBars\ =\ \%d\ WaitEvents\ =\ \%d\ SignalEvents\ =\ \%d}.
A wait or signal event is a fixed 24-byte record,
\texttt{\{\ uint32\ type;\ uint32\ targetMask;\ uint64\ id;\ uint64\ value;\ \}},
recovered from the event log strings
\texttt{waitEvent{[}\%d{]}:\ type=\%d\ mask=\%x\ id=\%lld\ value=\%lld}
and the matching signal line.

The procedure-call type gate admits only certain content types:
\texttt{getProcedureCallType(programId,\ procedureId)} must return
\texttt{ContentType\_0}, \texttt{\_3}, or \texttt{\_4} for a plain
\texttt{PROCEDURE\_CALL}, and \texttt{\_0} or \texttt{\_3} on the
cache-request path, under
\texttt{procedureId\ \textless{}\ pProg-\textgreater{}pProgram-\textgreater{}procNbr}.
The \texttt{WITH\_CUSTOM\_BARS} variant adds a trailing custom
execute-order array, whose offset and length are bounded by
\texttt{callSize\ \textless{}=\ customExecuteOrderArrayOffset},
\texttt{0\ \textless{}\ nbrOfCustomExecuteOrder},
\texttt{nbrOfCustomExecuteOrder\ \textless{}=\ 128}, and
\texttt{(customExecuteOrderArrayOffset\ +\ nbrOfCustomExecuteOrder\ *\ sizeof(uint32\_t))\ \textless{}=\ sizeof(sCSneCmdProcedureCallCopyContainer)}.
Each custom bar has \texttt{type\ ==\ ANE\_CUSTOM\_BAR\_GENERIC} and
\texttt{config\ ==\ ANE\_CUSTOM\_BAR\_HW\_32BIT}, the on-wire count is
bounded \texttt{nbrOfCustomBars\ \textless{}=\ 32}, and a blob in the
compiled program may store up to 128. The \texttt{WITH\_SIGNAL\_EVENTS}
variant adds the \texttt{procCallWithSignalEvents} sub-struct, bounded
by
\texttt{nbrOfSignalEvents\ \textgreater{}\ 0\ \&\&\ nbrOfSignalEvents\ \textless{}=\ 16},
with the sub-network custom execute-order disallowed
(\texttt{0\ ==\ bSubNetworkCustomExecuteOrder}).

\Cref{tbl:c30-percall-limits} collects the per-call and per-request
numeric limits so a host can size a submission ahead of dispatch.

\begin{longtable}[]{@{}
  >{\raggedright\arraybackslash}p{(\linewidth - 4\tabcolsep) * \real{0.3333}}
  >{\raggedright\arraybackslash}p{(\linewidth - 4\tabcolsep) * \real{0.3333}}
  >{\raggedright\arraybackslash}p{(\linewidth - 4\tabcolsep) * \real{0.3333}}@{}}
\caption{The per-call and per-request numeric limits on the M1
generation.}\label{tbl:c30-percall-limits}\tabularnewline
\toprule\noalign{}
\begin{minipage}[b]{\linewidth}\raggedright
\textbf{Quantity}
\end{minipage} & \begin{minipage}[b]{\linewidth}\raggedright
\textbf{Limit}
\end{minipage} & \begin{minipage}[b]{\linewidth}\raggedright
\textbf{Firmware gate}
\end{minipage} \\
\midrule\noalign{}
\endfirsthead
\toprule\noalign{}
\begin{minipage}[b]{\linewidth}\raggedright
\textbf{Quantity}
\end{minipage} & \begin{minipage}[b]{\linewidth}\raggedright
\textbf{Limit}
\end{minipage} & \begin{minipage}[b]{\linewidth}\raggedright
\textbf{Firmware gate}
\end{minipage} \\
\midrule\noalign{}
\endhead
\bottomrule\noalign{}
\endlastfoot
trigger input buffers & 16 &
\texttt{nbrOfInputBuffers\ \textless{}=\ (16)} \\
\rowcolor{zebra}active shared events & fewer than 2 &
\texttt{sharedEventsActiveNbr\ \textless{}\ 2} \\
task-descriptor partitions on the M1 single path & 1 &
\texttt{1\ ==\ nbrOfTdPartition} \\
\rowcolor{zebra}ANE requests in list on the M1 single path & 1 &
\texttt{1\ ==\ pCacheReq-\textgreater{}nbrOfAneRequestInList} \\
\end{longtable}

The resident cache-request family is what keeps a buffer set on the
engine across calls. An install command at \texttt{0x45} returns the
64-bit \texttt{cacheHandler} and creates a long-lived request indexed
internally by \texttt{cacheReqIdx} in the range
\texttt{0\ \textless{}=\ cacheReqIdx\ \textless{}\ maxCacheRequest},
allocated from a \texttt{CIndexPool}. A trigger command at \texttt{0x46}
names that handle and fires it, holding an \texttt{execTimestamp} that
must be strictly monotone against
\texttt{pCacheReq-\textgreater{}lastTriggerExecTimestamp}, and checking
that the handle is live first. The firmware drops a trigger on a handle
not in the \texttt{ANE\_DATA\_CHAINING\_CACHE\_REQUEST\_IN\_USE} state
with
\texttt{cacheHandler\ (0x\%llx)\ in\ wrong\ state\ (\%d)\ for\ trigger.\ trigger\ dropped}.
The recycle command at \texttt{0x47} returns a consumed output buffer to
a resident request, the invalidate command at \texttt{0x48} tears it
down, and the group select at \texttt{0x4f} picks the active member of a
buffer-sharing group. The force-disable at \texttt{0x4d} is the global
disable, which the firmware admits only when set true. The firmware
constrains chained buffers to the low address window: a data-chaining
buffer descriptor must have
\texttt{buffer.type\ ==\ eCSneBufferDescriptorType\_OutputBuffer},
\texttt{outputBufferSetId\ \textless{}\ nbrOfOutputBufferSets}, and a
device address whose high 32 bits are zero
(\texttt{(buffer\ \textgreater{}\textgreater{}\ 32)\ ==\ 0}).

The inference call at \texttt{0x5a} is the higher-level submission path,
which the firmware expands into one or more
\texttt{CANE\_SUB\_PACKET\_CMD\_PROCEDURE\_CALL} sub-packets, holding a
pre-mapped property buffer and validating the buffer count against the
program descriptor:
\texttt{pMsg-\textgreater{}bufNbr\ ==\ 1\ +\ pAneProgramDesc2-\textgreater{}procedures{[}procedureId{]}.numIoBuffers}.
The \texttt{PREMAP\_BUFFER} command at \texttt{0x4a} pre-maps that
property buffer, hard-validated for exactly one operation and exactly
one buffer (\texttt{pPreMapOp-\textgreater{}nbrOfBuffers\ ==\ 1}) with
nonzero address and size. The procedure call at \texttt{0x41} is the
lower-level direct form addressing the program, process, and procedure
triple; both share the \texttt{sCSneCmdProcedureCall} body and the same
priority scheduler.

The property channel exposes a register-poke surface that the host uses
to read and write firmware registers directly. The read command at
\texttt{0x1f} and the write command at \texttt{0x1e} each contain a
two-field body, recovered from the
\texttt{IPC\ Writer\ regAddr\ 0x\%lx\ regValue\ 0x\%x} log line as
\texttt{\{\ uintptr\ regAddr;\ uint32\ regValue;\ \}}, and a write
checks \texttt{regValue\ !=\ 0}. The back-channel RPC at \texttt{0x5b}
is a firmware-initiated call into the host driver: the firmware client
posts directly into the direct-proc-call event pool, gated on
\texttt{GetDirectProcCallEventPoolAvailNum()\ \textgreater{}=\ 2},
tracks outstanding calls in \texttt{RPCEventMap{[}\%d{]}}, and reports a
stuck slot with
\texttt{RPCEventMap{[}\%d{]}\ is\ dirty.\ (No\ ack\ from\ driver)}.

\section{Recovered command-id enum}\label{recovered-command-id-enum}

The dispatched table above is the contiguous logger ordering, but the
firmware also keeps the canonical protocol enum as a packed
\texttt{\{char*\ name;\ u64\ id\}} array at vaddr \texttt{0xf5ba8} (file
offset \texttt{0xf9ba8}, in \texttt{\_\_DATA.\_\_const}). This is the
\texttt{ECSneCmdId} enum, 94 named entries plus a trailing zero-id
terminator, and the id is the literal 16-bit value held on the wire at
message \texttt{+0x4}. The array is reached through a fixed-up pointer
with no \texttt{adrp}+\texttt{add} cross-reference, consistent with a
shared name-lookup helper. The protocol enum and the dispatched logger
ordering are not the same numbering in the \texttt{0x2xx} region, so
this enum is the wire vocabulary while the table in the preceding
section is the firmware's own logger index.

The named ids partition by high byte into families, each a contiguous
range (\Cref{tbl:c30-id-families}).



\section{Firmware fast path}\label{firmware-fast-path}

The live on-chip dispatcher is \texttt{CAneEngineExeLoop::dispatch} at
vaddr \texttt{0x4a42c}, which reads the 16-bit id with
\texttt{ldrh\ w8,{[}x21,\#4{]}} and decodes it through a compare tree.
The tree implements only about ten ids, all on the procedure-call and
inference fast path, and every other id falls through to a default arm
at \texttt{0x4aa4c} that logs
\texttt{Cmd\ 0x\%x\ is\ not\ supported\ through\ ExeLoop\ cmd\ channel}
and then asserts, panicking the firmware. Sending a non-fast-path id on
this channel is thus a hard fault on the live device, consistent with
the unrecoverable behavior of a DART or fence fault.

The implemented arm of the compare tree dispatches each id to a
PAC-signed virtual method on the engine object, reached through the
\texttt{ldr}/\texttt{autda}/\texttt{blraa} pattern at a fixed vtable
offset. Each such slot is an arm64e auth-rebase chained pointer in
\texttt{\_\_DATA.\_\_const} whose low thirty-two bits give the raw
target address, matched to its call site by the \texttt{movk}
discriminator, so the seventy-six-slot table resolves statically. The
procedure-call slot at \texttt{+0x200} targets \texttt{0x7374c} and the
inference slot at \texttt{+0x190} targets \texttt{0x74510}.
\Cref{tbl:c30-exeloop-ids} gives the ids the ExeLoop dispatcher
implements, with each id's vtable offset and behavior.

\begin{longtable}[]{@{}
  >{\raggedright\arraybackslash}p{(\linewidth - 6\tabcolsep) * \real{0.2500}}
  >{\raggedright\arraybackslash}p{(\linewidth - 6\tabcolsep) * \real{0.2500}}
  >{\raggedright\arraybackslash}p{(\linewidth - 6\tabcolsep) * \real{0.2500}}
  >{\raggedright\arraybackslash}p{(\linewidth - 6\tabcolsep) * \real{0.2500}}@{}}
\caption{The ids the ExeLoop dispatcher
implements.}\label{tbl:c30-exeloop-ids}\tabularnewline
\toprule\noalign{}
\begin{minipage}[b]{\linewidth}\raggedright
\textbf{ExeLoop id}
\end{minipage} & \begin{minipage}[b]{\linewidth}\raggedright
\textbf{Meaning}
\end{minipage} & \begin{minipage}[b]{\linewidth}\raggedright
\textbf{Vtable offset}
\end{minipage} & \begin{minipage}[b]{\linewidth}\raggedright
\textbf{Behavior}
\end{minipage} \\
\midrule\noalign{}
\endfirsthead
\toprule\noalign{}
\begin{minipage}[b]{\linewidth}\raggedright
\textbf{ExeLoop id}
\end{minipage} & \begin{minipage}[b]{\linewidth}\raggedright
\textbf{Meaning}
\end{minipage} & \begin{minipage}[b]{\linewidth}\raggedright
\textbf{Vtable offset}
\end{minipage} & \begin{minipage}[b]{\linewidth}\raggedright
\textbf{Behavior}
\end{minipage} \\
\midrule\noalign{}
\endhead
\bottomrule\noalign{}
\endlastfoot
\texttt{0x2d} & \texttt{CH\_DATA\_FILE\_LOAD2} & \texttt{+0x250} &
guarded by the engine flag at \texttt{{[}x19,\#0x2bb{]}} \\
\rowcolor{zebra}\texttt{0x204} & procedure call & \texttt{+0x200} &
validate \texttt{(programId,\ procId)} then dispatch, the inference
trigger \\
\texttt{0x209} & cache-request trigger & \texttt{+0x218} & logs
\texttt{cacheHandler\ 0x\%llx} \\
\rowcolor{zebra}\texttt{0x20a} & cache-request submit & cache path &
cache-request submission \\
\texttt{0x20c} & procedure call with bars & \texttt{+0x208} region &
custom buffer-address-register call \\
\rowcolor{zebra}\texttt{0x211} & procedure call with events & events
path & event-signalled call, force-disable-cache arm \\
\texttt{0x212} & procedure-call variant & \texttt{+0x210} & shared reply
and verify path \\
\rowcolor{zebra}\texttt{0x404} & \texttt{RETURN\_PROCESS\_ID} &
\texttt{+0x190} & process-id handshake reply \\
\end{longtable}

The procedure-call arm at id \texttt{0x204} decodes a fixed message
layout, recovered from the debug line
\texttt{{[}EXELOOP{]}\ CMD=\%\#04x\ {[}PROC\ CALL{]}\ at\ \%lld\ :\ ProgId=\%d\ ProcId=\%d\ Proc=\%d\ Pri=\%d}:
a \texttt{u16} id at \texttt{+0x04}, then \texttt{u32}
\texttt{programId}, \texttt{procId}, \texttt{proc}, and
\texttt{priority} at \texttt{+0x08} through \texttt{+0x14}. Its argument
validator at \texttt{0x48644} rejects a \texttt{programId} of
\texttt{144} or more and a \texttt{procId} of \texttt{288} or more, then
indexes a per-program validity byte at
\texttt{engine\ +\ programId\ *\ 0x20\ +\ 0x9948}, so a 144-entry
program-slot table of stride \texttt{0x20} is at engine offset
\texttt{0x9948}. The back-channel RPC handler at \texttt{0x47818} caps
its payload length at \texttt{0x800000} and requires the magic
\texttt{0x55aa55aa} before processing, the same back-channel surface as
id \texttt{0x5b} in the dispatched table.

The \texttt{0x00xx} config, power, trace, and profile commands are
present in this image only as enum names and as the per-command struct
and assert strings, for example \texttt{sCSneCmdReset},
\texttt{sCSneCmdConfigGet}, \texttt{sCSneCmdTraceEnable}, and
\texttt{sCSneCmdPowerSupplyControl}. Exhaustive
\texttt{adrp}+\texttt{add} and constant-pointer scans find zero
references to any of those handler strings anywhere in \texttt{\_\_text}
or \texttt{\_\_DATA.\_\_const}, so the strings are linked in from a
shared protocol header while the handlers are not compiled into this M1
and H13 build. \texttt{AppleH11ANEInterface} and
\texttt{AppleANELoadBalancer}, the kext that owns power, IPC-endpoint,
and buffer management, service those config-command handlers host-side,
so their status is opaque by absence rather than undecoded. The
\texttt{0x01xx} notification family is firmware-to-host output emitted
on the channel, so it is not dispatcher input and does not appear in the
ExeLoop compare tree either.

\chapter{Power and thermal}\label{power-and-thermal}

\begin{summarybox}

The engine runs at one fixed clock, with no voltage or frequency
scaling. It manages power through a peak-power regulator, armed once at
power-up by five loop-free stores with the control word \texttt{0x11}.
Idle reads 0 W with rails off. A sustained convolution holds flat at
about 1.66 W over 176 seconds, and the densest peak reaches about 4.7 W
at 6.96 TFLOP/s. The firmware has no thermal logic and reads no
temperature; thermal protection is off-engine in the system power
manager.

\end{summarybox}

The engine runs at one fixed operating point and manages power through a
peak-power regulator, not a frequency sequence.

\section{Boot and device tree}\label{boot-and-device-tree}

The engine appears in the device tree as a single node
\texttt{ane0@84000000}, class \texttt{AppleARMIODevice},
\texttt{compatible\ =\ "ane,t8020"}, \texttt{device\_type\ =\ "ane"}.
The match key \texttt{ane,t8020} is the kext bind key even on a later
die, since the engine block has the t8020-generation identity string
across the whole M1 family.

The node's properties decode as \Cref{tbl:c31-devicetree}.

\begin{longtable}[]{@{}
  >{\raggedright\arraybackslash}p{(\linewidth - 4\tabcolsep) * \real{0.3333}}
  >{\raggedright\arraybackslash}p{(\linewidth - 4\tabcolsep) * \real{0.3333}}
  >{\raggedright\arraybackslash}p{(\linewidth - 4\tabcolsep) * \real{0.3333}}@{}}
\caption{The properties of the \protect\texttt{ane0} device-tree node,
with decoded value and
meaning.}\label{tbl:c31-devicetree}\tabularnewline
\toprule\noalign{}
\begin{minipage}[b]{\linewidth}\raggedright
\textbf{Property}
\end{minipage} & \begin{minipage}[b]{\linewidth}\raggedright
\textbf{Decoded value}
\end{minipage} & \begin{minipage}[b]{\linewidth}\raggedright
\textbf{Meaning}
\end{minipage} \\
\midrule\noalign{}
\endfirsthead
\toprule\noalign{}
\begin{minipage}[b]{\linewidth}\raggedright
\textbf{Property}
\end{minipage} & \begin{minipage}[b]{\linewidth}\raggedright
\textbf{Decoded value}
\end{minipage} & \begin{minipage}[b]{\linewidth}\raggedright
\textbf{Meaning}
\end{minipage} \\
\midrule\noalign{}
\endhead
\bottomrule\noalign{}
\endlastfoot
\texttt{compatible} & \texttt{"ane,t8020"} & the kext match key \\
\rowcolor{zebra}\texttt{ane-type} & \texttt{0x60} & engine variant
identifier \\
\texttt{reg} bank 0 & \texttt{0x2\_8400\_0000}, length
\texttt{0x200\_0000} & the 32 MB engine control aperture \\
\rowcolor{zebra}\texttt{reg} bank 1 & \texttt{0x2\_8E08\_0000}, length
\texttt{0xC02C} & the power-manager slice \\
\texttt{interrupts} & \texttt{0x302} & the engine interrupt number \\
\rowcolor{zebra}\texttt{clock-ids} & \texttt{0x13e}, \texttt{0x13f},
\texttt{0x140}, \texttt{0x141} & four clock identifiers, one per compute
cluster \\
\texttt{clock-gates}, \texttt{power-gates} & \texttt{0x1cf} & the single
clock and power-gate index for the whole block \\
\rowcolor{zebra}\texttt{asc-dram-mask} & \texttt{0x1F0\_0000\_0000} &
the high-address window the coprocessor may reach in DRAM \\
\texttt{segment-names} & \texttt{\_\_TEXT;\_\_DATA} & the firmware
segments the loader places \\
\rowcolor{zebra}\texttt{pre-loaded} & \texttt{1} & the firmware is
resident in DRAM before the kernel attaches \\
\end{longtable}

Bank 0 is the engine control aperture at absolute physical
\texttt{0x2\_8400\_0000}, 32 MB, holding the coprocessor registers,
mailbox doorbells, and engine register file. Bank 1 is a small
power-manager slice at \texttt{0x2\_8E08\_0000}, length \texttt{0xC02C},
whose base matches the base of the system power-manager node
\texttt{pmgr@8E080000}, tying the engine's power-gate \texttt{0x1cf}
directly to the power-management register neighborhood. The four clock
identifiers \texttt{0x13e} through \texttt{0x141} are one per compute
cluster, matching the four-cluster geometry of the power model.

The firmware is not read from disk at first boot. The boot loader
verifies the signed firmware image and pre-loads its segments into the
carve-out named by the node's \texttt{segment-ranges}, marked by the
property \texttt{pre-loaded\ =\ 1}. The kernel driver then attaches to
\texttt{ane0}, hands the already-trusted image to the engine's
coprocessor over the mailbox, and the coprocessor boots. The image
itself is a bare uncompressed and unencrypted preload Mach-O wrapped in
an Image4 container with the payload tag \texttt{anef}, and it has no
embedded manifest, keybag, or code-signature load command of its own.
The per-device personalized boot ticket enforces its authenticity
externally, listing the \texttt{anef} object digest alongside the
kernel, boot-loader, and device-tree digests and binding to the device
by its unique chip identifier. A firmware image cannot be transplanted
to a different machine.

The power and peak-power tunables the firmware consumes are not on
\texttt{ane0}. They are on \texttt{pmgr@8E080000}: the master enable
\texttt{ane-dpe\ =\ 1}, calibration vector \texttt{dpe-ane-data} (ten
32-bit words), and per-operating-point current ceiling table
\texttt{ifane-max}. The ten calibration words decode to
\texttt{{[}9384,\ 661,\ 1323,\ 2641,\ 2395,\ 84489,\ 163813,\ 51491,\ 178500,\ 51491{]}},
the per-platform coefficients that convert dynamic-power-estimation
activity counts to energy, with the repeated value \texttt{51491} as a
shared scale factor. The \texttt{ifane-max} table is rows of an index
key and three current ceilings: the head rows are index \texttt{0x1f8}
with \texttt{\{0x18000,\ 0x17333,\ 0x17333\}}, index \texttt{0x2e8} with
\texttt{\{0x1fd70,\ 0x1fd70,\ 0x1fd70\}}, index \texttt{0x498} with
\texttt{\{0x2deb8,\ 0x2deb8,\ 0x2deb8\}}, index \texttt{0x6a8} with
\texttt{\{0x428f5,\ 0x428f5,\ 0x428f5\}}, and index \texttt{0x858} with
\texttt{\{0x628c5,\ 0x65666,\ 0x65666\}}. The first column is a monotone
activity-keyed operating point, and the three trailing columns are the
peak-current ceilings, one per energy-accumulator partition, that the
peak-power regulator clamps against. A parameter-push call at init
delivers these, and the firmware arms the power blocks from them.

\section{Power model}\label{power-model}

\Cref{tbl:c31-power-facts} collects the decoded boot, power, and
clocking facts with their values and sources.

\begin{longtable}[]{@{}
  >{\raggedright\arraybackslash}p{(\linewidth - 4\tabcolsep) * \real{0.3333}}
  >{\raggedright\arraybackslash}p{(\linewidth - 4\tabcolsep) * \real{0.3333}}
  >{\raggedright\arraybackslash}p{(\linewidth - 4\tabcolsep) * \real{0.3333}}@{}}
\caption{The decoded boot, power, and clocking facts; the register-init
constants are decoded in Appendix
\hyperref[appendix-c.-decoded-reference-tables]{C}.}\label{tbl:c31-power-facts}\tabularnewline
\toprule\noalign{}
\begin{minipage}[b]{\linewidth}\raggedright
\textbf{Fact}
\end{minipage} & \begin{minipage}[b]{\linewidth}\raggedright
\textbf{Value}
\end{minipage} & \begin{minipage}[b]{\linewidth}\raggedright
\textbf{Source}
\end{minipage} \\
\midrule\noalign{}
\endfirsthead
\toprule\noalign{}
\begin{minipage}[b]{\linewidth}\raggedright
\textbf{Fact}
\end{minipage} & \begin{minipage}[b]{\linewidth}\raggedright
\textbf{Value}
\end{minipage} & \begin{minipage}[b]{\linewidth}\raggedright
\textbf{Source}
\end{minipage} \\
\midrule\noalign{}
\endhead
\bottomrule\noalign{}
\endlastfoot
Engine node & \texttt{ane0@84000000}, \texttt{compatible\ "ane,t8020"} &
device tree \\
\rowcolor{zebra}Engine aperture & \texttt{0x2\_8400\_0000}, 32 MB &
device tree \\
Power-manager slice & \texttt{0x2\_8E08\_0000}, length \texttt{0xC02C} &
device tree \\
\rowcolor{zebra}Power-block base & \texttt{0x2\_6b8f\_0000} & firmware
disassembly \\
Voltage base (opaque) & \texttt{0x2\_3b70\_c008} & firmware
disassembly \\
\rowcolor{zebra}Power-block arm & 5 fixed stores, no loop & firmware
disassembly \\
Peak-power control word & \texttt{0x11} & firmware disassembly \\
\rowcolor{zebra}Compute clusters & 4, gated independently & device tree
and firmware \\
Idle power & 0 W, rails off & M1/H13 measured \\
\rowcolor{zebra}Sustained convolution & \(\approx 1.66\) W flat over 176
s & M1/H13 measured \\
Densest peak & \(\approx 4.7\) W, \(6.96\) TFLOP/s & M1/H13 measured \\
\end{longtable}

The engine runs at a single fixed clock with no local voltage or
frequency scaling of its own; any frequency change is the system power
manager's to make externally. The only clock string in the firmware is
the boot-time timebase report; the image has no frequency table, no
enumerated operating points, and no voltage string. Voltage is the
system power manager's concern, reached only through an opaque
power-management base address (\texttt{0x2\_3b70\_c008}) that the
firmware never interprets.

Performance is modulated by two controls, neither of them frequency. The
first is how many of the four compute clusters are powered. The second
is a peak-power regulator that limits activity under a fixed budget.
That regulator is the engine's substitute for a frequency and voltage
sequence, and it has three parts the firmware arms once at power-up.
Dynamic power estimation counts switching activity to estimate
instantaneous dynamic power. Peak-power tracking watches that estimate
against the budget and applies back-pressure on issue when the work
would exceed it, throttling activity within the fixed clock. Leakage and
energy estimation wires the result into the system power manager. The
running energy total that the energy-model telemetry channel reports is
held in three accumulator partitions.

Arming this model is loop-free and idempotent: five memory-mapped stores
at the fixed base \texttt{0x2\_6b8f\_0000}, gated on two per-instance
enable flags at offsets \texttt{+0x95} and \texttt{+0x96}, with no ramp
and no per-frequency programming (\Cref{lst:c31-arm-seq}).

\begin{listing}[H]
\caption{The five-store arm sequence for the power-estimation and peak-power blocks, with the fixed control words at their decoded memory-mapped offsets.}\label{lst:c31-arm-seq}

\begin{Shaded}
\begin{Highlighting}[]
\NormalTok{this+0x95 (byte)  : dynamic{-}power{-}estimation mode enabled?   (tested at entry)}
\NormalTok{this+0x96 (byte)  : peak{-}power{-}tracking mode enabled?}

\NormalTok{store  0x11             {-}\textgreater{} [0x2\_6b8f\_0000]   ; peak{-}power{-}tracking control word}
\NormalTok{store  (this+0x96 ? 1 : 0) {-}\textgreater{} [0x2\_6b8f\_0004] ; dynamic{-}power{-}estimation control word}
\NormalTok{reg = [0x2\_6b8f\_4000]; reg |= 1; store {-}\textgreater{} [0x2\_6b8f\_4000]  ; leakage{-}and{-}energy enable bit}
\NormalTok{store  0x3fff           {-}\textgreater{} [0x2\_6b8e\_c42c]   ; trailing control word}
\NormalTok{store  0xf              {-}\textgreater{} [0x2\_6b8e\_c5dc]   ; trailing control word (= 0x2\_6b8e\_c42c + 0x1b0)}
\end{Highlighting}
\end{Shaded}

\end{listing}

The leakage-and-energy block is enabled by setting bit 0 of the register
at \texttt{0x2\_6b8f\_4000}, and two fixed trailing control words close
the sequence. This estimation aperture at \texttt{0x2\_6b8f\_0000} is
distinct from the power-gating base \texttt{0x2\_3b70\_c008}, so power
gating and power estimation are separate hardware blocks. The
system-on-chip leakage-and-energy enable bit folds the result into the
system power manager. The rails drop fully when the engine goes idle, so
the estimation block loses state and the firmware re-applies the
retained calibration on every wake.

The four compute clusters gate independently through a stride-eight
status array, polled by the host driver until each domain's low byte
reads \texttt{0xff}, meaning all eight power straps have settled
(\Cref{tbl:c31-power-domains}).

\begin{longtable}[]{@{}
  >{\raggedright\arraybackslash}p{(\linewidth - 4\tabcolsep) * \real{0.3333}}
  >{\raggedright\arraybackslash}p{(\linewidth - 4\tabcolsep) * \real{0.3333}}
  >{\raggedright\arraybackslash}p{(\linewidth - 4\tabcolsep) * \real{0.3333}}@{}}
\caption{The power-domain status registers, a stride-eight array at
\protect\texttt{0xc000\ +\ domain*8} polled until each low byte reads
\protect\texttt{0xff}.}\label{tbl:c31-power-domains}\tabularnewline
\toprule\noalign{}
\begin{minipage}[b]{\linewidth}\raggedright
\textbf{Aperture-relative offset}
\end{minipage} & \begin{minipage}[b]{\linewidth}\raggedright
\textbf{Poll}
\end{minipage} & \begin{minipage}[b]{\linewidth}\raggedright
\textbf{Domain}
\end{minipage} \\
\midrule\noalign{}
\endfirsthead
\toprule\noalign{}
\begin{minipage}[b]{\linewidth}\raggedright
\textbf{Aperture-relative offset}
\end{minipage} & \begin{minipage}[b]{\linewidth}\raggedright
\textbf{Poll}
\end{minipage} & \begin{minipage}[b]{\linewidth}\raggedright
\textbf{Domain}
\end{minipage} \\
\midrule\noalign{}
\endhead
\bottomrule\noalign{}
\endlastfoot
\texttt{0x2c8} & \texttt{expect\ 0xff,\ mask\ 0xff,\ retries\ 5000} &
top-level ready and fabric gate, validated first \\
\rowcolor{zebra}\texttt{0xc000} &
\texttt{expect\ 0xff,\ mask\ 0xff,\ retries\ 5000} & base domain, the
always-on control fabric \\
\texttt{0xc008} & \texttt{expect\ 0xff,\ mask\ 0xff,\ retries\ 5000} &
compute cluster 1 \\
\rowcolor{zebra}\texttt{0xc010} &
\texttt{expect\ 0xff,\ mask\ 0xff,\ retries\ 5000} & compute cluster
2 \\
\texttt{0xc018} & \texttt{expect\ 0xff,\ mask\ 0xff,\ retries\ 5000} &
compute cluster 3 \\
\rowcolor{zebra}\texttt{0xc020} &
\texttt{expect\ 0xff,\ mask\ 0xff,\ retries\ 5000} & compute cluster
4 \\
\end{longtable}

The five domains are one always-on base domain and four independently
gated compute clusters; there is no fifth cluster. The firmware brings
the base domain up first and adds a compute cluster only when work needs
it, which is the mechanism behind the measured idle of 0 W. The engine
is in the all-off state until a procedure call arrives, then brings the
base domain up to a wait state and gates in the clusters, so the idle
floor is rail-off rather than clock-gated. A free-running firmware
timestamp pair is in the same aperture at \texttt{0x1170000} for the low
word and \texttt{0x1170004} for the high word, distinct from the
engine-internal timebase, and the firmware zeroes four scratch registers
at \texttt{0x1840048} through \texttt{0x1840054} at init.

For a workload held at the operating point for time \(t\), the energy is

\[E = P \, t\]

with measured constants from a sustained saturating loop on the M1: a
dense convolution at \(P \approx 1.66 \text{ W}\) held flat over
\(t = 176 \text{ s}\), and a densest-packed peak of
\(P \approx 4.7 \text{ W}\). At that peak the engine delivers \(6.96\)
TFLOP/s, which gives an efficiency of about \(1.5\) TFLOP/s per watt in
fp16, rising to about \(2.6\) TFLOP/s per watt on weight-reuse
convolutions. The power scales with cluster engagement and lane density;
no clock change is involved. Relative to a single cluster, throughput
rises \(1.99\times\), \(3.02\times\), and \(4.00\times\) at the second,
third, and fourth clusters. The rail then climbs smoothly from \(779\)
mW to \(1429\) mW as channels pack the four clusters, with no discrete
steps.

\section{Thermal behavior and the operating
point}\label{thermal-behavior-and-the-operating-point}

A scan of the image for temperature, throttle, junction, and similar
terms returns nothing. The engine does not read temperature, does not
throttle on temperature, and emits no thermal event of its own. Thermal
protection is entirely off-engine: the peak-power regulator bounds
power, which indirectly bounds heat, and package-level thermal control
is in the system power manager, which can delay or refuse dispatch but
does not reach into any engine counter.

A 176-second saturating loop sampled thermal pressure every two seconds
and read nominal on every one of the 89 samples, with power and
throughput flat to within half a percent and a slightly negative drift.
A frequency-scaling engine would show a ramp or multiple power modes as
it settled; a thermally limited engine would decay. The engine does
neither; power and throughput hold flat. The single outbound power
signal the firmware emits is a normalized margin level for the host to
react to, holding no thermal field.

The first dispatch after idle pays a fixed power-up cost because the
engine reaches a true rails-off state between jobs. That cost grows from
near zero at back-to-back dispatch to about \(0.5\) ms once the idle gap
reaches roughly 100 ms, then plateaus, which sets the residency window
for keeping the engine warm.

\subsection{Operating points and the absence of local
DVFS}\label{operating-points-and-the-absence-of-local-dvfs}

The engine has no frequency-or-voltage sequence of its own. The kext
delegates operating-point selection to the system power manager: the SoC
CLPC within ApplePMGR. The kext has no clock register addresses. A
disassembly of \texttt{enableAneSysClock} shows it performs no
memory-mapped writes itself; it loads an ApplePMGR service object,
authenticates the vtable pointer, and calls a PMGR method with the
enable argument set, with the only literal in the path being the string
\texttt{"ApplePMGR"}. The \texttt{ane0} device-tree node publishes the
PMGR clock and power-gate identifiers but has no \texttt{dvfm-states},
\texttt{voltage-states}, or \texttt{perf-states} property, so the
per-state frequency in Hz and voltage in mV stay held privately by
ApplePMGR behind the power-manager base and are not recoverable from the
engine binaries.

What is on the engine side is the firmware \texttt{H13TunableManager}
register-init table, where \texttt{H13} is the M1 family. A
\texttt{TunableManager} descriptor at base \texttt{0x2\_6b8f\_4000}
(\texttt{aneDpePpt\_soc\_dpe\_lee}, the SoC dynamic-power-estimation
control block) contains a 7-step monotonic credit sequence, written once
at firmware bring-up and gated by chip revision
(\Cref{tbl:c31-tunable}).

\begin{longtable}[]{@{}lllr@{}}
\caption{The 7-step monotonic DPE/PPT credit sequence in the
\protect\texttt{aneDpePpt\_soc\_dpe\_lee} block, with the decoded value
of each 9-bit field.}\label{tbl:c31-tunable}\tabularnewline
\toprule\noalign{}
\textbf{Offset} & \textbf{Mask} & \textbf{Value} & \textbf{Decoded} \\
\midrule\noalign{}
\endfirsthead
\toprule\noalign{}
\textbf{Offset} & \textbf{Mask} & \textbf{Value} & \textbf{Decoded} \\
\midrule\noalign{}
\endhead
\bottomrule\noalign{}
\endlastfoot
\texttt{+0x18} & \texttt{0x1ff} & \texttt{0x19} & 25 \\
\rowcolor{zebra}\texttt{+0x1c} & \texttt{0x1ff} & \texttt{0x32} & 50 \\
\texttt{+0x20} & \texttt{0x1ff} & \texttt{0x46} & 70 \\
\rowcolor{zebra}\texttt{+0x24} & \texttt{0x1ff} & \texttt{0x55} & 85 \\
\texttt{+0x28} & \texttt{0x1ff} & \texttt{0x5f} & 95 \\
\rowcolor{zebra}\texttt{+0x2c} & \texttt{0x1ff} & \texttt{0x69} & 105 \\
\texttt{+0x30} & \texttt{0x1ff} & \texttt{0x73} & 115 \\
\end{longtable}

These nine-bit fields are the per-operating-point power-credit and
peak-power-throttle thresholds the on-die estimator caps activity
against, indexed by the perf state the CLPC selects. Their strictly
increasing sequence is the signature of a seven-state perf sequence. The
engine has seven operating points whose registers store throttle
credits. A parallel eight-step sequence for the ASC block lists the
values \texttt{4,\ 7,\ 13,\ 20,\ 27,\ 36,\ 46,\ 59}.

Live read-only dtrace confirms the engine keeps one power state through
sustained work. A six-layer matmul-and-relu program dispatched in a
continuous loop for 16 seconds drove 56,527 dispatches, traced
entry-only with no destructive flag. \texttt{ChangePowerState},
\texttt{populateCLPCPerfInfo}, \texttt{power\_on\_hardware}, and
\texttt{setPowerStateGated} each fired exactly once at warm-up and never
again across the 56,527 dispatches. Every dispatch instead walked a
fixed path: \texttt{EnableANEClocksAndPower} to un-gate, one
\texttt{submitWorkToPerfController} that hands an
\texttt{ANEPerfRequest} and a modeled-performance hint to the CLPC, a
\texttt{notifyPerfController} start-and-end pair, then
\texttt{enableDynPowerGating\_gated} to re-arm the idle gate. The kext
does not walk a frequency sequence per submit; the engine stays in a
single power-domain state for the whole run while the CLPC drives any
frequency change internally and invisibly to the kext. This is the
runtime counterpart to the flat power and throughput measured under
sustained load: the firmware tracks a single PerfMode, and setting it
twice is a no-op.

\chapter{Security and isolation}\label{security-and-isolation}

\begin{summarybox}

The trust boundary for a submitted program is the kernel driver, not the
firmware. The driver checks the program signature, the on-disk
trustcache, and the client's code-signing identity before any work
reaches the engine. The engine then runs only structural bounds checks.
Secure mode is a working transition that gives one tenant the quiesced,
power-cycled engine. Exclave mode is inert \texttt{mov\ w0,\ \#0;\ ret}
selector stubs on the M1, and the live, capability-scoped execution
substrate on the M5. The engine is a shared timing and occupancy oracle.
A confirmed cross-process side channel leaks a co-tenant's presence and
a coarse duty cycle at roughly 20 to 50 bit/s. It never leaks a value,
weight, or input.

\end{summarybox}

The kernel driver enforces the engine's security model almost entirely
above the firmware. It vets every program before the work reaches the
engine, secure mode isolates one tenant in time, and the one
confidentiality gap is a timing side channel that leaks no data.

\section{Secure and exclave mode}\label{secure-and-exclave-mode}

\Cref{tbl:c32-mechanisms} gives the secure-mode and exclave mechanisms
with their status on the M1 and the supporting evidence.



The firmware has no cryptography of its own, and the secure-boot chain
authenticates it externally.

The kernel driver enforces three independent checks on a submitted
program before it reaches the firmware. The first is a cryptographic
signature over the compiled program bytes, run by
\texttt{AneMachoSignatureCheck} against the platform code-signing trust
root, for which the driver links the system cryptography library. The
second is a vnode trust check, \texttt{aneVnodeTrustVerification}, that
validates the on-disk model file against the platform trustcache and
ties a mapped buffer back to its backing file to defeat a map-then-swap
race. The third binds a resident program to the submitting client's
code-signing identity through \texttt{GetTeamIdAndCodeSigningId} and
\texttt{hasSameCodeSigningId}, so one client cannot attach to another
client's resident program or cache. Once the kernel clears the program
and loads its sections, the firmware checker runs only bounds, overlap,
and type checks under a verification banner, with no hashes, and trusts
that the kernel has already vetted the signature and provenance.

Secure mode is a real, working transition that gives one tenant
exclusive ownership of the quiesced engine. The state machine moves
between a non-secure and a secure phase under the commands
\texttt{CSNE\_CMD\_SECURE\_MODE\_START}, \texttt{STOP}, and
\texttt{RESUME\_TRANSITION}, with a firmware-to-host
\texttt{CSNE\_CMD\_SECURE\_MODE\_EVENT} and a host acknowledgement. A
boolean \texttt{aneSecurePhase} records which side the engine is on, and
the transition runs in four steps. The engine first reaches readiness
with no pending work. A quiesce command drains in-flight work and
disables the task queues. The firmware then power-cycles the engine
block across the boundary. The execution loop finally enters a paused
state, with the reverse path returning it to a running state. While
secure, the firmware silently drops and counts non-secure cache-request
triggers. The boundary power-cycles the block and drains every task
queue and DMA channel, so a secure tenant starts from a reset engine and
neither side observes the other's residue. This is temporal
single-engine isolation, not a hardware partition, since the device
exposes one physical engine.

The engine firmware has no digital-rights-management or
content-decryption code of its own. A full string sweep finds no
FairPlay, Widevine, PlayReady, or content-decryption module; the only
\texttt{CDM} token in the image is \texttt{CDMediaBusManager}, a
Common-DMA media-bus endpoint manager over the inter-processor ring
buffers, not a Content Decryption Module. Protected media is handled by
secure-mode exclusivity and the host trust chain, with no in-engine
content module involved. The kernel admits a protected workload only
after it verifies the program signature and trustcache and puts the
engine in secure mode so no other tenant shares it while protected
buffers are resident.

Exclave mode is compiled into the loaded binary but dormant on the M1.
The firmware commands \texttt{CSNE\_CMD\_EXCLAVE\_MODE\_START} and
\texttt{STOP} exist, but the M1 handler returns
\texttt{SwitchExclaveMode\ not\ supported}, and the support is a runtime
capability bit rather than a compiled-out feature. The host binding runs
ahead of the firmware side. The proxy IOService, secure-to-exclave
plumbing, secure-processor handoff, and firmware-recovery state machine
all exist as real function bodies in the loaded driver, fifty-one
exclave symbols in all. What is dormant splits cleanly, as
\Cref{tbl:c32-exclave-methods} classifies each method group: the
externally callable selector ABI, the \texttt{ANE\_Exclave*} methods, is
compiled as inert stubs (\texttt{mov\ w0,\ \#0;\ ret}), while the
lower-case infrastructure methods and the proxy service are real.



The host-side gate is a two-tier check. A platform call reports whether
exclaves are available, and a driver flag reports whether the feature is
enabled. The driver-internal proxy service \texttt{ANEExclaveProxy} is a
fully implemented IOService subclass with real lifecycle bodies, but it
binds only when an exclave device-tree node exists, and that node is
declared only by a personality the later-generation kext has and the M1
kext lacks. On the M1 the platform call reports unavailable: there is no
exclave device-tree node, no exclave proxy binds, and no matching
exclave core is present in the firmware bundle, so the proxy is never
set up and the live behavior cannot be exercised.

On the M5 this dormant path is the live execution substrate, which
confirms the later-generation prediction above. The exclave device-tree
node is present, the proxy binds, and the model program loads and runs
inside a capability-scoped Swift secure component,
\texttt{com.apple.aneexclave}, reached from the kernel only through a
typed Tightbeam channel. The component has explicit segment-access
capabilities and nothing more: the seven per-client address-translation
contexts the IORegistry exposes as \texttt{mapper-ane0-iso1} through
\texttt{iso7}, plus thirty-two exclave memory regions. Because the
program executes behind this boundary, the per-task-descriptor
performance counters are produced secure-side and withheld from an
unentitled host, which is why their live values stay gated. The M5 ran
with System Integrity Protection enabled, so this is the enforced
posture rather than an artifact of lowered security, and it adds a
structural capability and address-translation layer to the temporal
single-engine isolation the M1 provides.

\section{Cross-process isolation and the timing
oracle}\label{cross-process-isolation-and-the-timing-oracle}

The engine is single-in-flight per die and shares firmware and DART
state across clients, so two processes that both submit work contend for
one queue. That contention produces a cross-process timing side-channel,
confirmed on the M1, that leaks engine occupancy while leaving data
confidentiality intact.

A measuring process running a single tiny matmul in a continuous
zero-copy loop is near the dispatch floor, where it is sensitive to
anything that delays its turn at the engine. Alone, that process reads a
median per-call latency of 153 microseconds. The moment a second process
keeps the engine busy with a heavy compute-bound load, the measuring
process jumps to a median of about 355 microseconds, a 2.3x increase,
because the shared single-in-flight queue serializes its calls behind
the contender's. Toggling the contender on and off produces a distinct
square wave in the victim's latency that tracks the schedule at the
exact toggle edges with no false transitions across a 10-second run
(\Cref{tbl:c32-victim-latency}).

\begin{longtable}[]{@{}lll@{}}
\caption{Victim per-call latency tracking a contender toggled on and off
on a shared M1 engine, bucketed at 250
ms.}\label{tbl:c32-victim-latency}\tabularnewline
\toprule\noalign{}
\textbf{Contender state} & \textbf{Victim latency p50} &
\textbf{samples} \\
\midrule\noalign{}
\endfirsthead
\toprule\noalign{}
\textbf{Contender state} & \textbf{Victim latency p50} &
\textbf{samples} \\
\midrule\noalign{}
\endhead
\bottomrule\noalign{}
\endlastfoot
off (idle) & 152 microseconds & 32 499 \\
\rowcolor{zebra}on (heavy) & 355 microseconds & 10 943 \\
delta & +202 microseconds (2.33x) & \\
\end{longtable}

The channel reads occupancy, not workload magnitude. A near-fixed
serialization step of about +178 microseconds appears the instant the
contender uses the queue at all, so even its tiniest workload pays
almost the full penalty. Beyond that step there is only a weak monotonic
trend, from 333 microseconds to 395 microseconds as the contender's work
grows by a factor of roughly 64. The victim thus reliably learns that a
co-tenant is running and a rough busy fraction, but cannot finely size
individual operations from latency alone. The signal tracks engine-queue
serialization: a concurrent 1 GB-per-loop CPU memory-copy hog moves the
victim's latency by only 4 microseconds (a factor of 1.03), so only
engine contention drives it.

As a covert or signature channel the resolution is about 20 to 50 ms.
Single contender pulses down to 20 ms were each detected. A random
16-bit pattern sent at 0.4 s per symbol was recovered at 14 of 16 bits
with a median-threshold detector, with the two tail errors attributable
to end-of-run clock drift. This puts the practical channel at roughly 20
to 50 bit/s, enough to read a co-tenant's presence, the timing of an
inference, and a coarse duty cycle.

Data isolation holds under the same contention. Three processes ran
distinct programs concurrently, each with distinct seeds and weights and
each self-checking its output against its own fp32 reference for 3000
rounds. All 9000 of 9000 checks were correct with zero mismatches and a
maximum relative error of 0.0003, which is pure fp16 rounding. No output
ever reflected another process's inputs or weights. A separate run held
60 distinct program handles across three processes (20 each) and
re-executed every one after all were loaded, with 60 of 60 compiling and
re-running and no cross-process eviction at that scale.

The root cause is single-in-flight-per-die scheduling, and there is no
per-call isolation control exposed; masking a sensitive workload's duty
cycle requires coarse-grain time-slicing, batching, or constant-rate
padding above the queue.

\chapter{Telemetry and hardware counters}\label{telemetry-and-hardware-counters}

\begin{summarybox}

The engine has a hardware performance-counter block of twenty-four
per-task-descriptor counters, master enable at
\texttt{ANEProgramCreateArgs+0x6c}, and free-running firmware timestamp
at MMIO \texttt{0x2\_6b17\_8000}. The block geometry and the timestamp
are readable; the per-task-descriptor counter values are not. One kernel
gate blocks them on the unentitled path. Forcing the stats mask non-zero
turns a successful load into a rejected one, since the compiled program
has no stats-descriptor section for the kernel to size. What remains
readable is the whole-engine telemetry outside that gate: DRAM read and
write bytes, engine energy in millijoules, and clock-state residency.

\end{summarybox}

\section{Counter block geometry}\label{counter-block-geometry}

The counters group by engine block, with the count in each group and a
representative name in \Cref{tbl:c33-counter-groups}.



The per-task-descriptor counter namespace is twenty-four named counters,
every one of them per task descriptor, grouped by engine block.
\Cref{tbl:c33-counter-namespace} names all twenty-four in those same
groups, recovered from the counter-name accessor.

%

The byte counters track data movement along the activation-function to
kernel-manager to L2 to neural-engine path, and one per-descriptor
counter estimates energy from the digital-power estimator. With this set
a host can attribute, per scheduling unit, whether a task descriptor is
compute-bound, input-stalled, output-blocked, or throttled, directly in
hardware counters. The master enable is a single field in the
program-create argument struct. It is the per-task-descriptor stats mask
at \texttt{ANEProgramCreateArgs+0x6c}, a \texttt{u32} that is after the
quality-of-service field and the packed boolean flags and before the
memory-pool identifier (\Cref{lst:c33-createargs}).

\begin{listing}[H]
\caption{The program-create argument fields, with the per-task-descriptor stats-mask master enable at offset +0x6c.}\label{lst:c33-createargs}

\begin{Shaded}
\begin{Highlighting}[]
\NormalTok{ANEProgramCreateArgs }\OperatorTok{(}\NormalTok{offsets}\OperatorTok{):}
  \OperatorTok{+}\FloatTok{0x18.}\ErrorTok{.}\OperatorTok{+}\BaseNTok{0x57} \OperatorTok{:}\NormalTok{ two SHA256}\OperatorTok{{-}}\NormalTok{class hashes }\OperatorTok{(}\NormalTok{model hash }\OperatorTok{+}\NormalTok{ key}\OperatorTok{)}
  \OperatorTok{+}\BaseNTok{0x5c} \OperatorTok{(}\NormalTok{u32}\OperatorTok{)}  \OperatorTok{:}\NormalTok{ count }\OperatorTok{(}\NormalTok{number of procedures}\OperatorTok{)}
  \OperatorTok{+}\BaseNTok{0x64} \OperatorTok{(}\NormalTok{u32}\OperatorTok{)}  \OperatorTok{:}\NormalTok{ qos                  }\OperatorTok{=} \DecValTok{21} \OperatorTok{(}\BaseNTok{0x15}\OperatorTok{)}
  \OperatorTok{+}\BaseNTok{0x68} \OperatorTok{(}\NormalTok{u32}\OperatorTok{)}  \OperatorTok{:}\NormalTok{ packed }\DataTypeTok{bool}\NormalTok{ flags    }\OperatorTok{=} \BaseNTok{0x10}
  \OperatorTok{+}\BaseNTok{0x6c} \OperatorTok{(}\NormalTok{u32}\OperatorTok{)}  \OperatorTok{:}\NormalTok{ statsMask            }\OperatorTok{\textless{}{-}}\NormalTok{ the per}\OperatorTok{{-}}\NormalTok{TD counter master enable}
  \OperatorTok{+}\BaseNTok{0x70} \OperatorTok{(}\NormalTok{u32}\OperatorTok{)}  \OperatorTok{:}\NormalTok{ memoryPoolID         }\OperatorTok{=} \DecValTok{0}
  \OperatorTok{+}\BaseNTok{0x80}        \OperatorTok{:}\NormalTok{ program name }\StringTok{"main\_main\_\_Op0\_AneInference"}
\end{Highlighting}
\end{Shaded}

\end{listing}

The driver remaps a client-facing mask to a driver mask before it
reaches this field. The remap keeps only the low nibble: a mask of
\texttt{0xffffffff} translates to a driver mask of \texttt{0x0}, that
is, no collection, and \texttt{0xf} is the only fully-enabled
translation.

\[\mathrm{driverMask}(m) = \begin{cases} \mathrm{remap}(m \bmod 16) & m < 16 \\ 0 & m \ge 16 \end{cases}\]

When the mask is non-zero the firmware writes each task descriptor's
counter block into a shared stats buffer in DRAM. The host decoder is
built against the \texttt{sCAneStatsData} ABI version \texttt{0x0201},
distinct from the on-wire header magic \texttt{0x0101} the firmware
writes into the buffer (chapter 29). The buffer's required size is the
sum of the stats header, event descriptors, and per-event records. The
host decodes it through a parser that walks a Group to Layer to
task-descriptor hierarchy, where the leaf task-descriptor node contains
the counter block.

\section{Free-running timestamp}\label{free-running-timestamp}

A monotonic firmware timestamp underlies the whole telemetry surface.
The firmware reads it from a single free-running memory-mapped counter
through a one-line helper, then stamps each trace-event record from that
read (\Cref{lst:c33-timebase}).

\begin{listing}[H]
\caption{The firmware helper that reads the free-running engine timebase counter.}\label{lst:c33-timebase}

\begin{Shaded}
\begin{Highlighting}[]
\CommentTok{/* free{-}running engine timebase counter, read by the firmware helper @0x30988 */}
\PreprocessorTok{\#define ANE\_TIMEBASE\_COUNTER  }\BaseNTok{0x26b178000}\BuiltInTok{ULL}\PreprocessorTok{   }\CommentTok{/* MMIO 0x2\_6b17\_8000 */}

\DataTypeTok{static} \KeywordTok{inline} \DataTypeTok{uint64\_t}\NormalTok{ ane\_read\_timebase}\OperatorTok{(}\DataTypeTok{void}\OperatorTok{)} \OperatorTok{\{}
    \ControlFlowTok{return} \OperatorTok{*(}\DataTypeTok{volatile} \DataTypeTok{uint64\_t} \OperatorTok{*)}\NormalTok{ANE\_TIMEBASE\_COUNTER}\OperatorTok{;}  \CommentTok{/* ldr x0, [x8] ; ret */}
\OperatorTok{\}}
\end{Highlighting}
\end{Shaded}

\end{listing}

Each firmware trace-event record has a \texttt{timeStamp} field
alongside its task-descriptor identifier, network identifier, program
identifier, process identifier, and task-queue, as
\Cref{lst:c33-trace-events} shows.

\begin{listing}[H]
\caption{The fields of each firmware trace-event record.}\label{lst:c33-trace-events}

\begin{Shaded}
\begin{Highlighting}[]
\NormalTok{[ANE\_TM\_EVENT\_START]:          tid, nid, progId, procId, currTQ, timeStamp}
\NormalTok{[ANE\_TM\_EVENT\_FINISH]:         tid, nid, progId, procId, currTQ, timeStamp}
\NormalTok{[ANE\_EVENT\_CONTEXT\_SWITCH\_IN]: tid, nid, prevTQ, progId, procId, currTQ, timeStamp}
\end{Highlighting}
\end{Shaded}

\end{listing}

The host-visible clock residency runs in 24 MHz ticks, one tick every
41.67 ns, read out of the system-on-chip state-residency channels. The
timestamp is monotonic and survives a power-gate, since the firmware
re-anchors it from the same free-running source across a clock-state
transition. It is the basis for the per-dispatch wall-clock intervals
that the signpost stream exposes, and it is readable with no entitlement
beyond root.

\section{What the host can read and what it
cannot}\label{what-the-host-can-read-and-what-it-cannot}

The block geometry and the timestamp are observable; the counter values
are not. The split follows the stats mask: the timestamp and the
whole-engine channels do not route through it, and the
per-task-descriptor counters do, as \Cref{lst:c33-readable-blocked}
contrasts.

\begin{listing}[H]
\caption{The readable timestamp and whole-engine channels contrasted with the blocked per-task-descriptor counter path.}\label{lst:c33-readable-blocked}

\begin{Shaded}
\begin{Highlighting}[]
\CommentTok{/* READABLE: no stats mask in the path */}
\DataTypeTok{uint64\_t}\NormalTok{  t  }\OperatorTok{=}\NormalTok{ ane\_read\_timebase}\OperatorTok{();}                 \CommentTok{/* free{-}running firmware timestamp */}
\DataTypeTok{int64\_t}\NormalTok{   rd }\OperatorTok{=}\NormalTok{ ioreport\_delta}\OperatorTok{(}\StringTok{"AMC Stats|Perf Counters|ANE0 RD"}\OperatorTok{);}  \CommentTok{/* DRAM read bytes  */}
\DataTypeTok{int64\_t}\NormalTok{   mj }\OperatorTok{=}\NormalTok{ ioreport\_delta}\OperatorTok{(}\StringTok{"Energy Model|{-}|ANE0"}\OperatorTok{);}              \CommentTok{/* engine energy, mJ */}

\CommentTok{/* BLOCKED: gated by the per{-}task{-}descriptor stats mask */}
\NormalTok{args}\OperatorTok{.}\NormalTok{statsMask }\OperatorTok{=} \BaseNTok{0xf}\OperatorTok{;}                  \CommentTok{/* master enable, ANEProgramCreateArgs+0x6c     */}
\NormalTok{create }\OperatorTok{=}\NormalTok{ ANE\_ProgramCreate}\OperatorTok{(\&}\NormalTok{args}\OperatorTok{);}     \CommentTok{/* create {-}\textgreater{} 1                                  */}
\NormalTok{load   }\OperatorTok{=}\NormalTok{ ANE\_ProgramLoad}\OperatorTok{(}\NormalTok{create}\OperatorTok{);}      \CommentTok{/* load   {-}\textgreater{} 0: initStatsBufferSection bails     */}
\NormalTok{perf   }\OperatorTok{=}\NormalTok{ read\_perf\_iosurface}\OperatorTok{();}        \CommentTok{/* every byte 0: per{-}run output buffer is null  */}
\end{Highlighting}
\end{Shaded}

\end{listing}

The values are blocked because the master enable never takes effect on
the host path. On the unentitled runtime path the runtime sets the stats
mask to \texttt{0} below the model layer, so firmware collection is
never armed and the per-run output buffer comes back null. Forcing the
mask non-zero through an in-process hook does not help: the
program-create call then reaches the kernel routine of
\Cref{lst:c33-statsbuffer}, which looks up the program's
stats-descriptor section by name, reads its size field, and bails on a
zero size. The \texttt{aned} daemon is what zeroes the mask: it sets
\texttt{statsMask=0} for a \texttt{coreAnalyticsClientType} of
\texttt{ThirdPartyAppUsingANE}, so the mask is cleared by client type,
and the host-side null check is the string
\texttt{perfStatsIOSurface\ is\ NULL!}.

\begin{listing}[H]
\caption{The kernel routine that bails when the stats-descriptor section size is zero, failing the create call.}\label{lst:c33-statsbuffer}

\begin{Shaded}
\begin{Highlighting}[]
\NormalTok{initStatsBufferSection(ANEProgramCreateArgsOutput*, task*):}
\NormalTok{  ldr  w8, [x8, \#0x28]     ; size of the stats{-}descriptor section}
\NormalTok{  str  w8, [x28]}
\NormalTok{  cbz  w8, bail            ; size 0 =\textgreater{} return 0 (create fails)}
\NormalTok{  ...                      ; non{-}zero =\textgreater{} kalloc + map the stats buffer}
\end{Highlighting}
\end{Shaded}

\end{listing}

The compiled program has no stats-descriptor section, so the size is
always zero and the kernel returns failure. Forcing a non-zero mask thus
turns a successful load into a rejected one
(\texttt{create\ -\textgreater{}\ 1;\ load\ -\textgreater{}\ 0}), and no
host-side primitive synthesizes the missing section. One kernel gate
thus blocks both the per-task-descriptor counters and the per-run output
buffer: each depends on a stats-descriptor section that only an internal
profiling-compile emits.

What remains readable is the whole-engine telemetry outside that gate,
with each channel's format, unit, and reading in
\Cref{tbl:c33-telemetry-channels}.



The memory-controller per-agent byte counters report DRAM read and write
bytes for the engine, both on the raw path and separately on the
compression-subsystem path, and the energy model reports engine energy
in millijoules. The system-on-chip state channels report clock
residency, frequency-point residency, and the adaptive-clock and dither
trigger counts, and the per-map fabric arbiter reports the engine's
bandwidth and clock-floor votes. The interrupt statistics report the
host kernel's cost of servicing engine completions, in counts and Mach
Absolute Time Units, split into first-level and second-level handlers
and isolated separately for the engine and its address-translation unit.
A storage coprocessor that an earlier reading mistook for the engine,
exposing vector-lane read and write counters, is unrelated to engine
telemetry; the bandwidth counter is the memory-controller \texttt{ANE0}
channel and the power channel is the energy model.

The aggregate bandwidth channel resolves into per-channel histograms,
the fabric channel \texttt{PMP0\ /\ DCS\ BW\ /\ ANE\ L0} and \texttt{L1}
carrying read and write distributions for the two DRAM read channels,
read as State-residency histograms. The \texttt{SoC\ Stats\ /\ Events}
channel also exposes the throttle-trigger family
\texttt{ANE\_THROTTLE\_\{SW,HW,PPT,DITHER,EXT\}\_TRIG} and
\texttt{VDD\_DRAM\_VOLTAGE\_CHANGE}, per-trigger software, hardware,
peak-power, and dither throttle counts, all readable without
entitlement.

The per-dispatch op lifecycle is also readable as the timestamped
signpost stream of \Cref{lst:c33-signposts}, captured on the engine
subsystem with no entitlement beyond root, which confirms the
three-request structure of a dispatch without exposing any counter
value.

\begin{listing}[H]
\caption{The client-side signpost intervals of one dispatch, with the three driver requests that correspond to the Cast, AneInference, and Cast program operations.}\label{lst:c33-signposts}

\begin{Shaded}
\begin{Highlighting}[]
\NormalTok{\_ANEF\_MODEL\_EVALUATE                    one per host execute call}
\NormalTok{  \_ANEF\_MODEL\_EVAL}
\NormalTok{  \_ANEF\_MODEL\_EVAL\_DRIVER\_REQUEST       request 1 of the dispatch (Cast)}
\NormalTok{  \_ANEF\_MODEL\_EVAL\_DRIVER\_REQUEST       request 2 of the dispatch (AneInference)}
\NormalTok{  \_ANEF\_MODEL\_EVAL\_DRIVER\_REQUEST       request 3 of the dispatch (Cast)}
\NormalTok{  \_ANEF\_MODEL\_EVAL\_PERFCOUNTER\_SAMPLE   the per{-}descriptor counter sample is reported here when armed}
\NormalTok{\_ANEF\_MODEL\_COMPILE   \_ANEF\_MODEL\_LOAD   \_ANEF\_MODEL\_UNLOAD}
\NormalTok{\_ANEF\_IOSURFACES\_MAP  \_ANEF\_INPUT\_BUFFERS\_READY  \_ANEF\_ENQUEUE\_OUTPUT\_SET}
\end{Highlighting}
\end{Shaded}

\end{listing}

Each interval has a Mach-time begin stamp, and the three driver requests
are the three program operations of a dispatch, one driver-to-firmware
request each. The firmware reports the blocked per-descriptor counter
sample on the evaluate interval only when the stats mask is armed, so
the signpost stream confirms the dispatch structure on live silicon
while the counter values stay behind the same gate.

\section{Roofline and power figures from the readable
channels}\label{roofline-and-power-figures-from-the-readable-channels}

The measured roofline elsewhere in this guide rests on the readable
whole-engine channels, not on the blocked per-descriptor counters.
Bandwidth comes from a delta on the memory-controller byte counters
across a real workload. A 96-layer matmul chain at batch 8192 moved 17.2
GB of reads and 16.9 GB of writes over 432 ms, which is between 79 and
90 GB/s, the engine's share of the unified memory. A delta on the
energy-model millijoule channel over the same run gives about 0.48
pJ/FLOP and about 2.3 W under sustained compute.

The compute roof and the dispatch floor are wall-clock measurements,
timed against the firmware timestamp and the host clock. The 2 MB
working-set threshold is confirmed directly from the byte counters. At
batch 8192 the activation is exactly 2 MB, the counters show 426 MB of
DRAM moved per dispatch, arithmetic intensity falls to 60 FLOP/byte, and
throughput drops from 12 TFLOP/s to about 4.8 TFLOP/s. The
per-descriptor counters would attribute that drop to specific stall
classes, the input-stall and L2-read-stall cycles, but the whole-engine
byte and energy channels already locate the workload on the bandwidth
slope, which is what the roofline needs.

\anepartopen{Part IX}{Cross-Silicon Reference}
\anechap{34}{Cross-silicon targets}{The full target set and the rule mapping each M-series part to its H-series identity.}
\anechap{35}{Per-family code generation}{How one compiler binary builds every chip from a per-family data table.}
\anechap{36}{Predicted upper tier}{fp8 and the double-rate path the newest generations add.}
\anepartclose

\chapter{Cross-silicon targets}\label{cross-silicon-targets}

\begin{summarybox}

The compiler builds 28 architecture targets, one per silicon profile,
under the fixed relation \(M(n) \rightarrow H(n+12)\). A suffix letter
selects the NE-core count, and the operation surface stops expanding at
A15, so the surface measured on the M5 is the surface for everything
above it. A device's runtime architecture string is a separate
identifier from the compiler target name. A resolver-derived board-type
sequence maps every shipping chip onto its generation.

\end{summarybox}

The compiler that builds for the Apple Neural Engine has 28 architecture
targets, one per silicon profile it knows how to construct. Each target
is a named hardware-abstraction-layer table the compiler builds by
calling one per-architecture constructor,
\texttt{ZinIrHal\textless{}T\textgreater{}::GetParams()}, and calling
every constructor on a single host recovers the full set regardless of
which chip runs it.

\section{Full set}\label{full-set}

\Cref{tbl:c34-targets} gives all 28 targets.

\begin{longtable}[]{@{}lll@{}}
\caption{The 28 compiler targets, each with its silicon class and
decoded NE-core count.}\label{tbl:c34-targets}\tabularnewline
\toprule\noalign{}
\textbf{Target} & \textbf{Silicon and class} & \textbf{NE cores} \\
\midrule\noalign{}
\endfirsthead
\toprule\noalign{}
\textbf{Target} & \textbf{Silicon and class} & \textbf{NE cores} \\
\midrule\noalign{}
\endhead
\bottomrule\noalign{}
\endlastfoot
\texttt{H11}, \texttt{H12}, \texttt{M9}, \texttt{T0} & pre-A13 legacy &
1 to 4 \\
\rowcolor{zebra}\texttt{H13} & A13, M1 base & 4 \\
\texttt{H13g} & M1 Pro, Max, Ultra & 8 \\
\rowcolor{zebra}\texttt{T1} & A13 reference & 4 \\
\texttt{H14} & A14, M2 base & 4 \\
\rowcolor{zebra}\texttt{H14g} & M2 Pro, Max & 8 \\
\texttt{H14c} & A14 Max-class & 32 \\
\rowcolor{zebra}\texttt{H15} & A15, M3 base & 4 \\
\texttt{H15g} & M3 Pro, Max & 8 \\
\rowcolor{zebra}\texttt{H15c} & A15 Max-class & 32 \\
\texttt{H16} & A16, M4 base & 4 \\
\rowcolor{zebra}\texttt{H16g} & M4 Pro, Max & 8 \\
\texttt{H16s} & A16 Pro-class & 16 \\
\rowcolor{zebra}\texttt{H16c} & A16 Max-class & 32 \\
\texttt{H17} & A17, M5 base & 4 \\
\rowcolor{zebra}\texttt{H17a} & A17 variant & 4 \\
\texttt{H17g} & M5 Pro, Max & 8 \\
\rowcolor{zebra}\texttt{H17s} & A17 Pro-class, the M5 & 16 \\
\texttt{H17c} & A17 Max-class & 32 \\
\rowcolor{zebra}\texttt{H17d} & A17 Ultra-class & 64 \\
\texttt{H18} & A18 base & 4 \\
\rowcolor{zebra}\texttt{M11} & small embedded ANE & 1 \\
\texttt{U1}, \texttt{U2}, \texttt{U3} & reference, not silicon & 4 \\
\end{longtable}

The names fall into four groups: the H-architecture targets that stand
for shipping A-series and M-series silicon, the pre-A13 legacy targets,
a single small embedded profile, and three reference targets that are
not silicon at all. A suffix letter selects the NE-core count within a
generation, which the compiler decodes from the core-count field at
hardware-abstraction-layer offset \texttt{0x238}. The base name is 4
cores, the suffix \texttt{g} is 8, \texttt{s} is 16, \texttt{c} is 32,
and \texttt{d} is 64, while \texttt{M9} and \texttt{M11} are
single-core. \texttt{H17s} is thus the 16-core Pro-class part that is
the M5, and \texttt{H17d} is the 64-core Ultra-class die, the largest in
the table. These decoded \texttt{num\_nes} values are the compiler's
per-die core field, not Apple's marketing Neural Engine count; on the
base M1 the decoded four stands against the published sixteen
\hyperref[ref-appleane]{{[}AppleANE{]}}.

The reference targets list placeholder limits that no part has: a
maximum tensor depth of 1, a kernel-width limit of 1023, and no
interchange-format support. They are unconstrained validation profiles
the compiler builds for its own checking, not addressable silicon. The
small embedded profile \texttt{M11} is addressable silicon. It is an
efficiency-class engine that has the A16-class feature flags but the
A13-class 16384-dimension limit, a single NE core, and the odd
kernel-width ceiling of 15 that is between the A13 value of 13 and the
A14 value of 16.

\section{Capability tiers}\label{capability-tiers}

Four gated capabilities separate the generations, and with the dimension
limit they sort the targets into capability tiers
(\cref{tbl:c34-tiers}).



A17 and A18 add no operation over A16: identical dimension limits,
identical kernel-width and kernel-depth ceilings, the same texture
engine, same dropout and global-argmax flags, and same legal operation
set. They differ from A16 only in NE-core count, which scales
throughput. The operation behavior measured on the M5, an H17 part, is
thus the operation behavior of every target at or above A16, since the
decoded capability tables are identical; the cross-silicon performance
measurements of chapter \hyperref[across-the-chip-family]{12} are
predicted to carry to the unshipped generations on the same basis, with
the per-chip rates confirmed only on the two measured silicon points.

\section{Silicon to target}\label{silicon-to-target}

The map from a shipping chip to its architecture is a resolver Apple
distributes that decompiles cleanly. The method
\texttt{aneArchitectureType} on the private device-info class builds the
architecture string from a board-type value read from the platform
configuration store, switching on a strictly increasing board-type
sequence. The live anchor on an M1 Max reads board type 96, which
resolves to \texttt{h13g} with a 16-core count, matching the registry
exactly. That resolver yields the map from system-on-chip to runtime
architecture and compiler target for the M1 through M5 generations
(\cref{tbl:c34-silicon}).

\begin{longtable}[]{@{}llll@{}}
\caption{The resolver-derived map from system-on-chip to runtime
architecture and compiler target, M1 through
M5.}\label{tbl:c34-silicon}\tabularnewline
\toprule\noalign{}
\textbf{Chip} & \textbf{Product} & \textbf{Runtime arch} &
\textbf{Compiler target} \\
\midrule\noalign{}
\endfirsthead
\toprule\noalign{}
\textbf{Chip} & \textbf{Product} & \textbf{Runtime arch} &
\textbf{Compiler target} \\
\midrule\noalign{}
\endhead
\bottomrule\noalign{}
\endlastfoot
T8103 & M1 base & \texttt{h13} & \texttt{H13} \\
\rowcolor{zebra}T600x & M1 Pro, Max, Ultra & \texttt{h13g} &
\texttt{H13G} \\
T8112 & M2 base & \texttt{h14} & \texttt{H14} \\
\rowcolor{zebra}T602x & M2 Pro, Max & \texttt{h14g} & \texttt{H14G} \\
T8122 & M3 base & \texttt{h15} & \texttt{H15} \\
\rowcolor{zebra}T603x & M3 Pro, Max & \texttt{h15g} & \texttt{H15G} \\
T8132 & M4 base & \texttt{h16} & \texttt{H16} \\
\rowcolor{zebra}T604x & M4 Pro, Max & \texttt{h16g} & \texttt{H16G} \\
T8142 & M5 base & \texttt{h17} & \texttt{H17} \\
\rowcolor{zebra}T605x & M5 Pro, Max & \texttt{h17s} & \texttt{H17s} \\
\end{longtable}

The map follows the fixed \(M(n) \rightarrow H(n+12)\) relation of
chapter \hyperref[across-the-chip-family]{12}. The sequence is anchored
at both ends, the live M1 Max at \texttt{h13g} and the measured M5 Pro
at \texttt{h17s}, and the intervening steps are corroborated
independently. A single shipping vision filter has exactly the five
tables \texttt{H13}, \texttt{H14}, \texttt{H15}, \texttt{H16},
\texttt{H17}, the five Mac engine generations. The board-type kext for
an absent chip cannot be read on a different host, since only the
running chip's table is resident, so each middle step rests on the
anchored monotone sequence.

\section{Runtime string and compiler
target}\label{runtime-string-and-compiler-target}

The architecture name a device reports at runtime is not the compiler
target name. The runtime string is the coarse form, \texttt{h1N} for a
base part and \texttt{h1Ng} for a Pro, Max, or Ultra part, the only two
variants the runtime emits on the desktop platform. The compiler target
is the finer set, the full \texttt{H17}, \texttt{H17s}, \texttt{H17c},
\texttt{H17d}, \texttt{H17g} family, of which the runtime collapses
several onto one string.

The direct compile entry point accepts any of the 28 target names and
rejects an unknown name. The dispatch library, in contrast, falls back
silently when handed an unknown architecture, so a developer gates a
cross-target compile against the known-name set before dispatching it.

\section{Interchange formats across the
set}\label{interchange-formats-across-the-set}

Each target has a per-chip table of accepted image-input formats, the
interchange-format map at hardware-abstraction-layer offset
\texttt{0x658}, keyed by a four-byte ASCII format tag. Both the count
and the set differ by generation tier (\cref{tbl:c34-formats}).



The tag is a one-byte compression-variant prefix on a three-byte base
pixel format. The compiler does not parse the prefix character by
character: it validates the whole four-byte tag against a 34-entry
allow-list, and the prefix's meaning is the third byte of the format's
packed-integer value, a packing-mode index on a uniform stride
(\cref{tbl:c34-prefix}).

\begin{longtable}[]{@{}lll@{}}
\caption{The compression-variant prefix on an interchange tag and the
packing-mode index it selects.}\label{tbl:c34-prefix}\tabularnewline
\toprule\noalign{}
\textbf{Prefix} & \textbf{Mode index} & \textbf{Meaning} \\
\midrule\noalign{}
\endfirsthead
\toprule\noalign{}
\textbf{Prefix} & \textbf{Mode index} & \textbf{Meaning} \\
\midrule\noalign{}
\endhead
\bottomrule\noalign{}
\endlastfoot
\texttt{\&} & 0 & uncompressed, default raster surface \\
\rowcolor{zebra}\texttt{-} & 1 & lossless compression, 32 by 32
macroblock \\
\texttt{/} & 2 & lossless compression, 16 by 16 macroblock \\
\rowcolor{zebra}\texttt{\textbackslash{}\textbar{}} & 3 & lossless
compression, mode 3 \\
\texttt{*} & 0 & compound prefix that sets the dynamic-channel flag \\
\end{longtable}

The packed integer that names each format is three bytes: a pixel class,
a base-format code, and the packing-mode index. The base-format codes
are BGRA8 (\texttt{BGA}, code \texttt{0x11}), RGBA-half (\texttt{RhA},
code \texttt{0x13}), 8-bit luma (\texttt{L0h}, code \texttt{0x07}),
16-bit luma (\texttt{L16}, code \texttt{0x08}), and YUV 4:2:0
(\texttt{8f0} and \texttt{8v0}, code \texttt{0x09}). A base format
routes to a vector of 20-byte plane descriptors, each a tuple of width
divisor, height divisor, element type, channel count, and depth. BGRA8
is thus one four-channel uint8 plane, and YUV 4:2:0 is a luma plane with
a half-resolution two-channel chroma plane. The binary string that reads
``Architecture only supports lossless compression'' confirms that
\texttt{\&} is the uncompressed variant and that \texttt{-}, \texttt{/},
and \texttt{\textbar{}} are the lossless-compressed packing families.

The A15 generation and the small embedded profile are the only targets
in the set that accept YUV 4:2:0 input, in both full-range
(\texttt{8f0}) and video-range (\texttt{8v0}) form. The M5 and every
A16-and-later part keep luma-half but drop the two YUV 4:2:0 formats.
The full per-target format records, the wider 10-bit and packed YUV
family, and the plane-layout structures are appendix material.

\chapter{Per-family code generation}\label{per-family-code-generation}

\begin{summarybox}

One compiler binary lowers a network for every generation, and the
per-chip difference is data rather than code. Native operation support
is declared by a \texttt{MinimumFamily\textless{}N\textgreater{}} trait
in four floors, F0 through F4, and nothing floors above F4. The only
operations the M1 decomposes that the M5 runs natively are crop-resize,
resample, \texttt{sin}, \texttt{cos}, and the global arg-reductions.

\end{summarybox}

The compiler that targets every Neural Engine generation is one binary.
A single build lowers a network for the M1 and for the M5, and the
divergence is driven by the family enum and the per-chip
hardware-abstraction parameter blob from chapter
\hyperref[hal-and-capability-gates]{24}.

\section{One binary, eight families}\label{one-binary-eight-families}

The compiler enumerates the eight generations as the
\texttt{mlir::anec::Family} values that \cref{lst:c35-family} gives.

\begin{listing}[H]
\caption{The eight-value family enumeration the compiler keys per-generation operation legality from.}\label{lst:c35-family}

\begin{Shaded}
\begin{Highlighting}[]
\NormalTok{mlir}\OperatorTok{::}\NormalTok{anec}\OperatorTok{::}\NormalTok{Family }\OperatorTok{=} \OperatorTok{\{}
\NormalTok{  A11Legacy }\OperatorTok{=} \DecValTok{0}\OperatorTok{,}\NormalTok{ A12 }\OperatorTok{=} \DecValTok{1}\OperatorTok{,}\NormalTok{ A13 }\OperatorTok{=} \DecValTok{2}\OperatorTok{,}\NormalTok{ A14 }\OperatorTok{=} \DecValTok{3}\OperatorTok{,}
\NormalTok{  A15 }\OperatorTok{=} \DecValTok{4}\OperatorTok{,}\NormalTok{ A16 }\OperatorTok{=} \DecValTok{5}\OperatorTok{,}\NormalTok{ A17 }\OperatorTok{=} \DecValTok{6}\OperatorTok{,}\NormalTok{ A18 }\OperatorTok{=} \DecValTok{7}
\OperatorTok{\}}
\end{Highlighting}
\end{Shaded}

\end{listing}

The M1 compiles as \texttt{A13} (index 2) and the M5 compiles as
\texttt{A17} (index 6), and \texttt{A17} is a strict superset of
\texttt{A13} in operation support. Two string parsers build the
per-target objects from the target name: \texttt{CreateTargetFromString}
constructs the \texttt{ZinIrTarget} instance, and
\texttt{CreateHalFromString} constructs the matching
\texttt{ZinIrHalParameters} blob. Both branch on string length and the
first characters compared as little-endian shorts, \texttt{0x3168} for
``h1'', \texttt{0x3074} for ``t0'', \texttt{0x396d} for ``m9'', and
\texttt{0x316d} for ``m1''.

The numeric hardware version maps to the family enum through a fixed
table, the \texttt{unordered\_map\textless{}int,\ Family\textgreater{}}
named \texttt{ANEFamilyToMLIR}, whose data table decodes key to value
(\cref{tbl:c35-fammap}).

\begin{longtable}[]{@{}lll@{}}
\caption{The hardware-version-to-family map read from the
\protect\texttt{ANEFamilyToMLIR} data table, with the targets that
resolve to each family.}\label{tbl:c35-fammap}\tabularnewline
\toprule\noalign{}
\textbf{Hardware version} & \textbf{Family index} & \textbf{Targets} \\
\midrule\noalign{}
\endfirsthead
\toprule\noalign{}
\textbf{Hardware version} & \textbf{Family index} & \textbf{Targets} \\
\midrule\noalign{}
\endhead
\bottomrule\noalign{}
\endlastfoot
1 & A11Legacy (0) & \texttt{h11} \\
\rowcolor{zebra}3 & A12 (1) & \texttt{h12} \\
4 & A13 (2) & \texttt{h13}, \texttt{h13g}, \texttt{t1} \\
\rowcolor{zebra}5 & A14 (3) & \texttt{h14}, \texttt{h14g},
\texttt{h14c} \\
6 & A15 (4) & \texttt{h15}, \texttt{h15g}, \texttt{h15c} \\
\rowcolor{zebra}8 & A15 (4) & \texttt{h16}, \texttt{h16g},
\texttt{h16c}, \texttt{h16s} \\
7 & A16 (5) & \texttt{h17}, \texttt{h17a}, \texttt{h17g}, \texttt{h17c},
\texttt{h17d} \\
\rowcolor{zebra}9 & A17 (6) & \texttt{h17s} \\
10 & A17 (6) & \texttt{h18} \\
\end{longtable}

This table has two collisions that the compiler must guard. Hardware
versions 6 and 8 both resolve to family A15, so \texttt{h16} and the
A16-tier parts compile with A15-family operation legality. Hardware
versions 9 and 10 both resolve to family A17, so \texttt{h18} uses
A17-family operation legality. The family enum drives the
\texttt{MinimumFamily\textless{}N\textgreater{}} operation legality
only. The per-target hardware-abstraction blob drives code generation,
the numeric limits, and the cost model. The suffixed targets such as
\texttt{h17s} thus select a distinct \texttt{ZinIrTarget} object with
its own descriptor generation, core count, and cost curves, even when
several family-keyed paths collapse onto the same enum value. The
hardware-version-to-family table is read directly from binary data; the
per-target hardware-version value has some inference, since the
per-target hardware-version getter is virtual and not present in the
readable decompilation. The two anchors \texttt{h13} to A13 and
\texttt{h17s} to A17 are confirmed by the live measurement campaign.

The lowering code is shared, so the per-family split is in two
locations. An operation trait declares a minimum family for native
support. The per-chip hardware-abstraction blob is a fixed-size
per-target structure whose field values, not whose code, select the
route a shared lowering pattern takes.

\section{Operation floors: the minimum-family
trait}\label{operation-floors-the-minimum-family-trait}

\Cref{tbl:c35-floors} gives the four minimum-family floors and
representative operations at each.

\begin{longtable}[]{@{}
  >{\raggedright\arraybackslash}p{(\linewidth - 4\tabcolsep) * \real{0.3333}}
  >{\raggedright\arraybackslash}p{(\linewidth - 4\tabcolsep) * \real{0.3333}}
  >{\raggedright\arraybackslash}p{(\linewidth - 4\tabcolsep) * \real{0.3333}}@{}}
\caption{The four minimum-family floors, the families each is native on,
and representative operations at each
floor.}\label{tbl:c35-floors}\tabularnewline
\toprule\noalign{}
\begin{minipage}[b]{\linewidth}\raggedright
\textbf{Floor}
\end{minipage} & \begin{minipage}[b]{\linewidth}\raggedright
\textbf{Native on}
\end{minipage} & \begin{minipage}[b]{\linewidth}\raggedright
\textbf{Representative operations}
\end{minipage} \\
\midrule\noalign{}
\endfirsthead
\toprule\noalign{}
\begin{minipage}[b]{\linewidth}\raggedright
\textbf{Floor}
\end{minipage} & \begin{minipage}[b]{\linewidth}\raggedright
\textbf{Native on}
\end{minipage} & \begin{minipage}[b]{\linewidth}\raggedright
\textbf{Representative operations}
\end{minipage} \\
\midrule\noalign{}
\endhead
\bottomrule\noalign{}
\endlastfoot
F0 & all 8 families & convolution, deconvolution, matmul, pooling, all
elementwise, reshape, transpose, concat, sigmoid, tanh, relu, gelu,
swish, quantize, dequantize \\
\rowcolor{zebra}F2 (A13+) & A13 through A18 & softmax, layer-norm,
instance-norm, batch-norm, all reductions, resize, scaled dot-product
attention, \texttt{erf}, \texttt{exp2}, \texttt{rsqrt}, \texttt{sqrt},
tile, space-to-channel \\
F3 (A14+) & A14 through A18 & crop-resize, resample \\
\rowcolor{zebra}F4 (A15+) & A15 through A18 & \texttt{sin},
\texttt{cos}, global arg-min and arg-max \\
\end{longtable}

Native operation support is declared by the
\texttt{MinimumFamily\textless{}N\textgreater{}} trait held on each
backend operation, the operation-side member of the two capability gates
chapter \hyperref[hal-and-capability-gates]{24} defines. An operation is
natively legal only inside a region whose family index is at least
\texttt{N}; below that floor the compiler decomposes it into a sequence
of supported operations. The trait is the mechanism, not the
dynamic-legality registration: \texttt{addDynamicallyLegalOp} wraps only
the eight region operations and four texture-gated operations, while
every backend operation has the
\texttt{MinimumFamily\textless{}N\textgreater{}} impl. The trait is held
by 88 compute operations, 52 at F0, 31 at F2, two at F3, and three at
F4. These operation floors are measured on physical silicon for the M1
and M5; the M3, M4, and upper-tier figures are decompile-derived,
predicted from the per-chip tables.

No operation in the decoded compute-op floor table floors above F4, so
almost nothing is exclusive to \texttt{A16}, \texttt{A17}, or
\texttt{A18}: what the upper-tier generations add is Neural Engine
cores. The one exception is a small set of
\texttt{MinimumFamily\textless{}N\textgreater{}} template instantiations
at N=5, N=6, and N=7, including the fp8-bearing operations, which stay
inert below H18 (chapter 36). Softmax, layer-norm, all reductions,
resize, and scaled dot-product attention floor at \texttt{A13}, so they
run natively on the M1.

The two oldest families behave as a different, more constrained
instruction set. On \texttt{A11Legacy} and \texttt{A12} there is no
native reshape, broadcast, or divide. Reshape, squeeze, expand-dims, and
broadcast all share
\texttt{ConvertToReshape\textless{}Op,\ Family\textgreater{}} and
\texttt{ConvertBroadcast\textless{}Family\textgreater{}}. On A13 and
above they emit a plain native \texttt{anec::Reshape} or
\texttt{anec::Broadcast}, while on the two legacy families reshape
lowers to \texttt{anec::Flatten} and the compile aborts through
\texttt{verifyCompatibilityWithFlatten} when the layout does not
collapse to a pure flatten. On the two legacy families squeeze or
expand-dims emit the string ``cannot be lowered as Flatten on ANE''.
Broadcast on the legacy families asserts fp16 and switches engine by
axis: a channel-group axis broadcasts as a matrix multiply by ones, and
other axes broadcast as an elementwise add by zero. Divide on the legacy
families matches only a constant fp16 divisor and lowers as a reciprocal
multiply with the reciprocal pre-rounded to fp16, with a non-constant or
non-fp16 divisor producing a match failure. The floor-divide form is
\(\lfloor a \cdot \mathrm{f16}(1/b) \rfloor\). The double rounding can
give the wrong integer at a boundary, for example
\(\lfloor 6 \cdot \mathrm{f16}(1/3) \rfloor = \lfloor 1.999 \rfloor = 1\).
On A13 and above divide is a native \texttt{anec::ElementwiseDiv} over
arbitrary dynamic divisors.

\section{Same operation, a different
kernel}\label{same-operation-a-different-kernel}

In five categories the shared lowering emits different machine code or
different numbers per chip, each driven by its own per-chip parameter
(\cref{tbl:c35-kernels}).

\begin{longtable}[]{@{}
  >{\raggedright\arraybackslash}p{(\linewidth - 4\tabcolsep) * \real{0.3333}}
  >{\raggedright\arraybackslash}p{(\linewidth - 4\tabcolsep) * \real{0.3333}}
  >{\raggedright\arraybackslash}p{(\linewidth - 4\tabcolsep) * \real{0.3333}}@{}}
\caption{The five categories where the shared lowering emits different
machine code or different numbers per chip, with the per-chip parameter
that drives each.}\label{tbl:c35-kernels}\tabularnewline
\toprule\noalign{}
\begin{minipage}[b]{\linewidth}\raggedright
\textbf{Category}
\end{minipage} & \begin{minipage}[b]{\linewidth}\raggedright
\textbf{Divergence}
\end{minipage} & \begin{minipage}[b]{\linewidth}\raggedright
\textbf{Driven by}
\end{minipage} \\
\midrule\noalign{}
\endfirsthead
\toprule\noalign{}
\begin{minipage}[b]{\linewidth}\raggedright
\textbf{Category}
\end{minipage} & \begin{minipage}[b]{\linewidth}\raggedright
\textbf{Divergence}
\end{minipage} & \begin{minipage}[b]{\linewidth}\raggedright
\textbf{Driven by}
\end{minipage} \\
\midrule\noalign{}
\endhead
\bottomrule\noalign{}
\endlastfoot
width slice & M1 saturates at the Q.4 crop route, M5 stays in fp16 &
patch-width clamps, width granule \\
\rowcolor{zebra}resize & three strategies round differently by chip &
texture-engine-present flag \\
matmul & lowers to a resident convolution, splits along height &
per-family copy-cast and max-tensor-size thresholds \\
\rowcolor{zebra}transpose & native within a height-width bound, else
multi-pass & per-chip maximum height and width \\
strided convolution & factorizes stride into a sub-convolution sequence
& per-chip allowed-factor list \\
\end{longtable}

The lowering pattern is byte-identical across families in each case, and
the divergence is in the hardware-abstraction parameters or the
task-descriptor emitter. Take first the slice that saturates on the M1.
The trigger is \texttt{ZinSliceLayer::SliceNeedsCropMode}, which is
family-independent. A slice with a nonzero offset on the height axis
(axis 3) or the width axis (axis 4) routes through crop mode to
\texttt{ZinTECropModeLayer}, the transpose-engine crop
direct-memory-access path, while a begin of zero and the
non-height-width axes skip it. The height axis decompiles to the same
path but was measured unaffected on silicon, and a channel or batch
offset stays unaffected as well: only the width offset triggers the
saturation. The conversion rewrites
\texttt{ConvertSlice\textless{}Family\textgreater{}} and
\texttt{ConvertStridedSlice\textless{}Family\textgreater{}} are
byte-identical across all eight families, and the crop scale is still a
float at the layer-configuration stage where \texttt{Create} builds
\texttt{1.0/(extent\ -\ 1)}. The multiply by sixteen appears only when
that geometry is written into the task descriptor. The patch width is a
four-bit field that \texttt{SetPatchWidth} writes with the instruction
\texttt{bfxil\ w9,\ w8,\ \#0,\ \#4}, and
\texttt{IsFormatDMAConvertibleToFP16} and
\texttt{L2Allocate::ConvertFmt} make the fixed-point-versus-fp16 choice,
keyed to the format rather than to the chip. On the M1 the compiler
sends that copy through a fixed-point storage format with four
fractional bits, an implied multiply by sixteen. The stored value is the
source times sixteen and the storage clamps at the fp16 maximum, so any
source magnitude above the threshold overflows to \(\pm\infty\). The
threshold is exact:

\[V_{\max} = \frac{65504}{16} = 4094\]

\texttt{ZinMirL2Config::CalculateDMASrcBufferSizeAndStrides} reads the
width granule and clamps, taking a \texttt{ZinIrHalParameters\ const\&}
and reading \texttt{HAL+0x1c0} as both a divisor that recovers an
element count and a multiplier on the final byte stride. It also reads
the patch-width clamps at \texttt{HAL+0x3f8}, \texttt{HAL+0x400}, and
\texttt{HAL+0x410} through \texttt{ComputeMaxPatchWidth}, which returns
a \texttt{CeilLog2(width)} bounded by those clamps. A sixteen-wide
granule gives a log-base-two of four, the shift-left-by-four that is the
multiply by sixteen. On the M5 the route avoids the fixed-point format
and the slice stays in plain fp16, so the same slice that saturates on
the M1 is unaffected on the M5. The lowering pattern is the same
template on both chips; the route and format selection at the descriptor
and hardware-abstraction layer differs, driven by the patch-width clamps
and the width granule in the per-chip blob. The saturating path is
present on every patch-capable descriptor generation including the A17
generation. The M1-versus-M5 difference is thus the route and format
selection upstream. Any height-width-axis slice with a nonzero offset
whose fp16-convertible source can exceed 4094 is thus hazardous on the
whole patch-capable generation set, unless the route is known to avoid
crop mode.

The remaining four cases are decompositions gated by the per-chip blob
rather than the family enum. Resize is family-uniform at the conversion
level and diverges in \texttt{ZinResizeLayerUtils}:
\texttt{DecomposeResize} branches on the texture-engine-present flag at
\texttt{HAL{[}0x81d{]}}, and if it is absent and the no-texture
decomposition fails the compiler asserts ``failed to map resize layer on
this arch''. The no-texture path picks a deconvolution upsample against
a transpose-plus-convolution path through
\texttt{QualifiesforUpsampleDecomposition}, gated on the one-by-one
fast-path flag at \texttt{HAL{[}0x812{]}}, and
\texttt{GetMaxSmallKernelWidth}; the three strategies round differently,
so resize results differ by chip. Matmul lowers to a resident
convolution through \texttt{LowerNEMatMulToNEConv} when the right-hand
weight bytes meet the per-family copy-cast threshold at
\texttt{HAL{[}0x1b8{]}}, and \texttt{SliceMatMulAlongHeight} splits the
operation by the per-family maximum tensor size. Transpose runs natively
only when \texttt{IsValidNETransposeConfiguration} passes the maximum
height and width at \texttt{HAL+0x468} and \texttt{HAL+0x470} plus the
divisibility checks, and otherwise decomposes into multiple passes on
the more constrained families. A convolution with a large stride
factorizes the stride through \texttt{DecomposeConvWithLargeStride}
against the per-family allowed-factor list at \texttt{HAL{[}0x730{]}}
through \texttt{HAL{[}0x750{]}}, at most two factors per axis, so one
convolution becomes a different sub-convolution sequence per chip.

\section{What the M1 blocks and which chip enables
it}\label{what-the-m1-blocks-and-which-chip-enables-it}

The M1 blocks a specific set of operations and behaviors, each enabled
by a particular later family (\cref{tbl:c35-blocks}).



Some operations have no native form on any family, including
\texttt{tan}, \texttt{asin}, \texttt{acos}, \texttt{atan2},
\texttt{sinh}, \texttt{cosh}, \texttt{atanh}, \texttt{asinh},
\texttt{acosh}, the logical and-or-xor, the recurrent cells, scatter,
one-hot, non-zero, \texttt{mod}, band-part, and reverse-sequence. These
are decomposed on the host or in the graph on every chip and are not a
per-family difference. The slice saturation is the only axis that turns
a finite value into an infinity across chips; the wide accumulator and
the reduction route are uniform across families and are not a block.

\section{Kernel-data path: per-core replication against shared kernel
memory}\label{kernel-data-path-per-core-replication-against-shared-kernel-memory}

The weight layout splits into two code-generation functions that the
kernel-memory register at \texttt{HAL{[}0x48f{]}} selects.
\texttt{ZinMirBuildNEKernelData} is the per-core replicated path: on the
M1 the kernel coefficients are built per Neural-Engine core, walked
through the per-core byte stride that \texttt{CalculateNEOffsetJump}
computes, and must fit the 64 KB kernel-memory cap.
\texttt{ZinMirBuildNEKernelDataSharedKmem} is the A14-and-above shared
path, looped over core groups, selected by the kernel-memory
register-select byte and the \texttt{ExceedKmemSizeLimit} logic that
\cref{lst:c35-kmem} gives.

\begin{listing}[H]
\caption{The kernel-memory offset selection between the shared 16 MB path and the 64 KB per-core path.}\label{lst:c35-kmem}

\begin{Shaded}
\begin{Highlighting}[]
\NormalTok{off }\OperatorTok{=} \OperatorTok{(}\NormalTok{useShared }\OperatorTok{\&\&}\NormalTok{ HAL}\OperatorTok{[}\BaseNTok{0x48f}\OperatorTok{])} \OperatorTok{?} \BaseNTok{0x210} \OperatorTok{:} \BaseNTok{0x200}
\end{Highlighting}
\end{Shaded}

\end{listing}

On the M1 the register-select byte at \texttt{HAL{[}0x48f{]}} is set,
but the extended-kernel-memory-mode scalar at \texttt{0x288} reads zero
on every target in this compiler. The shared 16 MB path at offset
\texttt{0x210} is thus taken only when an operation explicitly requests
shared and the chip enables it. Otherwise the M1 falls to the 64 KB cap
at offset \texttt{0x200} and tiles. The A14 and later chips route the
large kernels through the shared builder instead of replicating per
core, which is the concrete kernel-memory architecture difference
between the M1 and the later families.

\section{Task-descriptor layer: per-generation field
offsets}\label{task-descriptor-layer-per-generation-field-offsets}

Below the family enum is the descriptor emitter
\texttt{ZinAneTd\textless{}Nu\textgreater{}}, instantiated for the
generations \(\{1, 4, 5, 6, 7, 8, 10, 11, 17, 19, 20\}\). Only the
generations \(\{7, 8, 10, 11, 17, 19, 20\}\) have the patch-width path;
the legacy generations \(\{1, 4, 5, 6\}\) use \texttt{HandleCcdmaLayer}
with no patch settings. The same logical field is at a different byte
offset per generation, so a field written at the wrong offset for the
target corrupts an unrelated descriptor word. The patch-width offset
moves with the generation (\cref{tbl:c35-descoffset}).

\begin{longtable}[]{@{}lll@{}}
\caption{The patch-width descriptor offset per task-descriptor
generation.}\label{tbl:c35-descoffset}\tabularnewline
\toprule\noalign{}
\textbf{Descriptor generation} & \textbf{Family} & \textbf{Patch-width
offset} \\
\midrule\noalign{}
\endfirsthead
\toprule\noalign{}
\textbf{Descriptor generation} & \textbf{Family} & \textbf{Patch-width
offset} \\
\midrule\noalign{}
\endhead
\bottomrule\noalign{}
\endlastfoot
7 & A13, M1 & \texttt{desc+0x200} \\
\rowcolor{zebra}8 & A13, A14 & \texttt{desc+0x224} \\
10 & A14 & \texttt{desc+0x12c} \\
\rowcolor{zebra}11 & A14, A15 & \texttt{desc+0x21c} \\
17 & A15, A16 & \texttt{desc+0x23c} \\
\rowcolor{zebra}19 & A16, A17 & \texttt{desc+0x248} \\
20 & A17, M5 & \texttt{desc+0x26c} \\
\end{longtable}

The primary and secondary source direct-memory-access fields and the
result direct-memory-access field follow the same per-generation-offset
pattern. A single compiler defect is in this layer.
\texttt{SetPatchSettings\textless{}7u\textgreater{}} targets
\texttt{ZinAneTd\textless{}7u\textgreater{}} on every setter except the
last, which calls
\texttt{ZinAneTd\textless{}8u\textgreater{}::SetTileOverlapPadReflect}
on a generation-7 pointer, writing the reflect flag at the generation-8
offset; a shipping target that uses generation 7 locates its
tile-overlap-pad-reflect bit in the wrong word.

\section{Hardware-abstraction parameter field
map}\label{hardware-abstraction-parameter-field-map}

\texttt{ZinIrHalParameters} is a 0x348-byte per-chip blob, copied by a
0x348-byte copy constructor. \Cref{tbl:c35-halmap} gives the fields the
shared lowering reads to drive the per-chip route, format, and limit
selection.

\begin{longtable}[]{@{}
  >{\raggedright\arraybackslash}p{(\linewidth - 2\tabcolsep) * \real{0.5000}}
  >{\raggedright\arraybackslash}p{(\linewidth - 2\tabcolsep) * \real{0.5000}}@{}}
\caption{The fields of the 0x348-byte hardware-abstraction parameter
blob and the route each drives.}\label{tbl:c35-halmap}\tabularnewline
\toprule\noalign{}
\begin{minipage}[b]{\linewidth}\raggedright
\textbf{Offset}
\end{minipage} & \begin{minipage}[b]{\linewidth}\raggedright
\textbf{Meaning}
\end{minipage} \\
\midrule\noalign{}
\endfirsthead
\toprule\noalign{}
\begin{minipage}[b]{\linewidth}\raggedright
\textbf{Offset}
\end{minipage} & \begin{minipage}[b]{\linewidth}\raggedright
\textbf{Meaning}
\end{minipage} \\
\midrule\noalign{}
\endhead
\bottomrule\noalign{}
\endlastfoot
\texttt{0x1b8} & matmul right-hand copy-cast byte threshold \\
\rowcolor{zebra}\texttt{0x1c0} & large-stride-conv enable and
direct-memory-access width granule \\
\texttt{0x328} & conv stride and unicast input-channel limit \\
\rowcolor{zebra}\texttt{0x30}, \texttt{0x38}, \texttt{0x48},
\texttt{0x50} & maximum small and large kernel width by tensor format \\
\texttt{0x88}, \texttt{0x90}, \texttt{0x98}, \texttt{0x138} through
\texttt{0x1a8} & maximum spatial and per-dimension tensor sizes \\
\rowcolor{zebra}\texttt{0x468}, \texttt{0x470} & maximum transpose
height and width \\
\texttt{0x730} through \texttt{0x750} & allowed kernel-width and stride
factor list \\
\rowcolor{zebra}\texttt{0x812} & one-by-one input upsample fast-path
flag \\
\texttt{0x81d} & resize and texture-engine-present flag \\
\rowcolor{zebra}\texttt{0x3f8}, \texttt{0x400}, \texttt{0x410} &
patch-width clamps for max-pool, floor, and non-pool \\
\texttt{0x228}, \texttt{0x238}, \texttt{0x240}, \texttt{0x248} &
cost-model frequency, cycles, and core count \\
\end{longtable}

The concrete numeric values at these offsets are in a constant data
section behind a packed selector the target constructors pass:
\texttt{H13} and \texttt{H17s} pass the identical selector, so both
share a base hardware-abstraction descriptor and diverge only in the
literals. The per-target literals are deferred; the offsets, consumers,
and mechanism are pinned here.

\section{Per-family cost model}\label{per-family-cost-model}

\texttt{getDeviceInfo} fills a device-information struct from the family
enum and the operation size. It writes a constant \texttt{0.0008} at
offset \texttt{0x30}, a size-scaled pair at offsets \texttt{0x14} and
\texttt{0x18}, a bandwidth clamp at offset \texttt{0x20}, and a
setup-latency and throughput pair at offsets \texttt{0x8} and
\texttt{0xc} from a data block. The dispatch reads family 5 and 6 as the
A16 and A17 tier, family 4 as the M1, and family 3 as A12, and buckets
the size at thresholds of 6, 10, 20, and 40. The bandwidth-clamp
sequence and the setup ramp both key off the family
(\cref{tbl:c35-costmodel}).

\begin{longtable}[]{@{}lll@{}}
\caption{The bandwidth-clamp sequence and setup ramp per family in the
analytic cost model.}\label{tbl:c35-costmodel}\tabularnewline
\toprule\noalign{}
\textbf{Family} & \textbf{Bandwidth clamp sequence by size bucket} &
\textbf{Setup ramp} \\
\midrule\noalign{}
\endfirsthead
\toprule\noalign{}
\textbf{Family} & \textbf{Bandwidth clamp sequence by size bucket} &
\textbf{Setup ramp} \\
\midrule\noalign{}
\endhead
\bottomrule\noalign{}
\endlastfoot
A12 & 50, 100, 200, 400, 800 & 1.84 to 14.72 \\
\rowcolor{zebra}A13, M1 & 50, 100, 200, 400, 800 & 1.84 to 7.36 \\
A16, A17, M5 & 62.5, 125, 250, 500 & 1.84 to 7.36 \\
\rowcolor{zebra}default (A11, A14, A15) & 34.1, 68.2, and on & 1.84 to
13.57 \\
\end{longtable}

The M5 is modeled with a higher per-bucket throughput but a clamp
ceiling of 500 against the M1 and A12 ceiling of 800.

\section{A15 codegen branch}\label{a15-codegen-branch}

The A15 family is its own code-generation tier, not a relabel of A14,
and the difference is visible in three independent locations in the
compiler. The following is decode-derived from the static compiler image
and is not measured on A15 silicon.

\texttt{getDeviceInfo} has a dedicated \texttt{anec::A15} branch. The
dispatch range-tests the family value and special-cases \texttt{A14} and
\texttt{A15} separately, so the \texttt{A15} arm fills the
device-information struct from its own cost-table block rather than
falling to the A14 arm or the shared default. A15 thus has distinct
cost-model constants, which is the data-side proof that the family is a
distinct capability tier.

The operation legality matches. Forty-five
\texttt{MinimumFamily\textless{}N\textgreater{}} template instantiations
have the F4 floor, the A15-and-above floor, against twenty-nine at the
A14 floor, so A15 adds a concrete operation set over A14. The distinct
compute operations at that floor are \texttt{sin}, \texttt{cos}, and the
global arg-reductions.

Beyond the H15 constructors in the chapter
\hyperref[cross-silicon-targets]{34} census, the compiler carries
recognized target strings for a wider H15 die family, \texttt{h15g},
\texttt{h15m}, \texttt{h15p}, \texttt{h15s}, \texttt{h15c}, and
\texttt{h15d}. The \texttt{h15m} suffix is an extra M-class target that
the h13 and h14 families do not list. A15 is also the only tier whose
per-chip interchange-format table accepts YUV420 image input, a 4CC
route that the surrounding families drop. The A15 cost-table constants
and the YUV420 acceptance are decoded from the compiler; the numeric
behavior of the A15 operation set, such as any accumulator or rounding
change against A13 and A14, is an A15-silicon measurement that this
guide does not have.

\section{Named validity gate}\label{named-validity-gate}

Only the M1 family has a named per-family validity gate.
\texttt{IsValidForH13} is the gather gate: it asserts the gather-axes
size is 3, the data batch and depth are 1, the index channel is 3 with
width and depth 1, plus an axis-pattern check.
\texttt{IsValidForH13GatherNCH} enforces the channel-height-width-layout
gather restriction. No equivalent named gate exists for any other
family; the rest gate numerically through the hardware-abstraction
clamps, and the compression gate \texttt{IsValidForCompression} is
hardware-abstraction-driven with the universal rule that only kernels of
at least eight bits are compressible.

\section{Cross-compiled across the
family}\label{cross-compiled-across-the-family}

The per-generation versioning that chapter
\hyperref[program-and-container-format]{23} reads out of the descriptor
accessors can be observed directly in the emitted program. Because the
compiler is one binary that lowers for any target from any machine, the
target is a command-line flag: a single \texttt{ANECompile} build, here
\texttt{9.509.0}, lowers the same network for \texttt{h13} through
\texttt{h17} while only the \texttt{-t} target varies. A diff across
those programs isolates the per-generation divergence from the compiler
version and the operating-system version, which a diff across physical
machines confounds. Lowering one convolution and one matmul for the five
base M-series targets, \texttt{h13} through \texttt{h17}, gives the
comparison in \cref{tbl:c35-crossgen}.



The cpusubtype at header offset \texttt{0x08}, the codegen revision
chapter \hyperref[program-and-container-format]{23} reads as
\texttt{0x4} on the M1, steps once per generation across the family:
\texttt{0x04}, \texttt{0x05}, \texttt{0x06}, \texttt{0x07}, and
\texttt{0x09}, the M5 base skipping \texttt{0x08}, with the A19 target
\texttt{h18} continuing to \texttt{0x0a}. The convolution op-config word
that chapter \hyperref[program-and-container-format]{23} decodes as the
version-7 value \texttt{0x5042a063} holds on the M1 and becomes
\texttt{0x5042a0c3} on the M2 and every later base target. The high
half-word \texttt{0x5042} is stable, as chapter
\hyperref[program-and-container-format]{23} states; it is the low byte
that carries a generation component, \texttt{0x63} on the version-7
image and \texttt{0xc3} from the A14 image on. The version-7 op-config
words in the \texttt{0x5000a0..} range, \texttt{0x5000a021} and
\texttt{0x5000a421}, appear only in the M1 image and are absent from the
M2 image onward, the byte-level form of the version bump chapter
\hyperref[program-and-container-format]{23} describes.

What does not move is the addressing aperture and the descriptor
prologue. The aperture virtual addresses, input \texttt{0x30004000} and
the \texttt{0x30008000}, \texttt{0x3000c000}, and \texttt{0x30010000}
that follow, are byte-identical across all five targets, as is the
descriptor header through its first \texttt{0x300} bytes. What moves is
the register program below them: the relocation records that carry the
aperture base addresses are packed at an 8-byte stride on the M1 and
move to a 16-byte stride from the M2 on, the concrete instance of the
rule that the byte offsets and field widths move between versions
(chapter 23) and of the per-generation descriptor-offset
\cref{tbl:c35-descoffset}. Both operations show the same stride shift at
the same boundary.

The largest single step is the M1 to the M2, the version-7 descriptor to
the A14 descriptor. The op-config word, the relocation stride, and the
version-7 config words all change there, while the M2 through the M5
share the op-config word and the stride and differ mainly in the
cpusubtype stamp and the kernel-section size, the convolution image
growing from 64 KB on the M1 through the M3 to 96 KB on the M4 and M5.
This is the program-format counterpart of chapter
\hyperref[cross-silicon-targets]{34}'s observation that the operation
surface stops expanding at A15 while the upper tiers add cores: the
format is largely settled after A14, and the later changes are
incremental. This comparison is the compiler's emitted program for each
target, read from the cross-compiled container, not an on-silicon run on
the M2, M3, and M4, which this guide does not have; it confirms the
decompile-derived per-generation versioning of chapter
\hyperref[program-and-container-format]{23} against the actual byte
stream, with the M1 and M5 endpoints also confirmed on silicon.

\section{Apple documents}\label{apple-documents}

Per-family code generation has no public counterpart, and this chapter
reports it as reverse-engineered across the target set. The public
conversion tools document the frontend operation set and the
palettization, quantization, and pruning passes that produce a model,
and the compute-unit selector that requests the engine
\hyperref[ref-applecoremltools]{{[}AppleCoreMLTools{]}}. They do not
document the family enum, minimum-family operation trait, per-chip
hardware-abstraction parameters, or route and format selection that
makes one slice saturate on the M1 and run clean on the M5.

\chapter{Predicted upper tier}\label{predicted-upper-tier}

\begin{summarybox}

Two datapaths are above the M5 in the compiler, and no part measured
here can run them. They are an fp8 weight and activation format, and a
multi-die collective communication layer. Both are present as decoded
structure, gated off on every family this guide reached, and stated as
predicted, not measured. The fp8 gate is the capability byte at offset
\texttt{0x52d}, set on H18 alone. The collective-enable byte at offset
\texttt{0x48b} reads zero on all 28 targets.

\end{summarybox}

\section{64-core ceiling}\label{core-ceiling}

The largest die this guide measured is the 16-core H17s, and the line
ends at the 64-core H17d that no part here could run. The decode shows
that nothing extra is family-gated for the 64-core case beyond the count
itself, and the following is a compiler decode, with no 64-core part on
hand to measure. Core count arrives as a runtime parameter, the
\texttt{ne\_core\_count} value sourced from the hardware-abstraction
blob, rather than as a hard-coded per-family constant.
\texttt{getDeviceInfo} takes that count as its second argument and
buckets it against the thresholds 7, 10, 20, and 40, which straddle the
four die sizes 8, 16, 32, and 64 held by the \texttt{g}, \texttt{s},
\texttt{c}, and \texttt{d} target suffixes. Several device-information
fields then scale linearly with the count, and the top bucket reached on
the A14-and-above branch stores the 800-unit peak. The compiler thus
emits a 64-core-scaled cost model for H17d purely from the count of 64,
and the geometry is decoded while the realized 64-core throughput stays
an on-silicon measurement.

The two datapaths above the M5 are the fp8 weight format and the
multi-die collective layer (\cref{tbl:c36-overview}).

\begin{longtable}[]{@{}
  >{\raggedright\arraybackslash}p{(\linewidth - 6\tabcolsep) * \real{0.2500}}
  >{\raggedright\arraybackslash}p{(\linewidth - 6\tabcolsep) * \real{0.2500}}
  >{\raggedright\arraybackslash}p{(\linewidth - 6\tabcolsep) * \real{0.2500}}
  >{\raggedright\arraybackslash}p{(\linewidth - 6\tabcolsep) * \real{0.2500}}@{}}
\caption{The decoded fields of the two fp8 formats and the collective
layer, with their gates and their state on the
M5.}\label{tbl:c36-overview}\tabularnewline
\toprule\noalign{}
\begin{minipage}[b]{\linewidth}\raggedright
\textbf{field}
\end{minipage} & \begin{minipage}[b]{\linewidth}\raggedright
\textbf{E4M3}
\end{minipage} & \begin{minipage}[b]{\linewidth}\raggedright
\textbf{E5M2}
\end{minipage} & \begin{minipage}[b]{\linewidth}\raggedright
\textbf{collective layer}
\end{minipage} \\
\midrule\noalign{}
\endfirsthead
\toprule\noalign{}
\begin{minipage}[b]{\linewidth}\raggedright
\textbf{field}
\end{minipage} & \begin{minipage}[b]{\linewidth}\raggedright
\textbf{E4M3}
\end{minipage} & \begin{minipage}[b]{\linewidth}\raggedright
\textbf{E5M2}
\end{minipage} & \begin{minipage}[b]{\linewidth}\raggedright
\textbf{collective layer}
\end{minipage} \\
\midrule\noalign{}
\endhead
\bottomrule\noalign{}
\endlastfoot
sign bits & 1 & 1 & not applicable \\
\rowcolor{zebra}exponent bits & 4 & 5 & not applicable \\
mantissa bits & 3 & 2 & not applicable \\
\rowcolor{zebra}exponent bias & 7 & 15 & not applicable \\
max finite magnitude & 448 & 57344 & not applicable \\
\rowcolor{zebra}role & gated weight and activation & fp16 conversion &
mesh reduce, gather, slice \\
family gate & \texttt{0x52d} set on H18 only & folds to fp16 broadly &
\texttt{0x48b} zero on all 28 \\
\rowcolor{zebra}register encoding & present from A17 encoder & output
slot on modern encoders & every setter a stub \\
state on M5 & off & conversion available & inert \\
\end{longtable}

\section{fp8 datapath}\label{fp8-datapath-1}

The compiler has two eight-bit floating-point formats, and they are not
symmetric in role. They live in the internal \texttt{ZinTensorFormat}
code-generation and direct-memory-access enumeration, which is separate
from the serialization data-type enumeration that the operation
attributes use. The serialization enumeration runs codes 0 through 10
with fp16 at 3, fp32 at 4, and int8 at 2, and has no fp8 codepoint at
all; the fp8 formats are recovered from the byte-size map
\texttt{\_ZinTensorFormatGetSizeInBytes} (\cref{tbl:c36-tensorfmt}).

\begin{longtable}[]{@{}lll@{}}
\caption{The fp8 codepoints in the internal tensor-format enumeration,
with the integer and half-precision neighbors that bound
them.}\label{tbl:c36-tensorfmt}\tabularnewline
\toprule\noalign{}
\textbf{\texttt{ZinTensorFormat} code} & \textbf{Format} &
\textbf{Bytes} \\
\midrule\noalign{}
\endfirsthead
\toprule\noalign{}
\textbf{\texttt{ZinTensorFormat} code} & \textbf{Format} &
\textbf{Bytes} \\
\midrule\noalign{}
\endhead
\bottomrule\noalign{}
\endlastfoot
1 & int8 & 1 \\
\rowcolor{zebra}2 & uint8 & 1 \\
3 & fp16 & 2 \\
\rowcolor{zebra}\texttt{0xb} (11) & fp32 & 4 \\
\texttt{0xc} (12) & E4M3 & 1 \\
\rowcolor{zebra}\texttt{0xd} (13) & E5M2 & 1 \\
\texttt{0xe} (14) & two-byte live-input format & 2 \\
\rowcolor{zebra}\texttt{0x10} (16) & int4 & under 1 \\
\end{longtable}

The compiler has the full set of fp8 variant types as C++ symbols,
\texttt{e4m3\_t} with 1273 references and \texttt{e5m2\_t} with 176,
alongside the finite-and-NaN and unsigned-zero variant type classes. The
shipping format is the E4M3FN variant, proven by the \(\pm 448\)
saturation bound that
\texttt{ZinPadLayer::ValidateBackgroundPaddingValue} enforces with the
string ``in {[}\%u, \%u{]} for e4m3 format'' and the NaN-only special
case in \texttt{ZinE4M3Expand}.

E4M3 is the gated weight and activation format. A value decodes from a
sign bit, four-bit exponent, and three-bit mantissa as

\[x = (-1)^s \, 2^{e - 7} \left(1 + \frac{m}{2^3}\right)\]

for a normal value, with an exponent bias of \(7\) and a maximum finite
magnitude of \(448\). The shipping variant is E4M3FN, the finite-and-NaN
form: it has no infinity, the all-ones exponent with a full mantissa
encodes NaN, and a value past \(\pm 448\) either maps to NaN or clamps
to the maximum normal. A one-bit mode held per direct-memory-access
source controls the overflow behavior, \(0\) for NaN and \(1\) for
saturate, so a narrowing cast picks the bit pattern \(\mathtt{0x7e}\)
for the saturated maximum or \(\mathtt{0x7f}\) for NaN. The narrower
exponent and the \(\pm 448\) ceiling cap the representable range below
half precision.

The other format, E5M2, serves conversion. It decodes as

\[x = (-1)^s \, 2^{e - 15} \left(1 + \frac{m}{2^2}\right)\]

with five exponent bits, a two-bit mantissa, and the same exponent bias
of \(15\) as half precision. That shared exponent is why E5M2 is the
high byte of an fp16 value: the conversion to fp16 is a left shift of
eight bits, and the inverse is the top byte with round-to-nearest and an
infinity clamp. E5M2 folds to fp16 by a direct-memory-access conversion,
so it is broadly available wherever that conversion path exists, and the
upscaler layer accepts it as input. The fold asymmetry is exact in the
conversion functions. \texttt{ZinE5M2ToF16} is a left shift of eight
bits, the literal high byte, and \texttt{ZinF16ToE5M2} is the top byte
with round-to-nearest and an infinity clamp at \texttt{0x7c00}. E4M3 is
not so trivial. \texttt{ZinE4M3Expand} is a bit-surgery decode with sign
at bit 7, the four-bit exponent at bits 6 through 3, and the three-bit
mantissa at bits 2 through 0, handling the subnormal and NaN encodings.
\texttt{ZinF32ToE4M3} narrows and selects the overflow byte,
\texttt{0x7e} for the saturated maximum and \texttt{0x7f} for NaN. The
fold predicate \texttt{IsFormatDMAConvertibleToFP16} returns
\texttt{(fmt\ \textless{}\ 0xe)\ \&\ (0x2ff0\ \textgreater{}\textgreater{}\ fmt)},
and the mask \texttt{0x2ff0} includes code 13 and excludes code 12. E5M2
thus folds by direct-memory-access conversion and E4M3 cannot, which is
the binary-level reason E4M3 requires the native datapath that is gated.

The family gate is a single hardware-abstraction-layer capability byte
at offset \texttt{0x52d}, the E4M3 direct-memory-access and
kernel-format capability. It is clear on the M1 generation, clear on the
M5 generation, and clear on the intermediate families, and it reads set
on the H18 family alone of the twenty-eight compiler targets. The master
per-format direct-memory-access validity function
\texttt{CheckValidDMAFormat} takes the four capability bytes
\texttt{HAL{[}0x52c{]}}, \texttt{HAL{[}0x52d{]}},
\texttt{HAL{[}0x52e{]}}, and \texttt{HAL{[}0x685{]}}. It validates fp32
against \texttt{0x52c}, E4M3 against \texttt{0x52d}, E5M2 against
\texttt{0x52e}, and the two-byte live formats against \texttt{0x685},
with every code below \texttt{0xb} always valid. Three
operation-semantics sites re-check the byte and abort with ``E4M3 is not
supported on this architecture'': the dequantize validator at line
129078, the quantize validator at line 227466, and a third semantics
validator at line 3407368. Below that runtime gate is a harder
compile-time limit on the M1 generation. The M1 task-descriptor encoder
packs the source element format into a two-bit field, which stores only
the integer and half-precision codes and has no bit space for the E4M3
codepoint. Feeding fp8 to the M1 encoder thus aborts the compile rather
than refusing the operation at runtime. The encoder that does have the
E4M3 codepoint widens that field to three bits and adds a distinct
output slot, and that wider encoder is present from the A17 generation
onward even though the runtime capability at \texttt{0x52d} fires only
on H18.

The operation-side gate matches the encoder, decoded from the compiler
and not measured on an H18 part. The
\texttt{MinimumFamily\textless{}N\textgreater{}} trait that chapter
\hyperref[per-family-code-generation]{35} reads as the
operation-legality floor has a high-N set, a few operations at N=5, N=6,
and N=7, the A16, A17, and A18 floors, and the fp8-bearing operations
are in that set. The fp8 converters and the quant-unit storage are
compiled into the one byte-identical image, so an fp8 operation parses
and type-checks on any target. Native execution depends on both that
high-N family floor and the \texttt{0x52d} capability byte, which is why
the datapath is present yet inert below H18.

The format-register delta between the M1 and the H18 encoder is exact.
The M1 generation builds the generation-5 and generation-7 descriptors,
whose \texttt{SetCommonInFmt} writes a two-bit field at
\texttt{this+0x48} holding only int8, uint8, and fp16, and the code
\texttt{0xc} aborts with ``Error: Invalid Common InFmt E4M3''. The
generation-20 descriptor that the A17 generation builds packs all three
format fields into a 32-bit word at \texttt{this+0x228}, widening each
lane to three bits. \texttt{SetCommonInFmt} gives E4M3 the source-one
code 4 at bits 2 through 0, \texttt{SetCommonSrc2InFmt} gives it the
source-two code \texttt{0x20} at bits 5 through 3, and
\texttt{SetCommonOutFmt} gives it a distinct output slot \texttt{0x100}
at bits 8 through 6. \Cref{tbl:c36-fmtreg} gives the task-descriptor
format-register delta.

\begin{longtable}[]{@{}
  >{\raggedright\arraybackslash}p{(\linewidth - 4\tabcolsep) * \real{0.3333}}
  >{\raggedright\arraybackslash}p{(\linewidth - 4\tabcolsep) * \real{0.3333}}
  >{\raggedright\arraybackslash}p{(\linewidth - 4\tabcolsep) * \real{0.3333}}@{}}
\caption{The task-descriptor format-register delta across the upper-tier
generations.}\label{tbl:c36-fmtreg}\tabularnewline
\toprule\noalign{}
\begin{minipage}[b]{\linewidth}\raggedright
\textbf{Field}
\end{minipage} & \begin{minipage}[b]{\linewidth}\raggedright
\textbf{M1 generation-5 at \texttt{this+0x48}}
\end{minipage} & \begin{minipage}[b]{\linewidth}\raggedright
\textbf{H18 generation-20 at \texttt{this+0x228}}
\end{minipage} \\
\midrule\noalign{}
\endfirsthead
\toprule\noalign{}
\begin{minipage}[b]{\linewidth}\raggedright
\textbf{Field}
\end{minipage} & \begin{minipage}[b]{\linewidth}\raggedright
\textbf{M1 generation-5 at \texttt{this+0x48}}
\end{minipage} & \begin{minipage}[b]{\linewidth}\raggedright
\textbf{H18 generation-20 at \texttt{this+0x228}}
\end{minipage} \\
\midrule\noalign{}
\endhead
\bottomrule\noalign{}
\endlastfoot
source-one format & two-bit lane, int8, uint8, fp16 only & three-bit
lane, plus E4M3 as 4 \\
\rowcolor{zebra}source-two format & folded into the two-bit space, no
distinct E4M3 code & three-bit lane, E4M3 as \texttt{0x20} \\
output format & two-bit lane, E5M2 reuses the fp16 slot \texttt{0x20} &
three-bit lane, E4M3 as \texttt{0x100} \\
\rowcolor{zebra}E4M3 codepoint & absent, a compile-time assert & present
as 4, \texttt{0x20}, \texttt{0x100} \\
E4M3 overflow register & stub assert on the generation-8 setter &
\texttt{this+0x2fc} and \texttt{this+0x300}, bit 24 \\
\end{longtable}

The overflow mode is a per-source register field on the generation-20
encoder, \texttt{SetTileDmaSrc1E4M3Overflow} writing bit 24 of
\texttt{this+0x2fc} and \texttt{SetTileDmaSrc2E4M3Overflow} writing bit
24 of \texttt{this+0x300}, where 1 selects saturate and 0 selects NaN.
On the older encoders the same setter is a stub that asserts
``E4M3Overflow is not supported'' only when the overflow option is
engaged.

In the multiply array fp8 is an input width, not an accumulator width.
An E4M3 weight expands to fp16 going into the multiply, the products
accumulate in the same wide register every family uses, and the output
port rounds the result to fp16. No fp8 accumulator type, register field,
or symbol exists in the compiler: the generation-20 format word encodes
only the source-one, source-two, and output element formats, with no
accumulator-format field. E4M3 is a native kernel format, code 6 in the
kernel-format set that \texttt{ZinSetFormat} admits through the mask
\texttt{0x3b}, alongside int8, uint8, fp16, and fp32; E5M2 is absent
from that set, so it is a conversion and upscaler-input format rather
than a kernel format. Throughput runs on the same double-rate path int8
uses, gated by
\texttt{ZinDoubleMacMode::CanUseDoubleMacModeBasedOnFormats}. A one-byte
activation against a one-byte kernel of the same numeric class is
eligible for the double-multiply mode, the eligibility turning on the
exclusive-or of the activation-float and kernel-float bits being zero,
which holds only when the two bits agree. An E4M3 activation against an
E4M3 kernel is both one-byte and both float, so it runs at twice the
per-element rate into the fp16 accumulator, while a mixed
int8-against-E4M3 pair is not eligible. E4M3 is symmetric-only, with the
zero point forced to zero, the same constraint the M1 imposes on int8.
The quantize and dequantize validators reject a zero point with ``Zero
point is not supported for quant with E4M3 output format'', and the same
rule reaches the palette layer. E4M3 weights stream as a palette through
\texttt{ZinIrWeight::DePalettizeWeightData\textless{}e4m3\_t\textgreater{}},
a codebook of E4M3 values indexed by a packed stream at bit-widths 1, 2,
3, 4, 6, and 8. This is the identical machinery the int4, int8, and fp16
palettes use under template specialization, dequantized as scale times
E4M3 with no zero point.

\section{Multi-die collective layer}\label{multi-die-collective-layer}

Each collective operation in the multi-die dialect has a backend
operation class and a unit-type code (\cref{tbl:c36-collops}).

\begin{longtable}[]{@{}
  >{\raggedright\arraybackslash}p{(\linewidth - 6\tabcolsep) * \real{0.2500}}
  >{\raggedright\arraybackslash}p{(\linewidth - 6\tabcolsep) * \real{0.2500}}
  >{\raggedright\arraybackslash}p{(\linewidth - 6\tabcolsep) * \real{0.2500}}
  >{\raggedright\arraybackslash}p{(\linewidth - 6\tabcolsep) * \real{0.2500}}@{}}
\caption{The collective operations of the multi-die dialect, their
backend operation classes, and the unit-type code of
each.}\label{tbl:c36-collops}\tabularnewline
\toprule\noalign{}
\begin{minipage}[b]{\linewidth}\raggedright
\textbf{silc mnemonic}
\end{minipage} & \begin{minipage}[b]{\linewidth}\raggedright
\textbf{C++ operation class}
\end{minipage} & \begin{minipage}[b]{\linewidth}\raggedright
\textbf{role}
\end{minipage} & \begin{minipage}[b]{\linewidth}\raggedright
\textbf{unit-type code}
\end{minipage} \\
\midrule\noalign{}
\endfirsthead
\toprule\noalign{}
\begin{minipage}[b]{\linewidth}\raggedright
\textbf{silc mnemonic}
\end{minipage} & \begin{minipage}[b]{\linewidth}\raggedright
\textbf{C++ operation class}
\end{minipage} & \begin{minipage}[b]{\linewidth}\raggedright
\textbf{role}
\end{minipage} & \begin{minipage}[b]{\linewidth}\raggedright
\textbf{unit-type code}
\end{minipage} \\
\midrule\noalign{}
\endhead
\bottomrule\noalign{}
\endlastfoot
\texttt{silc.all\_reduce} & \texttt{SilcAllReduceOp} & reduce a tensor
in place across a mesh axis & 78 \\
\rowcolor{zebra}\texttt{silc.all\_gather} & \texttt{SilcAllGatherOp} &
concatenate shards into a replicated tensor & 76 \\
\texttt{silc.all\_slice} & \texttt{SilcAllSliceOp} & scatter a
replicated tensor into shards & 75 \\
\rowcolor{zebra}\texttt{silc.mesh} & \texttt{SilcMeshOp} & declare the
device mesh & none \\
\texttt{silc.call} & \texttt{SilcCallOp} & per-die single-program call &
79 \\
\end{longtable}

The collective layer is a cross-die data path, not the independent
per-job steering a multi-die part otherwise does. The kernel
load-balancer steers whole independent submissions to the least-busy
engine die and never exchanges tensor data between them, and the driver
supports up to four dies. The M1 and the M1 Max each register a single
engine die, so this steering engages only on a multi-die part such as
the Ultra. The compiler's collective instead splits one tensor across
dies and exchanges partials, threaded through an
\texttt{optional\textless{}ZinIrDeviceMesh\textgreater{}} that
\texttt{GetAndValidateSpmdDeviceMesh} and
\texttt{ZinParseDeviceMeshAttributes} build, so a program compiled with
a device-mesh attribute gets a real all-reduce, all-gather, or all-slice
across the mesh. The cross-die move itself is the
\texttt{HandleCcdmaLayer} direct-memory-access primitive below, and
\texttt{ValidateMeshAxesInTensorFamily} is the family gate on whether a
given die admits the mesh. This contrast is decode-derived; whether a
two-die Ultra in any reached family accepts the mesh path is a multi-die
measurement this guide does not have.

It is a distinct intermediate-language dialect, \texttt{mlir::silc},
whose closed operation set is the three collectives above over a device
mesh, plus the mesh declaration and the per-die call. The
operation-class list is exactly these, with no reduce-scatter and no
broadcast, the all-slice operation standing in for the scatter. The
collective operations have the attributes \texttt{mesh},
\texttt{mesh\_axes}, \texttt{sharding}, the \texttt{members} membership
list, and \texttt{reduce\_op}.

The reduction kind is an enumerated attribute decoded directly from the
packed-string compare in \texttt{symbolizeReductionKind}, token by token
(\cref{tbl:c36-redkind}).

\begin{longtable}[]{@{}ll@{}}
\caption{The reduction-kind enumeration the all-reduce reduce-operation
attribute holds.}\label{tbl:c36-redkind}\tabularnewline
\toprule\noalign{}
\textbf{Token} & \textbf{Integer} \\
\midrule\noalign{}
\endfirsthead
\toprule\noalign{}
\textbf{Token} & \textbf{Integer} \\
\midrule\noalign{}
\endhead
\bottomrule\noalign{}
\endlastfoot
\texttt{sum} & 1 \\
\rowcolor{zebra}\texttt{max} & 2 \\
\texttt{min} & 3 \\
\rowcolor{zebra}\texttt{product} & 4 \\
\texttt{mean} & 5 \\
\rowcolor{zebra}miss & 0, invalid \\
\end{longtable}

The mesh is an N-dimensional grid of dies by engines-per-die, a vector
of per-axis extents held in \texttt{ZinIrDeviceMesh}, which exposes the
die count, total engine count, and engines-per-die count. A device
identifier maps to a mesh coordinate and an engine index by a
mixed-radix split at the die axis, which
\texttt{ZinSPMDUtils::AneIndexFromDeviceId} computes:

\[\mathrm{ane} = \sum_{i \ge d} \mathrm{id}[i] \prod_{j > i,\, j \ge d} \mathrm{ext}[j] \; + \; \left( \sum_{i < d} \mathrm{id}[i] \prod_{j < i} \mathrm{ext}[j] \right) \cdot \mathrm{anesPerDie}\]

where \(d\) is the die-axis split index: the axes above the die axis
fold into the within-die engine offset, and the axes below fold into the
die index, multiplied by engines-per-die. The inverse decode is
\texttt{DeviceIdFromAneIndex}. A layer runs on a die through
\texttt{ZinEngineLayer::RunsOnDeviceId}. A layer not assigned to the
single-program path runs on all dies. Otherwise it emits on a die only
when the engine index for that die is a key in the layer's engine-set
map, which is how the \texttt{members} attribute becomes a per-die emit
decision.

A sharding attribute maps each tensor axis to a mesh axis, splitting
that axis into one shard per device along the axis. The split requires
even divisibility, which the strings ``Input tensor dimension must be
divisible by the number of shards along the tensor dimension'' and
``Kernel dimension must be divisible by number of shards'' enforce. The
compiler rejects sharding the same mesh dimension twice and allows a
replicated section only on a multi-die network.

The reduce-across-the-mesh operation lowers to a collective
direct-memory-access that contains a hardware atomic read-modify-write:
the reduction happens in memory as the direct-memory-access writes into
a shared buffer, which is why the operation requires in-memory-reduction
support that the single-die families lack. The reduction-to-atomic map
is decoded from the \texttt{ZinIrReductionTypeToZinAtomicOpType} switch
(\cref{tbl:c36-atomic}).

\begin{longtable}[]{@{}ll@{}}
\caption{The reduction-type-to-atomic-operation register map for the
all-reduce collective, with the reductions that have no hardware atomic
and are rejected.}\label{tbl:c36-atomic}\tabularnewline
\toprule\noalign{}
\textbf{Reduction type} & \textbf{Atomic-op register value} \\
\midrule\noalign{}
\endfirsthead
\toprule\noalign{}
\textbf{Reduction type} & \textbf{Atomic-op register value} \\
\midrule\noalign{}
\endhead
\bottomrule\noalign{}
\endlastfoot
1 & 4 \\
\rowcolor{zebra}2 & 3 \\
4 & 1 \\
\rowcolor{zebra}5 & 2 \\
8 & 5 \\
\rowcolor{zebra}9 & 6 \\
10 & 7 \\
\rowcolor{zebra}3, 6, 7, 11 & assert, no hardware atomic \\
other & 0 \\
\end{longtable}

The atomic configuration packs as
\texttt{(atomicOp\ \&\ 0xff)\ \textbar{}\ (atomicDataType\ \textless{}\textless{}\ 8)},
and the input supplied to the collective must arrive through a bypass
pass-through, asserted by ``Input to inter-die AllReduce should be
NEBypass''. The all-reduce lowering itself further asserts ``AllReduce
is currently not supported for architectures that do not support
in-memory reductions''.

\texttt{HandleCcdmaLayer\textless{}Nu\textgreater{}} programs the
collective direct-memory-access into the task descriptor, instantiated
for the generations \(\{1, 4, 5, 6, 7, 8, 10, 11, 17, 19, 20\}\). Each
opens with the collective-enable gate and the per-die predicate before
driving the setters in a fixed order. That order is the source mode, the
counter mode, the data size, five shape words, four destination strides,
the destination base address, the optional constant, four source
strides, the source base address, the wait-event address from the mesh
symbol, the counter address, the atomic data type, the atomic operation,
the counter amount, and the wait-event value. The base addresses are all
emitted through \texttt{ZinSPMDUtils::GetSymbolOffsetToBaseAddr}, the
extended-addressing path that needs the cross-die address-reach byte.

The layer is decoded but inert in the compiler this guide reads. Every
collective direct-memory-access register setter is a stub that asserts
``CCDMA is not supported for this arch'', in every task-descriptor
generation present including generation 20, the would-be Ultra path: the
body calls the full setter sequence, but the setters abort. The
collective-enable capability byte at offset \texttt{0x48b} reads zero on
all twenty-eight targets including the M-class and Ultra-reference
targets. The cross-die capability bytes sit alongside it
(\cref{tbl:c36-capbytes}).



No family in this compiler arms the register encoding that would drive
the engine, though the front end is there in full: the operation set,
mesh and sharding math, reduction-to-atomic map, and per-die dispatch.

A second-layer gate is in the source-direct-memory-access emitters,
where the single-program flag drives an assert ``This target does not
allow sharding or SPMD functions''. A third gate in the tasklet emitter
asserts ``No tasklet for given architecture'' when the per-section
tasklet bit and the single-program predicate are not both set. On the M1
all three gates fail, so the M1 is single-die.

The pieces the single-die M1 silicon does touch are the on-die ordering
bits the same machinery shares. Those bits are the layer-two barrier,
event masks, and remote-dependency bookkeeping, none of which is the
cross-die reduction itself. On the M1 generation-10 descriptor the
layer-two barrier sets bit 23 of \texttt{TD+0x134} and the forward
barrier sets bit 30. The event setter writes a 26-bit signal mask into
\texttt{TD+0x10} and a 26-bit wait mask into \texttt{TD+0x18}, and a
direct-memory-backed event is rejected with ``DRAM Events not supported
for architecture''. The distributed unit is the tasklet, the multi-die
variant of the descriptor instruction, one tasklet per participating
die.

\section{What requires newer parts}\label{what-requires-newer-parts}

The fp8 datapath requires an H18-class part to set the capability byte
and exercise the native E4M3 multiply; the collective layer requires a
multi-die part and a newer compiler that arms the collective-enable byte
and replaces the register stubs with a real encoding. Until that
hardware and that compiler are in hand, the encodings here stand as the
predicted upper tier of the line.

\anepartopen{}{Back Matter}
\anechap{}{Methodology}{The measured silicon, the tools, and what each technique can and cannot observe.}
\anechap{}{Open questions}{The findings that remain unconfirmed and what would settle each.}
\anechap{}{Statements}{Provenance, reproduction, and the work's declarations.}
\anechap{}{Revision history}{}
\anepartclose
\setcounter{figure}{0}\setcounter{table}{0}\renewcommand{\thefigure}{\arabic{figure}}\renewcommand{\thetable}{\arabic{table}}

\chapter*{Methodology}\label{methodology}
\addcontentsline{toc}{chapter}{Methodology}

\begin{summarybox}

Every finding rests on one of four techniques: a direct private-runtime
path, static decompilation of the stack, live read-only instrumentation,
and compile-and-run probing. The results come from two measured silicon
points: M1/H13 as the primary host and M5/H17s as the second. Claims are
marked as measured on a named generation, or as predicted.

\end{summarybox}

\begin{figure}[H]\centering
\includegraphics[width=4.360in]{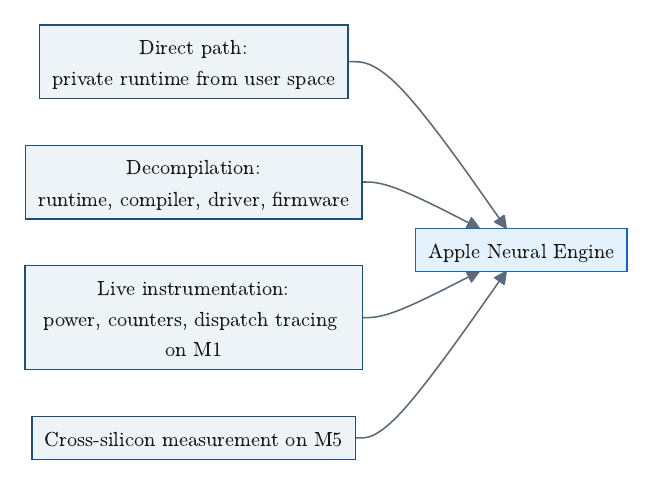}
\caption{The four lines of evidence behind the guide findings.}
\label{fig:meth-evidence}
\end{figure}

The direct private-runtime path, static decompilation, live
instrumentation, and cross-silicon measurement come at the engine from
different sides (\cref{fig:meth-evidence}). Every quantitative claim in
this guide is either measured on running silicon, read out of a binary
or table, or marked as predicted from those artifacts.

\section*{Reaching the engine}\label{reaching-the-engine}
\addcontentsline{toc}{section}{Reaching the engine}

The engine was reached without the public model framework. The same
private Espresso and dispatch runtime that the system's own dispatchers
use is callable from ordinary user space. No high-level framework is in
the path, and the operations the compiler accepts need no special
entitlement. The compiler lowers a graph to the engine's program format,
and the runtime loads it and drives it through an execution stream
directly. This route bypasses the public compute-unit selector
\hyperref[ref-applecoreml]{{[}AppleCoreML{]}}, and every measured result
in the guide was taken on it.

In-process instrumentation mapped the dispatch path: a library inserted
into the dispatching process interposed on the kernel-driver calls. The
mapping showed the dispatch as input and output surfaces passed to an
asynchronous driver selector, and it located the boundary below which
user space cannot see.

\section*{Static analysis of the
stack}\label{static-analysis-of-the-stack}
\addcontentsline{toc}{section}{Static analysis of the stack}

The structure of the stack was read out of its decompiled binaries
(\cref{tbl:meth-artifacts}).

\begin{longtable}[]{@{}
  >{\raggedright\arraybackslash}p{(\linewidth - 2\tabcolsep) * \real{0.5000}}
  >{\raggedright\arraybackslash}p{(\linewidth - 2\tabcolsep) * \real{0.5000}}@{}}
\caption{The four decompiled artifacts of the stack and what static
analysis read out of each.}\label{tbl:meth-artifacts}\tabularnewline
\toprule\noalign{}
\begin{minipage}[b]{\linewidth}\raggedright
\textbf{Artifact}
\end{minipage} & \begin{minipage}[b]{\linewidth}\raggedright
\textbf{What was read out}
\end{minipage} \\
\midrule\noalign{}
\endfirsthead
\toprule\noalign{}
\begin{minipage}[b]{\linewidth}\raggedright
\textbf{Artifact}
\end{minipage} & \begin{minipage}[b]{\linewidth}\raggedright
\textbf{What was read out}
\end{minipage} \\
\midrule\noalign{}
\endhead
\bottomrule\noalign{}
\endlastfoot
the dispatch and compiler runtime & the operation vocabulary, the
compiler passes, and the program bundle format \\
\rowcolor{zebra}the kernel driver & the user-to-kernel boundary and the
expanded program binary it lowers \\
the engine firmware & the on-engine execution model and the
host-to-firmware command protocol \\
\rowcolor{zebra}the intermediate-language operation set & the lowering
from the public operation set to the engine's own \\
\end{longtable}

Apple distributes the firmware on the M1 unencrypted, an ARM64
real-time-kernel image behind an image wrapper. Its execution loop, its
driver, and its ninety-three-command host protocol were read straight
out of that image. The compiler lowers a graph to a
ninety-seven-operation internal vocabulary that is a superset of the
public set. Static reading of those passes separated the hidden internal
operations from the user-reachable ones.

\section*{Live instrumentation}\label{live-instrumentation}
\addcontentsline{toc}{section}{Live instrumentation}

The running values that static reading cannot recover came from
instruments that escalate in privilege, each one read-only or
recoverable on a dedicated host. A user-space counter interface exposed
live hardware counters: DRAM bytes moved, energy in millijoules, and
clock frequency. Sampling those counters around a hot loop gave the
roofline directly. That fixed the M1 at about 12 fp16 TFLOP/s of
compute, about 85 GB/s of DRAM bandwidth, and near 0.5 pJ per FLOP
sustained, 0.37 at the compute optimum. It also fixed a 2 MB on-chip
working-set threshold, confirmed by the counters falling off exactly
where the working set crosses 2 MB. A signpost trace on the engine
subsystem gave the op-level event sequence and independently confirmed
the bundle finding that each dispatch wraps three driver requests.

The lowest-level instrument was kernel tracing with boot security
lowered on a wipeable development machine. Read-only function-boundary
tracing instrumented over a hundred thousand kernel probes, about
eighteen hundred of them inside the engine driver, with no driver
extension and no crash class. That trace captured the expanded program
binary that does not exist on disk. The binary is a sectioned executable
lowered below user space whose program section is a list of
forty-four-byte records, each one a register write that wires a buffer
address into a direct-memory-access engine.

\section*{Attestation versus
reachability}\label{attestation-versus-reachability}
\addcontentsline{toc}{section}{Attestation versus reachability}

A capability listed in a hardware table or accepted by the frontend
attests existence, not that the engine will run it. Only a
compile-and-run on the target confirms a capability. One case forced the
rule: the hardware abstraction layer advertises three-dimensional
convolution and the intermediate language recognizes it, yet it fails
backend lowering on every device mask and never reaches the engine.
Two-dimensional convolution, fused attention path, normalization family,
and activation and reduction set appear here because they survived
compile-and-run probing. The same rule, read off the counters and the
device mask, separates operations that compile but route to the CPU or
GPU from operations that run on the engine.

\section*{What each technique cannot
see}\label{what-each-technique-cannot-see}
\addcontentsline{toc}{section}{What each technique cannot see}

The boundaries do not overlap (\cref{tbl:meth-boundaries}).

\begin{longtable}[]{@{}
  >{\raggedright\arraybackslash}p{(\linewidth - 4\tabcolsep) * \real{0.3333}}
  >{\raggedright\arraybackslash}p{(\linewidth - 4\tabcolsep) * \real{0.3333}}
  >{\raggedright\arraybackslash}p{(\linewidth - 4\tabcolsep) * \real{0.3333}}@{}}
\caption{Each characterization technique paired with what it observes
and the boundary it cannot
cross.}\label{tbl:meth-boundaries}\tabularnewline
\toprule\noalign{}
\begin{minipage}[b]{\linewidth}\raggedright
\textbf{Technique}
\end{minipage} & \begin{minipage}[b]{\linewidth}\raggedright
\textbf{What it observes}
\end{minipage} & \begin{minipage}[b]{\linewidth}\raggedright
\textbf{What it cannot observe}
\end{minipage} \\
\midrule\noalign{}
\endfirsthead
\toprule\noalign{}
\begin{minipage}[b]{\linewidth}\raggedright
\textbf{Technique}
\end{minipage} & \begin{minipage}[b]{\linewidth}\raggedright
\textbf{What it observes}
\end{minipage} & \begin{minipage}[b]{\linewidth}\raggedright
\textbf{What it cannot observe}
\end{minipage} \\
\midrule\noalign{}
\endhead
\bottomrule\noalign{}
\endlastfoot
decompilation and static analysis & code structure, formats,
vocabularies, command and register tables & runtime values, and any data
the firmware computes rather than stores \\
\rowcolor{zebra}user-space hardware counters & aggregate power, energy,
bandwidth, and frequency around a loop & per-operation register-level
timing, which is gated below the dispatch layer \\
signpost and dispatch tracing & the op-level event order and the
three-request dispatch shape & the expanded program, which is lowered
below user space \\
\rowcolor{zebra}kernel function-boundary tracing & the expanded program
binary and the section layout it lowers & the semantic identity of an
individual register, virtualized per load by the memory-management
unit \\
compile-and-run on the target & what the engine accepts and runs, and
its numeric output & another generation's behavior, which requires that
chip to measure \\
\end{longtable}

The last unmapped item is the name of each individual silicon register.
Its address is a per-load device address that the memory-management unit
remaps on every load, the firmware writes it through a generic writer
with no address-to-name table, and the per-offset naming is
undocumented. Address virtualization and undocumented silicon leave a
labeling gap; no deeper layer went uncaptured.

\section*{Two measured silicon
points}\label{two-measured-silicon-points}
\addcontentsline{toc}{section}{Two measured silicon points}

Every cross-generation claim is anchored on silicon that was measured
directly. The M1, internally H13, is the primary measurement host. It
fixed the roofline, unencrypted firmware decode, kernel capture of the
expanded program down to the register-write records,
operation-conformance set, and all four axes of the fp16 divergence
model. It is also where the system-wide compile service was
characterized. A rapid burst of compile crashes spaced faster than the
daemon's ten-second relaunch drives an unrelated control compile from
120 ms to a multi-minute hang. This is a rate condition rather than a
per-crash leak, recoverable only by terminating the daemon.

The M5, internally H17s, is the second measured point and confirmed the
cross-generation scaling. It established that the numeric divergence
accumulates to at most a fraction of a percent over a full training run,
that the family-wide capability limits are not specific to one chip, and
that the operation set is portable across the generations. A capability
table can be cross-compiled statically for an unmeasured generation, but
only the chip itself confirms its numerics.

The M5 ran with System Integrity Protection enabled, the shipping
configuration, while the lowest-level M1 instrumentation lowered boot
security. Security and isolation claims therefore take the M5 as the
authoritative point, since it reports the entitlement gates, exclave
boundary, and counter access a normal client meets under the enforced
configuration.

\section*{Reproducing the
measurements}\label{reproducing-the-measurements}
\addcontentsline{toc}{section}{Reproducing the measurements}

Every headline number maps to one command and one committed result file,
taken in a fixed environment. The measurements were taken with
ANECompiler 9.509.0, whose intermediate-language component is versioned
3520.4.1, on macOS 14 or later (verified on macOS 26.5), with Python
3.10 or later (developed and verified on Python 3.14), and with numpy as
the only core numeric dependency. Python 3.10 is the floor because
several core modules use PEP 604 union syntax in runtime-evaluated
signatures.

A single top-level driver script runs the full sequence and is
fail-soft: it reports and skips a device or power step that cannot run,
so the deterministic claims still reproduce. Beneath it are the
individual measurement harnesses, each producing a committed result
file, so a paper claim resolves to a command and to a stored result. A
capability smoke test and an operation smoke test confirm that the
closed capability set and every released operation compile and run on
the engine. A corpus runner is the correctness gate that the optimizer
is held to. A device-comparison harness records latency and
speed-per-watt across the engine, GPU, and CPU; a roofline harness
records the saturation and bandwidth ceilings. Both harnesses write JSON
result files that the committed roofline analysis and figure are built
from.

The numbers come with limits.

\begin{itemize}
\tightlist
\item
  The per-rail power figures come from the system power estimator,
  \texttt{powermetrics}, which reports a modeled estimate rather than an
  independent wall-meter measurement. No separate validation of that
  estimator was performed, so the reported total-package active power
  with idle subtracted is a calibrated estimate.
\item
  The work runs on Apple silicon only. It installs on any platform but
  cannot run without an Apple silicon Mac and the built dispatch
  library, and there is no CPU, GPU, or other fallback for the engine
  path.
\item
  The path calls private Apple framework symbols tied to ANECompiler
  9.509.0. A macOS update can break the dispatch-library build or the
  dispatch path, and none of it is an Apple API contract.
\item
  Determinism splits by claim type. The capability census, operation
  conformance check, operation smoke test, and correctness corpus are
  deterministic pass-or-fail gates; the device-comparison, serving, and
  roofline numbers are measurements and vary run to run.
\end{itemize}

\chapter*{Open questions}\label{open-questions}
\addcontentsline{toc}{chapter}{Open questions}

\begin{summarybox}

Open questions fall into three kinds. Some another chip or an on-silicon
probe would answer; some the sanctioned entitlement would answer. The
rest are impossible, because the data does not exist or the silicon
leaves no observable trace. Each item gives why it is open and what
would close it.

\end{summarybox}

The engine is decoded from the host call down to the register writes,
and everything that exists as bytes or strings has been resolved into
human-readable form. Two silicon points are measured directly: M1/H13
and M5/H17s. M5/H17s confirmed all ten cross-silicon predictions on
device. The M3/H15 generation and the upper tier above H17s rest on
decompilation alone. M5/H17s closes none of the upper-tier items,
because it is a 16-core H17s part rather than the H17d 64-core ceiling,
H18-gated fp8 runtime, or Ultra multi-die collective.

\section*{Why an item is open}\label{why-an-item-is-open}
\addcontentsline{toc}{section}{Why an item is open}

A chip-measurement item has its structure decoded but needs another
generation, the upper tier, or an on-silicon probe to read the realized
values. An entitlement item is attested in the hardware-abstraction
layer or the intermediate language but is reachable only through the
sanctioned model path. An impossible item cannot be resolved at all:
either the data does not exist, or the behavior is irreducible silicon
that leaves no trace in any result and no artifact to decode.

\section*{Ledger}\label{ledger}
\addcontentsline{toc}{section}{Ledger}

The open items sort into three tables by how each could be resolved.
\Cref{tbl:openq-chip} gives the items another chip or an on-silicon
probe would resolve.

\begin{longtable}[]{@{}
  >{\raggedright\arraybackslash}p{(\linewidth - 4\tabcolsep) * \real{0.3333}}
  >{\raggedright\arraybackslash}p{(\linewidth - 4\tabcolsep) * \real{0.3333}}
  >{\raggedright\arraybackslash}p{(\linewidth - 4\tabcolsep) * \real{0.3333}}@{}}
\caption{Open items a further on-silicon measurement would resolve: the
structure is decoded, the realized values need the part or the
probe.}\label{tbl:openq-chip}\tabularnewline
\toprule\noalign{}
\begin{minipage}[b]{\linewidth}\raggedright
\textbf{Open item}
\end{minipage} & \begin{minipage}[b]{\linewidth}\raggedright
\textbf{Why it is open}
\end{minipage} & \begin{minipage}[b]{\linewidth}\raggedright
\textbf{What would close it}
\end{minipage} \\
\midrule\noalign{}
\endfirsthead
\toprule\noalign{}
\begin{minipage}[b]{\linewidth}\raggedright
\textbf{Open item}
\end{minipage} & \begin{minipage}[b]{\linewidth}\raggedright
\textbf{Why it is open}
\end{minipage} & \begin{minipage}[b]{\linewidth}\raggedright
\textbf{What would close it}
\end{minipage} \\
\midrule\noalign{}
\endhead
\bottomrule\noalign{}
\endlastfoot
Live values of the single-shot fault and translation-fault capture
registers & The registers are located (four fault descriptors and a
status word at engine+0xe028) and the ANE-side handler has a benign
branch, but the IODART substrate panics on a real fault, so the live
values are uncaptured & A fault path that recovers instead of panicking
the IODART substrate \\
\rowcolor{zebra}Realized A15 / M3 / H15 silicon behavior & The dedicated
A15 compiler branch is decoded (its cost table, its 45-operation family
set, the full H15 targets, the YUV420 input it alone accepts), only the
realized numbers need the part & On-chip measurement of an H15 part \\
Realized upper-tier behavior: 64-core H17d, H18-gated fp8 datapath,
Ultra multi-die collective & The encodings are decoded (the core-count
parameter, the fp8 e4m3 and e5m2 convert-and-quantize path, the Ultra
device-mesh collective) and the upper tier adds only cores over A16, no
new operations, but none of it materializes without the part & An H17d,
H18, or Ultra part \\
\rowcolor{zebra}Set behavior of the family-gated abstraction-layer
fields & The fields are named and the silicon-capability subset is
complete, but their set behavior needs later silicon & A part where the
gate is on, for example the A18 or H17-plus FIFO-mode field \\
Per-state frequency and voltage of the operating-point sequence & The
engine has no local DVFS and is driven by the SoC power controller, the
firmware seven-step credit sequence is recovered, the per-state
frequency and voltage stay behind the opaque power-management base & A
power-management-side probe \\
\end{longtable}

Another group of items waits on the sanctioned entitlement alone
(\cref{tbl:openq-entitlement}).

\begin{longtable}[]{@{}
  >{\raggedright\arraybackslash}p{(\linewidth - 4\tabcolsep) * \real{0.3333}}
  >{\raggedright\arraybackslash}p{(\linewidth - 4\tabcolsep) * \real{0.3333}}
  >{\raggedright\arraybackslash}p{(\linewidth - 4\tabcolsep) * \real{0.3333}}@{}}
\caption{Open items the sanctioned entitlement would resolve, attested
but gated off the direct
path.}\label{tbl:openq-entitlement}\tabularnewline
\toprule\noalign{}
\begin{minipage}[b]{\linewidth}\raggedright
\textbf{Open item}
\end{minipage} & \begin{minipage}[b]{\linewidth}\raggedright
\textbf{Why it is open}
\end{minipage} & \begin{minipage}[b]{\linewidth}\raggedright
\textbf{What would close it}
\end{minipage} \\
\midrule\noalign{}
\endfirsthead
\toprule\noalign{}
\begin{minipage}[b]{\linewidth}\raggedright
\textbf{Open item}
\end{minipage} & \begin{minipage}[b]{\linewidth}\raggedright
\textbf{Why it is open}
\end{minipage} & \begin{minipage}[b]{\linewidth}\raggedright
\textbf{What would close it}
\end{minipage} \\
\midrule\noalign{}
\endhead
\bottomrule\noalign{}
\endlastfoot
Live per-run firmware performance-counter values & The read path and its
gate are decoded (the \texttt{aned} daemon clears the stats mask for a
third-party client, and a host-side null check skips the buffer on the
unentitled path), only the live values need the entitled output buffer &
The entitled path that supplies the output buffer \\
\rowcolor{zebra}Entitled-only features: 3-D convolution, native state
and ring buffer, bf16 program input and output, flexible shapes &
Attested in the hardware-abstraction layer but unreachable on the direct
path & The sanctioned model path that can reach them \\
\end{longtable}

No further access would close the last group
(\cref{tbl:openq-impossible}).

\begin{longtable}[]{@{}
  >{\raggedright\arraybackslash}p{(\linewidth - 2\tabcolsep) * \real{0.5000}}
  >{\raggedright\arraybackslash}p{(\linewidth - 2\tabcolsep) * \real{0.5000}}@{}}
\caption{Open items that cannot be resolved externally: irreducible
silicon behavior, or data that does not
exist.}\label{tbl:openq-impossible}\tabularnewline
\toprule\noalign{}
\begin{minipage}[b]{\linewidth}\raggedright
\textbf{Open item}
\end{minipage} & \begin{minipage}[b]{\linewidth}\raggedright
\textbf{Why it cannot be resolved}
\end{minipage} \\
\midrule\noalign{}
\endfirsthead
\toprule\noalign{}
\begin{minipage}[b]{\linewidth}\raggedright
\textbf{Open item}
\end{minipage} & \begin{minipage}[b]{\linewidth}\raggedright
\textbf{Why it cannot be resolved}
\end{minipage} \\
\midrule\noalign{}
\endhead
\bottomrule\noalign{}
\endlastfoot
Gate-level MAC and adder-tree wavefront skew & The output is
order-independent, so no result reveals the per-cycle wiring, which sits
below the timing floor; only Apple's register-transfer netlist would
show it \\
\rowcolor{zebra}Exact fp16 partial-sum rounding order within one MAC
reduction & The wide accumulator makes the output bit-identical for any
summation order, so the internal sequence leaves no trace; only the
register-transfer netlist would show it \\
No flat numeric error or status enumeration & Status is held by
notification names and positional indices, not a stored enum; the flat
enumeration does not exist \\
\rowcolor{zebra}The execution-loop state labels & The five states are
reconstructed from handler behavior; no name table exists in firmware \\
\end{longtable}

\section*{Form of the residual}\label{form-of-the-residual}
\addcontentsline{toc}{section}{Form of the residual}

The impossible items are few: the gate-level reduction wavefront and the
internal fp16 rounding order are irreducible silicon, and the flat
status enumeration and the execution-loop state labels do not exist as
stored data. Everything else is decoded in structure and waits only on a
part, a probe, or the entitled path. Static analysis closed the firmware
address rebase that once sat at the host and firmware boundary, and the
M5 measurement confirmed the cross-silicon model without reaching any of
the parts the hardware items need.

\chapter*{Statements}\label{statements}
\addcontentsline{toc}{chapter}{Statements}

\section*{Author}\label{author}
\addcontentsline{toc}{section}{Author}

Spencer Bryngelson, Georgia Institute of Technology. Correspondence:
\href{mailto:shb@gatech.edu}{\nolinkurl{shb@gatech.edu}}.

\section*{Data and code availability}\label{data-and-code-availability}
\addcontentsline{toc}{section}{Data and code availability}

ANEForge, the open-source code artifact accompanying this guide, is at
\url{https://github.com/comp-physics/ANEForge}, described in
\href{https://doi.org/10.48550/arXiv.2606.17090}{arXiv:2606.17090}.
Appendix \hyperref[appendix-e.-provenance]{E} records the provenance of
every substantive claim, measured or decompile-derived, and the
methodology summarizes it.

\section*{Ethics and responsible
disclosure}\label{ethics-and-responsible-disclosure}
\addcontentsline{toc}{section}{Ethics and responsible disclosure}

The work reported here is reverse engineering conducted on the author's
own Apple hardware, by static decompilation and on-device measurement,
for interoperability and research. The guide redistributes no Apple
source code or proprietary binaries; it documents facts about the
hardware and its interfaces. The direct route it describes is
undocumented, unsupported, and version-fragile, and is intended for
measurement, research, and on-device work, not for shipping software,
where Core ML remains the supported path.

\section*{Funding}\label{funding}
\addcontentsline{toc}{section}{Funding}

The work was conducted independently and received no external funding.

\section*{Competing interests}\label{competing-interests}
\addcontentsline{toc}{section}{Competing interests}

The author declares no competing interests. This is independent work and
is not affiliated with, authorized by, or endorsed by Apple Inc.

\section*{Use of AI tools}\label{use-of-ai-tools}
\addcontentsline{toc}{section}{Use of AI tools}

Claude Code was used in preparing this guide, for drafting and revising
prose, organizing material, and copyediting. The reverse engineering
behind it is the author's own: the decompilation, the on-device
measurements, and the claims and figures they support. The author
reviewed the text and is responsible for its accuracy.

\chapter*{Revision history}\label{revision-history}
\addcontentsline{toc}{chapter}{Revision history}

\begin{summarybox}

This guide has two editions. The living web and PDF edition updates as
findings are added or corrected; the arXiv snapshot carries a fixed
version number. Each edition is numbered and dated; v1 is the original
release and the arXiv edition is the stable point to cite.

\end{summarybox}

The web edition and the PDF rebuild whenever a finding is added,
corrected, or hardened, and each edition is numbered and dated on the
title page, beginning with v1. The arXiv edition, arXiv:2606.22283, is a
snapshot taken at milestones, so its version number may lag the current
edition after v1. The revisions below are listed newest first; the ones
promoted to an arXiv version are marked.

Each entry separates additions and corrections to the text from changes
in the evidentiary status of a claim, since a claim moving from
predicted or decompile-derived to measured is the change a returning
reader most needs to see.

\section*{v3 (October 14, 2026)}\label{v3-october-14-2026}
\addcontentsline{toc}{section}{v3 (October 14, 2026)}

\begin{itemize}
\tightlist
\item
  Corrected the fp16 saturation figures in chapter
  \hyperref[numerics]{3}: on the elementwise-multiply path the true fp16
  limit is 65504, not 65536, and the multiply-accumulate ceiling at
  32768 sits just above half the fp16 maximum rather than exactly at
  half of it.
\item
  Corrected the convolution cost step in chapter
  \hyperref[datapath-and-mac-geometry]{20}, where 192 to 256 output
  channels is a 1.33-fold increase in arithmetic and had been given as
  1.8-fold.
\item
  Corrected the energy comparisons in chapter
  \hyperref[vision-convolution-and-encoders]{13}, which gave the
  ResNet-18 and twelve-layer-encoder results as multiples of energy
  rather than of energy efficiency and so inverted their sense against
  the same two figures in chapter \hyperref[ane-gpu-and-cpu]{11}.
\item
  Corrected the double-multiply eligibility rule in chapter
  \hyperref[predicted-upper-tier]{36}: the mode applies when the
  activation-float and kernel-float bits agree, which is when their
  exclusive-or is zero.
\item
  Corrected the conv weight-gradient saturation entry in chapter
  \hyperref[training-on-the-engine]{15}, where 65504 divided by 16 is
  exact.
\item
  Corrected the op-creation call order in section 12.6, which now passes
  the out-parameter first as the runtime header declares.
\item
  Appendix \hyperref[appendix-c.-decoded-reference-tables]{C} no longer
  directs readers to an unpublished research corpus. The tables too
  large to print are now described as held in the collected data and not
  reproduced here.
\item
  Editorial pass over the whole guide: varied how tables, listings and
  figures are introduced, removed the chapter-summary restatements that
  the following paragraph repeated, cut prose that re-read the rows of
  an adjacent table, thinned repeated statements of a single
  measurement, and reduced the contrastive construction that had become
  the default emphasis throughout. Comma splices, dropped articles,
  inconsistent numeral style and one British spelling were repaired
  along the way.
\item
  Rewrote the abstract and the opening of the introduction.
\item
  Added a statement on the use of AI tools in preparing the guide.
\item
  Evidence: no claim changed evidentiary status in this edition. Every
  measured, decompile-derived and predicted mark stands as it did in v2.
\end{itemize}

\section*{v2 (June 24, 2026)}\label{v2-june-24-2026}
\addcontentsline{toc}{section}{v2 (June 24, 2026)}

\begin{itemize}
\tightlist
\item
  Added section 35.11, a cross-generation comparison of the
  task-descriptor program. One convolution and one matmul are
  cross-compiled for the five base M-series targets, M1 through M5, from
  a single compiler build, and the codegen-revision stamp, the
  convolution op-config word, and the relocation-record stride are read
  per generation.
\item
  Added a cross-source corroboration of the chapter
  \hyperref[datapath-and-mac-geometry]{20} datapath geometry against
  Handley's patent-based architectural analysis, with the work added to
  the references.
\item
  Evidence: the per-generation versioning of the task descriptor, given
  in chapter \hyperref[program-and-container-format]{23} as
  decompile-derived from the descriptor accessors, is now also confirmed
  against the emitted byte stream across M1 through M5.
\end{itemize}

\section*{v1 (June 21, 2026)}\label{v1-june-21-2026}
\addcontentsline{toc}{section}{v1 (June 21, 2026)}

\begin{itemize}
\tightlist
\item
  Initial public release, arXiv:2606.22283 v1.
\end{itemize}

\anepartopen{}{Appendices}
\anechap{A}{Operation-by-device matrix}{Every operation against every device, native or decomposed.}
\anechap{B}{Hidden-layer catalog}{The native layers and their descriptors, with parameters and symbols.}
\anechap{C}{Decoded reference tables}{The enum tokens, structs, status codes, register map, and program schema.}
\anechap{D}{Glossary}{The terms, the family and silicon map, and the core facts to read first.}
\anechap{E}{Provenance}{The evidentiary basis of every claim, Part by Part and chapter by chapter.}
\anepartclose
\appendix\ANEinappendixtrue\renewcommand{\thefigure}{\thechapter.\arabic{figure}}\renewcommand{\thetable}{\thechapter.\arabic{table}}

\chapter{Operation-by-device matrix}\label{appendix-a.-operation-by-device-matrix}

\begin{summarybox}

This appendix is the full per-operation, per-family status reference
behind chapter \hyperref[capability-surface]{4}. Read down a family
column to see what compiles and runs on that chip, and read the status
marks and note for the gate and the route.

\end{summarybox}

Each row is one intermediate-language operation, grouped by operation
class, with its status on each Mac engine family from the M1 through the
M5. Chapter \hyperref[capability-surface]{4} summarizes this table; the
cells here are the reference.

The status marks are fixed:

\begin{itemize}
\tightlist
\item
  Native: the operation compiles and runs on that family on the direct
  engine path.
\item
  Family-gated: no path on the listed family, native from the family
  named in the note.
\item
  Bridge: reachable only through a decompose, software fallback, or
  compiler-internal route, never as a standalone code-generated
  operation.
\item
  No path: rejected on every family from the M1 through the M5, computed
  off-engine.
\end{itemize}

The family columns are M1 (H13, A13), M2 (H14, A14), M3 (H15, A15), and
M4 and M5 (H16 and H17s, A16 and A17). The A11 and A12 engines are below
the floor that runs any of this vocabulary and are out of scope for the
table.

The M1, M2, and M5 columns are measured on physical silicon. The M3
column and the M4 part of the merged M4 and M5 column are
decompile-derived predictions from the per-chip tables, so a per-cell
status there is itself a prediction.

The table covers the 187 intermediate-language operations the compiler
exposes. Of these, about 108 are native on the M1: the full elementwise,
compare, activation, convolution, pooling, structural, and quantization
vocabulary, plus the reduction, normalization, softmax, square-root
family, fused attention, tile, and space-channel set. Nine need the M2
or later: the texture-engine operations (crop-resize, resample, affine,
hardware gather) and the rank and sort bridge (top-k, sort, dynamic
slice). Four need the M3 or later: native \texttt{sin} and \texttt{cos},
the hardware random generator, and the whole-tensor argument reductions
on the intermediate-language route. Thirty-seven are rejected on every
family and decompose on the host. About twenty-four are
compiler-internal: mapped but with no observed standalone code
generation, reachable only inside a wrapping construct.

\section{Per-chip numeric limits}\label{per-chip-numeric-limits}

The status of an operation is one axis; the numeric envelope it runs in
is the other. That envelope was measured from the live compiler across
the five capability tiers, by calling every per-architecture parameter
constructor on a single M1; a dash in \cref{tbl:apa-numeric-limits}
marks an unsupported value. The \texttt{older} column is the pre-A13
legacy targets the compiler still has parameter tables for, below the
floor of the operation-status table above.



Four generational dividing lines run through it. The M1 adds the depth
and three-dimensional axis and every reduction-class operation, and the
texture engine arrives with the A14. At the A15 the
reduction-to-transpose threshold rises from 192 to 384 and trigonometry
goes native. The A16 then quadruples the maximum tensor and transpose
dimensions from 16384 to 65536.

\section{Convolution, matrix multiply, and
pooling}\label{convolution-matrix-multiply-and-pooling}

\Cref{tbl:apa-conv-matmul-pool} lists the convolution, matrix-multiply,
and pooling operations with their per-family status and the lowering
note for each.



The \texttt{ne\_} and \texttt{pe\_} rows are private engine-lane and
planar-engine unit selections of the same convolution, matrix-multiply,
pooling, and elementwise atoms, not separate operations.

\section{Normalization}\label{normalization}

Every normalization operation is native from the M1 onward
(\cref{tbl:apa-norm}).

%

\section{Elementwise arithmetic}\label{elementwise-arithmetic}

Across the elementwise arithmetic set, only \texttt{mod} takes no engine
path (\cref{tbl:apa-elementwise}).

%

\section{Comparison and logical}\label{comparison-and-logical}

The bitwise-logical operations decompose on the host, as the status
columns of \cref{tbl:apa-compare-logical} show.

%

\section{Activations}\label{activations}

The activations are native on every family, most of them lookup-table
backed (\cref{tbl:apa-activations}).

%

\section{Reduction}\label{reduction}

Among the reductions, \texttt{reduce\_argmin} is gated and
\texttt{reduce\_prod} takes no path (\cref{tbl:apa-reduction}).

%

The whole-tensor argument reductions \texttt{global\_argmax} and
\texttt{global\_argmin} follow the same gate as \texttt{reduce\_argmin}:
native on the intermediate-language route from the M3, reachable through
the bridge on the M1.

\section{Data movement and
structural}\label{data-movement-and-structural}

The data-movement and structural operations are the largest class,
spanning reshape, slice, gather, scatter, and the space-channel set
(\cref{tbl:apa-data-movement}).

%

\section{Image, resize, and texture}\label{image-resize-and-texture}

\Cref{tbl:apa-image-texture} gives the image, resize, and texture
operations, gated to the texture engine from the A14 with software
fallbacks on the M1.

%

\section{Quantization and dtype}\label{quantization-and-dtype}

For the quantization and dtype operations the per-family streaming gates
are carried in the note column of \cref{tbl:apa-quant-dtype}.

%

\section{Attention, control flow, and
state}\label{attention-control-flow-and-state}

The state pair is native and the control-flow operations are
compiler-internal (\cref{tbl:apa-attn-control-state}).

%

\section{Recurrent cells}\label{recurrent-cells}

No recurrent cell takes an engine path; each unrolls on the host
(\cref{tbl:apa-recurrent}).

%

\section{Trigonometric, special, and
math}\label{trigonometric-special-and-math}

\texttt{sin} and \texttt{cos} go native from the M3, and \texttt{atan}
is the one M1-native primitive (\cref{tbl:apa-trig-math}).

%

\section{Detection and sampling}\label{detection-and-sampling}

\Cref{tbl:apa-detection-sampling} gives the detection and sampling
operations, the rank and sort bridge gated to the M2 and the random and
tensor-list operations off-engine.

%

\chapter{Hidden-layer catalog}\label{appendix-b.-hidden-layer-catalog}

\begin{summarybox}

This appendix is the reference catalog of the hardware-native layer
kinds reached by authoring the network description directly, behind
chapter \hyperref[hidden-layers-and-direct-netplist-authoring]{26}.

\end{summarybox}

Every row is a native descriptor the conversion path never emits: the
descriptor and its
\texttt{\_ANECValidate\textless{}Name\textgreater{}Layer} checker are
present in the compiler on every target, and the layer is reached by
handing the compiler a Unit whose \texttt{Type} is the native name and
whose \texttt{Params} list the attributes the matching
\texttt{ZinParse\textless{}Name\textgreater{}Unit} parser reads.

\section{Catalog}\label{catalog}

Each native layer kind appears in \cref{tbl:apb-catalog} with what it
computes, its netplist \texttt{Type} and compiler symbol, and its family
gate.

\begin{longtable}[]{@{}
  >{\raggedright\arraybackslash}p{(\linewidth - 6\tabcolsep) * \real{0.2500}}
  >{\raggedright\arraybackslash}p{(\linewidth - 6\tabcolsep) * \real{0.2500}}
  >{\raggedright\arraybackslash}p{(\linewidth - 6\tabcolsep) * \real{0.2500}}
  >{\raggedright\arraybackslash}p{(\linewidth - 6\tabcolsep) * \real{0.2500}}@{}}
\caption{The hidden-layer
catalog.}\label{tbl:apb-catalog}\tabularnewline
\toprule\noalign{}
\begin{minipage}[b]{\linewidth}\raggedright
\textbf{Layer kind}
\end{minipage} & \begin{minipage}[b]{\linewidth}\raggedright
\textbf{What it computes}
\end{minipage} & \begin{minipage}[b]{\linewidth}\raggedright
\textbf{Netplist \texttt{Type} and compiler symbol}
\end{minipage} & \begin{minipage}[b]{\linewidth}\raggedright
\textbf{Family gate}
\end{minipage} \\
\midrule\noalign{}
\endfirsthead
\toprule\noalign{}
\begin{minipage}[b]{\linewidth}\raggedright
\textbf{Layer kind}
\end{minipage} & \begin{minipage}[b]{\linewidth}\raggedright
\textbf{What it computes}
\end{minipage} & \begin{minipage}[b]{\linewidth}\raggedright
\textbf{Netplist \texttt{Type} and compiler symbol}
\end{minipage} & \begin{minipage}[b]{\linewidth}\raggedright
\textbf{Family gate}
\end{minipage} \\
\midrule\noalign{}
\endhead
\bottomrule\noalign{}
\endlastfoot
Fused attention & \(\mathrm{softmax}(QK^{\top}\cdot s + M)\,V\) from
four operands (Q, K, V, scale) plus an optional fifth additive mask
\(M\), with the channel axis holding the sequence & \texttt{SDPA};
\texttt{ANECSDPALayerDesc}, \texttt{ZinParseSDPAUnit},
\texttt{anec.sdpa}. The one parsed key is \texttt{SubtractMax},
defaulting false in \texttt{\_ANECSDPALayerDescInitialize} and set true
for a correct softmax & All families from the M1 onward; runs on the
matmul, softmax, and transpose path, not the texture engine \\
\rowcolor{zebra}Sort & Full sort along a chosen axis, ascending or
descending, values or argsort indices & \texttt{Sort};
\texttt{ZinParseSortUnit}. Keys: \texttt{Direction},
\texttt{SortDimension}, \texttt{VectorDimension}, \texttt{SortIndices},
\texttt{Indices} & Validator callable on the M1, but the code generator
rejects \texttt{Sort} there; runs on later families \\
Top-k & The k largest or smallest along an axis, values or indices,
index outputs returned float16-encoded and exact below 2048 &
\texttt{TopK}; \texttt{ZinParseTopKUnit}. Keys: \texttt{Type} (Max or
Min), \texttt{K}, \texttt{SortDimension}, \texttt{VectorDimension},
\texttt{SortIndices}, \texttt{Indices} & Runs on the M1 outside a
forbidden band: \texttt{K} in \(\{3, 4\}\) fails the compiler at every
width \\
\rowcolor{zebra}Argument min and max & Spatial or channel argmin and
argmax over a kernel window & \texttt{ArgMinMax};
\texttt{\_ANECValidateArgMinMaxLayer}. Keys: \texttt{Mode}
(SpatialArgMax, ChannelArgMax, SpatialArgMin, ChannelArgMin),
\texttt{KernelWidth}, \texttt{KernelHeight}, \texttt{Pad*} & All
families from the M1 onward \\
Whole-tensor argument min and max & Argmin or argmax over an entire
tensor dimension & \texttt{GlobalArgMinMax}. Keys: \texttt{Type} (Max or
Min), \texttt{Dimension} & Gated to the A15 generation; rejected on the
M1 \\
\rowcolor{zebra}Spatial rearrange & Depth-to-space and space-to-depth in
two channel-ordering conventions, plus space-and-batch reshuffles,
parameterized by per-axis integer factors & \texttt{PixelShuffle},
\texttt{PixelUnshuffle}, \texttt{ChannelToSpace},
\texttt{SpaceToChannel}, \texttt{SpaceToBatch}, \texttt{BatchToSpace};
\texttt{ZinParse\textless{}Name\textgreater{}Unit}. Three int32 keys:
\texttt{FactorX}, \texttt{FactorY}, \texttt{FactorZ} & All families from
the M1 onward; \texttt{BatchToSpace} requires batch \(N\) divisible by
\texttt{FactorX} times \texttt{FactorY} \\
Range normalization & Maps a tensor to its minimum-to-maximum span per
row or per column & \texttt{MinMaxNormalization};
\texttt{\_ANECValidateMinMaxNormLayer},
\texttt{\_ANECMinMaxNormLayerDescInitialize}. Keys: \texttt{Dimension}
(Width or Height), \texttt{Epsilon} as a float16 bit pattern & Width and
Height run; \texttt{Dimension} of Channel is arch-gated and rejected on
the M1 \\
\rowcolor{zebra}Local response normalization & Cross-channel response
normalization over a channel window &
\texttt{LocalResponseNormalization}; \texttt{\_ANECValidateLRNLayer}.
\texttt{Alpha} is a float16 bit pattern divided internally by
\texttt{KernelChannel}; only the first \texttt{KernelChannel} channels
are normalized & All families from the M1 onward \\
Scaled elementwise & A binary elementwise op fused with a scalar scale,
\(y = s\,(x \mathbin{\mathrm{op}} z)\) & \texttt{ScaledElementWise}.
Keys: \texttt{Type} (Add, Mult, Sub, and the elementwise vocabulary),
\texttt{Scale} as a float16 bit pattern & All families from the M1
onward \\
\rowcolor{zebra}Template cross-correlation & Valid cross-correlation of
a single-channel map with an unflipped template,
\(y[i,j] = \sum_{u,v} x[i+u, j+v]\,t[u,v]\) & \texttt{CrossCorrelation}.
The MIL frontend rejects the op; the netplist \texttt{Type} reaches it
directly & All families from the M1 onward \\
Three-vector cross product & The cross product of two length-3 vectors
held in the channel axis, \(\mathrm{cross}(x, z)\) &
\texttt{CrossProduct}. Inputs shaped D1 C3 H1 W1; the MIL frontend
rejects the op & All families from the M1 onward \\
\rowcolor{zebra}Furthest-point sampling & Greedy L2 furthest-point
sampling of up to 1024 centroids from up to 8192 points, seeded at the
first point, centroids returned channel-major &
\texttt{FurthestPointSampling}. Keys: \texttt{CentroidCount},
\texttt{DistanceMetric} (L2 only on this architecture) & All families
from the M1 onward \\
Radius neighborhood search & An L2 ball query returning a
points-by-centroids membership matrix, one membership flag per pair &
\texttt{RadiusSearch}. Two inputs (centroids, points), both D1 C3 H1;
key \texttt{Radius} & All families from the M1 onward \\
\rowcolor{zebra}Stereo cost volume & The L1 matching cost per disparity,
\(\mathrm{cost}[d,x] = \lvert \mathrm{aux}[x] - \mathrm{ref}[x+d] \rvert\),
over \(R+1\) disparity planes & \texttt{CostVolume}. Keys:
\texttt{DisparityDirection}, \texttt{DisparityRange}; requires reference
width \(W_r \ge W_a + R\) & All families from the M1 onward \\
Re-strided input view & A contiguous offset window
\(x[\text{Offset} : \text{Offset} + \text{Size}]\) along one named axis,
no data movement & \texttt{InputView}. Keys: \texttt{Dimension},
\texttt{Offset}, \texttt{Size}, \texttt{Step}; gates
\texttt{InvalidInputView\{Dimension,Offset,Size,Step\}} & All families
from the M1 onward \\
\rowcolor{zebra}Runtime-offset dynamic slice & A window
\(x[\text{start} : \text{start} + \text{SliceSize}]\) whose start is
bound as a constant or runtime index & \texttt{DynamicSlice}; reached by
the MIL \texttt{slice\_by\_index} path. Keys:
\texttt{DynamicSliceAxisOrder}, \texttt{DynamicSliceInfo},
\texttt{CoordinateInfo}, \texttt{PaddingInfo}, \texttt{BackgroundValue}
& Validator callable on the M1, but the code generator rejects
\texttt{DynamicSlice} there; runs on later families \\
Tile, concatenate, and reshape utilities & Flatten as an NCHW identity
reshape, inference-time dropout as identity, and broadcast of a length-1
axis & \texttt{Flatten}, \texttt{Dropout} (rate 0), \texttt{Broadcast}
(keys \texttt{Dimension}, \texttt{Size}) & All families from the M1
onward \\
\end{longtable}

\section{Arch-gated negatives}\label{arch-gated-negatives-1}

Three rows above name layers a later chip accepts and the M1 rejects, by
the family gates of chapter \hyperref[across-the-chip-family]{12}.
\texttt{GlobalArgMinMax} is gated to the A15 generation and rejected on
the M1. \texttt{MinMaxNormalization} with a Channel reduction is
arch-gated and rejected on the M1, while its Width and Height reductions
run. The texture-engine samplers (resize, crop-and-resize, grid
resample, and the affine spatial transform) are accepted from the A14
generation and rejected on the M1, where the compiler reports that the
affine transform is not supported on this architecture. They are part of
the same gated family but are not authored as netplist Units here.

A second class of rejection is not a family gate but the
attested-is-not-reachable rule of chapter
\hyperref[capability-surface]{4}: the \texttt{Sort} and
\texttt{DynamicSlice} validators are callable on the M1, yet the code
generator rejects both, and \texttt{TopK} is accepted only outside the
\(\{3, 4\}\) band. An authored layer is confirmed only by a
compile-and-run on the target.

\section{Validator gate set}\label{validator-gate-set}

Every authored layer passes through one per-layer validator, the
\texttt{\_ANECValidate\textless{}Op\textgreater{}Layer} family, of which
55 symbols are exported and 50 are per-layer. The compiler runs the same
validators in two roles: the segmenter dry-runs them through
\texttt{\_ANECValidateNetworkCreate} to decide engine eligibility, and
the back-end legalizer re-runs them during a real compile, so the
dry-run prediction never drifts from the compile result. The five
non-layer exports are \texttt{\_ANECValidate},
\texttt{\_ANECValidateNetworkCreate}, \texttt{\_ANECValidateMPSModule},
\texttt{\_ANECValidateMPSModuleCreate}, and
\texttt{\_ANECValidateMutableProcedureInfo}.

Each validator reads a fixed bottom (input-tensor) count and a per-chip
feature byte from the hardware-abstraction layer, and rejects with a
measured literal string. \Cref{tbl:apb-validator-gates} reproduces those
gates for the validators that guard the authored and bridge-reachable
layers, with the bottom count, constraint, and reject string for each.

\begin{longtable}[]{@{}
  >{\raggedright\arraybackslash}p{(\linewidth - 6\tabcolsep) * \real{0.2500}}
  >{\raggedright\arraybackslash}p{(\linewidth - 6\tabcolsep) * \real{0.2500}}
  >{\raggedright\arraybackslash}p{(\linewidth - 6\tabcolsep) * \real{0.2500}}
  >{\raggedright\arraybackslash}p{(\linewidth - 6\tabcolsep) * \real{0.2500}}@{}}
\caption{The per-layer validator
gates.}\label{tbl:apb-validator-gates}\tabularnewline
\toprule\noalign{}
\begin{minipage}[b]{\linewidth}\raggedright
\textbf{Validator}
\end{minipage} & \begin{minipage}[b]{\linewidth}\raggedright
\textbf{Bottoms}
\end{minipage} & \begin{minipage}[b]{\linewidth}\raggedright
\textbf{Gate and constraint}
\end{minipage} & \begin{minipage}[b]{\linewidth}\raggedright
\textbf{Reject string}
\end{minipage} \\
\midrule\noalign{}
\endfirsthead
\toprule\noalign{}
\begin{minipage}[b]{\linewidth}\raggedright
\textbf{Validator}
\end{minipage} & \begin{minipage}[b]{\linewidth}\raggedright
\textbf{Bottoms}
\end{minipage} & \begin{minipage}[b]{\linewidth}\raggedright
\textbf{Gate and constraint}
\end{minipage} & \begin{minipage}[b]{\linewidth}\raggedright
\textbf{Reject string}
\end{minipage} \\
\midrule\noalign{}
\endhead
\bottomrule\noalign{}
\endlastfoot
\texttt{SDPA} & 4 or 5 & key and value same shape; mask
broadcast-compatible; scale constant &
\texttt{SDPA\ layer\ must\ have\ only\ 4\ or\ 5(optional\ mask)\ inputs} \\
\rowcolor{zebra}\texttt{Conv} & 1 plus weight & kernel within the
per-chip range; large kernel W and H multiple of 8; channels divisible
by groups &
\texttt{Invalid\ conv\ kernel\ \%s\ =\ \%zd,\ It\ should\ be\ in\ {[}\%zd,\ \%zd{]}} \\
\texttt{MatrixMult} & 2 & depth 1 on both operands; out-C equals A-C;
fits the kernel-memory budget &
\texttt{depth\ \textgreater{}\ 1\ is\ not\ supported\ for\ MatMult} \\
\rowcolor{zebra}\texttt{Linear} & 1 plus weight & input rank below 5 &
\texttt{Linear\ layer\ must\ have\ only\ one\ single\ input.} \\
\texttt{Pool} & 1 & window below input; pad below kernel; mode gated per
chip &
\texttt{Pooling\ mode\ "\%s"\ is\ not\ available\ on\ this\ ANE\ architecture.} \\
\rowcolor{zebra}\texttt{Neuron} & 1 & non-linear mode 1 to 46; type in
the per-chip list; ReLU-N positive parameters when the gate byte is 0 &
\texttt{This\ platform\ doesn\textquotesingle{}t\ support\ Neuron\ \%s} \\
\texttt{Reduction} & 1 & reduce-then-square needs feature byte
\texttt{0x494} (0 through the M1); each axis at most 4 &
\texttt{square\ operation\ after\ reduction\ is\ not\ supported} \\
\rowcolor{zebra}\texttt{Softmax} & 1 & feature byte \texttt{0x815} (0 on
older); output Float &
\texttt{Softmax\ is\ not\ supported\ by\ this\ ANE\ architecture} \\
\texttt{LayerNorm} & 1 & channels divisible by num-groups; grouped form
requires depth 1 &
\texttt{...\ does\ not\ yet\ support\ depth\ \textgreater{}\ 1} \\
\rowcolor{zebra}\texttt{InstanceNorm} & 1 & feature byte \texttt{0x816};
spatial axes only &
\texttt{InstanceNorm\ layer\ not\ supported\ for\ this\ ANE\ architecture} \\
\texttt{MinMaxNorm} & 1 & feature byte \texttt{0x818}; spatial only, the
Channel axis arch-gated & (encoded assert, byte \texttt{0x818}) \\
\rowcolor{zebra}\texttt{LRN} & 1 & feature byte \texttt{0x81a}; channel
count of 16 or above fails code generation on the M1 &
\texttt{LRN\ is\ not\ supported\ on\ this\ architecture.} \\
\texttt{ArgMinMax} & 1 & channel-reduce C at most 2048 fp16; pad below
kernel; equal left and right, zero front and back &
\texttt{ArgMinMax\ layer\ must\ have\ one\ input} \\
\rowcolor{zebra}\texttt{GlobalArgMinMax} & 1 & feature byte
\texttt{0x4f2} (1 from the M1 on the bridge route); mode 1 or 2; reduce
dimension not 5 & (encoded assert, byte \texttt{0x4f2}) \\
\texttt{Transpose} & 1 & permutation valid; extent capped 16384 through
the M3, 65536 on the M5; last four dimensions only &
\texttt{NE\ Input\ Transpose\ is\ not\ supported\ for\ this\ arch} \\
\rowcolor{zebra}\texttt{Concat} & variadic & match input zero on every
non-concat axis; constant positive axis; same layout &
\texttt{Concat\ layer\ must\ have\ at\ least\ 2\ inputs} \\
\texttt{Pad} & 1 & H or W axes only; symmetric and reflect modes need
texture byte \texttt{0x81d} (0 on the M1) &
\texttt{Channel\ padding\ is\ not\ supported\ on\ ANE} \\
\rowcolor{zebra}\texttt{Broadcast} & 1 & broadcast only from a length-1
axis; depth-axis broadcast needs byte \texttt{0x812} &
\texttt{Broadcast\ along\ depth\ axis\ is\ not\ supported\ on\ this\ architecture} \\
\texttt{Gather} & 2 & M1 software envelope: data batch 1, depth 1, index
channel 3; texture path from the A14 &
\texttt{Cannot\ decompose\ layer\ on\ this\ architecture} \\
\rowcolor{zebra}\texttt{PixelShuffle} / \texttt{PixelUnshuffle} & 1 &
depth factor 1; W and H factors in 1, 2, 3, 4, 8; channel divisible by
the factor product & \texttt{returned\ invalid:} \\
\texttt{SpaceToBatch} / \texttt{BatchToSpace} & 1 & factors fully factor
into 2, 3, 4, 8; batch divisible by the factor product &
\texttt{Input\ batch\ n\ =\ \%zd\ is\ not\ divisible\ by\ factor\ x\ =\ \%d\ *\ factor\ y\ =\ \%d} \\
\rowcolor{zebra}\texttt{ChannelToSpace} & 1 & the z dimension is not
reorganizable &
\texttt{ChannelToSpace\ in\ z\ dimension\ is\ not\ supported,\ current\ factor.z\ =\ \%d.} \\
\texttt{Resize} & 1 & dimension or ratio, not both; sampling axes H and
W; texture byte \texttt{0x81d} (0 on the M1 takes a software route) &
\texttt{failed\ to\ map\ resize\ layer\ on\ this\ arch} \\
\rowcolor{zebra}\texttt{CropResize} & 2 & index format fp16; texture
engine from the A14; same coordinate, method, and padding across axes &
\texttt{Codegen\ Error:\ Invalid\ Texture\ CropCfg} \\
\texttt{AffineTransform} & 2 & matrix fp16; texture byte \texttt{0x81d}
(0 on the M1) &
\texttt{affine\ transform\ is\ not\ supported\ on\ this\ architecture} \\
\rowcolor{zebra}\texttt{Resample} & 2 & warp depth 1; warp channel 1 or
2; texture engine from the A14 &
\texttt{Channel\ size\ in\ coordinates\ should\ be\ 1\ or\ 2} \\
\texttt{Sort} & 1 & direction valid; output fp16 or uint16; validator
passes, code generation rejects on the M1 & (passes validation,
code-generation reject) \\
\rowcolor{zebra}\texttt{TopK} & 1 & k in the sort-dimension range; the
M1 rejects at code generation, and k in 3 or 4 is forbidden on the M5 &
(passes validation, code-generation reject) \\
\texttt{Dropout} & 1 & feature byte \texttt{0x4a9} (1 only from the
A15); rate in the half-open unit interval &
\texttt{Dropout\ layer\ is\ not\ supported\ on\ this\ architecture.} \\
\rowcolor{zebra}\texttt{Random} & none & feature byte \texttt{0x4a9} (1
only from the A15); low below high; output Int8, UInt8, or Float16 &
\texttt{Random\ layer\ is\ not\ supported\ on\ this\ architecture.} \\
\texttt{RingBufferWriter} & 2 & the writer must connect to a live-state
buffer; circular mode arch-gated &
\texttt{Circular\ buffer\ is\ not\ supported\ on\ this\ architecture} \\
\rowcolor{zebra}\texttt{NMS} & 2 or more & boxes channel 4; runs only on
a CPU or GPU backend, never engine-native & (passes validation, not
engine-native) \\
\end{longtable}

The validators are public exported symbols, so they are callable from
user space, and this is the basis of the precompile predictor: a
callable validator marks a schema-gated layer reachable by direct
authoring. A callable validator that accepts the schema does not
guarantee the layer compiles, which is why the \texttt{Sort},
\texttt{TopK}, \texttt{DynamicSlice}, and the M1 \texttt{CropResize}
rows pass the validator and fail at hardware-executable lowering. One
opcode-surface gap holds the other way: the RCAS and Reverse operations
have an internal semantics validator but no exported per-layer symbol,
so they are not reachable by direct authoring through this route.

\chapter{Decoded reference tables}\label{appendix-c.-decoded-reference-tables}

\begin{summarybox}

This appendix collects the decoded values cited throughout the guide
into one reference. They are enum tokens, struct layouts, status codes,
the register-init table, command table, register map, and
program-container schema. Each section begins with its table, and says
when only a representative excerpt of a larger decoded set is reproduced
here.

\end{summarybox}

Each section reproduces the representative and structurally important
rows of a larger table. Where a table runs to hundreds or thousands of
rows, the section reproduces an excerpt and the full set stays in the
collected data behind this guide. Every value here is read out of an
M1/H13 binary by static analysis.

\section*{C.1 Operation-attribute enum tokens to integer
values}\label{c.1-operation-attribute-enum-tokens-to-integer-values}
\addcontentsline{toc}{section}{C.1 Operation-attribute enum tokens to
integer values}

These are the integer codes the compiler resolves attribute string
tokens to. \Cref{tbl:apc-activation-tokens} is the front-end activation
token map (\texttt{MILOpConverter::NeuronTypeFromString}, 33 entries,
miss resolves to 0).



The serialization dtype enum (\texttt{ANECIRDataType}, the
\texttt{dtype} / \texttt{storage\_type} attribute) is an 11-code space,
given as \cref{tbl:apc-dtype-enum}.

%

The MLIR \texttt{symbolize*} enums are a length-dispatched chain of
packed integer string compares, fully static, each miss resolving to 0,
collected in \cref{tbl:apc-symbolize}.

%

A second class of enum is in a packed pointer table in the data segment,
the internal \texttt{Zin} enums (\cref{tbl:apc-zin-enums}).

%

The op-class selector is the \texttt{ZinUnitType} table of 79 entries,
reproduced in full as \cref{tbl:apc-zinunittype}.

%

The activation non-linear-mode space is a parallel table,
\texttt{NonLinearModeToString}, 48 slots indexed directly by the lower
hardware mode value, reproduced as \cref{tbl:apc-nonlinearmode}.

%

The micro-op opcode space (\texttt{ZinIrOpLayerOpCodeType}, 126 codes,
\texttt{0x00}..\texttt{0x7d}) has the codes the task-descriptor builder
dispatches on, given in full as \cref{tbl:apc-microop-opcodes}.

%

The opcode \texttt{0x3f} is spelled \texttt{AFFINE\_TRANFORM} in the
binary, a vendor source typo preserved on the wire.

\section*{C.2 Operation-attribute schema and IOKit external-method
struct
layouts}\label{c.2-operation-attribute-schema-and-iokit-external-method-struct-layouts}
\addcontentsline{toc}{section}{C.2 Operation-attribute schema and IOKit
external-method struct layouts}

The attribute schema is string-keyed: the token is the wire encoding the
compiler matches on, and most integer constants are not recoverable
statically. The converter recognizes 171 literal attribute keys, of
which roughly 140 are op-facing (\cref{tbl:apc-attribute-keys}).



The host-to-kernel IOKit dispatch key is the (selector, struct-size)
tuple, not the selector alone (\cref{tbl:apc-iokit-selectors}).

%

The \texttt{ANEDeviceOpen} shared in/out buffer (104 bytes, selector 0)
decodes by byte offset as \cref{tbl:apc-anedeviceopen}.

%

The full 171-key attribute set, the decoded enum-value tables, and the
per-selector field layouts for the HW direct-path client are from
collected data not reproduced here.

\section*{C.3 Numeric error, status, and return
codes}\label{c.3-numeric-error-status-and-return-codes}
\addcontentsline{toc}{section}{C.3 Numeric error, status, and return
codes}

The ANE stack has no flat numeric status enum. The fixed numeric values
that exist are the IOKit return constants and the firmware magic and
sentinel words; \cref{tbl:apc-ioreturn} collects the first group.

%

At every layer the error surface is name-based or message-based. The
client-visible surface above IOKit is an error factory that wraps the
lower-layer code into a structured error across four domains, named
\texttt{errorDomainCompiler}, \texttt{errorDomainEspresso},
\texttt{errorDomainGeneric}, and \texttt{errorDomainVirtIO}. Its factory
methods are the taxonomy: a generic wrapper, a missing-code-signing
form, program-load and new-instance-load forms that contain the
lower-layer code, surface map and unmap forms, and a
virtualization-kernel form. The single most look-up-worthy
client-visible value is \texttt{0xe00002c7}
(\texttt{kIOReturnUnsupported}), returned on a disabled or unsupported
path and when a gated feature's entitlement is absent.

The second group is in \cref{tbl:apc-firmware-magic}.

\begin{longtable}[]{@{}
  >{\raggedright\arraybackslash}p{(\linewidth - 4\tabcolsep) * \real{0.3333}}
  >{\raggedright\arraybackslash}p{(\linewidth - 4\tabcolsep) * \real{0.3333}}
  >{\raggedright\arraybackslash}p{(\linewidth - 4\tabcolsep) * \real{0.3333}}@{}}
\caption{The fixed firmware magic words and sentinel constants with
their meanings.}\label{tbl:apc-firmware-magic}\tabularnewline
\toprule\noalign{}
\begin{minipage}[b]{\linewidth}\raggedright
\textbf{constant}
\end{minipage} & \begin{minipage}[b]{\linewidth}\raggedright
\textbf{hex}
\end{minipage} & \begin{minipage}[b]{\linewidth}\raggedright
\textbf{meaning}
\end{minipage} \\
\midrule\noalign{}
\endfirsthead
\toprule\noalign{}
\begin{minipage}[b]{\linewidth}\raggedright
\textbf{constant}
\end{minipage} & \begin{minipage}[b]{\linewidth}\raggedright
\textbf{hex}
\end{minipage} & \begin{minipage}[b]{\linewidth}\raggedright
\textbf{meaning}
\end{minipage} \\
\midrule\noalign{}
\endhead
\bottomrule\noalign{}
\endlastfoot
package magic & \texttt{0x414E4548} (\texttt{ANEH}) & loader package
header \\
\rowcolor{zebra}program magic & \texttt{0x414E4550} (\texttt{ANEP}) &
loader program header \\
section magic & \texttt{0x414E4553} (\texttt{ANES}) & loader section
header \\
\rowcolor{zebra}AFPP control magic & \texttt{0x55AA55AA} & AFPP control
struct \\
checksum-valid sentinel & \texttt{0xFFFFFFFF} & command checksum
initialized and valid \\
\rowcolor{zebra}invalid id & \texttt{0xFFFFFFFF} &
\texttt{ECSneCmdId\_Invalid}, unbound program or process \\
padding & \texttt{0x00000000} & command padding must be zero \\
\rowcolor{zebra}power-status byte & \texttt{0xFF} / \texttt{0x00} &
fully on / fully off \\
\end{longtable}

This section shows the three loader magic words as 32-bit integers; on
disk the bytes are little-endian, so a raw byte scan finds
\texttt{HENA}, \texttt{PENA}, and \texttt{SENA} (the characters of
\texttt{ANEH}, \texttt{ANEP}, \texttt{ANES} reversed). The
firmware-to-host notification names, inline \texttt{status=0x\%x} print
sites, AArch64 and L2C fault-register dump fields, and compiler
diagnostic categories are from collected data not reproduced here.

\section*{C.4 The tunable register-init
table}\label{c.4-the-tunable-register-init-table}
\addcontentsline{toc}{section}{C.4 The tunable register-init table}

The per-chip register init is a sequence of 12-byte
\texttt{(offset,\ mask,\ value)} records, each applied as a masked
read-modify-write,
\texttt{reg\ =\ (reg\ \&\ \textasciitilde{}mask)\ \textbar{}\ value},
where \texttt{reg} is the block MMIO base plus the offset. The M1 (ASC
AscChinook) firmware has 1994 records across 10 named MMIO blocks, each
block reached through a 32-byte descriptor of name, MMIO base, record
pointer, and count (\cref{tbl:apc-reginit-blocks}).



One or more records from each block are reproduced in
\cref{tbl:apc-reginit-records}.

%

The 32 \texttt{mask=1\ value=1} records at a regular stride are direct
evidence of the 32-tile MAC array geometry, each tile individually clock
and power gateable. The DPE system block has seven ascending sampling
thresholds (25, 50, 70, 85, 95, 105, 115), and the SoC
leakage-estimation block has eight (10, 22, 39, 64, 89, 121, 164, 189):
the firmware-side breakpoints of the power model. All 1994 decoded
records are from collected data not reproduced here.

\section*{\texorpdfstring{C.5 The \texttt{CSNE\_CMD\_*} numeric command
table}{C.5 The CSNE\_CMD\_* numeric command table}}\label{c.5-the-csne_cmd_-numeric-command-table}
\addcontentsline{toc}{section}{C.5 The \texttt{CSNE\_CMD\_*} numeric
command table}

The host-to-firmware command set is 93 entries, numbered \texttt{0x00}
through \texttt{0x5c}, indexed by \texttt{eCSneCmdId} into the firmware
command-name string table, whose index is the numeric command
identifier. \texttt{0xFFFFFFFF} is the no-command sentinel. The full set
follows in \cref{tbl:apc-command-table}, whose \texttt{dir} column reads
\texttt{H-\textgreater{}FW} for a host request and
\texttt{FW-\textgreater{}H} for a firmware notification; the subsystem
codes are lifecycle, power, secure, program, execution, cache, ipc,
buffer, property, and stats.



Two strings the prior corpus counted have no numeric identifier:
\texttt{CSNE\_CMD\_START} is a standalone lifecycle log alias, and
\texttt{CSNE\_CMD\_IPC\_ENDPOINT\_TYPE\_DATA\_CHAINING} is an
endpoint-type enum value rather than a command. The per-call numeric
limits the command bodies enforce on the M1 are fixed. The dispatch caps
are at most 16 signal events per call, at most 32 custom barriers on the
wire (128 in the program container), at most 128 custom execute-order
entries, at most 16 trigger input buffers, and fewer than 2 active
shared events. Priority levels run 0 through 7, split into a privileged
band of 0 and 1 and a normal band of 2 through 7. The single dispatch
path takes exactly one output buffer set, one task-descriptor partition,
and one engine request per list.

A fixed header, \texttt{sCSneControllerCmdHdr}, prefixes every ring
message (\cref{tbl:apc-command-header}).



The full 93-entry table with file offsets, the decoded request structs,
and the per-call numeric limits are from collected data not reproduced
here.

\section*{C.6 The task-descriptor hardware register
map}\label{c.6-the-task-descriptor-hardware-register-map}
\addcontentsline{toc}{section}{C.6 The task-descriptor hardware register
map}

A captured \texttt{ane\_reg} record is a \texttt{(regAddr,\ regValue)}
pair whose low address selects one of 7 aperture groups. The aperture
map converts a raw address to a group (\cref{tbl:apc-aperture-map}).

%

The G1 dimension fields also pack the format and control bits, a
representative set of them in \cref{tbl:apc-g1-fields}.

%

To invert a raw value: DMA strides are 26-bit signed at bits {[}31:6{]};
L2-result base and strides are 17-bit at bits {[}20:4{]}; a full device
address is \texttt{(hi\ \textless{}\textless{}\ 32)\ \textbar{}\ lo}
with \texttt{lo} 64-byte aligned and \texttt{hi} 10 bits, capped at 42
bits. The complete inventory of roughly 190 register fields across the 7
groups, 11 reloc slots, and on-M1 stubbed engines (CCDMA, atomic
scatter, LDTID) is from collected data not reproduced here.

Each operation descriptor in the task-descriptor stream has a 32-bit
opcode word. Three such words were decoded from a live M1 program
(\cref{tbl:apc-opcode-words}).

\begin{longtable}[]{@{}ll@{}}
\caption{The version-7 / H13 codegen opcode words for three operations,
with the high half-word shared and the low 16 bits selecting the
operation.}\label{tbl:apc-opcode-words}\tabularnewline
\toprule\noalign{}
\textbf{operation} & \textbf{opcode word} \\
\midrule\noalign{}
\endfirsthead
\toprule\noalign{}
\textbf{operation} & \textbf{opcode word} \\
\midrule\noalign{}
\endhead
\bottomrule\noalign{}
\endlastfoot
convolution & \texttt{0x5042a063} \\
\rowcolor{zebra}reduce-mean & \texttt{0x5000a021} \\
matrix multiply & \texttt{0x5000b021} \\
\end{longtable}

\section*{\texorpdfstring{C.7 The \texttt{.e5} FlatBuffer
schema}{C.7 The .e5 FlatBuffer schema}}\label{c.7-the-.e5-flatbuffer-schema}
\addcontentsline{toc}{section}{C.7 The \texttt{.e5} FlatBuffer schema}

The program container is a FlatBuffer whose root table has four fields,
the schema given as \cref{lst:apc-e5-schema}. The schema reconstructs
from the serializer method set and the wire bytes, since the binary
strips the reflection schema, and round-trips cleanly through the
FlatBuffers tool to 23 tables, 4 enums, and 1 union. The data-type and
op-type enums are recovered by name; the numeric ordinals are inferred.

\par\addvspace{\medskipamount}\noindent\captionof{listing}{The reconstructed program-container FlatBuffer schema: root table, type enums, tensor descriptor, section tables, and operation structure.}\label{lst:apc-e5-schema}\nopagebreak\par\vspace{1pt}

\begin{Shaded}
\begin{Highlighting}[]
\KeywordTok{namespace}\NormalTok{ E5RT.fb;}

\KeywordTok{enum}\NormalTok{ TensorDataType : }\DataTypeTok{int}\NormalTok{ \{}
\NormalTok{  Invalid = }\DecValTok{0}\NormalTok{, Float16 = }\DecValTok{1}\NormalTok{, Float32 = }\DecValTok{2}\NormalTok{, Int8 = }\DecValTok{3}\NormalTok{, UInt8 = }\DecValTok{4}\NormalTok{,}
\NormalTok{  Int16 = }\DecValTok{5}\NormalTok{, Int32 = }\DecValTok{6}\NormalTok{, Int4 = }\DecValTok{7}\NormalTok{, Bool = }\DecValTok{8}\NormalTok{, E4M3 = }\DecValTok{9}\NormalTok{, E5M2 = }\DecValTok{10}
\NormalTok{\}}

\KeywordTok{enum}\NormalTok{ OpType : }\DataTypeTok{int}\NormalTok{ \{}
\NormalTok{  Cast = }\DecValTok{0}\NormalTok{, AneInference = }\DecValTok{1}\NormalTok{, EirInference = }\DecValTok{2}\NormalTok{, CpuInference = }\DecValTok{3}\NormalTok{,}
\NormalTok{  BnnsCpuInference = }\DecValTok{4}\NormalTok{, MlcCpuInference = }\DecValTok{5}\NormalTok{, MpsGraphInference = }\DecValTok{6}\NormalTok{,}
\NormalTok{  E5MinimalCpu = }\DecValTok{7}\NormalTok{, Quant = }\DecValTok{8}\NormalTok{, Dequant = }\DecValTok{9}\NormalTok{, Barrier = }\DecValTok{10}\NormalTok{, JitCall = }\DecValTok{11}
\NormalTok{\}}

\KeywordTok{table}\NormalTok{ TensorDescriptor \{}
\NormalTok{  dim:[}\DataTypeTok{ulong}\NormalTok{];}
\NormalTok{  stride:[}\DataTypeTok{ulong}\NormalTok{];}
\NormalTok{  width:}\DataTypeTok{ulong}\NormalTok{;}
\NormalTok{  height:}\DataTypeTok{ulong}\NormalTok{;}
\NormalTok{  channels:}\DataTypeTok{ulong}\NormalTok{;}
\NormalTok{  batch\_number:}\DataTypeTok{ulong}\NormalTok{;}
\NormalTok{  sequence\_length:}\DataTypeTok{ulong}\NormalTok{;}
\NormalTok{  stride\_width:}\DataTypeTok{ulong}\NormalTok{;}
\NormalTok{  stride\_height:}\DataTypeTok{ulong}\NormalTok{;}
\NormalTok{  stride\_channels:}\DataTypeTok{ulong}\NormalTok{;}
\NormalTok{  stride\_batch\_number:}\DataTypeTok{ulong}\NormalTok{;}
\NormalTok{  stride\_sequence\_length:}\DataTypeTok{ulong}\NormalTok{;}
\NormalTok{  storage\_type:TensorDataType;}
\NormalTok{  component\_pack:}\DataTypeTok{int}\NormalTok{;}
\NormalTok{\}}

\KeywordTok{table}\NormalTok{ BuildInfoEntry \{ key:}\DataTypeTok{string}\NormalTok{; value:}\DataTypeTok{string}\NormalTok{; \}}
\KeywordTok{table}\NormalTok{ BuildInfo \{ entries:[BuildInfoEntry]; \}   }\CommentTok{// 7 key{-}value pairs in the sample}

\KeywordTok{table}\NormalTok{ AliasSymbol \{ name:}\DataTypeTok{string}\NormalTok{; symbol\_index:}\DataTypeTok{uint}\NormalTok{; addr\_offset:}\DataTypeTok{uint}\NormalTok{; \}}

\KeywordTok{table}\NormalTok{ IOPort \{ name:}\DataTypeTok{string}\NormalTok{; byte\_size:}\DataTypeTok{ulong}\NormalTok{; aperture\_va:}\DataTypeTok{ulong}\NormalTok{; \}}

\KeywordTok{table}\NormalTok{ Operand \{ descriptor:TensorDescriptor; \}}

\KeywordTok{table}\NormalTok{ CastAttrs \{ src\_dtype:TensorDataType; dst\_dtype:TensorDataType; component\_pack:}\DataTypeTok{int}\NormalTok{; \}}

\KeywordTok{table}\NormalTok{ AneInferenceAttrs \{}
\NormalTok{  procedure\_name:}\DataTypeTok{string}\NormalTok{;}
\NormalTok{  anehash:}\DataTypeTok{string}\NormalTok{;}
\NormalTok{  program\_symbol:}\DataTypeTok{string}\NormalTok{;}
\NormalTok{  intermediate\_buffer\_handle:}\DataTypeTok{uint}\NormalTok{;}
\NormalTok{  compiler\_options:}\DataTypeTok{string}\NormalTok{;}
\NormalTok{\}}

\KeywordTok{table}\NormalTok{ Operation \{}
\NormalTok{  name:}\DataTypeTok{string}\NormalTok{;}
\NormalTok{  op\_type:OpType;}
\NormalTok{  inputs:[}\DataTypeTok{uint}\NormalTok{];}
\NormalTok{  outputs:[}\DataTypeTok{uint}\NormalTok{];}
\NormalTok{  arg\_frame:}\DataTypeTok{string}\NormalTok{;        }\CommentTok{// \_\_arg\_frame section reference}
\NormalTok{  attrs\_section:}\DataTypeTok{string}\NormalTok{;    }\CommentTok{// \_\_op\_attrs section reference}
\NormalTok{\}}

\KeywordTok{table}\NormalTok{ Block \{ name:}\DataTypeTok{string}\NormalTok{; operations:[Operation]; \}}

\KeywordTok{table}\NormalTok{ Function \{ name:}\DataTypeTok{string}\NormalTok{; anehash\_path:}\DataTypeTok{string}\NormalTok{; blocks:[Block]; \}}

\KeywordTok{table}\NormalTok{ Section \{ name:}\DataTypeTok{string}\NormalTok{; kind:}\DataTypeTok{int}\NormalTok{; \}}

\KeywordTok{table}\NormalTok{ E5Program \{}
\NormalTok{  symbol\_names:[}\DataTypeTok{string}\NormalTok{];   }\CommentTok{// field[0] name vector}
\NormalTok{  build\_info:BuildInfo;    }\CommentTok{// field[1] 7{-}entry sub{-}table}
\NormalTok{  sections:[Section];      }\CommentTok{// field[2] 6{-}entry section vector}
\NormalTok{  format\_version:}\DataTypeTok{int}\NormalTok{;      }\CommentTok{// field[3] inline scalar == 4}
\NormalTok{\}}

\KeywordTok{root\_type}\NormalTok{ E5Program;}
\end{Highlighting}
\end{Shaded}

\addvspace{\medskipamount}

The whole fused graph collapses to a single \texttt{AneInference}
operation, with the surrounding \texttt{Cast} operations holding the
input and output dtype conversion. The validation against the round-9
\texttt{H13C.e5} sample reads the four root fields, the seven build-info
pairs (\texttt{built-for-profiling}, \texttt{input-file-path}, the
component versions, and \texttt{on-device-compilation}), and the
operation chain \texttt{Cast}, \texttt{AneInference}, \texttt{Cast}
straight out of the bytes with no contradiction. The enum ordinals, the
\texttt{e4m3} and \texttt{e5m2} dtypes, segment-chaining fields, and
field sets of the ten op-attribute tables other than \texttt{CastAttrs}
and \texttt{AneInferenceAttrs} are inferred rather than byte-confirmed
in this single-segment sample.

\chapter{Glossary}\label{appendix-d.-glossary}

\begin{summarybox}

This appendix defines the acronyms, proper nouns, and symbols the guide
uses across all nine parts.

\end{summarybox}

The term \cref{tbl:apd-terms} is the reference the rest of the guide
leans on.

\section{Core facts}\label{core-facts}

Read these four corrections first; several entries in the table below
depend on them. The M5 base part is H17, specifically the \texttt{h17s}
compiler target, not H16s; H16s and H17s are separate targets
distinguished only by the generation-tag byte. The multiply-accumulate
datapath uses a single wide accumulator of the fp32 class, supplied by
radix-4 fp16-rounded input tiles; the accumulator width is fixed
hardware on every device and is never a per-chip parameter, so it is not
a recursive fp16 reduction tree. On M1/H13 two weight-compression forms
stream natively, the int4 palette (int4-LUT) and the sparse form whose
mask and values have at least 50 percent zeros; int8 and
blockwise-affine weights fold into the descriptor on that generation and
only stream natively from later families. The firmware task-queue
notification identifier (NID) is 8-bit, taking values 1 through 255.

\section{Terms}\label{terms}

\begin{longtable}[]{@{}
  >{\raggedright\arraybackslash}p{(\linewidth - 2\tabcolsep) * \real{0.5000}}
  >{\raggedright\arraybackslash}p{(\linewidth - 2\tabcolsep) * \real{0.5000}}@{}}
\caption{The acronyms, proper nouns, and symbols used across the guide
with their definitions.}\label{tbl:apd-terms}\tabularnewline
\toprule\noalign{}
\begin{minipage}[b]{\linewidth}\raggedright
\textbf{Term}
\end{minipage} & \begin{minipage}[b]{\linewidth}\raggedright
\textbf{Definition}
\end{minipage} \\
\midrule\noalign{}
\endfirsthead
\toprule\noalign{}
\begin{minipage}[b]{\linewidth}\raggedright
\textbf{Term}
\end{minipage} & \begin{minipage}[b]{\linewidth}\raggedright
\textbf{Definition}
\end{minipage} \\
\midrule\noalign{}
\endhead
\bottomrule\noalign{}
\endlastfoot
A11 through A18, M1 through M5 & Apple system-on-chip marketing names,
each with an ANE of a specific H-generation, related by
\(M(n) = H(n+12)\). \\
\rowcolor{zebra}AFPP & The on-device firmware program container: a
three-level big-endian FourCC package, \texttt{ANEH} then \texttt{ANEP}
then sections. \\
\texttt{aned} & The system ANE broker daemon at
\texttt{/usr/libexec/aned} that holds the IOKit access gate, so every
unentitled client reaches the engine by sending it a request. \\
\rowcolor{zebra}\texttt{aneuserd} & The per-user sibling of
\texttt{aned}, the other holder of the IOKit access gate. \\
ANEC, \texttt{anec.*} & The compiler backend intermediate-language
dialect of 97 operations, the target of front-end lowering and the input
to task-descriptor codegen. \\
\rowcolor{zebra}ANECCompile & The backend compile entry point that
direct netplist authoring supplies rather than bypasses. \\
ANECompiler & The single compiler binary that lowers the front IR to the
backend IR and then to task descriptors for every target, so one host
can construct any of the 28 targets' hardware-abstraction blobs. \\
\rowcolor{zebra}ANECompilerService & The out-of-process compile service;
repeated failed compiles in quick succession can stall it, so pace
compiles after a failure by about 15 seconds, covered in Part V. \\
ANEServices & The user-space framework layer beneath the runtime that
marshals requests into IOKit calls. \\
\rowcolor{zebra}AppleH11ANEInterface & The kernel driver for the engine,
decompiled at version 9.511.3, that holds the IOKit class hierarchy and
the user client. \\
ASC\_CHINOOK & The firmware's internal chip codename string for the H13
ANE coprocessor. \\
\rowcolor{zebra}bridge ops & Hidden backend layer kinds reached by
direct netplist authoring, such as fused attention, fused rank, and
fused rearrange, each paired with one frontend bridge module, described
in Part II and cataloged in Appendix
\hyperref[appendix-b.-hidden-layer-catalog]{B}. \\
CCDMA & The cross-chip and cross-engine DMA and event-sync engine; on
M1/H13 it is folded, and it is present natively on A15 and later and on
M5, where it enables resident state. \\
\rowcolor{zebra}CSNE, \texttt{CSNE\_CMD\_*} & The host-to-firmware
command protocol, where \texttt{CSNE\_CMD\_*} are the numeric command
opcodes the host mailbox issues to the firmware. \\
DART & The ANE's IOMMU, a 16 KB-page, 3.5 GiB-window unit that maps host
physical RAM into the engine's device address space. \\
\rowcolor{zebra}DPE & Dynamic Power Estimation: a firmware
activity-counter power estimate calibrated by 10 device-tree
coefficients and bounded by the peak-power ceiling. \\
DVA & Device Virtual Address: the address the engine issues, resolved by
DART to physical RAM, used interchangeably with IOVA. \\
\rowcolor{zebra}dispatch floor & The fixed per-dispatch latency, about
0.23 ms on the M1 anchor and a fitted 0.11 ms on the M5, below which a
kernel cannot run regardless of its size, covered in Part III. \\
\texttt{.e5} & The compiled-program dispatch-layer container whose size
tracks the segment and dispatch count rather than the operation
count. \\
\rowcolor{zebra}e5rt & The E5 runtime, the C API that the frontend uses
to load and stream programs and the unentitled reachable surface, which
still relays to \texttt{aned} underneath. \\
EIR, NitroIR & The runtime's lowered IR, a Lisp-style S-expression node
tree serialized on disk, whose pivot type is the fp16
\texttt{ndarray\textless{}half\textgreater{}}. \\
\rowcolor{zebra}Espresso & Apple's cross-backend neural-network runtime
and scheduler that hosts the E5 execution engine and the cost-model
placement segmenter. \\
ExeLoop & The firmware's main control execution loop and finite-state
machine that fetches and dispatches task descriptors through the RUN,
IDLE, and EXEC states. \\
\rowcolor{zebra}fp16 & Half-precision IEEE float, the engine's native
compute and storage type, with a maximum finite magnitude of 65504. \\
fp16 accumulator & A shorthand tag for the MAC numeric behavior: a
single wide accumulator of the fp32 class supplied by radix-4
fp16-rounded input tiles, where the only quantization is fp16 input and
partial rounding and the fp16 output grid. \\
\rowcolor{zebra}fp8 (E4M3, E5M2) & The 8-bit float weight and activation
datapath, present only on H18, which decodes to fp16 before the MAC. \\
generation-tag & The hardware-abstraction blob's generation byte at
offset \texttt{0x0}, holding the hex H-number, the decisive
discriminator between near-identical targets such as H16s and H17s. \\
\rowcolor{zebra}GOC, DynamicGOC & Generate-Output-Channels, the dynamic
unit that generates the output-channel-group kernel tiles from a runtime
weight, present from M1 onward. \\
HAL & The Hardware Abstraction Layer: the per-target scalar and byte
blob that data-drives nearly all per-family behavior and is the source
of truth for capability, limit, and cost, detailed in Part IX. \\
\rowcolor{zebra}\texttt{.hwx} & The fully lowered hardware-executable
container, the counterpart to the \texttt{.e5}. \\
int4-LUT, palette weights & Palettized 4-bit weight compression, one of
the two formats that stream natively on M1/H13 (alongside the sparse
form) at about 2.37 times, where int8 and blockwise-affine instead fold
into the descriptor. \\
\rowcolor{zebra}IOVA & IO Virtual Address: the device virtual address
DART produces from host physical RAM, on a 16 KB page over a 3.5 GiB
window. \\
KMEM & The on-chip working and scratch buffer the task descriptor sizes
for weights and tiles, gated at 64 KB at legalization and the basis of
the working-set threshold. \\
\rowcolor{zebra}KV-cache (resident) & An on-device key and value cache
that stays resident across dispatches for decode, built on M1 with
\texttt{share\_buffer} rather than native state, covered in Part
VIII. \\
LUT & Lookup table, used both for piecewise-linear activation
approximation and for the palette of int4 weight compression. \\
\rowcolor{zebra}MAC & Multiply-accumulate, the compute primitive whose
datapath is an fp16 multiply, radix-4 fp16-rounded input tiles, and one
wide accumulator of the fp32 class. \\
MIL & The Model Intermediate Language, the front IR in single-assignment
form that the compiler segments and lowers to the backend IR. \\
\rowcolor{zebra}MLComputePlan & The model-framework introspection
surface that reports per-operation device assignment and a cost weight,
the readable view of the placement segmenter. \\
\texttt{.mlmodelc} & The compiled model bundle that pairs the runtime
net, shapes, weights, and the \texttt{.hwx}. \\
\rowcolor{zebra}multi-die, AllReduce, AllGather & The multi-die
collective-communication layer present on the multi-die H14 through H18
Max and Ultra-class dies, not a base-class feature. \\
NE core & A neural-engine compute core; the per-family count decodes
from the HAL as base 4, then 8, 16, 32, or 64 by suffix, described in
Part IX. \\
\rowcolor{zebra}NID & The firmware task-queue notification identifier
owned by the state machine, 8-bit and taking values 1 through 255. \\
OCG & The Output-Channel Group, the compiler's output-channel tiling
unit sized to the accumulator file, where a larger group means fewer DMA
re-bases. \\
\rowcolor{zebra}OC/cycle & The per-cycle output-channel throughput of
the MAC array, the roofline unit in the cost model. \\
Path-A & Direct netplist authoring, hand-writing the backend netplist to
reach hidden layer kinds, which supplies \texttt{ANECCompile} and so
cannot reach a lowering the backend rejects, covered in Part II. \\
\rowcolor{zebra}palettization & Weight compression that maps each weight
to a lookup-table index, the int4 form of which streams natively on
M1/H13. \\
power domains & The five independently gated power domains of the H13
ANE, by which the engine modulates power through the number it
energizes. \\
\rowcolor{zebra}PPL & Page Protection Layer: the kernel page-table
protection layer through which the ANE's DART leaf writes go. \\
PPT & The peak-power ceiling and throttle under which the Dynamic Power
Estimation values are bounded. \\
\rowcolor{zebra}\texttt{pushTDList} & The firmware function that hands a
task descriptor to the hardware and re-enters with an already-built
descriptor for resident chains. \\
roofline & The performance bound that takes the smaller of
compute-limited and bandwidth-limited rates as a function of arithmetic
intensity, the basis of the cost model in Part III. \\
\rowcolor{zebra}RTBuddy & Apple's coprocessor real-time-OS runtime
framework, the substrate the ANE firmware app runs on. \\
RTKit & The real-time-OS substrate beneath the ANE firmware, providing
the task and thread model and the synchronization primitives. \\
\rowcolor{zebra}\texttt{share\_buffer} & The runtime primitive that
aliases an output buffer to an input buffer after compile, giving a
zero-copy resident cache without native state. \\
slice ×16 saturation & An H13 codegen defect in which a slice with a
nonzero last-axis begin lowers to a scaled kernel that silently sends
values above 4094, which is \(65504 / 16\), to infinity; H13-only and
clean on H17. \\
\rowcolor{zebra}SoC T-number & The SoC part number, such as T8103 for
the M1 base and T8142 for the M5 base, which maps to an ANE H-generation
through the board-type sequence. \\
sparse-binary, SparseFmt & The sparse-weight compute format, where the
weight has a binary sparsity mask; the mask-and-values form streams
natively on M1/H13, while the packed sparse-binary palette-index form is
absent from the M1 version-5 descriptor and present from A15 and M5. \\
\rowcolor{zebra}SPTM & Secure Page Table Monitor: the kernel monitor
that, with the Page Protection Layer, governs page-table edits and
physical-frame ownership. \\
stack layers & The top-to-bottom software path from the frontend through
the runtime, the framework, \texttt{aned}, ANEServices, IOKit, and
firmware to silicon, in Part VIII. \\
\rowcolor{zebra}styx & The firmware and chip codename for the M1/H13
ANE. \\
TD, task descriptor & The hardware work unit the firmware loads and the
engine executes: a register-image descriptor of DMA sub-blocks and
framing, emitted per generation from the compiler's descriptor struct,
detailed in Part VII. \\
\rowcolor{zebra}TileDMA, KernelDMA & The conv datapath DMA engines: the
kernel source streams weight coefficients, the tile source streams input
activation tiles, and the tile destination writes outputs. \\
TM, Tensor-Mover & The firmware tile-manager driver that moves tiles and
drives the texture layers. \\
\rowcolor{zebra}TQ, task queue & The firmware queue the state machine
enqueues a program's task-descriptor partitions into. \\
wide accumulator & The fp32-class running sum of the MAC, which holds
small addends rather than dropping them, so a sum of representable terms
stays near-exact, covered in Part III. \\
\rowcolor{zebra}Winograd & The Winograd fast-convolution transform the
compiler can emit for small kernels, trading multiplies for
transforms. \\
Zin, ZinIr, ZinMir & The compiler's internal class namespaces: the
IR-object layer, the mid-IR build layer, and the task-descriptor codegen
layer. \\
\end{longtable}

\section{Family and silicon map}\label{family-and-silicon-map}

The relation \(M(n) = H(n+12)\) is anchored at both ends, with the live
M1 reporting \texttt{h13g} and the M5 cost-model trees decompiled on the
M5 host reporting \texttt{H17C} and \texttt{H17S}. The compiler-family
index drives operation legality, and the per-target HAL drives codegen,
limits, and cost. \Cref{tbl:apd-family-map} maps each marketing name to
its ANE generation, architecture string, compiler family,
generation-tag, and core counts; Part IX gives the full table.

\begin{longtable}[]{@{}
  >{\raggedright\arraybackslash}p{(\linewidth - 10\tabcolsep) * \real{0.1667}}
  >{\raggedright\arraybackslash}p{(\linewidth - 10\tabcolsep) * \real{0.1667}}
  >{\raggedright\arraybackslash}p{(\linewidth - 10\tabcolsep) * \real{0.1667}}
  >{\raggedright\arraybackslash}p{(\linewidth - 10\tabcolsep) * \real{0.1667}}
  >{\raggedright\arraybackslash}p{(\linewidth - 10\tabcolsep) * \real{0.1667}}
  >{\raggedright\arraybackslash}p{(\linewidth - 10\tabcolsep) * \real{0.1667}}@{}}
\caption{The Apple chip marketing names mapped to ANE generation,
architecture string, compiler family, generation-tag, and core
counts.}\label{tbl:apd-family-map}\tabularnewline
\toprule\noalign{}
\begin{minipage}[b]{\linewidth}\raggedright
\textbf{Marketing name}
\end{minipage} & \begin{minipage}[b]{\linewidth}\raggedright
\textbf{ANE H-gen}
\end{minipage} & \begin{minipage}[b]{\linewidth}\raggedright
\textbf{OS arch string}
\end{minipage} & \begin{minipage}[b]{\linewidth}\raggedright
\textbf{Compiler family}
\end{minipage} & \begin{minipage}[b]{\linewidth}\raggedright
\textbf{generation-tag}
\end{minipage} & \begin{minipage}[b]{\linewidth}\raggedright
\textbf{NE cores by suffix}
\end{minipage} \\
\midrule\noalign{}
\endfirsthead
\toprule\noalign{}
\begin{minipage}[b]{\linewidth}\raggedright
\textbf{Marketing name}
\end{minipage} & \begin{minipage}[b]{\linewidth}\raggedright
\textbf{ANE H-gen}
\end{minipage} & \begin{minipage}[b]{\linewidth}\raggedright
\textbf{OS arch string}
\end{minipage} & \begin{minipage}[b]{\linewidth}\raggedright
\textbf{Compiler family}
\end{minipage} & \begin{minipage}[b]{\linewidth}\raggedright
\textbf{generation-tag}
\end{minipage} & \begin{minipage}[b]{\linewidth}\raggedright
\textbf{NE cores by suffix}
\end{minipage} \\
\midrule\noalign{}
\endhead
\bottomrule\noalign{}
\endlastfoot
A13, M1 & H13 & \texttt{h13}, \texttt{h13g} & A13 & \texttt{0x0d} & 4, 8
(g) \\
\rowcolor{zebra}A14, M2 & H14 & \texttt{h14}, \texttt{h14g},
\texttt{h14c} & A14 & \texttt{0x0e} & 4, 8, 32 \\
A15, M3 & H15 & \texttt{h15}, \texttt{h15g}, \texttt{h15c} & A15 &
\texttt{0x0f} & 4, 8, 32 \\
\rowcolor{zebra}A16, M4 & H16 & \texttt{h16}, \texttt{h16s} & A16 &
\texttt{0x10} & 4, 8, 16, 32 \\
A17, M5 & H17 & \texttt{h17}, \texttt{h17s} & A17 & \texttt{0x11} & 4,
8, 16 (M5), 32, 64 \\
\rowcolor{zebra}A18 & H18 & \texttt{h18} & A18 & \texttt{0x12} & 4 \\
\end{longtable}

The M5 base is the \texttt{h17} runtime arch and the \texttt{H17s}
compiler target, the 16-core variant, not H16s. The suffixes \texttt{g},
\texttt{s}, \texttt{c}, and \texttt{d} decode to NE-core counts of 8,
16, 32, and 64 from a single HAL field. A17 and A18 add no new operation
capabilities over A16 and scale only the core count; the A13 to A16 jump
was the last capability expansion. The fp8 datapath is H18 only.

\chapter{Provenance}\label{appendix-e.-provenance}

This appendix records the evidentiary basis of every substantive claim
in the guide, Part by Part and chapter by chapter. The method is the one
given in the Methodology chapter: the engine was reached directly below
Core ML, the stack was read by static decompilation of the runtime,
compiler, kernel driver, and firmware, and values were taken by live
read-only instrumentation and by compile-and-run probing. Two silicon
points were measured directly, M1/H13 (Apple M1) as the primary host and
M5/H17s (Apple M5) as the second, with an A14-class part (Apple M2, H14)
as a middle point where a claim required one. Each claim below is marked
one of three ways: measured on a named generation, decompile-derived
from a named binary, or predicted from the per-chip tables and not yet
confirmed on silicon.

\section*{Part I. The Machine}\label{part-i.-the-machine}
\addcontentsline{toc}{section}{Part I. The Machine}

Unless a chapter below notes otherwise, every figure in Part I was taken
on M1/H13 (Apple M1).

\subsection*{\texorpdfstring{Chapter \hyperref[what-the-ane-is]{1}. What
the ANE
is}{Chapter 1. What the ANE is}}\label{chapter-1.-what-the-ane-is}
\addcontentsline{toc}{subsection}{Chapter {1}. What the ANE is}

Apple documents the engine only as the \texttt{MLComputeUnits}
compute-unit selector at
developer.apple.com/documentation/coreml/mlcomputeunits, a placement
hint with no direct device API and no way to confirm which unit ran a
segment; this chapter extends that account by reaching the engine
directly below the selector. The roofline figures and the convolution
advantage over the GPU (3.8x faster, 9x more energy efficient) are
M1/H13 measured. The fp16-product and wide-accumulator result is
decompile-derived from the firmware and the compiler and confirmed by
the M1/H13 cancellation probe. The Core ML placement planner and the
direct Espresso route are decompile-derived.

\subsection*{\texorpdfstring{Chapter \hyperref[execution-model]{2}.
Execution
model}{Chapter 2. Execution model}}\label{chapter-2.-execution-model}
\addcontentsline{toc}{subsection}{Chapter {2}. Execution model}

Apple documents only the load-and-predict surface (\texttt{MLModel},
\texttt{prediction(from:)}); the autonomous-coprocessor model in this
chapter, with its command mailbox, walked segment graph, and resident
state across dispatches, has no public counterpart and is reported as
reverse-engineered. The compile-once, dispatch-many split and the
disk-cached program are decompile-derived and confirmed by M1/H13
dispatch tracing. The mailbox-and-doorbell command channel, the
autonomous controller, operand mapping through the address-translation
unit, and the walked segment graph with its static-control-flow
consequence are decompile-derived from the firmware, kernel-driver, and
program-format work. The resident-state mechanism through
output-to-input buffer aliasing is M1/H13 measured, observed to persist
and update an accumulator and a key-value cache across successive
dispatches with no host resubmission.

\subsection*{\texorpdfstring{Chapter \hyperref[numerics]{3}.
Numerics}{Chapter 3. Numerics}}\label{chapter-3.-numerics}
\addcontentsline{toc}{subsection}{Chapter {3}. Numerics}

The fp16 datapath and the type limits are decompile-derived from the
firmware and the compiler, with the activation-table coefficients read
out of the compiler binary constant section. The wide accumulator and
its radix-4 first stage, activation-table behavior and the
twenty-three-op accuracy sweep, NaN coercion and the gelu and swish
origin biases, round-half-to-even output grid, MAC saturation at exactly
\(2^{15}\), denormal handling, softmax max-subtraction, and
bit-deterministic output for a fixed graph and input are all M1/H13
measured. The slice saturation threshold at 4094 is M1/H13 measured and
reproduced on A14/H14 (Apple M2), where 4094 stays finite and 4096
overflows; the clean route arrives at A15 and later. The
cross-generation accumulator behavior holds family-wide, the M5/H17s
difference being a one-unit-in-the-last-place scheduling and tiling
reorder rather than an accumulator-width change. Denormal preservation
inside the M5/H17s accumulator, where the M1 flushes to zero inside the
multiply-accumulate, is M5/H17s measured and generational. The int8
relation here agrees with the affine form \(w = s\,(q - z)\) that
Apple's conversion tooling documents for low-precision compression at
apple.github.io/coremltools; nothing public covers the wide-accumulator
and per-product cancellation results, which are reported as
reverse-engineered.

\subsection*{\texorpdfstring{Chapter \hyperref[capability-surface]{4}.
Capability
surface}{Chapter 4. Capability surface}}\label{chapter-4.-capability-surface}
\addcontentsline{toc}{subsection}{Chapter {4}. Capability surface}

The native classes, type limits, compile-legal envelope, and M1 limits
are M1/H13 measured by operation-conformance runs, each native operation
compiled and run and each unsupported one rejected on device. The
attested-is-not-reachable rule is M1/H13 measured: three-dimensional
convolution fails backend lowering on every device mask despite its
capability byte, and the top-k, sort, and dynamic-slice validators are
callable but code-generation-rejected on the M1. The cumsum result is
M1/H13 measured through the curated runtime path, correcting the earlier
no-path status. The unsupported-on-every-family set and the per-family
unlock points for the texture-engine operations, sin and cos, and the
rank and sort bridge are decompile-derived from the operation floors and
validators and confirmed on the M1 only at its boundary; the chip that
first runs each is predicted from the floor table. The public account is
the convertible operation set at apple.github.io/coremltools and
developer.apple.com, which this chapter extends and partly corrects by
reporting what compiles and runs on the direct path, since some
operations the converter accepts, such as three-dimensional convolution,
do not lower to the engine.

\section*{Part II. Reaching the ANE}\label{part-ii.-reaching-the-ane}
\addcontentsline{toc}{section}{Part II. Reaching the ANE}

The same M1/H13 (Apple M1) default applies through Part II, except in
chapter \hyperref[weights-and-compression]{7}, which also reports
figures from M2/H14 (Apple M2 Pro) and M5/H17s (Apple M5).

\subsection*{\texorpdfstring{Chapter \hyperref[software-stack]{5}.
Software
stack}{Chapter 5. Software stack}}\label{chapter-5.-software-stack}
\addcontentsline{toc}{subsection}{Chapter {5}. Software stack}

Apple documents the compute-unit selector, model loading, and the
placement read-out (\texttt{MLComputePlan.load},
\texttt{deviceUsage(for:)}, \texttt{estimatedCost(of:)}) at
developer.apple.com/documentation/coreml, and that read-out agrees with
the segmenter decision reported here; the layered runtime beneath it and
its internal cost graph are not documented and are reported as
reverse-engineered. The layering and the runtime's ownership of compile,
library, stream, and descriptors; the shortest-path placement segmenter
with its per-operation-by-backend cost graph, learned decision trees,
and launch and transfer penalties; the broker model with its single
privileged device gate, content-hashed program cache, and time-shared
request queue; the 292-export runtime surface and its options
dictionary; and the per-inference submit through
\texttt{IOConnectCallAsyncMethod} selector 2 with the daemon's lifecycle
selectors 3 through 6 are decompile-derived from the framework, runtime,
and daemon binaries. The reachability of the runtime below the
framework, with no placement planner and no entitlement for accepted
operations, is decompile-derived and confirmed by M1/H13 dispatch
tracing.

\subsection*{\texorpdfstring{Chapter
\hyperref[dispatching-without-core-ml]{6}. Dispatching without Core
ML}{Chapter 6. Dispatching without Core ML}}\label{chapter-6.-dispatching-without-core-ml}
\addcontentsline{toc}{subsection}{Chapter {6}. Dispatching without Core
ML}

Apple documents only the indirect \texttt{MLModel} and
\texttt{prediction(from:)} route; the direct compile, load, bind, and
dispatch route here is reported as reverse-engineered and extends that
account with a planner-free path that targets the engine on purpose. The
five-step workflow and the absence of a placement planner are
decompile-derived from the runtime and the format pipeline. The absence
of an entitlement requirement for accepted operations is
decompile-derived and confirmed M1/H13 measured from ordinary user
space, and the single-submission multi-step drive with its
performance-neutral result and the resident-state buffer aliasing it
rests on are M1/H13 measured.

\subsection*{\texorpdfstring{Chapter
\hyperref[weights-and-compression]{7}. Weights and
compression}{Chapter 7. Weights and compression}}\label{chapter-7.-weights-and-compression}
\addcontentsline{toc}{subsection}{Chapter {7}. Weights and compression}

This chapter measures across three endpoints: M1/H13 (Apple M1), A14 on
M2/H14 (Apple M2 Pro), and M5/H17s (Apple M5). The int4 and sparse
streaming speedups and byte ratios, the int8 matmul latency and
weight-byte halving, and the M1 folding of the int8 and blockwise forms
are M1/H13 measured. The A14 int8 and sparse stream and the A14
blockwise fold are M2/H14 measured, and the streaming of all four forms
is M5/H17s measured. The weight-reconstruction codecs and the per-format
hardware-abstraction-layer streaming gates are decompile-derived from
the ANE compiler; the A15 floor at which the blockwise form first
streams is predicted from the gate pattern and not yet
silicon-confirmed. Palettization, quantization, and pruning appear in
Apple's conversion tools at apple.github.io/coremltools as a model-size
feature; the finding here is that the same forms stream on the
unentitled engine for a bandwidth gain.

\subsection*{\texorpdfstring{Chapter \hyperref[entitlement-boundary]{8}.
Entitlement
boundary}{Chapter 8. Entitlement boundary}}\label{chapter-8.-entitlement-boundary}
\addcontentsline{toc}{subsection}{Chapter {8}. Entitlement boundary}

The four gated features and the layer each gate is at are M1/H13
measured: three-dimensional convolution fails backend lowering on every
device mask, a native-state program fails the compile with the
counter-and-event engine stubbed in the M1 descriptor, a bf16 input or
output fails with an unsupported-dtype rejection and is absent from the
eleven-code program-I/O enumeration, and a symbolic dimension parses but
fails to lower. The framework cast of bf16 to fp16 and the bucketed
fixed-shape handling of flexible shapes are decompile-derived from the
framework and system bundles, and the load-time signature boundary is
decompile-derived from the kernel driver's corecrypto signature and
trustcache vnode-trust checks, with the load rejection code
\texttt{0xe00002e2} observed on the M1. The arrival of native resident
state on a later generation is predicted from the per-family hardware
descriptor and not measured on that silicon. Apple documents these
capabilities (flexible shapes, the \texttt{MLState} type) at the
conversion and runtime layer; the finding shows they are not reachable
on the unentitled direct path, correcting the impression that a
documented capability runs on the engine directly.

\section*{Part III. Performance and
Fit}\label{part-iii.-performance-and-fit}
\addcontentsline{toc}{section}{Part III. Performance and Fit}

Part III uses two passes, the first on M1/H13 (Apple M1; M1 Max) and the
second on M5/H17s (Apple M5), with an A14-class middle point on M2/H14
where a cross-device or cross-generation claim required one. Apple
publishes no roofline, no per-workload power or efficiency figures, and
no cross-processor comparison for the engine, so the figures across this
Part stand on their own measurements.

\subsection*{\texorpdfstring{Chapter \hyperref[roofline]{9}.
Roofline}{Chapter 9. Roofline}}\label{chapter-9.-roofline}
\addcontentsline{toc}{subsection}{Chapter {9}. Roofline}

The M1 compute and bandwidth ceilings, the 141 FLOP-per-byte ridge
point, 2 MB working-set threshold, 0.23 ms dispatch floor, and
conv-throughput scaling are M1/H13 measured from the memory-controller
and energy counters and end-to-end timing. The saturating-method figures
(the large-matmul compute ceiling, effective peak, wall-clock
weight-stream bandwidth matching the compiler's internal 50 GB/s
constant, int8 rate, and full-call floor) are M1/H13 measured live in a
read-only dtrace session, and are presented alongside the counter-based
and slope-based figures with the methodology difference noted. The
cross-device ridge points, the standalone and weight-stream bandwidths,
and fused-block rate are M5/H17s measured. The fusion-and-floor model is
decompile-derived from the analytic cost model and fit to the measured
M1 convs within plus or minus 17 percent; cross-chip scaling of the
ceilings is predicted from that model, not asserted as a measured M1
fact.

\subsection*{\texorpdfstring{Chapter
\hyperref[power-and-efficiency]{10}. Power and
efficiency}{Chapter 10. Power and efficiency}}\label{chapter-10.-power-and-efficiency}
\addcontentsline{toc}{subsection}{Chapter {10}. Power and efficiency}

The convolution-stack efficiency figures, the absolute-power comparison
on the 4096 matrix multiply, the 2-to-14-times efficiency range, and
sustained-load behavior are M1/H13 measured, with package power from
hardware instrumentation. The power-utilization model (the zero idle
rail, dispatch floor, fp16 and int8 compute-bound draw, regime-dependent
points, and operations-per-watt optimum near 0.37 pJ per FLOP) is M1 Max
measured with the root power sampler, which exposes only a power reading
and a binary on-or-off state, with no frequency or voltage telemetry.
The A14-class middle generation is measured on that silicon, and the M5
efficiency figures are M5/H17s measured.

\subsection*{\texorpdfstring{Chapter \hyperref[ane-gpu-and-cpu]{11}.
ANE, GPU, and
CPU}{Chapter 11. ANE, GPU, and CPU}}\label{chapter-11.-ane-gpu-and-cpu}
\addcontentsline{toc}{subsection}{Chapter {11}. ANE, GPU, and CPU}

The per-class speed and energy verdicts are M1/H13 and M5/H17s measured
by a single sixteen-class harness recording minimum latency,
idle-subtracted total-package power, and fp16 relative error per class.
The saturation peaks, the large-N matrix-multiply falloff, and serving
crossovers are M5/H17s (Apple M5 Pro) measured; the per-eval overhead
floor and the M1 power gap are M1/H13 measured.

\subsection*{\texorpdfstring{Chapter
\hyperref[across-the-chip-family]{12}. Across the chip
family}{Chapter 12. Across the chip family}}\label{chapter-12.-across-the-chip-family}
\addcontentsline{toc}{subsection}{Chapter {12}. Across the chip family}

The naming rule \(M(n) \rightarrow H(n+12)\) is decompile-derived from
the per-family device tables and confirmed on the measured parts: M1/H13
by the live h13g architecture string, M2/H14g on the live A14 host, and
M5/H17s by the resolved target. The family-wide operation limits and the
core-and-clock scaling are decompile-derived from the operation floors
and the per-target scalar tables. The M5 throughput and working-set
threshold, the ten-of-ten cross-silicon prediction pass, the
M1-versus-M5 training and inference parity (0.9080 against 0.9070,
deterministic across runs), and the one-unit-in-the-last-place fp16
divergence bound are M5/H17s and M1/H13 measured; the M2 campaign
measured training accuracy, the four fp16 axes, peak throughput,
compression, and the per-op max-dim caps on A14 silicon. The A15 and A16
generations and their M3 and M4 counterparts are decompile-derived from
the device tables and not individually measured, the A15/M3 rail being
the one generation that remains unmeasured. Apple's product
specifications publish a marketing core count per chip, for example a
16-core engine, a different quantity from the four physical compute sets
measured on the M1; the H-architecture naming, the mapping, and
per-family gates are reported as reverse-engineered. The M5
single-program matmul peak of about 9.5 fp16 TFLOP/s and the
weight-stream bandwidth of about 145 GB/s over two DRAM read channels
are M5/H17s measured.

\section*{Part IV. Workloads}\label{part-iv.-workloads}
\addcontentsline{toc}{section}{Part IV. Workloads}

Part IV pairs M1/H13 (Apple M1 Max) with M5/H17s (Apple M5 Pro) and adds
an A14-class point on M2/H14 where noted, each figure marked for the
generation it was taken on.

\subsection*{\texorpdfstring{Chapter
\hyperref[vision-convolution-and-encoders]{13}. Vision, convolution, and
encoders}{Chapter 13. Vision, convolution, and encoders}}\label{chapter-13.-vision-convolution-and-encoders}
\addcontentsline{toc}{subsection}{Chapter {13}. Vision, convolution, and
encoders}

The convolution speed and efficiency, the convolution-stack and roofline
figures, the power gap, and per-eval floor are M1/H13 measured; the M5
convolution-stack, ResNet-18, twelve-layer-encoder, and
single-sentence-encoder ratios and the serving crossovers are M5/H17s
measured. The convolution lowering, Winograd gate, 2 MB working-set
constant, per-output-channel fold, and texture-engine operation set with
its single A14 family gate are decompile-derived from the ANE compiler
and the per-chip parameter table. The Q.4 crop-scale saturation
threshold, where \(4094 \times 16 = 65504\) reaches the fp16 ceiling, is
decompile-derived and confirmed by the fp16 range probe on M1/H13 and on
A14 (M2); the clean route arrives on A15 and later. For vision and image
models on the engine the public surface is the compute-unit selector
alone (developer.apple.com/documentation/coreml,
developer.apple.com/documentation/vision), with no direct datapath API
and no cross-processor figures; the measured economics against the GPU
are new here, and so is the texture-engine preprocessing path, which
that surface does not expose.

\subsection*{\texorpdfstring{Chapter \hyperref[llm-case-study]{14}. LLM
case
study}{Chapter 14. LLM case study}}\label{chapter-14.-llm-case-study}
\addcontentsline{toc}{subsection}{Chapter {14}. LLM case study}

The per-eval dispatch floor, int8-hybrid result, and resident-cache step
are M1/H13 measured, the last by a two-proof resident-buffer probe; the
batched-decode comparison, the per-projection placement table and
position rule, and the speculative-decoding and batched-prefill serving
controls are M5/H17s measured. The dispatch and resident-state machinery
of the direct path is decompile-derived from the runtime, including the
output-to-input buffer aliasing that keeps the cache resident, the
in-flight cap of 127, and per-process loaded-program cap near 128 (the
next load returns \texttt{GetANEFModel:\ must\ re-compile}). Apple does
not publish engine-versus-GPU decode measurements, so the per-batch
decode verdict is reported as measured.

\subsection*{\texorpdfstring{Chapter
\hyperref[training-on-the-engine]{15}. Training on the
engine}{Chapter 15. Training on the engine}}\label{chapter-15.-training-on-the-engine}
\addcontentsline{toc}{subsection}{Chapter {15}. Training on the engine}

On-device model update, documented at
developer.apple.com/documentation/coreml, is a limited fine-tuning
surface whose backward pass runs off the engine; the full forward,
backward, and optimizer loop reported here runs as engine graph
operations, optimizer state resident across steps. The gradient audit,
differentiable-vocabulary correctness to a cosine of \(1.0000\), conv
weight-gradient saturation threshold, M1 training accuracy and
loss-scale curves, and two-generation parity (0.9080 on M1 against
0.9070 on M5 for the identical seeded network) are M1/H13 and M5/H17s
measured. The width-axis slice saturation, exact at loss scale 384 and
first overflowing at 512 above \(65504 / 16 \approx 4094\), was
reproduced on A14 (M2), locating the clean route on A15 and later. The
absence of an engine-native backward operation is decompile-derived from
the shared compiler, which has gradient operations only in the
graphics-processor dialect.

\subsection*{\texorpdfstring{Chapter
\hyperref[numerical-and-scientific-computing]{16}. Numerical and
scientific
computing}{Chapter 16. Numerical and scientific computing}}\label{chapter-16.-numerical-and-scientific-computing}
\addcontentsline{toc}{subsection}{Chapter {16}. Numerical and scientific
computing}

The iterative-solver envelope, the size bounds on the unrolled
factorizations, the full-spectral-decomposition relative errors, and
wide-accumulator behavior are M1/H13 measured, with the fused five-point
stencil margin measured on M5/H17s and the DFT-as-matmul throughput on
the M2 generation at \(N = 1024\). The static-dataflow constraint and
the absence of data-dependent control flow are decompile-derived from
the operation set and the compiler. The fp16-clean DFT bound near
\(N = 2048\) is predicted from the wide-accumulator reduction and the
fp16 rounding of the matrix entries, an edge of the representable range
consistent with the accumulator measurements. Apple documents dense
linear algebra and signal processing through the accelerate framework at
developer.apple.com/documentation/accelerate, all of it CPU-targeted;
this chapter maps which of those kernels fit the engine and which are
architecture-limited, a mapping the public documentation does not
provide.

\section*{Part V. Practice}\label{part-v.-practice}
\addcontentsline{toc}{section}{Part V. Practice}

For Part V the host was M1/H13 (Apple M1), with M5/H17s (Apple M5)
standing in as the cross-chip reference where a second generation is
needed.

\subsection*{\texorpdfstring{Chapter \hyperref[model-design-rules]{17}.
Model-design
rules}{Chapter 17. Model-design rules}}\label{chapter-17.-model-design-rules}
\addcontentsline{toc}{subsection}{Chapter {17}. Model-design rules}

The tensor-dimension boundaries, the convolution kernel and stride
limits, the arg-min and arg-max 2048 cap, and pooling-window behavior
are M1/H13 measured, swept until compile flipped from accept to reject.
The per-operation validators, the kernel-format and maximum-dimension
fields, the divisibility checks, and working-set and kernel-memory
budgets are decompile-derived and joined to those sweeps; the per-family
unlock points for the texture-engine padding modes, the
square-after-reduction mode, and sin and cos are decompile-derived from
the per-chip support flags, confirmed on the M1 only at its boundary and
predicted for the chip that first enables each. The conversion-time
constraints and supported converter configurations are documented at
apple.github.io/coremltools; what this chapter adds is the argument,
shape, and mode limits the on-device validators enforce, which the
public converter does not enumerate.

\subsection*{\texorpdfstring{Chapter
\hyperref[optimization-and-the-cost-model]{18}. Optimization and the
cost
model}{Chapter 18. Optimization and the cost model}}\label{chapter-18.-optimization-and-the-cost-model}
\addcontentsline{toc}{subsection}{Chapter {18}. Optimization and the
cost model}

The convolution latency fit and the dispatch floor are M1/H13 measured
(Apple M1; M1 Max), and the cross-chip bandwidth, floor, and peak re-fit
are M5/H17s measured, with the core-scaled bandwidth confirmed by direct
streaming at 57.5 GB/s against the M1's 10.4 GB/s. The compiler's
analytic cost functions (the cycles, roofline, and wall-time chain) are
decompile-derived with the per-chip parameters walked live from the
hardware-abstraction table; the cross-chip scaling of unmeasured targets
is predicted from core-count and clock ratios. The cost-model fidelity
is M1/H13 measured: a median error near 31 percent with 11 of 68 shapes
within plus or minus 17 percent, sound as an ordinal placement tool
rather than an absolute-latency oracle, with attention unmodeled and the
9.0 GB/s bandwidth anchor held below the roughly 40 GB/s effective rate
because it is jointly calibrated for the convolution fit. Apple does not
publish the engine compiler's cost model or its per-chip parameters, so
the model is reported here as decompile-derived and validated against
M1/H13 and M5/H17s measured latency.

\subsection*{\texorpdfstring{Chapter \hyperref[pitfalls-and-limits]{19}.
Pitfalls and
limits}{Chapter 19. Pitfalls and limits}}\label{chapter-19.-pitfalls-and-limits}
\addcontentsline{toc}{subsection}{Chapter {19}. Pitfalls and limits}

The slice-saturation threshold, dynamic-weight convolution batch
boundary, compile-failure back-off, and four-character-code lowering
limit are M1/H13 measured, and the A14 generation (M2) was measured to
saturate on the same slice, locating the clean route on A15 and above
and confirmed clean on the M5. The per-target slice-lowering template
and its DMA source-path and patch-width routines are decompile-derived,
giving the times-16 fixed-point DMA format and the target-keyed slice
route; the clean A15 route is decompile-derived from the per-family
lowering and confirmed clean on the M5, not measured on A15 silicon
directly. These are failure modes of the private compiler and have no
public counterpart: they are reported as measured on the M1 and
reverse-engineered from the compiler binary.

\section*{Part VI. The Silicon}\label{part-vi.-the-silicon}
\addcontentsline{toc}{section}{Part VI. The Silicon}

Part VI was measured on M1/H13 (Apple M1; M1 Max, live architecture
string \texttt{h13g}); the datapath geometry and memory hierarchy are
decompile-derived from the per-chip hardware-abstraction table and the
engine compiler, calibrated against the M1 anchors.

\subsection*{\texorpdfstring{Chapter
\hyperref[datapath-and-mac-geometry]{20}. Datapath and MAC
geometry}{Chapter 20. Datapath and MAC geometry}}\label{chapter-20.-datapath-and-mac-geometry}
\addcontentsline{toc}{subsection}{Chapter {20}. Datapath and MAC
geometry}

The core count of four, per-core throughput scaling of 1 to 4,
power-rail step per core, radix-4 fan-in, wide accumulator, and
output-channel-group pass-doubling threshold at 192 to 256 are M1/H13
measured. The accumulator budget of eight and the lane widths of four
and eight are decompile-derived and uniform across chips per the
hardware-abstraction table, which was carved from the H13 firmware blob
and calibrated at the core-count, cycle-divisor, and working-set
offsets; the convolution-lowering and performance-model functions are
decompile-derived from the ANE compiler
(\texttt{GetNumOutputChannelsPerCycle},
\texttt{GetNumOutputChannelsPerAccumulator}, \texttt{ComputeMaxOcgSize},
\texttt{ZinMirNECoreAssignment}, \texttt{GetNumNeededNEsNextPow2}).
Apple publishes a marketing core count per chip, for example a 16-core
M1, a different quantity from the decoded \texttt{num\_nes} of four; the
multiply-accumulate geometry has no public counterpart and is reported
as reverse-engineered and measured. The int8 compile flag is weight-only
quantization that leaves the multiply-accumulate in fp16, neutral on
compute-bound work and about 1.5 times faster only on
weight-bandwidth-bound matmuls near a 4096-by-4096 weight, M5/H17s
measured.

\subsection*{\texorpdfstring{Chapter \hyperref[memory-hierarchy]{21}.
Memory
hierarchy}{Chapter 21. Memory hierarchy}}\label{chapter-21.-memory-hierarchy}
\addcontentsline{toc}{subsection}{Chapter {21}. Memory hierarchy}

The 2.28 to 2.34 MB threshold and the 64-byte throughput period are
M1/H13 measured on the dispatch path, and the absence of a runtime
replacement policy is an M1/H13 measured negative, with sequential and
random re-reference order identical at every footprint. The field values
are decompile-derived: \texttt{0x1b8} (2 MB operand working set),
\texttt{0x1c8} (64 banks), \texttt{0x1c0} (16-byte granule),
\texttt{0x1f8} (2 MB stride ceiling), \texttt{0x1f0} (residency
threshold, 0 on M1), and \texttt{0x288} (64 KB kernel store); the
operand-size comparator, the bank function and conflict model, and
inverted residency-buffer gate are decompile-derived from the named
compiler routines, with the 2 MB boundary itself the operand-size
comparator. The A15-class, A16-class, and M5 values of field
\texttt{0x1f0} are read from the same table by offset but predicted for
those parts, the gate behavior confirmed on the M1 only. The memory
hierarchy, the bank function, and residency threshold have no public
counterpart and are reported as reverse-engineered and measured. The
compiler name for field \texttt{0x1b8}, \texttt{MemCacheSize} (also
\texttt{L2Size}) with its \texttt{fl2-size} override, is
decompile-derived; on the M5 the working-set crossing is smooth in
throughput and shows instead in DRAM energy per operation, bottoming
near a 2 MB operand, M5/H17s measured.

\section*{Part VII. The Toolchain and
Encoding}\label{part-vii.-the-toolchain-and-encoding}
\addcontentsline{toc}{section}{Part VII. The Toolchain and Encoding}

Part VII shares that M1/H13 host (Apple M1; M1 Max, live string
\texttt{h13g}), and chapter \hyperref[compression-internals]{25} extends
to M2/H14 (Apple M2 Pro) and M5/H17s (Apple M5). This Part is mostly
decompile-derived from the engine compiler decompile and its
constraint-string corpus, the per-chip hardware-abstraction table read
by byte offset across 28 target entries, and the runtime serializers and
task-descriptor setters; unless a chapter says otherwise, scalar
offsets, capability-byte offsets, struct fields, and symbol names are
decompile-derived.

\subsection*{\texorpdfstring{Chapter \hyperref[compiler]{22}.
Compiler}{Chapter 22. Compiler}}\label{chapter-22.-compiler}
\addcontentsline{toc}{subsection}{Chapter {22}. Compiler}

The four-phase pipeline, task-descriptor partition budget,
allocation-type set, \texttt{anec.matmul} and \texttt{anec.convolution}
lowerings, and fusion rules (the GOC fused unit, seven-slot epilogue,
fusable epilogues and the two-live-input, concat, and attention-cut
barriers) are decompile-derived from the engine compiler framework (the
9.509 build, a 4.1-million-line decompile) and the constraint-string
corpus. The validator export set, the per-layer reject strings, and the
code-generation rejections for top-k, sort, dynamic-slice, and
three-dimensional convolution are M1/H13 measured. The compiler
internals, the backend \texttt{anec.*} dialect, and the
\texttt{\_ANECValidate*} surface have no public counterpart and are
reported as reverse-engineered; the public tools document only the
frontend operation set and conversion passes.

\subsection*{\texorpdfstring{Chapter
\hyperref[program-and-container-format]{23}. Program and container
format}{Chapter 23. Program and container format}}\label{chapter-23.-program-and-container-format}
\addcontentsline{toc}{subsection}{Chapter {23}. Program and container
format}

The two-layer split, four-field root table, cast-inference-cast
operation shape, seven register groups, 15-bit and 17-bit dimension
widths, relocation-slot model, 44-byte sparse record, and the HWX
on-disk layout (the \texttt{0xbeefface} Mach-O variant, the segment set,
the \texttt{ZinAneTd} linked list, the per-lane weight tiles, and shape
descriptors) are decompile-derived from the serializer and
task-descriptor symbols and cross-confirmed against the vendor's
task-descriptor symbol table. The dispatch-descriptor schema was
validated by a round-trip through the schema compiler (version
25.12.19), which regenerated an object-API header and a binary
reflection schema without error. The decoded identity-linear program
with its segment map, port descriptors, register records, and weight
bank are M1/H13 measured, parsed byte for byte from real on-disk files
in the runtime caches, with the resolved tensor frame read from the
post-compile status sidecar. The format has no public counterpart and is
reported as reverse-engineered; the full FlatBuffer schema is Appendix
\hyperref[appendix-c.-decoded-reference-tables]{C}. The custom-bar
bit-field relocation, which patches a resolved value into a named
descriptor field by bit offset and width, and the range-checked live-in
shape and stride parameters that let one program serve a range of input
shapes, are decompile-derived from the program loader.

\subsection*{\texorpdfstring{Chapter
\hyperref[hal-and-capability-gates]{24}. HAL and capability
gates}{Chapter 24. HAL and capability gates}}\label{chapter-24.-hal-and-capability-gates}
\addcontentsline{toc}{subsection}{Chapter {24}. HAL and capability
gates}

The scalar offsets, the capability-byte offsets, and family floors are
decompile-derived, with every per-target constructor invoked on this
host (live string \texttt{h13g}) to read byte-exact values for all 28
targets, joined to the minimum-family operation trait and tier
assignment read from the same binary. The per-family unlock generations
for the texture engine, sin and cos, and the dimension and format-count
steps are decompile-derived from the per-target tables and confirmed on
the M1 only at its boundary; the generation that first enables each is
predicted from the floor table. The attested-is-not-reachable rule is
M1/H13 measured: three-dimensional convolution has its HAL kernel-depth
attestation at offset \texttt{0x70} and fails backend lowering on every
device mask, and the top-k, sort, and dynamic-slice validators are
callable but code-generation-rejected. The packed-bitfield struct
measures \texttt{0x938} bytes; its non-flag residual is the cost-model
coefficient block at \texttt{0x580} through \texttt{0x7f0} plus about
two soft fp64 coefficients, and the offsets past \texttt{0x938}, read in
an earlier round as an A12 operation-emulation catalog at \texttt{0xa30}
through \texttt{0xe84}, are a read into zeroed memory beyond the struct
and have no table; the capability-flag offsets are decompile-derived
from the \texttt{ZinIrHalParameters} reader symbols. The HAL table, the
capability-byte region, and minimum-family trait have no public
counterpart; Apple documents only the model framework and the
convertible operation set.

\subsection*{\texorpdfstring{Chapter
\hyperref[compression-internals]{25}. Compression
internals}{Chapter 25. Compression internals}}\label{chapter-25.-compression-internals}
\addcontentsline{toc}{subsection}{Chapter {25}. Compression internals}

The bit-layout and address detail are M1/H13 (table-descriptor codegen
version five, family two, A13) unless another version or family is
named. The public account of palettization, quantization, and pruning at
apple.github.io/coremltools treats them as a model-size feature; the
on-device codec arithmetic and the per-family streaming datapath go
beyond it. The sparse and int4 speedups and byte ratios, the
byte-identical compiled program, and the M1 fold of the int8 and
blockwise forms are M1/H13 measured; the A14 int8 stream and blockwise
fold are M2/H14 (Apple M2 Pro) measured; the all-forms-stream endpoint
is M5/H17s (Apple M5) measured. The kernel-format helper tables, affine
and palette dequantization codecs, dequantize-to-dense fold path,
streaming and palette gates, on-chip-memory budget caps, and Winograd
eligibility gate are decompile-derived, with the on-device sparse-format
field read from the register map; the A15 floor at which the blockwise
form first streams is predicted from the gate pattern, and the resident
Winograd transform matrices are predicted, their textbook forms matching
the engine's behavior but not byte-confirmable from the binary.

\subsection*{\texorpdfstring{Chapter
\hyperref[hidden-layers-and-direct-netplist-authoring]{26}. Hidden
layers and direct netplist
authoring}{Chapter 26. Hidden layers and direct netplist authoring}}\label{chapter-26.-hidden-layers-and-direct-netplist-authoring}
\addcontentsline{toc}{subsection}{Chapter {26}. Hidden layers and direct
netplist authoring}

The fused-attention, ranking, and spatial-rearrange layers were
authored, compiled, and dispatched on the M5/H17s byte-exact against a
host reference, and the M1 gates and the top-k forbidden band were
confirmed on the M1 (measured). The native layer-descriptor catalog, its
per-layer \texttt{ZinParse\textless{}Name\textgreater{}Unit} parsers and
\texttt{\_ANECValidate\textless{}Name\textgreater{}Layer} checkers, and
the constant-string constraint corpus are decompile-derived, joined to
the netplist schema read out of the runtime framework. The model
converter and its intermediate-language operation set, documented at
apple.github.io/coremltools, emit none of these native layer kinds; this
chapter authors them directly, the attention, ranking,
spatial-rearrange, geometry, and normalization descriptors being present
in the compiler and reachable through the network description even
though the converter never produces them. The 33-knot activation-LUT
format and the gated \texttt{NeuronCustom} netplist path (a parser that
requires and then rejects the same field sets) are decompile-derived;
the rectifier-basis reproduction of an arbitrary pointwise function is
M5/H17s measured.

\section*{Part VIII. System
Internals}\label{part-viii.-system-internals}
\addcontentsline{toc}{section}{Part VIII. System Internals}

Part VIII rests on M1/H13 (Apple M1; M1 Max, and the T6000-generation
engine where a multi-die part is needed), with an M2-class kernel cache
as the cross-generation reference where cited. Much of it is
decompile-derived static analysis with no firmware executed: the
unencrypted real-time-kernel preload executable is carved from the
on-package firmware image and read for its strings, asserts, and
disassembled handlers, and the kernel cache is read for its symbols,
dispatch arrays, and call sites.

\subsection*{\texorpdfstring{Chapter
\hyperref[kernel-driver-and-iokit-abi]{27}. Kernel driver and IOKit
ABI}{Chapter 27. Kernel driver and IOKit ABI}}\label{chapter-27.-kernel-driver-and-iokit-abi}
\addcontentsline{toc}{subsection}{Chapter {27}. Kernel driver and IOKit
ABI}

The two \texttt{IOExternalMethodDispatch2022} arrays are M1/H13
measured, read byte for byte from the kernel cache's read-only data
section and corroborated on an M2-class cache where all 26 size tuples
are byte-identical, and the control-client open and submit struct sizes
are cross-validated against captured user-space call blobs. The selector
handlers, four-layer call path, doorbell register write,
entitlement-check call sites for
\texttt{com.apple.ane.iokit-user-access} and
\texttt{com.apple.ane.allow-dataChaining-access}, driver class hierarchy
and device properties, and broker model are decompile-derived from the
unstripped kernel-cache symbols, kext property lists, live device
registry, and a system-wide entitlement sweep. The driver's user-client
ABI has no public counterpart and is reported as reverse-engineered; the
IOKit user-client framework and the
\texttt{IOExternalMethodDispatch2022} structure are public, but this
driver's selector numbers, struct sizes, and handler set are not. On the
M5 the client-creation gate \texttt{ANEClientInfo::create}, its
\texttt{copyClientEntitlement} stamp of \texttt{isPrivileged} and
\texttt{allowDataChaining}, and six further driver-enforced
\texttt{com.apple.ane} and \texttt{com.apple.private.ane} entitlements
are decompile-derived from the M5 kernel driver.

\subsection*{\texorpdfstring{Chapter
\hyperref[address-translation-and-the-dart]{28}. Address translation and
the
DART}{Chapter 28. Address translation and the DART}}\label{chapter-28.-address-translation-and-the-dart}
\addcontentsline{toc}{subsection}{Chapter {28}. Address translation and
the DART}

The leaf word \texttt{phys\ \textbar{}\ 0x8000000000000000}, 16 KB
granule, active stream set \texttt{\{0,\ 1,\ 2\}}, translation-table
base \texttt{0x90022320}, and host-to-firmware rebase to the
\texttt{0x1bc4} aperture are M1/H13 measured read-only on the live
dispatch path (Apple M1 Pro, T6000-generation DART): the granule and
stream set from the live device tree, the base register from
function-boundary probes, and the leaf-word template, segment structure,
and protection classes from probes across 26178 leaf-map events. The
leaf-word bit layout, the fault-register offset map, the
panic-terminated fault path (panic confirmed M1/H13 from the disassembly
of every fault-path function), the \texttt{IODARTErrorInfo} descriptor
layout, the \texttt{{[}engine+0xe028{]}} status predicate, and firmware
rebase arithmetic with its three aperture-config offsets are
decompile-derived, the rebase arithmetic unicorn-verified. The
fault-capture register decode is predicted from the published controller
field layout and not measured, because a fault panics the machine. The
address-translation unit, its leaf entry format, the rebase boundary,
and the fault-capture block have no public counterpart and are reported
as reverse-engineered. The per-client isolation contexts, the eight
\texttt{mapper-ane0} translation mappers in the live IORegistry and the
\texttt{ANEIsoID1} through \texttt{ID7} exclave capabilities that bind
them, are M5/H17s measured with System Integrity Protection enabled.

\subsection*{\texorpdfstring{Chapter \hyperref[firmware]{29}.
Firmware}{Chapter 29. Firmware}}\label{chapter-29.-firmware}
\addcontentsline{toc}{subsection}{Chapter {29}. Firmware}

The preload executable is the M1-generation real-time-kernel image,
build identity \texttt{RTKit-3255.120.11.release}, chip tag
\texttt{ASC\_CHINOOK}. The task roster, priority bands, heap and pool
model, execution-loop command set, scheduler deadline, and fault
post-mortem layout; the command-record classes and their
\texttt{sCSneCmdProcedureCall*} invariants; the doorbell-emit sequence
around bit 39 of \texttt{S3\_3\_C15\_C8\_0}, host-notify site
\texttt{@0x4c890}, and engine-to-graphics-processor doorbell at
\texttt{0x2\_0646\_8000}; the bring-up order, seven MMIO banks, RTBuddy
endpoint, and three scratch handshakes; and the per-run statistics
buffer with its header, per-engine descriptors, and host-side null gate
are decompile-derived from the embedded strings, assertion expressions,
and disassembled handlers. The firmware has no public counterpart and is
reported as reverse-engineered from the unencrypted image; Apple
documents only the model framework and conversion tools above this
layer. The CHINOOK control-CPU register map, its eleven thread contexts,
level-two cache, and pipeline error-capture and power-down-save
registers, is decompile-derived from the kernel driver.

\subsection*{\texorpdfstring{Chapter
\hyperref[host-to-firmware-command-protocol]{30}. Host-to-firmware
command
protocol}{Chapter 30. Host-to-firmware command protocol}}\label{chapter-30.-host-to-firmware-command-protocol}
\addcontentsline{toc}{subsection}{Chapter {30}. Host-to-firmware command
protocol}

This chapter is static analysis of the unencrypted M1 firmware image, an
ARM64e real-time-kernel Mach-O, with no firmware executed. The command
vocabulary, numeric identifiers, header layout, and body bounds are read
from the ordered command-name string table, struct-size asserts, and log
format strings; header byte offsets are inferred from field order and
alignment, while field presence, widths, and bounds are read directly
from in-binary asserts. The 94-entry \texttt{CSNE\_CMD} enumeration (93
dispatched command identifiers plus the invalid sentinel), the roughly
ten fast-path ids, and the seventy-six-slot dispatch vtable (arm64e
auth-rebase chained pointers in \texttt{\_\_DATA.\_\_const}, low
thirty-two bits giving the target, the procedure-call slot at
\texttt{+0x200} reaching \texttt{0x7374c} and the inference slot at
\texttt{+0x190} reaching \texttt{0x74510}) are decompile-derived. The
protocol, the command header, and \texttt{CSNE\_CMD\_*} vocabulary have
no public counterpart and are reported as reverse-engineered; the public
model framework reaches only application-level model loading. The full
numeric command table and the decoded request structs are Appendix
\hyperref[appendix-c.-decoded-reference-tables]{C}.

\subsection*{\texorpdfstring{Chapter \hyperref[power-and-thermal]{31}.
Power and
thermal}{Chapter 31. Power and thermal}}\label{chapter-31.-power-and-thermal}
\addcontentsline{toc}{subsection}{Chapter {31}. Power and thermal}

The clean 0 mW idle, the 176-second saturating loop holding flat power
and throughput under nominal thermal pressure, the first-op power-up tax
near 0.5 ms past a 100 ms idle gap, and the single held power state
across 56,527 dispatches are M1/H13 measured by read-only tracing. The
power-block base \texttt{0x2\_6b8f\_0000}, the opaque voltage base
\texttt{0x2\_3b70\_c008}, the five-store power-block arm, the
\texttt{0x11} peak-power control word, the engine and power-manager
device-tree nodes (\texttt{ane0@84000000},
\texttt{compatible\ "ane,t8020"}, the \texttt{0x2\_8400\_0000} aperture,
the \texttt{0x2\_8E08\_0000} power-manager slice), the absence of local
DVFS, and firmware seven-step credit sequence are decompile-derived from
the H13 firmware Mach-O and live device-tree enumeration. Apple does not
document the power model, the fixed operating point, or the thermal
behavior; with nothing public to check against, this account is reported
as reverse-engineered and measured.

\subsection*{\texorpdfstring{Chapter
\hyperref[security-and-isolation]{32}. Security and
isolation}{Chapter 32. Security and isolation}}\label{chapter-32.-security-and-isolation}
\addcontentsline{toc}{subsection}{Chapter {32}. Security and isolation}

The cross-process timing side-channel (a 2.3 times latency jump under
contention and a 20 to 50 bit/s occupancy channel) and the intact data
isolation across 9000 concurrent results are M1/H13 measured. The
kernel-side trust boundary and its three program checks (code signature,
vnode trustcache, client code-signing identity), the firmware's
structural-only check, and the secure and exclave method bodies (the
secure-mode transition state machine, the
\texttt{SwitchExclaveMode\ not\ supported} stub, the inert
\texttt{mov\ w0,\ \#0;\ ret} exclave selectors) are decompile-derived
from the loaded kernel driver and the firmware image. This chapter is
reported as reverse-engineered and measured throughout, because Apple
does not document the secure and exclave transition internals or the
cross-process isolation behavior. On the M5 the exclave is live rather
than stubbed: the secure component \texttt{com.apple.aneexclave}, its
capability-scoped segment access, and the Tightbeam submit path are
decompile-derived from the M5 exclave bundle and boot kernelcache, with
System Integrity Protection enabled.

\subsection*{\texorpdfstring{Chapter
\hyperref[telemetry-and-hardware-counters]{33}. Telemetry and hardware
counters}{Chapter 33. Telemetry and hardware counters}}\label{chapter-33.-telemetry-and-hardware-counters}
\addcontentsline{toc}{subsection}{Chapter {33}. Telemetry and hardware
counters}

On the Apple M1 and M1 Max, the readable whole-engine channels (DRAM
bytes, energy, clock residency), the signpost lifecycle, the all-zero
per-run output buffer, and forced-mask load rejection are M1/H13
measured. The \texttt{ANEProgramCreateArgs} layout, the \texttt{+0x6c}
stats-mask offset, the twenty-four per-descriptor counter namespace, the
stats-buffer ABI \texttt{0x0201}, the \texttt{initStatsBufferSection}
bail branch, and the free-running engine timebase counter at MMIO
\texttt{0x2\_6b17\_8000} (read by the firmware helper \texttt{@0x30988})
are decompile-derived from the runtime dylibs and the kernel driver.
None of the hardware performance-counter block, per-task-descriptor
namespace, stats-mask enable, or firmware timestamp is documented
anywhere public, so the block geometry, master enable,
readable-versus-walled split, and kernel gate are reported as
reverse-engineered and measured. On the M5 the gate actor (the
\texttt{aned} daemon forcing \texttt{statsMask=0} for a
\texttt{ThirdPartyAppUsingANE} client), the per-channel
\texttt{DCS\ BW\ /\ ANE\ L0} and \texttt{L1} State-residency bandwidth
histograms readable on the unentitled path, and the fuller
\texttt{ANE\_THROTTLE\_*} and \texttt{VDD\_DRAM\_VOLTAGE\_CHANGE}
trigger family are M5/H17s measured with System Integrity Protection
enabled.

\section*{Part IX. Cross-Silicon
Reference}\label{part-ix.-cross-silicon-reference}
\addcontentsline{toc}{section}{Part IX. Cross-Silicon Reference}

Part IX is decompilation and static analysis of the one engine compiler
binary across its full target set, with boundaries reproduced on silicon
where a part was in hand: M1/H13 (Apple M1; M1 Max) measured, and
M5/H17s (Apple M5) as the cross-chip reference.

\subsection*{\texorpdfstring{Chapter
\hyperref[cross-silicon-targets]{34}. Cross-silicon
targets}{Chapter 34. Cross-silicon targets}}\label{chapter-34.-cross-silicon-targets}
\addcontentsline{toc}{subsection}{Chapter {34}. Cross-silicon targets}

The 28-target set is measured, extracted by invoking each
per-architecture builder on the M1 (host chip irrelevant) and resolving
the M5 to \texttt{H17s} and the M1 Max to \texttt{H13G} by
fixed-build-directory compile. The target names, the suffix-to-core
decode at HAL offset \texttt{0x238}, the interchange-format tables at
HAL offset \texttt{0x658}, and four-byte format decode are
decompile-derived; the A14, A15, A16, and A18 targets and their M-series
counterparts are decompile-derived from the per-target tables and not
individually measured, and the \(M(n) \rightarrow H(n+12)\) mapping is
confirmed only at the M1 and M5 ends with the middle generations
predicted. Apple's product specifications publish a marketing core count
per chip, for example a 16-core engine; the decoded \texttt{num\_nes} is
a different quantity, counting per-die compute sets, four on the base M1
against the published sixteen. The 28-target compiler set, the
H-architecture naming, the suffix-to-core decode, and interchange tables
are not publicly documented and are reported as reverse-engineered.

\subsection*{\texorpdfstring{Chapter
\hyperref[per-family-code-generation]{35}. Per-family code
generation}{Chapter 35. Per-family code generation}}\label{chapter-35.-per-family-code-generation}
\addcontentsline{toc}{subsection}{Chapter {35}. Per-family code
generation}

The slice-saturation threshold, the top-k, sort, and dynamic-slice
code-generation rejections, and the operation decompositions are M1/H13
measured; the clean slice route and the native crop-resize, resample,
and trig operations are M5/H17s measured. The family enum,
\texttt{MinimumFamily\textless{}N\textgreater{}} trait and its four
operation tiers, per-chip hardware-abstraction offsets, task-descriptor
patch-width path, and
\texttt{ConvertSlice\textless{}Family\textgreater{}} lowering are
decompile-derived from the engine compiler framework (the 9.509 build,
87,874 functions); the A14 and A15 unlock points, M-series families
above the M1, and dedicated A15 code-generation branch with its cost
table, 45-operation floor, full H15 targets, and YUV420 input are
decompile-derived and not measured on A15 silicon. The family enum,
minimum-family trait, per-chip parameters, and per-family route
selection have no public counterpart; the public conversion tools
document the frontend operation set, optimization passes, and
compute-unit selector, and stop above the backend per-family lowering
reported here.

\subsection*{\texorpdfstring{Chapter
\hyperref[predicted-upper-tier]{36}. Predicted upper
tier}{Chapter 36. Predicted upper tier}}\label{chapter-36.-predicted-upper-tier}
\addcontentsline{toc}{subsection}{Chapter {36}. Predicted upper tier}

The fp8 format converters, E4M3 overflow enumeration, format-register
encoders, double-multiply gate, collective dialect operation set,
device-mesh and sharding lowering, reduction-to-atomic map, and
collective direct-memory-access emitter are decompile-derived, with the
family gates read from the 28-target capability bytes; the 64-core
ceiling, fp8 \texttt{e4m3} and \texttt{e5m2} datapath, and Ultra
device-mesh collective are decompile-derived and not measured on the
upper-tier parts. The E4M3 native multiply, accumulation, and saturation
(inferred from the encoders and the H18-only capability byte at offset
\texttt{0x52d}), the fp16 accumulation of an fp8 multiply (from the
absence of any fp8 accumulator field), and the running collective (the
enable byte at offset \texttt{0x48b} zero on all 28 targets and the
register encoding stubbed on every family) are predicted, with no
current family materializing the collective. That the load-balancer
supports up to four engine dies while the M1 and M1 Max each register a
single engine, so cross-die steering engages only on a multi-die part
such as the Ultra, is M1 Max measured by the device registry. The fp8
datapath and the multi-die collective layer have no public counterpart
in Apple's documentation and are reported as reverse-engineered and
explicitly unmeasured.

\section*{Back matter}\label{back-matter}
\addcontentsline{toc}{section}{Back matter}

The back-matter chapters rest on M1/H13 as the primary host and M5/H17s
for cross-generation scaling.

\subsection*{Methodology}\label{methodology-1}
\addcontentsline{toc}{subsection}{Methodology}

The only public account is the compute-unit selector at
developer.apple.com/documentation/coreml/mlcomputeunits, with no direct
device API; this chapter goes below the selector, reaching the engine
directly and characterizing it by static analysis and live
instrumentation. The direct dispatch route, the four static-analysis
artifacts, and the program-binary capture are decompile-derived and
confirmed by the kernel trace; the roofline figures and the
compile-service rate condition are M1/H13 measured, and the
cross-generation scaling and bounded numeric drift are M5/H17s measured.

\subsection*{Open questions}\label{open-questions-1}
\addcontentsline{toc}{subsection}{Open questions}

The decoded baseline and the boundary limits are M1/H13,
decompile-derived from the compiler, hardware-abstraction tables, kernel
driver, and firmware and joined to live instrumentation; M5/H17s
confirmed the cross-family predictions. The M3/H15 and upper-tier
runtime behavior is predicted, decompile-derived from the gates and not
confirmed on silicon.

\section*{Appendices}\label{appendices}
\addcontentsline{toc}{section}{Appendices}

The reference tables are decompile-derived from static, read-only
analysis of the M1/H13 binaries: the ANE compiler
(\texttt{ANECompiler\ 9.509}), its per-family operation-floor tables,
its per-layer validators and parsers, the host runtime, and the
unencrypted firmware image (\texttt{h13\_ane\_fw\_styx\_j5x.im4p}), with
no firmware executed and no engine jobs run. Where a value or status is
measured rather than decompile-derived, it was confirmed on physical
silicon, primarily M5/H17s and M1/H13, by compiling and dispatching
against a host reference.

\subsection*{\texorpdfstring{Appendix
\hyperref[appendix-a.-operation-by-device-matrix]{A}. The
operation-by-device
matrix}{Appendix A. The operation-by-device matrix}}\label{appendix-a.-the-operation-by-device-matrix}
\addcontentsline{toc}{subsection}{Appendix {A}. The operation-by-device
matrix}

The M1, M2, and M5 columns are measured by operation-conformance runs,
each native operation compiled and run and each no-path one rejected on
device; the M3 column and the M4 part of the merged M4-and-M5 column are
decompile-derived predictions from the per-chip tables. The per-family
unlock points for the texture-engine operations, sin and cos, rank and
sort bridge, argument reductions, and weight-stream gates are
decompile-derived from the operation floors, validators, and
symbol-resolution map and confirmed on the M1 only at its boundary; the
family that first runs each is predicted from the floor table. Apple
documents the convertible operation set at apple.github.io/coremltools
and developer.apple.com; this table extends and partly corrects that
account, reporting what compiles and runs on the direct engine path,
since some accepted operations, such as three-dimensional convolution,
do not lower to the engine.

\subsection*{\texorpdfstring{Appendix
\hyperref[appendix-b.-hidden-layer-catalog]{B}. The hidden-layer
catalog}{Appendix B. The hidden-layer catalog}}\label{appendix-b.-the-hidden-layer-catalog}
\addcontentsline{toc}{subsection}{Appendix {B}. The hidden-layer
catalog}

Each layer's \texttt{Type} tag, descriptor symbol, and \texttt{Params}
key set are decompile-derived from the compiler export table, parser
disassembly, constant-string key atlas, per-layer
\texttt{ZinParse\textless{}Name\textgreater{}Unit} parsers and
\texttt{\_ANECValidate\textless{}Name\textgreater{}Layer} checkers, and
descriptor-initializer routines
\texttt{\_ANEC\textless{}Name\textgreater{}LayerDescInitialize}, joined
to the netplist schema read out of the runtime framework. The
fused-attention operand contract, the spatial-rearrange channel
ordering, the float16-bit-pattern convention for \texttt{Alpha},
\texttt{Epsilon}, and \texttt{Scale}, and point-cloud output contracts
are M5/H17s measured by authoring and dispatching the layers, and the M1
arch gates, the \texttt{Sort} and \texttt{DynamicSlice} rejections, and
the top-k \(\{3, 4\}\) forbidden band are M1/H13 measured. Apple
documents the model converter and its intermediate-language operation
set at apple.github.io/coremltools, which does not emit these native
layer kinds; this catalog authors the native descriptors directly
through the network description.

\subsection*{\texorpdfstring{Appendix
\hyperref[appendix-c.-decoded-reference-tables]{C}. Decoded reference
tables}{Appendix C. Decoded reference tables}}\label{appendix-c.-decoded-reference-tables-1}
\addcontentsline{toc}{subsection}{Appendix {C}. Decoded reference
tables}

Every value is read out of an M1/H13 binary by static analysis, with no
firmware executed: the attribute and opcode integers and IOKit struct
layouts from the compiler decompile (\texttt{ANECompiler\ 9.509}) and
the host runtime; the error constants, command table, and tunable table
from the unencrypted firmware image
(\texttt{h13\_ane\_fw\_styx\_j5x.im4p}) and the standard IOKit return
macros; the register map from the compiler's task-descriptor setters and
getters; and the \texttt{.e5} schema from the runtime serializer
symbols, validated byte-for-byte against a captured sample. The runtime,
firmware, compiler, and ABI surface consolidated here is private and
undocumented and is reported as reverse-engineered; the public model
framework, conversion tools, and intermediate-language reference
describe none of these numeric tables.

\chapter*{References}\label{references}
\addcontentsline{toc}{chapter}{References}

This guide is archived on arXiv as
\href{https://arxiv.org/abs/2606.22283}{arXiv:2606.22283}. To cite it:

\begin{Shaded}
\begin{Highlighting}[]
\VariableTok{@misc}\NormalTok{\{}\OtherTok{bryngelson2026ane}\NormalTok{,}
  \DataTypeTok{author}\NormalTok{        = \{Bryngelson, Spencer H.\},}
  \DataTypeTok{title}\NormalTok{         = \{Apple Neural Engine: Architecture, Programming, and Performance\},}
  \DataTypeTok{year}\NormalTok{          = \{2026\},}
  \DataTypeTok{archivePrefix}\NormalTok{ = \{arXiv\},}
  \DataTypeTok{eprint}\NormalTok{        = \{2606.22283\},}
  \DataTypeTok{doi}\NormalTok{           = \{10.48550/arXiv.2606.22283\},}
  \DataTypeTok{url}\NormalTok{           = \{https://arxiv.org/abs/2606.22283\},}
\NormalTok{\}}
\end{Highlighting}
\end{Shaded}

\begin{center}\rule{0.5\linewidth}{0.5pt}\end{center}

\begin{enumerate}
\def\labelenumi{\arabic{enumi}.}
\tightlist
\item
  \protect\phantomsection\label{ref-libane}{AmiraniLabs.} ``libane: a
  native Apple Neural Engine runtime.'' Repository,
  \url{https://github.com/AmiraniLabs/libane}.
\item
  \protect\phantomsection\label{ref-appleaccelerate}{Apple.} Accelerate
  and BNNS documentation.
  \url{https://developer.apple.com/documentation/accelerate}.
\item
  \protect\phantomsection\label{ref-appleactivedevices2026}{Apple.}
  Active installed base of 2.5 billion devices, reported by T. Cook on
  the first-quarter fiscal 2026 earnings call, January 29, 2026.
  apple.com.
\item
  \protect\phantomsection\label{ref-appleane}{Apple.} Apple silicon
  technical specifications. \url{https://www.apple.com/mac/compare/}.
\item
  \protect\phantomsection\label{ref-applecoreml}{Apple.} Core ML
  framework documentation.
  \url{https://developer.apple.com/documentation/coreml}.
\item
  \protect\phantomsection\label{ref-applecoremltools}{Apple.} Core ML
  Tools (coremltools) documentation.
  \url{https://apple.github.io/coremltools}.
\item
  \protect\phantomsection\label{ref-applevision}{Apple.} Vision
  framework documentation.
  \url{https://developer.apple.com/documentation/vision}.
\item
  \protect\phantomsection\label{ref-appleanetransformers}{Apple Machine
  Learning Research.} ``Deploying Transformers on the Apple Neural
  Engine.'' Apple Machine Learning Research article, 2022.
\item
  \protect\phantomsection\label{ref-benazir2026}{Benazir, A., and Lin,
  F. X.} ``Efficient Mixture-of-Experts LLM Inference with Apple Silicon
  NPUs.'' Preprint,
  \href{https://arxiv.org/abs/2604.18788}{arXiv:2604.18788}, 2026.
\item
  \protect\phantomsection\label{ref-bi2026}{Bi, Z., Chen, X., Sun, L.,
  Yao, Y., Shen, Q., Lou, J., and Deng, C.} ``RooflineBench: A
  Benchmarking Framework for On-Device LLMs via Roofline Analysis.''
  Preprint, \href{https://arxiv.org/abs/2602.11506}{arXiv:2602.11506},
  2026.
\item
  \protect\phantomsection\label{ref-aneforge2026}{Bryngelson, S. H.}
  ``ANEForge: Python for direct computation on the Apple Neural
  Engine.'' Preprint,
  \href{https://arxiv.org/abs/2606.17090}{arXiv:2606.17090}, 2026.
\item
  \protect\phantomsection\label{ref-chen2025}{Chen, L., Feng, D., Feng,
  E., Wang, Y., Zhao, R., Xia, Y., Xu, P., and Chen, H.}
  ``Characterizing Mobile SoC for Accelerating Heterogeneous LLM
  Inference.'' ACM SIGOPS Symposium on Operating Systems Principles
  (SOSP), 2025.
  \href{https://arxiv.org/abs/2501.14794}{arXiv:2501.14794}, DOI
  10.1145/3731569.3764808.
\item
  \protect\phantomsection\label{ref-choi2013}{Choi, J. W., Bedard, D.,
  Fowler, R., and Vuduc, R.} ``A Roofline Model of Energy.'' IEEE
  International Symposium on Parallel and Distributed Processing
  (IPDPS), 661-672, 2013. DOI 10.1109/IPDPS.2013.77.
\item
  \protect\phantomsection\label{ref-communityane}{Community Apple Neural
  Engine reverse-engineering repositories.} johnmai-dev/ANE-LM,
  mechramc/Orion, skyfallsin/apple-neural-engine-field-guide, and
  dmaynor/apple-vuln-research. Repositories.
\item
  \protect\phantomsection\label{ref-ding2019}{Ding, N., and Williams,
  S.} ``An Instruction Roofline Model for GPUs.'' IEEE/ACM Performance
  Modeling, Benchmarking and Simulation of High Performance Computer
  Systems (PMBS), 7-18, 2019. DOI 10.1109/PMBS49563.2019.00007.
\item
  \protect\phantomsection\label{ref-fanariotis2025}{Fanariotis, A.,
  Orphanoudakis, T., and Fotopoulos, V.} ``Evaluating the Energy
  Efficiency of NPU-Accelerated Machine Learning Inference on Embedded
  Microcontrollers.'' Preprint,
  \href{https://arxiv.org/abs/2509.17533}{arXiv:2509.17533}, 2025.
\item
  \protect\phantomsection\label{ref-whispercpp}{Gerganov, G.}
  ``whisper.cpp: Whisper inference in C/C++ with Core ML Neural Engine
  support.'' Repository, \url{https://github.com/ggml-org/whisper.cpp}.
\item
  \protect\phantomsection\label{ref-handley}{Handley, M.} ``Apple Neural
  Engine.'' AArch64-Explore, vol.~7. Patent-based architectural analysis
  of the engine, \url{https://github.com/name99-org/AArch64-Explore}.
\item
  \protect\phantomsection\label{ref-hollemans}{Hollemans, M.} ``The
  Neural Engine: What Do We Know About It?'' Community-maintained
  repository, \url{https://github.com/hollance/neural-engine}.
\item
  \protect\phantomsection\label{ref-tinygrad}{Hotz, G., and the tinygrad
  authors.} ``tinygrad.'' Repository,
  \url{https://github.com/tinygrad/tinygrad}.
\item
  \protect\phantomsection\label{ref-hubner2025}{Hübner, P., Hu, A.,
  Peng, I., and Markidis, S.} ``Apple vs.~Oranges: Evaluating the Apple
  Silicon M-Series SoCs for HPC Performance and Efficiency.'' Preprint,
  \href{https://arxiv.org/abs/2502.05317}{arXiv:2502.05317}, 2025.
\item
  \protect\phantomsection\label{ref-ignatov2019}{Ignatov, A., Timofte,
  R., Kulik, A., Yang, S., Wang, K., Baum, F., Wu, M., Xu, L., and Van
  Gool, L.} ``AI Benchmark: All About Deep Learning on Smartphones in
  2019.'' Preprint,
  \href{https://arxiv.org/abs/1910.06663}{arXiv:1910.06663}, 2019.
\item
  \protect\phantomsection\label{ref-ilic2014}{Ilic, A., Pratas, F., and
  Sousa, L.} ``Cache-Aware Roofline Model: Upgrading the Loft.'' IEEE
  Computer Architecture Letters, 13(1), 21-24, 2014. DOI
  10.1109/L-CA.2013.6.
\item
  \protect\phantomsection\label{ref-jayanth2024}{Jayanth, R., Gupta, N.,
  and Prasanna, V.} ``Benchmarking Edge AI Platforms for
  High-Performance ML Inference.'' Preprint,
  \href{https://arxiv.org/abs/2409.14803}{arXiv:2409.14803}, 2024.
\item
  \protect\phantomsection\label{ref-jouppi2017}{Jouppi, N. P., Young,
  C., Patil, N., Patterson, D. A., et al.} ``In-Datacenter Performance
  Analysis of a Tensor Processing Unit.'' International Symposium on
  Computer Architecture (ISCA), 1-12, 2017. Also
  \href{https://arxiv.org/abs/1704.04760}{arXiv:1704.04760}.
\item
  \protect\phantomsection\label{ref-orion2026}{Kumaresan, R.} ``Orion:
  Characterizing and Programming Apple's Neural Engine for LLM Training
  and Inference.'' Preprint,
  \href{https://arxiv.org/abs/2603.06728}{arXiv:2603.06728}, 2026.
\item
  \protect\phantomsection\label{ref-zeus2025}{ML.ENERGY / Zeus.}
  ``Programmatic Energy Consumption Measurement on Apple Silicon
  (macOS).'' Project issue report (\#159), 2025.
\item
  \protect\phantomsection\label{ref-moon2025}{Moon, S., Cha, J., Park,
  H., and Kim, J.} ``Hybe: GPU-NPU Hybrid System for Efficient LLM
  Inference with Million-Token Context Window.'' International Symposium
  on Computer Architecture (ISCA), 808-820, 2025. DOI
  10.1145/3695053.3731051.
\item
  \protect\phantomsection\label{ref-plyenkov2019}{Plyenkov, B.}
  ``Decoupling Machine Intelligence from Application in IoT Devices.''
  Master's thesis, Aalto University, 2019.
\item
  \protect\phantomsection\label{ref-prashanthi2025}{Prashanthi, S. K.,
  Sahoo, K. K., Saikia, A. R., Gupta, P., Joshi, A. V., Pansari, P., and
  Simmhan, Y.} ``Pagoda: An Energy and Time Roofline Study for DNN
  Workloads on Edge Accelerators.'' Preprint,
  \href{https://arxiv.org/abs/2509.20189}{arXiv:2509.20189}, 2025.
\item
  \protect\phantomsection\label{ref-singh2026}{Singh, M.} ``Inside the
  M4 Apple Neural Engine, Part 1: Reverse Engineering.'' Blog post and
  repository, 2026, \url{https://github.com/maderix/ANE}.
\item
  \protect\phantomsection\label{ref-tummalapalli2026}{Tummalapalli, P.,
  Arayakandy, S., Pal, R., and Kundan, K.} ``LLM Inference at the Edge:
  Mobile, NPU, and GPU Performance Efficiency Trade-offs Under Sustained
  Load.'' Preprint,
  \href{https://arxiv.org/abs/2603.23640}{arXiv:2603.23640}, 2026.
\item
  \protect\phantomsection\label{ref-verhelst2025}{Verhelst, M., Benini,
  L., and Verma, N.} ``How to Keep Pushing ML Accelerator Performance?
  Know Your Rooflines!'' IEEE Journal of Solid-State Circuits, 2025. DOI
  10.1109/JSSC.2025.3553765.
\item
  \protect\phantomsection\label{ref-williams2009}{Williams, S.,
  Waterman, A., and Patterson, D. A.} ``Roofline: An Insightful Visual
  Performance Model for Multicore Architectures.'' Communications of the
  ACM, 52(4), 65-76, 2009. DOI 10.1145/1498765.1498785.
\item
  \protect\phantomsection\label{ref-xu2025}{Xu, D., Zhang, H., Yang, L.,
  Liu, R., Huang, G., Xu, M., and Liu, X.} ``Fast On-device LLM
  Inference with NPUs.'' ACM International Conference on Architectural
  Support for Programming Languages and Operating Systems (ASPLOS),
  2025. \href{https://arxiv.org/abs/2407.05858}{arXiv:2407.05858}, DOI
  10.1145/3669940.3707239.
\item
  \protect\phantomsection\label{ref-yang2020}{Yang, C., Kurth, T., and
  Williams, S.} ``Hierarchical Roofline Analysis for GPUs: Accelerating
  Performance Optimization for the NERSC-9 Perlmutter System.''
  Concurrency and Computation: Practice and Experience, 32(20), e5547,
  2020. DOI 10.1002/cpe.5547.
\item
  \protect\phantomsection\label{ref-eilnane}{Yoon, E.} ``ane: a
  reverse-engineered Linux driver for the Apple Neural Engine, with
  anecc.'' Repository, 2022, \url{https://github.com/eiln/ane}.
\end{enumerate}

\end{document}